\documentclass[10pt]{article}    
\usepackage{url}  		
\usepackage{ifthen}  	
\usepackage{multicol}   
\usepackage[utf8]{inputenc} 
\usepackage{amsmath,amssymb}
\usepackage{tikz,times}
\usepackage{epstopdf}
\usepackage{caption}
\usepackage{appendix}
\usepackage{varwidth}
\usepackage{subcaption}
\usepackage{multirow}
\usepackage{amsthm}
\usepackage{float}
\usepackage{enumerate}
\usepackage{algorithm}
\usepackage{algpseudocode}
\usepackage{graphicx,color}
\usepackage[a4paper,margin=1cm]{geometry}
\usepackage{hyperref}
\hypersetup{colorlinks=true,%
    		citecolor=oucrimsonred,%
    		filecolor=oucrimsonred,%
    		linkcolor=oucrimsonred,%
    		urlcolor=oucrimsonred 
}
\newtheorem{theorem}{Theorem}[section]
\newtheorem{corollary}{Corollary}[theorem]

\usetikzlibrary{positioning,shapes,shadows,arrows}
\usetikzlibrary{mindmap,backgrounds}
\newboolean{publ}
\newenvironment{bmcformat}{\baselineskip20pt\sloppy\setboolean{publ}{false}}{\baselineskip20pt\sloppy}

\begin{document}
\begin{bmcformat}
\definecolor{darkorchid}{rgb}{0.6, 0.2, 0.8}
\definecolor{brightgreen}{rgb}{0.4, 1.0, 0.0}
\definecolor{americanrose}{rgb}{1.0, 0.01, 0.24}
\definecolor{darkorange}{rgb}{1.0, 0.55, 0.0}
\definecolor{jazzberryjam}{rgb}{0.65, 0.04, 0.37}
\definecolor{harvardcrimson}{rgb}{0.79, 0.0, 0.09}
\definecolor{mahogany}{rgb}{0.75, 0.25, 0.0}
\definecolor{chromeyellow}{rgb}{1.0, 0.65, 0.0}
\definecolor{cinnabar}{rgb}{0.89, 0.26, 0.2}
\definecolor{cobalt}{rgb}{0.0, 0.28, 0.67}
\definecolor{darkmagenta}{rgb}{0.55, 0.0, 0.55}
\definecolor{patriarch}{rgb}{0.5, 0.0, 0.5}
\definecolor{oucrimsonred}{rgb}{0.6, 0.0, 0.0}
\definecolor{tyrianpurple}{rgb}{0.4, 0.01, 0.24}
\definecolor{darkgreen}{rgb}{0.0, 0.2, 0.13}
\definecolor{darkpastelgreen}{rgb}{0.01, 0.75, 0.24}
\definecolor{forestgreenweb}{rgb}{0.13, 0.55, 0.13}
\definecolor{americanrose}{rgb}{1.0, 0.01, 0.24}
\definecolor{alizarin}{rgb}{0.82, 0.1, 0.26}
\definecolor{awesome}{rgb}{1.0, 0.13, 0.32} 	
\definecolor{bostonuniversityred}{rgb}{0.8, 0.0, 0.0} 	
\definecolor{brightpink}{rgb}{1.0, 0.0, 0.5} 	
\definecolor{byzantine}{rgb}{0.74, 0.2, 0.64} 	
\definecolor{candyapplered}{rgb}{1.0, 0.03, 0.0} 	
\definecolor{carminered}{rgb}{1.0, 0.0, 0.22} 	
\definecolor{deepmagenta}{rgb}{0.8, 0.0, 0.8}
\definecolor{deeppink}{rgb}{1.0, 0.08, 0.58}
\definecolor{hollywoodcerise}{rgb}{0.96, 0.0, 0.63}
\definecolor{brightpink}{rgb}{1.0, 0.0, 0.5}
\definecolor{redviolet}{rgb}{0.78, 0.08, 0.52}
\definecolor{carmine}{rgb}{0.59, 0.0, 0.09}
\definecolor{sangria}{rgb}{0.57, 0.0, 0.04}
\definecolor{tyrianpurple}{rgb}{0.4, 0.01, 0.24}
\definecolor{wildwatermelon}{rgb}{0.99, 0.42, 0.52}
\definecolor{wildstrawberry}{rgb}{1.0, 0.26, 0.64}
\definecolor{vividcerise}{rgb}{0.85, 0.11, 0.51}
\definecolor{venetianred}{rgb}{0.78, 0.03, 0.08}
\definecolor{utahcrimson}{rgb}{0.83, 0.0, 0.25}
\definecolor{tractorred}{rgb}{0.99, 0.05, 0.21}
\definecolor{persianrose}{rgb}{1.0, 0.16, 0.64}
\definecolor{ferrarired}{rgb}{1.0, 0.11, 0.0}
\definecolor{folly}{rgb}{1.0, 0.0, 0.31}
\definecolor{ao}{rgb}{0.0, 0.0, 1.0}
\definecolor{persianblue}{rgb}{0.11, 0.22, 0.73}
\definecolor{brandeisblue}{rgb}{0.0, 0.44, 1.0}
\definecolor{hanpurple}{rgb}{0.32, 0.09, 0.98}
\definecolor{indigoweb}{rgb}{0.29, 0.0, 0.51} 	
\definecolor{internationalkleinblue}{rgb}{0.0, 0.18, 0.65} 	
\definecolor{mediumblue}{rgb}{0.0, 0.0, 0.8} 	
\definecolor{ultramarine}{rgb}{0.07, 0.04, 0.56} 	
\definecolor{blue-violet}{rgb}{0.54, 0.17, 0.89}
\definecolor{coolblack}{rgb}{0.0, 0.18, 0.39}
\definecolor{cyanprocess}{rgb}{0.0, 0.72, 0.92}
\definecolor{richelectricblue}{rgb}{0.03, 0.57, 0.82}
\definecolor{royalazure}{rgb}{0.0, 0.22, 0.66}
\definecolor{tiffanyblue}{rgb}{0.04, 0.73, 0.71}
\definecolor{trueblue}{rgb}{0.0, 0.45, 0.81}
\definecolor{britishracinggreen}{rgb}{0.0, 0.26, 0.15}
\definecolor{greenhtmlcssgreen}{rgb}{0.0, 0.5, 0.0}
\definecolor{greenpigment}{rgb}{0.0, 0.65, 0.31}
\definecolor{olive}{rgb}{0.5, 0.5, 0.0}
\definecolor{tangelo}{rgb}{0.98, 0.3, 0.0}
\definecolor{orange-red}{rgb}{1.0, 0.27, 0.0}
\def\bdoc{\begin{document}}
\def\edoc{\end{document}}
\def\babs{\begin{abstract}}
\def\eabs{\end{abstract}}
\def\bcc{\begin{center}}
\def\ecc{\end{center}}
\def\ben{\begin{enumerate}}
\def\een{\end{enumerate}}
\def\bit{\begin{itemize}}
\def\eit{\end{itemize}}
\def\bpic{\begin{picture}}
\def\epic{\end{picture}}
\def\beq{\begin{equation}}
\def\eeq{\end{equation}}
\def\barr{\begin{array}}
\def\earr{\end{array}}
\def\btab{\begin{tabular}}
\def\etab{\end{tabular}}
\def\btabu{\begin{tabular}}
\def\etabu{\end{tabular}}
\def\bcc{\begin{center}}
\def\ecc{\end{center}}
\def\beqn{\begin{eqnarray}}
\def\eeqn{\end{eqnarray}}
\def\beqnx{\begin{eqnarray*}}
\def\eeqnx{\end{eqnarray*}}
\def\bfig{\begin{figure}}
\def\efig{\end{figure}}
\def\bver{\begin{verb}}
\def\ever{\end{verb}}

\def\blue#1{{\color{blue}#1}}
\def\red#1{{\color{red}#1}}
\def\green#1{{\color{green}#1}}

\def\bm#1{\mbox{\boldmath $#1$}}
\def\ab{\mbox{\boldmath $a$}}
\def\bb{\mbox{\boldmath $b$}}
\def\cb{\mbox{\boldmath $c$}}
\def\db{\mbox{\boldmath $d$}}
\def\eb{\mbox{\boldmath $e$}}
\def\fb{\mbox{\boldmath $f$}}
\def\gb{\mbox{\boldmath $g$}}
\def\hb{\mbox{\boldmath $h$}}
\def\ib{\mbox{\boldmath $i$}}
\def\jb{\mbox{\boldmath $j$}}
\def\kb{\mbox{\boldmath $k$}}
\def\lb{\mbox{\boldmath $l$}}
\def\mb{\mbox{\boldmath $m$}}
\def\nb{\mbox{\boldmath $n$}}
\def\ob{\mbox{\boldmath $o$}}
\def\pb{\mbox{\boldmath $p$}}
\def\qb{\mbox{\boldmath $q$}}
\def\rb{\mbox{\boldmath $r$}}
\def\sb{\mbox{\boldmath $s$}}
\def\tb{\mbox{\boldmath $t$}}
\def\ub{\mbox{\boldmath $u$}}
\def\vb{\mbox{\boldmath $v$}}
\def\wb{\mbox{\boldmath $w$}}
\def\xb{\mbox{\boldmath $x$}}
\def\yb{\mbox{\boldmath $y$}}
\def\zb{\mbox{\boldmath $z$}}
\def\Ab{\mbox{\boldmath $A$}}
\def\Bb{\mbox{\boldmath $B$}}
\def\Cb{\mbox{\boldmath $C$}}
\def\Db{\mbox{\boldmath $D$}}
\def\Eb{\mbox{\boldmath $E$}}
\def\Fb{\mbox{\boldmath $F$}}
\def\Gb{\mbox{\boldmath $G$}}
\def\Hb{\mbox{\boldmath $H$}}
\def\Ib{\mbox{\boldmath $I$}}
\def\Jb{\mbox{\boldmath $J$}}
\def\Kb{\mbox{\boldmath $K$}}
\def\Lb{\mbox{\boldmath $L$}}
\def\Mb{\mbox{\boldmath $M$}}
\def\Nb{\mbox{\boldmath $N$}}
\def\Ob{\mbox{\boldmath $O$}}
\def\Pb{\mbox{\boldmath $P$}}
\def\Qb{\mbox{\boldmath $Q$}}
\def\Rb{\mbox{\boldmath $R$}}
\def\Sb{\mbox{\boldmath $S$}}
\def\Tb{\mbox{\boldmath $T$}}
\def\Ub{\mbox{\boldmath $U$}}
\def\Vb{\mbox{\boldmath $V$}}
\def\Wb{\mbox{\boldmath $W$}}
\def\Xb{\mbox{\boldmath $X$}}
\def\Yb{\mbox{\boldmath $Y$}}
\def\Zb{\mbox{\boldmath $Z$}}
\def\aht{\widehat{a}}		\def\ati{\widetilde{a}}
\def\bht{\widehat{b}}		\def\bti{\widetilde{b}}		
\def\cht{\widehat{c}}		\def\cti{\widetilde{c}}		
\def\dht{\widehat{d}}		\def\dti{\widetilde{d}}		
\def\eht{\widehat{e}}		\def\eti{\widetilde{e}}		
\def\fht{\widehat{f}}		\def\fti{\widetilde{f}}		
\def\ght{\widehat{g}}		\def\gti{\widetilde{g}}		
\def\hht{\widehat{h}}		\def\hti{\widetilde{h}}		
\def\iht{\widehat{i}}		\def\iti{\widetilde{i}}		
\def\jht{\widehat{j}}		\def\jti{\widetilde{j}}		
\def\kht{\widehat{k}}		\def\kti{\widetilde{k}}		
\def\lht{\widehat{l}}		\def\lti{\widetilde{l}}		
\def\mht{\widehat{m}}		\def\mti{\widetilde{m}}		
\def\nht{\widehat{n}}		\def\nti{\widetilde{n}}		
\def\oht{\widehat{o}}		\def\oti{\widetilde{o}}		
\def\pht{\widehat{p}}		\def\pti{\widetilde{p}}		
\def\qht{\widehat{q}}		\def\qti{\widetilde{q}}		
\def\rht{\widehat{r}}		\def\rti{\widetilde{r}}		
\def\sht{\widehat{s}}		\def\sti{\widetilde{s}}		
\def\tht{\widehat{t}}		\def\tti{\widetilde{t}}		
\def\uht{\widehat{u}}		\def\uti{\widetilde{u}}		
\def\vht{\widehat{v}}		\def\vti{\widetilde{v}}		
\def\wht{\widehat{w}}		\def\wti{\widetilde{w}}		
\def\xht{\widehat{x}}		\def\xti{\widetilde{x}}		
\def\yht{\widehat{y}}		\def\yti{\widetilde{y}}		
\def\zht{\widehat{z}}		\def\zti{\widetilde{z}}		
\def\alphah{\widehat{\alpha}}		\def\alphat{\widetilde{\alpha}}
\def\betah{\widehat{\beta}}			\def\betat{\widetilde{\beta}}
\def\deltah{\widehat{\delta}}		\def\deltat{\widetilde{\delta}}
\def\epsilonh{\widehat{\epsilon}} 	\def\epsilont{\widetilde{\epsilon}}
\def\gammah{\widehat{\gamma}}		\def\gammat{\widetilde{\gamma}}
\def\omegah{\widehat{\omega}}		\def\omegat{\widetilde{\omega}}	
\def\thetah{\widehat{\theta}}		\def\thetat{\widetilde{\theta}}	
\def\xih{\widehat{\xi}}				\def\xit{\widetilde{\xi}}	
\def\lambdah{\widehat{\lambda}}		\def\lambdat{\widetilde{\lambda}}	
\def\tauh{\widehat{\tau}}			\def\taut{\widetilde{\tau}}	
\def\phih{\widehat{\phi}}			\def\phit{\widetilde{\phi}}	
\def\muh{\widehat{\mu}}				\def\mut{\widetilde{\mu}}	
\def\psih{\widehat{\psi}}			\def\psit{\widetilde{\psi}}	
\def\chih{\widehat{\chi}}			\def\chit{\widetilde{\chi}}	
\def\sigmah{\widehat{\sigma}}		\def\sigmat{\widetilde{\sigma}}	
\def\zetah{\widehat{\zeta}}			\def\zetat{\widetilde{\zeta}}	
\def\etah{\widehat{\eta}}			\def\etat{\widetilde{\eta}}	
\def\iotah{\widehat{\iota}}			\def\iotat{\widetilde{\iota}}	
\def\kappah{\widehat{\kappa}}		\def\kappat{\widetilde{\kappa}}	
\def\nuh{\widehat{\nu}}				\def\nut{\widetilde{\nu}}	
\def\rhoh{\widehat{\rho}}			\def\rhot{\widetilde{\rho}}	
\def\upsilonh{\widehat{\upsilon}}	\def\upsilont{\widetilde{\upsilon}}	
\def\alphab{\mbox{\boldmath $\alpha$}}
\def\betab{\mbox{\boldmath $\beta$}}
\def\deltab{\mbox{\boldmath $\delta$}}
\def\epsilonb{\mbox{\boldmath $\epsilon$}}
\def\gammab{\mbox{\boldmath $\gamma$}}
\def\omegab{\mbox{\boldmath $\omega$}}
\def\thetab{\mbox{\boldmath $\theta$}}
\def\xib{\mbox{\boldmath $\xi$}}
\def\lambdab{\mbox{\boldmath $\lambda$}}
\def\taub{\mbox{\boldmath $\tau$}}
\def\phib{\mbox{\boldmath $\phi$}}
\def\mub{\mbox{\boldmath $\mu$}}
\def\psib{\mbox{\boldmath $\psi$}}
\def\chib{\mbox{\boldmath $\chi$}}
\def\sigmab{\mbox{\boldmath $\sigma$}}
\def\zetab{\mbox{\boldmath $\zeta$}}
\def\etab{\mbox{\boldmath $\eta$}}
\def\iotab{\mbox{\boldmath $\iota$}}
\def\kappab{\mbox{\boldmath $\kappa$}}
\def\nub{\mbox{\boldmath $\nu$}}
\def\rhob{\mbox{\boldmath $\rho$}}
\def\upsilonb{\mbox{\boldmath $\upsilon$}}
\def\alphabh{\widehat{\alphab}}		\def\alphabt{\widetilde{\alphab}}
\def\betabh{\widehat{\betab}}		\def\betabt{\widetilde{\betab}}
\def\deltabh{\widehat{\deltab}}		\def\deltabt{\widetilde{\deltab}}
\def\epsilonbh{\widehat{\epsilonb}} \def\epsilonbt{\widetilde{\epsilonb}}
\def\gammabh{\widehat{\gammab}}		\def\gammabt{\widetilde{\gammab}}
\def\omegabh{\widehat{\omegab}}		\def\omegabt{\widetilde{\omegab}}	
\def\thetabh{\widehat{\thetab}}		\def\thetabt{\widetilde{\thetab}}	
\def\xibh{\widehat{\xib}}			\def\xibt{\widetilde{\xib}}	
\def\lambdabh{\widehat{\lambdab}}	\def\lambdabt{\widetilde{\lambdab}}	
\def\taubh{\widehat{\taub}}			\def\taubt{\widetilde{\taub}}	
\def\phibh{\widehat{\phib}}			\def\phibt{\widetilde{\phib}}	
\def\mubh{\widehat{\mub}}			\def\mubt{\widetilde{\mub}}	
\def\psibh{\widehat{\psib}}			\def\psibt{\widetilde{\psib}}	
\def\chibh{\widehat{\chib}}			\def\chibt{\widetilde{\chib}}	
\def\sigmabh{\widehat{\sigmab}}		\def\sigmabt{\widetilde{\sigmab}}	
\def\zetabh{\widehat{\zetab}}		\def\zetabt{\widetilde{\zetab}}	
\def\etabh{\widehat{\etab}}			\def\etabt{\widetilde{\etab}}	
\def\iotabh{\widehat{\iotab}}		\def\iotabt{\widetilde{\iotab}}	
\def\kappabh{\widehat{\kappab}}		\def\kappabt{\widetilde{\kappab}}	
\def\nubh{\widehat{\nub}}			\def\nubt{\widetilde{\nub}}	
\def\rhobh{\widehat{\rhob}}			\def\rhobt{\widetilde{\rhob}}	
\def\upsilonbh{\widehat{\upsilonb}}	\def\upsilonbt{\widetilde{\upsilonb}}	
\def\ralpha{\red{\alpha}}
\def\rbeta{\red{\beta}}
\def\rdelta{\red{\delta}}
\def\repsilon{\red{\epsilon}}
\def\rgamma{\red{\gamma}}
\def\romega{\red{\omega}}
\def\rtheta{\red{\theta}}
\def\rxi{\red{\xi}}
\def\rlambda{\red{\lambda}}
\def\rtau{\red{\tau}}
\def\rphi{\red{\phi}}
\def\rmu{\red{\mu}}
\def\rpsi{\red{\psi}}
\def\rchi{\red{\chi}}
\def\rsigma{\red{\sigma}}
\def\rzeta{\red{\zeta}}
\def\reta{\red{\eta}}
\def\riota{\red{\iota}}
\def\rkappa{\red{\kappa}}
\def\rnu{\red{\nu}}
\def\rrho{\red{\rho}}
\def\rupsilon{\red{\upsilon}}
\def\ralphabh{\red{\widehat{\alphab}}}	  	\def\ralphabt{\red{\widetilde{\alphab}}}
\def\rbetabh{\red{\widehat{\betab}}}	    \def\rbetabt{\red{\widetilde{\betab}}}
\def\rdeltabh{\red{\widehat{\deltab}}}	  	\def\rdeltabt{\red{\widetilde{\deltab}}}
\def\repsilonbh{\red{\widehat{\epsilonb}}}  \def\repsilonbt{\red{\widetilde{\epsilonb}}}
\def\rgammabh{\red{\widehat{\gammab}}}	  	\def\rgammabt{\red{\widetilde{\gammab}}}
\def\romegabh{\red{\widehat{\omegab}}}	  	\def\romegabt{\red{\widetilde{\omegab}}}
\def\rthetabh{\red{\widehat{\thetab}}}	  	\def\rthetabt{\red{\widetilde{\thetab}}}
\def\rxibh{\red{\widehat{\xib}}}		    \def\rxibt{\red{\widetilde{\xib}}}
\def\rlambdabh{\red{\widehat{\lambdab}}}    \def\rlambdabt{\red{\widetilde{\lambdab}}}
\def\rtaubh{\red{\widehat{\taub}}}		  	\def\rtaubt{\red{\widetilde{\taub}}}
\def\rphibh{\red{\widehat{\phib}}}		  	\def\rphibt{\red{\widetilde{\phib}}}	
\def\rmubh{\red{\widehat{\mub}}}		    \def\rmubt{\red{\widetilde{\mub}}}
\def\rpsibh{\red{\widehat{\psib}}}		  	\def\rpsibt{\red{\widetilde{\psib}}}
\def\rchibh{\red{\widehat{\chib}}}		  	\def\rchibt{\red{\widetilde{\chib}}}	
\def\rsigmabh{\red{\widehat{\sigmab}}}	  	\def\rsigmabt{\red{\widetilde{\sigmab}}}
\def\rzetabh{\red{\widehat{\zetab}}}	    \def\rzetabt{\red{\widetilde{\zetab}}}
\def\retabh{\red{\widehat{\etab}}}		  	\def\retabt{\red{\widetilde{\etab}}}
\def\riotabh{\red{\widehat{\iotab}}}	    \def\riotabt{\red{\widetilde{\iotab}}}
\def\rkappabh{\red{\widehat{\kappab}}}	  	\def\rkappabt{\red{\widetilde{\kappab}}}
\def\rnubh{\red{\widehat{\nub}}}		    \def\rnubt{\red{\widetilde{\nub}}}
\def\rrhobh{\red{\widehat{\rhob}}}		  	\def\rrhobt{\red{\widetilde{\rhob}}}
\def\rupsilonbh{\red{\widehat{\upsilonb}}}  \def\rupsilonbt{\red{\widetilde{\upsilonb}}}

\def\balpha{\blue{\alpha}}
\def\bbeta{\blue{\beta}}
\def\bdelta{\blue{\delta}}
\def\bepsilon{\blue{\epsilon}}
\def\bgamma{\blue{\gamma}}
\def\bomega{\blue{\omega}}
\def\btheta{\blue{\theta}}
\def\bxi{\blue{\xi}}
\def\blambda{\blue{\lambda}}
\def\btau{\blue{\tau}}
\def\bphi{\blue{\phi}}
\def\bmu{\blue{\mu}}
\def\bpsi{\blue{\psi}}
\def\bchi{\blue{\chi}}
\def\bsigma{\blue{\sigma}}
\def\bzeta{\blue{\zeta}}
\def\biota{\blue{\iota}}
\def\bkappa{\blue{\kappa}}
\def\bnu{\blue{\nu}}
\def\brho{\blue{\rho}}
\def\bupsilon{\blue{\upsilon}}
\def\balphabh{\blue{\widehat{\alphab}}}	  	\def\balphabt{\blue{\widetilde{\alphab}}}
\def\bbetabh{\blue{\widehat{\betab}}}	    \def\bbetabt{\blue{\widetilde{\betab}}}
\def\bdeltabh{\blue{\widehat{\deltab}}}	  	\def\bdeltabt{\blue{\widetilde{\deltab}}}
\def\bepsilonbh{\blue{\widehat{\epsilonb}}} \def\bepsilonbt{\blue{\widetilde{\epsilonb}}}
\def\bgammabh{\blue{\widehat{\gammab}}}	  	\def\bgammabt{\blue{\widetilde{\gammab}}}
\def\bomegabh{\blue{\widehat{\omegab}}}	  	\def\bomegabt{\blue{\widetilde{\omegab}}}
\def\bthetabh{\blue{\widehat{\thetab}}}	  	\def\bthetabt{\blue{\widetilde{\thetab}}}
\def\bxibh{\blue{\widehat{\xib}}}		    \def\bxibt{\blue{\widetilde{\xib}}}
\def\blambdabh{\blue{\widehat{\lambdab}}}   \def\blambdabt{\blue{\widetilde{\lambdab}}}
\def\btaubh{\blue{\widehat{\taub}}}		  	\def\btaubt{\blue{\widetilde{\taub}}}
\def\bphibh{\blue{\widehat{\phib}}}		  	\def\bphibt{\blue{\widetilde{\phib}}}	
\def\bmubh{\blue{\widehat{\mub}}}		    \def\bmubt{\blue{\widetilde{\mub}}}
\def\bpsibh{\blue{\widehat{\psib}}}		  	\def\bpsibt{\blue{\widetilde{\psib}}}
\def\bchibh{\blue{\widehat{\chib}}}		  	\def\bchibt{\blue{\widetilde{\chib}}}	
\def\bsigmabh{\blue{\widehat{\sigmab}}}	  	\def\bsigmabt{\blue{\widetilde{\sigmab}}}
\def\bzetabh{\blue{\widehat{\zetab}}}	    \def\bzetabt{\blue{\widetilde{\zetab}}}
\def\biotabh{\blue{\widehat{\iotab}}}	    \def\biotabt{\blue{\widetilde{\iotab}}}
\def\bkappabh{\blue{\widehat{\kappab}}}	  	\def\bkappabt{\blue{\widetilde{\kappab}}}
\def\bnubh{\blue{\widehat{\nub}}}		    \def\bnubt{\blue{\widetilde{\nub}}}
\def\brhobh{\blue{\widehat{\rhob}}}		  	\def\brhobt{\blue{\widetilde{\rhob}}}
\def\bupsilonbh{\blue{\widehat{\upsilonb}}} \def\bupsilonbt{\blue{\widetilde{\upsilonb}}}
\def\ralphab{\red{\alphab}}
\def\rbetab{\red{\betab}}
\def\rdeltab{\red{\deltab}}
\def\repsilonb{\red{\epsilonb}}
\def\rgammab{\red{\gammab}}
\def\romegab{\red{\omegab}}
\def\rthetab{\red{\thetab}}
\def\rxib{\red{\xib}}
\def\rlambdab{\red{\lambdab}}
\def\rtaub{\red{\taub}}
\def\rphib{\red{\phib}}
\def\rmub{\red{\mub}}
\def\rpsib{\red{\psib}}
\def\rchib{\red{\chib}}
\def\rsigmab{\red{\sigmab}}
\def\rzetab{\red{\zetab}}
\def\retab{\red{\etab}}
\def\riotab{\red{\iotab}}
\def\rkappab{\red{\kappab}}
\def\rnub{\red{\nub}}
\def\rrhob{\red{\rhob}}
\def\rupsilonb{\red{\upsilonb}}
\def\balphab{\blue{\alphab}}
\def\bbetab{\blue{\betab}}
\def\bdeltab{\blue{\deltab}}
\def\bepsilonb{\blue{\epsilonb}}
\def\bgammab{\blue{\gammab}}
\def\bomegab{\blue{\omegab}}
\def\bthetab{\blue{\thetab}}
\def\bxib{\blue{\xib}}
\def\blambdab{\blue{\lambdab}}
\def\btaub{\blue{\taub}}
\def\bphib{\blue{\phib}}
\def\bmub{\blue{\mub}}
\def\bpsib{\blue{\psib}}
\def\bchib{\blue{\chib}}
\def\bsigmab{\blue{\sigmab}}
\def\bzetab{\blue{\zetab}}
\def\biotab{\blue{\iotab}}
\def\bkappab{\blue{\kappab}}
\def\bnub{\blue{\nub}}
\def\brhob{\blue{\rhob}}
\def\bupsilonb{\blue{\upsilonb}}
\def\Gammab{\mbox{\boldmath $\Gamma$}}
\def\Lambdab{\mbox{\boldmath $\Lambda$}}
\def\Sigmab{\mbox{\boldmath $\Sigma$}}
\def\Psib{\mbox{\boldmath $\Psi$}}
\def\Deltab{\mbox{\boldmath $\Delta$}}
\def\Xib{\mbox{\boldmath $\Xi$}}
\def\Upsilonb{\mbox{\boldmath $\Upsilon$}}
\def\Omegab{\mbox{\boldmath $\Omega$}}
\def\Thetab{\mbox{\boldmath $\Theta$}}
\def\Pib{\mbox{\boldmath $\Pi$}}
\def\Phib{\mbox{\boldmath $\Phi$}}
\def\Gammah{\widehat{\Gamma}}		\def\Gammat{\widetilde{\Gamma}}
\def\Lambdah{\widehat{\Lambda}}		\def\Lambdat{\widetilde{\Lambda}}
\def\Sigmah{\widehat{\Sigma}}		\def\Sigmat{\widetilde{\Sigma}}
\def\Psih{\widehat{\Psi}}			\def\Psit{\widetilde{\Psi}}
\def\Deltah{\widehat{\Delta}}		\def\Deltat{\widetilde{\Delta}}
\def\Xih{\widehat{\Xi}}				\def\Xit{\widetilde{\Xi}}
\def\Upsilonh{\widehat{\Upsilon}}	\def\Upsilont{\widetilde{\Upsilon}}
\def\Omegah{\widehat{\Omega}}		\def\Omegat{\widetilde{\Omega}}	
\def\Thetah{\widehat{\Theta}}		\def\Thetat{\widetilde{\Theta}}
\def\Pih{\widehat{\Pi}}				\def\Pit{\widetilde{\Pi}}
\def\Phih{\widehat{\Phi}}			\def\Phit{\widetilde{\Phi}}
\def\Gammabh{\widehat{\Gammab}}		\def\Gammabt{\widetilde{\Gammab}}
\def\Lambdabh{\widehat{\Lambdab}}		\def\Lambdabt{\widetilde{\Lambdab}}
\def\Sigmabh{\widehat{\Sigmab}}		\def\Sigmabt{\widetilde{\Sigmab}}
\def\Psibh{\widehat{\Psib}}			\def\Psibt{\widetilde{\Psib}}
\def\Deltabh{\widehat{\Deltab}}		\def\Deltabt{\widetilde{\Deltab}}
\def\Xibh{\widehat{\Xib}}				\def\Xibt{\widetilde{\Xib}}
\def\Upsilonbh{\widehat{\Upsilonb}}	\def\Upsilonbt{\widetilde{\Upsilonb}}
\def\Omegabh{\widehat{\Omegab}}		\def\Omegabt{\widetilde{\Omegab}}	
\def\Thetabh{\widehat{\Thetab}}		\def\Thetabt{\widetilde{\Thetab}}
\def\Pibh{\widehat{\Pib}}				\def\Pibt{\widetilde{\Pib}}
\def\Phibh{\widehat{\Phib}}			\def\Phibt{\widetilde{\Phib}}
\def\bGammabh{\blue{\widehat{\Gammab}}}		\def\bGammabt{\blue{\widetilde{\Gammab}}}
\def\bLambdabh{\blue{\widehat{\Lambdab}}}	\def\bLambdabt{\blue{\widetilde{\Lambdab}}}
\def\bSigmabh{\blue{\widehat{\Sigmab}}}		\def\bSigmabt{\blue{\widetilde{\Sigmab}}}
\def\bPsibh{\blue{\widehat{\Psib}}}			\def\bPsibt{\blue{\widetilde{\Psib}}}	
\def\bDeltabh{\blue{\widehat{\Deltab}}}		\def\bDeltabt{\blue{\widetilde{\Deltab}}}
\def\bXibh{\blue{\widehat{\Xib}}}			\def\bXibt{\blue{\widetilde{\Xib}}}
\def\bUpsilonbh{\blue{\widehat{\Upsilonb}}}	\def\bUpsilonbt{\blue{\widetilde{\Upsilonb}}}
\def\bOmegabh{\blue{\widehat{\Omegab}}}		\def\bOmegabt{\blue{\widetilde{\Omegab}}}	
\def\bThetabh{\blue{\widehat{\Thetab}}}		\def\bThetabt{\blue{\widetilde{\Thetab}}}
\def\bPibh{\blue{\widehat{\Pib}}}			\def\bPibt{\blue{\widetilde{\Pib}}}	
\def\bPhibh{\blue{\widehat{\Phib}}}			\def\bPhibt{\blue{\widetilde{\Phib}}}	
\def\rGammabh{\red{\widehat{\Gammab}}}		\def\rGammabt{\red{\widetilde{\Gammab}}}
\def\rLambdabh{\red{\widehat{\Lambdab}}}	\def\rLambdabt{\red{\widetilde{\Lambdab}}}
\def\rSigmabh{\red{\widehat{\Sigmab}}}		\def\rSigmabt{\red{\widetilde{\Sigmab}}}
\def\rPsibh{\red{\widehat{\Psib}}}			\def\rPsibt{\red{\widetilde{\Psib}}}	
\def\rDeltabh{\red{\widehat{\Deltab}}}		\def\rDeltabt{\red{\widetilde{\Deltab}}}
\def\rXibh{\red{\widehat{\Xib}}}			\def\rXibt{\red{\widetilde{\Xib}}}
\def\rUpsilonbh{\red{\widehat{\Upsilonb}}}	\def\rUpsilonbt{\red{\widetilde{\Upsilonb}}}
\def\rOmegabh{\red{\widehat{\Omegab}}}		\def\rOmegabt{\red{\widetilde{\Omegab}}}	
\def\rThetabh{\red{\widehat{\Thetab}}}		\def\rThetabt{\red{\widetilde{\Thetab}}}
\def\rPibh{\red{\widehat{\Pib}}}			\def\rPibt{\red{\widetilde{\Pib}}}	
\def\rPhibh{\red{\widehat{\Phib}}}			\def\rPhibt{\red{\widetilde{\Phib}}}	
\def\rGammab{\red{\Gammab}}
\def\rLambdab{\red{\Lambdab}}
\def\rSigmab{\red{\Sigmab}}
\def\rPsib{\red{\Psib}}
\def\rDeltab{\red{\Deltab}}
\def\rXib{\red{\Xib}}
\def\rUpsilonb{\red{\Upsilonb}}
\def\rOmegab{\red{\Omegab}}
\def\rThetab{\red{\Thetab}}
\def\rPib{\red{\Pib}}
\def\rPhib{\red{\Phib}}
\def\bGammab{\blue{\Gammab}}
\def\bLambdab{\blue{\Lambdab}}
\def\bSigmab{\blue{\Sigmab}}
\def\bPsib{\blue{\Psib}}
\def\bDeltab{\blue{\Deltab}}
\def\bXib{\blue{\Xib}}
\def\bUpsilonb{\blue{\Upsilonb}}
\def\bOmegab{\blue{\Omegab}}
\def\bThetab{\blue{\Thetab}}
\def\bPib{\blue{\Pib}}
\def\bPhib{\blue{\Phib}}
\def\Ac{{\cal A}}
\def\Bc{{\cal B}}
\def\Cc{{\cal C}}
\def\Dc{{\cal D}}
\def\Ec{{\cal E}}
\def\Fc{{\cal F}}
\def\Gc{{\cal G}}
\def\Hc{{\cal H}}
\def\Ic{{\cal I}}
\def\Jc{{\cal J}}
\def\Kc{{\cal K}}
\def\Lc{{\cal L}}
\def\Mc{{\cal M}}
\def\Nc{{\cal N}}
\def\Oc{{\cal O}}
\def\Pc{{\cal P}}
\def\Qc{{\cal Q}}
\def\Rc{{\cal R}}
\def\Sc{{\cal S}}
\def\Tc{{\cal T}}
\def\Uc{{\cal U}}
\def\Vc{{\cal V}}
\def\Wc{{\cal W}}
\def\Xc{{\cal X}}
\def\Yc{{\cal Y}}
\def\Zc{{\cal Z}}
\def\ba{\blue{a}}
\def\bb{\blue{b}}
\def\bc{\blue{c}}
\def\bd{\blue{d}}
\def\be{\blue{e}}
\def\bf{\blue{f}}
\def\bg{\blue{g}}
\def\bh{\blue{h}}
\def\bi{\blue{i}}
\def\bj{\blue{j}}
\def\bk{\blue{k}}
\def\bl{\blue{l}}
\def\bm{\blue{m}}
\def\bn{\blue{n}}
\def\bo{\blue{o}}
\def\bp{\blue{p}}
\def\bq{\blue{q}}
\def\br{\blue{r}}
\def\bs{\blue{s}}
\def\bt{\blue{t}}
\def\bu{\blue{u}}
\def\bv{\blue{v}}
\def\bw{\blue{w}}
\def\bx{\blue{x}}
\def\by{\blue{y}}
\def\bz{\blue{z}}
\def\ra{\red{a}}
\def\rb{\red{b}}
\def\rc{\red{c}}
\def\rd{\red{d}}
\def\re{\red{e}}
\def\rf{\red{f}}
\def\rg{\red{g}}
\def\rh{\red{h}}
\def\ri{\red{i}}
\def\rj{\red{j}}
\def\rk{\red{k}}
\def\rl{\red{l}}
\def\rm{\red{m}}
\def\rn{\red{n}}
\def\ro{\red{o}}
\def\rp{\red{p}}
\def\rq{\red{q}}
\def\rr{\red{r}}
\def\rs{\red{s}}
\def\rt{\red{t}}
\def\ru{\red{u}}
\def\rv{\red{v}}
\def\rw{\red{w}}
\def\rx{\red{x}}
\def\ry{\red{y}}
\def\rz{\red{z}}
\def\rah{\red{\widehat{a}}}		\def\rat{\red{\widetilde{a}}}
\def\rbh{\red{\widehat{b}}}		\def\rbt{\red{\widetilde{b}}}		
\def\rch{\red{\widehat{c}}}		\def\rct{\red{\widetilde{c}}}		
\def\rdh{\red{\widehat{d}}}		\def\rdt{\red{\widetilde{d}}}		
\def\reh{\red{\widehat{e}}}		\def\ret{\red{\widetilde{e}}}		
\def\rfh{\red{\widehat{f}}}		\def\rft{\red{\widetilde{f}}}		
\def\rgh{\red{\widehat{g}}}		\def\rgt{\red{\widetilde{g}}}		
\def\rhh{\red{\widehat{h}}}		\def\rht{\red{\widetilde{h}}}		
\def\rih{\red{\widehat{i}}}		\def\rit{\red{\widetilde{i}}}		
\def\rjh{\red{\widehat{j}}}		\def\rjt{\red{\widetilde{j}}}		
\def\rkh{\red{\widehat{k}}}		\def\rkt{\red{\widetilde{k}}}		
\def\rlh{\red{\widehat{l}}}		\def\rlt{\red{\widetilde{l}}}		
\def\rmh{\red{\widehat{m}}}		\def\rmt{\red{\widetilde{m}}}		
\def\rnh{\red{\widehat{n}}}		\def\rnt{\red{\widetilde{n}}}		
\def\roh{\red{\widehat{o}}}		\def\rot{\red{\widetilde{o}}}		
\def\rph{\red{\widehat{p}}}		\def\rpt{\red{\widetilde{p}}}		
\def\rqh{\red{\widehat{q}}}		\def\rqt{\red{\widetilde{q}}}		
\def\rrh{\red{\widehat{r}}}		\def\rrt{\red{\widetilde{r}}}		
\def\rsh{\red{\widehat{s}}}		\def\rst{\red{\widetilde{s}}}		
\def\rth{\red{\widehat{t}}}		\def\rtt{\red{\widetilde{t}}}		
\def\ruh{\red{\widehat{u}}}		\def\rut{\red{\widetilde{u}}}		
\def\rvh{\red{\widehat{v}}}		\def\rvt{\red{\widetilde{v}}}		
\def\rwh{\red{\widehat{w}}}		\def\rwt{\red{\widetilde{w}}}		
\def\rxh{\red{\widehat{x}}}		\def\rxt{\red{\widetilde{x}}}		
\def\ryh{\red{\widehat{y}}}		\def\ryt{\red{\widetilde{y}}}		
\def\rzh{\red{\widehat{z}}}		\def\rzt{\red{\widetilde{z}}}		
\def\bah{\blue{\widehat{a}}}	\def\bat{\blue{\widetilde{a}}}	
\def\bbh{\blue{\widehat{b}}}	\def\bbt{\blue{\widetilde{b}}}	
\def\bch{\blue{\widehat{c}}}	\def\bct{\blue{\widetilde{c}}}	
\def\bdh{\blue{\widehat{d}}}	\def\bdt{\blue{\widetilde{d}}}	
\def\beh{\blue{\widehat{e}}}	\def\bet{\blue{\widetilde{e}}}	
\def\bfh{\blue{\widehat{f}}}	\def\bft{\blue{\widetilde{f}}}	
\def\bgh{\blue{\widehat{g}}}	\def\bgt{\blue{\widetilde{g}}}	
\def\bhh{\blue{\widehat{h}}}	\def\bht{\blue{\widetilde{h}}}	
\def\bih{\blue{\widehat{i}}}	\def\bit{\blue{\widetilde{i}}}	
\def\bjh{\blue{\widehat{j}}}	\def\bjt{\blue{\widetilde{j}}}	
\def\bkh{\blue{\widehat{k}}}	\def\bkt{\blue{\widetilde{k}}}	
\def\blh{\blue{\widehat{l}}}	\def\blt{\blue{\widetilde{l}}}	
\def\bmh{\blue{\widehat{m}}}	\def\bmt{\blue{\widetilde{m}}}	
\def\bnh{\blue{\widehat{n}}}	\def\bnt{\blue{\widetilde{n}}}	
\def\boh{\blue{\widehat{o}}}	\def\bot{\blue{\widetilde{o}}}	
\def\bph{\blue{\widehat{p}}}	\def\bpt{\blue{\widetilde{p}}}	
\def\bqh{\blue{\widehat{q}}}	\def\bqt{\blue{\widetilde{q}}}	
\def\brh{\blue{\widehat{r}}}	\def\brt{\blue{\widetilde{r}}}	
\def\bsh{\blue{\widehat{s}}}	\def\bst{\blue{\widetilde{s}}}	
\def\bth{\blue{\widehat{t}}}	\def\btt{\blue{\widetilde{t}}}	
\def\buh{\blue{\widehat{u}}}	\def\but{\blue{\widetilde{u}}}	
\def\bvh{\blue{\widehat{v}}}	\def\bvt{\blue{\widetilde{v}}}	
\def\bwh{\blue{\widehat{w}}}	\def\bwt{\blue{\widetilde{w}}}	
\def\bxh{\blue{\widehat{x}}}	\def\bxt{\blue{\widetilde{x}}}	
\def\byh{\blue{\widehat{y}}}	\def\byt{\blue{\widetilde{y}}}	
\def\bzh{\blue{\widehat{z}}}	\def\bzt{\blue{\widetilde{z}}}	
\def\bab{\blue{\ab}}
\def\bbb{\blue{\bb}}
\def\bcb{\blue{\cb}}
\def\bdb{\blue{\db}}
\def\beb{\blue{\eb}}
\def\bfb{\blue{\fb}}
\def\bgb{\blue{\gb}}
\def\bhb{\blue{\hb}}
\def\bib{\blue{\ib}}
\def\bjb{\blue{\jb}}
\def\bkb{\blue{\kb}}
\def\blb{\blue{\lb}}
\def\bmb{\blue{\mb}}
\def\bnb{\blue{\nb}}
\def\bob{\blue{\ob}}
\def\bpb{\blue{\pb}}
\def\bqb{\blue{\qb}}
\def\brb{\blue{\rb}}
\def\bsb{\blue{\sb}}
\def\btb{\blue{\tb}}
\def\bub{\blue{\ub}}
\def\bvb{\blue{\vb}}
\def\bwb{\blue{\wb}}
\def\bxb{\blue{\xb}}
\def\byb{\blue{\yb}}
\def\bzb{\blue{\zb}}
\def\rab{\red{\ab}}
\def\rbb{\red{\bb}}
\def\rcb{\red{\cb}}
\def\rdb{\red{\db}}
\def\reb{\red{\eb}}
\def\rfb{\red{\fb}}
\def\rfbone{\rfb_{\red{1}}}
\def\rfbtwo{\rfb_{\red{2}}}
\def\rgb{\red{\gb}}
\def\rhb{\red{\hb}}
\def\rib{\red{\ib}}
\def\rjb{\red{\jb}}
\def\rkb{\red{\kb}}
\def\rlb{\red{\lb}}
\def\rmb{\red{\mb}}
\def\rnb{\red{\nb}}
\def\rob{\red{\ob}}
\def\rpb{\red{\pb}}
\def\rqb{\red{\qb}}
\def\rrb{\red{\rb}}
\def\rsb{\red{\sb}}
\def\rtb{\red{\tb}}
\def\rub{\red{\ub}}
\def\rvb{\red{\vb}}
\def\rwb{\red{\wb}}
\def\rxb{\red{\xb}}
\def\ryb{\red{\yb}}
\def\rzb{\red{\zb}}
\def\rabh{\red{\widehat{\ab}}}	\def\rabt{\red{\widetilde{\ab}}}	
\def\rbbh{\red{\widehat{\bb}}}	\def\rbbt{\red{\widetilde{\bb}}}	
\def\rcbh{\red{\widehat{\cb}}}	\def\rcbt{\red{\widetilde{\cb}}}	
\def\rdbh{\red{\widehat{\db}}}	\def\rdbt{\red{\widetilde{\db}}}	
\def\rebh{\red{\widehat{\eb}}}	\def\rebt{\red{\widetilde{\eb}}}	
\def\rfbh{\red{\widehat{\fb}}}	\def\rfbt{\red{\widetilde{\fb}}}	
\def\rgbh{\red{\widehat{\gb}}}	\def\rgbt{\red{\widetilde{\gb}}}	
\def\rhbh{\red{\widehat{\hb}}}	\def\rhbt{\red{\widetilde{\hb}}}	
\def\ribh{\red{\widehat{\ib}}}	\def\ribt{\red{\widetilde{\ib}}}	
\def\rjbh{\red{\widehat{\jb}}}	\def\rjbt{\red{\widetilde{\jb}}}	
\def\rkbh{\red{\widehat{\kb}}}	\def\rkbt{\red{\widetilde{\kb}}}	
\def\rlbh{\red{\widehat{\lb}}}	\def\rlbt{\red{\widetilde{\lb}}}	
\def\rmbh{\red{\widehat{\mb}}}	\def\rmbt{\red{\widetilde{\mb}}}	
\def\rnbh{\red{\widehat{\nb}}}	\def\rnbt{\red{\widetilde{\nb}}}	
\def\robh{\red{\widehat{\ob}}}	\def\robt{\red{\widetilde{\ob}}}	
\def\rpbh{\red{\widehat{\pb}}}	\def\rpbt{\red{\widetilde{\pb}}}	
\def\rqbh{\red{\widehat{\qb}}}	\def\rqbt{\red{\widetilde{\qb}}}	
\def\rrbh{\red{\widehat{\rb}}}	\def\rrbt{\red{\widetilde{\rb}}}	
\def\rsbh{\red{\widehat{\sb}}}	\def\rsbt{\red{\widetilde{\sb}}}	
\def\rtbh{\red{\widehat{\tb}}}	\def\rtbt{\red{\widetilde{\tb}}}	
\def\rubh{\red{\widehat{\ub}}}	\def\rubt{\red{\widetilde{\ub}}}	
\def\rvbh{\red{\widehat{\vb}}}	\def\rvbt{\red{\widetilde{\vb}}}	
\def\rwbh{\red{\widehat{\wb}}}	\def\rwbt{\red{\widetilde{\wb}}}	
\def\rxbh{\red{\widehat{\xb}}}	\def\rxbt{\red{\widetilde{\xb}}}	
\def\rybh{\red{\widehat{\yb}}}	\def\rybt{\red{\widetilde{\yb}}}	
\def\rzbh{\red{\widehat{\zb}}}	\def\rzbt{\red{\widetilde{\zb}}}	
\def\babh{\blue{\widehat{\ab}}}	\def\babt{\blue{\widetilde{\ab}}}	
\def\bbbh{\blue{\widehat{\bb}}}	\def\bbbt{\blue{\widetilde{\bb}}}	
\def\bcbh{\blue{\widehat{\cb}}}	\def\bcbt{\blue{\widetilde{\cb}}}	
\def\bdbh{\blue{\widehat{\db}}}	\def\bdbt{\blue{\widetilde{\db}}}	
\def\bebh{\blue{\widehat{\eb}}}	\def\bebt{\blue{\widetilde{\eb}}}	
\def\bfbh{\blue{\widehat{\fb}}}	\def\bfbt{\blue{\widetilde{\fb}}}	
\def\bgbh{\blue{\widehat{\gb}}}	\def\bgbt{\blue{\widetilde{\gb}}}	
\def\bhbh{\blue{\widehat{\hb}}}	\def\bhbt{\blue{\widetilde{\hb}}}	
\def\bibh{\blue{\widehat{\ib}}}	\def\bibt{\blue{\widetilde{\ib}}}	
\def\bjbh{\blue{\widehat{\jb}}}	\def\bjbt{\blue{\widetilde{\jb}}}	
\def\bkbh{\blue{\widehat{\kb}}}	\def\bkbt{\blue{\widetilde{\kb}}}	
\def\blbh{\blue{\widehat{\lb}}}	\def\blbt{\blue{\widetilde{\lb}}}	
\def\bmbh{\blue{\widehat{\mb}}}	\def\bmbt{\blue{\widetilde{\mb}}}	
\def\bnbh{\blue{\widehat{\nb}}}	\def\bnbt{\blue{\widetilde{\nb}}}	
\def\bobh{\blue{\widehat{\ob}}}	\def\bobt{\blue{\widetilde{\ob}}}	
\def\bpbh{\blue{\widehat{\pb}}}	\def\bpbt{\blue{\widetilde{\pb}}}	
\def\bqbh{\blue{\widehat{\qb}}}	\def\bqbt{\blue{\widetilde{\qb}}}	
\def\brbh{\blue{\widehat{\rb}}}	\def\brbt{\blue{\widetilde{\rb}}}	
\def\bsbh{\blue{\widehat{\sb}}}	\def\bsbt{\blue{\widetilde{\sb}}}	
\def\btbh{\blue{\widehat{\tb}}}	\def\btbt{\blue{\widetilde{\tb}}}	
\def\bubh{\blue{\widehat{\ub}}}	\def\bubt{\blue{\widetilde{\ub}}}	
\def\bvbh{\blue{\widehat{\vb}}}	\def\bvbt{\blue{\widetilde{\vb}}}	
\def\bwbh{\blue{\widehat{\wb}}}	\def\bwbt{\blue{\widetilde{\wb}}}	
\def\bxbh{\blue{\widehat{\xb}}}	\def\bxbt{\blue{\widetilde{\xb}}}	
\def\bybh{\blue{\widehat{\yb}}}	\def\bybt{\blue{\widetilde{\yb}}}	
\def\bzbh{\blue{\widehat{\zb}}}	\def\bzbt{\blue{\widetilde{\zb}}}	
\def\bA{\blue{A}}
\def\bB{\blue{B}}
\def\bC{\blue{C}}
\def\bD{\blue{D}}
\def\bE{\blue{E}}
\def\bF{\blue{F}}
\def\bG{\blue{G}}
\def\bH{\blue{H}}
\def\bI{\blue{I}}
\def\bJ{\blue{J}}
\def\bK{\blue{K}}
\def\bL{\blue{L}}
\def\bM{\blue{M}}
\def\bN{\blue{N}}
\def\bO{\blue{O}}
\def\bP{\blue{P}}
\def\bQ{\blue{Q}}
\def\bR{\blue{R}}
\def\bS{\blue{S}}
\def\bT{\blue{T}}
\def\bU{\blue{U}}
\def\bV{\blue{V}}
\def\bW{\blue{W}}
\def\bX{\blue{X}}
\def\bY{\blue{Y}}
\def\bZ{\blue{Z}}
\def\rA{\red{A}}
\def\rB{\red{B}}
\def\rC{\red{C}}
\def\rD{\red{D}}
\def\rE{\red{E}}
\def\rF{\red{F}}
\def\rG{\red{G}}
\def\rH{\red{H}}
\def\rI{\red{I}}
\def\rJ{\red{J}}
\def\rK{\red{K}}
\def\rL{\red{L}}
\def\rM{\red{M}}
\def\rN{\red{N}}
\def\rO{\red{O}}
\def\rP{\red{P}}
\def\rQ{\red{Q}}
\def\rR{\red{R}}
\def\rS{\red{S}}
\def\rT{\red{T}}
\def\rU{\red{U}}
\def\rV{\red{V}}
\def\rW{\red{W}}
\def\rX{\red{X}}
\def\rY{\red{Y}}
\def\rZ{\red{Z}}
\def\rAh{\red{\widehat{A}}}	\def\rAt{\red{\widetilde{A}}}	
\def\rBh{\red{\widehat{B}}}	\def\rBt{\red{\widetilde{B}}}	
\def\rCh{\red{\widehat{C}}}	\def\rCt{\red{\widetilde{C}}}	
\def\rDh{\red{\widehat{D}}}	\def\rDt{\red{\widetilde{D}}}	
\def\rEh{\red{\widehat{E}}}	\def\rEt{\red{\widetilde{E}}}	
\def\rFh{\red{\widehat{F}}}	\def\rFt{\red{\widetilde{F}}}	
\def\rGh{\red{\widehat{G}}}	\def\rGt{\red{\widetilde{G}}}	
\def\rHh{\red{\widehat{H}}}	\def\rHt{\red{\widetilde{H}}}	
\def\rIh{\red{\widehat{I}}}	\def\rIt{\red{\widetilde{I}}}	
\def\rJh{\red{\widehat{J}}}	\def\rJt{\red{\widetilde{J}}}	
\def\rKh{\red{\widehat{K}}}	\def\rKt{\red{\widetilde{K}}}	
\def\rLh{\red{\widehat{L}}}	\def\rLt{\red{\widetilde{L}}}	
\def\rMh{\red{\widehat{M}}}	\def\rMt{\red{\widetilde{M}}}	
\def\rNh{\red{\widehat{N}}}	\def\rNt{\red{\widetilde{N}}}	
\def\rOh{\red{\widehat{O}}}	\def\rOt{\red{\widetilde{O}}}	
\def\rPh{\red{\widehat{P}}}	\def\rPt{\red{\widetilde{P}}}	
\def\rQh{\red{\widehat{Q}}}	\def\rQt{\red{\widetilde{Q}}}	
\def\rRh{\red{\widehat{R}}}	\def\rRt{\red{\widetilde{R}}}	
\def\rSh{\red{\widehat{S}}}	\def\rSt{\red{\widetilde{S}}}	
\def\rTh{\red{\widehat{T}}}	\def\rTt{\red{\widetilde{T}}}	
\def\rUh{\red{\widehat{U}}}	\def\rUt{\red{\widetilde{U}}}	
\def\rVh{\red{\widehat{V}}}	\def\rVt{\red{\widetilde{V}}}	
\def\rWh{\red{\widehat{W}}}	\def\rWt{\red{\widetilde{W}}}	
\def\rXh{\red{\widehat{X}}}	\def\rXt{\red{\widetilde{X}}}	
\def\rYh{\red{\widehat{Y}}}	\def\rYt{\red{\widetilde{Y}}}	
\def\rZh{\red{\widehat{Z}}}	\def\rZt{\red{\widetilde{Z}}}	
\def\bAh{\blue{\widehat{A}}}	\def\bAt{\blue{\widetilde{A}}}	
\def\bBh{\blue{\widehat{B}}}	\def\bBt{\blue{\widetilde{B}}}	
\def\bCh{\blue{\widehat{C}}}	\def\bCt{\blue{\widetilde{C}}}	
\def\bDh{\blue{\widehat{D}}}	\def\bDt{\blue{\widetilde{D}}}	
\def\bEh{\blue{\widehat{E}}}	\def\bEt{\blue{\widetilde{E}}}	
\def\bFh{\blue{\widehat{F}}}	\def\bFt{\blue{\widetilde{F}}}	
\def\bGh{\blue{\widehat{G}}}	\def\bGt{\blue{\widetilde{G}}}	
\def\bHh{\blue{\widehat{H}}}	\def\bHt{\blue{\widetilde{H}}}	
\def\bIh{\blue{\widehat{I}}}	\def\bIt{\blue{\widetilde{I}}}	
\def\bJh{\blue{\widehat{J}}}	\def\bJt{\blue{\widetilde{J}}}	
\def\bKh{\blue{\widehat{K}}}	\def\bKt{\blue{\widetilde{K}}}	
\def\bLh{\blue{\widehat{L}}}	\def\bLt{\blue{\widetilde{L}}}	
\def\bMh{\blue{\widehat{M}}}	\def\bMt{\blue{\widetilde{M}}}	
\def\bNh{\blue{\widehat{N}}}	\def\bNt{\blue{\widetilde{N}}}	
\def\bOh{\blue{\widehat{O}}}	\def\bOt{\blue{\widetilde{O}}}	
\def\bPh{\blue{\widehat{P}}}	\def\bPt{\blue{\widetilde{P}}}	
\def\bQh{\blue{\widehat{Q}}}	\def\bQt{\blue{\widetilde{Q}}}	
\def\bRh{\blue{\widehat{R}}}	\def\bRt{\blue{\widetilde{R}}}	
\def\bSh{\blue{\widehat{S}}}	\def\bSt{\blue{\widetilde{S}}}	
\def\bTh{\blue{\widehat{T}}}	\def\bTt{\blue{\widetilde{T}}}	
\def\bUh{\blue{\widehat{U}}}	\def\bUt{\blue{\widetilde{U}}}	
\def\bVh{\blue{\widehat{V}}}	\def\bVt{\blue{\widetilde{V}}}	
\def\bWh{\blue{\widehat{W}}}	\def\bWt{\blue{\widetilde{W}}}	
\def\bXh{\blue{\widehat{X}}}	\def\bXt{\blue{\widetilde{X}}}	
\def\bYh{\blue{\widehat{Y}}}	\def\bYt{\blue{\widetilde{Y}}}	
\def\bZh{\blue{\widehat{Z}}}	\def\bZt{\blue{\widetilde{Z}}}	
\def\bAb{\blue{\Ab}}
\def\bBb{\blue{\Bb}}
\def\bCb{\blue{\Cb}}
\def\bDb{\blue{\Db}}
\def\bEb{\blue{\Eb}}
\def\bFb{\blue{\Fb}}
\def\bGB{\blue{\Gb}}
\def\bHb{\blue{\Hb}}
\def\bIb{\blue{\Ib}}
\def\bJb{\blue{\Jb}}
\def\bKb{\blue{\Kb}}
\def\bLb{\blue{\Lb}}
\def\bMb{\blue{\Mb}}
\def\bNb{\blue{\Nb}}
\def\bOb{\blue{\Ob}}
\def\bPb{\blue{\Pb}}
\def\bQb{\blue{\Qb}}
\def\bRb{\blue{\Rb}}
\def\bSb{\blue{\Sb}}
\def\bTb{\blue{\Tb}}
\def\bUb{\blue{\Ub}}
\def\bVb{\blue{\Vb}}
\def\bWb{\blue{\Wb}}
\def\bXb{\blue{\Xb}}
\def\bYb{\blue{\Yb}}
\def\bZb{\blue{\Zb}}
\def\rAbh{\red{\widehat{\Ab}}}	\def\rAbt{\red{\widetilde{\Ab}}}	
\def\rBbh{\red{\widehat{\Bb}}}	\def\rBbt{\red{\widetilde{\Bb}}}	
\def\rCbh{\red{\widehat{\Cb}}}	\def\rCbt{\red{\widetilde{\Cb}}}	
\def\rDbh{\red{\widehat{\Db}}}	\def\rDbt{\red{\widetilde{\Db}}}	
\def\rEbh{\red{\widehat{\Eb}}}	\def\rEbt{\red{\widetilde{\Eb}}}	
\def\rFbh{\red{\widehat{\Fb}}}	\def\rFbt{\red{\widetilde{\Fb}}}	
\def\rGbh{\red{\widehat{\Gb}}}	\def\rGbt{\red{\widetilde{\Gb}}}	
\def\rHbh{\red{\widehat{\Hb}}}	\def\rHbt{\red{\widetilde{\Hb}}}	
\def\rIbh{\red{\widehat{\Ib}}}	\def\rIbt{\red{\widetilde{\Ib}}}	
\def\rJbh{\red{\widehat{\Jb}}}	\def\rJbt{\red{\widetilde{\Jb}}}	
\def\rKbh{\red{\widehat{\Kb}}}	\def\rKbt{\red{\widetilde{\Kb}}}	
\def\rLbh{\red{\widehat{\Lb}}}	\def\rLbt{\red{\widetilde{\Lb}}}	
\def\rMbh{\red{\widehat{\Mb}}}	\def\rMbt{\red{\widetilde{\Mb}}}	
\def\rNbh{\red{\widehat{\Nb}}}	\def\rNbt{\red{\widetilde{\Nb}}}	
\def\rObh{\red{\widehat{\Ob}}}	\def\rObt{\red{\widetilde{\Ob}}}	
\def\rPbh{\red{\widehat{\Pb}}}	\def\rPbt{\red{\widetilde{\Pb}}}	
\def\rQbh{\red{\widehat{\Qb}}}	\def\rQbt{\red{\widetilde{\Qb}}}	
\def\rRbh{\red{\widehat{\Rb}}}	\def\rRbt{\red{\widetilde{\Rb}}}	
\def\rSbh{\red{\widehat{\Sb}}}	\def\rSbt{\red{\widetilde{\Sb}}}	
\def\rTbh{\red{\widehat{\Tb}}}	\def\rTbt{\red{\widetilde{\Tb}}}	
\def\rUbh{\red{\widehat{\Ub}}}	\def\rUbt{\red{\widetilde{\Ub}}}	
\def\rVbh{\red{\widehat{\Vb}}}	\def\rVbt{\red{\widetilde{\Vb}}}	
\def\rWbh{\red{\widehat{\Wb}}}	\def\rWbt{\red{\widetilde{\Wb}}}	
\def\rXbh{\red{\widehat{\Xb}}}	\def\rXbt{\red{\widetilde{\Xb}}}	
\def\rYbh{\red{\widehat{\Yb}}}	\def\rYbt{\red{\widetilde{\Yb}}}	
\def\rZbh{\red{\widehat{\Zb}}}	\def\rZbt{\red{\widetilde{\Zb}}}	
\def\bAbh{\blue{\widehat{\Ab}}}		\def\bAbt{\blue{\widetilde{\Ab}}}		
\def\bBbh{\blue{\widehat{\Bb}}}		\def\bBbt{\blue{\widetilde{\Bb}}}		
\def\bCbh{\blue{\widehat{\Cb}}}		\def\bCbt{\blue{\widetilde{\Cb}}}		
\def\bDbh{\blue{\widehat{\Db}}}		\def\bDbt{\blue{\widetilde{\Db}}}		
\def\bEbh{\blue{\widehat{\Eb}}}		\def\bEbt{\blue{\widetilde{\Eb}}}		
\def\bFbh{\blue{\widehat{\Fb}}}		\def\bFbt{\blue{\widetilde{\Fb}}}		
\def\bGbh{\blue{\widehat{\Gb}}}		\def\bGbt{\blue{\widetilde{\Gb}}}		
\def\bHbh{\blue{\widehat{\Hb}}}		\def\bHbt{\blue{\widetilde{\Hb}}}		
\def\bIbh{\blue{\widehat{\Ib}}}		\def\bIbt{\blue{\widetilde{\Ib}}}		
\def\bJbh{\blue{\widehat{\Jb}}}		\def\bJbt{\blue{\widetilde{\Jb}}}		
\def\bKbh{\blue{\widehat{\Kb}}}		\def\bKbt{\blue{\widetilde{\Kb}}}		
\def\bLbh{\blue{\widehat{\Lb}}}		\def\bLbt{\blue{\widetilde{\Lb}}}		
\def\bMbh{\blue{\widehat{\Mb}}}		\def\bMbt{\blue{\widetilde{\Mb}}}		
\def\bNbh{\blue{\widehat{\Nb}}}		\def\bNbt{\blue{\widetilde{\Nb}}}		
\def\bObh{\blue{\widehat{\Ob}}}		\def\bObt{\blue{\widetilde{\Ob}}}		
\def\bPbh{\blue{\widehat{\Pb}}}		\def\bPbt{\blue{\widetilde{\Pb}}}		
\def\bQbh{\blue{\widehat{\Qb}}}		\def\bQbt{\blue{\widetilde{\Qb}}}		
\def\bRbh{\blue{\widehat{\Rb}}}		\def\bRbt{\blue{\widetilde{\Rb}}}		
\def\bSbh{\blue{\widehat{\Sb}}}		\def\bSbt{\blue{\widetilde{\Sb}}}		
\def\bTbh{\blue{\widehat{\Tb}}}		\def\bTbt{\blue{\widetilde{\Tb}}}		
\def\bUbh{\blue{\widehat{\Ub}}}		\def\bUbt{\blue{\widetilde{\Ub}}}		
\def\bVbh{\blue{\widehat{\Vb}}}		\def\bVbt{\blue{\widetilde{\Vb}}}		
\def\bWbh{\blue{\widehat{\Wb}}}		\def\bWbt{\blue{\widetilde{\Wb}}}		
\def\bXbh{\blue{\widehat{\Xb}}}		\def\bXbt{\blue{\widetilde{\Xb}}}		
\def\bYbh{\blue{\widehat{\Yb}}}		\def\bYbt{\blue{\widetilde{\Yb}}}		
\def\bZbh{\blue{\widehat{\Zb}}}		\def\bZbt{\blue{\widetilde{\Zb}}}		
\def\rAb{\red{\Ab}}
\def\rBb{\red{\Bb}}
\def\rCb{\red{\Cb}}
\def\rDb{\red{\Db}}
\def\rEb{\red{\Eb}}
\def\rFb{\red{\Fb}}
\def\rGB{\red{\Gb}}
\def\rHb{\red{\Hb}}
\def\rIb{\red{\Ib}}
\def\rJb{\red{\Jb}}
\def\rKb{\red{\Kb}}
\def\rLb{\red{\Lb}}
\def\rMb{\red{\Mb}}
\def\rNb{\red{\Nb}}
\def\rOb{\red{\Ob}}
\def\rPb{\red{\Pb}}
\def\rQb{\red{\Qb}}
\def\rRb{\red{\Rb}}
\def\rSb{\red{\Sb}}
\def\rTb{\red{\Tb}}
\def\rUb{\red{\Ub}}
\def\rVb{\red{\Vb}}
\def\rWb{\red{\Wb}}
\def\rXb{\red{\Xb}}
\def\rYb{\red{\Yb}}
\def\rZb{\red{\Zb}}
\def\zerob{\mbox{\boldmath $0$}}
\def\oneb{\mbox{\boldmath $1$}}
\def\twob{\mbox{\boldmath $2$}}
\def\threeb{\mbox{\boldmath $3$}}
\def\fourb{\mbox{\boldmath $4$}}
\def\fiveb{\mbox{\boldmath $5$}}
\def\sixb{\mbox{\boldmath $6$}}
\def\sevenb{\mbox{\boldmath $7$}}
\def\eightb{\mbox{\boldmath $8$}}
\def\nineb{\mbox{\boldmath $9$}}

\def\rzero{\red{0}}
\def\rone{\red{1}}
\def\rtwo{\red{2}}
\def\rthree{\red{3}}
\def\rfour{\red{4}}
\def\rfive{\red{5}}
\def\rsix{\red{6}}
\def\rseven{\red{7}}
\def\reight{\red{8}}
\def\rnine{\red{9}}
\def\rzerob{\red{\zerob}}
\def\roneb{\red{\oneb}}
\def\rtwob{\red{\twob}}
\def\rthreeb{\red{\threeb}}
\def\rfourb{\red{\fourb}}
\def\rfiveb{\red{\fiveb}}
\def\rsixb{\red{\sixb}}
\def\rsevenb{\red{\sevenb}}
\def\reightb{\red{\eightb}}
\def\rnineb{\red{\nineb}}
\def\bzero{\blue{0}}
\def\bone{\blue{1}}
\def\btwo{\blue{2}}
\def\bthree{\blue{3}}
\def\bfour{\blue{4}}
\def\bfive{\blue{5}}
\def\bsix{\blue{6}}
\def\bseven{\blue{7}}
\def\beight{\blue{8}}
\def\bnine{\blue{9}}
\def\bzerob{\blue{\zerob}}
\def\boneb{\blue{\oneb}}
\def\btwob{\blue{\twob}}
\def\bthreeb{\blue{\threeb}}
\def\bfourb{\blue{\fourb}}
\def\bfiveb{\blue{\fiveb}}
\def\bsixb{\blue{\sixb}}
\def\bsevenb{\blue{\sevenb}}
\def\beightb{\blue{\eightb}}
\def\bnineb{\blue{\nineb}}
\def\diag#1{\mbox{diag}\left[#1\right]}
\def\Prob#1{\mbox{Pr}\left\{#1\right\}}
\def\var#1{\mbox{Var}\left\{#1\right\}}
\def\cov#1{\mbox{Cov}\left\{#1\right\}}
\def\Cov#1{\mbox{Cov}\left[#1\right]}
\def\corr#1{\mbox{Corr}\left\{#1\right\}}
\def\trace#1{\mbox{Tr}\left\{#1\right\}}
\def\Tr#1{\mbox{Tr}\left[#1\right]}
\def\rang#1{\mbox{rang}\left\{#1\right\}}
\def\det#1{\mbox{det}\left\{#1\right\}}
\def\d{\mbox{d}}


\title{Sparsity enforcing priors in inverse problems via Normal variance mixtures: model selection, algorithms and applications}
\date{}
\author{Mircea Dumitru$^{1}$\\[10pt]
  \multicolumn{1}{p{.7\textwidth}}{\centering\emph{$^1$Laboratoire des signaux et syst\`emes (L2S),\\ CNRS -- CentraleSup\'elec -- Universit\'e Paris-Sud, \\ CentraleSup\'elec, Plateau de Moulon, 91192 Gif-sur-Yvette, France}}}
\maketitle

\begin{abstract}
The sparse structure of the solution for an inverse problem can be modelled using different sparsity enforcing priors when the Bayesian approach is considered. Analytical expression for the unknowns of the model can be obtained by building hierarchical models based on sparsity enforcing distributions expressed via conjugate priors. We consider heavy tailed distributions with this property: the Student-t distribution, which is expressed as a Normal scale mixture, with the mixing distribution the Inverse Gamma distribution, the Laplace distribution, which can also be expressed as a Normal scale mixture, with the mixing distribution the Exponential distribution or can be expressed as a Normal inverse scale mixture, with the mixing distribution the Inverse Gamma distribution, the Hyperbolic distribution, the Variance-Gamma distribution, the Normal-Inverse Gaussian distribution, all three expressed via conjugate distributions using the Generalized Hyperbolic distribution. For all distributions iterative algorithms are derived based on hierarchical models that account for the uncertainties of the forward model. For estimation,  Maximum A Posterior (MAP) and Posterior Mean (PM) via variational Bayesian approximation (VBA) are used. The performances of resulting algorithm are compared in applications in 3D computed tomography (3D-CT) and chronobiology.
Finally, a theoretical study is developed for comparison between sparsity enforcing algorithms obtained via the Bayesian approach and the sparsity enforcing algorithms issued from regularization techniques, like LASSO and some others. 
\\[5pt]
{Keywords:} inverse problems, sparsity enforcing priors, conjugate priors, Student-t prior model (StPM), Laplace prior model (LPM), uncertainties model, estimation, variational Bayesian approximation (VBA) 3D computed tomography, chronobiology
\end{abstract}
\newpage
\tableofcontents

\section{Introduction}
\label{Sec:Introduction}
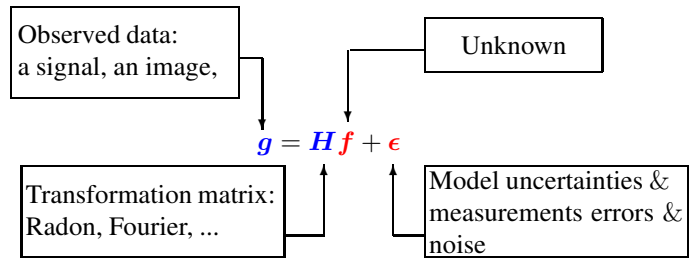
\begin{figure}[!htb]
\center
\begin{tabular}{c}
\begin{picture}(220,95)

\put(1,62){\framebox(85,34){\parbox{80\unitlength}{Observed data: \\ a signal, an image,}}}
\put(86,77){\line(1,0){9}}
\put(95,77){\vector(0,-1){27}}

\put(156,72){\framebox(66,20){Unknown}}
\put(156,80){\line(-1,0){28}}
\put(127,80){\vector(0,-1){27}}

\put(120,45){\makebox(0,0){$\bgb = \bHb \rfb + \repsilonb$}}

\put(5,2){\framebox(98,34){\parbox{95\unitlength}{Transformation matrix: \\ Radon, Fourier, ... }}}
\put(103,10){\line(1,0){15}}
\put(118,10){\vector(0,1){27}}

\put(156,2){\framebox(98,34){\parbox{95\unitlength}{Model uncertainties $\&$ \\ measurements errors $\&$ noise}}}
\put(156,10){\line(-1,0){12}}
\put(144,10){\vector(0,1){27}}

\end{picture}
\end{tabular}
\caption{Interpretation of the forward linear model, Equation~\eqref{Eq:LinearModel}}
\label{Fig:InterpretationOfTheForwardModel}
\end{figure}

\begin{figure}[!htb]
\center
\begin{tabular}{c}
\begin{picture}(220,75)

\put(30,-8){\framebox(36,16){$\psi_{\theta_\epsilon}$}}
\put(66,0){\vector(1,0){24}}

\put(100,0){\circle{20}}
\put(100,0){\makebox(0,0){$\rthetab_\repsilon$}}
\put(110,0){\vector(1,0){20}}

\put(140,0){\circle{20}}
\put(140,0){\makebox(0,0){$\repsilonb$}}
\put(150,0){\vector(1,0){20}}

\put(180,0){\circle{20}}
\put(180,0){\makebox(0,0){$\bgb$}}

\put(62,37){$\bgb = \bHb \rfb + \repsilon$}

\put(180,40){\circle{20}}
\put(180,40){\makebox(0,0){$\rfb$}}
\put(180,30){\vector(0,-1){20}}
\put(188,22){\makebox(0,0){$\blue{\bHb}$}}

\put(30,72){\framebox(36,16){$\psi_{\theta_f}$}}
\put(66,80){\vector(1,0){104}}

\put(180,80){\circle{20}}
\put(180,80){\makebox(0,0){$\rthetab_\rf$}}
\put(180,70){\vector(0,-1){20}}

\end{picture}
\end{tabular}
\caption{From linear forward model, Equation~\eqref{Eq:LinearModel} to the Hierarchical Model: Direct sparsity, i.e. $\rfb$ is sparse}
\label{Fig:FromForwardModelHierarchicalModel}
\end{figure}
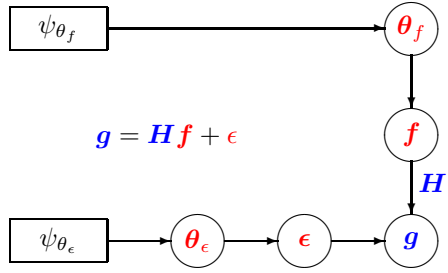

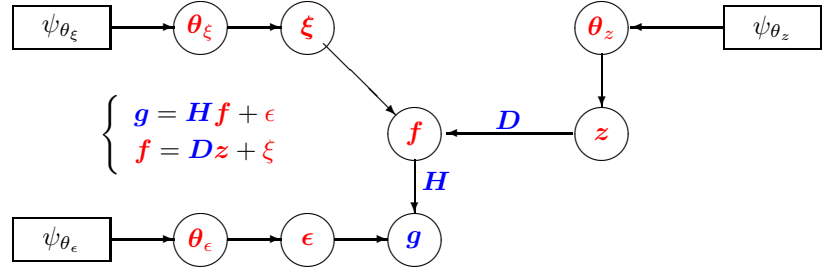
\begin{figure}[!htb]
\center
\begin{tabular}{c}
\begin{picture}(330,75)

\put(30,-8){\framebox(36,16){$\psi_{\theta_\epsilon}$}}
\put(66,0){\vector(1,0){24}}

\put(100,0){\circle{20}}
\put(100,0){\makebox(0,0){$\rthetab_\repsilon$}}
\put(110,0){\vector(1,0){20}}

\put(140,0){\circle{20}}
\put(140,0){\makebox(0,0){$\repsilonb$}}
\put(150,0){\vector(1,0){20}}

\put(180,0){\circle{20}}
\put(180,0){\makebox(0,0){$\bgb$}}

\put(180,40){\circle{20}}
\put(180,40){\makebox(0,0){$\rfb$}}
\put(180,30){\vector(0,-1){20}}
\put(188,22){\makebox(0,0){$\blue{\bHb}$}}

\put(215,45){\makebox(0,0){$\bDb$}}
\put(250,40){\circle{20}}
\put(250,40){\makebox(0,0){$\rzb$}}
\put(239,40){\vector(-1,0){47}}

\put(250,80){\circle{20}}
\put(250,80){\makebox(0,0){$\rthetab_\rz$}}
\put(250,70){\vector(0,-1){20}}

\put(296,72){\framebox(36,16){$\psi_{\theta_z}$}}
\put(296,80){\vector(-1,0){35}}

\put(62,37){$\left\{\barr{ll}
\bgb = \bHb \rfb + \repsilon
\\[2pt]
\rfb = \bDb \rzb + \rxi
\\[1pt]
\earr\right.$}

\put(30,72){\framebox(36,16){$\psi_{\theta_\xi}$}}
\put(66,80){\vector(1,0){24}}

\put(100,80){\circle{20}}
\put(100,80){\makebox(0,0){$\rthetab_\rxi$}}
\put(110,80){\vector(1,0){20}}

\put(140,80){\circle{20}}
\put(140,80){\makebox(0,0){$\rxib$}}
\put(146,74){\vector(1,-1){27}}


\end{picture}
\end{tabular}
\caption{From linear forward model, Equation~\eqref{Eq:LinearModel} to the Hierarchical Model: Sparsity via a transformation, i.e. $\bDb\rfb$ is sparse}
\label{Fig:FromForwardModelHierarchicalModel2}
\end{figure}

In many applications, the prior information concerning the unknown(s) of the model, namely the \textit{classical} linear forward model, Equation~\eqref{Eq:LinearModel}
\beq
\bgb = \bHb \rfb + \repsilonb
\label{Eq:LinearModel}
\eeq
used in inverse problems,  can be translated as the \textit{sparse} structure of the unknown(s) i.e. the $\rfb$ in Equation~\eqref{Eq:LinearModel}. 
In particular, the linear forward model expressed in Equation~\eqref{Eq:LinearModel} corresponds to many application such as signal deconvolution, image restoration, Computed Tomography (CT) image reconstruction, Fourier Synthesis (FS) inversion, microwave imaging~\cite{Nguyen1994} and~\cite{Feron2007}, ultrasound echography, seismic imaging, radio astronomy ~\cite{kuruoglu2004source} fluorescence imaging, inverse scattering~\cite{Carfantan1997},~\cite{Feron2005c},~\cite{Ayasso2010b} and~\cite{Gharsalli2013b}, Eddy current non destructive testing~\cite{Nikolova1996} or SAR imaging ~\cite{Achim2006}.  
In all these examples the common inverse problem is to estimate $\rfb$ from the observations of $\bgb$. In general, the inverse problems are ill-posed~\cite{Hadamard1901}, since the conditioning number of the matrix $\bHb$ is very high. This means that, in practice, the data $\bgb$ alone is not sufficient to define an unique and satisfactory solution. The interpretation of the linear forward model, Equation~\eqref{Eq:LinearModel} is presented in Figure~\eqref{Fig:InterpretationOfTheForwardModel}.
\vspace{+0.3cm}
When the Bayesian approach is considered, one way to build hierarchical models that are \textit{favouring} a sparse solution is to consider distributions that are known to enforce sparsity for the prior. Such an approach gives the possibility to estimate the hyperparameters of the hierarchical model, i.e. the associated variances for $\rfb$ and $\repsilonb$. A typical hierarchical model associated to the forward model Equation~\eqref{Eq:LinearModel} is presented in Figure~\eqref{Fig:FromForwardModelHierarchicalModel}.
\vspace{+0.3cm}
However, Figure~\eqref{Fig:FromForwardModelHierarchicalModel} presents an hierarchical model for direct sparsity, i.e. an hierarchical model that asumes the sparse structure of $\rfb$. In many applications, $\rfb$ is not sparse but can be expressed via a transformation $\bDb$ on a sparse structure $\rzb$. Evidently, when considering the transformation on the sparse structure, the uncertainties and modelling errors have to accounted, Equation~\eqref{Eq:SparsityViaTransformation}: 
\beq
\rfb = \bDb \rzb + \rxi
\label{Eq:SparsityViaTransformation}
\eeq
In this cases, a more general general Hierarchical model is presented in  Figure~\eqref{Fig:FromForwardModelHierarchicalModel2}.
\vspace{+0.3cm}
When referring to the strategy used in the Bayesian approach for searching sparse solution in the inverse problem context, we have used the word \textit{favouring}. It is important to mention that generally, the linear forward model, Equation~\eqref{Eq:LinearModel}, may have an infinite number of solutions. Using a sparsity enforcing prior to model $\rfb$ results in algorithms selecting sparse solutions, but this is possible only when the linear forward model, Equation~\eqref{Eq:LinearModel} is allowing such solutions. Therefore, for those type of algorithms there is \textit{no guarantee} for the sparse structure of the solution.  
\\
In this work we present three classes of sparsity enforcing priors and show how a hierarchical model can be build using these kind of priors. We discuss then the mechanism of sparsity enforcing and present the advantages of iterative algorithms using sparsity enforcing priors that can be expressed via conjugate priors. e place this work in the context of such heavy tailed distributions that in particular can be expressed via \textit{conjugate priors}.

\section{Sparsity enforcing priors via conjugate priors}
\label{Sec:SparsityEnforcingPriors}
This section presents some sparsity enforcing priors, namely heavy tailed distributions that can be expressed via conjugate priors. First, a brief presentation of the Bayesian approach in inverse problems, Subsection~\eqref{Subsec:GeneralPerspective}. The sparsity mechanism and its key factors that are to be considered when selecting a good sparsity enforcing prior is presented in Subsection~\eqref{Subsec:SparsityMechanism}. It is shown that not only the heavy-tailed form of the distribution is of great interest for enforcing sparsity but also the associated variance vector of the sparse structure plays a crucial role, for which we aim a specific behaviour: small variances associated with the zero or close to zero values and important variances for the other values. Subsection~\eqref{Subsec:StudentPrior} presents the Student-t distribution, expressed via a Normal variance mixture, using as the mixing distribution the Inverse Gamma distribution. In particular, it is shown that via those two conjugate priors a two parameters version of the Student-t distribution is obtained when no condition is imposed for the scale and shape parameters of the Inverse Gamma distribution, for which the corresponding variance can be decreased to any positive value, which is a crucial fact in the sparsity mechanism. The same Normal variance mixture is obtained if the Student-t distribution is viewed as a generalized Hyperbolic distribution. In Subsection~\eqref{Subsec:LaplacePrior} the Laplace distribution is expressed via two conjugate distributions as a Normal variance mixture where the mixing distribution is the Exponential distribution. The same Normal variance mixture is obtained if the Laplace distribution is viewed as a generalized Hyperbolic distribution. Another way to express the Laplace distribution using conjugate distributions is via a Normal inverse-variance mixture where the mixing distribution is the Inverse Gamma distribution. In Subsection~\eqref{Subsec:VarianceGammaPrior} the Variance-Gamma distribution is expressed via two conjugate distributions as a Normal mean-variance mixture where the mixing distribution is the Inverse Gamma distribution. The expression is obtained using the generalized Hyperbolic distribution. In Subsection~\eqref{Subsec:NormalInverseGaussianPrior} the Normal-Inverse Gaussian distribution is expressed via two conjugate distributions as a Normal mean-variance mixture where the mixing distribution is the Inverse Gaussian distribution. 

\subsection{General Perspective: from the forward model to inversion via a Bayesian approach}
\label{Subsec:GeneralPerspective}
\begin{figure}[!htb]
\center
\tikzset{level 1 concept/.append style={font=\sf, sibling angle=120, level distance = 5.2cm, scale=0.9}}
\tikzset{level 2 concept/.append style={font=\sf, sibling angle=60,level distance = 2.6cm, scale=0.9}}
\begin{tikzpicture}[mindmap, grow cyclic, every node/.style=concept, concept color=orange!90, text=white]
\node{\textbf{Hierarchical Model}}
   child [concept color=blue!80] { node {\textbf{Likelihood: $p(\gb|\fb)$ from $p(\epsilonb)$}}
    }
   child [concept color=hollywoodcerise!80] { node {\textbf{Hyperparam: $p(\thetab)$}}
    }
   child [concept color=purple!80] { node {\textbf{Sparsity Prior: $p(\fb)$}}
    };
\end{tikzpicture}
        \caption{Hierarchical Model: probability density functions assigned for the unknowns}
\label{Fig:MindMapHierarchicalModel}
\end{figure}
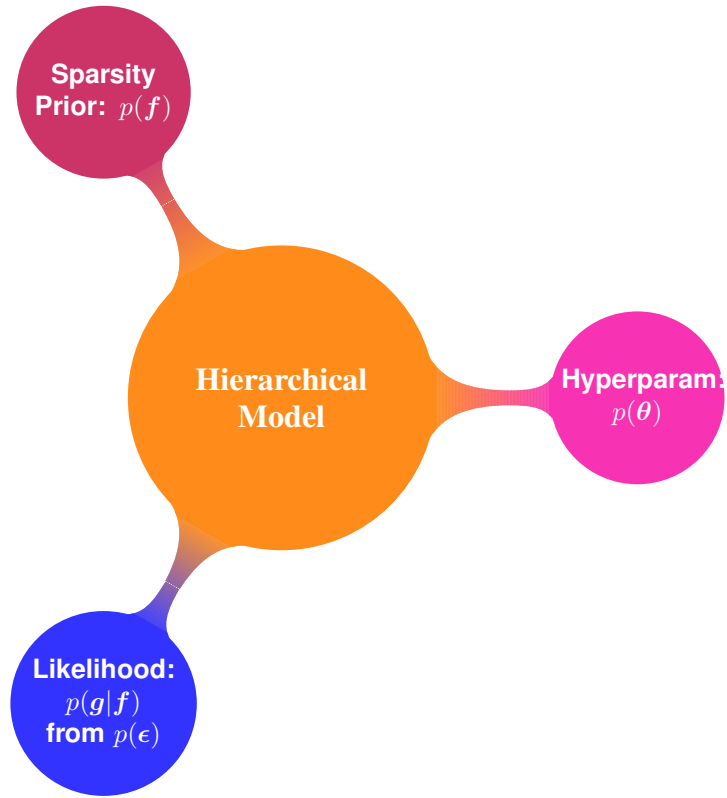

The strategy adopted for doing the inversion in Equation~\eqref{Eq:LinearModel} (or an equivalent one) is to build an hierarchical model, accounting for the available prior informations (i.e. accounting for the sparsity when the prior information concerning $\rfb$ is its sparse structure) and also accounting for the particularities of the errors and uncertainties of the model, modelled via $\repsilonb$ based on a Bayesian approach. Considering the linear forward model, Equation~\eqref{Eq:LinearModel}, the Bayesian inference is based on the fundamental relation given by the Bayes rule:
\beq
p(\rfb|\bgb,\thetab_\epsilon,\thetab_f)
=\frac{p(\bgb|\rfb,\thetab_\epsilon)\, p(\rfb|\thetab_f)}{p(\bgb|\thetab_\epsilon,\thetab_f)}, \thetab=(\thetab_\epsilon,\thetab_f)
\label{Eq:BayesRule}
\eeq
where $\thetab$ represents the hyperparameters appearing in the hierarchical model (namely the variances $\rvb_\rf$ and $\rvb_\repsilon$ associated with the unknowns of the linear forward model, $\rfb$ and $\repsilonb$. The Bayes rule can be interpreted as a proportionality relation between the posterior law and the product of the prior law (the prior information, sparsity in our case) and the likelihood:
\beq
p(\rfb|\bgb, \thetab_\epsilon, \thetab_f)
\propto p(\bgb|\rfb,\thetab_\epsilon)\, p(\rfb|\thetab_f)
\label{BayesProportionality}
\eeq
Generally, the likelihood is obtained via the linear forward model, Equation~\eqref{Eq:LinearModel} and the distribution considered for modelling the errors and the uncertainties $\repsilonb$. More details on this, in Section~\eqref{Sec:UncertaintiesModels}. An extension of Equation~\eqref{BayesProportionality} is the general Bayesian Inference, where the hyperparameters $\thetab=(\thetab_\epsilon,\thetab_f)$ from Equation~\eqref{BayesProportionality} are considered to be unknown and are to be estimated, along with the unknowns of the forward model, Equation~\eqref{Eq:LinearModel}:
\beq
p(\rfb,\rthetab_\repsilon, \rthetab_\rf|\bgb)
\propto p(\bgb|\rfb,\rthetab_\repsilon)\, p(\rfb|\rthetab_\rf)\, p(\rthetab_\repsilon|\psi_{\theta_\epsilon})\, p(\rthetab_\rf|\psi_{\theta_f})\,p(\psi_{\theta_\epsilon})\, p(\psi_{\theta_f})
\label{Bayes2}
\eeq
For the case when the sparsity appears via a transformation, the forward linear model Equation~\eqref{Eq:LinearModel} and Equation~\eqref{Eq:SparsityViaTransformation} are considered (Figure~\eqref{Fig:FromForwardModelHierarchicalModel2}):
\beq
p(\rfb,\rthetab_\repsilon, \rthetab_\rxi, \rthetab_\rz | \bgb)
\propto
p(\bgb | \rfb, \rthetab_\repsilon)\, 
p(\rfb | \rzb, \rthetab_\rxi)\,
p(\rzb | \rthetab_\rz)\,
p(\rthetab_\repsilon | \psi_{\theta_\epsilon})\,
p(\rthetab_\rxi | \psi_{\theta_\xi})\,
p(\rthetab_\rz | \psi_{\theta_z})\,
p(\psi_{\theta_\epsilon})\,
p(\psi_{\theta_\xi})\,
p(\psi_{\theta_z})
\tag{\ref{Bayes2}bis}
\label{Bayes2_IS}
\eeq
Evidently, considering the general Bayesian Inference implies assigning distributions for the hyperparameters, i.e., assigning $p(\rthetab|\psi)$. The set of distribution assigned for the prior, the likelihood and for the hyperparameters represents the hierarchical model, Figure~\eqref{Fig:MindMapHierarchicalModel}. 
The choices of the distributions are done in accordance with the application and with the available prior informations. In particular, when the prior information is the sparse structure of the unknown $\rfb$ of the forward model Equation~\eqref{Eq:LinearModel} sparsity enforcing priors will be used. From the hierarchical model the posterior distribution is obtained via Equation~\eqref{Bayes2} from which the unknowns and the hyperparameters can be estimated. Another representation of an hierarchical model built on the linear forward model, Equation~\eqref{Eq:LinearModel} is presented in Figure~\eqref{Fig:MindMapHierarchicalModel}.
We will see that when sparsity enforcing priors that can be expressed via conjugate distributions laws are considered, analytical expressions can be obtained for the unknowns of the hierarchical model. This is a great advantage and the fundamental reason for the great interest of such priors.

\subsection{Iterative algorithms sparsity mechanism}
\label{Subsec:SparsityMechanism}
In those type of approaches, the mechanism of sparsity is based not only on the heavy tailed property of the prior distribution (or its property to induce sparsity) but also on a particular behaviour of the associated variances. In such an approach a bivariate prior is set for the unknown of the model that needs to be estimated  and for the corresponding variance. The algorithm that results is an iterative one, updating at every iteration both the unknown of the model and the corresponding variance. In order to obtain a sparse solution for the unknowns, the structure of the variance must be sparse itself. In particular the variances associated with the zero or close to zero points from the unknown of the model must be small, and the variances associated with the non-zero elements of the sparse unknowns of the model must be significant.   
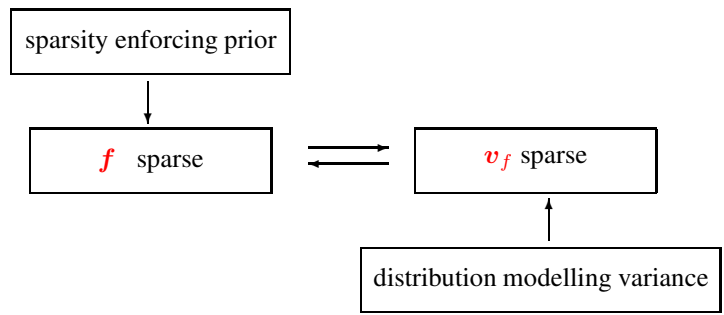
\begin{figure}[!htb]
\center
\begin{picture}(450,130)
\put(99,96){\framebox(104,24){sparsity enforcing prior}}
\put(150,93){\vector(0,-1){16}}
\put(106,51){\framebox(90,24){$\rfb$ \; sparse}}
\put(210,68){\vector(1,0){30}}
\put(240,62){\vector(-1,0){30}}
\put(250,51){\framebox(90,24){$\rvb_{\rf}$ sparse}}
\put(300,33){\vector(0,1){16}}
\put(230,6){\framebox(134,24){distribution modelling variance}}
\end{picture}
        \caption{Sparsity Mechanism: For direct sparsity applications}
\label{Fig:SparsityMechanism}
\end{figure}
Therefore, the parameters of the distribution modelling the variance (for example the shape and scale parameters of the Inverse Gamma distribution appearing in the conjugate prior models for sparsity enforcing via Student-t or Laplace) must be chosen such that the variance vector $\rvb_{\rf}$ is sparse, i.e. the expected value of the elements $\rv_{\rf_\rj}$ is close to zero, $\mbox{E}\left[ \rv_{\rf_\rj} \right] \searrow 0$. Furthermore, in a Bayesian approach, we may have a prior knowledge for the numerical value associated with the variance of the distribution modelling $\rv_{\rf_\rj}$, i.e. $\mbox{Var}\left[ \rv_{\rf_\rj} \right] = w$, where $w$ is the numerical value obtained via prior knowledge.  Evidently, depending on the parameters corresponding to the distribution modelling the variance, the behaviour of the marginal, i.e. the sparsity enforcing prior distribution. We note that in order to select a prior model that enforces sparsity, those parameters must be chosen such that the prior distribution expressed via conjugate prior distributions, modelling the unknown of the linear forward model, Equation~\eqref{Eq:LinearModel}, is concentrated around zero, i.e. it's variance is very small, $\mbox{Var}_{Prior}\left[\rf_\red{kj}\right] \searrow 0$.

\subsection{Student-t prior: expressed via conjugate priors: Normal and Inverse Gamma}
\label{Subsec:StudentPrior}
\begin{figure}[!htb]
        \centering
        \begin{subfigure}{0.49\textwidth}
          		\centering        
                \includegraphics[width=7.4cm,height=4cm]{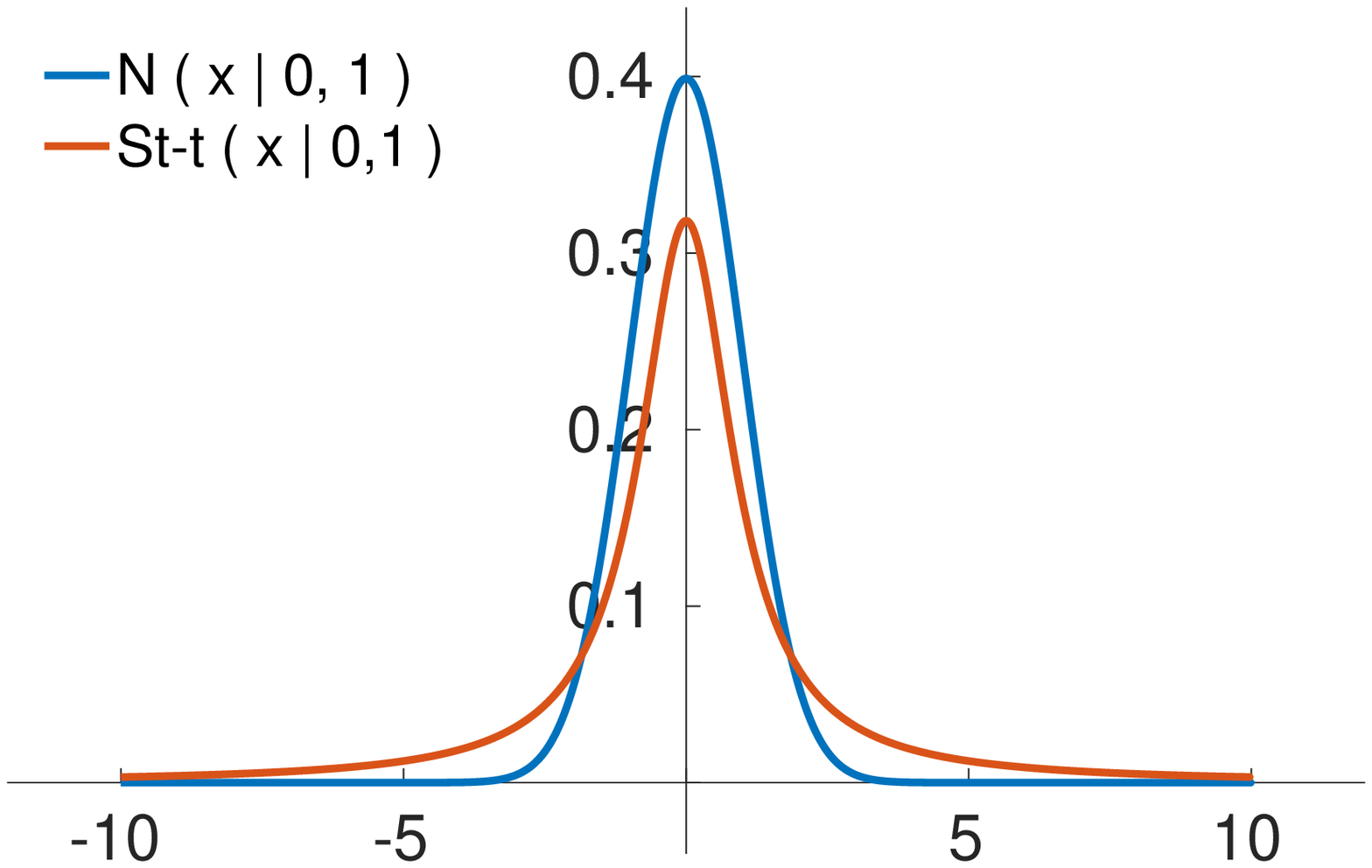}
                \caption{\textcolor{cobalt}{Normal} vs. \textcolor{cinnabar}{Student-t} distribution}
                \label{fig:StudentVsNormal2a}
        \end{subfigure}                               
        \begin{subfigure}{0.49\textwidth}
          		\centering        
                \includegraphics[width=7.4cm,height=4cm]{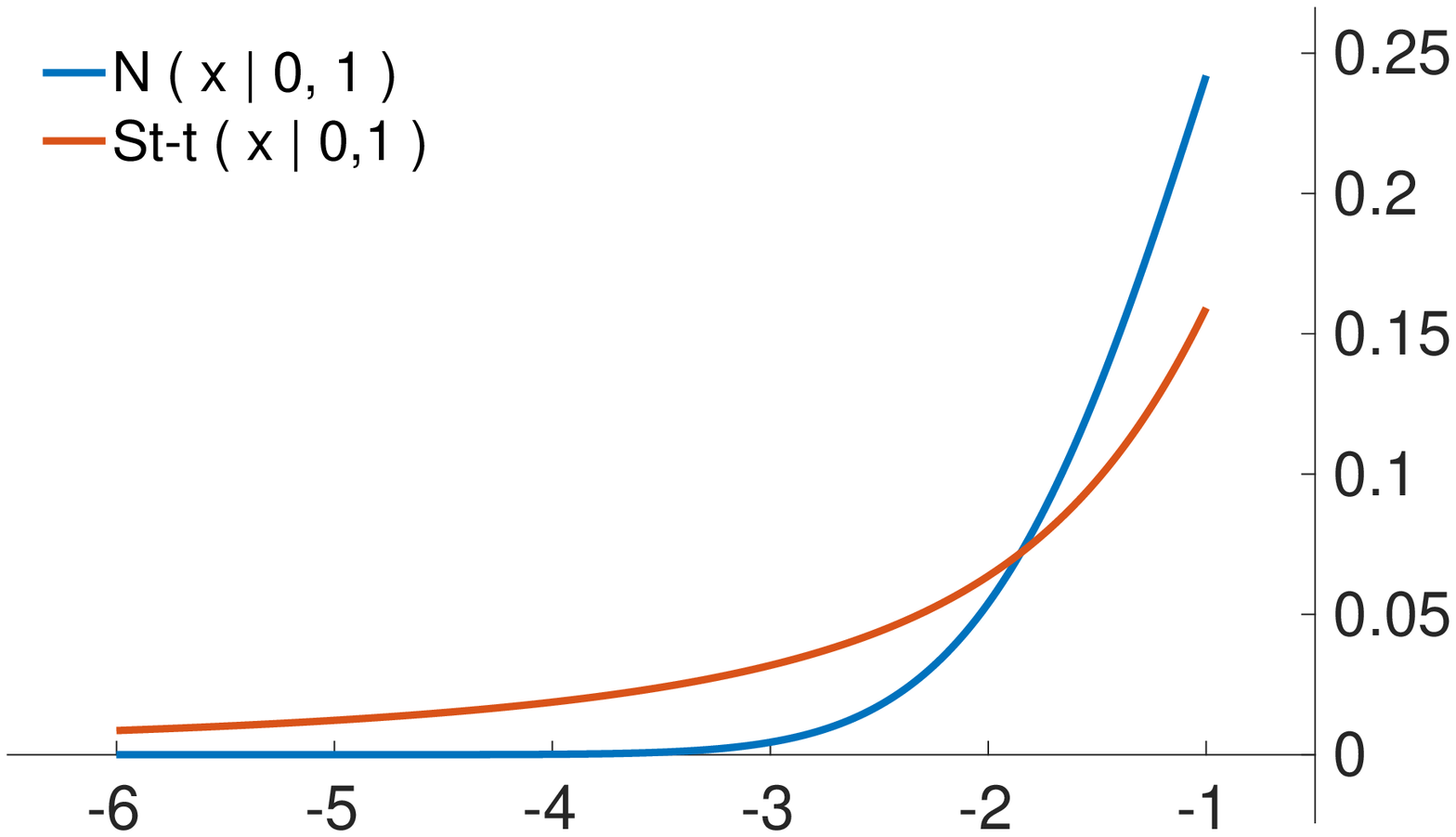}
                \caption{Heavy tailed property of the Student distribution}
                \label{fig:StudentVsNormal2b}
        \end{subfigure}                                         
        \caption{Comparison between the \textcolor{cobalt}{Normal} distribution $\Nc \left( x | 0, 1 \right)$ and \textcolor{cinnabar}{Student-t} distribution $\Sc t-t \left( x | 0, 1 \right)$.}
        \label{fig:StudentVsNormal2}
\end{figure}

In the following, the Student-t distribution is discussed. It is a sparsity enforcing prior because of its heavy-tailed form. It can be expressed via conjugate priors as the marginal of a Normal variance mixture distribution, with the mixing distribution an Inverse Gamma distribution. The standard form, with one parameter $\nu$ representing the degrees of freedom is obtained when shape and scale parameters corresponding to the Inverse Gamma distribution are considered equal, $\alpha = \beta = \frac{\nu}{2}$. When this equality is not imposed, a two parameters Student-t distribution is obtained, which is of great importance in the context of sparsity enforcing since in this case, the corresponding variance (which generally needs to be small) can take any positive values.
In probability and statistics, Student's t-distribution (or simply the t-distribution) is any member of a family of continuous probability distributions that arises when estimating the mean of a normally distributed population in situations where the sample size is small and population standard deviation is unknown. It was developed by William Sealy Gosset under the pseudonym Student. Whereas a normal distribution describes a full population, t-distributions describe samples drawn from a full population; accordingly, the t-distribution for each sample size is different, and the larger the sample, the more the distribution resembles a normal distribution. The t-distribution plays a role in a number of widely used statistical analyses, including Student's t-test for assessing the statistical significance of the difference between two sample means, the construction of confidence intervals for the difference between two population means, and in linear regression analysis. The Student's t-distribution also arises in the Bayesian analysis of data from a normal family.\\
A random variable has a Student-t$(x|\nu)$ distribution if its probability density function is:
\beq
\Sc t-t(x|\nu) = \frac{\Gamma\left(\frac{\nu+1}{2} \right)}{\sqrt{\pi\nu}\Gamma\left(\frac{\nu}{2} \right)} \left( 1 + \frac{x^2}{\nu} \right)^{\left(-\frac{\nu + 1}{2}\right)}, \; \nu \in \mathbb{R_+}, \; \Gamma \textrm{ representing the Gamma function};\eeq
The comparison between the standard Normal distribution $\Nc \left( x | 0, 1 \right)$ and the standard Student-t distribution $\Sc t-t \left( x | 0, 1 \right)$ is presented in Figure~\eqref{fig:StudentVsNormal2}.
\subsubsection{Student-t distribution via Normal variance mixture}
\label{Subsubsec:StudenttViaIGSM}
For the linear model expressed in Equation~\eqref{Eq:LinearModel}, the Student-t distribution can be used as a prior via the Student-t Prior Model (StPM), Equation~\eqref{Eq:StPM1}, which considers a zero-mean Normal distribution for $\rf_\rj|\rv_{\rf_\rj}$, with the variance $\rv_{\rf_\rj}|\alpha, \beta$  modelled as an Inverse Gamma distribution, with the corresponding shape and scale parameters, $\alpha$ and $\beta$:
\beq
\textbf{StPM:}\;\;\;\;\;\;\;
\left\{\barr{ll}
p(\rf_\rj|0,\rv_{\rf_\rj}) = \Nc(\rf_\rj|0, \; \rv_{\rf_\rj}) = \left( 2 \pi \right)^{-\frac{1}{2}} \; \rv_{\rf_\rj}^{-\frac{1}{2}} \; \exp \left\lbrace -\frac{\rf_\rj^2}{2\rv_{\rf_\rj}} \right\rbrace
\\[7pt]
p(\rv_{\rf_\rj}|\alpha, \beta) = \Ic\Gc(\rv_{\rf_\rj}|\alpha, \beta) = \frac{\beta^{\alpha}}{\Gamma(\alpha)} \; \rv_{\rf_\rj}^{-\alpha-1} \; \exp \left\lbrace -\frac{\beta}{\rv_{\rf_\rj}} \right\rbrace,\\[4pt]
\earr\right.
\label{Eq:StPM1}
\eeq
The expression of the joint probability distribution $\rf_\rj,\rv_{\rf_\rj}|\alpha, \beta$ is a bivariate Normal-Inverse Gamma distribution, Equation~\eqref{Eq:JointStPM1},
\beq
p(\rf_\rj,\rv_{\rf_\rj}|\alpha, \beta) = \Nc(\rf_\rj|0, \; \rv_{\rf_\rj}) \; \Ic\Gc(\rv_{\rf_\rj}|\alpha, \beta),
\label{Eq:JointStPM1}
\eeq 
and the marginal $p(\rf_\rj|\alpha, \beta)$ is, Equation~\eqref{Eq:MargStPM1}:
\beq
p(\rf_\rj|\alpha, \beta)
=
\left( 2 \pi \right)^{-\frac{1}{2}} \;
\frac{\beta^{\alpha}}{\Gamma(\alpha)} \;
\int \rv_{\rf_\rj}^{-\frac{1}{2}} \; \exp \left\lbrace -\frac{\rf_\rj^2}{2\rv_{\rf_\rj}} \right\rbrace  \rv_{\rf_\rj}^{-\alpha-1} \; \exp \left\lbrace -\frac{\beta}{\rv_{\rf_\rj}} \right\rbrace  \d \rv_{\rf_\rj}.
\label{Eq:MargStPM1}
\eeq 
In Equation~\eqref{Eq:MargStPM1} an $\Ic\Gc$ distribution can be identified inside the integral, 
\beq
I = \int \rv_{\rf_\rj}^{-\left(\alpha+\frac{1}{2}\right)-1} \; \exp \left\lbrace -\frac{\beta+\frac{\rf_\rj^2}{2}}{\rv_{\rf_\rj}}  \right\rbrace \; \d \rv_{\rf_\rj}
= 
\frac{\Gamma(\alpha+\frac{1}{2})}{{\left( \beta + \frac{\rf_\rj^2}{2} \right)}^{\left(\alpha+\frac{1}{2}\right)}}
\underbrace{\int \Ic\Gc\left( \rv_{\rf_\rj} | \alpha+\frac{1}{2}, \beta + \frac{\rf_\rj^2}{2} \right) \; \d \rv_{\rf_\rj}}_{1},
\label{Eq:StInt}
\eeq 
so, the marginal from Equation~\eqref{Eq:MargStPM1} is two-parameters Student-t distribution:
\beq
p(\rf_\rj|\alpha, \beta) = \left( 2 \pi \right)^{-\frac{1}{2}} \; \frac{\beta^{\alpha}}{\Gamma(\alpha)} \; \frac{\Gamma(\alpha+\frac{1}{2})}{{\left( \beta + \frac{\rf_\rj^2}{2} \right)}^{\left(\alpha+\frac{1}{2}\right)}} = \left( 2 \pi \beta \right)^{-\frac{1}{2}} \; \frac{\Gamma(\alpha+\frac{1}{2})}{\Gamma(\alpha)} \; \left( 1 +\frac{\rf_\rj^2}{2\beta} \right)^{-\left(\alpha + \frac{1}{2} \right)} 
\label{Eq:MargStPM1_1}
\eeq 
If $\alpha = \beta = \frac{\nu}{2}$, from Equation~\eqref{Eq:MargStPM1_1} we can conclude that the marginal is the Student-t distribution, $p(\rf_\rj|\alpha = \frac{\nu}{2}, \beta = \frac{\nu}{2}) = \Sc t \left( \rf_\rj | \nu \right)$:
\beq
p(\rf_\rj|\alpha=\frac{\nu}{2},\beta=\frac{\nu}{2}) 
=
\frac{\Gamma\left(\frac{\nu+1}{2} \right)}{\sqrt{\pi\nu}\Gamma\left(\frac{\nu}{2} \right)} \left( 1 + \frac{\rf_\rj^2}{\nu} \right)^{\left(-\frac{\nu + 1}{2}\right)}
=
\Sc t \left( \rf_\rj | \nu \right)
\eeq
Using the Student-t Prior Model (StPM), Equation~\eqref{Eq:StPM1}, the Student-t distribution can be used as a sparsity enforcing prior for the forward linear model, Equation~\eqref{Eq:LinearModel}. The interest of expressing the prior distribution via prior models using conjugate priors is the possibility to compute analytical expressions for the unknown estimates.
\\
Section~\eqref{Sec:HierarchicalModelsWithStudent-tPriorViaStPM} presents the developments of hierarchical models based on Student-t prior distribution for the linear forward models Figure~\eqref{Fig:FromForwardModelHierarchicalModel} and Figure~\eqref{Fig:FromForwardModelHierarchicalModel2}.
\subsubsection{Student-t distribution via the Generalized Hyperbolic distribution}
\label{Subsec:StudenttViaGH}
\begin{figure}[!htb]
        \centering
        \begin{subfigure}{0.49\textwidth}
          		\centering
                \includegraphics[width=7.4cm,height=4cm]{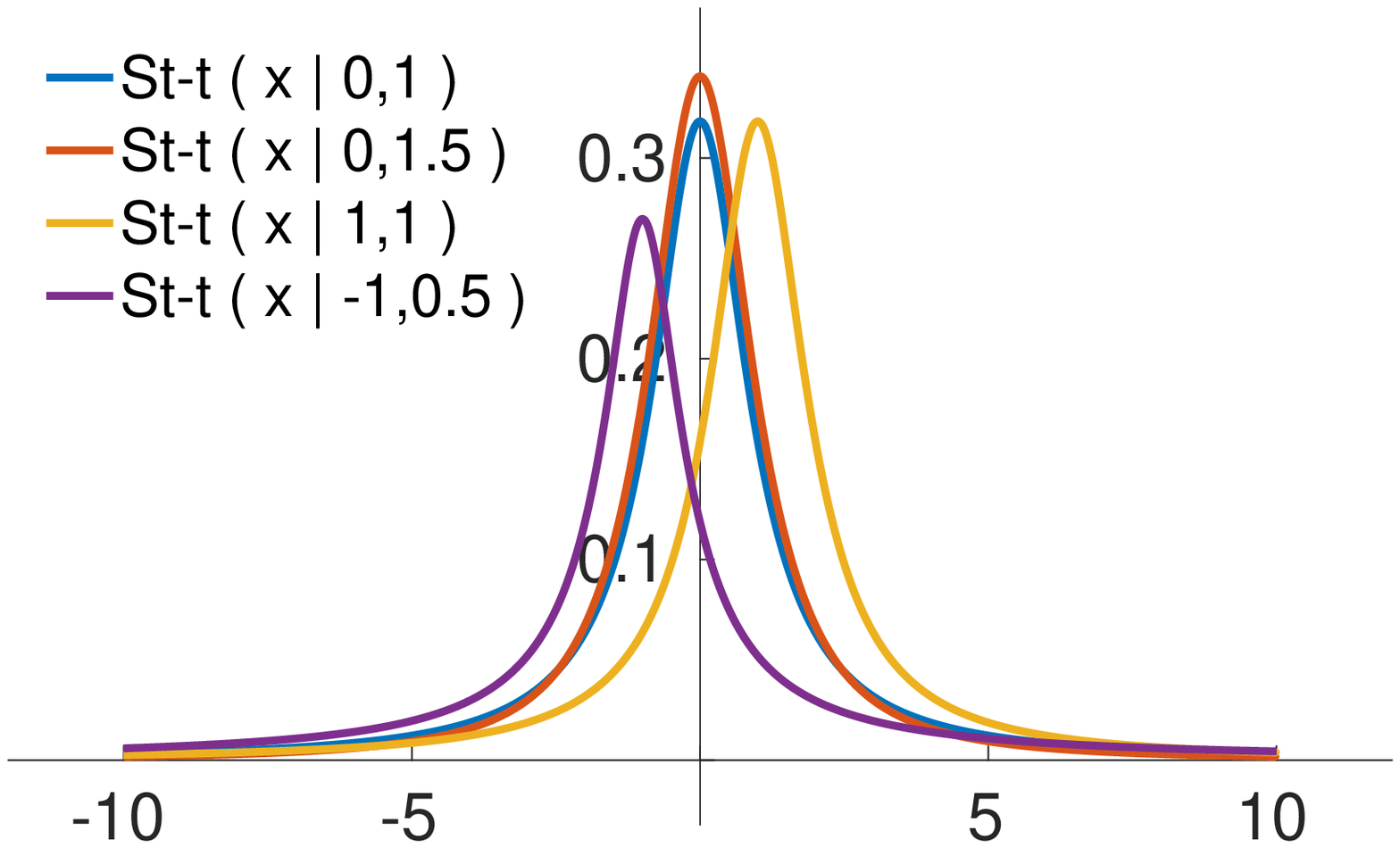}
                \caption{$ \Sc t-t \left( x | \mu , \nu \right) $}
                \label{fig:St_GenHypAsMStu}
        \end{subfigure}
        \begin{subfigure}{0.49\textwidth}
          		\centering        
                \includegraphics[width=7.4cm,height=4cm]{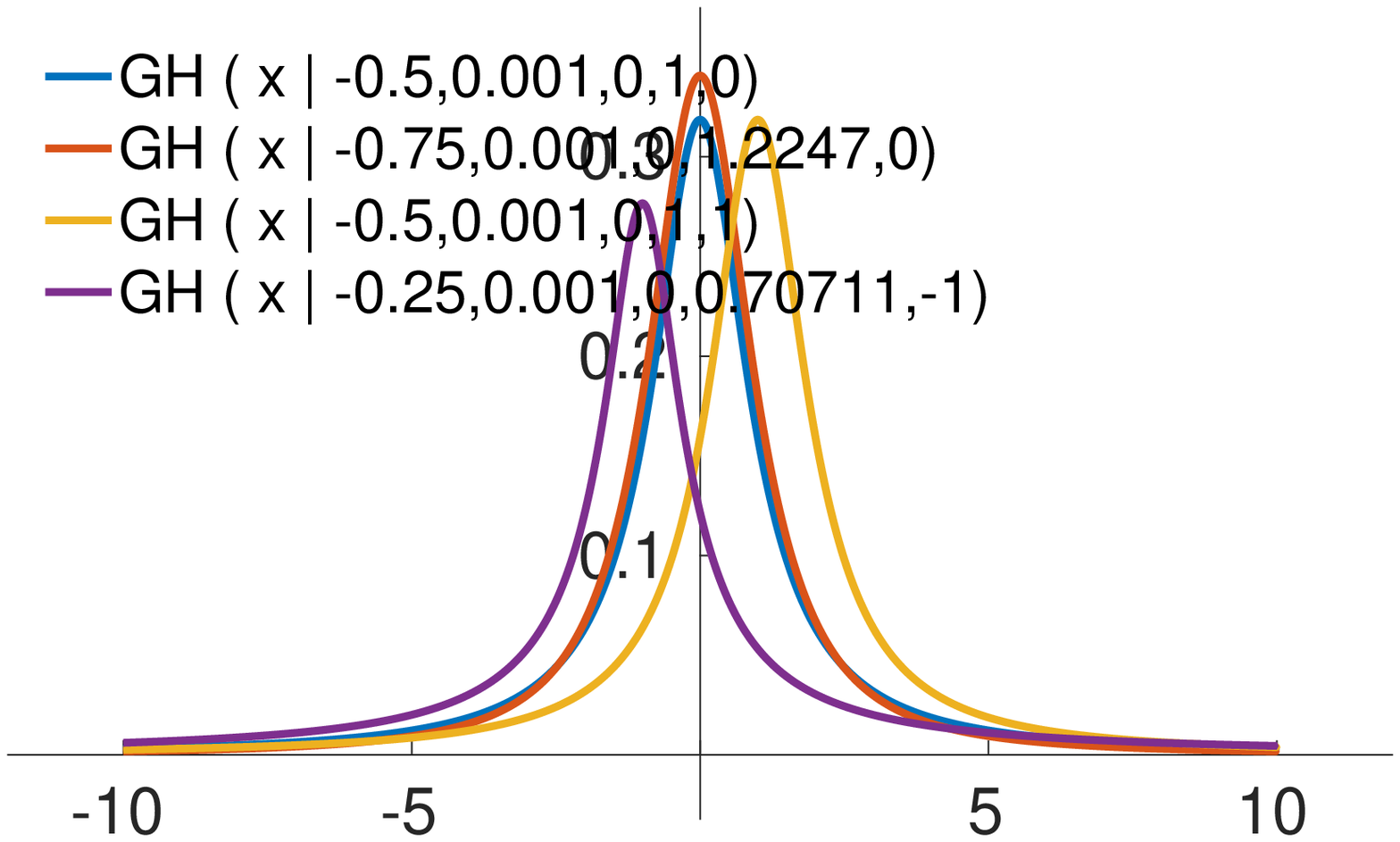}
                \caption{\small{$ \Gc\Hc \left( x | \lambda = -\frac{\nu}{2}, \alpha = 0.001, \beta = 0, \delta = \sqrt{\nu}, \mu=0 \right) $}}
                \label{fig:GHSt_GenHypAsMStu}
        \end{subfigure}        
\par\bigskip       
        \begin{subfigure}{0.49\textwidth}
          		\centering        
                \includegraphics[width=7.4cm,height=4cm]{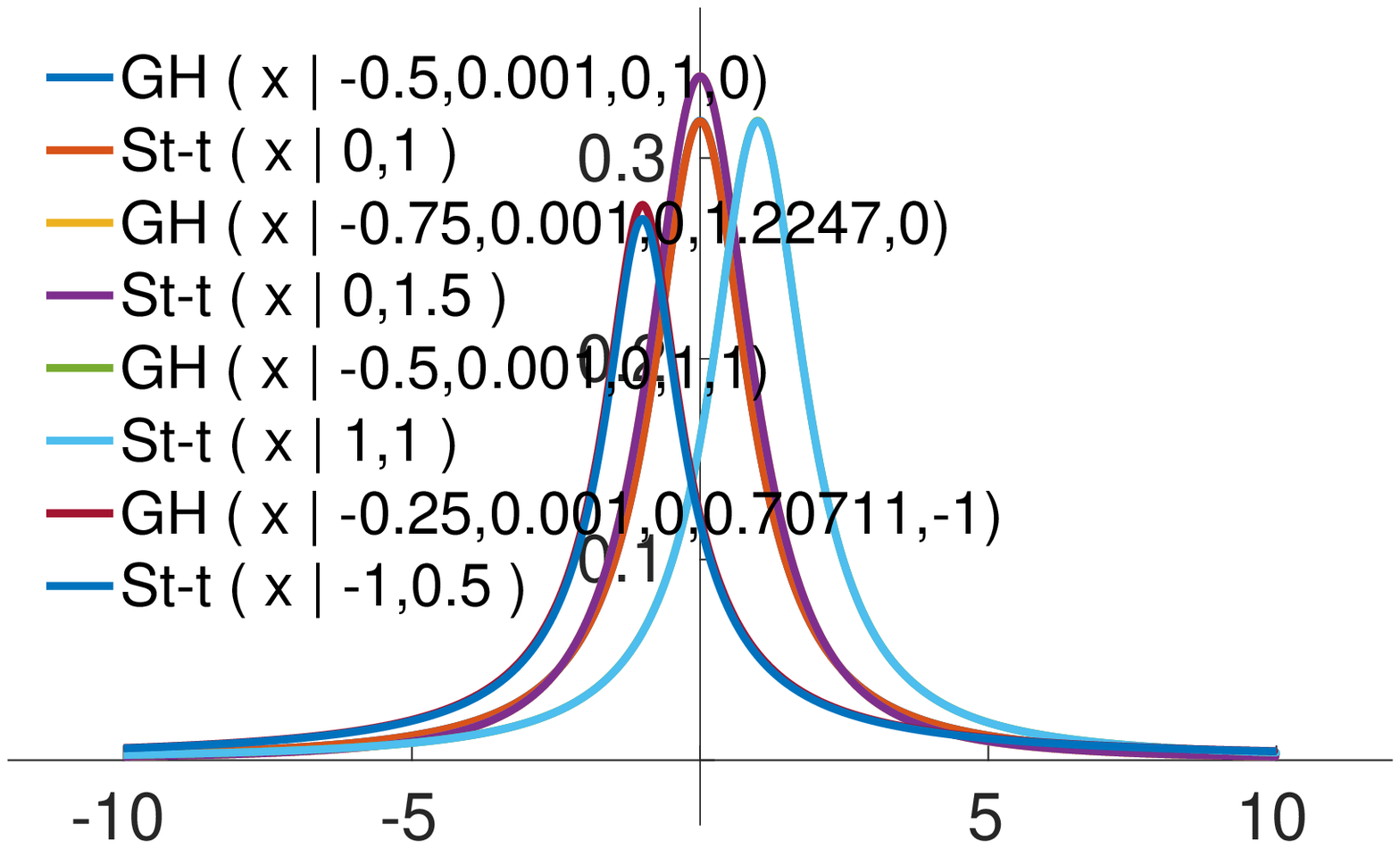}
                \caption{$ \Sc t-t $ vs. $ \Gc\Hc $}
                \label{fig:GHStVsSt_GenHypAsMStu}
        \end{subfigure}                               
        \begin{subfigure}{0.49\textwidth}
          		\centering        
                \includegraphics[width=7.4cm,height=4cm]{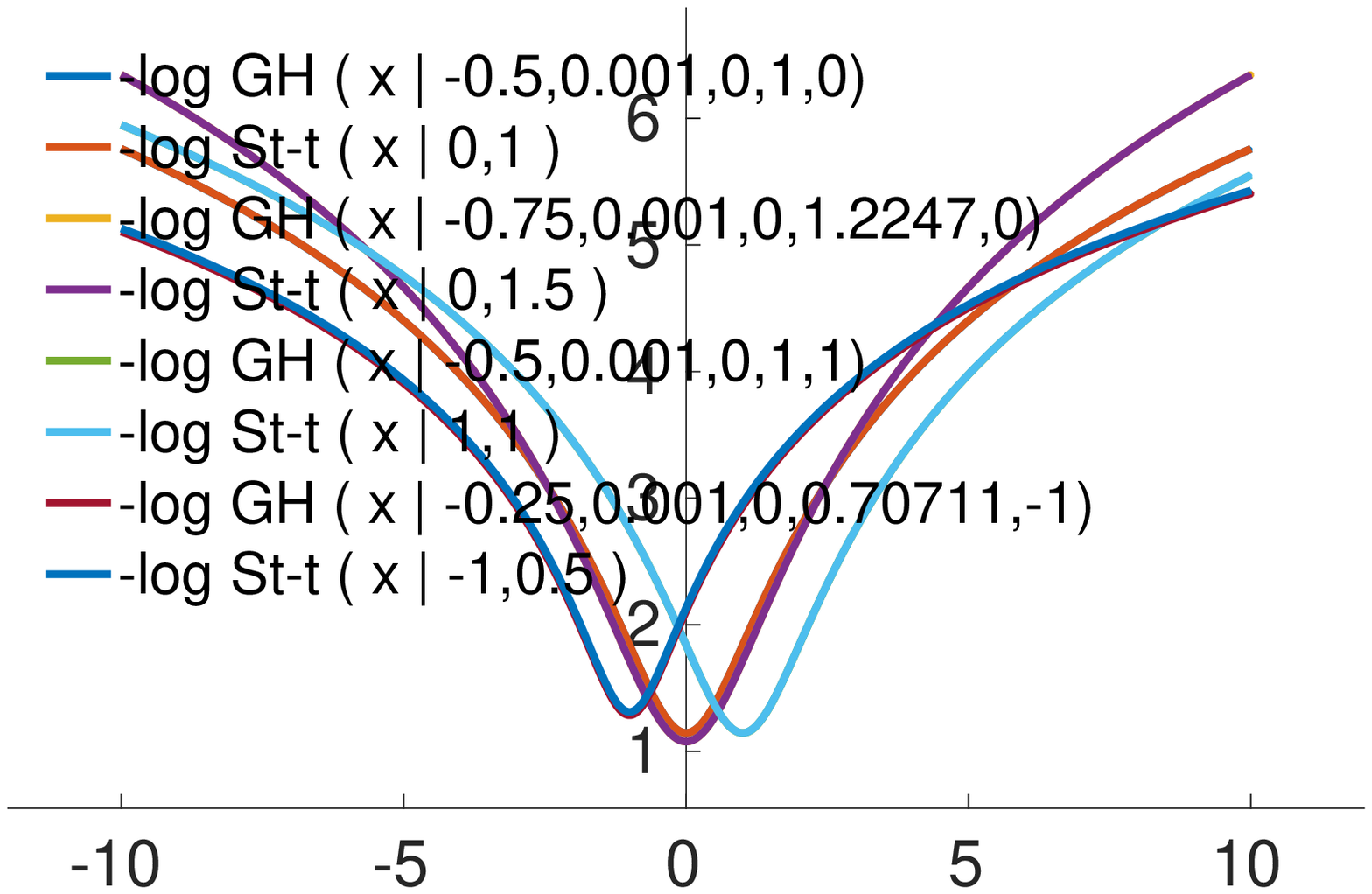}
                \caption{$ - \log \Sc t-t $ vs. $ - \log \Gc\Hc $}
                \label{fig:Log_GHStVsSt_GenHypAsMStu}
        \end{subfigure}                                         
        \caption{Four Student-t probability density functions with different means ($\mu=0$, blue and red, $\mu=1$, yellow and $\mu=-1$, violet) and different degrees of freedom ($\nu=1$, blue and yellow, $\nu=1,5$, red and $\nu=0.5$, violet), Figure~\eqref{fig:St_GenHypAsMStu} and the corresponding Generalized Hyperbolic probability density functions, with parameters set as in Equation~\eqref{Eq:GenHyperbolicDistributionAsStudent3}, with $\alpha = 0.001$, Figure~\eqref{fig:GHSt_GenHypAsStu}. Comparison between the distributions, $\Sc t-t$ vs. $\Gc\Hc$, Figure~\eqref{fig:GHStVsSt_GenHypAsStu} and between the logarithm of the distributions $\log \Sc t-t$ vs. $\log \Gc\Hc$, Figure~\eqref{fig:Log_GHStVsSt_GenHypAsStu}.}
        \label{fig:Comp_GenHypAsMStu}
\end{figure}

\begin{figure}[!htb]
        \centering
        \begin{subfigure}{0.49\textwidth}
          		\centering
                \includegraphics[width=7.4cm,height=4cm]{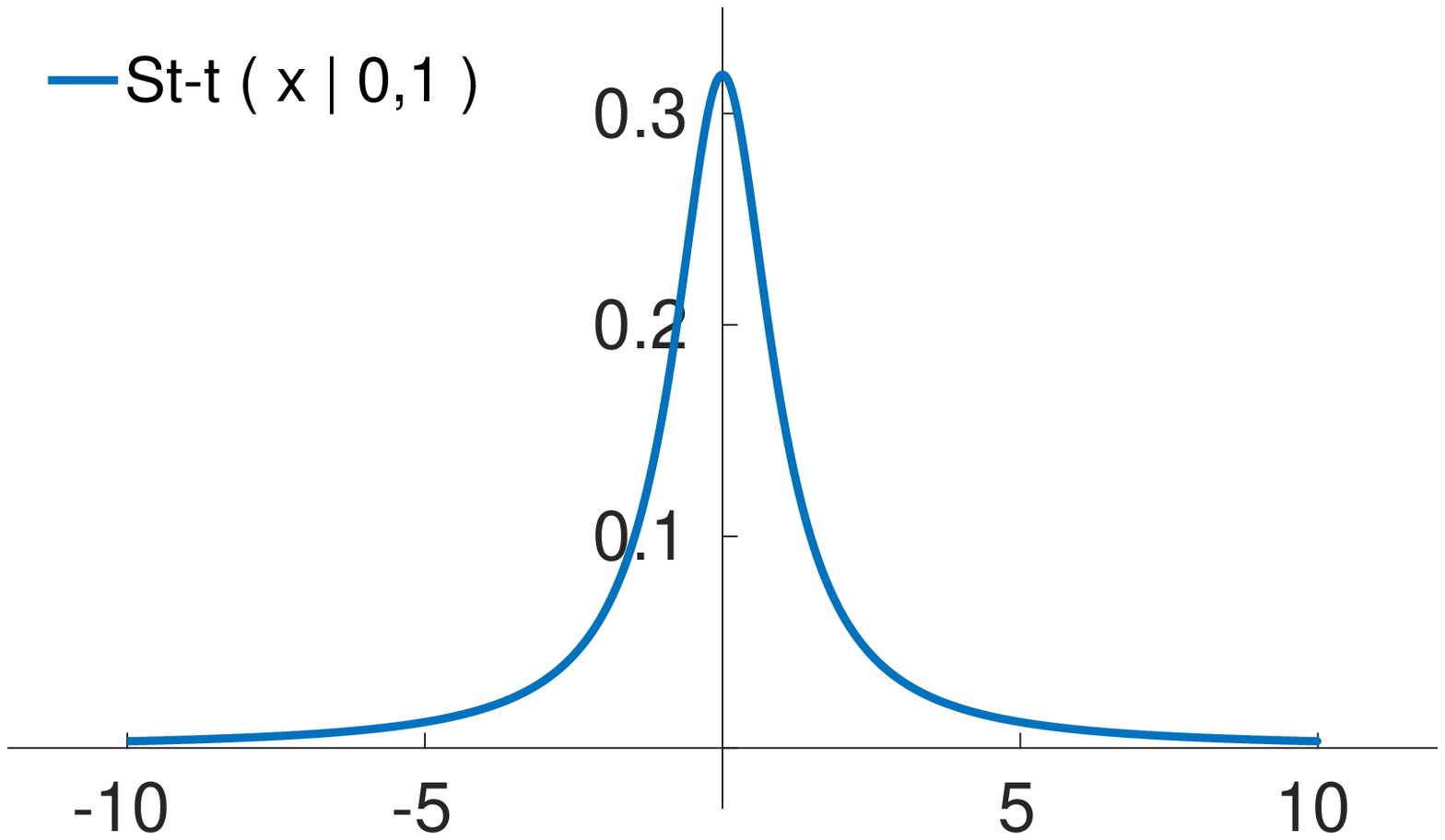}
                \caption{$ \Sc t-t \left( x | \mu = 0 , \nu = 1 \right) $}
                \label{fig:St_GenHypAsStu}
        \end{subfigure}
        \begin{subfigure}{0.49\textwidth}
          		\centering        
                \includegraphics[width=7.4cm,height=4cm]{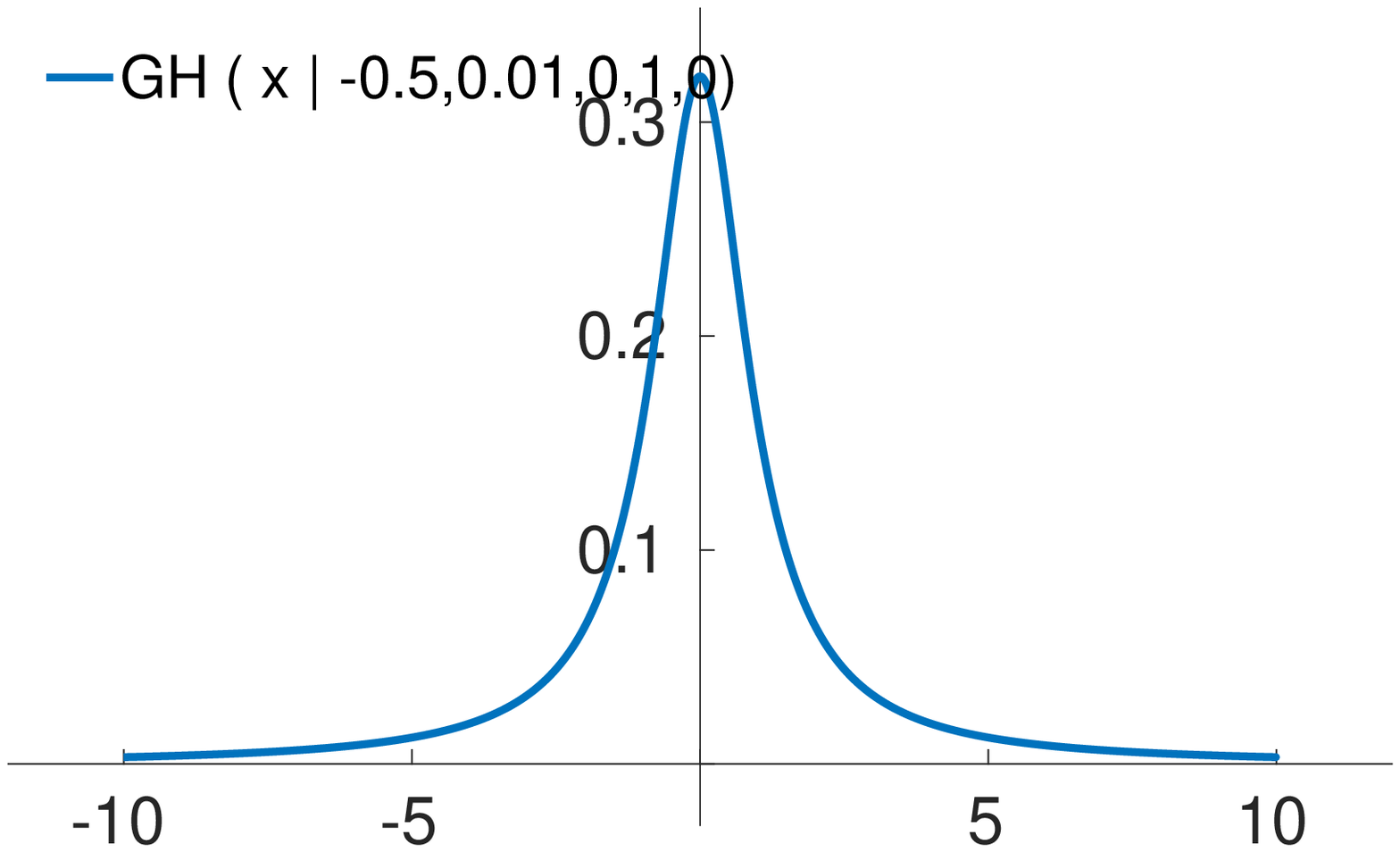}
                \caption{\small{$ \Gc\Hc \left( x | \lambda = -\frac{\nu}{2}, \alpha = 0.01, \beta = 0, \delta = \sqrt{\nu}, \mu=0 \right) $}}
                \label{fig:GHSt_GenHypAsStu}
        \end{subfigure}        
\par\bigskip       
        \begin{subfigure}{0.49\textwidth}
          		\centering        
                \includegraphics[width=7.4cm,height=4cm]{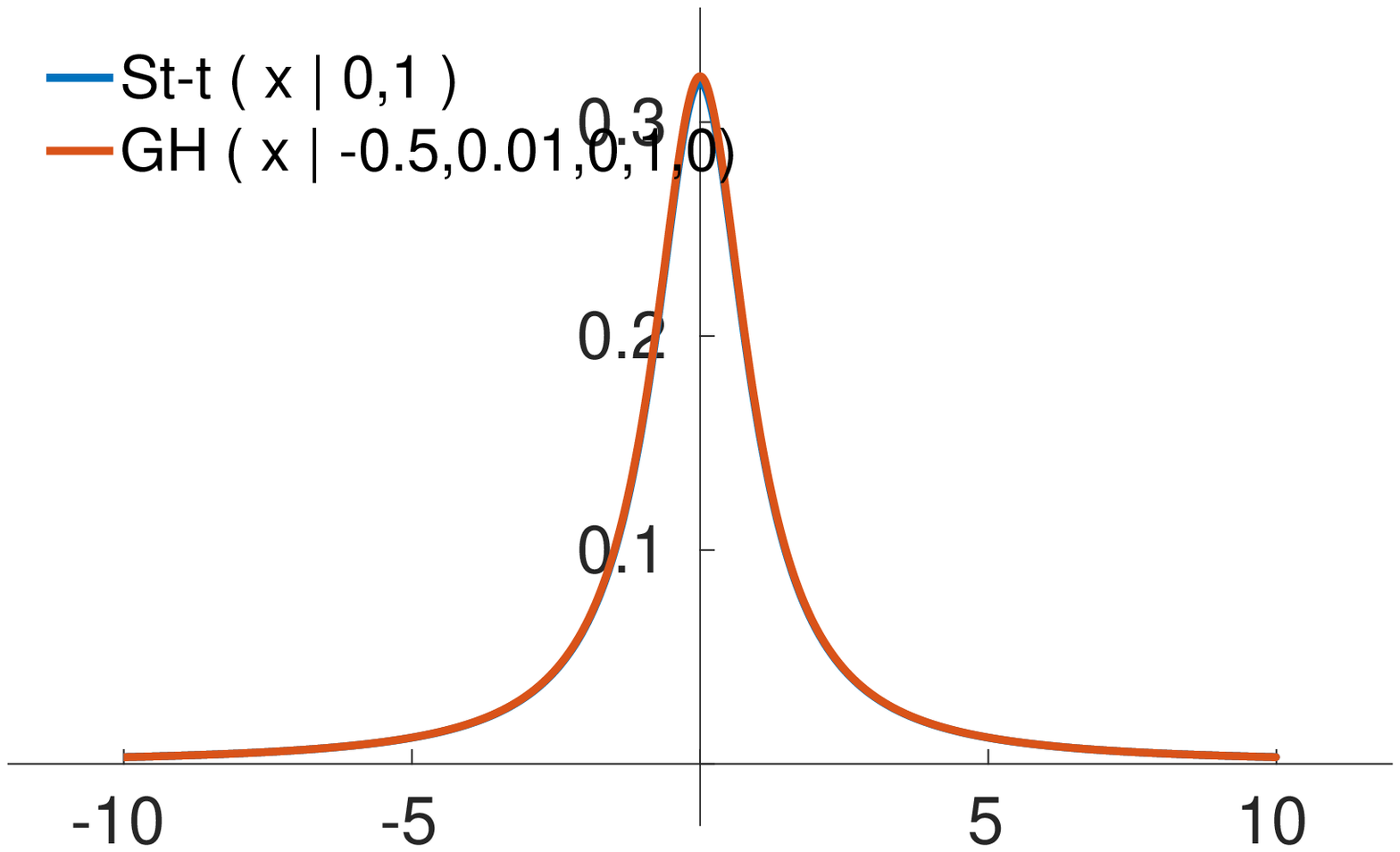}
                \caption{$ \Sc t-t $ vs. $ \Gc\Hc $}
                \label{fig:GHStVsSt_GenHypAsStu}
        \end{subfigure}                               
        \begin{subfigure}{0.49\textwidth}
          		\centering        
                \includegraphics[width=7.4cm,height=4cm]{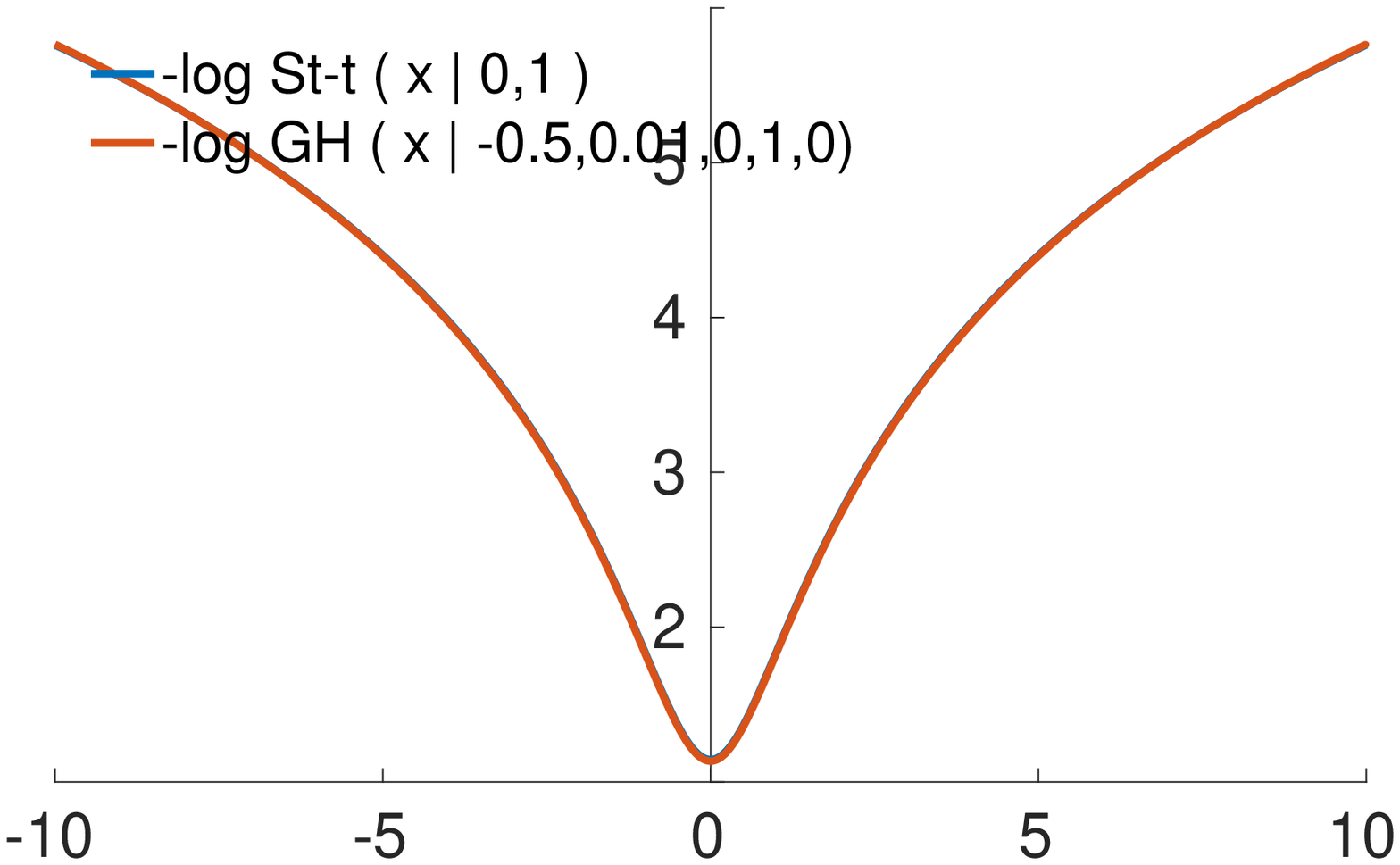}
                \caption{$ -\log \Sc t-t $ vs. $ -\log \Gc\Hc $}
                \label{fig:Log_GHStVsSt_GenHypAsStu}
        \end{subfigure}                                         
        \caption{The standard Student-t distribution ($\nu=1$), Figure~\eqref{fig:St_GenHypAsStu} and the Generalized Hyperbolic distribution, with parameters set as in Equation~\eqref{Eq:GenHyperbolicDistributionAsStudent3}, Figure~\eqref{fig:GHSt_GenHypAsStu}. Comparison between the two distribution, $\Sc t-t$ vs. $\Gc\Hc$, Figure~\eqref{fig:GHStVsSt_GenHypAsStu} and between the logarithm of the distributions $-\log \Sc t-t$ vs. $-\log \Gc\Hc$, Figure~\eqref{fig:Log_GHStVsSt_GenHypAsStu}.}
        \label{fig:Comp_GenHypAsStu}
\end{figure}

\begin{figure}[!htb]
        \centering
        \begin{subfigure}{0.49\textwidth}
          		\centering        
                \includegraphics[width=7.4cm,height=4cm]{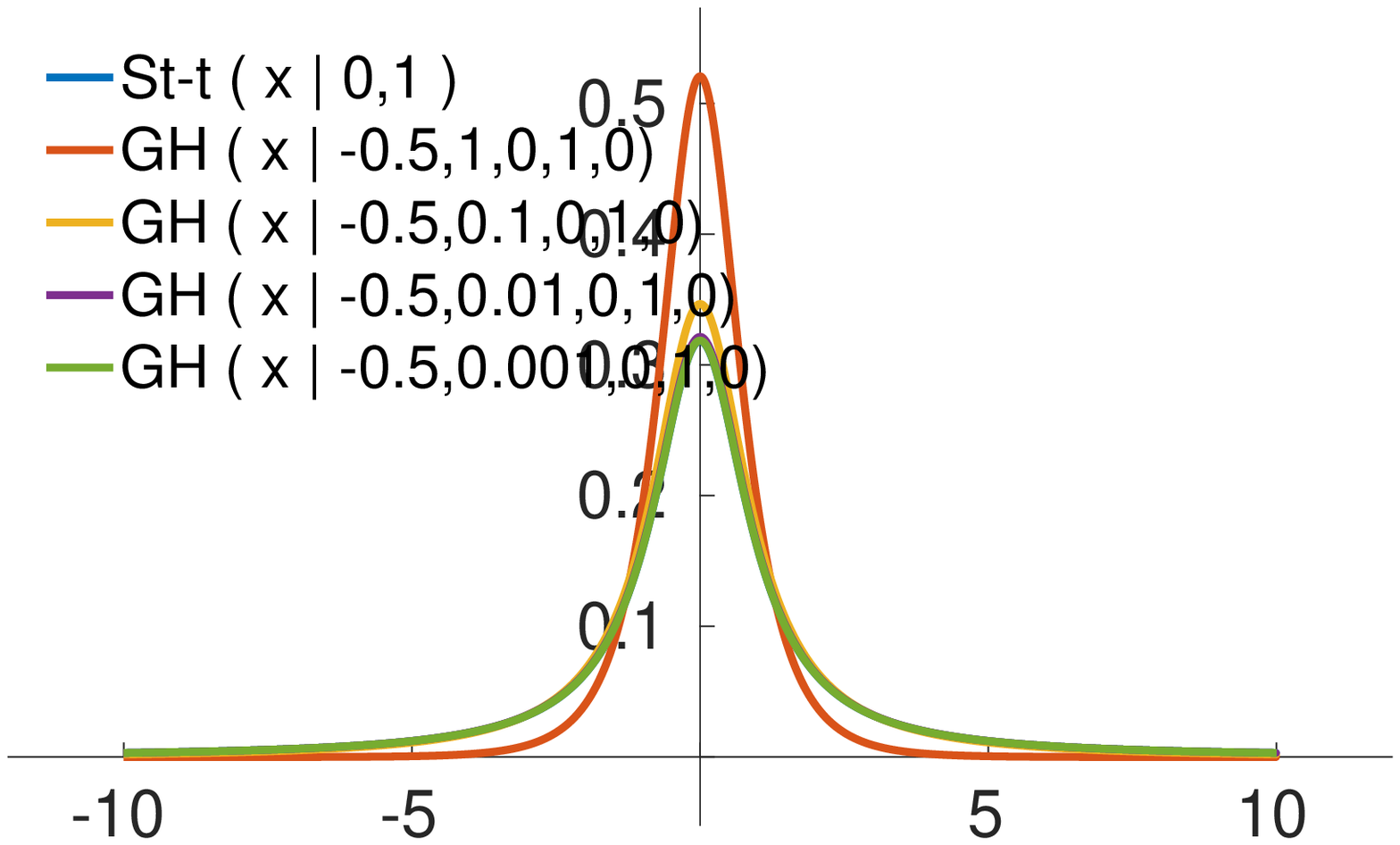}
                \caption{$ \Sc t-t $ vs. $ \Gc\Hc$ with $\alpha \in \left\lbrace 1, 0.1, 0.01, 0.001 \right\rbrace$}
                \label{fig:GHStVsSt_GenHypAsStu1}
        \end{subfigure}                               
        \begin{subfigure}{0.49\textwidth}
          		\centering        
                \includegraphics[width=7.4cm,height=4cm]{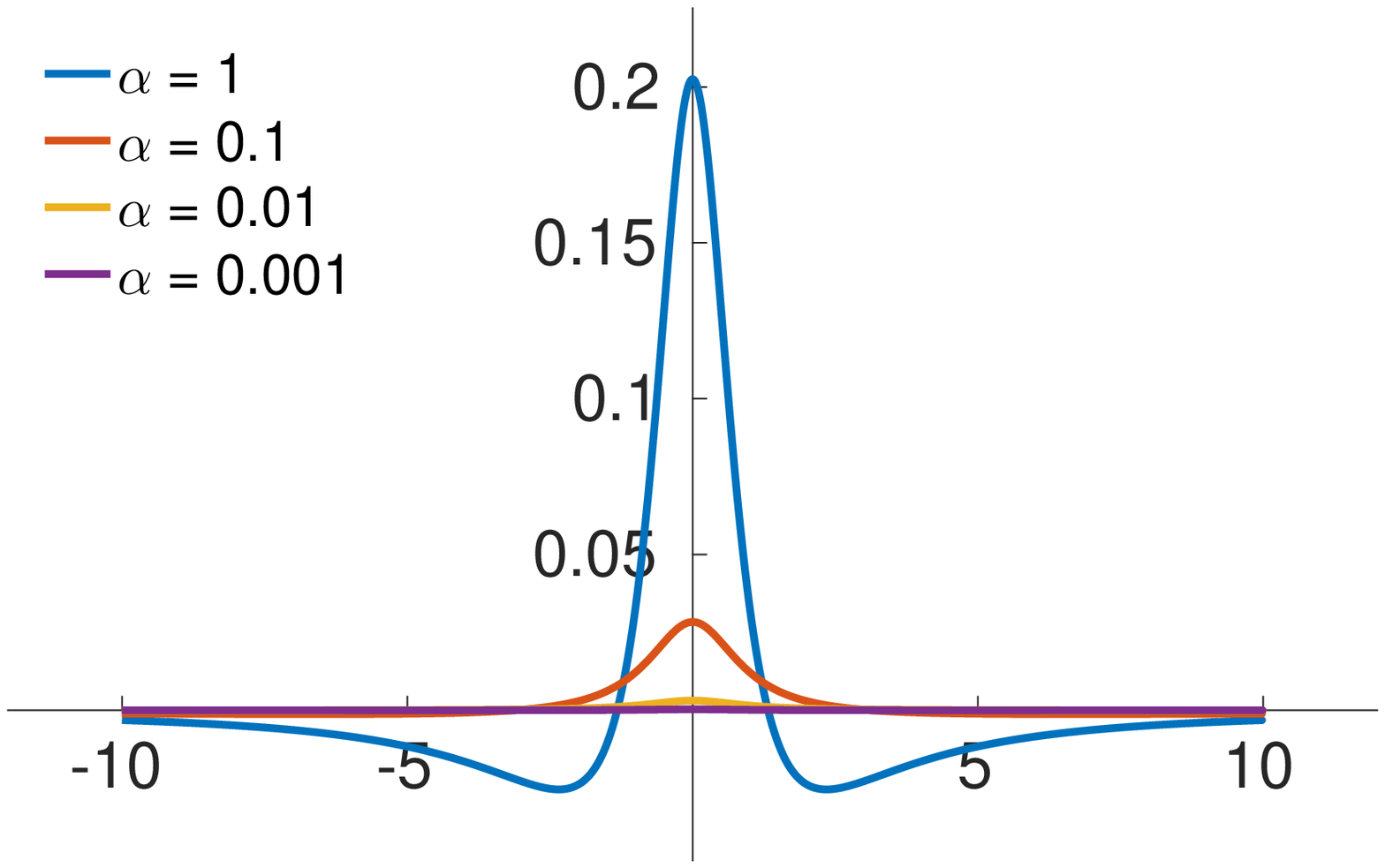}
                \caption{$ \Gc\Hc - \Sc t-t $ for $\alpha \in \left\lbrace 1, 0.1, 0.01, 0.001 \right\rbrace$}
                \label{fig:GHStMinusSt_GenHypAsStu1}
        \end{subfigure}                                         
        \caption{The behaviour of the Generalized Hyperbolic density function $\Gc\Hc (x | \lambda = -\nu/2, \alpha \searrow 0, \beta = 0, \delta = \sqrt{\nu}, \mu=0)$ depending on $\alpha$.}
        \label{fig:Comp_GenHypAsStu1}
\end{figure}

In the following, a short introduction of the Generalized Hyperbolic distribution. The interest of this distribution is its heavy tailed form for different parameters and the fact that it can be expressed via conjugate priors, namely the Normal distribution and the generalized Inverse Gaussian distribution.
\\
The generalised Hyperbolic distribution ($\Gc\Hc$), introduced by Ole Barndorff-Nielsen, is a continuous probability distribution defined as the normal variance-mean mixture where the mixing distribution is the generalized Inverse Gaussian ($\Gc\Ic\Gc$) distribution. Its probability density function is given in terms of modified Bessel function of the second kind, $\Kc_{\lambda}$. A random variable has a generalized Hyperbolic $\Gc\Hc(x|\lambda,\alpha,\beta,\delta,\mu,\gamma)$ distribution if its probability density function is:
\beq
\Gc\Hc(x|\lambda,\alpha,\beta,\delta,\mu) 
= 
\frac{\left(\frac{\gamma}{\delta}\right)^{\lambda}}{\sqrt{2\pi}\Kc_{\lambda} \left( \delta \gamma \right)}
\;
\frac{\Kc_{\lambda-\frac{1}{2}} \left( \alpha \sqrt{ \delta^2 + \left( x - \mu \right)^2} \right)}{\left( \sqrt{ \delta^2 + \left( x - \mu \right)^2} / \alpha \right)^{\frac{1}{2}-\lambda}}
\;
\exp \left\lbrace {\beta \left( x - \mu \right)} \right\rbrace
, 
\lambda, \alpha, \beta, \delta, \mu \in \mathbb{R}, \gamma = \sqrt{\alpha^2-\beta^2};
\label{Eq:GenHyperbolicDistribution}
\eeq
For the linear model expressed in Equation~\eqref{Eq:LinearModel}, the Generalized Hyperbolic Prior Model (GHPM), Equation~\eqref{Eq:GHPM1}, considers a  Normal distribution for $\rf_\rj|\rv_{\rf_\rj}$ with the mean $\mu + \beta \rv_{\rf_\rj}$ and variance $\rv_{\rf_\rj}$. The variance $\rv_{\rf_\rj}| \gamma^2, \; \delta^2, \; \lambda$ is modelled as a generalized Inverse Gaussian distribution, with the corresponding parameters $\gamma^2$, $\delta^2$, and $\lambda$:
\beq
\textbf{GHPM:}\;\;\;\;\;\;\;
\left\{\barr{ll}
p(\rf_\rj | \mu, \; \rv_{\rf_\rj}, \; \beta)
=
\Nc\left(\rf_\rj|\mu + \beta \rv_{\rf_\rj}, \; \rv_{\rf_\rj}\right) 
=
\left( 2 \pi \right)^{-\frac{1}{2}} \; \rv_{\rf_\rj}^{-\frac{1}{2}} \; \exp \left\lbrace -\frac{1}{2}\frac{\left( x - \mu - \beta \rv_{\rf_\rj} \right)^2}{\rv_{\rf_\rj}} \right\rbrace
\\[7pt]
p(\rv_{\rf_\rj}|\gamma^2, \; \delta^2, \; \lambda) 
=
\Gc\Ic\Gc(\rv_{\rf_\rj}|\gamma^2, \; \delta^2, \; \lambda)
=
\frac{\left(\frac{\gamma}{\delta}\right)^{\lambda}}{2\Kc_{\lambda} \left( \delta \gamma \right)}
\;
\rv_{\rf_\rj}^{\lambda-1}
\;
\exp \left\lbrace -\frac{1}{2} \left( \gamma^2 \rv_{\rf_\rj} + \delta^2 \rv_{\rf_\rj}^{-1} \right) \right\rbrace,\\[4pt]
\earr\right.
\label{Eq:GHPM1}
\eeq
The expression of the joint probability distribution $\rf_\rj,\rv_{\rf_\rj}| \mu, \; \beta, \; \gamma^2, \; \delta^2, \; \lambda$ is given by Equation~\eqref{Eq:JointGHPM1},
\beq
p\left( \rf_\rj,\rv_{\rf_\rj}| \mu, \; \beta, \; \gamma^2, \; \delta^2, \; \lambda \right) = \Nc\left(\rf_\rj|\mu + \beta \rv_{\rf_\rj}, \; \rv_{\rf_\rj}\right) \; \Gc\Ic\Gc(\rv_{\rf_\rj}|\gamma^2, \; \delta^2, \; \lambda),
\label{Eq:JointGHPM1}
\eeq 
so the marginal is, Equation~\eqref{Eq:MargGHPM1}:
\beq
p(\rf_\rj|\mu, \beta, \gamma^2, \delta^2, \lambda)
=
\frac{\left(\frac{\gamma}{\delta}\right)^{\lambda}}{2 \sqrt{2 \pi} \Kc_{\lambda} \left( \delta \gamma \right)}
\;
\int
\rv_{\rf_\rj}^{\lambda-\frac{3}{2}}
\;
\exp \left\lbrace -\frac{1}{2} \left( \frac{\left( \rf_\rj - \mu - \beta \rv_{\rf_\rj} \right)^2}{\rv_{\rf_\rj}} + \gamma^2 \rv_{\rf_\rj} + \delta^2 \rv_{\rf_\rj}^{-1} \right) \right\rbrace
\;
\d \rv_{\rf_\rj}
\label{Eq:MargGHPM1}
\eeq 
For solving Equation~\eqref{Eq:MargGHPM1} we can identify a $\Gc\Ic\Gc$ inside the integral: 
\beq
\begin{split}
I 
=&
\exp \left\lbrace \beta \left( \rf_\rj - \mu \right) \right\rbrace 
\int
\rv_{\rf_\rj}^{\left( \lambda-\frac{1}{2} \right) -1} 
\exp \left\lbrace -\frac{1}{2} \left(\left[ \beta^{2} + \gamma^{2} \right] \rv_{\rf_\rj} + \left[ \delta^2 + \left( \rf_\rj-\mu \right)^{2} \right] \rv_{\rf_\rj}^{-1} \right) \right\rbrace \d \rv_{\rf_\rj}
\\
=&
\exp \left\lbrace \beta \left( \rf_\rj - \mu \right) \right\rbrace  
\frac{2\Kc_{\lambda-\frac{1}{2}}\sqrt{\left( \beta^{2} + \gamma^{2} \right) \left( \delta^2 + \left(\rf_\rj-\mu \right)^{2} \right)}}{\left( \frac{ \beta^{2} + \gamma^{2} }{ \delta^2 + \left(\rf_\rj-\mu \right)^{2}}\right)^{\frac{\lambda}{2}-\frac{1}{4}}} 
\underbrace{\int \Gc\Ic\Gc \left( \rv_{\rf_\rj} | \lambda-\frac{1}{2}, \left[ \beta^{2} + \gamma^{2} \right], \left[ \delta^2 + \left(\rf_\rj-\mu \right)^{2} \right] \right) \d \rv_{\rf_\rj}}_{1}
\end{split}
\label{Eq:LapIntegral2}
\eeq 
Introducing the notations:
\beq
\alpha^{2} = \beta^{2} + \gamma^{2} 
\;\;\;\; ; \;\;\;\;
q(\rf_\rj)^{2} = \delta^{2} + \left(\rf_\rj-\mu \right)^{2},
\label{Eq:Notat_GHPM1} 
\eeq
the marginal from Equation~\eqref{Eq:MargGHPM1} can be written as:
\beq
p(\rf_\rj|\mu, \beta, \gamma^2, \delta^2, \lambda)
= 
\frac{\left(\frac{\gamma}{\delta}\right)^{\lambda}}{\sqrt{2 \pi} \Kc_{\lambda} \left( \delta \gamma \right)}
\;
\frac{\Kc_{\lambda-\frac{1}{2}}\sqrt{ \alpha q(\rf_\rj) }}{\left( \frac{q(\rf_\rj)}{\alpha}\right)^{\frac{1}{2}-\lambda}} 
\;
\exp \left\lbrace \beta \left( \rf_\rj - \mu \right) \right\rbrace 
\label{Eq:MargGHPM1_1}
\eeq 
Introducing between the parameters the parameter $\alpha$ and excluding the parameter $\gamma$, considering as parameters $\delta$ instead of $\delta^2$ and reordering the parameters we obtain:
\beq
p(\rf_\rj|\lambda, \alpha, \beta, \delta, \mu)
= 
\frac{\left(\frac{\gamma}{\delta}\right)^{\lambda}}{\sqrt{2 \pi} \Kc_{\lambda} \left( \delta \gamma \right)}
\;
\frac{\Kc_{\lambda-\frac{1}{2}}\sqrt{ \alpha q(\rf_\rj) }}{\left( \frac{q(\rf_\rj)}{\alpha}\right)^{\frac{1}{2}-\lambda}} 
\;
\exp \left\lbrace \beta \left( \rf_\rj - \mu \right) \right\rbrace,  \gamma = \sqrt{\alpha^{2} - \beta^{2}}  
\label{Eq:MargGHPM1_2}
\eeq 
The interest of this distribution is that is of a very general form, being the superclass of, among others, the Student-t ($\Sc t - t$) distribution for the particular set of parameters $\Gc\Hc \left( -\frac{\nu}{2}, 0, 0, \sqrt{\nu}, \mu \right)$, the Hyperbolic ($\Hc$) distribution for the particular set of parameters $\Gc\Hc(x|\lambda = 1,\alpha,\beta,\delta,\mu) $ - Subsection~\eqref{Subsubsec:GenHyperbolicAsHyp}, the Laplace ($\Lc$) distribution for the particular set of parameters $\Gc\Hc(x|\lambda = 1, \alpha = b^{-1}, \beta = 0, \delta \searrow  0, \mu) $ - Subsection~\eqref{Subsubsec:GenHyperbolicAsLap}, the variance-gamma ($\Vc\Gc$) distribution for the particular set of parameters $\Gc\Hc(x|\lambda,\alpha,\beta,\delta \searrow 0,\mu)$ - Subsection~\eqref{Subsubsec:GenHyperbolicAsVarGam}, the Normal-Inverse Gaussian ($\Nc\Ic\Gc$) distribution $\Gc\Hc(x|\lambda,\alpha=-\frac{1}{2},\beta,\delta,\mu)$ - Subsection~\eqref{Subsubsec:GenHyperbolicAsNIG}.\\ 
In the following we consider the Student-t distribution expressed via the generalized Hyperbolic distribution:\\
$X \sim \Gc\Hc(x|\lambda = \frac{-\nu}{2},\alpha \searrow 0,\beta = 0, \delta = \sqrt{\nu},\mu)$ has a Student's t-distribution with $\nu$ degrees of freedom, $\Sc t-t \left( x | \mu, \nu \right)$.\\ 
Fixing the asymmetry parameter $\beta = 0$, as an immediate consequence $\gamma=\alpha$ and $\exp \left\lbrace \beta \left( x - \mu \right) \right\rbrace = 1$. The particular case of the probability density function is:
\beq
\Gc\Hc(x|\lambda,\alpha,\beta=0,\delta,\mu) 
= 
\frac{\left(\frac{\alpha}{\delta}\right)^{\lambda}}{\sqrt{2\pi}\Kc_{\lambda} \left( \alpha \delta \right)}
\;
\frac{\Kc_{\lambda-\frac{1}{2}} \left( \alpha \delta \sqrt{ 1 + \frac{ \left( x - \mu \right)^2}{\delta^2}} \right)}{\left( \delta \sqrt{ 1 + \frac{\left( x - \mu \right)^2}{\delta^2}} / \alpha \right)^{\frac{1}{2}-\lambda}}.
\label{Eq:GenHyperbolicDistributionAsStudent}
\eeq 
Fixing $\delta^2 = \nu$ and considering $\alpha \searrow 0$, the particular case of the probability density function is:
\beq
\begin{split}
\Gc\Hc(x|\lambda,\alpha \searrow 0,\beta = 0, \delta = \sqrt{\nu},\mu) 
&= 
\frac{1}{\sqrt{2\pi}}
\left(\frac{\alpha}{\sqrt{\nu}}\right)^{\lambda}
\left( \frac{\sqrt{\nu}}{\alpha} \sqrt{ 1 + \frac{\left( x - \mu \right)^2}{\nu}} \right)^{\lambda-\frac{1}{2}}
\;
\frac{\Kc_{\lambda-\frac{1}{2}} \left( \alpha \sqrt{\nu} \sqrt{ 1 + \frac{ \left( x - \mu \right)^2}{\nu}} \right)}{\Kc_{\lambda} \left( \alpha \sqrt{\nu} \right)}
\\
&= 
\frac{1}{\sqrt{2\pi}}
\left(\frac{\alpha}{\sqrt{\nu}}\right)^{\frac{1}{2}}
\left( 1 + \frac{\left( x - \mu \right)^2}{\nu} \right)^{\frac{1}{2} \left( \lambda - \frac{1}{2} \right)}
\;
\frac{\Kc_{\lambda-\frac{1}{2}} \left( \alpha \sqrt{\nu} \sqrt{ 1 + \frac{ \left( x - \mu \right)^2}{\nu}} \right)}{\Kc_{\lambda} \left( \alpha \sqrt{\nu} \right)}.
\end{split}
\label{Eq:GenHyperbolicDistributionAsStudent1}
\eeq
Since we considered $\alpha \searrow 0 $, for both arguments of the modified Bessel functions of the second kind appearing in Equation~\eqref{Eq:GenHyperbolicDistributionAsStudent1} we have $ \alpha \sqrt{\nu} \sqrt{ 1 + \frac{ \left( x - \mu \right)^2}{\nu}} \searrow 0 $ and $ \alpha \sqrt{\nu} \searrow 0 $ so for computing the values of the two modified Bessel functions of the second kind we can use the asymptotic relations for small arguments:
\beq
\begin{split}
\Kc_{\lambda}\left( x \right) &\sim \Gamma \left( \lambda \right) \; 2^{\lambda-1} \; x^{-\lambda}, \; \mbox{as} \; x \searrow 0 \; \mbox{and} \; \lambda > 0
\\
\Kc_{\lambda}\left( x \right) &\sim \Gamma \left( - \lambda \right) \; 2^{-\lambda-1} \; x^{\lambda}, \; \mbox{as} \; x \searrow 0 \; \mbox{and} \; \lambda < 0.
\end{split}
\label{Eq:ModifiedBesselAsymptoticRelation}
\eeq
We have:
\beq
\begin{split}
\Kc_{\lambda-\frac{1}{2}} \left( \alpha \sqrt{\nu} \sqrt{ 1 + \frac{ \left( x - \mu \right)^2}{\nu}} \right) &= \Gamma \left( \frac{1}{2} - \lambda \right) 2^{-\lambda + \frac{1}{2} - 1} \left( \alpha \sqrt{\nu} \right)^{\lambda - \frac{1}{2}} \left( 1 + \frac{\left( x - \mu \right)^2}{\nu} \right)^{\frac{1}{2}\left(\lambda - \frac{1}{2}\right)}
\\
\Kc_{\lambda} \left( \alpha \sqrt{\nu} \right) &= \Gamma \left( - \lambda \right) 2^{-\lambda - 1} \left( \alpha \sqrt{\nu} \right)^{\lambda} 
\end{split}
\label{Eq:ModifiedBesselValues}
\eeq
Using Equation~\eqref{Eq:ModifiedBesselValues} in Equation~\eqref{Eq:GenHyperbolicDistributionAsStudent1}, the particular case of the probability density function becomes:
\beq
\begin{split}
\Gc\Hc(x|\lambda,\alpha \searrow 0,\beta = 0, \delta = \sqrt{\nu},\mu) 
&= 
\frac{1}{\sqrt{2\pi}}
\left(\frac{\alpha}{\sqrt{\nu}}\right)^{\frac{1}{2}}
\sqrt{2}
\left( \alpha \sqrt{\nu} \right)^{-\frac{1}{2}}
\;
\frac{\Gamma \left( \frac{1}{2} - \lambda \right) }{\Gamma \left( - \lambda \right)}
\;
\left( 1 + \frac{\left( x - \mu \right)^2}{\nu} \right)^{\lambda - \frac{1}{2}}
\end{split}
\label{Eq:GenHyperbolicDistributionAsStudent2}
\eeq
Finally, fixing $\lambda = -\frac{\nu}{2}$, Equation~\eqref{Eq:GenHyperbolicDistributionAsStudent2} becomes:
\beq
\begin{split}
\Gc\Hc(x|\lambda = \frac{-\nu}{2},\alpha \searrow 0,\beta = 0, \delta = \sqrt{\nu},\mu) 
&= 
\frac{1}{\sqrt{\nu\pi}}
\;
\frac{\Gamma \left( \frac{\nu+1}{2} \right) }{\Gamma \left( \frac{\nu}{2} \right)}
\;
\left( 1 + \frac{\left( x - \mu \right)^2}{\nu} \right)^{-\frac{\nu + 1}{2}}
=
\Sc t-t \left( x | \mu, \nu \right).
\end{split}
\label{Eq:GenHyperbolicDistributionAsStudent3}
\eeq
Figure~\eqref{fig:St_GenHypAsMStu} presents four Student-t probability density functions with different means ($\mu=0$, blue and red, $\mu=1$, yellow and $\mu=-1$, violet) and different degrees of freedom ($\nu=1$, blue and yellow, $\nu=1,5$, red and $\nu=0.5$, violet). 
Figure~\eqref{fig:GHSt_GenHypAsMStu} presents the Generalized Hyperbolic probability density function for parameters $\Gc\Hc (x | \lambda = -\nu/2, \alpha \searrow 0, \beta = 0, \delta = \sqrt{\nu}, \mu)$ with $\nu$ and $\mu$ set with the same numerical values as the corresponding ones from Figure~\eqref{fig:St_GenHypAsMStu}. For $\alpha$, the numerical values is $0.001$.
Figure~\eqref{fig:GHStVsSt_GenHypAsMStu} presents the comparison between the four Student-t probability density functions and the corresponding Generalized Hyperbolic probability density functions, $ \Sc t-t $ vs. $ \Gc\Hc $. In all forth cases the probability density functions are superposed. 
Figure~\eqref{fig:Log_GHStVsSt_GenHypAsMStu} presents the comparison between the logarithm of distributions, $ - \log \Sc t-t$ vs. $ - \log \Gc\Hc$.  
Figure~\eqref{fig:GHStVsSt_GenHypAsStu} presents the comparison between the standard Student-t probability density function, $\Sc t-t \left( x | \mu=0, \nu=1 \right)$ (reported in Figure~\eqref{fig:St_GenHypAsStu}) and the Generalized Hyperbolic density function for parameters $\Gc\Hc (x | \lambda = -\frac{\nu}{2} = \frac{1}{2}, \alpha \searrow 0, \beta = 0, \delta = \sqrt{\nu} = 1, \mu=0)$ (reported in Figure~\eqref{fig:GHSt_GenHypAsStu}), showing that the two probability density functions are almost superposed. Figure~\eqref{fig:Log_GHStVsSt_GenHypAsStu} presents the comparison between the logarithm of the two distributions, $ - \log \Sc t-t$ vs. $ - \log \Gc\Hc$. In this case, the numerical value for $\alpha$ is $0.01$. 
The behaviour of the Generalized Hyperbolic density function $\Gc\Hc (x | \lambda = -\nu/2, \alpha \searrow 0, \beta = 0, \delta = \sqrt{\nu}, \mu=0)$ depending on $\alpha$ is presented in Figure~\eqref{fig:GHStVsSt_GenHypAsStu1}: a comparison between the standard Student-t probability density function $\Sc t-t \left( x | \mu=0, \nu=1 \right)$ (in blue) and the Generalized Hyperbolic density function $\Gc\Hc (x | \lambda = - \frac{\nu}{2} = -\frac{1}{2}, \alpha \searrow 0, \beta = 0, \delta = \sqrt{\nu} = 1, \mu=0)$ for $\alpha=1$ (in red), $\alpha=0.1$ (in yellow), $\alpha=0.01$ (in violet) and $\alpha=0.001$ (in green) is presented in Figure~\eqref{fig:GHStVsSt_GenHypAsStu1}. The difference between those four Generalized Hyperbolic density functions and the Student-t density function is presented in Figure~\eqref{fig:GHStMinusSt_GenHypAsStu1}. 
Considering the Generalized Hyperbolic Prior Model (GHPM), Equation~\eqref{Eq:GHPM1}, for the structure of the parameters corresponding to the Student-t distribution $ \lambda = \frac{-\nu}{2},\alpha \searrow 0,\beta = 0, \delta = \sqrt{\nu},\mu $ we obtain a Normal variance mixture (since $\beta=0$) with the mixing distribution the Inverse Gamma distribution (since $\Gc\Ic\Gc( \rv_{\rf_\rj} | \gamma^2 \searrow 0, \; \nu, \; \frac{-\nu}{2}) = \Ic\Gc(\rv_{\rf_\rj}| \frac{\nu}{2} \; \frac{\nu}{2})$). Indeed, if $ \alpha \searrow 0, \beta = 0$ and $ \gamma = \sqrt{ \alpha^2 + \beta^2 }$ we obtain $\gamma \searrow 0$. So $ p(\rv_{\rf_\rj} | \gamma^2, \; \delta^2, \; \lambda) = p(\rv_{\rf_\rj} | \gamma^2 \searrow 0, \; \nu, \; \frac{-\nu}{2}) = \Gc\Ic\Gc(\rv_{\rf_\rj}|\gamma^2 \searrow 0, \; \nu, \; \frac{-\nu}{2})$. Since $\gamma^2 \searrow 0$, for the modified Bessel function of the second kind appearing in the expression of the generalized Inverse Gaussian distribution the asymptotic relation for small arguments can be used, $ \Kc_{\lambda} \left( \delta \gamma \right) \sim \Gamma \left( \lambda \right) 2^{-\lambda-1} \left( \delta \gamma \right)^{\lambda} $, so the generalized Inverse Gaussian can be written as an Inverse Gamma with equal shape and scale parameters $ \Gc\Ic\Gc(\rv_{\rf_\rj}|\gamma^2 \searrow 0, \; \nu, \; \frac{-\nu}{2}) \sim \frac{\left(\frac{\nu}{2}\right)^{\frac{\nu}{2}}}{\Gamma \left( \frac{\nu}{2} \right)} \;\rv_{\rf_\rj}^{-\frac{\nu}{2}-1} \;\exp \left\lbrace - \frac{\nu}{2} \rv_{\rf_\rj}^{-1} \right\rbrace = \Ic\Gc(\rv_{\rf_\rj}|\frac{\nu}{2}, \; \frac{\nu}{2})$. We obtain the same conjugate prior model for the Student-t distribution, the Normal variance mixture with the mixing distribution an Inverse Gamma distribution. If the condition $\delta = \sqrt{\nu}$ is not imposed in the Generalized Hyperbolic distribution, the two parameters Student-t distribution form is obtained.

\subsection{Hyperbolic prior: expressed via conjugate priors}
\label{Subsec:HyperbolicPrior}
The interest is given by the fact that the Hyperbolic distribution is a \textit{heavy-tailed} distribution and therefore can be successfully used as a sparsity enforcing prior. The Hyperbolic distribution is a continuous probability distribution, for the logarithm of the probability density function is a hyperbola, therefore the distribution decreases exponentially, which is more slowly than the Normal distribution. It is suitable to model phenomena where numerically large values are more probable than is the case for the Normal distribution. The origin of the distribution is the observation by Ralph Alger Bagnold, that the logarithm of the histogram of the empirical size distribution of sand deposits tends to form a hyperbola. The Hyperbolic distribution has the following probability density function:
\beq
\Hc \left( x | \alpha, \beta, \delta, \mu \right)
=
\frac{\gamma}{2\alpha\delta \Kc_{1} \left( \delta \gamma \right)}
\exp \left\lbrace - \alpha \sqrt{ \delta^2 + \left( x - \mu \right)^2 } + \beta \left( x - \mu \right) \right\rbrace,
\gamma = \sqrt{ \alpha^2 - \beta^2 }
\label{Eq:HyperbolicDistribution}
\eeq
\subsubsection{Hyperbolic prior: via Generalized Hyperbolic distribution}
\label{Subsubsec:GenHyperbolicAsHyp}
\begin{figure}[!htb]
        \centering
        \begin{subfigure}{0.49\textwidth}
          		\centering        
                \includegraphics[width=7.4cm,height=4cm]{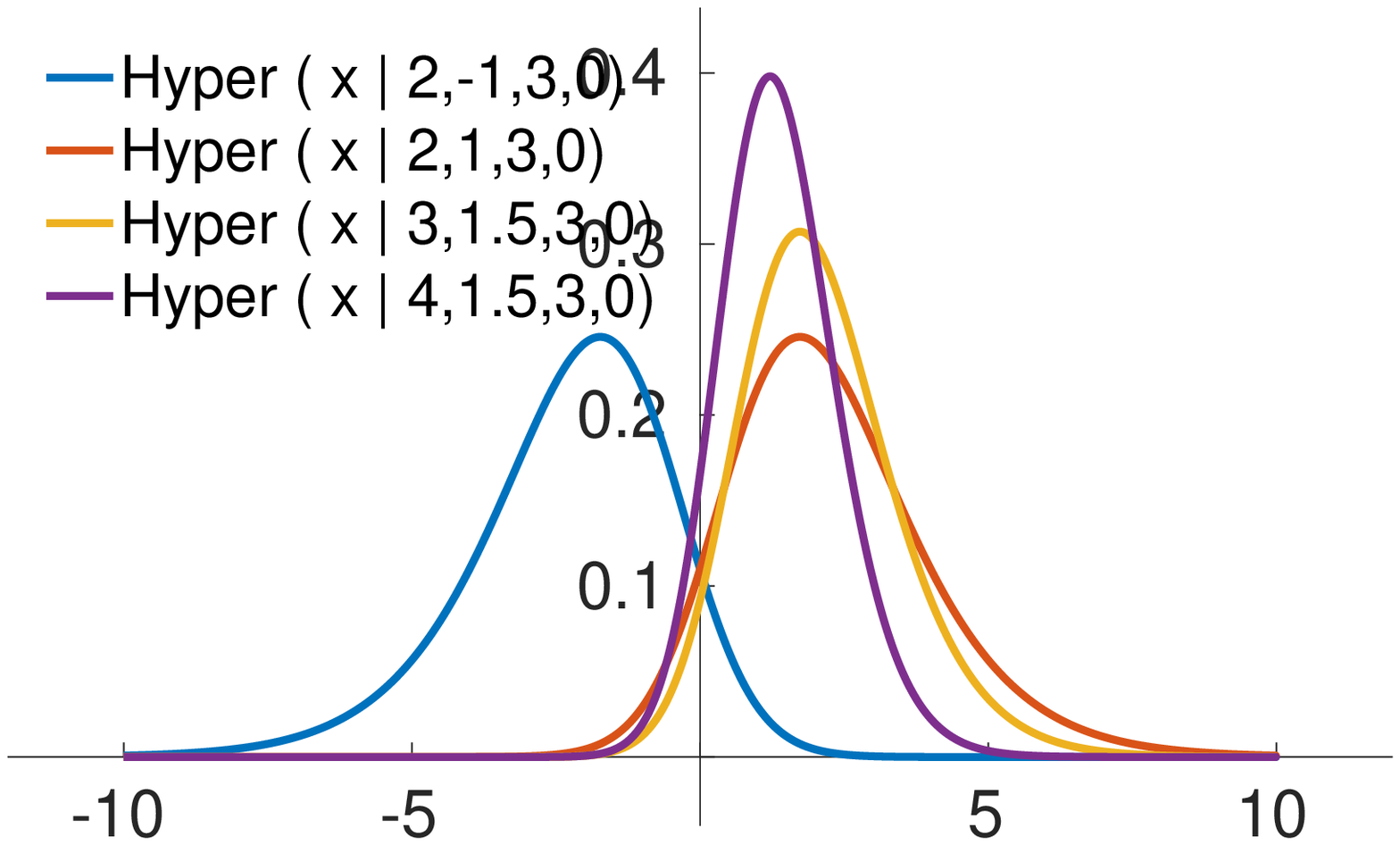}
                \caption{$ \Hc \left( x | \alpha,\beta, \delta, \mu \right) $}
                \label{fig:Hyp_GenHypAsHyp}
        \end{subfigure}
        \begin{subfigure}{0.49\textwidth}
          		\centering        
                \includegraphics[width=7.4cm,height=4cm]{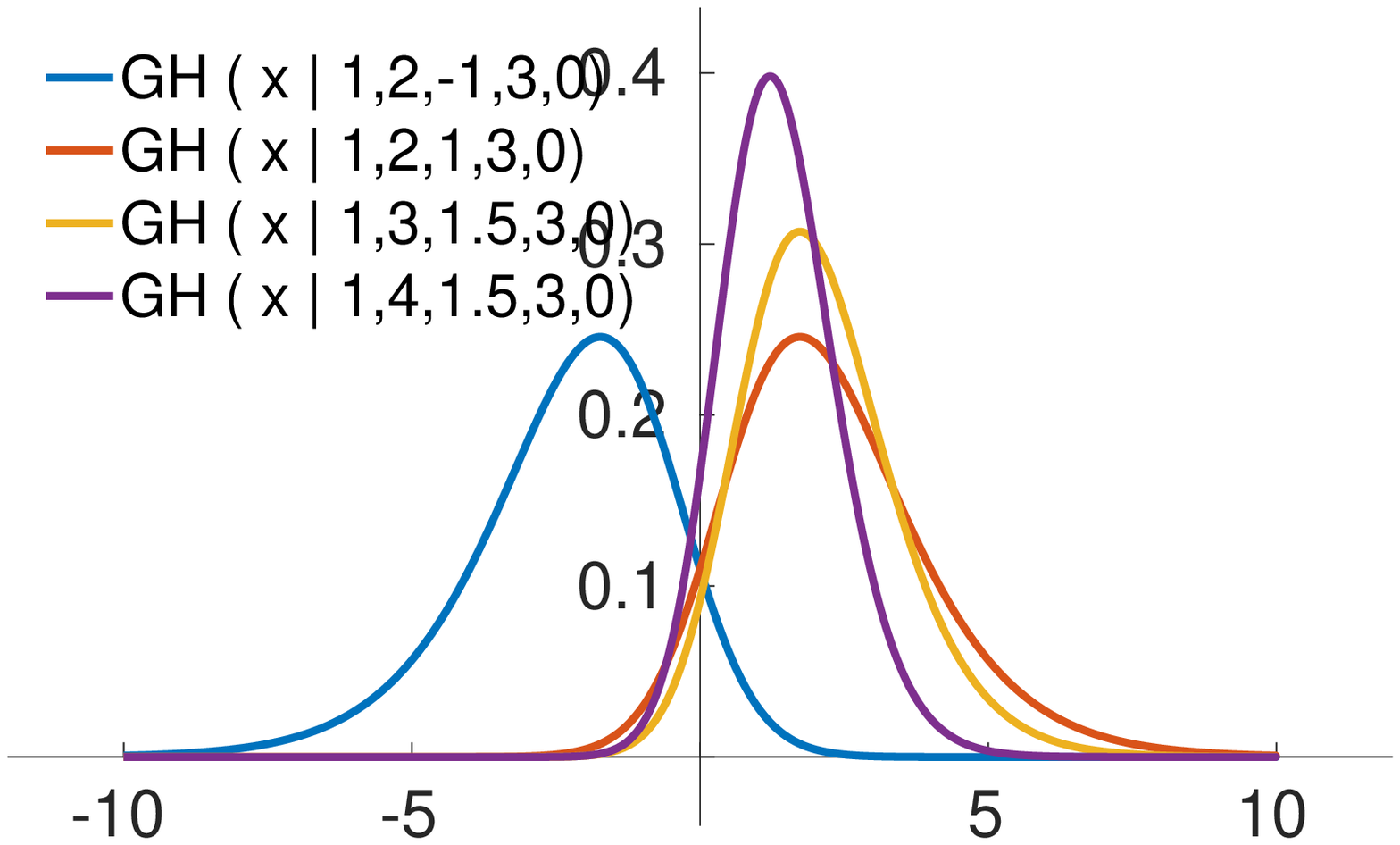}
                \caption{\small{$ \Gc\Hc \left( x | \lambda = 1, \alpha, \beta, \delta, \mu \right) $}}
                \label{fig:GHHyp_GenHypAsHyp}
        \end{subfigure}        
\par\bigskip        
        \begin{subfigure}{0.49\textwidth}
          		\centering        
                \includegraphics[width=7.4cm,height=4cm]{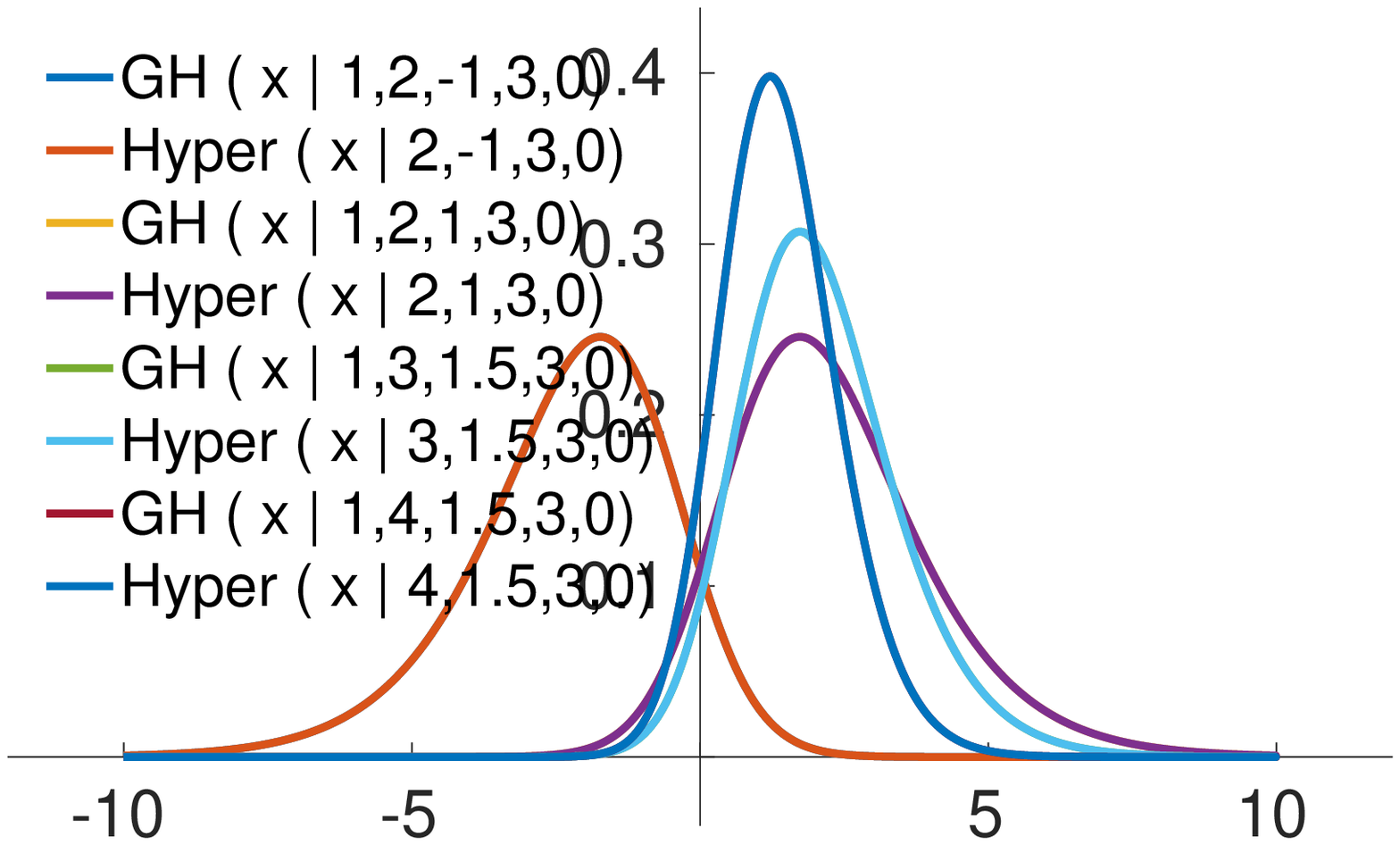}
                \caption{$ \Hc $ vs. $ \Gc\Hc $}
                \label{fig:GHHypVsHyp_GenHypAsHyp}
        \end{subfigure}                               
        \begin{subfigure}{0.49\textwidth}
          		\centering        
                \includegraphics[width=7.4cm,height=4cm]{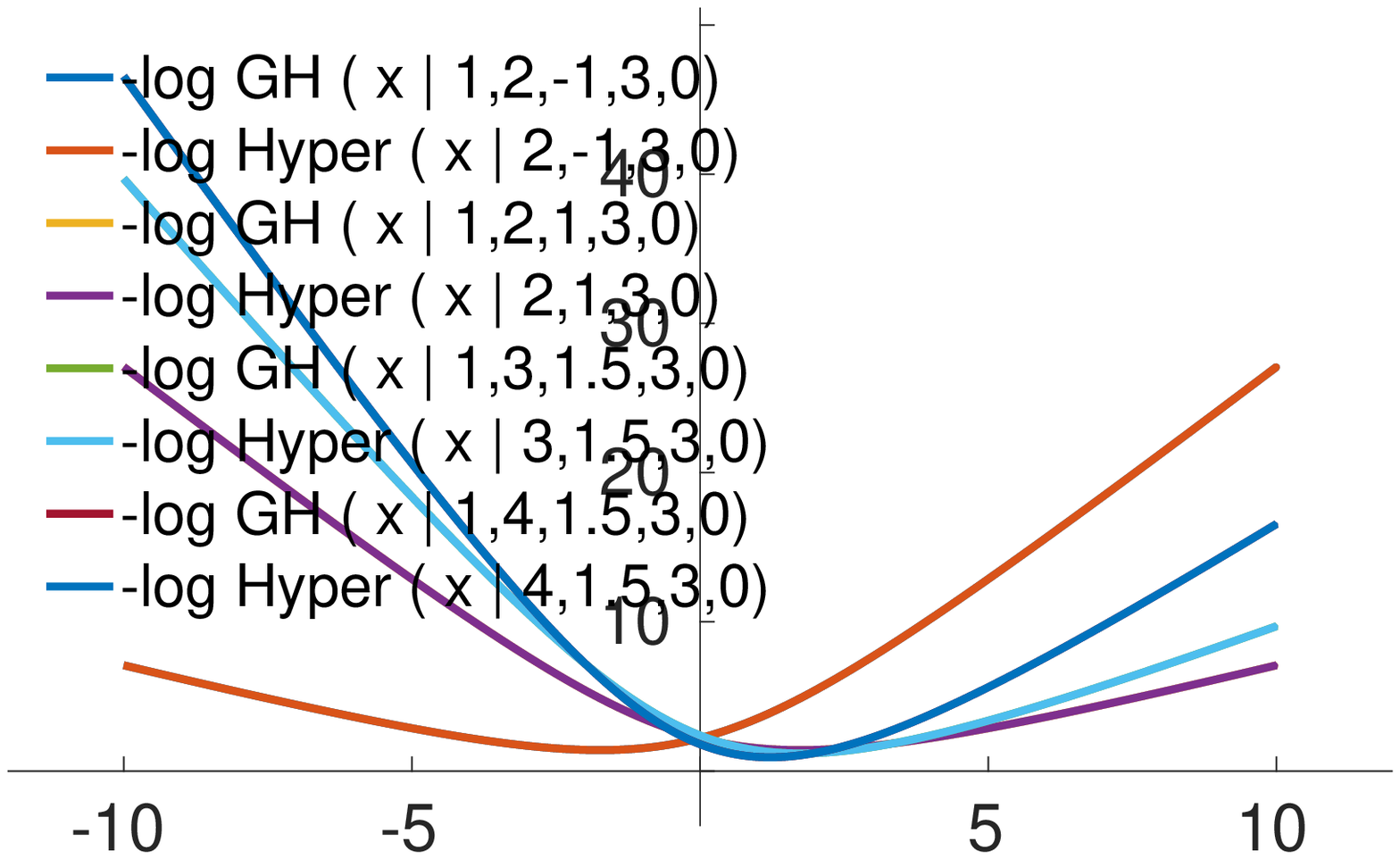}
                \caption{$ -\log \Hc $ vs. $ -\log \Gc\Hc$}
                \label{fig:Log_GHHypVsHyp_GenHypAsHyp}
        \end{subfigure}                                       
        \caption{The Hyperbolic distribution, Figure~\eqref{fig:Hyp_GenHypAsHyp} and the Generalized Hyperbolic distribution, with parameters set as in Equation~\eqref{Eq:GenHyperbolicDistributionAsHyp2}, Figure~\eqref{fig:GHHyp_GenHypAsHyp}. Comparison between the two distribution, $\Hc$ vs. $\Gc\Hc$, Figure~\eqref{fig:GHHypVsHyp_GenHypAsHyp} and between the logarithm of the distributions $-\log \Hc$ vs. $-\log \Gc\Hc$, Figure~\eqref{fig:Log_GHHypVsHyp_GenHypAsHyp}.}
        \label{fig:Comp_GenHypAsHyp}
\end{figure}

$X \sim \Gc\Hc(x|\lambda = 1,\alpha,\beta,\delta,\mu) $ has a Hyperbolic distribution, $\Hc \left( x | \alpha, \beta, \delta, \mu \right)$.\\
The goal of this section is to derive the Hyperbolic distribution from the Generalized Hyperbolic distribution. 
For the particular case of the Generalized Hyperbolic distribution with $\lambda = 1$, the probability density function is:
\beq
\Gc\Hc(x|\lambda = 1,\alpha,\beta,\delta,\mu) 
= 
\frac{1}{\sqrt{2\pi}}
\;
\frac{\gamma}{\delta}
\;
\left( \frac{\sqrt{ \delta^2 + \left( x - \mu \right)^2}}{ \alpha} \right)^{\frac{1}{2}}
\;
\frac{\Kc_{\frac{1}{2}} \left( \alpha \sqrt{ \delta^2 + \left( x - \mu \right)^2} \right)}{\Kc_{1} \left( \delta \gamma \right)}
\;
\exp\left\lbrace\beta \left( x - \mu \right)\right\rbrace,
\label{Eq:GenHyperbolicDistributionAsHyp}
\eeq
For $\lambda=\frac{1}{2}$, the modified Bessel function of the second kind $\Kc_{\lambda} \left( x \right)$ can be stated explicitly with:
\beq
\Kc_{\frac{1}{2}} \left( x \right) = \Kc_{-\frac{1}{2}} \left( x \right) = \sqrt{\frac{\pi}{2x}} \exp \left\lbrace -x \right\rbrace, x > 0,
\label{Eq:ModifiedBesselExplicitly}
\eeq
so:
\beq
\Kc_{\frac{1}{2}} \left( \alpha \sqrt{ \delta^2 + \left( x - \mu \right)^2} \right) 
=
\left(\frac{\pi}{2\alpha \sqrt{ \delta^2 + \left( x - \mu \right)^2}}\right)^{\frac{1}{2}} 
\exp \left\lbrace -\alpha \sqrt{ \delta^2 + \left( x - \mu \right)^2} \right\rbrace. 
\label{Eq:GenHyperbolicDistributionAsHyp1}
\eeq
Plugging Equation~\eqref{Eq:GenHyperbolicDistributionAsHyp1} in Equation~\eqref{Eq:GenHyperbolicDistributionAsHyp}:
\beq
\Gc\Hc(x|\lambda = 1,\alpha,\beta,\delta,\mu) 
= 
\frac{\gamma}{2 \alpha \delta \Kc_{1} \left( \delta \gamma \right)}
\;
\exp \left\lbrace -\alpha \sqrt{ \delta^2 + \left( x - \mu \right)^2} + \beta \left( x - \mu \right) \right\rbrace
=
\Hc \left( x | \alpha, \beta, \delta, \mu \right).
\label{Eq:GenHyperbolicDistributionAsHyp2}
\eeq
Figure~\eqref{fig:Hyp_GenHypAsHyp} presents four Hyperbolic probability density functions $\Hc \left( x |\alpha,\beta,\delta,\mu \right)$ with different parameters.  
Figure~\eqref{fig:GHHyp_GenHypAsHyp} the corresponding Generalized Hyperbolic density functions for parameters $\Gc\Hc (x | \lambda = 1,\alpha,\beta,\delta,\mu)$.
The Hyperbolic probability density functions and the corresponding Generalized Hyperbolic density functions are superposed, Figure~\eqref{fig:GHHypVsHyp_GenHypAsHyp}.
Figure~\eqref{fig:Log_GHHypVsHyp_GenHypAsHyp} presents the comparison between the logarithm of the two distributions, $-\log \Hc$ vs. $-\log \Gc\Hc$.
The Hyperbolic distribution can be expressed using the Generalized Hyperbolic Prior Model (GHPM), Equation~\eqref{Eq:GHPM1} by fixing $\lambda =1$, . 
\beq
\textbf{HypPM:}\;\;\;\;\;\;\;
\left\{\barr{ll}
p(\rf_\rj | \mu, \; \rv_{\rf_\rj}, \; \beta)
=
\Nc\left(\rf_\rj|\mu + \beta \rv_{\rf_\rj}, \; \rv_{\rf_\rj}\right) 
=
\left( 2 \pi \right)^{-\frac{1}{2}} \; \rv_{\rf_\rj}^{-\frac{1}{2}} \; \exp \left\lbrace -\frac{1}{2}\frac{\left( x - \mu - \beta \rv_{\rf_\rj} \right)^2}{\rv_{\rf_\rj}} \right\rbrace
\\[7pt]
p(\rv_{\rf_\rj}|\gamma^2, \; \delta^2, \; \lambda) 
=
\Gc\Ic\Gc(\rv_{\rf_\rj}|\gamma^2, \; \delta^2, \; \lambda = 1)
=
\frac{\left(\frac{\gamma}{\delta}\right)}{2\Kc_{1} \left( \delta \gamma \right)}
\;
\exp \left\lbrace -\frac{1}{2} \left( \gamma^2 \rv_{\rf_\rj} + \delta^2 \rv_{\rf_\rj}^{-1} \right) \right\rbrace,\\[4pt]
\earr\right.
\label{Eq:GHPM_Hyp}
\eeq

\subsection{Laplace prior: expressed via conjugate priors}
\label{Subsec:LaplacePrior}
\begin{figure}[!htb]
        \centering
        \begin{subfigure}{0.49\textwidth}
          		\centering        
                \includegraphics[width=7.4cm,height=4cm]{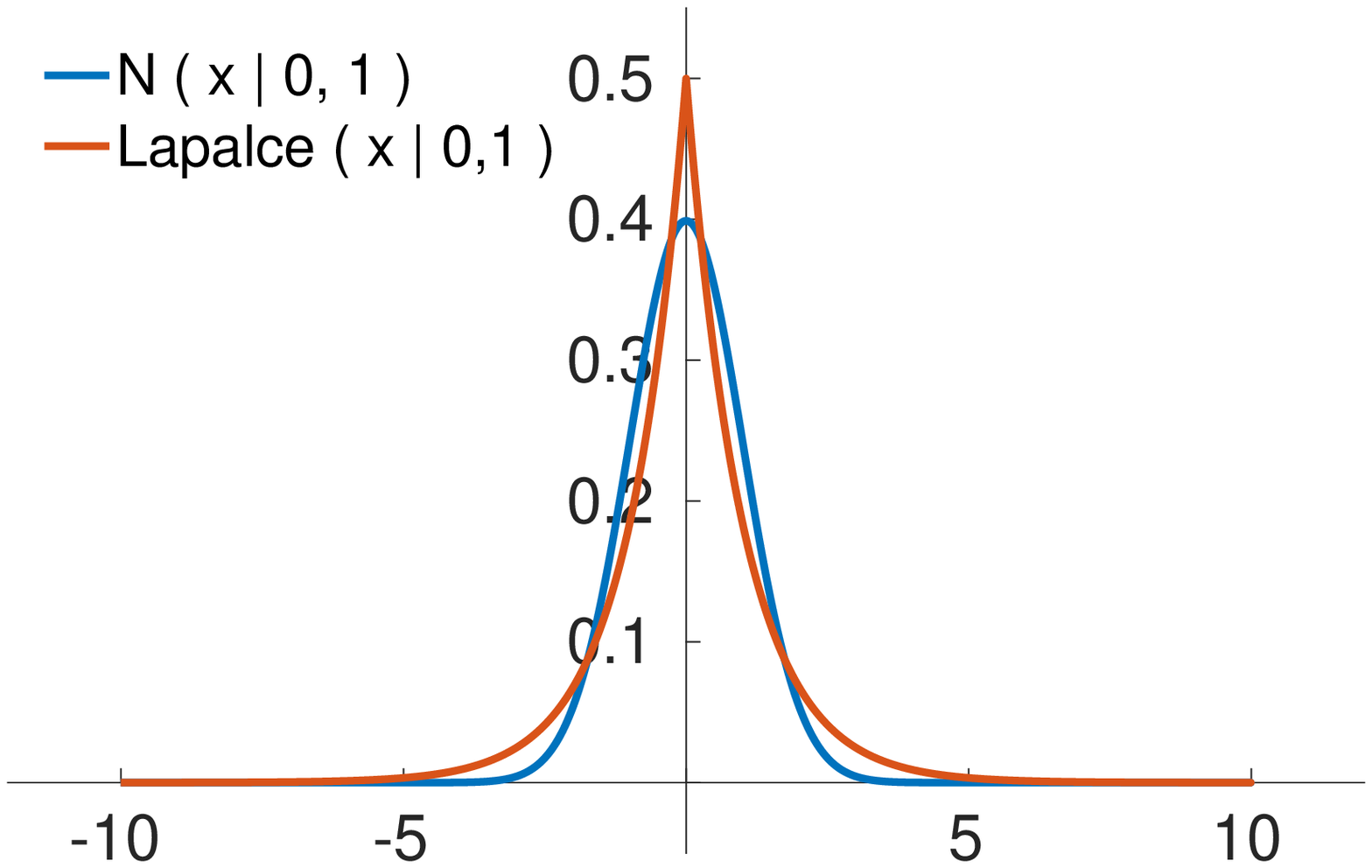}
                \caption{\textcolor{cobalt}{Normal} vs. \textcolor{cinnabar}{Laplace} distribution}
                \label{fig:LaplaceVsNormal2a}
        \end{subfigure}                               
        \begin{subfigure}{0.49\textwidth}
          		\centering        
                \includegraphics[width=7.4cm,height=4cm]{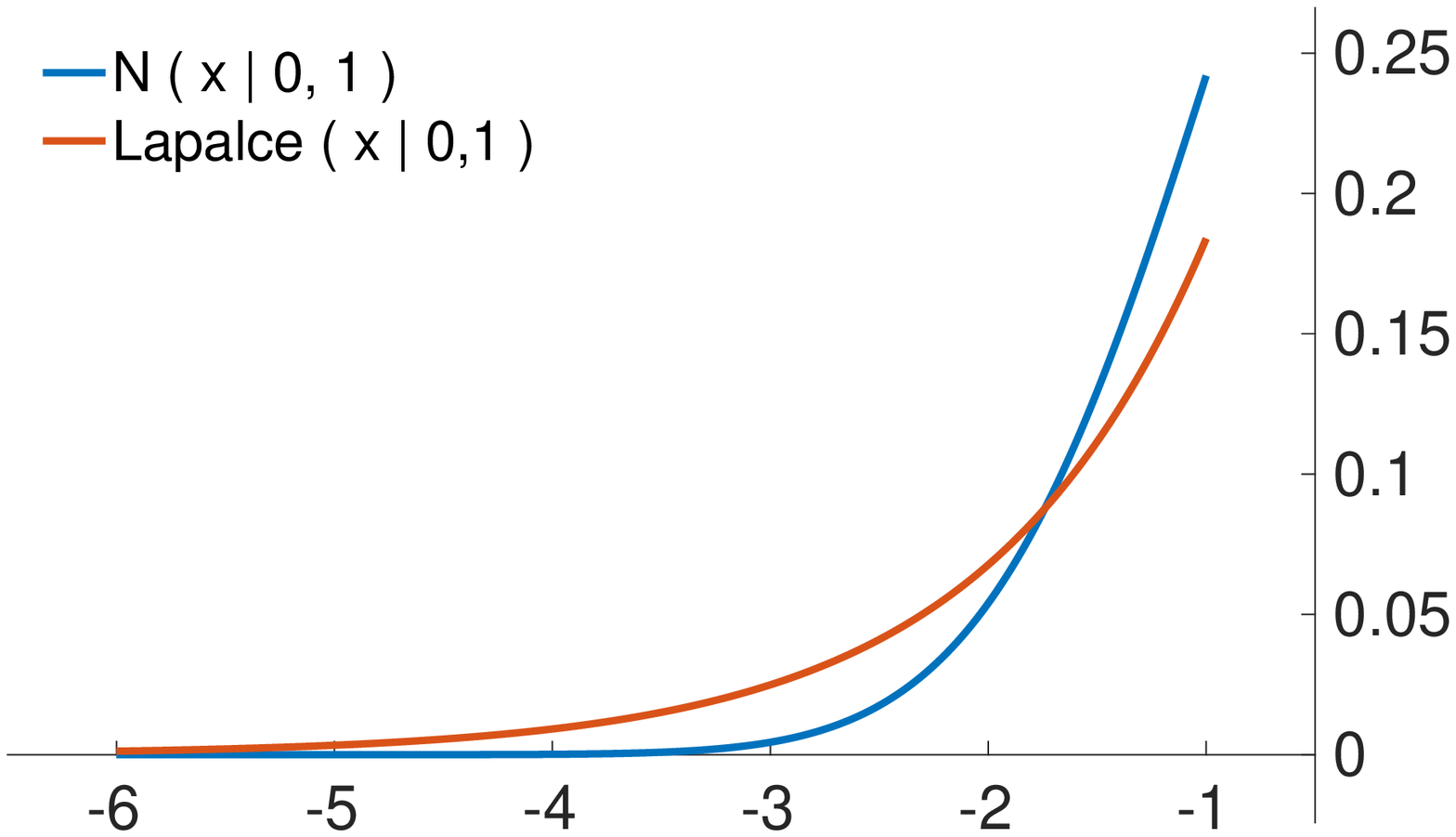}
                \caption{Heavy tailed property of the Laplace distribution}
                \label{fig:LaplaceVsNormal2b}
        \end{subfigure}                                         
        \caption{Comparison between the standard \textcolor{cobalt}{Normal} distribution and the \textcolor{cinnabar}{Laplace} distribution.}
        \label{fig:LaplaceVsNormal2}
\end{figure}

In the following, we present how the Laplace distribution, a sparsity enforcing prior because of its heavy-tailed form can be expressed via conjugate priors, namely the Normal distribution and the Exponential distribution.
\\
For this part of the work, the following resources were used:
\\
\textcolor{americanrose}{$\left[ 2 \right]$ \textbf{A variational Bayes framework for sparse adaptive estimation} - K. E. THEMELIS, A. A. RONTOGIANNIS, arvix: 1401.2771v1 [statML], 13 Jan 2014}.   
\\
In probability theory and statistics, the Laplace distribution is a continuous probability distribution named after Pierre-Simon Laplace. It is also sometimes called the double exponential distribution, because it can be thought of as two exponential distributions (with an additional location parameter) spliced together back-to-back, although the term 'double exponential distribution' is also sometimes used to refer to the Gumbel distribution. The difference between two independent identically distributed exponential random variables is governed by a Laplace distribution, as is a Brownian motion evaluated at an exponentially distributed random time. Increments of Laplace motion or a variance gamma process evaluated over the time scale also have a Laplace distribution.\\
A random variable has a Laplace$(\mu,b)$ distribution if its probability density function is:
\beq
p(x|\mu,b) 
= 
\frac{1}{2 b} \;
\exp \left\lbrace - \frac{|x-\mu|}{b} \right\rbrace
=
\frac{1}{2 b}
\left\{\barr{ll}
\exp \left\lbrace - \frac{\mu-x}{b} \right\rbrace, x<\mu
\\[7pt]
\exp \left\lbrace - \frac{x-\mu}{b} \right\rbrace, x\geq \mu\\[4pt]
\earr\right.
, b > 0;
\label{LaplaceDistributionBis}
\eeq
Figure~\eqref{fig:LaplaceVsNormal2} presents the comparison between the Normal distribution and the standard Laplace distribution.
We present two ways to obtain the Laplace distribution via conjugate priors. The first possibility is to consider a zero-mean Normal distribution for which the variance is modelled as an Exponential distribution. The second possibility considers an Inverse Gamma distribution for the inverse of the variance, with the shape parameter set at 1.
\subsubsection{Laplace prior: Normal and Exponential}
\label{Subsubsec:LaplacePriorBis}
We start with the linear model expressed in Equation~\eqref{Eq:LinearModel}. We consider the Laplace Prior Model (LPM), Equation~\eqref{Eq:LPMBis1}, which considers a zero-mean Normal distribution for $\rf_\rj|\rv_{\rf_\rj}$ while the variance $\rv_{\rf_\rj}|b$ is modelled as an Exponential distribution, with the corresponding parameter $\lambda$:
\beq
\textbf{LPM:}\;\;\;\;\;\;\;
\left\{\barr{ll}
p(\rf_\rj|0,\rv_{\rf_\rj}) = \Nc(\rf_\rj|0, \; \rv_{\rf_\rj}) = \left( 2 \pi \right)^{-\frac{1}{2}} \; \rv_{\rf_\rj}^{-\frac{1}{2}} \; \exp \left\lbrace -\frac{1}{2} \; \frac{\rf_\rj^2}{\rv_{\rf_\rj}} \right\rbrace
\\[7pt]
p(\rv_{\rf_\rj}|\lambda) = \Ec(\rv_{\rf_\rj}|\lambda) = \lambda  \; \exp \left\lbrace - \lambda \; \rv_{\rf_\rj} \right\rbrace,\\[4pt]
\earr\right.
\label{Eq:LPMBis1}
\eeq
The expression of the joint probability distribution $\rf_\rj,\rv_{\rf_\rj}|b$ is given by Equation~\eqref{Eq:JointLPMBis1},
\beq
p(\rf_\rj,\rv_{\rf_\rj}|\lambda) = \Nc(\rf_\rj|0, \; \rv_{\rf_\rj}) \; \Ec(\rv_{\rf_\rj}|\lambda),
\label{Eq:JointLPMBis1}
\eeq 
so the marginal is, Equation~\eqref{Eq:MargLPMBis1}:
\beq
p(\rf_\rj|\rv_{\rf_\rj},\lambda) = \left( 2 \pi \right)^{-\frac{1}{2}} \lambda \int \rv_{\rf_\rj}^{-\frac{1}{2}} \exp \left\lbrace -\frac{1}{2} \left( 2 \lambda \rv_{\rf_\rj} + \frac{\rf_\rj^{2}}{\rv_{\rf_\rj}} \right) \right\rbrace \d \rv_{\rf_\rj}
\label{Eq:MargLPMBis1}
\eeq 
The probability distribution of the generalized inverse Gaussian distribution is presented in Equation~\eqref{Eq:GenInvGauBis} 
\beq
\Gc\Ic\Gc(x|a,b,c) = \frac{\left(a/b\right)^{\frac{c}{2}}}{2\Kc_c(\sqrt{ab})} \; x^{c-1} \; \exp \left\lbrace -\frac{1}{2} \left( a x + b x^{-1} \right) \right\rbrace, \; a, b \geq 0, \; c \in \mathbb{R},
\label{Eq:GenInvGauBis}
\eeq 
and $\Kc_c(\circ)$ is the modified Bessel function of the second kind. So, for solving Equation~\eqref{Eq:MargLPMBis1} we can identify a $\Gc\Ic\Gc$ inside the integral: 
\beq
I = \int \rv_{\rf_\rj}^{-\frac{1}{2}} \exp \left\lbrace -\frac{1}{2} \left(2 \lambda \rv_{\rf_\rj} + \frac{\rf_\rj^{2}}{\rv_{\rf_\rj}} \right) \right\rbrace \d \rv_{\rf_\rj}
= 
2\Kc_{\frac{1}{2}}\left(\sqrt{2\lambda\rf_\rj^2}\right) \left( \frac{2\lambda}{\rf_\rj^2} \right)^{-\frac{1}{4}} \underbrace{\int \Gc\Ic\Gc \left( \rv_{\rf_\rj} | 2\lambda, \rf_\rj^2, \frac{1}{2} \right) \d \rv_{\rf_\rj}}_{1}
\label{Eq:LapIntegralBis}
\eeq 
So, the marginal from Equation~\eqref{Eq:MargLPMBis1} can be written as:
\beq
p(\rf_\rj|\rv_{\rf_\rj},b) = 
\left( 2 \pi \right)^{-\frac{1}{2}}
\;
\lambda
\;
\left( 2 \lambda \right)^{-\frac{1}{4}}
\;
\rf_\rj^{\frac{1}{2}}  
\;
2
\;
\Kc_{\frac{1}{2}} \left( \sqrt{2\lambda\rf_\rj^2} \right) 
=
2^{\frac{1}{4}}
\;
\pi^{-\frac{1}{2}}
\;
\lambda^{\frac{3}{4}}
\;
\rf_\rj^{\frac{1}{2}}  
\;
\Kc_{\frac{1}{2}} \left( \sqrt{2\lambda\rf_\rj^2} \right) 
\label{Eq:MargLPMBis1_1}
\eeq 
The following equality stands:
\beq
\Kc_{\frac{1}{2}}\left(\sqrt{2\lambda\rf_\rj^2}\right) = \sqrt{\frac{\pi}{2}} \; \left( 2\lambda \right)^{\frac{-1}{4}} \; \rf_\rj^{-\frac{1}{2}} \; \exp \left\lbrace -\sqrt{2\lambda\rf_\rj^2}  \right\rbrace
\label{Eq:BesselIdentBis}
\eeq 
From Equation~\eqref{Eq:MargLPMBis1_1} and Equation~\eqref{Eq:BesselIdentBis} the marginal from Equation~\eqref{Eq:MargLPMBis1} becomes:
\beq
p(\rf_\rj|\rv_{\rf_\rj},\lambda) 
= 
2^{\frac{1}{4}}
\;
\pi^{-\frac{1}{2}}
\;
\lambda^{\frac{3}{4}}
\;
\rf_\rj^{\frac{1}{2}}  
\;
\pi^{\frac{1}{2}}
\;
2^{-\frac{1}{2}}
\; 
2^{\frac{-1}{4}} 
\; 
\lambda^{\frac{-1}{4}} 
\; 
\rf_\rj^{-\frac{1}{2}} 
\;
\exp \left\lbrace -\sqrt{2\lambda\rf_\rj^2} \right\rbrace
=
2^{-\frac{1}{2}}
\lambda^{\frac{1}{2}} \;
\exp \left\lbrace -\sqrt{2\lambda\rf_\rj^2} \right\rbrace
\eeq
So, we conclude that the marginal is a Laplace distribution, $p(\rf_\rj|\rv_{\rf_\rj},\lambda) = \Lc \left( \rf_\rj | 0 \; \left( 2\lambda \right)^{-\frac{1}{2}} \right)$:
\beq
p(\rf_\rj|\rv_{\rf_\rj},\lambda) 
= 
\frac{1}{2 (2\lambda)^{-\frac{1}{2}}} \;
\exp \left\lbrace - \frac{|\rf_\rj|}{(2\lambda)^{-\frac{1}{2}}} \right\rbrace
=
\Lc \left( \rf_\rj | 0 \; \left( 2\lambda \right)^{-\frac{1}{2}} \right)
\eeq
\subsubsection{Laplace prior: Normal and Inverse Gamma}
\label{Subsubsec:LaplacePrior}
We start with the linear model expressed in Equation~\eqref{Eq:LinearModel}. We consider the Laplace Prior Model (LPM), Equation~\eqref{Eq:LPM1}, which considers a zero-mean Normal distribution for $\rf_\rj|\rv_{\rf_\rj}$ and the inverse of the variance. The variance $\rv_{\rf_\rj}|b$ is modelled as an Inverse Gamma distribution, with the corresponding shape parameter $\alpha$ equal to $1$:
\beq
\textbf{LPM:}\;\;\;\;\;\;\;
\left\{\barr{ll}
p(\rf_\rj|0,\rv_{\rf_\rj}) = \Nc(\rf_\rj|0, \; \rv_{\rf_\rj}^{-1}) = \left( 2 \pi \right)^{-\frac{1}{2}} \; \rv_{\rf_\rj}^{\frac{1}{2}} \; \exp \left\lbrace -\frac{\rv_{\rf_\rj}}{2} \; \rf_\rj^2 \right\rbrace
\\[7pt]
p(\rv_{\rf_\rj}|b) = \Ic\Gc(\rv_{\rf_\rj}|1, \; \frac{b}{2}) = \frac{b}{2} \; \rv_{\rf_\rj}^{-2} \; \exp \left\lbrace -\frac{b}{2} \; \rv_{\rf_\rj}^{-1} \right\rbrace,\\[4pt]
\earr\right.
\label{Eq:LPM1}
\eeq
The expression of the joint probability distribution $\rf_\rj,\rv_{\rf_\rj}|b$ is given by Equation~\eqref{Eq:JointLPM1},
\beq
p(\rf_\rj,\rv_{\rf_\rj}|b) = \Nc(\rf_\rj|0, \; \rv_{\rf_\rj}^{-1}) \; \Ic\Gc(\rv_{\rf_\rj}|1, \; \frac{b}{2}),
\label{Eq:JointLPM1}
\eeq 
so the marginal is, Equation~\eqref{Eq:MargLPM1}:
\beq
p(\rf_\rj|\rv_{\rf_\rj},b) = \left( 2 \pi \right)^{-\frac{1}{2}} \frac{b}{2} \int \rv_{\rf_\rj}^{-\frac{1}{2}-1} \exp \left\lbrace -\frac{1}{2} \left( \rf_\rj^{2} \rv_{\rf_\rj} + b \rv_{\rf_\rj}^{-1} \right) \right\rbrace \d \rv_{\rf_\rj}
\label{Eq:MargLPM1}
\eeq 
The probability distribution of generalized inverse Gaussian distribution is presented in Equation~\eqref{Eq:GenInvGau} 
\beq
\Gc\Ic\Gc(x|a,b,c) = \frac{\left(a/b\right)^{\frac{c}{2}}}{2\Kc_c(\sqrt{ab})} \; x^{c-1} \; \exp \left\lbrace -\frac{1}{2} \left( a x + b x^{-1} \right) \right\rbrace, \; a, b \geq 0, \; c \in \mathbb{R},
\label{Eq:GenInvGau}
\eeq 
and $\Kc_c(\circ)$ is the modified Bessel function of the second kind. So, for solving Equation~\eqref{Eq:MargLPM1} we can identify a $\Gc\Ic\Gc$ inside the integral: 
\beq
I = \int \rv_{\rf_\rj}^{-\frac{1}{2}-1} \exp \left\lbrace -\frac{1}{2} \left( \rf_\rj^{2} \rv_{\rf_\rj} + b \rv_{\rf_\rj}^{-1} \right) \right\rbrace \d \rv_{\rf_\rj}
= 
\frac{2\Kc_{-\frac{1}{2}}(\sqrt{b\rf_\rj^2})}{\left( \frac{\rf_\rj^2}{b} \right)^{-\frac{1}{4}}} \underbrace{\int \Gc\Ic\Gc \left( \rv_{\rf_\rj} | \rf_\rj^2, b, -\frac{1}{2} \right) \d \rv_{\rf_\rj}}_{1}
\label{Eq:LapIntegral}
\eeq 
So, the marginal from Equation~\eqref{Eq:MargLPM1} can be written as:
\beq
p(\rf_\rj|\rv_{\rf_\rj},b) = \left( 2 \pi \right)^{-\frac{1}{2}} \; \frac{b}{2} \; \rv_{\rf_\rj}^{-\frac{1}{2}-1} \; \left( \frac{\rf_\rj^2}{b} \right)^{\frac{1}{4}} \; 2 \; \Kc_{-\frac{1}{2}}\left(\sqrt{b\rf_\rj^2}\right) = 
\left( 2 \pi \right)^{-\frac{1}{2}}  \; b^{\frac{3}{4}}  \; \rf_\rj^{\frac{1}{2}} \; \Kc_{-\frac{1}{2}} \left( \sqrt{b\rf_\rj^2}\right)
\label{Eq:MargLPM1_1}
\eeq 
The following equality stands:
\beq
\Kc_{-\frac{1}{2}}\left(\sqrt{b\rf_\rj^2}\right) = \sqrt{\frac{\pi}{2}} \; b^{\frac{-1}{4}} \; \rf_\rj^{-\frac{1}{2}} \exp \left\lbrace -\sqrt{b \rf_\rj^2} \right\rbrace
\label{Eq:BesselIdent}
\eeq 
From Equation~\eqref{Eq:MargLPM1_1} and Equation~\eqref{Eq:BesselIdent} the marginal from Equation~\eqref{Eq:MargLPM1} becomes:
\beq
p(\rf_\rj|\rv_{\rf_\rj},b) = 
2^{-\frac{1}{2}} \; 
\pi^{-\frac{1}{2}} \;
b^{\frac{3}{4}} \; 
\rf_\rj^{\frac{1}{2}} \; 
\pi^{\frac{1}{2}} \;
2^{-\frac{1}{2}} \;
b^{-\frac{1}{4}} \; 
\rf_\rj^{-\frac{1}{2}} \; 
\exp \left\lbrace -\sqrt{b \rf_\rj^2} \right\rbrace
=
\frac{1}{2} \;
\sqrt{b} \;
\exp \left\lbrace -\sqrt{b \rf_\rj^2} \right\rbrace
\eeq
So, we conclude that the marginal is a Laplace distribution, $p(\rf_\rj|\rv_{\rf_\rj},b) = \Lc \left( \rf_\rj | 0, \; b^{-\frac{1}{2}} \right)$:
\beq
p(\rf_\rj|\rv_{\rf_\rj},b) 
= 
\frac{1}{2 b^{-\frac{1}{2}}} \;
\exp \left\lbrace - \frac{|\rf_\rj|}{b^{-\frac{1}{2}}} \right\rbrace
=
\Lc \left( \rf_\rj | 0, \; b^{-\frac{1}{2}} \right)
\eeq
\subsubsection{Laplace prior: via the Generalized Hyperbolic distribution}
\label{Subsubsec:GenHyperbolicAsLap}
\begin{figure}[!htb]
        \centering
        \begin{subfigure}{0.49\textwidth}
          		\centering
                \includegraphics[width=7.4cm,height=4cm]{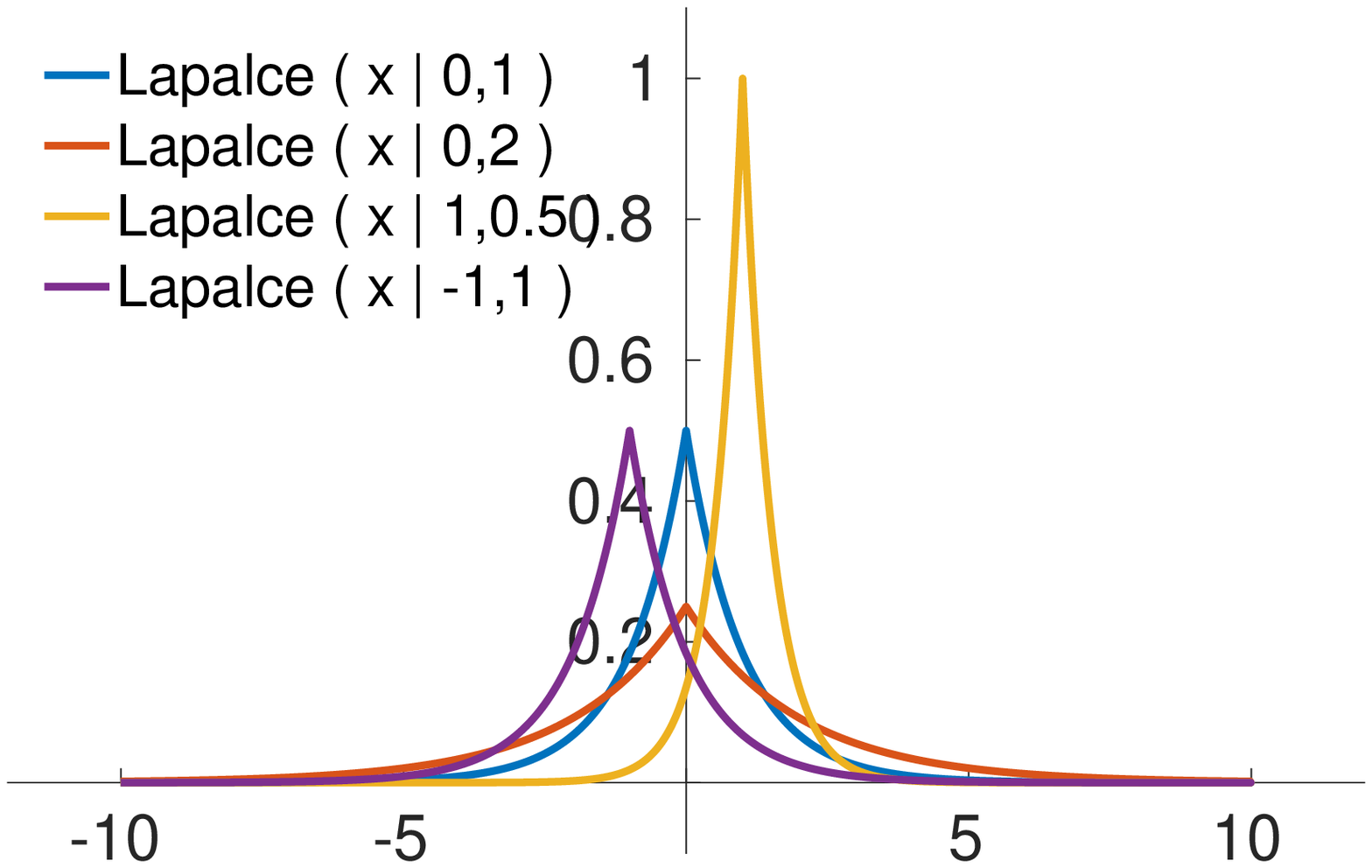}
                \caption{$ \Lc \left( x | \mu, b \right) $}
                \label{fig:L_GenHypAsMLap}
        \end{subfigure}
        \begin{subfigure}{0.49\textwidth}
          		\centering        
                \includegraphics[width=7.4cm,height=4cm]{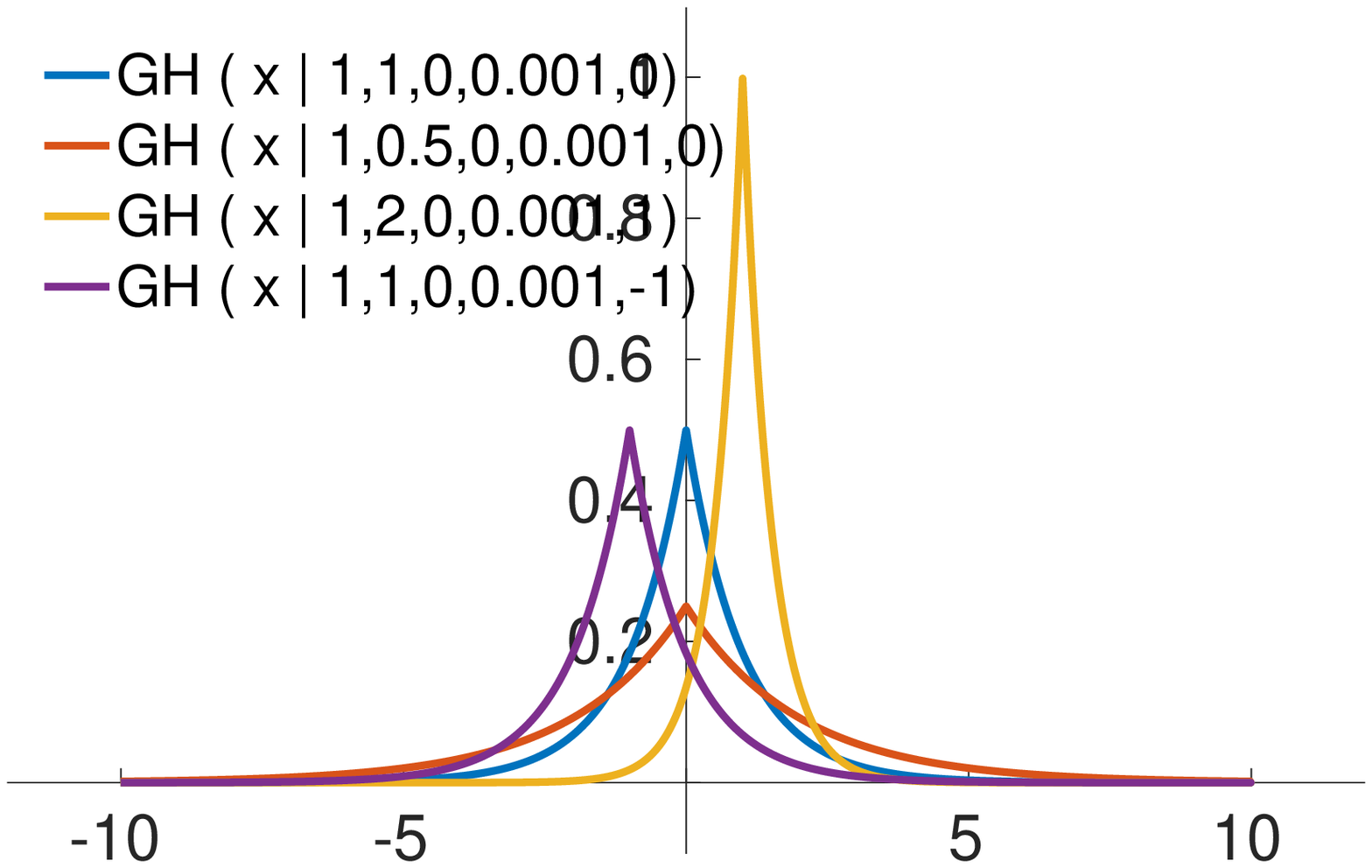}
                \caption{\small{$ \Gc\Hc(x|\lambda = 1, \alpha = b^{-1}, \beta = 0, \delta \searrow  0, \mu) $}}
                \label{fig:GHL_GenHypAsMLap}
        \end{subfigure}        
\par\bigskip       
        \begin{subfigure}{0.49\textwidth}
          		\centering        
                \includegraphics[width=7.4cm,height=4cm]{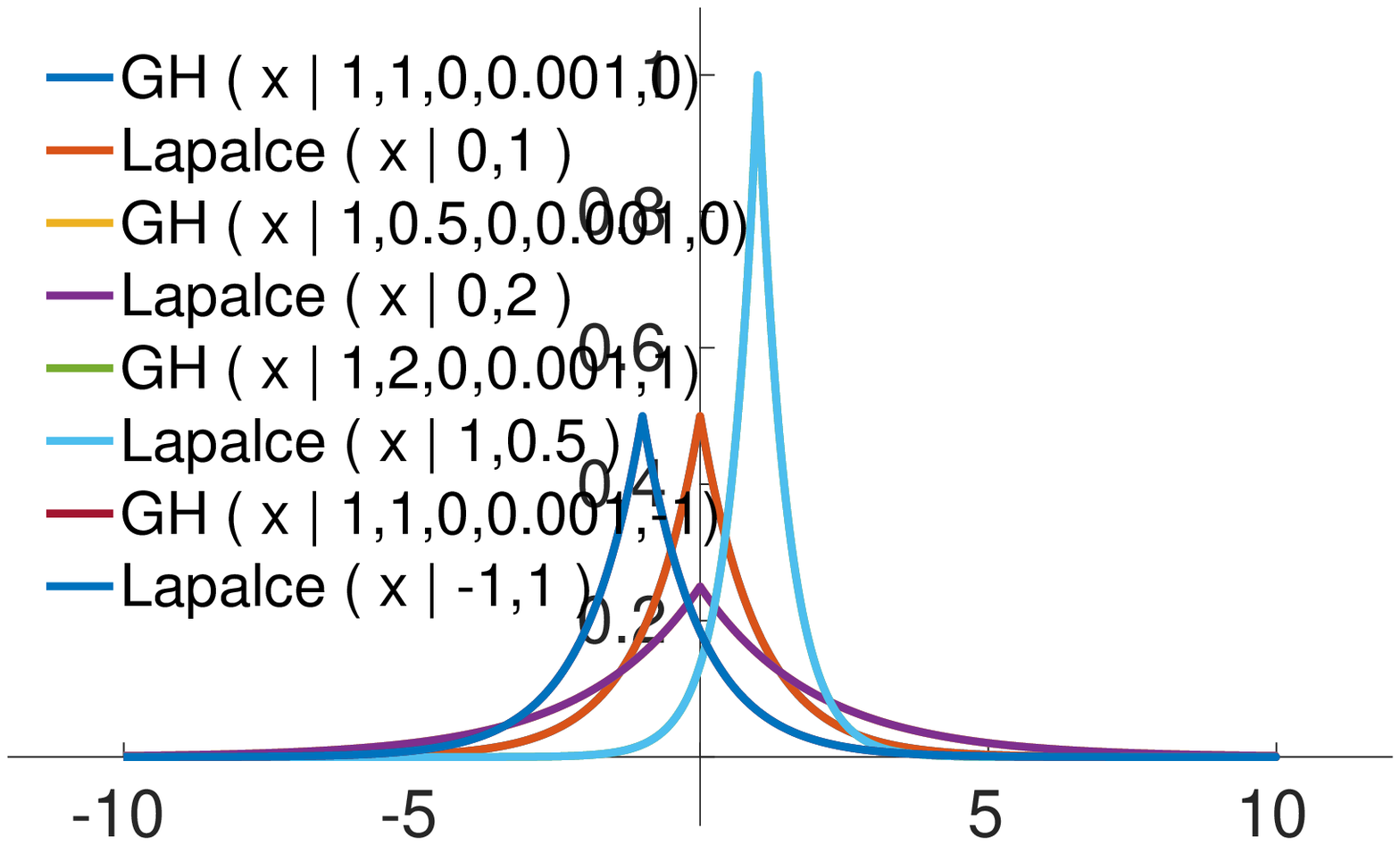}
                \caption{$ \Lc $ vs. $ \Gc\Hc $}
                \label{fig:GHLVsL_GenHypAsMLap}
        \end{subfigure}                               
        \begin{subfigure}{0.49\textwidth}
          		\centering        
                \includegraphics[width=7.4cm,height=4cm]{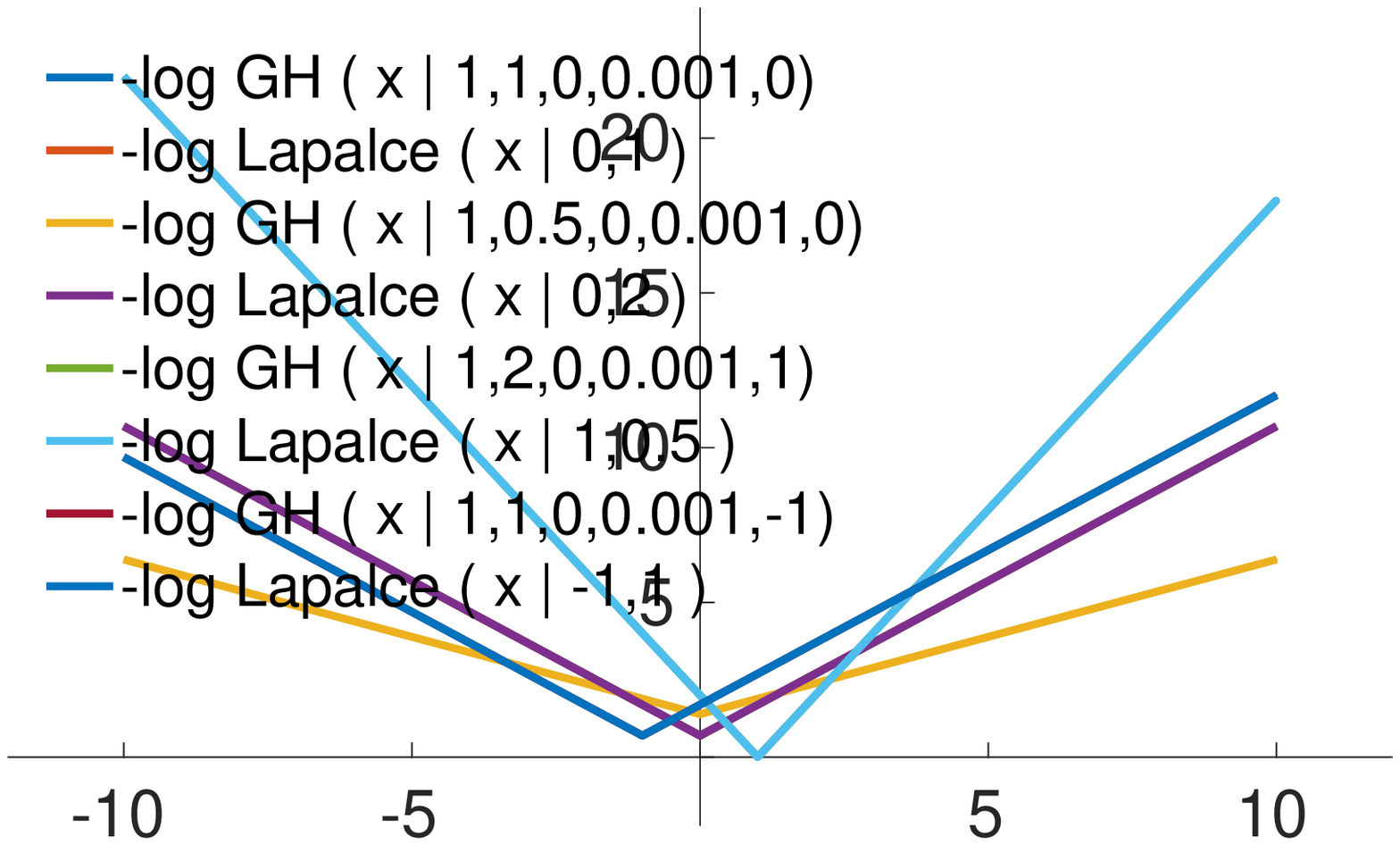}
                \caption{$ - \log \Lc $ vs. $ - \log \Gc\Hc $}
                \label{fig:Log_GHLVsL_GenHypAsMLap}
        \end{subfigure}                                         
        \caption{Four Laplace probability density functions with different means values ($\mu=0$, blue and red, $\mu=1$, yellow and $\mu=-1$, violet) and different scale parameter values ($b=1$, blue and violet, $b=2$, red and $b=0.5$, yellow), Figure~\eqref{fig:L_GenHypAsMLap} and the corresponding Generalized Hyperbolic probability density functions, with parameters set as in Equation~\eqref{Eq:GenHyperbolicDistributionAsLaplace4}, with $\delta = 0.001$, Figure~\eqref{fig:GHL_GenHypAsMLap}. Comparison between the distributions, $ \Lc $ vs. $\Gc\Hc$, Figure~\eqref{fig:GHLVsL_GenHypAsMLap} and between the logarithm of the distributions $\log \Lc $ vs. $\log \Gc\Hc$, Figure~\eqref{fig:Log_GHLVsL_GenHypAsMLap}.}
        \label{fig:Comp_GenHypAsMLap}
\end{figure}

\begin{figure}[!htb]
        \centering
        \begin{subfigure}{0.49\textwidth}
          		\centering        
                \includegraphics[width=7.4cm,height=4cm]{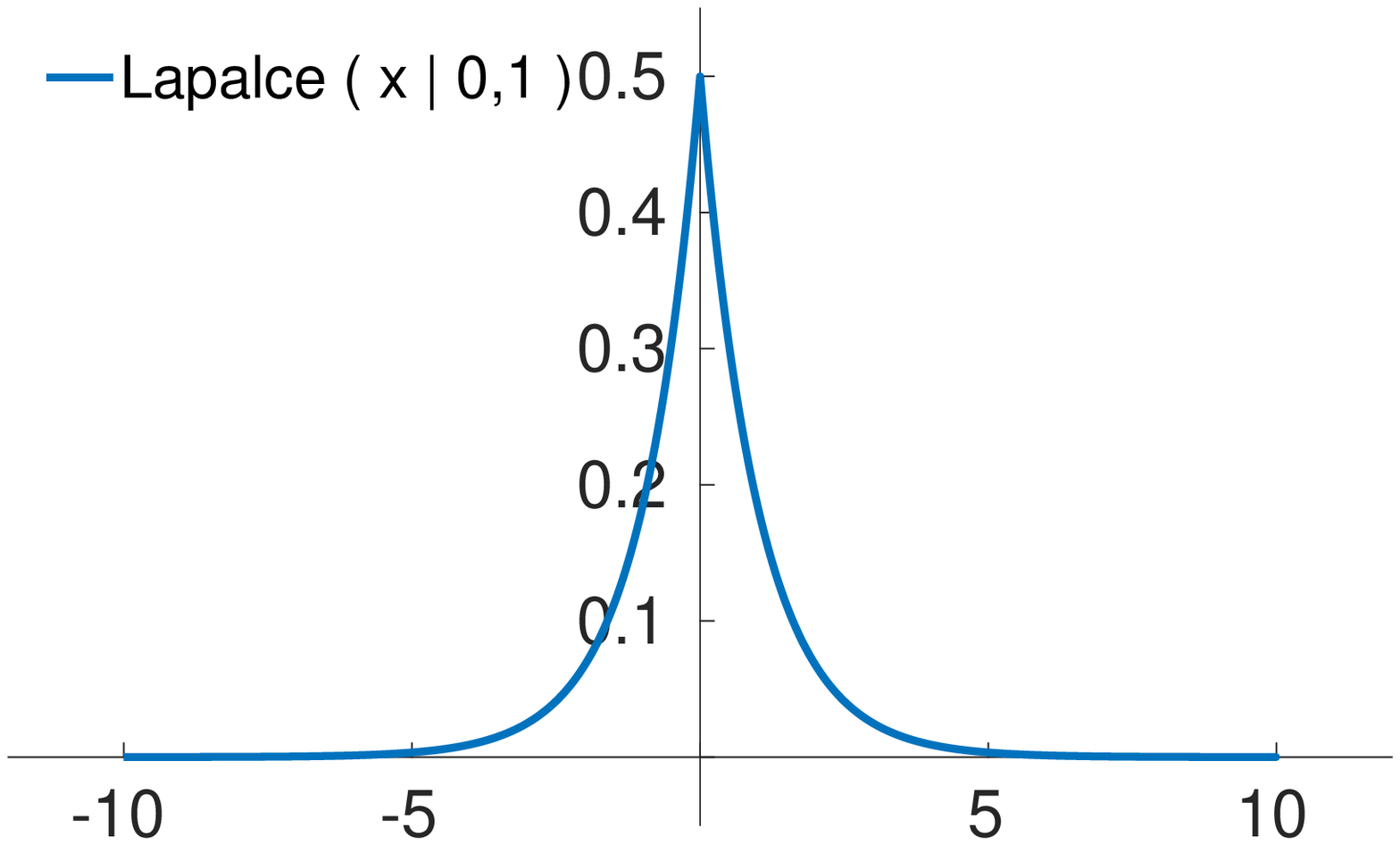}
                \caption{$ \Lc \left( x | \mu = 0, b =1 \right)$}
                \label{fig:L_GenHypAsL}
        \end{subfigure}
        \begin{subfigure}{0.49\textwidth}
          		\centering        
                \includegraphics[width=7.4cm,height=4cm]{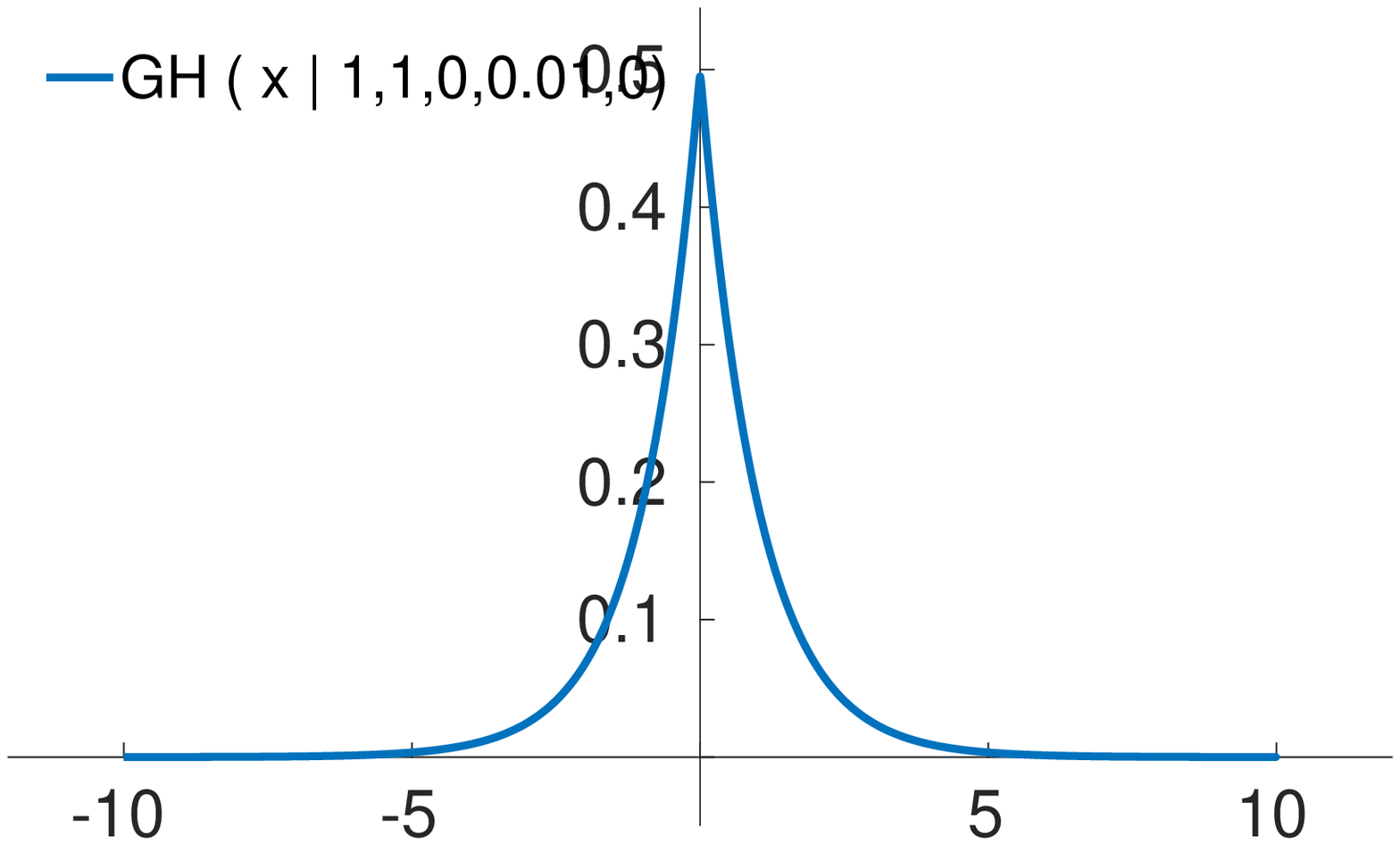}
                \caption{\small{$ \Gc\Hc(x | \lambda = 1, \alpha = b^{-1} = 1, \beta = 0, \delta = 0.01, \mu = 0) $}}
                \label{fig:GHL_GenHypAsL}
        \end{subfigure}        
\par\bigskip        
        \begin{subfigure}{0.49\textwidth}
          		\centering        
                \includegraphics[width=7.4cm,height=4cm]{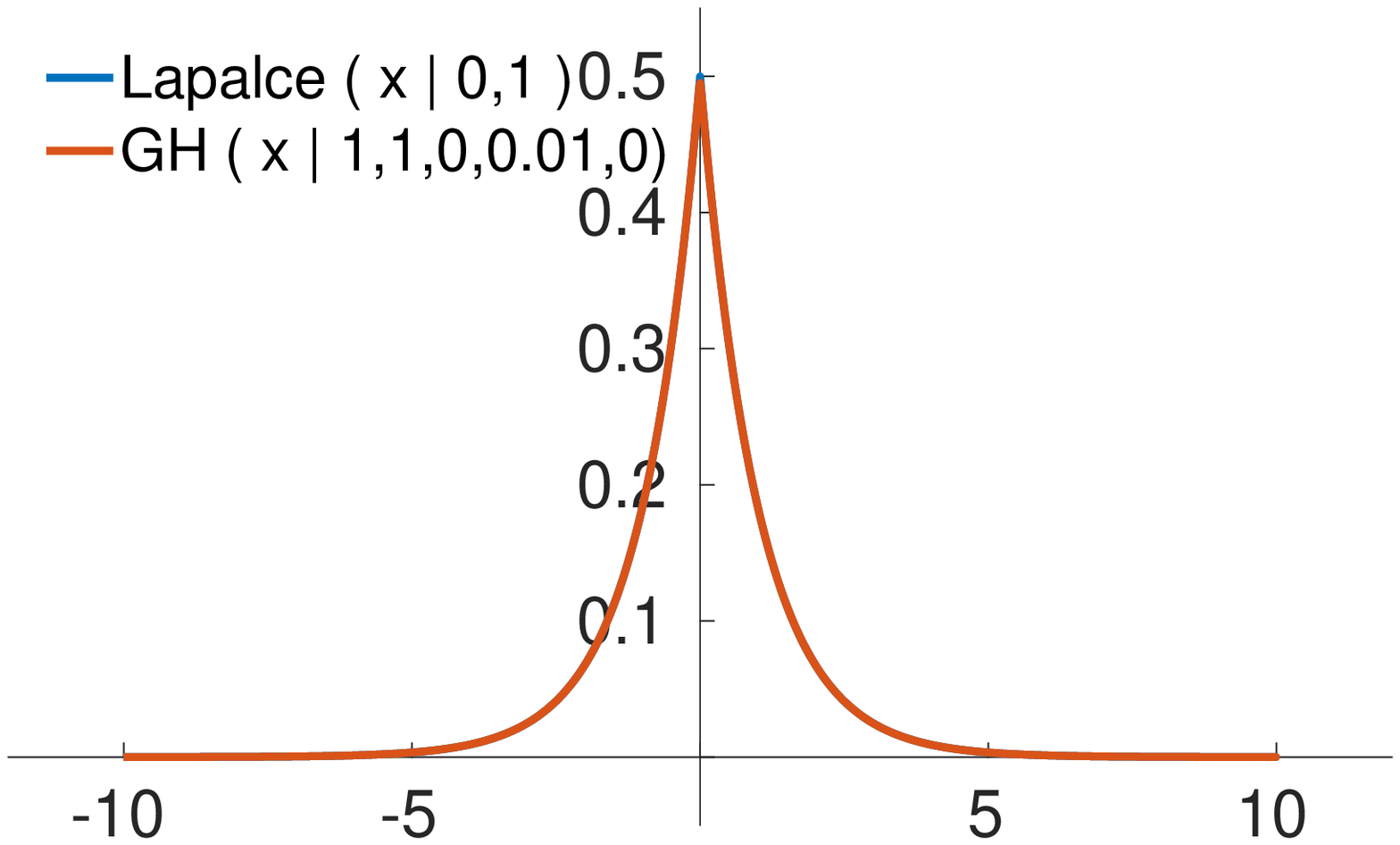}
                \caption{$ \Lc $ vs. $ \Gc\Hc $}
                \label{fig:GHLVsL_GenHypAsL}
        \end{subfigure}                               
        \begin{subfigure}{0.49\textwidth}
          		\centering        
                \includegraphics[width=7.4cm,height=4cm]{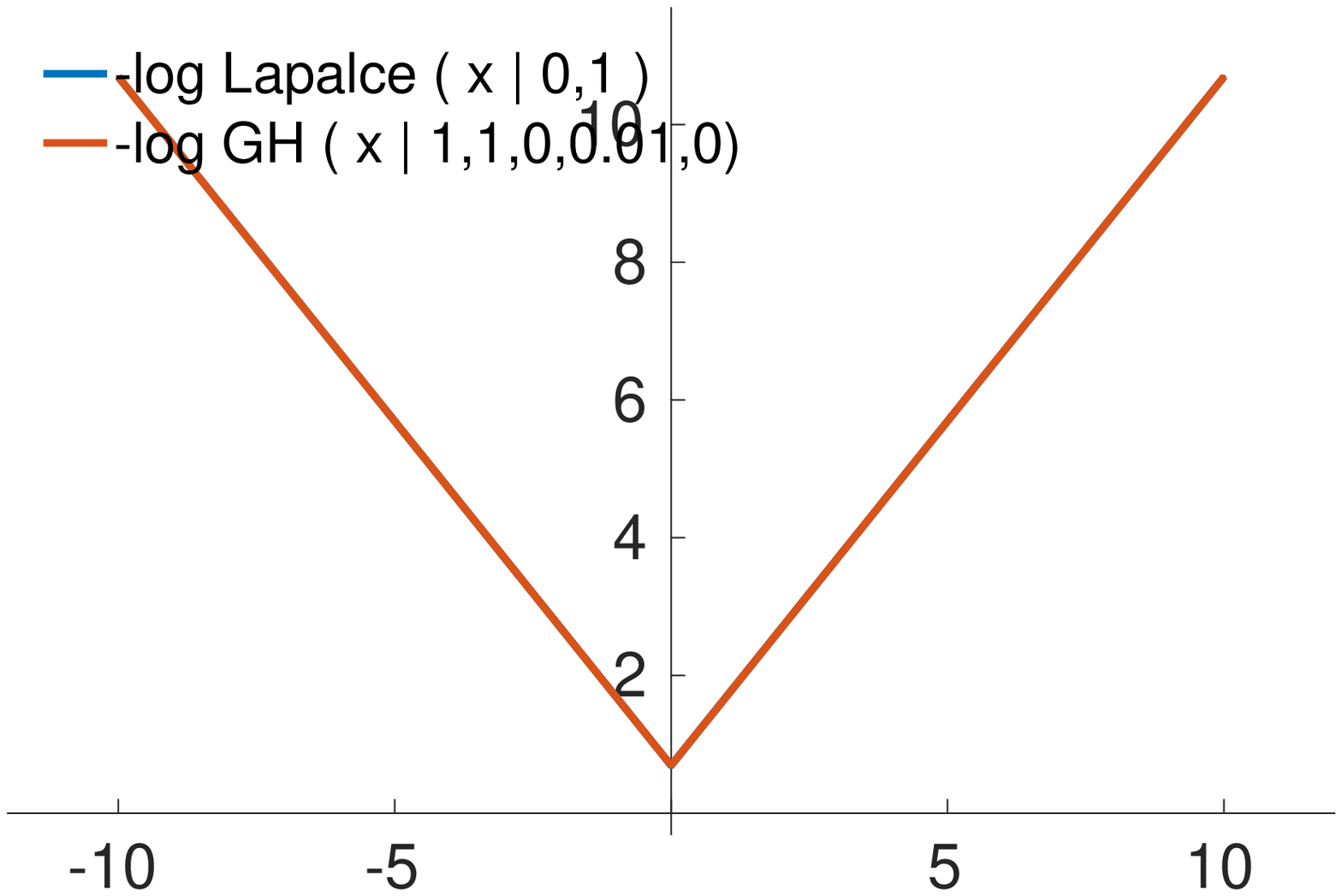}
                \caption{$- \log \Lc $ vs. $- \log \Gc\Hc $}
                \label{fig:Log_GHLVsL_GenHypAsL}
        \end{subfigure}                                       
        \caption{The standard Laplace distribution, Figure~\eqref{fig:L_GenHypAsL} and the Generalized Hyperbolic distribution, with parameters set as in Equation~\eqref{Eq:GenHyperbolicDistributionAsLaplace4}, Figure~\eqref{fig:GHL_GenHypAsL}. Comparison between the two distribution, $ \Lc $ vs. $ \Gc\Hc $, Figure~\eqref{fig:GHLVsL_GenHypAsL} and between the logarithm of the distributions $- \log \Lc $ vs. $- \log \Gc\Hc $, Figure~\eqref{fig:Log_GHLVsL_GenHypAsL}.}
        \label{fig:Comp_GenHypAsL}
\end{figure}

\begin{figure}[!htb]
        \centering
        \begin{subfigure}{0.49\textwidth}
          		\centering        
                \includegraphics[width=7.4cm,height=4cm]{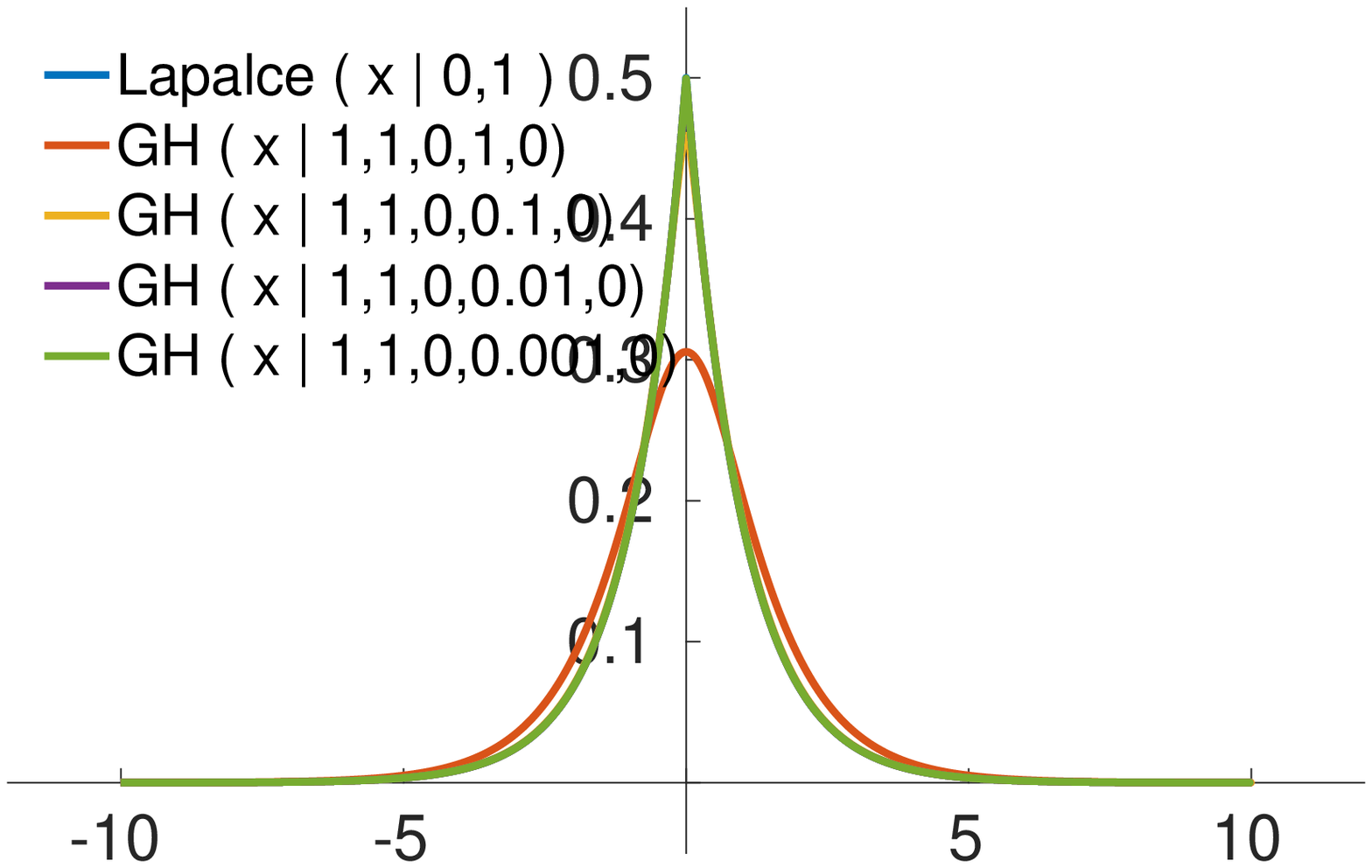}
                \caption{$ \Lc $ vs. $ \Gc\Hc$ for $\delta \in \left\lbrace 1, 0.1, 0.01, 0.001 \right\rbrace$}
                \label{fig:GHLVsL_GenHypAsL1}
        \end{subfigure}                               
        \begin{subfigure}{0.49\textwidth}
          		\centering        
                \includegraphics[width=7.4cm,height=4cm]{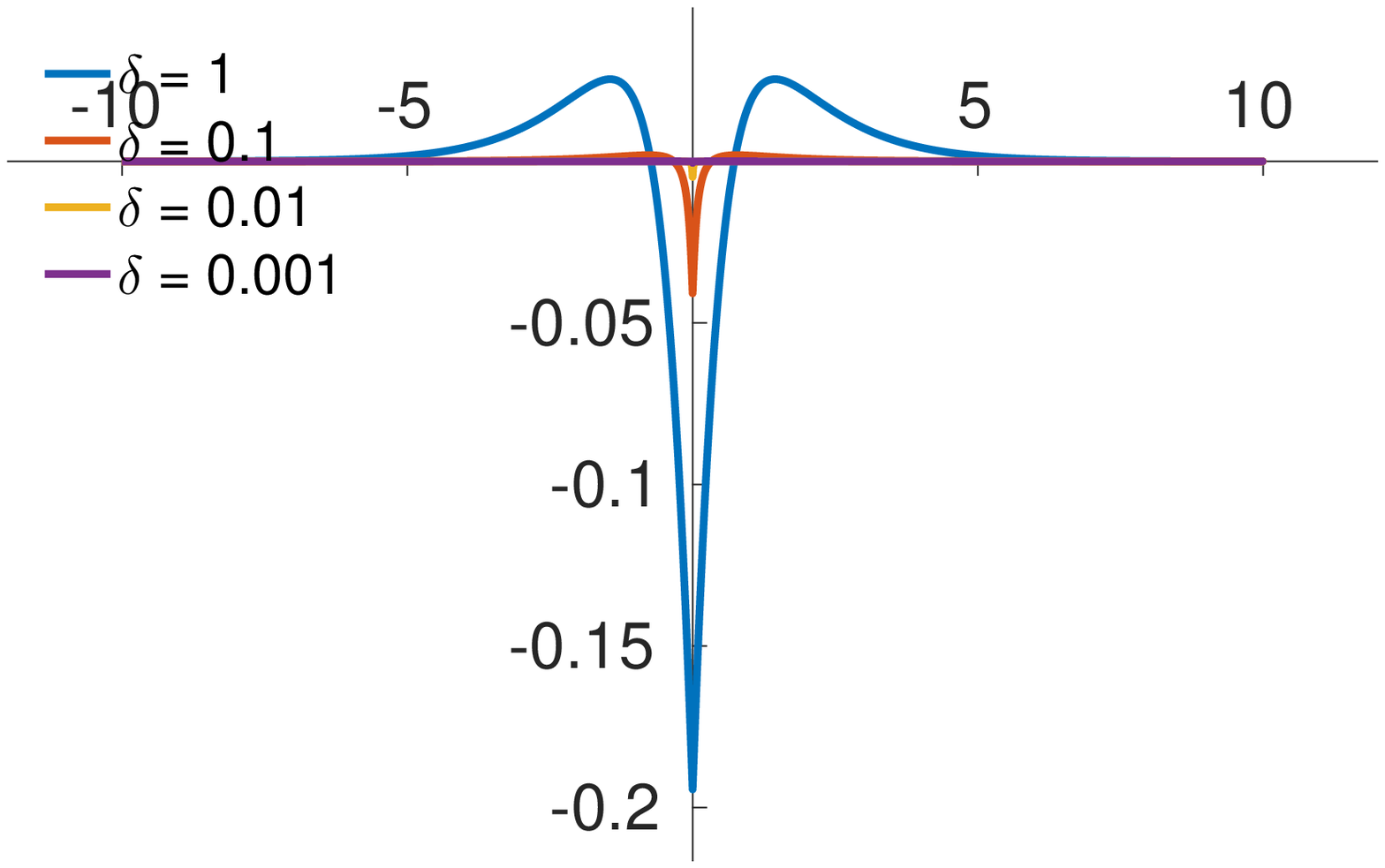}
                \caption{$ \Gc\Hc - \Lc $ for $\delta \in \left\lbrace 1, 0.1, 0.01, 0.001 \right\rbrace$}
                \label{fig:GHLMinusL_GenHypAsL1}
        \end{subfigure}                                         
        \caption{The behaviour of the Generalized Hyperbolic density function $\Gc\Hc (x | \lambda = 1, \alpha = b^{-1}, \beta = 0, \delta \searrow 0, \mu)$ depending on $\delta$.}
        \label{fig:Comp_GenHypAsL1}
\end{figure}

$X \sim \Gc\Hc(x|\lambda = 1, \alpha = b^{-1}, \beta = 0, \delta \searrow  0, \mu) $ has a Laplace distribution, $\Lc \left( x | \mu, b \right)$.\\
The goal of this section is to derive the Laplace distribution from the Generalized Hyperbolic distribution. The Laplace distribution has the following probability density function:
\beq
\Lc \left( x | \mu, b \right) = \frac{1}{2b} \exp \left\lbrace -\frac{\lvert x-\mu \rvert}{b} \right\rbrace
\eeq
Considering $\lambda=1$ in the expression of the Generalized Hyperbolic probability density function, a Hyperbolic distribution is obtained, Subsection~\eqref{Subsubsec:GenHyperbolicAsHyp}. Fixing $\beta=0$ implies $\gamma=\alpha$ and the expression of the probability density function becomes:
\beq
\begin{split}
\Gc\Hc(x|\lambda = 1, \alpha, \beta = 0, \delta ,\mu) 
= 
\Hc(x|\alpha, \beta = 0, \delta ,\mu) 
=
\frac{1}{2\delta\Kc_{1}\left( \delta \alpha \right)}
\exp 
\left\lbrace 
- \alpha \sqrt{\delta^2 + \left( x-\mu \right)^2} 
\right\rbrace
\end{split}
\label{Eq:GenHyperbolicDistributionAsLaplace}
\eeq
Further, considering $\delta \searrow  0$, for the expression of the modified Bessel function of the second degree $\Kc_{1}\left( \delta \alpha \right)$ the asymptotic relation for small arguments presented in Equation~\eqref{Eq:ModifiedBesselAsymptoticRelation} can be used, obtaining:
\beq
\Kc_{1}\left( \delta \alpha \right) = \alpha^{-1} \delta^{-1}, \mbox{ for } \delta \searrow  0.
\label{Eq:GenHyperbolicDistributionAsLaplace2}
\eeq
Using Equation~\eqref{Eq:GenHyperbolicDistributionAsLaplace2} in Equation~\eqref{Eq:GenHyperbolicDistributionAsLaplace}, the expression of the probability density function becomes:
\beq
\begin{split}
\Gc\Hc(x|\lambda = 1, \alpha, \beta = 0, \delta \searrow  0, \mu) 
= 
\Hc(x|\alpha, \beta = 0, \delta \searrow  0, \mu) 
=
\frac{1}{2\alpha^{-1}}
\exp 
\left\lbrace 
- \frac{\lvert x-\mu \rvert}{\alpha^{-1}}  
\right\rbrace.
\end{split}
\label{Eq:GenHyperbolicDistributionAsLaplace3}
\eeq
Finally, for $\alpha = b^{-1}$, the standard Laplace probability density function is obtained:
\beq
\begin{split}
\Gc\Hc(x|\lambda = 1, \alpha = b^{-1}, \beta = 0, \delta \searrow  0, \mu) 
= 
\Hc(x|\alpha = b^{-1}, \beta = 0, \delta \searrow  0, \mu) 
=
\frac{1}{2b}
\exp 
\left\lbrace 
- \frac{\lvert x-\mu \rvert}{b}  
\right\rbrace
=
\Lc \left( x | \mu, b \right)
\end{split}
\label{Eq:GenHyperbolicDistributionAsLaplace4}
\eeq
Figure~\eqref{fig:L_GenHypAsMLap} presents four Laplace probability density functions $\Lc \left( x | \mu, b \right)
$ with different means values ($\mu=0$, blue and red, $\mu=1$, yellow and $\mu=-1$, violet) and different scale parameter values ($b=1$, blue and violet, $b=2$, red and $b=0.5$, yellow).
Figure~\eqref{fig:GHL_GenHypAsMLap} presents the Generalized Hyperbolic probability density function for parameters $\Gc\Hc(x|\lambda = 1, \alpha = b^{-1}, \beta = 0, \delta \searrow  0, \mu)$ with $\mu$ and $b$ set with the same numerical values as the corresponding ones from Figure~\eqref{fig:L_GenHypAsMLap}. For $\delta$, the numerical values is $0.001$.
Figure~\eqref{fig:GHLVsL_GenHypAsMLap} presents the comparison between the four Laplace probability density functions and the corresponding Generalized Hyperbolic probability density functions, $ \Lc $ vs. $ \Gc\Hc $. In all forth cases the probability density functions are superposed. 
Figure~\eqref{fig:Log_GHLVsL_GenHypAsMLap} presents the comparison between the logarithm of distributions, $ - \log \Lc $ vs. $ - \log \Gc\Hc$.  
Figure~\eqref{fig:GHLVsL_GenHypAsL} presents the comparison between the standard Laplace probability density function, $\Lc \left( x | \mu = 0, b = 1 \right)$ (reported in Figure~\eqref{fig:L_GenHypAsL}) and the Generalized Hyperbolic density function for parameters $\Gc\Hc (x | \lambda = 1, \alpha = b^{-1} = 1, \beta = 0, \delta = 0.01, \mu = 0)$ (presented in Figure~\eqref{fig:GHL_GenHypAsL}), showing the two distributions superposed. Figure~\eqref{fig:Log_GHLVsL_GenHypAsL} presents the comparison between the logarithm of the two distributions, $- \log \Lc$ vs. $- \log \Gc\Hc$.
The behaviour of the Generalized Hyperbolic density function $\Gc\Hc (x | \lambda = 1, \alpha = b^{-1} = 1, \beta = 0, \delta \searrow 0, \mu = 0)$ depending on $\delta$ is presented in Figure~\eqref{fig:GHLVsL_GenHypAsL1}: a comparison between the Laplace probability density function (in blue) and the Generalized Hyperbolic density function $\Gc\Hc (x | \lambda = 1, \alpha = b^{-1} = 1, \beta = 0, \delta \searrow 0, \mu = 0)$ for $\delta=1$ (in red), $\delta=0.1$ (in yellow), $\delta=0.01$ (in violet) and $\delta=0.001$ (in green) is presented in Figure~\eqref{fig:GHLVsL_GenHypAsL1}. The difference between those four Generalized Hyperbolic density functions and the Laplace density function is presented in Figure~\eqref{fig:GHLMinusL_GenHypAsL1}. 
The Laplace distribution can be expressed using the Generalized Hyperbolic Prior Model (GHPM), Equation~\eqref{Eq:GHPM1} by fixing $ \lambda = 1, \alpha = b^{-1}, \beta = 0, \delta \searrow 0$: 
\beq
\textbf{LPM:}\;\;\;\;
\left\{\barr{ll}
p(\rf_\rj | \mu, \; \rv_{\rf_\rj}, \; \beta)
=
\Nc\left(\rf_\rj|\mu + \beta \rv_{\rf_\rj}, \; \rv_{\rf_\rj}\right) 
=
\left( 2 \pi \right)^{-\frac{1}{2}} \; \rv_{\rf_\rj}^{-\frac{1}{2}} \; \exp \left\lbrace -\frac{1}{2}\frac{\left( x - \mu - \beta \rv_{\rf_\rj} \right)^2}{\rv_{\rf_\rj}} \right\rbrace
\\[7pt]
p(\rv_{\rf_\rj}|\gamma^2, \; \delta^2 \searrow 0, \; \lambda = 1) 
=
\Gc\Ic\Gc(\rv_{\rf_\rj}|\gamma^2, \; \delta^2 \searrow 0, \; \lambda = 1)
=
\Ec(\rv_{\rf_\rj}|\frac{1}{2 b^2})
=
\frac{1}{2 b^2}
\;
\exp \left\lbrace -\frac{1}{2 b^2} \rv_{\rf_\rj} \right\rbrace,\\[4pt]
\earr\right.
\label{Eq:GHPM_Lap}
\eeq

\subsection{Variance-Gamma distribution: expressed via conjugate priors}
\label{Subsec:VarianceGammaPrior}
The Variance-Gamma distribution, or the generalized Laplace distribution or Bessel function distribution is a continuous probability distribution that is defined as the normal variance-mean mixture where the mixing density is the gamma distribution. The tails of the distribution decrease more slowly than the normal distribution. It is therefore suitable to model phenomena where numerically large values are more probable than is the case for the normal distribution. In particular, it can be successfully used for modelling sparse phenomena. Examples are returns from financial assets and turbulent wind speeds. The distribution was introduced in the financial literature by Madan and Seneta \textcolor{americanrose}{(\textit{D.B. Madan and E. Seneta (1990): The variance gamma (V.G.) model for share market returns, Journal of Business, 63, pp. 511–524.})}. The Variance-Gamma distributions form a subclass of the generalised hyperbolic distributions. The fact that there is a simple expression for the moment generating function implies that simple expressions for all moments are available. The Variance-Gamma distribution has the following probability density function:
\beq
\Vc-\Gc(x|\alpha,\beta,\lambda,\mu) 
= 
\frac{\gamma^{2\lambda} \lvert x - \mu \rvert^{\lambda-\frac{1}{2}}\Kc_{\lambda-\frac{1}{2}} \left( \alpha \lvert x - \mu \rvert \right)
}{\sqrt{\pi} \Gamma \left( \lambda \right) \left( 2 \alpha \right)^{\lambda-\frac{1}{2}}} 
\;
\exp \left\lbrace {\beta \left( x - \mu \right)} \right\rbrace
, 
\lambda > 0, \gamma = \sqrt{\alpha^2-\beta^2};
\label{Eq:VarianceGammaDistribution}
\eeq
\subsubsection{Variance-Gamma: via Generalized Hyperbolic distribution}
\label{Subsubsec:GenHyperbolicAsVarGam}
\begin{figure}[!htb]
        \centering
        \begin{subfigure}{0.49\textwidth}
          		\centering        
                \includegraphics[width=7.4cm,height=4cm]{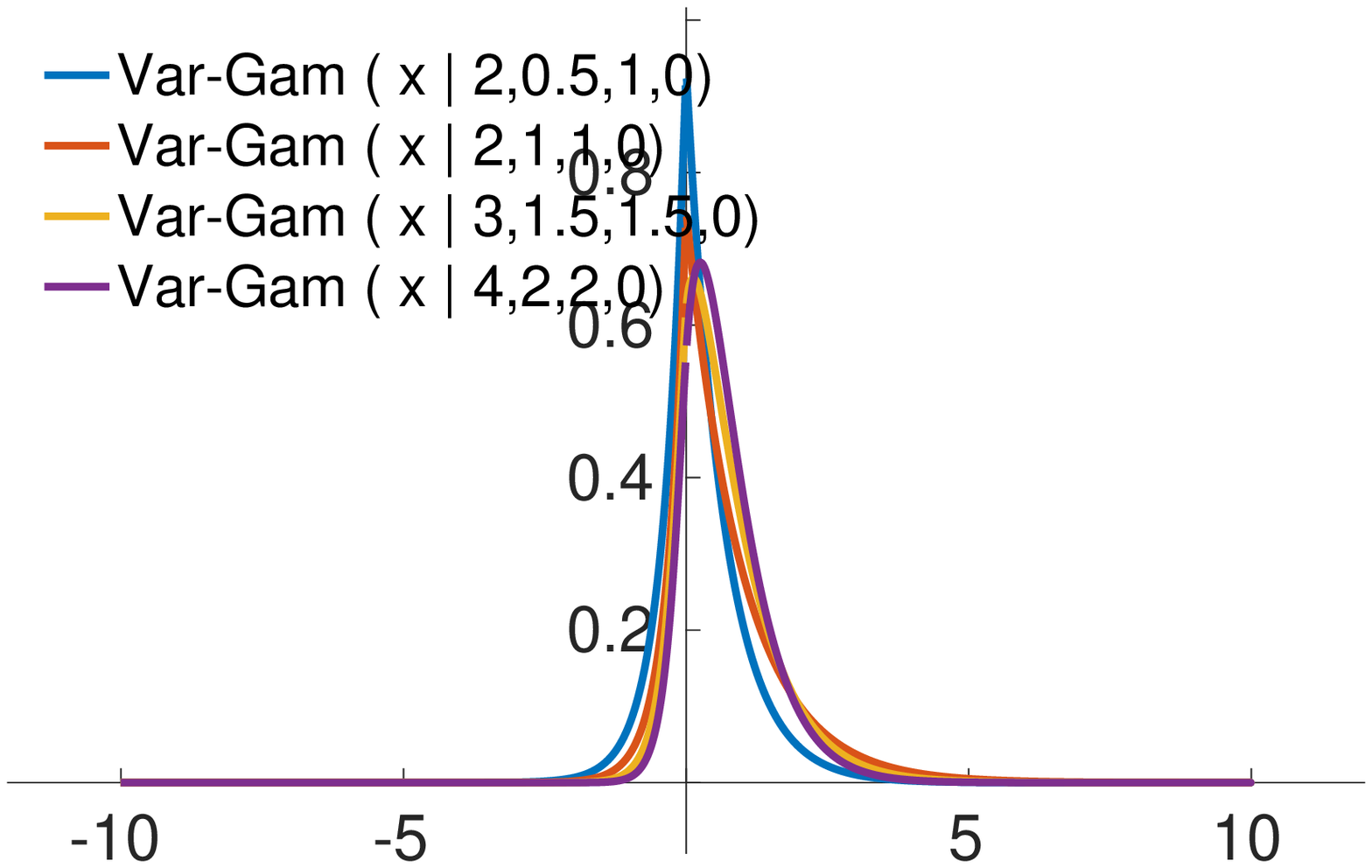}
                \caption{$ \Vc-\Gc(x|\alpha,\beta,\lambda,\mu) $}
                \label{fig:VG_GenHypAsVG}
        \end{subfigure}
        \begin{subfigure}{0.49\textwidth}
          		\centering        
                \includegraphics[width=7.4cm,height=4cm]{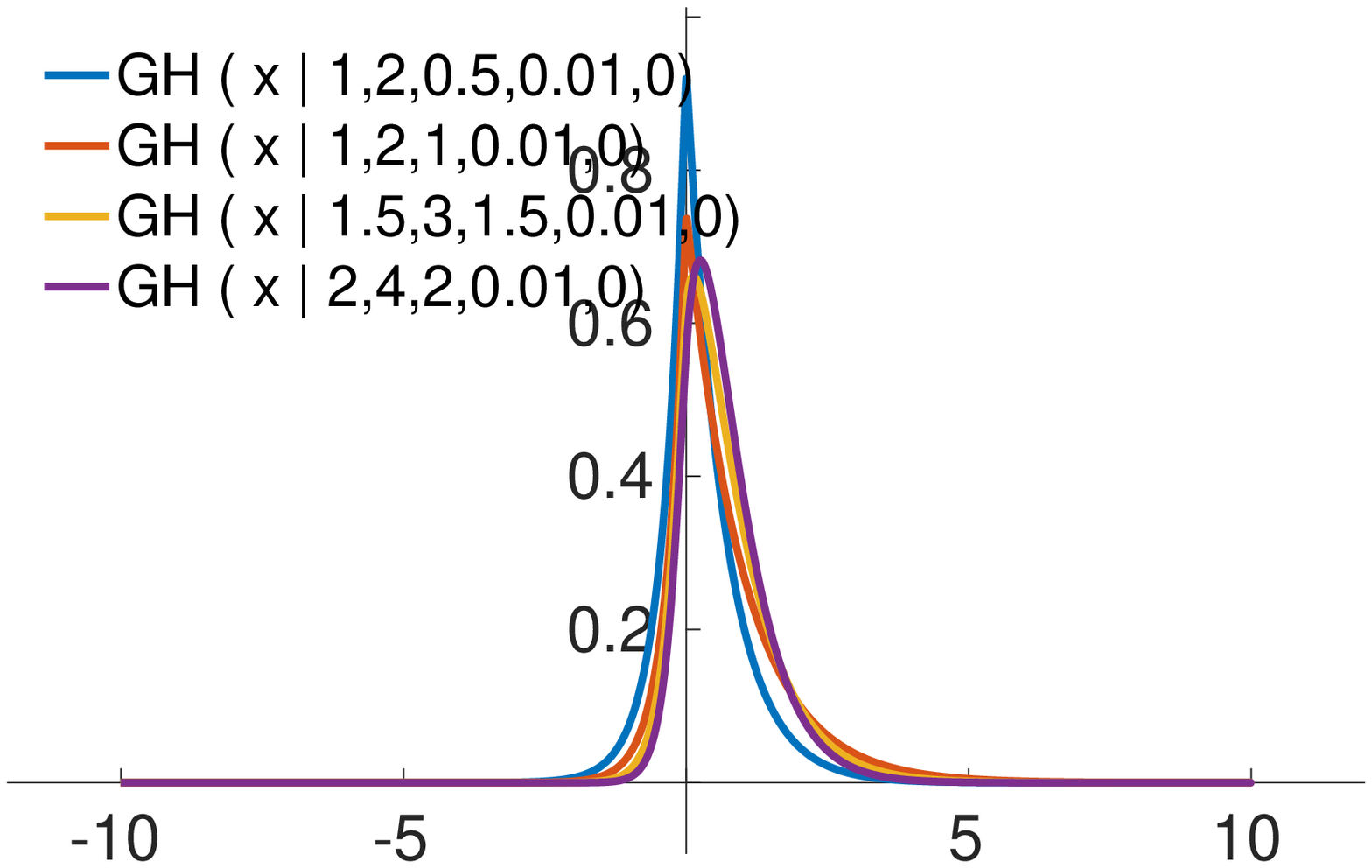}
                \caption{\small{$ \Gc\Hc(x|\lambda,\alpha,\beta,\delta \searrow 0,\mu) $}}
                \label{fig:GHVG_GenHypAsVG}
        \end{subfigure}        
\par\bigskip        
        \begin{subfigure}{0.49\textwidth}
          		\centering        
                \includegraphics[width=7.4cm,height=4cm]{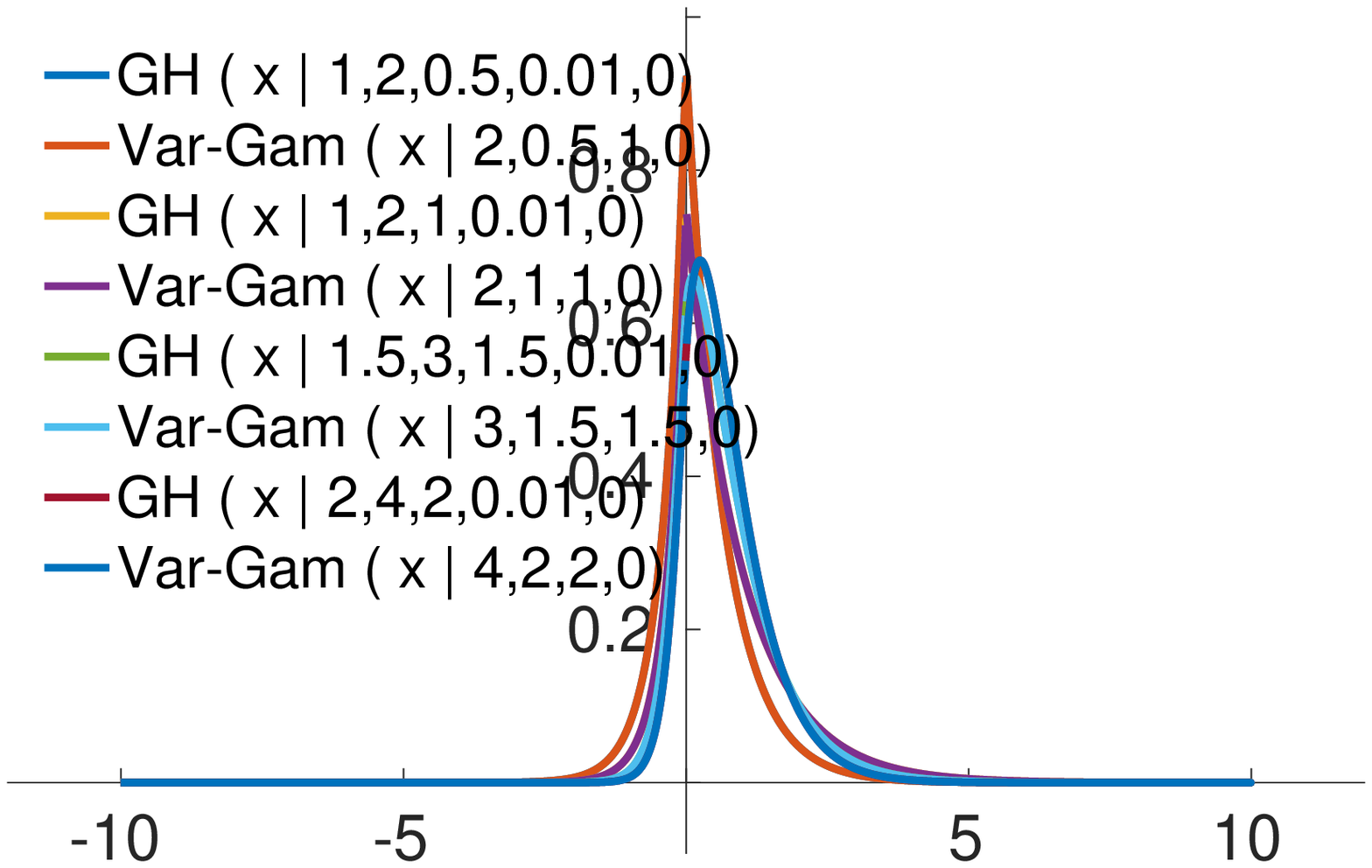}
                \caption{$ \Vc-\Gc $ vs. $ \Gc\Hc $}
                \label{fig:GHVGVsVG_GenHypAsVG}
        \end{subfigure}                               
        \begin{subfigure}{0.49\textwidth}
          		\centering        
                \includegraphics[width=7.4cm,height=4cm]{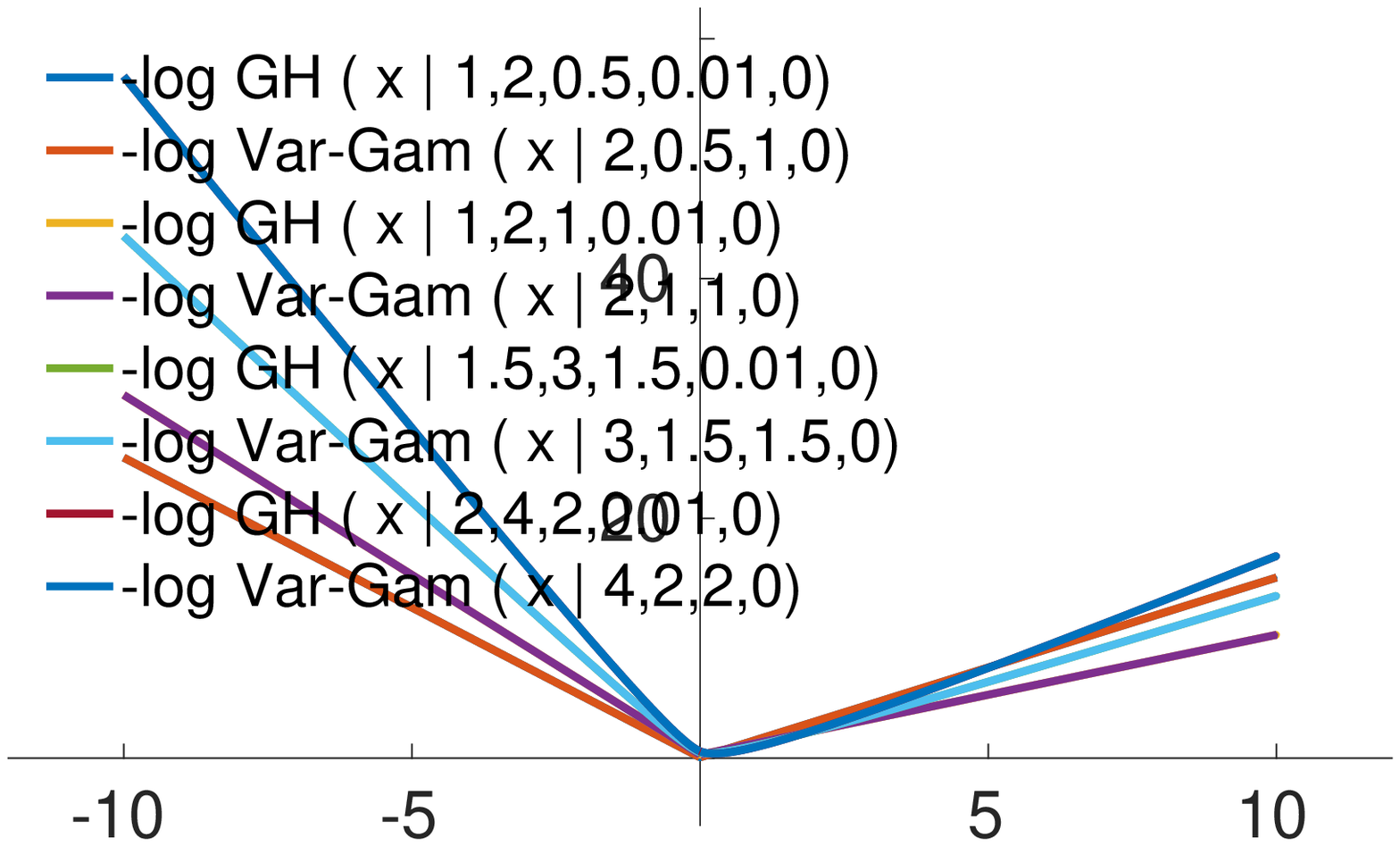}
                \caption{$ \log \Vc-\Gc $ vs. $ \log \Gc\Hc$}
                \label{fig:Log_GHVGVsVG_GenHypAsVG}
        \end{subfigure}                                       
        \caption{The Variance-Gamma distribution, Figure~\eqref{fig:VG_GenHypAsVG} and the Generalized Hyperbolic distribution, with parameters set as in Equation~\eqref{Eq:GenHyperbolicDistributionAsVarGam3}, Figure~\eqref{fig:GHVG_GenHypAsVG}. Comparison between the two distribution, $\Vc-\Gc$ vs. $\Gc\Hc$, Figure~\eqref{fig:GHVGVsVG_GenHypAsVG} and between the logarithm of the distributions $\log \Vc-\Gc$ vs. $\log \Gc\Hc$, Figure~\eqref{fig:Log_GHVGVsVG_GenHypAsVG}.}
        \label{fig:Comp_GenHypAsVG}
\end{figure}

The goal of this subsection is to derive the Variance-Gamma distribution from the Generalized Hyperbolic distribution.
\\
$X \sim \Gc\Hc(x|\lambda,\alpha,\beta,\delta \searrow 0,\mu)$ has a variance-gamma distribution, $\Vc-\Gc(x|\alpha,\beta,\lambda,\mu)$. 
Considering $\delta \searrow 0$, the particular form of the Generalized Hyperbolic distribution becomes:
\beq
\Gc\Hc(x|\lambda,\alpha,\beta,\delta \searrow 0,\mu) 
= 
\frac{{\gamma}^{\lambda}\lvert x - \mu \rvert^{\lambda - \frac{1}{2}}\Kc_{\lambda-\frac{1}{2}} \left( \alpha \lvert x - \mu \rvert \right)}{{\delta}^{\lambda} \sqrt{2\pi} {\alpha}^{\lambda - \frac{1}{2}} \Kc_{\lambda} \left( \delta \gamma \right)}
\;
\exp \left\lbrace {\beta \left( x - \mu \right)} \right\rbrace
\label{Eq:GenHyperbolicDistributionAsVarGam}
\eeq
Since $\delta \searrow  0$, for the expression of the modified Bessel function of the second degree $\Kc_{\lambda} \left( \delta \gamma \right)$ the asymptotic relation for small arguments presented in Equation~\eqref{Eq:ModifiedBesselAsymptoticRelation} can be used, obtaining:
\beq
\Kc_{\lambda} \left( \delta \gamma \right) = \Gamma \left( \lambda \right) 2^{\lambda-1} \left( \delta \gamma \right)^{-\lambda}, \mbox{ for } \delta \searrow  0.
\label{Eq:GenHyperbolicDistributionAsVarGam2}
\eeq
Using Equation~\eqref{Eq:GenHyperbolicDistributionAsVarGam2} in Equation~\eqref{Eq:GenHyperbolicDistributionAsVarGam}, the particular form of the Generalized Hyperbolic distribution becomes:
\beq
\Gc\Hc(x|\lambda,\alpha,\beta,\delta \searrow 0,\mu) 
= 
\frac{{\gamma}^{2\lambda}\lvert x - \mu \rvert^{\lambda - \frac{1}{2}}\Kc_{\lambda-\frac{1}{2}} \left( \alpha \lvert x - \mu \rvert \right)}{ \sqrt{\pi} \Gamma \left( \lambda \right) \left({2\alpha}\right)^{\lambda - \frac{1}{2}}}
\;
\exp \left\lbrace {\beta \left( x - \mu \right)} \right\rbrace
=
\Vc-\Gc(x|\alpha,\beta,\lambda,\mu)
\label{Eq:GenHyperbolicDistributionAsVarGam3}
\eeq
Figure~\eqref{fig:GHVGVsVG_GenHypAsVG} presents the comparison between the Variance-Gamma probability density function, $\Vc-\Gc(x|\alpha,\beta,\lambda,\mu)$ (presented in Figure~\eqref{fig:VG_GenHypAsVG}) and the Generalized Hyperbolic density function for parameters $\Gc\Hc(x|\lambda,\alpha,\beta,\delta \searrow 0,\mu)$ (presented in Figure~\eqref{fig:GHVG_GenHypAsVG}). Figure~\eqref{fig:Log_GHVGVsVG_GenHypAsVG} presents the comparison between the logarithm of the two distributions, $\log \Vc-\Gc$ vs. $\log \Gc\Hc$.
The Variance-Gamma distribution can be expressed using the Generalized Hyperbolic Prior Model (GHPM), Equation~\eqref{Eq:GHPM1} by fixing $\delta \searrow 0$: 
\beq
\textbf{V-GPM:}\;\;
\left\{\barr{ll}
p(\rf_\rj | \mu, \; \rv_{\rf_\rj}, \; \beta)
=
\Nc\left(\rf_\rj|\mu + \beta \rv_{\rf_\rj}, \; \rv_{\rf_\rj}\right) 
=
\left( 2 \pi \right)^{-\frac{1}{2}} \; \rv_{\rf_\rj}^{-\frac{1}{2}} \; \exp \left\lbrace -\frac{1}{2}\frac{\left( x - \mu - \beta \rv_{\rf_\rj} \right)^2}{\rv_{\rf_\rj}} \right\rbrace
\\[7pt]
p(\rv_{\rf_\rj}|\gamma^2, \; \delta^2 \searrow 0, \; \lambda = 1) 
=
\Gc\Ic\Gc(\rv_{\rf_\rj}|\gamma^2, \; \delta^2 \searrow 0, \; \lambda = 1)
=
\Ic\Gc(\rv_{\rf_\rj}|\lambda, \frac{\gamma^2}{2})
=
\frac{\left( \frac{\gamma^2}{2} \right)^{\lambda-1}}{\Gamma \left( \lambda \right)} \rv_{\rf_\rj}^{\lambda-1} \exp \left\lbrace -\frac{\gamma^2}{2} \rv_{\rf_\rj} \right\rbrace,\\[4pt]
\earr\right.
\label{Eq:GHPM_Lap}
\eeq

\subsection{Normal-Inverse Gaussian distribution: expressed via conjugate priors}
\label{Subsec:NormalInverseGaussianPrior}
The interest of the Normal-Inverse Gaussian distribution is given by the fact that it can be a \textit{heavy-tailed} distribution and therefore can be successfully used as a sparsity enforcing prior. The Normal-Inverse Gaussian distribution is a continuous probability distribution, defined as the Normal variance-mean mixture, where the mixing density is the Inverse Gaussian distribution. The parameters of the Normal-Inverse Gaussian distribution can be used to construct a heaviness and skewness plot called the NIG-triangle. The Normal-Inverse Gaussian distribution has the following probability density function:
\beq
\Nc\Ic\Gc \left( x | \alpha, \beta, \delta, \mu \right)
= \frac{\alpha \delta \Kc_{1} \left( \alpha \sqrt{ \delta^2 + \left( x - \mu \right)^2 } \right)}{\pi \sqrt{ \delta^2 + \left( x - \mu \right)^2 }}
\exp \left\lbrace \gamma \delta + \beta \left( x - \mu \right) \right\rbrace
\label{Eq:NIGDistribution}
\eeq
\subsubsection{Normal-Inverse Gaussian distribution: via Generalized Hyperbolic distribution}
\label{Subsubsec:GenHyperbolicAsNIG}
\begin{figure}[!htb]
        \centering
        \begin{subfigure}{0.49\textwidth}
          		\centering        
                \includegraphics[width=7.4cm,height=4cm]{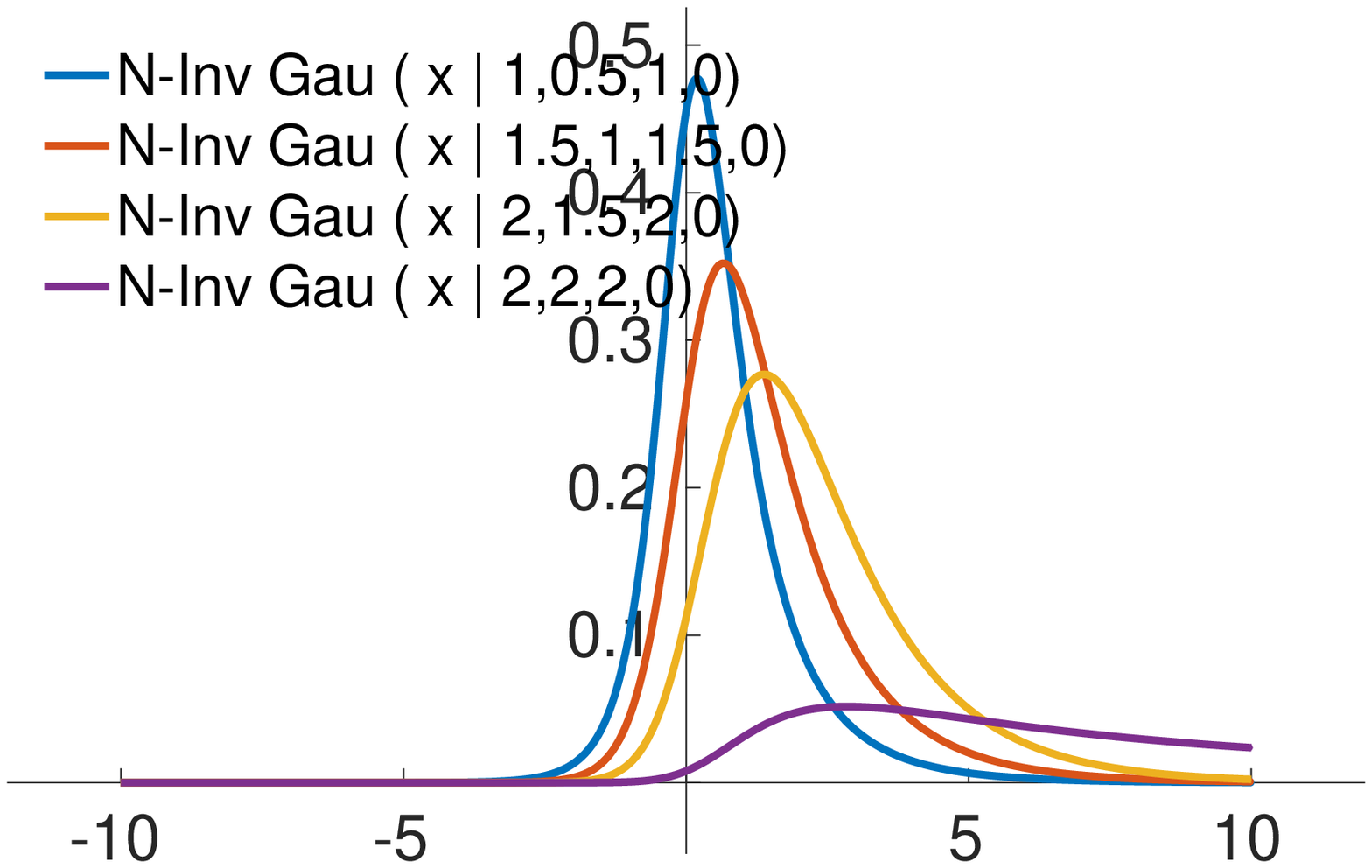}
                \caption{$ \Nc\Ic\Gc(x|\alpha,\beta,\delta,\mu) $}
                \label{fig:NIG_GenHypAsNIG}
        \end{subfigure}
        \begin{subfigure}{0.49\textwidth}
          		\centering        
                \includegraphics[width=7.4cm,height=4cm]{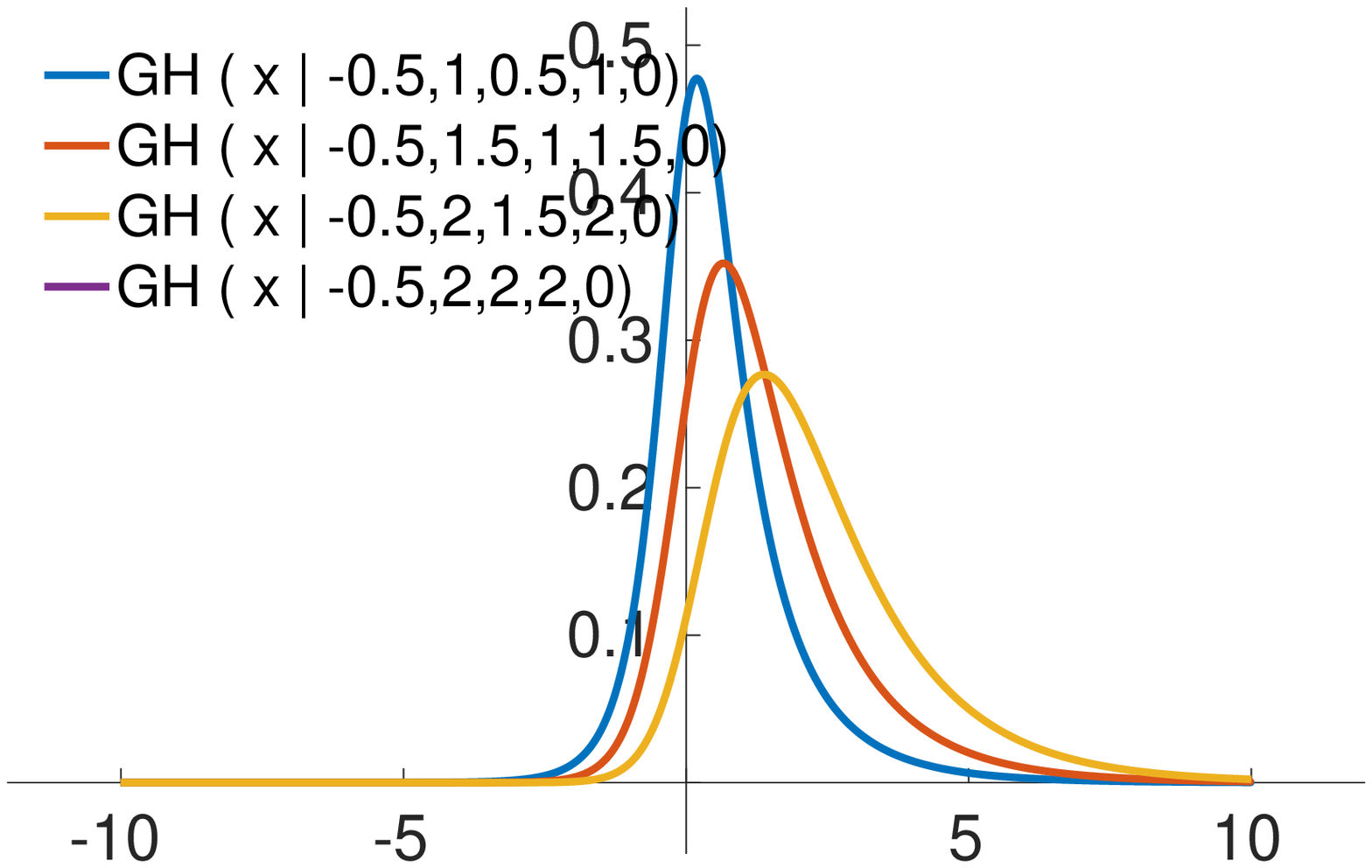}
                \caption{\small{$ \Gc\Hc(x|\lambda=-\frac{1}{2},\alpha,\beta,\delta,\mu) $}}
                \label{fig:GHNIG_GenHypAsNIG}
        \end{subfigure}        
\par\bigskip        
        \begin{subfigure}{0.49\textwidth}
          		\centering        
                \includegraphics[width=7.4cm,height=4cm]{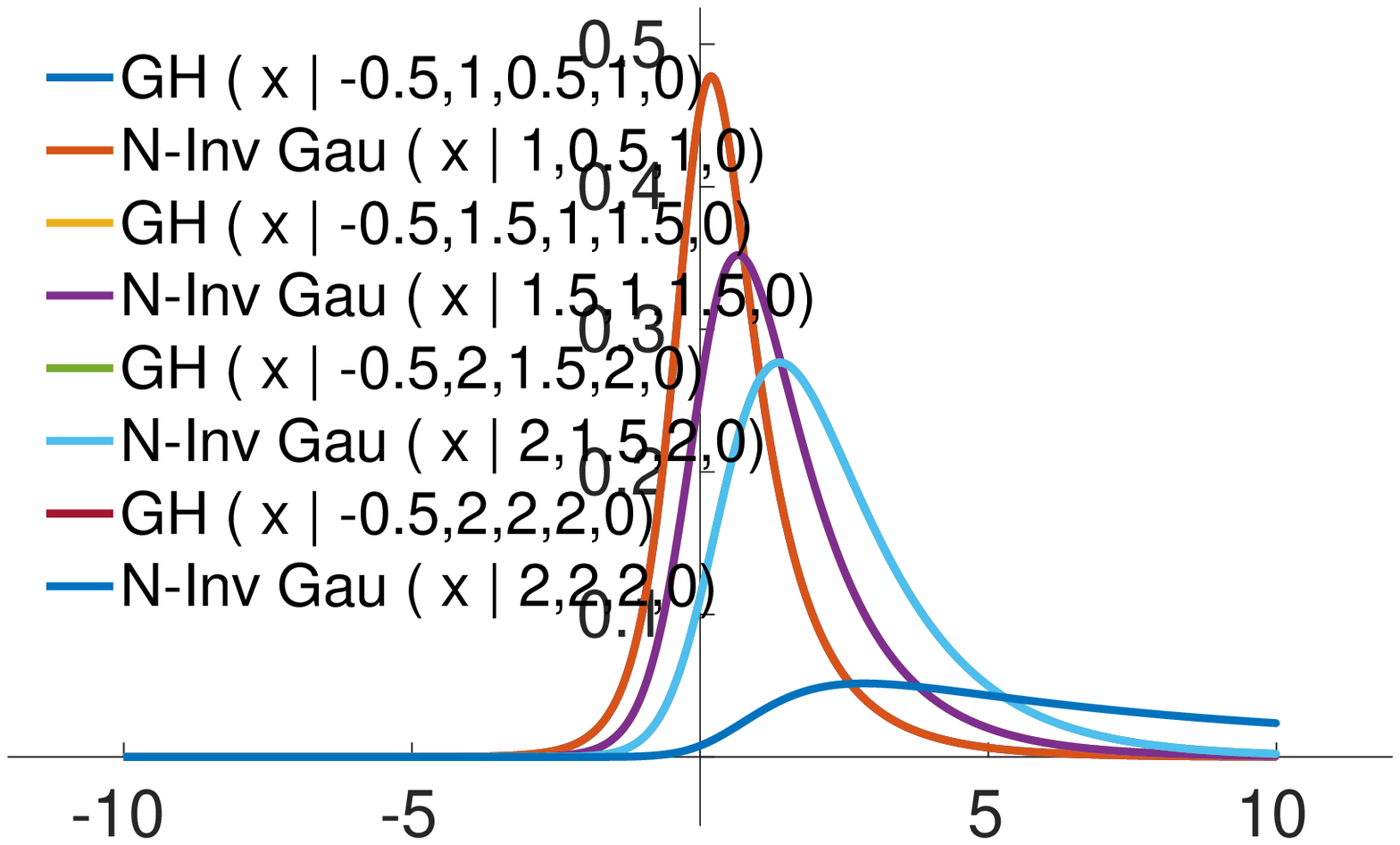}
                \caption{$ \Nc\Ic\Gc $ vs. $ \Gc\Hc $}
                \label{fig:GHNIGVsNIG_GenHypAsNIG}
        \end{subfigure}                               
        \begin{subfigure}{0.49\textwidth}
          		\centering        
                \includegraphics[width=7.4cm,height=4cm]{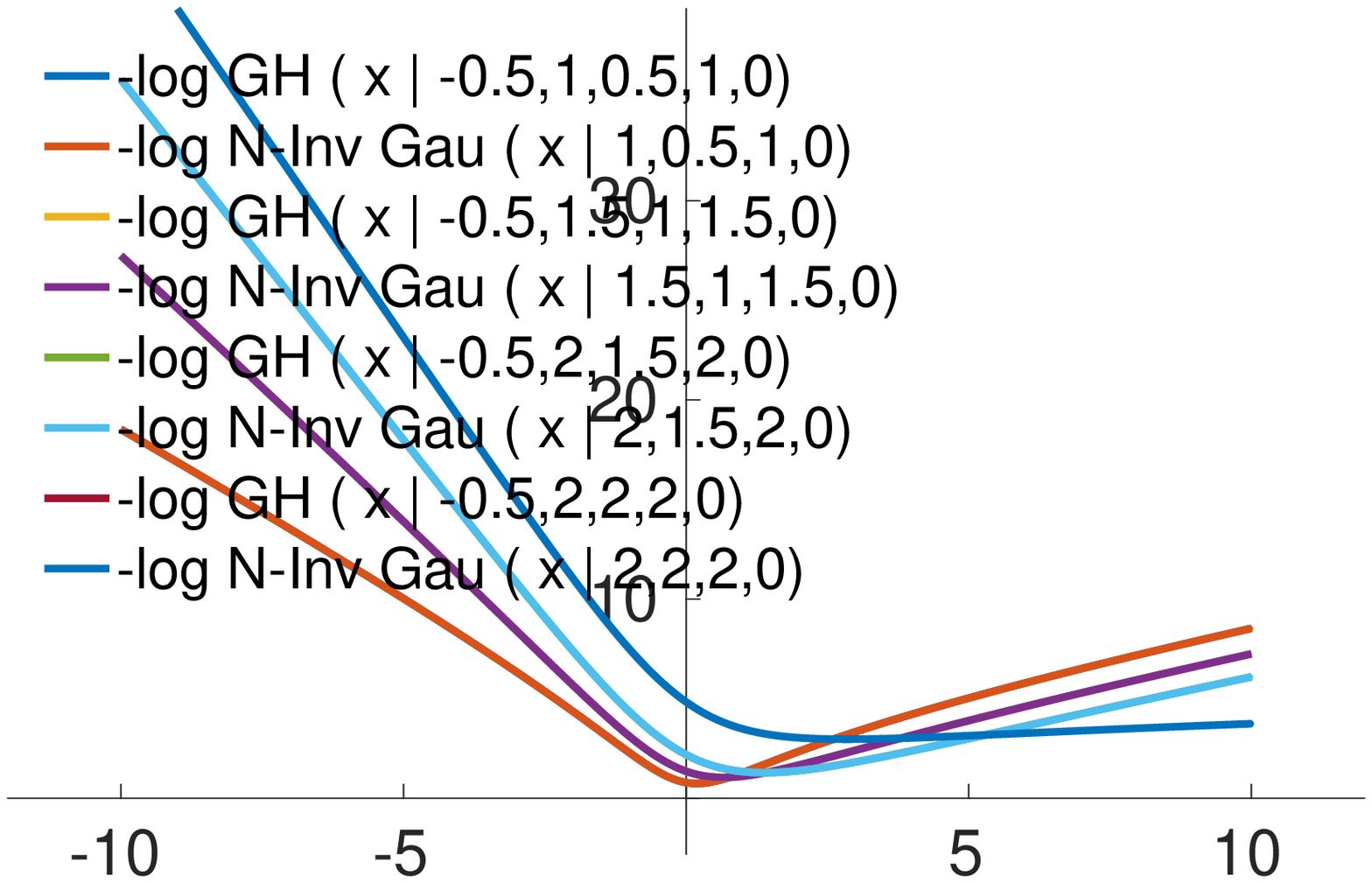}
                \caption{$ \log \Nc\Ic\Gc $ vs. $ \log \Gc\Hc$}
                \label{fig:Log_GHNIGVsNIG_GenHypAsNIG}
        \end{subfigure}                                       
        \caption{The Normal-Inverse Gaussian distribution, Figure~\eqref{fig:NIG_GenHypAsNIG} and the Generalized Hyperbolic distribution, with parameters set as in Equation~\eqref{Eq:GenHyperbolicDistributionAsNIG3}, Figure~\eqref{fig:GHNIG_GenHypAsNIG}. Comparison between the two distribution, $\Nc\Ic\Gc$ vs. $\Gc\Hc$, Figure~\eqref{fig:GHNIGVsNIG_GenHypAsNIG} and between the logarithm of the distributions $\log \Nc\Ic\Gc$ vs. $\log \Gc\Hc$, Figure~\eqref{fig:Log_GHNIGVsNIG_GenHypAsNIG}.}
        \label{fig:Comp_GenHypAsNIG}
\end{figure}

The goal of this subsection is to derive the Normal-Inverse Gaussian distribution from the Generalized Hyperbolic distribution.
\\
$X \sim \Gc\Hc(x|\lambda=-\frac{1}{2},\alpha,\beta,\delta,\mu)$ has a Normal-Inverse Gaussian distribution.
\\
Considering $\lambda = - \frac{1}{2}$, the Generalized Hyperbolic distribution is:
\beq
\Gc\Hc(x|\lambda,\alpha=-\frac{1}{2},\beta,\delta,\mu) 
= 
\frac{\left( \frac{\gamma}{\delta} \right)^{-\frac{1}{2}}}{\sqrt{2\pi}\Kc_{-\frac{1}{2}}\left( \gamma \delta \right)}
\frac{\Kc_{1} \left( \alpha \sqrt{ \delta^2 + \left( x - \mu \right)^2} \right)}{\left( \frac{\sqrt{ \delta^2 + \left( x - \mu \right)^2}}{\alpha} \right)^{\frac{1}{2}-\lambda}}
\exp \left\lbrace \beta \left( x - \mu \right) \right\rbrace
\label{Eq:GenHyperbolicDistributionAsNIG}
\eeq
Using the fact that for $\lambda=\frac{1}{2}$, the modified Bessel function of the second kind $\Kc_{\lambda} \left( x \right)$ can be stated explicitly, Equation~\eqref{Eq:ModifiedBesselExplicitly}, we express  $\Kc_{-\frac{1}{2}} \left( \gamma \delta \right)$:
\beq
\Kc_{\frac{1}{2}} \left( \gamma \delta \right) 
=
\left(\frac{\pi}{2\gamma \delta}\right)^{\frac{1}{2}} 
\exp \left\lbrace -\gamma \delta \right\rbrace. 
\label{Eq:GenHyperbolicDistributionAsNIG2}
\eeq
Plugging Equation~\eqref{Eq:GenHyperbolicDistributionAsNIG2} in Equation~\eqref{Eq:GenHyperbolicDistributionAsNIG}:
\beq
\Gc\Hc(x|\lambda=-\frac{1}{2},\alpha,\beta,\delta,\mu) 
= 
\frac{\alpha \delta \Kc_{1} \left( \alpha \sqrt{ \delta^2 + \left( x - \mu \right)^2} \right)}{\pi \sqrt{ \delta^2 + \left( x - \mu \right)^2}}
\exp \left\lbrace \gamma \delta + \beta \left( x - \mu \right) \right\rbrace
=
\Nc\Ic\Gc \left( x | \alpha, \beta, \delta, \mu \right)
\label{Eq:GenHyperbolicDistributionAsNIG3}
\eeq
Figure~\eqref{fig:GHNIGVsNIG_GenHypAsNIG} presents the comparison between the Normal-Inverse Gaussian probability density function, $\Nc\Ic\Gc(x|\alpha,\beta,\delta,\mu)$ (presented in Figure~\eqref{fig:NIG_GenHypAsNIG}) and the Generalized Hyperbolic density function for parameters $\Gc\Hc(x|\lambda=-\frac{1}{2},\alpha,\beta,\delta,\mu) $ (presented in Figure~\eqref{fig:GHNIG_GenHypAsNIG}). Figure~\eqref{fig:Log_GHNIGVsNIG_GenHypAsNIG} presents the comparison between the logarithm of the two distributions, $\log \Nc\Ic\Gc$ vs. $\log \Gc\Hc$.


\section{From uncertainties models to likelihood models}
\label{Sec:UncertaintiesModels}
Depending on the application the corresponding linear forward model variances might be unknown and are to be estimated.
For the linear forward model, Equation~\eqref{Eq:LinearModel}, the unknowns are $\rfb$, representing a signal or an image and $\repsilonb$, which accounts for measurements errors and uncertainties. When the corresponding variances are alos considered to be unknowns, we want to estimate the associated variances, i.e. $\rvb_{\rf}$ and $\rvb_{\repsilon}$. Typically, the construction of the hierarchical models is based on general Bayesian inference, which gives the possibility to derive the posterior distribution from the prior and likelihood. In this section, for the prior distribution we will use the Student-t distribution, in order to enforce sparsity on $\rfb$. The likelihood is obtained from the distribution proposed for modelling the uncertainties of the model, $\repsilonb$. Different distributions can be proposed for modelling the uncertainties, resulting in different likelihoods.
\subsection{Stationary Gaussian Uncertainties Model}
\label{Subsec:StationaryGaussianUncertaintiesModel}
A stationary Gaussian uncertainties model (sGUM) can be proposed, under the assumption that the associated uncertainties variances are equal, Equation~\eqref{Eq:StationaryNoise}:
\beq
\rv_{\repsilon_{\ri}} = \rv_{\repsilon_{\rj}} = \rv_{\repsilon}.
\label{Eq:StationaryNoise}
\eeq
The uncertainties vector elements are modelled using zero-mean Normal distributions, with the same associated variance, $\rv_{\repsilon}$, Equation~\eqref{Eq:StationaryModel}:
\beq
p\left( \repsilon_{\ri} | \rv_{\repsilon} \right)
=
\Nc\left( \repsilon_{\ri} | 0, \rv_{\repsilon} \right), i \in \left\lbrace 1, 2, \ldots, N \right\rbrace.
\label{Eq:StationaryModel}
\eeq
leading to the sGUM, a multivariate normal distribution, Equation~\eqref{Eq:sGUM}:
\beq
\textbf{sGUM:}\;\;\;\;\;\;\;
p\left( \repsilonb | \rv_{\repsilon} \right)
=
\Nc\left( \repsilonb | 0, \rv_{\repsilon} \Ib \right)
\label{Eq:sGUM}
\eeq
The likelihood $p\left( \bgb | \rfb, \rv_{\repsilon} \right)$ is obtained via the linear forward model, Equation~\eqref{Eq:LinearModel} and the stationary Gaussian uncertainties model (sGUM), Equation~\eqref{Eq:sGUM}. The distribution modelling the likelihood is also a multivariate normal distribution with the same covariance matrix $\rv_{\repsilon} \Ib$ and mean $\bHb\rfb$, stationary Gaussian likelihood (sGL), Equation~\eqref{Eq:sGL}:
\beq
\textbf{sGL:}\;\;\;\;\;\;\;
p\left( \bgb | \rfb, \rv_{\repsilon} \right)
=
\Nc\left( \bgb | \bHb\rfb, \rv_{\repsilon} \Ib \right) 
\label{Eq:sGL}
\eeq
\subsection{Non-stationary Gaussian Uncertainties Model}
\label{Subsec:NonStationaryGaussianUncertaintiesModel}
A non-stationary Gaussian uncertainties model (nsGUM) can be proposed, when the assumption that the associated uncertainties variances are equal, Equation~\eqref{Eq:StationaryNoise} is not imposed.\\
The uncertainties vector elements are modelled using zero-mean Normal distributions like in sGUM, but with different associated variances, $\rv_{\repsilon_{\ri}}$ for each uncertainties vector element $\repsilon_{\ri}$, Equation~\eqref{Eq:Non-StationaryModel}:
\beq
p\left( \repsilon_{\ri} | \rv_{\repsilon_{\ri}} \right)
=
\Nc\left( \repsilon_{\ri} | 0, \rv_{\repsilon_{\ri}} \right), i \in \left\lbrace 1, 2, \ldots, N \right\rbrace.
\label{Eq:Non-StationaryModel}
\eeq
leading to the nsGUM, a multivariate normal distribution, Equation~\eqref{Eq:nsGUM}:
\beq
\textbf{nsGUM:}\;\;\;\;\;\;\;
p\left( \repsilonb | \rvb_{\repsilon} \right)
=
\Nc\left( \repsilonb | 0, \rVb_{\repsilon} \right),
\label{Eq:nsGUM}
\eeq
where the following notations were used:
\beq
\rvb_{\repsilon} = 
\left[
\rv_{\repsilon_\red{1}}, 
\rv_{\repsilon_\red{2}}, 
\ldots , 
\rv_{\repsilon_{\rN}}
\right]
\;\;\; ; \;\;\;
\rVb_{\repsilon} = \diag{\rvb_{\repsilon}}.
\eeq
The likelihood $p\left( \bgb | \rfb, \rvb_{\repsilon} \right)$ is obtained via the linear forward model, Equation~\eqref{Eq:LinearModel} and the stationary Gaussian uncertainties model (nsGUM), Equation~\eqref{Eq:nsGUM}. The distribution modelling the likelihood is also a multivariate normal distribution with the same covariance matrix $\rVb_{\repsilon}$ and mean $\bHb\rfb$, non-stationary Gaussian likelihood (nsGL), Equation~\eqref{Eq:nsGL}:
\beq
\textbf{nsGL:}\;\;\;\;\;\;\;
p\left( \bgb | \rfb, \rv_{\repsilon} \right)
=
\Nc\left( \bgb | \bHb\rfb, \rVb_{\repsilon} \right) 
\label{Eq:nsGL}
\eeq

\subsection{Stationary Student-t Uncertainties Model}
\label{Subsec:StationaryStudentUncertaintiesModel}
A stationary Student-t uncertainties model (sStUM) can be proposed, under the assumption that the associated uncertainties variances are equal, Equation~\eqref{Eq:StationaryNoise}.\\
The uncertainties vector elements given the same associated variance $\rv_{\repsilon}$, are modelled using zero-mean Normal distributions, $ \repsilon | \rv_{\repsilon}  \sim \Nc \left( \repsilon | 0, \rv_{\repsilon} \right)$ and the variance  $\rv_{\repsilon}$ is modelled using an Inverse Gamma distribution, Equation~\eqref{Eq:UncVarAsInv}:
\beq
p\left( \rv_{\repsilon} | \alpha_{\epsilon}, \beta_{\epsilon} \right)
=
\Ic\Gc\left( \rv_{\repsilon} | \alpha_{\epsilon}, \beta_{\epsilon} \right)
=
\frac{\beta_{\epsilon}^{\alpha_{\epsilon}}}{\Gamma(\alpha_{\epsilon})} \; \rv_{\repsilon}^{-\alpha_{\epsilon}-1} \; \exp \left\lbrace -\frac{\beta_{\epsilon}}{\rv_{\repsilon}} \right\rbrace,
\label{Eq:UncVarAsInv}
\eeq
such that via the Student-t Prior Model, Equation~\eqref{Eq:StPM1}, the error vector $\repsilonb$ is modelled by a Student-t distribution, Equation~\eqref{Eq:UncAsSt}:
\beq
p(\repsilon_{\ri}|\alpha_{\epsilon}, \beta_{\epsilon}) = \left( 2 \pi \beta_{\epsilon} \right)^{-\frac{1}{2}} \; \frac{\Gamma( \alpha_{\epsilon} + \frac{1}{2} )}{\Gamma(\alpha_{\epsilon})} \; \left( 1 + \frac{\repsilon_{\ri}^2}{2 \beta_{\epsilon}} \right)^{-\left( \alpha_{\epsilon} + \frac{1}{2} \right)}. 
\label{Eq:UncAsSt}
\eeq 
The sStUM is represented by a multivariate Student-t distribution, Equation~\eqref{Eq:sStUMSt}:
\beq
p\left( \repsilonb | \alpha_{\epsilon}, \beta_{\epsilon} \right)
=
\left( 2 \pi \beta_{\epsilon} \right)^{-\frac{N}{2}} \;
\left( \frac{\Gamma( \alpha_{\epsilon} + \frac{1}{2} )}{\Gamma(\alpha_{\epsilon})} \right)^N \; \prod_{i=1}^{N} \left( 1 + \frac{\repsilon_{\ri}^2}{2 \beta_{\epsilon}} \right)^{-\left( \alpha_{\epsilon} + \frac{1}{2} \right)} 
\label{Eq:sStUMSt}
\eeq
and is expressed by a multivariate Normal distribution and an Inverse Gamma distribution, Equation~\eqref{Eq:sStUM}:
\beq
\textbf{sStUM:}\;\;\;\;\;\;\;
\left\{\barr{ll}
p\left( \repsilonb | \rv_{\repsilon} \right) = \Nc\left( \repsilonb | 0, \rv_{\repsilon} \Ib \right)
\\[7pt]
p\left( \rv_{\repsilon} | \alpha_{\epsilon}, \beta_{\epsilon} \right) = \Ic\Gc\left( \rv_{\repsilon} | \alpha_{\epsilon}, \beta_{\epsilon} \right),\\[4pt]
\earr\right.
\label{Eq:sStUM}
\eeq
The likelihood $p\left( \bgb | \rfb, \rv_{\repsilon} \right)$ is obtained via the linear forward model, Equation~\eqref{Eq:LinearModel} and the stationary Student-t uncertainties model (sStUM), Equation~\eqref{Eq:sStUM}. The distribution modelling the likelihood is also a multivariate normal distribution with the same covariance matrix $\rv_{\repsilon} \Ib$ and mean $\bHb\rfb$, stationary Student-t likelihood (sStL), Equation~\eqref{Eq:sStL}:
\beq
\textbf{sStL:}\;\;\;\;\;\;\;
\left\{\barr{ll}
p\left( \bgb | \rfb, \rv_{\repsilon} \right) = \Nc \left( \bgb | \bHb\rfb, \rv_{\repsilon} \Ib \right) 
\\[7pt]
p\left( \rv_{\repsilon} | \alpha_{\epsilon}, \beta_{\epsilon} \right) = \Ic\Gc\left( \rv_{\repsilon} | \alpha_{\epsilon}, \beta_{\epsilon} \right),\\[4pt]
\earr\right.
\label{Eq:sStL}
\eeq

\subsection{Non-stationary Student-t Uncertainties Model}
\label{Subsec:NonStationaryStudentUncertaintiesModel}
A non-stationary Student-t uncertainties model (nsStUM) can be proposed, when the assumption that the associated uncertainties variances are equal, Equation~\eqref{Eq:StationaryNoise} is not imposed.\\ 
The uncertainties vector elements given the associated variances $\rv_{\repsilon_\ri}$ are modelled using zero-mean Normal distributions, $ \repsilon_\ri | \rv_{\repsilon_\ri}  \sim \Nc \left( \repsilon_\ri | 0, \rv_{\repsilon_\ri} \right)$ and the variances $\rv_{\repsilon_\ri}$ are modelled using Inverse Gamma distributions, having the same shape and scale parameters, Equation~\eqref{Eq:unUncVarAsInv}: 
\beq
p\left( \rv_{\repsilon_\ri} | \alpha_{\epsilon}, \beta_{\epsilon} \right)
=
\Ic\Gc\left( \rv_{\repsilon_\ri} | \alpha_{\epsilon}, \beta_{\epsilon} \right)
=
\frac{\beta_{\epsilon}^{\alpha_{\epsilon}}}{\Gamma(\alpha_{\epsilon})} \; \rv_{\repsilon_\ri}^{-\alpha_{\epsilon}-1} \; \exp \left\lbrace -\frac{\beta_{\epsilon}}{\rv_{\repsilon_\ri}} \right\rbrace, i \in \left\lbrace 1, 2 \ldots, N \right\rbrace,
\label{Eq:unUncVarAsInv}
\eeq
such that via the Student-t Prior Model, Equation~\eqref{Eq:StPM1}, every element of the uncertainties vector $\repsilonb$ is modelled by a Student-t distribution, Equation~\eqref{Eq:UncAsSt}. The sStUM is represented by a multivariate Student-t distribution, Equation~\eqref{Eq:sStUMSt} and is expressed by a multivariate Normal distribution and a product of Inverse Gamma distributions, Equation~\eqref{Eq:nsStUM}:
\beq
\textbf{nsStUM:}\;\;\;\;\;\;\;
\left\{\barr{ll}
p\left( \repsilonb | \rvb_{\repsilon} \right) = \Nc\left( \repsilonb | 0, \rVb_{\repsilon} \right)
\\[7pt]
p\left( \rvb_{\repsilon} | \alpha_{\epsilon}, \beta_{\epsilon} \right) = \prod_{i=1}^{N} \Ic\Gc\left( \rv_{\repsilon_\ri} | \alpha_{\epsilon}, \beta_{\epsilon} \right),\\[4pt]
\earr\right.
\label{Eq:nsStUM}
\eeq
The likelihood $p\left( \bgb | \rfb, \rv_{\repsilon} \right)$ is obtained via the linear forward model, Equation~\eqref{Eq:LinearModel} and the non-stationary Student-t uncertainties model (nsStUM), Equation~\eqref{Eq:nsStUM}. The distribution modelling the likelihood is also a multivariate normal distribution with the same covariance matrix $\rv_{\repsilon} \Ib$ and mean $\bHb\rfb$, stationary Student-t likelihood (nsStL), Equation~\eqref{Eq:nsStL}:
\beq
\textbf{nsStL:}\;\;\;\;\;\;\;
\left\{\barr{ll}
p\left( \bgb | \rfb, \rvb_{\repsilon} \right) = \Nc \left( \bgb | \bHb\rfb, \rVb_{\repsilon} \right) 
\\[7pt]
p\left( \rvb_{\repsilon} | \alpha_{\epsilon}, \beta_{\epsilon} \right) = \prod_{i=1}^{N} \Ic\Gc\left( \rv_{\repsilon_\ri} | \alpha_{\epsilon}, \beta_{\epsilon} \right),\\[4pt]
\earr\right.
\label{Eq:nsStL}
\eeq

\subsection{Stationary Laplace Uncertainties Model}
\label{Subsec:StationaryLaplaceUncertaintiesModel}
A stationary Laplace uncertainties model (sLUM) can be proposed, under the assumption that the associated uncertainties variances are equal, Equation~\eqref{Eq:StationaryNoise}.\\ 
The uncertainties vector elements given the same associated variance $\rv_{\repsilon}$, are all modelled using zero-mean Normal distributions, $ \repsilon_\ri | \rv_{\repsilon}  \sim \Nc \left( \repsilon_\ri | 0, \rv_{\repsilon}^{-1} \right)$ and the variance  $\rv_{\repsilon}$ is modelled using an Inverse Gamma distribution, for which the shape parameter is set at $1$ and the considered scale parameter is $\frac{b_{\epsilon}}{2}$ Equation~\eqref{Eq:UncVarAsInvL}:
\beq
p\left( \rv_{\repsilon} | 1, \frac{b_{\epsilon}}{2} \right)
=
\Ic\Gc\left( \rv_{\repsilon} | 1, \frac{b_{\epsilon}}{2} \right)
=
\frac{b_{\epsilon}}{2} \; \rv_{\repsilon}^{-2} \; \exp \left\lbrace -\frac{b_{\epsilon}}{2 \rv_{\repsilon}} \right\rbrace,
\label{Eq:UncVarAsInvL}
\eeq
such that via the Laplace Prior Model, Equation~\eqref{Eq:LPM1}, the error vector $\repsilonb$ is modelled by a Laplace distribution, Equation~\eqref{Eq:UncAsL}:
\beq
p(\repsilon_{\ri}| b_{\epsilon}) =  \frac{1}{2 b_{\epsilon}^{-\frac{1}{2}}} \; \exp \left\lbrace -\frac{\repsilon_{\ri}}{b_{\epsilon}^{-\frac{1}{2}}} \right\rbrace .
\label{Eq:UncAsL}
\eeq 
The sLUM is represented by a multivariate Laplace distribution, Equation~\eqref{Eq:sLUML}:
\beq
p\left( \repsilonb | b_{\epsilon} \right)
=
\prod_{i=1}^{N} p(\repsilon_{\ri}| b_{\epsilon}) =  \frac{1}{2^N b_{\epsilon}^{-\frac{N}{2}}} \; \exp \left\lbrace -\frac{ \sum_{i=1}^{N} \repsilon_{\ri}}{b_{\epsilon}^{-\frac{1}{2}}} \right\rbrace
\label{Eq:sLUML}
\eeq
and is expressed by a multivariate Normal distribution and an Inverse Gamma distribution, Equation~\eqref{Eq:sLUM}:
\beq
\textbf{sLUM:}\;\;\;\;\;\;\;
\left\{\barr{ll}
p\left( \repsilonb | \rv_{\repsilon} \right) = \Nc\left( \repsilonb | 0, \rv_{\repsilon}^{-1} \Ib \right)
\\[7pt]
p\left( \rv_{\repsilon} | 1, \frac{b_{\epsilon}}{2} \right)
= \Ic\Gc\left( \rv_{\repsilon} | 1, \frac{b_{\epsilon}}{2} \right),\\[4pt]
\earr\right.
\label{Eq:sLUM}
\eeq
The likelihood $p\left( \bgb | \rfb, \rv_{\repsilon} \right)$ is obtained via the linear forward model, Equation~\eqref{Eq:LinearModel} and the stationary Laplace uncertainties model (sLUM), Equation~\eqref{Eq:sLUM}. The distribution modelling the likelihood is also a multivariate normal distribution with the same covariance matrix $\rv_{\repsilon}^{-1} \Ib$ and mean $\bHb\rfb$, stationary Student-t likelihood (sLL), Equation~\eqref{Eq:sLL}:
\beq
\textbf{sLL:}\;\;\;\;\;\;\;
\left\{\barr{ll}
p\left( \bgb | \rfb, \rv_{\repsilon} \right) = \Nc \left( \bgb | \bHb\rfb, \rv_{\repsilon}^{-1} \Ib \right) 
\\[7pt]
p\left( \rv_{\repsilon} | 1, \frac{b_{\epsilon}}{2} \right)
= \Ic\Gc\left( \rv_{\repsilon} | 1, \frac{b_{\epsilon}}{2} \right),\\[4pt]
\earr\right.
\label{Eq:sLL}
\eeq

\subsection{Non-stationary Laplace Uncertainties Model}
\label{Subsec:NonStationaryLaplaceUncertaintiesModel}
A non-stationary Laplace uncertainties model (nsLUM) can be proposed, when the assumption that the associated uncertainties variances are equal, Equation~\eqref{Eq:StationaryNoise} is not imposed.\\ 
The uncertainties vector elements given the associated variance $\rv_{\repsilon_\ri}$, are modelled using zero-mean Normal distributions, $ \repsilon_\ri | \rv_{\repsilon_\ri}  \sim \Nc \left( \repsilon_\ri | 0, \rv_{\repsilon_\ri}^{-1} \right)$ and the variances $\rv_{\repsilon_\ri}$ are modelled using Inverse Gamma distributions, for which the shape parameter is set at $1$ and the considered scale parameter is $\frac{b_{\epsilon}}{2}$, Equation~\eqref{Eq:unUncVarAsInvL}: 
\beq
p\left( \rv_{\repsilon_\ri} | 1, \frac{b_{\epsilon}}{2} \right)
=
\Ic\Gc\left( \rv_{\repsilon_\ri} | 1, \frac{b_{\epsilon}}{2} \right)
=
\frac{b_{\epsilon}}{2} \; \rv_{\repsilon_\ri}^{-2} \; \exp \left\lbrace -\frac{b_{\epsilon}}{2 \rv_{\repsilon_\ri}} \right\rbrace, i \in \left\lbrace 1, 2 \ldots, N \right\rbrace,
\label{Eq:unUncVarAsInvL}
\eeq
such that via the Laplace Prior Model, Equation~\eqref{Eq:LPM1}, every element of the uncertainties vector $\repsilonb$ is modelled by a Laplace distribution, Equation~\eqref{Eq:UncAsL}. The nsLUM is represented by a multivariate Laplace distribution, Equation~\eqref{Eq:sLUML} and is expressed by a multivariate Normal distribution and a product of Inverse Gamma distributions, Equation~\eqref{Eq:nsLUM}:
\beq
\textbf{nsLUM:}\;\;\;\;\;\;\;
\left\{\barr{ll}
p\left( \repsilonb | \rvb_{\repsilon} \right) = \Nc\left( \repsilonb | 0, \rVb_{\repsilon}^{-1} \right)
\\[7pt]
p\left( \rvb_{\repsilon} | 1, \frac{b_{\epsilon}}{2} \right)
= \prod_{i=1}^{N} \Ic\Gc\left( \rv_{\repsilon} | 1, \frac{b_{\epsilon}}{2} \right),\\[4pt]
\earr\right.
\label{Eq:nsLUM}
\eeq
The likelihood $p\left( \bgb | \rfb, \rvb_{\repsilon} \right)$ is obtained via the linear forward model, Equation~\eqref{Eq:LinearModel} and the non-stationary Laplace uncertainties model (nsLUM), Equation~\eqref{Eq:nsLUM}. The distribution modelling the likelihood is also a multivariate normal distribution with the same covariance matrix $\rVb_{\repsilon}^{-1} \Ib$ and mean $\bHb\rfb$, non-stationary Laplace likelihood (nsLL), Equation~\eqref{Eq:nsLL}:
\beq
\textbf{nsLL:}\;\;\;\;\;\;\;
\left\{\barr{ll}
p\left( \bgb | \rfb, \rvb_{\repsilon} \right) = \Nc \left( \bgb | \bHb\rfb, \rVb_{\repsilon}^{-1} \right) 
\\[7pt]
p\left( \rvb_{\repsilon} | 1, \frac{b_{\epsilon}}{2} \right)
= \prod_{i=1}^{N} \Ic\Gc\left( \rv_{\repsilon_\ri} | 1, \frac{b_{\epsilon}}{2} \right),\\[4pt]
\earr\right.
\label{Eq:nsLL}
\eeq

\section{Hierarchical Models with Student-t prior via StPM}
\label{Sec:HierarchicalModelsWithStudent-tPriorViaStPM}
Considering the Student-t distribution for modelling the sparse structure of $\rfb$ in linear forward model, Equation~\eqref{Eq:LinearModel}, expressed via the conjugate priors StPM, depending on the model proposed for the uncertainties of the model, $\repsilonb$, and implicitly the corresponding likelihood, sGL or nsGL, different hierarchical models can be proposed:

\subsection{Student-t hierarchical model: stationary Gaussian uncertainties model, known uncertainties variance}
\label{Subsec:St_sGL_k}
\begin{itemize}
\item the hierarchical model is using as a \textbf{prior} the \textbf{Student-t} distribution;
\item the Student-t prior distribution is expressed via \textbf{StPM}, Equation~\eqref{Eq:StPM1}, considering the variance $\rvb_\rf$ unknown;
\item the \textbf{likelihood} is derived from the distribution proposed for modelling the uncertainties vector $\repsilonb$;
\item for the uncertainties vector $\repsilonb$ a \textbf{stationary Gaussian uncertainties model} is proposed, i.e. a multivariate Gaussian distribution is used under the following two assumptions:\\
a) each element of the uncertainties vector has the same \textbf{variance}, $v_{\epsilon}$;\\
b) the variance $v_{\epsilon}$ is \textbf{known}; 
\end{itemize}
\beq
\rotatebox{90}{\hspace{-1.1cm}Student-t sGL K} \;
\vline
\left.\barr{ll}
Likelihood: \textbf{sGL:}\;\;\;\;\;\;\;
p\left( \bgb | \rfb, v_{\epsilon} \right)
=
\Nc\left( \bgb | \bHb\rfb, v_{\epsilon} \Ib \right)
\propto v_{\epsilon}^{-\frac{N}{2}} \; \exp \left\lbrace - \frac{1}{2 v_{\epsilon}}\| \left( \bgb - \bHb\rfb \right) \| \right\rbrace.
\\[7pt]
Prior:\;\;\;\;\;\;\;\textbf{StPM:}\;\;
\left\{\barr{ll}
p(\rfb|0,\rvb_{\rf}) = \Nc(\rfb|0, \; \rVb_{\rf}) \propto \det{\rVb_{\rf}}^{-\frac{1}{2}} \; \exp \left\lbrace - \frac{1}{2}\| \rVb_{\rf}^{-\frac{1}{2}} \rfb \| \right\rbrace,
\\[7pt]
p(\rvb_{\rf}|\alpha_{f}, \beta_{f}) = \prod_{j=1}^{M} \Ic\Gc(\rv_{\rf_\rj}|\alpha_{f}, \beta_{f}) \propto \prod_{j=1}^{M}\rv_{\rf_\rj}^{-\alpha_{f}-1} \; \exp \left\lbrace -\sum_{j=1}^{M}\frac{\beta_{f}}{\rv_{\rf_\rj}} \right\rbrace,\\[7pt]
\hspace{5.6cm}
\rVb_{\rf} = \diag{\rvb_{\rf}},\; \rvb_{\rf} = \left[\ldots,\rv_{\rf_{\rj}}, \ldots\right].\\[4pt]
\earr\right.
\earr\right.
\label{Eq:St_sGL_k}
\eeq

\subsection{Student-t hierarchical model: non-stationary Gaussian uncertainties model, known uncertainties variances}
\label{Subsec:St_nsGL_k}
\begin{itemize}
\item the hierarchical model is using as a \textbf{prior} the \textbf{Student-t} distribution;
\item the Student-t prior distribution is expressed via \textbf{StPM}, Equation~\eqref{Eq:StPM1}, considering the variance $\rvb_\rf$ unknown;
\item the \textbf{likelihood} is derived from the distribution proposed for modelling the uncertainties vector $\repsilonb$;
\item for the uncertainties vector $\repsilonb$ a \textbf{stationary Gaussian uncertainties model} is proposed, i.e. a multivariate Gaussian distribution is used under the following assumption:\\
a) the variance vector $\vb_{\epsilon}$ is \textbf{known}; 
\end{itemize}
\beq
\rotatebox{90}{\hspace{-1.15cm}Student-t nsGL K} \;
\vline
\left.\barr{ll}
Likelihood: \textbf{nsGL:}\;
\left\{\barr{ll}
\begin{split}
p\left( \bgb | \rfb, \vb_{\epsilon} \right) 
=
\Nc\left( \bgb | \bHb\rfb, \Vb_{\epsilon} \right) 
\propto 
\prod_{i=1}^{N} & v_{\epsilon_i}^{-\frac{1}{2}}
\exp \left\lbrace - \frac{1}{2}\| \Vb_{\epsilon}^{-\frac{1}{2}} \left( \bgb - \bHb\rfb \right) \| \right\rbrace,
\\
&
\Vb_{\epsilon} = \diag{\vb_{\epsilon}},\; \vb_{\epsilon} = \left[\ldots v_{\epsilon_{i}}, \ldots\right]. 
\end{split}
\earr\right.
\\[25pt]
Prior:\;\;\;\;\;\;\;\textbf{StPM:}\;\;\;
\left\{\barr{ll}
p(\rfb|0,\rvb_{\rf}) = \Nc(\rfb|0, \; \rVb_{\rf}) \propto \det{\rVb_{\rf}}^{-\frac{1}{2}} \; \exp \left\lbrace - \frac{1}{2}\| \rVb_{\rf}^{-\frac{1}{2}} \rfb \| \right\rbrace,
\\[6.7pt]
p(\rvb_{\rf}|\alpha_{f}, \beta_{f}) = \prod_{j=1}^{M} \Ic\Gc(\rv_{\rf_\rj}|\alpha_{f}, \beta_{f}) \propto  \prod_{j=1}^{M} \rv_{\rf_\rj}^{-\alpha_{f}-1} \; \exp \left\lbrace - \sum_{j=1}^{M}\frac{\beta_{f}}{\rv_{\rf_\rj}} \right\rbrace,\\[7pt]
\hspace{5.6cm}
\rVb_{\rf} = \diag{\rvb_{\rf}},\; \rvb_{\rf} = \left[\ldots,\rv_{\rf_{\rj}}, \ldots\right].\\[4pt]
\earr\right.
\earr\right.
\label{Eq:St_nsGL_k}
\eeq

\subsection{Student-t hierarchical model: stationary Gaussian uncertainties model, unknown uncertainties variance}
\label{Subsec:St_sGL_un}
\begin{itemize}
\item the hierarchical model is using as a \textbf{prior} the \textbf{Student-t} distribution;
\item the Student-t prior distribution is expressed via \textbf{StPM}, Equation~\eqref{Eq:StPM1}, considering the variance $\rvb_\rf$ as unknown;
\item the \textbf{likelihood} is derived from the distribution proposed for modelling the uncertainties vector $\repsilonb$;
\item for the uncertainties vector $\repsilonb$ a \textbf{stationary Gaussian uncertainties model} is proposed, i.e. a multivariate Gaussian distribution is used under the following two assumptions:\\
a) each element of the uncertainties vector has the same \textbf{variance}, $\rv_{\repsilon}$;\\
b) the variance $\rv_{\repsilon}$ is \textbf{unknown}; 
\end{itemize}
\beq
\rotatebox{90}{\hspace{-1.35cm}Student-t sGL UNK} \;
\vline
\left.\barr{ll}
Likelihood: \textbf{sGL:}\;\;\;\;\;\;\;
p\left( \bgb | \rfb, \rv_{\repsilon} \right)
=
\Nc\left( \bgb | \bHb\rfb, \rv_{\repsilon} \Ib \right)
\propto \rv_{\repsilon}^{-\frac{N}{2}} \; \exp \left\lbrace - \frac{1}{2\rv_{\repsilon}}\| \left( \bgb - \bHb\rfb \right) \| \right\rbrace.
\\[7pt]
Prior:\;\;\;\;\;\;\;\textbf{StPM:}\;\;
\left\{\barr{ll}
p(\rfb|0,\rvb_{\rf}) = \Nc(\rfb|0, \; \rVb_{\rf}) \propto \det{\rVb_{\rf}}^{-\frac{1}{2}} \; \exp \left\lbrace - \frac{1}{2}\| \rVb_{\rf}^{-\frac{1}{2}} \rfb \| \right\rbrace,
\\[7pt]
p(\rvb_{\rf}|\alpha_{f}, \beta_{f}) = \prod_{j=1}^{M} \Ic\Gc(\rv_{\rf_\rj}|\alpha_{f}, \beta_{f}) \propto \prod_{j=1}^{M}\rv_{\rf_\rj}^{-\alpha_{f}-1} \; \exp \left\lbrace -\sum_{j=1}^{M}\frac{\beta_{f}}{\rv_{\rf_\rj}} \right\rbrace,\\[7pt]
\hspace{5.6cm}
\rVb_{\rf} = \diag{\rvb_{\rf}},\; \rvb_{\rf} = \left[\ldots,\rv_{\rf_{\rj}}, \ldots\right].\\[4pt]
\earr\right.
\earr\right.
\label{Eq:St_sGL_un}
\eeq

\subsection{Student-t hierarchical model: non-stationary Gaussian uncertainties model, unknown uncertainties variances}
\label{Subsec:St_nsGL_un}
\begin{itemize}
\item the hierarchical model is using as a \textbf{prior} the \textbf{Student-t} distribution;
\item the Student-t prior distribution is expressed via \textbf{StPM}, Equation~\eqref{Eq:StPM1}, considering the variance $\rvb_\rf$ as unknown;
\item the \textbf{likelihood} is derived from the distribution proposed for modelling the uncertainties vector $\repsilonb$;
\item for the uncertainties vector $\repsilonb$ a \textbf{stationary Gaussian uncertainties model} is proposed, i.e. a multivariate Gaussian distribution is used under the following assumption:\\
a) the variance vector $\rvb_{\repsilon}$ is \textbf{unknown}; 
\end{itemize}
\beq
\rotatebox{90}{\hspace{-1.55cm}Student-t nsGL UNK} \;
\vline
\left.\barr{ll}
Likelihood: \textbf{nsGL:}\;
\left\{\barr{ll}
\begin{split}
p\left( \bgb | \rfb, \rv_{\repsilon} \right) 
=
\Nc\left( \bgb | \bHb\rfb, \rVb_{\repsilon} \right) 
\propto 
\prod_{i=1}^{N} & \rv_{\repsilon_\ri}^{-\frac{1}{2}}
\exp \left\lbrace - \frac{1}{2}\| \rVb_{\repsilon}^{-\frac{1}{2}} \left( \bgb - \bHb\rfb \right) \| \right\rbrace,
\\
&
\rVb_{\repsilon} = \diag{\rvb_{\repsilon}},\; \rvb_{\repsilon} = \left[\ldots,\rv_{\repsilon_{\ri}}, \ldots\right]. 
\end{split}
\earr\right.
\\[25pt]
Prior:\;\;\;\;\;\;\;\textbf{StPM:}\;\;\;
\left\{\barr{ll}
p(\rfb|0,\rvb_{\rf}) = \Nc(\rfb|0, \; \rVb_{\rf}) \propto \det{\rVb_{\rf}}^{-\frac{1}{2}} \; \exp \left\lbrace - \frac{1}{2}\| \rVb_{\rf}^{-\frac{1}{2}} \rfb \| \right\rbrace,
\\[6.7pt]
p(\rvb_{\rf}|\alpha_{f}, \beta_{f}) = \prod_{j=1}^{M} \Ic\Gc(\rv_{\rf_\rj}|\alpha_{f}, \beta_{f}) \propto  \prod_{j=1}^{M} \rv_{\rf_\rj}^{-\alpha_{f}-1} \; \exp \left\lbrace - \sum_{j=1}^{M}\frac{\beta_{f}}{\rv_{\rf_\rj}} \right\rbrace,\\[7pt]
\hspace{5.6cm}
\rVb_{\rf} = \diag{\rvb_{\rf}},\; \rvb_{\rf} = \left[\ldots,\rv_{\rf_{\rj}}, \ldots\right].\\[4pt]
\earr\right.
\earr\right.
\label{Eq:St_nsGL_un}
\eeq

\subsection{Student-t hierarchical model: stationary Student-t uncertainties model, unknown uncertainties variance}
\label{Subsec:St_sStL_un}
\begin{itemize}
\item the hierarchical model is using as a \textbf{prior} the \textbf{Student-t} distribution;
\item the Student-t prior distribution is expressed via \textbf{StPM}, Equation~\eqref{Eq:StPM1}, considering the variance $\rvb_\rf$ as unknown;
\item the \textbf{likelihood} is derived from the distribution proposed for modelling the uncertainties vector $\repsilonb$;
\item for the uncertainties vector $\repsilonb$ a \textbf{stationary Student-t uncertainties model} is proposed, i.e. a multivariate Student distribution expressed via \textbf{StPM} is used under the following two assumptions:\\
a) each element of the uncertainties vector has the same \textbf{variance}, $\rv_{\repsilon}$;\\
b) the variance $\rv_{\repsilon}$ is \textbf{unknown}; 
\end{itemize}
\beq
\rotatebox{90}{\hspace{-1.35cm}Student-t sStL UNK} \;
\vline
\left.\barr{ll}
Likelihood: \textbf{sStL:}\;
\left\{\barr{ll}
p\left( \bgb | \rfb, \rv_{\repsilon} \right)
=
\Nc\left( \bgb | \bHb\rfb, \rv_{\repsilon} \Ib \right)
\propto \rv_{\repsilon}^{-\frac{N}{2}} \; \exp \left\lbrace - \frac{1}{2\rv_{\repsilon}}\| \left( \bgb - \bHb\rfb \right) \| \right\rbrace,
\\[7pt]
p\left( \rv_{\repsilon} | \alpha_{\epsilon}, \beta_{\epsilon} \right) = 
\Ic\Gc\left( \rv_{\repsilon} | \alpha_{\epsilon}, \beta_{\epsilon} \right) \propto \rv_{\repsilon}^{-\alpha_{\epsilon}-1} \; \exp \left\lbrace \frac{\beta_{\epsilon}}{\rv_{\repsilon}} \right\rbrace,\\[4pt]
\earr\right.
\\[25pt]
Prior:\;\;\;\;\;\;\;\textbf{StPM:}\;\;
\left\{\barr{ll}
p(\rfb|0,\rvb_{\rf}) = \Nc(\rfb|0, \; \rVb_{\rf}) \propto \det{\rVb_{\rf}}^{-\frac{1}{2}} \; \exp \left\lbrace - \frac{1}{2}\| \rVb_{\rf}^{-\frac{1}{2}} \rfb \| \right\rbrace,
\\[7pt]
p(\rvb_{\rf}|\alpha_{f}, \beta_{f}) = \prod_{j=1}^{M} \Ic\Gc(\rv_{\rf_\rj}|\alpha_{f}, \beta_{f}) \propto \prod_{j=1}^{M}\rv_{\rf_\rj}^{-\alpha_{f}-1} \; \exp \left\lbrace -\sum_{j=1}^{M}\frac{\beta_{f}}{\rv_{\rf_\rj}} \right\rbrace,\\[7pt]
\hspace{5.6cm}
\rVb_{\rf} = \diag{\rvb_{\rf}},\; \rvb_{\rf} = \left[\ldots,\rv_{\rf_{\rj}}, \ldots\right].\\[4pt]
\earr\right.
\earr\right.
\label{Eq:St_sStL_un}
\eeq

\subsection{Student-t hierarchical model: non-stationary Student-t uncertainties model, unknown uncertainties variances}
\label{Subsec:St_nsStL_un}
\begin{itemize}
\item the hierarchical model is using as a \textbf{prior} the \textbf{Student-t} distribution;
\item the Student-t prior distribution is expressed via \textbf{StPM}, Equation~\eqref{Eq:StPM1}, considering the variance $\rvb_\rf$ as unknown;
\item the \textbf{likelihood} is derived from the distribution proposed for modelling the uncertainties vector $\repsilonb$;
\item for the uncertainties vector $\repsilonb$ a \textbf{non-stationary Student-t uncertainties model} is proposed, i.e. a multivariate Student-t distribution expressed via \textbf{StPM} is used under the following assumption:\\
a) the variance vector $\rvb_{\repsilon}$ is \textbf{unknown}; 
\end{itemize}
\beq
\rotatebox{90}{\hspace{-1.55cm}Student-t nsStL UNK} \;
\vline
\left.\barr{ll}
Likelihood: \textbf{nsStL:}\;
\left\{\barr{ll}
\begin{split}
p\left( \bgb | \rfb, \rvb_{\repsilon} \right) 
=
\Nc\left( \bgb | \bHb\rfb, \rVb_{\repsilon} \right) 
\propto 
\prod_{i=1}^{N} & \rv_{\repsilon_\ri}^{-\frac{1}{2}}
\exp \left\lbrace - \frac{1}{2}\| \rVb_{\repsilon}^{-\frac{1}{2}} \left( \bgb - \bHb\rfb \right) \| \right\rbrace, 
\end{split} 
\\[7pt]
p\left( \rvb_{\repsilon} | \alpha_{\epsilon}, \beta_{\epsilon} \right)
= \prod_{i=1}^{N} \Ic\Gc\left( \rv_{\repsilon_\ri} | \alpha_{\epsilon}, \beta_{\epsilon} \right) \propto  \prod_{i=1}^{N} \rv_{\repsilon_\ri}^{-\alpha_{\epsilon}-1} \; \exp \left\lbrace - \sum_{i=1}^{N}\frac{\beta_{\epsilon}}{\rv_{\repsilon_\ri}} \right\rbrace,\\[7pt]
\hspace{5.55cm}
\rVb_{\repsilon} = \diag{\rvb_{\repsilon}},\; \rvb_{\repsilon} = \left[\ldots,\rv_{\repsilon_{\ri}}, \ldots\right],\\[4pt]
\earr\right.
\\[33pt]
Prior:\;\;\;\;\;\;\;\textbf{StPM:}\;\;\;
\left\{\barr{ll}
p(\rfb|0,\rvb_{\rf}) = \Nc(\rfb|0, \; \rVb_{\rf}) \propto \det{\rVb_{\rf}}^{-\frac{1}{2}} \; \exp \left\lbrace - \frac{1}{2}\| \rVb_{\rf}^{-\frac{1}{2}} \rfb \| \right\rbrace,
\\[6.7pt]
p(\rvb_{\rf}|\alpha_{f}, \beta_{f}) = \prod_{j=1}^{M} \Ic\Gc(\rv_{\rf_\rj}|\alpha_{f}, \beta_{f}) \propto  \prod_{j=1}^{M} \rv_{\rf_\rj}^{-\alpha_{f}-1} \; \exp \left\lbrace - \sum_{j=1}^{M}\frac{\beta_{f}}{\rv_{\rf_\rj}} \right\rbrace,\\[7pt]
\hspace{5.65cm}
\rVb_{\rf} = \diag{\rvb_{\rf}},\; \rvb_{\rf} = \left[\ldots,\rv_{\rf_{\rj}}, \ldots\right].\\[4pt]
\earr\right.
\earr\right.
\label{Eq:St_nsStL_un}
\eeq
The hierarchical model build over the linear forward model, Equation~\eqref{Eq:LinearModel}, using as a prior for $\rfb$ a Student-t distribution, expressed via the Student-t Prior Model (StPM), Equation~\eqref{Eq:StPM1} and modelling the uncertainties of the model $\repsilonb$ using the non-stationary Student-t Uncertainties Model (nsStUM), Equation~\eqref{Eq:nsStUM}, is presented in Equation~\eqref{Eq:St_nsStL_un}. The posterior distribution is obtained via the Bayes rule, Equation~\eqref{Eq:Post_St_nsStL_un}:
\beq
\begin{split}
p(\rfb, \rvb_\rf, \rvb_\repsilon | \bgb, \alpha_{f}, \beta_{f}, \alpha_{\epsilon}, \beta_{\epsilon})
& \propto 
\; 
p\left( \bgb | \rfb, \rvb_{\repsilon} \right) 
\;
p\left( \rvb_{\repsilon} | \alpha_{\epsilon}  \beta_{\epsilon} \right) 
\;
p(\rfb|0,\rvb_{\rf}) 
\;
p(\rvb_{\rf}|\alpha_{f}, \beta_{f})\\
& \propto 
\;
\prod_{i=1}^{N}  \rv_{\repsilon_\ri}^{-\frac{1}{2}}
\exp \left\lbrace - \frac{1}{2}\| \rVb_{\repsilon}^{-\frac{1}{2}} \left( \bgb - \bHb\rfb \right) \| \right\rbrace 
\;
\prod_{i=1}^{N} \rv_{\repsilon_\ri}^{-\alpha_{\epsilon}-1} \; \exp \left\lbrace - \sum_{i=1}^{N}\frac{\beta_{\epsilon}}{\rv_{\repsilon_\ri}} \right\rbrace\\
&
\;\;\;\;\;\prod_{j=1}^{M}  \rv_{\rf_\ri}^{-\frac{1}{2}} \; \exp \left\lbrace - \frac{1}{2}\| \rVb_{\rf}^{-\frac{1}{2}} \rfb \| \right\rbrace
\;
\prod_{j=1}^{M} \rv_{\rf_\rj}^{-\alpha_{f}-1} \; \exp \left\lbrace - \sum_{j=1}^{M}\frac{\beta_{f}}{\rv_{\rf_\rj}} \right\rbrace
\end{split}
\label{Eq:Post_St_nsStL_un}
\eeq
The goal is to estimate the unknowns of the hierarchical model, namely $\rfb$, the main unknown of the linear forward model, Equation~\eqref{Eq:LinearModel} which was suppose sparse, and consequently modelled via the Student-t distribution and the two variances appearing in the hierarchical model, Equation~\eqref{Eq:St_nsStL_un}, the variance corresponding to the sparse structure $\rfb$, namely $\rvb_\rf$ and the variance corresponding to uncertainties of model $\repsilonb$, namely $\rvb_\repsilon$. 
\subsubsection{Joint MAP estimation}
\label{Subsubsec:JMAP_St_nsStL_un}
First, the Joint Maximum A Posterior (JMAP) estimation is considered: the unknowns are estimated on the basis of the available data $\bgb$, by maximizing the posterior distribution:
\beq
\left( \rfbh, \; \rvbh_\rf, \; \rvbh_\repsilon \right)
=
\operatorname*{arg\,max}_{\left( \rfb, \; \rvb_\rf, \; \rvb_\repsilon \right)} p(\rfb, \; \rvb_\rf, \; \rvb_\repsilon | \bgb, \; \alpha_{f}, \; \beta_{f}, \; \alpha_{\epsilon}, \; \beta_{\epsilon})
=
\operatorname*{arg\,min}_{\left( \rfb, \; \rvb_\rf, \; \rvb_\repsilon \right)} \Lc(\rfb, \; \rvb_\rf, \; \rvb_\repsilon),
\eeq 
where for the second equality the criterion $\Lc(\rfb, \rvb_{\rf}, \rvb_{\repsilon})$ is defined as:
\beq
\Lc(\rfb, \rvb_{\rf}, , \rvb_{\repsilon})= -\ln  p(\rfb, \; \rvb_\rf, \; \rvb_\repsilon | \bgb, \; \alpha_{f}, \; \beta_{f}, \; \alpha_{\epsilon}, \; \beta_{\epsilon})
\label{Eq:Def_L_Criterion}
\eeq
The MAP estimation corresponds to the solution minimizing the criterion $\Lc(\rfb, \rvb_{\rf}, \rvb_{\repsilon})$.
From the analytical expression of the posterior distribution, Equation~\eqref{Eq:Post_St_nsStL_un} and the definition of the criterion $\Lc$,  Equation~\eqref{Eq:Def_L_Criterion}, we obtain:
\beq
\begin{split}
\Lc(\rfb, \rvb_{\repsilon}, \rvb_{\rf})
=
-\ln  p(\rfb, \rvb_{\repsilon}, \rvb_{\rf}|\bgb)
&
=
\frac{1}{2}\| \rVb_{\repsilon}^{-\frac{1}{2}} \left( \bgb - \bHb\rfb \right) \|
+
\left( \alpha_{\epsilon} + \frac{3}{2} \right) \sum_{i=1}^{N} \ln \rv_{\repsilon_\ri} +
\sum_{i=1}^{N}\frac{\beta_{\epsilon}}{\rv_{\repsilon_\ri}} 
\\
&
+
\frac{1}{2}\| \rVb_{\rf}^{-\frac{1}{2}} \rfb \|
+
\left(\alpha_{f} + \frac{3}{2} \right) \sum_{j=1}^{M} \ln \rv_{\rf_\rj}
+
\sum_{j=1}^{M}\frac{\beta_{f}}{\rv_{\rf_\rj}}
\end{split}
\label{Eq:L_Criterion}
\eeq
One of the simplest optimisation algorithm that can be used is an alternate optimization of the criterion $\Lc(\rfb, \rvb_{\repsilon}, \rvb_{\rf})$ with respect to the each unknown:
\begin{itemize}
\item With respect to $\rfb$:
\beq
\begin{split}
\frac{\partial \Lc(\rfb, \rvb_{\rf}, \rvb_{\repsilon})}{\partial \rfb}=0
&\Leftrightarrow
\frac{\partial}{\partial \rfb}
\left(
\| \rVb_\repsilon^{-\frac{1}{2}} \left( \bgb - \bHb\rfb \right) \|^2
+
\|\rVb_\rf^{-\frac{1}{2}}\rfb\|^2
\right)
=
-\bHb^T \rVb_\repsilon^{-1} \left( \bgb - \bHb \rfb \right) + \rVb_\rf^{-1} \rfb = 0
\\
&
\Leftrightarrow 
\left( \bHb^T \rVb_\repsilon^{-1} \bHb + \rVb_\rf^{-1} \right) \rfb  = 
\bHb^T \rVb_\repsilon^{-1} \bgb
\\
&
\Rightarrow
\rfbh 
=
\left( \bHb^T \rVb_\repsilon^{-1} \bHb + \rVb_\rf^{-1} \right)^{-1} \bHb^T \rVb_\repsilon^{-1} \bgb
\end{split}
\nonumber
\eeq
\item With respect to $\rv_\rf$, $j \in \left\lbrace 1,2,\ldots, M \right\rbrace$:
\beq
\begin{split}
\frac{\partial \Lc(\rfb, \rvb_{\rf}, \rvb_{\rf})}{\partial \rv_{\rf_\rj}}=0
&
\Leftrightarrow
\frac{\partial}{\partial \rv_{\rf_\rj}}
\left[
\left( \alpha_f + \frac{3}{2}  \right)\ln \rv_{\rf_\rj}
+
\left( \beta_f + \frac{1}{2} \rf_\rj^2 \right)
\rv_{\rf_\rj}^{-1}
\right]
=0
\\
&
\Leftrightarrow
\left( \alpha_f + \frac{3}{2}  \right) \rv_{\rf_\rj}
-
\left( \beta_f + \frac{1}{2} \rf_\rj^2 \right)
=0
\\
&
\Rightarrow
\rvh_{\rf_\rj} = \frac{\beta_f + \frac{1}{2} \rf_\rj^2}{\alpha_f + \frac{3}{2}}
\end{split}
\nonumber
\eeq
\item With respect to $\rv_{\repsilon_\ri}$, $i \in \left\lbrace 1,2,\ldots, N \right\rbrace$:\\
First, we develop the norm $\| \rVb_\repsilon^{-\frac{1}{2}} \left( \bgb - \bHb\rfb \right) \|$:
\beq
\begin{split}
\| \rVb_\repsilon^{-\frac{1}{2}} \left( \bgb - \bHb\rfb \right) \|
&= 
\bgb^{T} \rVb_\repsilon^{-1} \bgb
-
2 \bgb^{T} \rVb_\repsilon^{-1} \bHb \rfb
+
\bHb^{T} \rfb^{T} \rVb_\repsilon^{-1} \bHb \rfb
\\
&=
\sum_{i=1}^{N} \rv_{\repsilon_\ri}^{-1} \bg_\bi^{2}
-
2 \sum_{i=1}^{N} \rv_{\repsilon_\ri}^{-1} \bg_\bi \bH_\bi \rfb 
+
\sum_{i=1}^{N} \rv_{\repsilon_\ri}^{-1} \rfb^{T} \bH_\bi^{T} \bH_\bi \rfb,
\nonumber
\end{split}
\eeq
where $\bHb_\bi$ denotes the line i of the matrix $\bHb$, $i\in \left\lbrace 1,2,\ldots,N \right\rbrace$, i.e. $\bHb_\bi = \left[ \bh_{\bi\blue{1}}, \bh_{\bi\blue{1}}, \ldots, \bh_{\bi\bM} \right]$. 
\beq
\begin{split}
\frac{\partial \Lc(\rfb, \rvb_{\rf}, \rvb_{\repsilon})}{\partial \rv_{\repsilon_\ri}}=0
&
\Leftrightarrow
\frac{\partial}{\partial \rv_{\repsilon_\ri}}
\left[
\left( \alpha_{\epsilon} + \frac{3}{2} \right)\ln \rv_{\repsilon_\ri}
+
\left(\beta_{\epsilon} + \frac{1}{2} \left( \bg_\bi^{2} - 2 \bg_\bi \bHb_\bi \rfb  + \rfb^{T} \bHb_\bi^{T} \bHb_\bi \rfb \right)
\right) \rv_{\repsilon_\ri}^{-1}
\right]
=0
\\
&
\Leftrightarrow
\left( \alpha_{\epsilon} + \frac{3}{2} \right) \rv_{\repsilon_\ri}
-
\left(\beta_{\epsilon} + \frac{1}{2} \left( \bg_\bi - \bHb_\bi \rfb  \right)^2 \right)=0
\\
&
\Rightarrow
\rvh_{\repsilon_\ri} = \frac{\beta_{\epsilon} + \frac{1}{2} \left( \bg_\bi - \bHb_\bi \rfb \right)^2}{\alpha_{\epsilon} + \frac{3}{2}}
\end{split}
\nonumber
\eeq
\end{itemize}
The iterative algorithm obtained via JMAP estimation is presented Figure~\eqref{Fig:IA_JMAP_St_nsStL_un}.
\tikzstyle{cloud} = [draw=blue!30!green,fill=orange!12, ellipse, thick, node distance=6em, text width=12em, text centered, minimum height=4em, minimum width=12em]
\tikzstyle{boxbig} = [draw=blue!30!green, fill=orange!12, rectangle, rounded corners, thick, node distance=6.5em, text width=25em, text centered, minimum height=5em, minimum width=25em]
\tikzstyle{box} = [draw=blue!30!green, fill=orange!12, rectangle, rounded corners, thick, node distance=6.5em, text width=21em, text centered, minimum height=5em, minimum width=21em]
\tikzstyle{boxsmall} = [draw=blue!30!green, fill=orange!12, rectangle, rounded corners, thick, node distance=8.5em, text width=9em, text centered, minimum height=5em, minimum width=9em]
\tikzstyle{line} = [draw=blue!30!green, -latex']
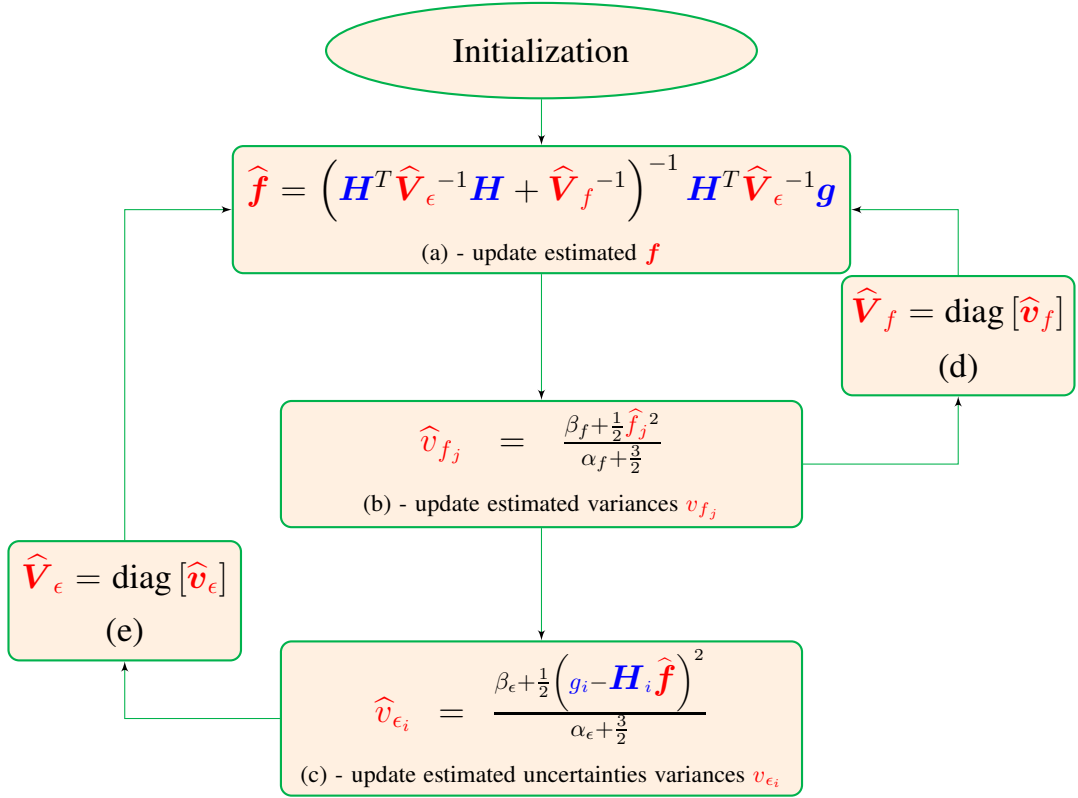
\begin{figure}[!htb]
\centering
\begin{center}
\begin{tikzpicture}[auto]
    \node [boxbig, scale=0.9] (f) {\Large{$\rfbh  = \left( \bHb^T \rVbh_\repsilon^{-1} \bHb + \rVbh_\rf^{-1} \right)^{-1} \bHb^T \rVbh_\repsilon^{-1} \bgb$}
\\[8pt]
\normalsize{(a) - update estimated $\rfb$} 
};

    \node [box, below of=f, node distance=9.6em, scale=0.9] (vf)
{\Large{$ \rvh_{\rf_\rj} = \frac{\beta_f + \frac{1}{2} \rfh_\rj^2}{\alpha_f + \frac{3}{2}} $}
\\[8pt]
\normalsize{(b) - update estimated variances $\rv_{\rf_\rj}$} 
};

    \node [box, below of=vf, node distance=9.6em, scale=0.9] (veps) {\Large{$\rvh_{\repsilon_\ri} = \frac{\beta_{\epsilon} + \frac{1}{2} \left( \bg_\bi - \bHb_\bi \rfbh \right)^2}{\alpha_{\epsilon} + \frac{3}{2}}$}
\\[8pt]
\normalsize{(c) - update estimated uncertainties variances $\rv_{\repsilon_\ri}$}     
};

    \node [boxsmall, above right = 0.05cm and 0.5cm of vf, node distance=17em, scale=0.9] (Vf) {\Large{$\rVbh_\rf = \diag {\rvbh_{\rf}}$}
\\[8pt]
(d) 
};

    \node [boxsmall, below left = -3.4cm and 0.5cm of veps, node distance=17em, scale=0.9] (Veps) {\Large{$\rVbh_\repsilon = \diag {\rvbh_{\repsilon}}$}
\\[8pt]
(e) 
};

\node [cloud, above of=f, scale=0.9] (init)
{\Large{Initialization}};  

    \path [line] (f) -- (vf)[near start];
    \path [line] (Vf) |- (f);
    \path [line] (vf) -| (Vf);
    \path [line] (vf) -- (veps);
    \path [line] (Veps) |- (f);
    \path [line] (veps) -| (Veps);
    \path [line] (init) -- (f);
    
\end{tikzpicture}
\end{center}
\caption{Iterative algorithm corresponding to Joint MAP estimation for Student-t hierarchical model, non-stationary Student-t uncertainties model}
\label{Fig:IA_JMAP_St_nsStL_un}
\end{figure}

\begin{algorithm}
\caption{Joint MAP - Student-t hierarchical model, non-stationary Student-t uncertainties model}
\begin{algorithmic}[1]
\Ensure INITIALIZATION $\rfbh^{(0)}$ \Comment{Where $\rfbh^{(0)} = \rfbh$ via a basic estimation method or $\rfbh^{(0)} = \zerob$}
\Function{JMAP}{$\alpha_{\epsilon},\beta_{\epsilon},\alpha_{f},\beta_{f},\bgb,\bHb,\rfb^{(0)},M,N,NoIter$} \Comment{$\alpha_{\epsilon}, \beta_{\epsilon}, \alpha_{f}, \beta_{f}$ are set in modeling phase, Eq.~\eqref{Eq:St_nsStL_un}}

\For{$n = 0$ to ${IterNumb}$}

\For{$j = 1$ to ${M}$}

\State $\rvh_{\rf_\rj}^{(n)} = \frac{\beta_f + \frac{1}{2} (\rfh_\rj^{(n)})^2}{\alpha_f + \frac{3}{2}}$ \Comment{$\rvh_{\rf_\rj}^{(n)}$ is computed using $\alpha_{f},\beta_{f},\rfb^{(n)}$}

\EndFor

\State $\rVbh_\rf^{(n)} = \diag {\rvbh_{\rf}^{(n)}}$ \Comment{The diagonal matrix $\rVbh_\rf^{(n)}$ is build using $\rvh_{\rf_\rj}^{(n)}$, $j \in \left\lbrace 1,2,\ldots, M\right\rbrace$}

\For{$i = 1$ to ${N}$}

\State $\rvh_{\repsilon_\ri}^{(n)} = \frac{\beta_{\epsilon} + \frac{1}{2} \left( \bg_\bi - \bHb_\bi \rfbh^{(n)} \right)^2}{\alpha_{\epsilon} + \frac{3}{2}}$ \Comment{$\rvh_{\repsilon_\ri}^{(n)}$ is computed using $\alpha_{\epsilon},\beta_{\epsilon}, \bg_\bi, \bHb_\bi, \rfb^{(n)}$} 

\EndFor

\State $\rVbh_\repsilon^{(n)} = \diag {\rvbh_{\repsilon}^{(n)}}$ \Comment{The diagonal matrix $\rVbh_\repsilon^{(n)}$ is build using $\rvh_{\repsilon_\ri}^{(n)}$, $i \in \left\lbrace 1,2,\ldots, N \right\rbrace$}

\State $\rfbh^{(n+1)}  = \left( \bHb^T \left(\rVbh_\repsilon^{(n)}\right)^{-1} \bHb + \left( \rVbh_\rf^{(n)}\right)^{-1} \right)^{-1} \bHb^T \left(\rVbh_\repsilon^{(n)}\right)^{-1} \bgb$ \Comment{New value $\rfbh^{(n+1)}$}

\EndFor

\Return $\left( \rfbh^{(n+1)}, \rvbh_\rf^{(n)}, \rvbh_{\repsilon}^{(n)} \right)$ , $n = NoIter$
\EndFunction
\end{algorithmic}
\end{algorithm}
\subsubsection{Posterior Mean estimation via VBA, partial separability}
\label{Subsubsec:PM_PS_St_nsStL_un}
In this subsection, the Posterior Mean (PM) estimation is considered. The Joint MAP computes the mod of the posterior distribution. The PM computes the mean of the posterior distribution. One of the advantages of this estimator is that it minimizes the Mean Square Error (MSE). Computing the posterior means of any unknown needs great dimensional integration. For example, the mean corresponding to $\rfb$ can be computed from the posterior distribution using Equation~\eqref{Eq:PM_Computation_F},
\beq
E_p \left\lbrace \rfb \right\rbrace = \iiint \rfb \; p(\rfb, \rvb_\rf, \rvb_\repsilon | \bgb, \alpha_{f}, \beta_{f}, \alpha_{\epsilon}, \beta_{\epsilon}) \d\rfb \d\rvb_{\rf} \d\rvb_{\repsilon}.
\label{Eq:PM_Computation_F}
\eeq
In general, these computations are not easy. One way to obtain approximate estimates is to approximate $p(\rfb, \rvb_\rf, \rvb_\repsilon | \bgb, \alpha_{f}, \beta_{f}, \alpha_{\epsilon}, \beta_{\epsilon})$ by a separable one $q(\rfb, \rvb_\rf, \rvb_\repsilon | \bgb, \alpha_{f}, \beta_{f}, \alpha_{\epsilon}, \beta_{\epsilon}) = q_1(\rfb) \; q_2(\rvb_{\rf}) \; q_3(\rvb_{\repsilon}) $, then computing the posterior means using the separability. The mean corresponding to $\rfb$ is computed using the corresponding separable distribution $q_1(\rfb)$, Equation~\eqref{Eq:PM_Computation_F_Separable}, 
\beq
E_{q_{1}} \left\lbrace \rfb \right\rbrace = \int \rfb \; q_1(\rfb) \d\rfb.
\label{Eq:PM_Computation_F_Separable}
\eeq
If the approximation of the posterior distribution with a separable one can be done in such a way that conserves the mean, i.e. Equation~\eqref{Eq:PM_Conservation},
\beq
E_q \left\lbrace x \right\rbrace = E_p \left\lbrace x \right\rbrace, 
\label{Eq:PM_Conservation}
\eeq
for all the unknowns of the model, a great amount of computational cost is gained. In particular, for the proposed hierarchical model, Equation~\eqref{Eq:St_nsStL_un}, the posterior distribution, Equation~\eqref{Eq:Post_St_nsStL_un}, is not a separable one, making the analytical computations of the PM very difficult. One way the compute the PM in this case is to first approximate the posterior law $p(\rfb, \rvb_\rf, \rvb_\repsilon | \bgb, \alpha_{f}, \beta_{f}, \alpha_{\epsilon}, \beta_{\epsilon})$ with a separable law $q(\rfb, \rvb_\rf, \rvb_\repsilon | \bgb, \alpha_{f}, \beta_{f}, \alpha_{\epsilon}, \beta_{\epsilon})$, Equation~\eqref{Eq:Posterior_Approximation},
\beq
p(\rfb, \rvb_\rf, \rvb_\repsilon | \bgb, \alpha_{f}, \beta_{f}, \alpha_{\epsilon}, \beta_{\epsilon}) 
\approx 
q(\rfb, \rvb_\rf, \rvb_\repsilon | \bgb, \alpha_{f}, \beta_{f}, \alpha_{\epsilon}, \beta_{\epsilon})
=
q_1(\rfb) \; q_2(\rvb_{\rf}) \; q_3(\rvb_{\repsilon})\label{Eq:Posterior_Approximation}
\eeq
where the notations from Equation~\eqref{Eq:Posterior_Approximation_Notations1} are used
\beq
q_{2}(\rvb_{\rf}) = \prod_{j=1}^{M} q_{2j}(\rv_{\rf_\rj}),
\;\;;\;\;
q_{3}(\rvb_{\repsilon}) = \prod_{i=1}^{N} q_{3i}(\rv_{\repsilon_\ri})
\label{Eq:Posterior_Approximation_Notations1}
\eeq
by minimizing of the Kullback-Leibler divergence, defined as:
\beq
\begin{split}
\mbox{KL} & \left( q(\rfb, \rvb_\rf, \rvb_\repsilon | \bgb, \alpha_{f}, \beta_{f}, \alpha_{\epsilon}, \beta_{\epsilon})
 : p(\rfb, \rvb_\rf, \rvb_\repsilon | \bgb, \alpha_{f}, \beta_{f}, \alpha_{\epsilon}, \beta_{\epsilon}) \right) =
\\
&=\iint \ldots \int q(\rfb, \rvb_\rf, \rvb_\repsilon | \bgb, \alpha_{f}, \beta_{f}, \alpha_{\epsilon}, \beta_{\epsilon}) \; \ln\frac{q(\rfb, \rvb_\rf, \rvb_\repsilon | \bgb, \alpha_{f}, \beta_{f}, \alpha_{\epsilon}, \beta_{\epsilon})} {p(\rfb, \rvb_\rf, \rvb_\repsilon | \bgb, \alpha_{f}, \beta_{f}, \alpha_{\epsilon}, \beta_{\epsilon})} \d \rfb \d \rvb_{\repsilon} \d \rvb_{\rf}
\label{Eq:Kullback-Leibler}
\end{split}
\eeq
where the notations from Equation~\eqref{Eq:Posterior_Approximation_Notations2} are used
\beq
\d \rvb_{\rf} = \prod_{j=1}^{M} \d \rv_{\rf_\rj}
\;\;;\;\;
\d \rvb_{\repsilon} = \prod_{i=1}^{N} \d \rv_{\repsilon_\ri}.
\label{Eq:Posterior_Approximation_Notations2}
\eeq
Equation~\eqref{Eq:Posterior_Approximation_Notations1} is selecting a partial separability for the approximated posterior distribution $q(\rfb, \rvb_\rf, \rvb_\repsilon | \bgb, \alpha_{f}, \beta_{f}, \alpha_{\epsilon}, \beta_{\epsilon})$ in the sense that a total separability is imposed for the distributions corresponding to the two variances appearing in the hierarchical model, $q_2\left( \rvb_{\rf} \right)$ and $q_3\left( \rvb_{\repsilon} \right)$ but not for the distribution corresponding to $\rfb$. Evidently, a full separability can be imposed, by adding the supplementary condition $q_{1}(\rfb) = \prod_{j=1}^{M} q_{1j}(\rf_\rj)$ in Equation~\eqref{Eq:Posterior_Approximation_Notations1}. This case is considered in Subsection~\eqref{Subsubsec:PM_FS_St_nsStL_un}. The minimization can be done via alternate optimization resulting the following proportionalities from Equations~\eqref{Eq:VBA_Proportionalities1},~\eqref{Eq:VBA_Proportionalities2} and ~\eqref{Eq:VBA_Proportionalities3},  
\begin{subequations}
\begin{align}
&q_1(\rfb) \;\;\; \propto \; \exp \biggl\lbrace \left\langle \ln p(\rfb, \rvb_\rf, \rvb_\repsilon | \bgb, \alpha_{f}, \beta_{f}, \alpha_{\epsilon}, \beta_{\epsilon}) \biggr\rangle_{ q_2(\rvb_{\rf}) \; q_3(\rvb_{\repsilon})} \right\rbrace,
\label{Eq:VBA_Proportionalities1}
\\[4pt]
&q_{2j}(\rv_{\rf_\red{j}}) \propto \; \exp \biggl\lbrace \left\langle \ln p(\rfb, \rvb_\rf, \rvb_\repsilon | \bgb, \alpha_{f}, \beta_{f}, \alpha_{\epsilon}, \beta_{\epsilon}) \biggr\rangle_{q_1(\rfb) \; q_{2-j}(\rv_{\rf_\red{j}}) \; q_3(\rvb_{\repsilon})} \right\rbrace,\;\;j \in \left\lbrace 1,2 \ldots, M \right\rbrace,
\label{Eq:VBA_Proportionalities2}
\\[4pt]
&q_{3i}(\rv_{\repsilon_\ri}) \; \propto \; \exp \biggl\lbrace \left\langle \ln p(\rfb, \rvb_\rf, \rvb_\repsilon | \bgb, \alpha_{f}, \beta_{f}, \alpha_{\epsilon}, \beta_{\epsilon}) \biggr\rangle_{q_1(\rfb) \; q_2(\rvb_{\rf}) \; q_{3-i}(\rv_{\repsilon_\ri})} \right\rbrace,\;\;i \in \left\lbrace 1,2 \ldots, N \right\rbrace,
\label{Eq:VBA_Proportionalities3}
\end{align}
\end{subequations}
using the notations:
\beq
q_{2-j}(\rv_{\rf_\red{j}})=\displaystyle \prod_{k=1,k \neq j}^{M} q_{2k}(\rv_{\rf_\red{k}})
\;\;\;;\;\;\;
q_{3-i}(\rv_{\repsilon_\ri})=\displaystyle \prod_{k=1,k \neq i}^{N} q_{3k}(\rv_{\repsilon_\rk})
\label{Eq:Def_q_Minus}
\eeq
and also
\beq
\biggl\langle u(x) \biggr\rangle_{v(y)}= \displaystyle \int u(x) v(y) \d y. 
\label{Eq:Def_Integral}
\eeq
Via Equation~\eqref{Eq:Def_L_Criterion} and Equation~\eqref{Eq:L_Criterion}, the analytical expression of logarithm of the posterior distribution is obtained, Equation~\eqref{Eq:Log_Posterior}:
\beq
\begin{split}
\ln  p(\rfb, \rvb_{\repsilon}, \rvb_{\rf}|\bgb)
=
&
-\frac{1}{2}\| \rVb_{\repsilon}^{-\frac{1}{2}} \left( \bgb - \bHb\rfb \right) \|
-
\left( \alpha_{\epsilon} + \frac{3}{2} \right) \sum_{i=1}^{N} \ln \rv_{\repsilon_\ri} -
\sum_{i=1}^{N}\frac{\beta_{\epsilon}}{\rv_{\repsilon_\ri}} 
\\
&
-
\frac{1}{2}\| \rVb_{\rf}^{-\frac{1}{2}} \rfb \|
-
\left(\alpha_{f} + \frac{3}{2} \right) \sum_{j=1}^{M} \ln \rv_{\rf_\rj}
-
\sum_{j=1}^{M}\frac{\beta_{f}}{\rv_{\rf_\rj}}
\end{split}
\label{Eq:Log_Posterior}
\eeq
\paragraph{Computation of the analytical expression of $q_1(\rfb)$.}
The proportionality relation corresponding to $q_1(\rfb)$ is presented in established in Equation~\eqref{Eq:VBA_Proportionalities1}. In the expression of $\ln p\left(\rfb, \rvb_\rf, \rvb_\repsilon | \bgb, \alpha_{f}, \beta_{f}, \alpha_{\epsilon}, \beta_{\epsilon} \right)$ all the terms free of $\rfb$ can be regarded as constants. Via Equation~\eqref{Eq:Log_Posterior} the integral $\left\langle \right\rangle$ defined in Equation~\eqref{Eq:Def_Integral} becomes:
\beq
\begin{split}
\biggl\langle \ln p(\rfb, \rvb_\rf, \rvb_\repsilon | \bgb, \alpha_{f}, \beta_{f}, \alpha_{\epsilon}, \beta_{\epsilon}) \biggr\rangle_{q_2(\rvb_{\rf}) \; q_3(\rvb_{\repsilon})}
&
=
-\frac{1}{2}
\biggl\langle 
\|\rVb_{\repsilon}^{-\frac{1}{2}} \left(\bgb - \bHb\rfb\right) \|
\biggr\rangle_{ q_3(\rvb_{\repsilon}) }
-\frac{1}{2}
\biggl\langle 
\| \rVb_{\rf}^{-\frac{1}{2}} \rfb \|
\biggr\rangle_{ q_2(\rvb_{\rf}) }.
\end{split}
\label{Eq:q1_Integral_1}
\eeq
Introducing the notations:
\beq
\begin{split}
\rvt_{\rf_\rj} = \biggl\langle \rv_{\rf_\rj}^{-1} \biggr\rangle_{q_{4j}\left( \rv_{\rf_\rj} \right)}
\;\;;\;\;
\rvbt_{\rf}\;&=\;
\begin{bmatrix}
\rvt_{\rf_\red{1}} \ldots 
\rvt_{\rf_\rj} \ldots 
\rvt_{\rf_\rM}
\end{bmatrix}^T
\;\;;\;\;
\rVbt_{\rf} = \diag {\rvbt_{\rf}}\;
\\
\rvt_{\repsilon_\ri}
=
\biggl\langle \rv_{\repsilon_\ri}^{-1} \biggr\rangle_{q_{3i}\left( \rv_{\repsilon_\ri} \right)}
\;\;;\;\;
\rvbt_\repsilon\;&=\;
\begin{bmatrix}
\rvt_{\repsilon_\red{1}}
\ldots 
\rvt_{\repsilon_\ri}
\ldots 
\rvt_{\repsilon_\rN}
\end{bmatrix}^T
\;\;;\;\;
\rVbt_\repsilon = \diag {\rvbt_\repsilon}
\end{split}
\label{Eq:q1_Integral_Not1}
\eeq
the integral from Equation~\eqref{Eq:q1_Integral_1} becomes:
\beq
\begin{split}
\biggl\langle \ln p(\rfb, \rvb_\rf, \rvb_\repsilon | \bgb, \alpha_{f}, \beta_{f}, \alpha_{\epsilon}, \beta_{\epsilon}) \biggr\rangle_{q_2(\rvb_{\rf}) \; q_3(\rvb_{\repsilon})}
&
= -\frac{1}{2} \|\rVbt_{\repsilon}^{-\frac{1}{2}} \left(\bgb - \bHb\rfb\right) \| - \frac{1}{2} \| \rVbt_{\rf}^{-\frac{1}{2}} \rfb \|.
\end{split}
\eeq
Noting that $\biggl\langle \ln p(\rfb, \rvb_\rf, \rvb_\repsilon | \bgb, \alpha_{f}, \beta_{f}, \alpha_{\epsilon}, \beta_{\epsilon}) \biggr\rangle_{q_2(\rvb_{\rf}) \; q_3(\rvb_{\repsilon})}
$ is a quadratic criterion and considering the proportionality from Equation~\eqref{Eq:VBA_Proportionalities1} it can be concluded that $q_1 \left( \rfb \right)$  is a multivariate Normal distribution. Minimizing the criterion leads to the analytical expression of the corresponding mean. The variance is obtained by identification: 
\beq
q_1(\rfb) = \Nc\left( \rfb | \rfbh, \Sigmabh \right),
\left\{\barr{ll}
\rfbh = 
\left( \bHb^T  \rVbt_\repsilon \bHb + \rVbt_{\rf} \right)^{-1}
\bHb^T \rVbt_\repsilon \bgb,
\\[10pt]
\Sigmabh = \left( \bHb_\blue{1}^T  \rVbt_\repsilon \bHb + \rVbt_{\rf} \right)^{-1}.
\earr\right.
\label{Eq:q1_Analytical_1}
\eeq
We note that both the expressions of the mean and variance depend on expectancies corresponding to two variances of the hierarchical model.
\paragraph{Computation of the analytical expression of $q_{2j}(\rv_{\rf_\rj})$.}
The proportionality relation corresponding to $q_{2j}(\rv_{\rf_\rj})$ is presented in established in Equation~\eqref{Eq:VBA_Proportionalities2}. In the expression of $\ln p\left(\rfb, \rvb_\rf, \rvb_\repsilon | \bgb, \alpha_{f}, \beta_{f}, \alpha_{\epsilon}, \beta_{\epsilon} \right)$ all the terms free of $\rv_{\rf_\rj}$ can be regarded as constants. Via Equation~\eqref{Eq:Log_Posterior} the integral defined in Equation~\eqref{Eq:Def_Integral} becomes:
\beq
\begin{split}
\biggl\langle \ln p(\rfb, \rvb_\rf, \rvb_\repsilon | \bgb, \alpha_{f}, \beta_{f}, \alpha_{\epsilon}, \beta_{\epsilon}) \biggr\rangle_{q_1(\rfb) \; q_{2-j}(\rv_{\rf_\red{j}}) \; q_3(\rvb_{\repsilon})}
=
&
-
\frac{1}{2}
\biggl\langle
\| \rVb_{\rf}^{-\frac{1}{2}} \rfb \|
\biggr\rangle_{q_1(\rfb) \; q_{2-j}(\rv_{\rf_\red{j}})}
\\
&
-
\left(\alpha_{f} + \frac{3}{2} \right) \ln \rv_{\rf_\rj}
-
\frac{\beta_{f}}{\rv_{\rf_\rj}}
\end{split}
\label{Eq:q2_Integral_1}
\eeq
Introducing the notations:
\beq
\rvbt_{\rf_{-\ri}}^{-1}=
\begin{bmatrix}
\rvt_{\rf_\red{1}}^{-1} \;
\ldots \;
\rvt_{\rf_{\ri-\red{1}}}^{-1} \;
\rv_{\rf_\ri}^{-1}  \;
\rvt_{\rf_{\ri+\red{1}}}^{-1}  \;
\ldots \;
\rvt_{\rf_\rN}^{-1}
\end{bmatrix}^T
\;\;;\;\;
\rVbt_{\rf_{-\ri}}^{-1}=
\mbox{diag}\left(\rvbt_{\rf_{-\ri}}^{-1} \right)
\label{Eq:q2_Integral_Not1}
\eeq
the integral $\biggl\langle
\| \rVb_{\rf}^{-\frac{1}{2}} \rfb \|
\biggr\rangle_{q_1(\rfb) \; q_{2-j}(\rv_{\rf_\red{j}})}
$ can be written:
\beq
\biggl\langle
\| \rVb_{\rf}^{-\frac{1}{2}} \rfb \|
\biggr\rangle_{q_1(\rfb) \; q_{2-j}(\rv_{\rf_\red{j}})}
= 
\biggl\langle \| \rVbt_{\rf_{-\ri}}^{-\frac{1}{2}} \rfb \|^2 \biggr\rangle_{q_1(\rfb)}
\eeq 
Considering that $q_1(\rfb)$ is a multivariate Normal distribution, Equation~\eqref{Eq:q1_Analytical_1}:
\beq
\biggl\langle \|  \rVbt_{\rf_{-\ri}}^{-\frac{1}{2}} \rfb \|^2 \biggr\rangle_{q_1(\rfb)}
=
\| \rVbt_{\rf_{-\ri}}^{-\frac{1}{2}} \rfbh \|^2 + \mbox{Tr}\left(\rVbt_{\rf_{-\ri}}^{-1} \Sigmabh \right) 
=
C + \rv_{\rf_\ri}^{-1} \left( \rfh_\rj^2 + \Sigmabh_{jj} \right)
\label{Eq:q2_Integral_2}
\eeq
From Equation~\eqref{Eq:q2_Integral_1} and Equation~\eqref{Eq:q2_Integral_2}: 
\beq
\begin{split}
\biggl\langle \ln p(\rfb, \rvb_\rf, \rvb_\repsilon | \bgb, \alpha_{f}, \beta_{f}, \alpha_{\epsilon}, \beta_{\epsilon}) \biggr\rangle_{q_1(\rfb) \; q_{2-j}(\rv_{\rf_\red{j}}) \; q_3(\rvb_{\repsilon})}
=
- \left(\alpha_{f} + \frac{3}{2} \right) \ln \rv_{\rf_\rj}
- \left( \beta_{f} + \frac{1}{2} \left( \rfh_\rj^2 + \Sigmabh_{jj} \right) \right) \rv_\rf^{-1}
\end{split}
\eeq
from which it can establish the proportionality corresponding to
$q_{2j}(\rv_{\rf_\rj})$:
\beq
q_{2j}(\rv_{\rf_\rj})
\propto
\rv_{\rf_\rj}^{-\left(\alpha_{f} + \frac{3}{2} \right)}
\exp
\left\lbrace 
-
\frac{\beta_{f} +
\frac{1}{2}
\left(\rfh_{\rj}^2 + \Sigmabh_{jj} \right) }{\rv_\rf}
\right\rbrace,
\eeq
leading to the conclusion that $q_{2j}(\rv_{\rf_\rj})$ is an Inverse Gamma distribution with the following shape and scale parameters:
\beq
q_{2j}(\rv_{\rf_\rj})=\Ic\Gc\left(\rv_{\rf_\rj}|\alphah_{f_j},\betah_{f_j}\right),
\left\{\barr{ll}
\alphah_{f_j} = \alpha_f + \frac{1}{2}
\\[10pt]
\betah_{f_j} = \beta_f + \frac{1}{2}
\left(\rfh_{\rj}^2 + \Sigmabh_{jj} \right)
\earr\right.
\label{Eq:q2_Analytical_1}
\eeq
\paragraph{Computation of the analytical expression of $q_{3i}(\rv_{\repsilon_\ri})$.}
The proportionality relation corresponding to $q_{3i}(\rv_{\repsilon_\ri})$ is presented in established in Equation~\eqref{Eq:VBA_Proportionalities3}. In the expression of $\ln p\left(\rfb, \rvb_\rf, \rvb_\repsilon | \bgb, \alpha_{f}, \beta_{f}, \alpha_{\epsilon}, \beta_{\epsilon} \right)$ all the terms free of $\rv_{\repsilon_\ri}$ can be regarded as constants. Via Equation~\eqref{Eq:Log_Posterior} the integral defined in Equation~\eqref{Eq:Def_Integral} becomes:
\beq
\begin{split}
\biggl\langle \ln p(\rfb, \rvb_\rf, \rvb_\repsilon | \bgb, \alpha_{f}, \beta_{f}, \alpha_{\epsilon}, \beta_{\epsilon}) \biggr\rangle_{q_1(\rfb) \; q_{2}(\rvb_{\rf}) \; q_{3-i}(\rv_{\repsilon_\ri})}
=
&
-\frac{1}{2} \left\langle \| \rVb_{\repsilon}^{-\frac{1}{2}} \left( \bgb - \bHb \rfb \right) \| \right\rangle_{q_1(\rfb) \; q_{3-i}(\rv_{\repsilon_\ri})}
\\
&
-
\left(\alpha_{\epsilon} + \frac{3}{2} \right) \ln \rv_{\repsilon_\ri}
-
\frac{\beta_{\epsilon}}{\rv_{\repsilon_\ri}}
\end{split}
\label{Eq:q3_Integral_1}
\eeq
Introducing the notations:
\beq
\rvbt_{\repsilon_{-\ri}}^{-1}=
\begin{bmatrix}
\rvt_{\repsilon_\red{1}}^{-1} \;
\ldots \;
\rvt_{\repsilon_{\ri-\red{1}}}^{-1} \;
\rv_{\repsilon_\ri}^{-1}  \;
\rvt_{\repsilon_{\ri+\red{1}}}^{-1}  \;
\ldots \;
\rvt_{\repsilon_\rN}^{-1}
\end{bmatrix}^T
\;\;;\;\;
\rVbt_{\repsilon_{-\ri}}^{-1}=
\mbox{diag}\left(\rvbt_{\repsilon_{-\ri}}^{-1} \right)
\label{Eq:q3_Integral_Not1}
\eeq
the integral $\biggl\langle \| \rVb_{\repsilon}^{-\frac{1}{2}} \left( \bgb - \bHb \rfb \right) \| \biggr\rangle_{q_1(\rfb) \; q_{3-i}(\rv_{\repsilon_\ri})}
$ can be written:
\beq
\biggl\langle \| \rVb_{\repsilon}^{-\frac{1}{2}} \left( \bgb - \bHb \rfb \right) \| \biggr\rangle_{q_1(\rfb) \; q_{3-i}(\rv_{\repsilon_\ri})}
= 
\biggl\langle \| \left(\rVbt_{\repsilon_{-\ri}}^{-1}\right)^{\frac{1}{2}} \left( \bgb - \bHb \rfb \right) \|^2 \biggr\rangle_{q_1(\rfb)}
\eeq 
Considering that $q_1(\rfb)$ is a multivariate Normal distribution, Equation~\eqref{Eq:q1_Analytical_1}:
\beq
\biggl\langle
\| \left( \rVbt_{\repsilon_{-\ri}}^{-1} \right)^{\frac{1}{2}} \left( \bgb - \bHb \rfb \right) \|^2
\biggr\rangle_{q_1(\rfb)}
=
\| \left( \rVbt_{\repsilon_{-\ri}}^{-1}\right)^{\frac{1}{2}}
\left( \bgb - \bHb \rfbh \right) \|^2
+
\mbox{Tr}\left( \bHb^T \rVb_{\repsilon_{-\ri}}^{-1} \bHb \Sigmabh \right)
\label{Eq:q3_Integral_2}
\eeq
and considering as constants all terms free of $\rv_{\repsilon_\ri}$:
\beq
\| \left( \rVbt_{\repsilon_{-\ri}}^{-1}\right)^{\frac{1}{2}}
\left( \bgb - \bHb \rfbh \right) \|^2
=
C +
\rv_{\repsilon_\ri}^{-1} \left( \bg_\bi - \bHb_\bi \rfbh \right)^2
\;\;;\;\;
\mbox{Tr}\left( \bHb^T \rVbt_{\repsilon_{-\ri}}^{-1} \bHb \Sigmabh \right)
=
C + 
\rv_{\repsilon_\ri}^{-1}\bHb_\bi \Sigmabh \bHb_\bi^T
\eeq
where $\bHb_\bi$ is the line i of the matrix $\bHb$, so we can conclude:
\beq
\biggl\langle
\| \rVb_{\repsilon}^{-\frac{1}{2}}\left( \bgb - \bHb\rfb \right)\|^2
\biggr\rangle_{q_1(\rfb) \; q_{3-i}(\rv_{\repsilon_\ri})}
=
C + 
\left(
\bHb_\bi \Sigmabh \bHb_\bi^T
+
\left( \bg_\bi - \bHb_\bi \rfbh \right)^2
\right)
\rv_{\repsilon_\ri}^{-1}
\label{Eq:q3_Integral_3}
\eeq
From Equation~\eqref{Eq:q3_Integral_1} and Equation~\eqref{Eq:q3_Integral_3}: 
\beq
\begin{split}
\left\langle \ln p(\rfb, \rvb_\rf, \rvb_\repsilon | \bgb, \alpha_{f}, \beta_{f}, \alpha_{\epsilon}, \beta_{\epsilon}) \right\rangle_{q_1(\rfb) \; q_{2}(\rvb_{\rf}) \; q_{3-i}(\rv_{\repsilon_\ri})}
=
&
- \left(\alpha_{\epsilon} + \frac{3}{2} \right) \ln \rv_{\repsilon_\ri}
\\
&
- 
\left( \beta_{\epsilon} + \frac{1}{2} \left(
\bHb_\bi \Sigmabh \bHb_\bi^T
+
\left( \bg_\bi - \bHb_\bi \rfbh \right)^2
\right)
 \right) \rv_{\repsilon_\ri}^{-1}
\end{split}
\eeq
from which it can establish the proportionality corresponding to
$q_{3i}(\rv_{\repsilon_\ri})$:
\beq
q_{3i}(\rv_{\repsilon_\ri})
\propto
\rv_{\repsilon_\ri}^{-\left(\alpha_{\epsilon} + \frac{3}{2} \right)}
\exp
\left\lbrace 
-
\frac{\beta_{\epsilon} +
\frac{1}{2}
\left(
\bHb_\bi \Sigmabh \bHb_\bi^T
+
\left( \bg_\bi - \bHb_\bi \rfbh \right)^2
\right)}{\rv_\repsilon}
\right\rbrace,
\eeq
leading to the conclusion that $q_{3i}(\rv_{\repsilon_\ri})$ is an Inverse Gamma distribution with the following shape and scale parameters:
\beq
q_{3i}(\rv_{\repsilon_\ri})=\Ic\Gc\left(\rv_{\repsilon_\ri}|\alphah_{\epsilon_i},\betah_{\epsilon_i}\right),
\left\{\barr{ll}
\alphah_{\epsilon_j} = \alpha_\epsilon + \frac{1}{2}
\\[10pt]
\betah_{\epsilon_j} = \beta_\epsilon +
\frac{1}{2}
\left(
\bHb_\bi \Sigmabh \bHb_\bi^T
+
\left( \bg_\bi - \bHb_\bi \rfbh \right)^2
\right)
\earr\right.
\label{Eq:q3_Analytical_1}
\eeq
Equations~\eqref{Eq:q1_Analytical_1},~\eqref{Eq:q2_Analytical_1} and~\eqref{Eq:q3_Analytical_1} resume the distributions families and the corresponding parameters for $q_1(\rfb)$, a multivariate Normal distribution and  $q_{2j}(\rv_{\rf_\rj})$, $j\in\left\lbrace 1, 2, \ldots, M \right\rbrace$ and $q_{3i}(\rv_{\repsilon_\ri})$, $i\in\left\lbrace 1, 2, \ldots, N \right\rbrace$, Inverse Gamma distributions. However, the parameters corresponding to the multivariate Normal distribution are expressed via $\rVbt_{\repsilon}^{-1}$ and $\rVbt_{\rf}^{-1}$ (and by extension all elements forming the three matrices $\rvt_{\repsilon_\ri}^{-1}$, $i\in\left\lbrace 1, 2, \ldots, N \right\rbrace$ and $\rvt_{\rf_\rj}^{-1}$, $j\in\left\lbrace 1, 2, \ldots, M \right\rbrace$).
\paragraph{Computation of the analytical expressions of $\rVbt_{\repsilon}^{-1}$ and $\rVbt_{\rf}^{-1}$.}
For an Inverse Gamma distribution with scale and shape parameters $\alpha$ and $\beta$, $\Ic\Gc\left(x|\alpha, \beta \right)$, the following relation holds:
\beq
\biggl\langle x^{-1} \biggr\rangle_{\Ic\Gc(x|\alpha,\beta)} 
=
\frac{\alpha}{\beta}
\label{Eq:Inverse_Gamma_Integral}
\eeq
The prove of the above relation is done by direct computation, using the analytical expression of the Inverse Gamma Distribution: 
\beq
\begin{split}
\biggl\langle x^{-1} \biggr\rangle_{\Ic\Gc(x|\alpha,\beta)} 
&= 
\int 
x^{-1}
\frac{{\beta}^{\alpha}}{\Gamma(\alpha)}
x^{-\alpha-1}
\exp
\left\lbrace
- \frac{\beta}{x}
\right\rbrace
\d x
=
\frac{{\beta}^{\alpha}}{\Gamma(\alpha)}
\frac{\Gamma(\alpha+1)}{{\beta}^{\alpha+1}}
\int
\frac{{\beta}^{\alpha+1}}{\Gamma(\alpha+1)}
x^{-(\alpha+1)-1}
\exp\left\lbrace -\frac{\beta}{x}\right\rbrace
\d x
=\\
&=
\frac{\alpha}{\beta}
\underbrace{
\int \Ic\Gc(x|\alpha+1,\beta)
}_{1}
\d x
=
\frac{\alpha}{\beta}
\end{split}
\nonumber
\eeq
Since $q_{2j}(\rv_{\rf_\rj})$, $j\in\left\lbrace 1, 2, \ldots, M \right\rbrace$ and $q_{3i}(\rv_{\repsilon_\ri})$, $i\in\left\lbrace 1, 2, \ldots, N \right\rbrace$ are Inverse Gamma distributions, with the corresponding parameters $\alphah_{f_j}$ and $\betah_{f_j}$, $j\in\left\lbrace 1, 2, \ldots, M \right\rbrace$ respectively $\alphah_{\epsilon_i}$ and $\betah_{\epsilon_i}$, $i\in\left\lbrace 1, 2, \ldots, N \right\rbrace$ the expectancies $\rvt_{\rf_\rj}^{-1}$ and $\rvt_{\repsilon_\ri}^{-1}$ can be expressed via the parameters of the two Inverse Gamma distributions using Equation~\eqref{Eq:Inverse_Gamma_Integral}:
\beq
\rvt_{\rf}^{-1}
=
\frac{\alphah_{f}}{\betah_{f}}
\;\;\;;\;\;\;
\rvt_{\repsilon_\ri}^{-1}
=
\frac{\alphah_{\epsilon_i}}{\betah_{\epsilon_i}}
\label{Eq:VepsVfExpectanciesIGSMGen}
\eeq
Using the notation introduced in \eqref{Eq:q1_Integral_Not1}:
\beq
\rVbt_{\rf}^{-1}=
\begin{bmatrix}
\frac{\alphah_{f_1}}{\betah_{f_1}} \ldots 0 \ldots 0 \\
\vdots \ddots \vdots \ddots \vdots \\
0 \ldots \frac{\alphah_{f_j}}{\betah_{f_j}} \ldots 0 \\
\vdots \ddots \vdots \ddots \vdots \\
0 \ldots 0 \ldots \frac{\alphah_{f_M}}{\betah_{f_M}} \\
\end{bmatrix}
=
\rVbh_{\rf}^{-1}
\;\;;\;\;
\rVbt_{\repsilon}^{-1}=
\begin{bmatrix}
\frac{\alphah_{\epsilon_1}}{\betah_{\epsilon_1}} \ldots 0 \ldots 0 \\
\vdots \ddots \vdots \ddots \vdots \\
0 \ldots \frac{\alphah_{\epsilon_i}}{\betah_{\epsilon_i}} \ldots 0 \\
\vdots \ddots \vdots \ddots \vdots \\
0 \ldots 0 \ldots \frac{\alphah_{\epsilon_N}}{\betah_{\epsilon_N}} \\
\end{bmatrix}
=
\rVbh_{\repsilon}^{-1}
\label{Eq:V_Expectancies}
\eeq
In Equation~\eqref{Eq:V_Expectancies} other notations are introduced for $\rVbt_{\rf}^{-1}$ and $\rVbt_{\repsilon}^{-1}$. Both values were expressed during the model via unknown expectancies, but via Equation~\eqref{Eq:V_Expectancies} those values don't contain any more integrals to be computed. Therefore, the new notations represent the final analytical expressions used for expressing the density functions $q_i$.
Using Equation~\eqref{Eq:V_Expectancies} and Equations~\eqref{Eq:q1_Analytical_1},~\eqref{Eq:q2_Analytical_1} and~\eqref{Eq:q3_Analytical_1}, the final analytical expressions of the separable distributions $q_i$ are presented in Equations~\eqref{Eq:q1_Analytical_2},~\eqref{Eq:q2_Analytical_2} and~\eqref{Eq:q3_Analytical_2}.
\begin{subequations}
\begin{align}
&q_1(\rfb) = \Nc\left( \rfb | \rfbh, \Sigmabh \right),
\;\;\;\;\;\;\;\;\;\;\;\;
\left\{\barr{ll}
\rfbh = 
\left( \bHb^T  \rVbh_\repsilon \bHb + \rVbh_{\rf} \right)^{-1}
\bHb^T \rVbh_\repsilon \bgb,
\\[10pt]
\Sigmabh = \left( \bHb_\blue{1}^T  \rVbh_\repsilon \bHb + \rVbh_{\rf} \right)^{-1}
\earr\right.,
\label{Eq:q1_Analytical_2}
\\[4pt]
&q_{2j}(\rv_{\rf_\rj})=\Ic\Gc\left(\rv_{\rf_\rj}|\alphah_{f_j},\betah_{f_j}\right),
\left\{\barr{ll}
\alphah_{f_j} = \alpha_f + \frac{1}{2}
\\[10pt]
\betah_{f_j} = \beta_f + \frac{1}{2}
\left(\rfh_{\rj}^2 + \Sigmabh_{jj} \right)
\earr\right.
,j\in\left\lbrace 1, 2, \ldots, M \right\rbrace,
\label{Eq:q2_Analytical_2}
\\[4pt]
&q_{3i}(\rv_{\repsilon_\ri})=\Ic\Gc\left(\rv_{\repsilon_\ri}|\alphah_{\epsilon_i},\betah_{\epsilon_i}\right),
\;\;\;
\left\{\barr{ll}
\alphah_{\epsilon_j} = \alpha_\epsilon + \frac{1}{2}
\\[10pt]
\betah_{\epsilon_j} = \beta_\epsilon +
\frac{1}{2}
\left(
\bHb_\bi \Sigmabh \bHb_\bi^T
+
\left( \bg_\bi - \bHb_\bi \rfbh \right)^2
\right)
\earr\right.
,i\in\left\lbrace 1, 2, \ldots, N \right\rbrace.
\label{Eq:q3_Analytical_2}
\end{align}
\end{subequations}
Equation~\eqref{Eq:q1_Analytical_2} establishes the dependency between the parameters corresponding to the multivariate Normal distribution $q_1(\rfb)$ and the others parameters involved in the hierarchical model: the mean $\rfbh$ and the covariance matrix $\Sigmabh$ depend on $\rVbh_{\repsilon}^{-1}$ and $\rVbh_{\rf}^{-1}$ which, via Equation~\eqref{Eq:V_Expectancies} are defined using $\left\lbrace \alphah_{f_j},\betah_{f_j}\right\rbrace, j\in \left\lbrace 1, 2, \ldots, M \right\rbrace $ and $\left\lbrace \alphah_{\epsilon_i},\betah_{\epsilon_i}\right\rbrace, i\in \left\lbrace 1, 2, \ldots, N \right\rbrace $. The dependency between the parameters of the multivariate Normal distribution $q_1(\rfb)$ and the parameters of the Inverse Gamma distributions $q_{2j}(\rvb_{\rf_\rj}) ,j \in \left\lbrace 1,2,\ldots,M \right\rbrace$ and $q_{3i}(\rvb_{\repsilon_\ri}), i \in \left\lbrace 1,2,\ldots,N \right\rbrace$ is presented in Figure~\eqref{Fig:Dependency_Scheme_1}.
\begin{figure}[!htb]
\center
\begin{tabular}{c}
\begin{picture}(180,30)
\put(0,0){\framebox(100,26){$\left\lbrace \alphah_{f_j},\betah_{f_j}\right\rbrace,\left\lbrace \alphah_{\epsilon_j},\betah_{\epsilon_j}\right\rbrace$}}
\put(102,13){\vector(1,0){26}}
\put(128,0){\framebox(56,26){$\rfbh \; , \; \Sigmabh$}}
\end{picture}
\end{tabular}
\caption{Dependency between $q_1(\rfb)$ parameters and $q_{2j}(\rvb_{\rf_\rj})$ and $q_{3i}(\rvb_{\repsilon_\ri})$ parameters}
\label{Fig:Dependency_Scheme_1}
\end{figure}
Equation~\eqref{Eq:q2_Analytical_2} establishes the dependency between the parameters corresponding to the Inverse Gamma distributions $q_{2j}(\rvb_{\rf_\rj}) ,j \in \left\lbrace 1,2,\ldots,M \right\rbrace$ and the others parameters involved in the hierarchical model: the shape and scale parameters $\left\lbrace \alphah_{f_j}, \betah_{f_j} \right\rbrace, j \in \left\lbrace 1,2,\ldots,M \right\rbrace$ depend on the mean $\rfbh$ and the covariance matrix $\Sigmabh$ of the multivariate Normal distribution $q_1(\rfb)$, Figure~\eqref{Fig:Dependency_Scheme_2}.
\begin{figure}[!htb]
\center
\begin{tabular}{c}
\begin{picture}(120,30)
\put(0,0){\framebox(46,26){$ \rfbh \; , \; \Sigmabh$}}
\put(48,13){\vector(1,0){26}}
\put(76,0){\framebox(56,26){$\left\lbrace \alphah_{f_j},\betah_{f_j}\right\rbrace$}}
\end{picture}
\end{tabular}
\caption{Dependency between $q_{2j}(\rvb_{\rf_\rj})$ parameters and $q_1(\rfb)$ and $q_{3i}(\rvb_{\repsilon_\ri})$ parameters}
\label{Fig:Dependency_Scheme_2}
\end{figure}
Equation~\eqref{Eq:q3_Analytical_2} establishes the dependency between the parameters corresponding to the Inverse Gamma distributions $q_{3i}(\rvb_{\repsilon_\ri}) ,i \in \left\lbrace 1,2,\ldots,N \right\rbrace$ and the others parameters involved in the hierarchical model: the shape and scale parameters $\left\lbrace \alphah_{\epsilon_i}, \betah_{\epsilon_i} \right\rbrace, i \in \left\lbrace 1,2,\ldots,N \right\rbrace$ depend on the mean $\rfbh$ and the covariance matrix $\Sigmabh$ of the multivariate Normal distribution $q_1(\rfb)$, Figure~\eqref{Fig:Dependency_Scheme_3}.
\begin{figure}[!htb]
\center
\begin{tabular}{c}
\begin{picture}(120,30)
\put(0,0){\framebox(46,26){$ \rfbh \; , \; \Sigmabh $}}
\put(48,13){\vector(1,0){26}}
\put(76,0){\framebox(56,26){$\left\lbrace \alphah_{\epsilon_j},\betah_{\epsilon_j}\right\rbrace $}}
\end{picture}
\end{tabular}
\caption{Dependency between $q_{3i}(\rvb_{\repsilon_\ri})$ parameters and $q_{2j}(\rvb_{\rf_\rj})$ and $q_1(\rfb)$ parameters}
\label{Fig:Dependency_Scheme_3}
\end{figure}
\newline
The iterative algorithm obtained via PM estimation is presented Figure~\eqref{Fig:IA_PM_St_nsStL_un}.
\tikzstyle{cloud} = [draw=blue!30!green,fill=orange!12, ellipse, thick, node distance=6em, text width=12em, text centered, minimum height=4em, minimum width=12em]
\tikzstyle{boxbig} = [draw=blue!30!green, fill=orange!12, rectangle, rounded corners, thick, node distance=8em, text width=25em, text centered, minimum height=5em, minimum width=25em]
\tikzstyle{box} = [draw=blue!30!green, fill=orange!12, rectangle, rounded corners, thick, node distance=6.5em, text width=21em, text centered, minimum height=5em, minimum width=21em]
\tikzstyle{boxsmall} = [draw=blue!30!green, fill=orange!12, rectangle, rounded corners, thick, node distance=8.5em, text width=11em, text centered, minimum height=5em, minimum width=11em]
\tikzstyle{line} = [draw=blue!30!green, -latex']
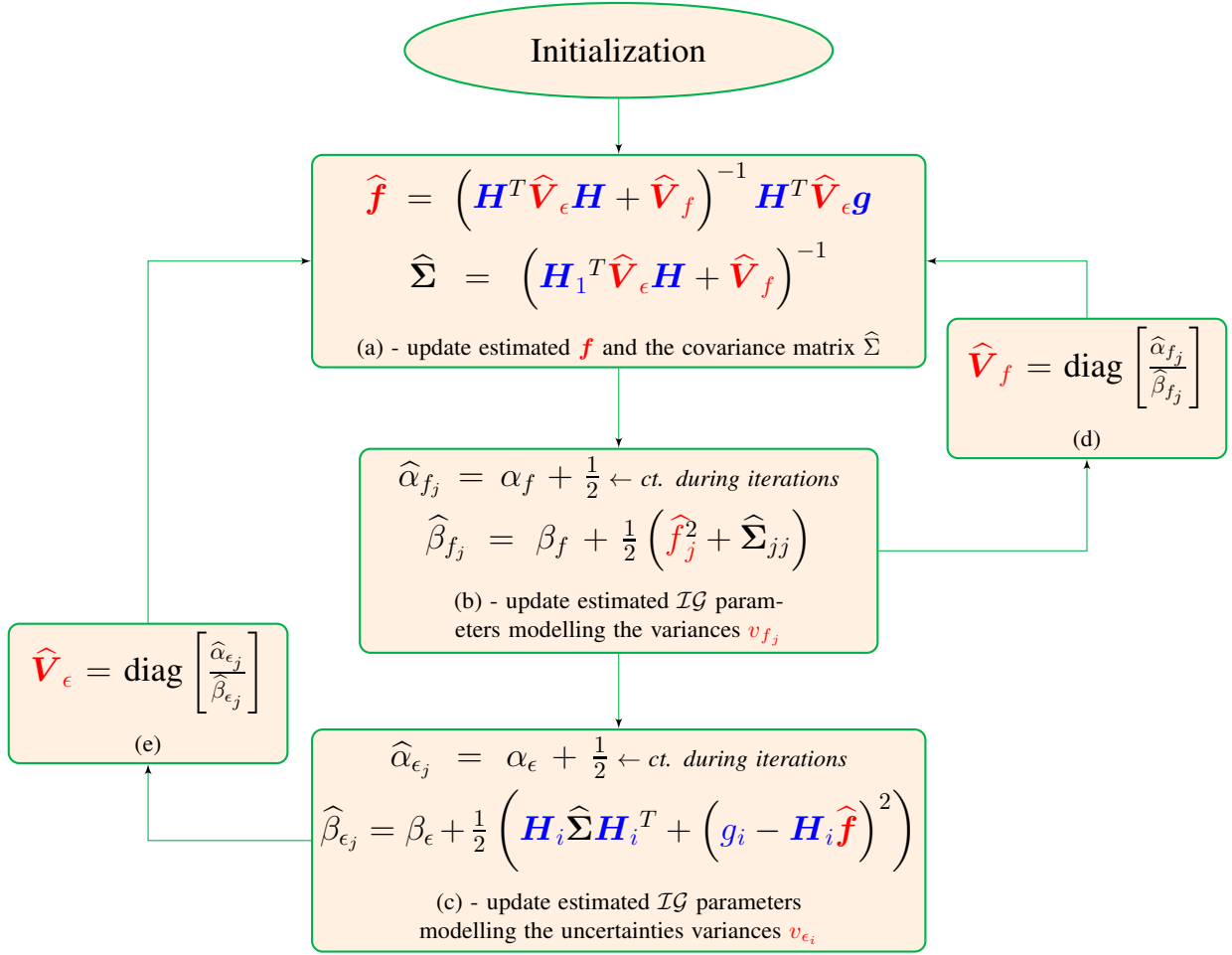
\begin{figure}[!htb]
\centering
\begin{center}
\begin{tikzpicture}[auto]
    \node [boxbig, scale=0.9] (f) {\Large{$\rfbh = 
\left( \bHb^T  \rVbh_\repsilon \bHb + \rVbh_{\rf} \right)^{-1}
\bHb^T \rVbh_\repsilon \bgb$}
\\[4pt]
\Large{$\Sigmabh = \left( \bHb_\blue{1}^T  \rVbh_\repsilon \bHb + \rVbh_{\rf} \right)^{-1}$}
\\[8pt]
\normalsize{(a) - update estimated $\rfb$ and the covariance matrix $\Sigmah$} 
};

    \node [box, below of=f, node distance=11em, scale=0.9] (vf)
{\Large{$ \alphah_{f_j} = \alpha_f + \frac{1}{2}$}
\normalsize{$\leftarrow$ \textit{ct. during iterations}}
\\[4pt]
\Large{$\betah_{f_j} = \beta_f + \frac{1}{2}
\left(\rfh_{\rj}^2 + \Sigmabh_{jj} \right)$}
\\[8pt]
\normalsize{(b) - update estimated $\Ic\Gc$ parameters modelling the variances $\rv_{\rf_\rj}$} 
};

    \node [boxbig, below of=vf, node distance=11em, scale=0.9] (veps) {\Large{$\alphah_{\epsilon_j} = \alpha_\epsilon + \frac{1}{2}$}
\normalsize{$\leftarrow$ \textit{ct. during iterations}}
\\[4pt]
\Large{$\betah_{\epsilon_j} = \beta_\epsilon + \frac{1}{2} \left( \bHb_\bi \Sigmabh \bHb_\bi^T + \left( \bg_\bi - \bHb_\bi \rfbh \right)^2 \right)$}
\\[8pt]
\normalsize{(c) - update estimated $\Ic\Gc$ parameters modelling the uncertainties variances $\rv_{\repsilon_\ri}$}     
};

    \node [boxsmall, above right = -0.15cm and 0.9cm of vf, node distance=17em, scale=0.9] (Vf) {\Large{$\rVbh_\rf = \diag {\frac{\alphah_{f_j}}{\betah_{f_j}}}$}
\\[8pt]
\normalsize{(d)} 
};

    \node [boxsmall, below left = -4.4cm and 0.3cm of veps, node distance=17em, scale=0.9] (Veps) {\Large{$\rVbh_\repsilon = \diag {\frac{\alphah_{\epsilon_j}}{\betah_{\epsilon_j}}}$}
\\[8pt]
\normalsize{(e)} 
};

\node [cloud, above of=f, node distance=8em, scale=0.9] (init)
{\Large{Initialization}};  

    \path [line] (f) -- (vf)[near start];
    \path [line] (Vf) |- (f);
    \path [line] (vf) -| (Vf);
    \path [line] (vf) -- (veps);
    \path [line] (Veps) |- (f);
    \path [line] (veps) -| (Veps);
    \path [line] (init) -- (f);
    
\end{tikzpicture}
\end{center}
\caption{Iterative algorithm corresponding to PM estimation via VBA - partial separability for Student-t hierarchical model, non-stationary Student-t uncertainties model}
\label{Fig:IA_PM_St_nsStL_un}
\end{figure}

\begin{algorithm}
\caption{PM via VBA partial sep. - Student-t hierarchical model, non-stationary Student-t uncertainties model}
\begin{algorithmic}[1]
\Ensure INITIALIZATION $\rfbh^{(0)}$ \Comment{Where $\rfbh^{(0)} = \rfbh$ via a basic estimation method or $\rfbh^{(0)} = \zerob$}
\Function{JMAP}{$\alpha_{\epsilon},\beta_{\epsilon},\alpha_{f},\beta_{f},\bgb,\bHb,\rfb^{(0)},M,N,NoIter$} \Comment{$\alpha_{\epsilon}, \beta_{\epsilon}, \alpha_{f}, \beta_{f}$ are set in modeling phase, Eq.~\eqref{Eq:St_nsStL_un}}

\For{$n = 0$ to ${IterNumb}$}

\For{$j = 1$ to ${M}$}

\State $\rvh_{\rf_\rj}^{(n)} = \frac{\beta_f + \frac{1}{2} (\rfh_\rj^{(n)})^2}{\alpha_f + \frac{3}{2}}$ \Comment{$\rvh_{\rf_\rj}^{(n)}$ is computed using $\alpha_{f},\beta_{f},\rfb^{(n)}$}

\EndFor

\State $\rVbh_\rf^{(n)} = \diag {\rvbh_{\rf}^{(n)}}$ \Comment{The diagonal matrix $\rVbh_\rf^{(n)}$ is build using $\rvh_{\rf_\rj}^{(n)}$, $j \in \left\lbrace 1,2,\ldots, M\right\rbrace$}

\For{$i = 1$ to ${N}$}

\State $\rvh_{\repsilon_\ri}^{(n)} = \frac{\beta_{\epsilon} + \frac{1}{2} \left( \bg_\bi - \bHb_\bi \rfbh^{(n)} \right)^2}{\alpha_{\epsilon} + \frac{3}{2}}$ \Comment{$\rvh_{\repsilon_\ri}^{(n)}$ is computed using $\alpha_{\epsilon},\beta_{\epsilon}, \bg_\bi, \bHb_\bi, \rfb^{(n)}$} 

\EndFor

\State $\rVbh_\repsilon^{(n)} = \diag {\rvbh_{\repsilon}^{(n)}}$ \Comment{The diagonal matrix $\rVbh_\repsilon^{(n)}$ is build using $\rvh_{\repsilon_\ri}^{(n)}$, $i \in \left\lbrace 1,2,\ldots, N \right\rbrace$}

\State $\rfbh^{(n+1)}  = \left( \bHb^T \left(\rVbh_\repsilon^{(n)}\right)^{-1} \bHb + \left( \rVbh_\rf^{(n)}\right)^{-1} \right)^{-1} \bHb^T \left(\rVbh_\repsilon^{(n)}\right)^{-1} \bgb$ \Comment{New value $\rfbh^{(n+1)}$}

\EndFor

\Return $\left( \rfbh^{(n+1)}, \rvbh_\rf^{(n)}, \rvbh_{\repsilon}^{(n)} \right)$ , $n = NoIter$
\EndFunction
\end{algorithmic}
\end{algorithm}
\subsubsection{Posterior Mean estimation via VBA, full separability}
\label{Subsubsec:PM_FS_St_nsStL_un}
In this subsection, the Posterior Mean (PM) estimation is again considered, but via a full separable approximation. The posterior distribution is approximated by a full separable distribution $q\left( \rfb, \rvb_{\rf},\rvb_{\repsilon} \right)$, i.e. a supplementary condition is added in Equation~\eqref{Eq:Posterior_Approximation_Notations1}:
\beq
q_{1}(\rfb) = \prod_{j=1}^{M} q_{1j}(\rf_{\rj}),
\;\;;\;\;
q_{2}(\rvb_{\rf}) = \prod_{j=1}^{M} q_{2j}(\rv_{\rf_\rj}),
\;\;;\;\;
q_{3}(\rvb_{\repsilon}) = \prod_{i=1}^{N} q_{3i}(\rv_{\repsilon_\ri})
\label{Eq:Posterior_Approximation_Notations1bis}
\eeq
As in Subsection~\eqref{Subsubsec:PM_PS_St_nsStL_un}, the approximation is done by minimizing of the Kullback-Leibler divergence, Equation~\eqref{Eq:Kullback-Leibler}, via alternate optimization resulting the following proportionalities from Equations~\eqref{Eq:VBA_Proportionalities1bis},~\eqref{Eq:VBA_Proportionalities2bis} and ~\eqref{Eq:VBA_Proportionalities3bis},  
\begin{subequations}
\begin{align}
&q_1(\rf_\red{j}) \;\;\; \propto \; \exp \biggl\lbrace \left\langle \ln p(\rfb, \rvb_\rf, \rvb_\repsilon | \bgb, \alpha_{f}, \beta_{f}, \alpha_{\epsilon}, \beta_{\epsilon}) \biggr\rangle_{ q_{1-j}(\rf_\rj) q_2(\rvb_{\rf}) \; q_3(\rvb_{\repsilon})} \right\rbrace,\;\;j \in \left\lbrace 1,2 \ldots, M \right\rbrace,
\label{Eq:VBA_Proportionalities1bis}
\\[4pt]
&q_{2j}(\rv_{\rf_\red{j}}) \propto \; \exp \biggl\lbrace \left\langle \ln p(\rfb, \rvb_\rf, \rvb_\repsilon | \bgb, \alpha_{f}, \beta_{f}, \alpha_{\epsilon}, \beta_{\epsilon}) \biggr\rangle_{q_1(\rfb) \; q_{2-j}(\rv_{\rf_\red{j}}) \; q_3(\rvb_{\repsilon})} \right\rbrace,\;\;j \in \left\lbrace 1,2 \ldots, M \right\rbrace,
\label{Eq:VBA_Proportionalities2bis}
\\[4pt]
&q_{3i}(\rv_{\repsilon_\ri}) \; \propto \; \exp \left\lbrace \biggl\langle \ln p(\rfb, \rvb_\rf, \rvb_\repsilon | \bgb, \alpha_{f}, \beta_{f}, \alpha_{\epsilon}, \beta_{\epsilon}) \biggr\rangle_{q_1(\rfb) \; q_2(\rvb_{\rf}) \; q_{3-i}(\rv_{\repsilon_\ri})} \right\rbrace,\;\;i \in \left\lbrace 1,2 \ldots, N \right\rbrace,
\label{Eq:VBA_Proportionalities3bis}
\end{align}
\end{subequations}
using the notations introduced in Equation~\eqref{Eq:Def_q_Minus}, Equation~\eqref{Eq:Def_Integral} and Equation~\eqref{Eq:Def_q_Minusbis}.
\beq
q_{1-j}(\rf_{\rj})=\displaystyle \prod_{k=1,k \neq j}^{M} q_{1k}(\rf_\rk)
\label{Eq:Def_q_Minusbis}
\eeq
The analytical expression of logarithm of the posterior distribution $\ln p(\rfb, \rvb_\rf, \rvb_\repsilon | \bgb, \alpha_{f}, \beta_{f}, \alpha_{\epsilon}, \beta_{\epsilon})$ is obtained in Equation~\eqref{Eq:Log_Posterior}.
\paragraph{Computation of the analytical expression of $q_{1j}(\rf_\rj)$.}
The proportionality relation corresponding to $q_1(\rfb)$ is presented in established in Equation~\eqref{Eq:VBA_Proportionalities1bis}. In the expression of $\ln p\left(\rfb, \rvb_\rf, \rvb_\repsilon | \bgb, \alpha_{f}, \beta_{f}, \alpha_{\epsilon}, \beta_{\epsilon} \right)$ all the terms free of $\rf_\ri$ can be regarded as constants:
\beq
\begin{split}
\biggl\langle \ln p(\rfb, \rvb_\rf, \rvb_\repsilon | \bgb, \alpha_{f}, \beta_{f}, \alpha_{\epsilon}, \beta_{\epsilon}) \biggr\rangle_{q_{1-j}(\rf_\rj) \; q_2(\rvb_{\rf}) \; q_3(\rvb_{\repsilon})}
=
&
-\frac{1}{2}
\biggl\langle 
\|\rVb_{\repsilon}^{-\frac{1}{2}} \left(\bgb - \bHb\rfb\right) \|
\biggr\rangle_{q_{1-j}(\rf_\rj) \; q_3(\rvb_{\repsilon}) }
\\
&
-\frac{1}{2}
\biggl\langle 
\| \rVb_{\rf}^{-\frac{1}{2}} \rfb \|
\biggr\rangle_{q_{1-j}(\rf_\rj) \; q_2(\rvb_{\rf}) }.
\end{split}
\label{Eq:q1_Integral_1bis}
\eeq
Using Equation~\eqref{Eq:q1_Integral_Not1}
the integral from Equation~\eqref{Eq:q1_Integral_1bis} becomes:
\beq
\begin{split}
\biggl\langle \ln p(\rfb, \rvb_\rf, \rvb_\repsilon | \bgb, \alpha_{f}, \beta_{f}, \alpha_{\epsilon}, \beta_{\epsilon}) \biggr\rangle_{q_{1-j}(\rf_\rj) \; q_2(\rvb_{\rf}) \; q_3(\rvb_{\repsilon})}
&
= 
-
\frac{1}{2} 
\biggl\langle 
\|\rVbt_{\repsilon}^{-\frac{1}{2}} \left(\bgb - \bHb\rfb\right) \| 
\biggr\rangle_{q_{1-j}(\rf_\rj)} 
-
\frac{1}{2} 
\biggl\langle 
\| \rVbt_{\rf}^{-\frac{1}{2}} \rfb \| 
\biggr\rangle_{q_{1-j}(\rf_\rj)}.
\label{Eq:q1_Integral_1bis1}
\end{split}
\eeq
Considering all the $\rf_\rj$ free terms as constants, the first norm can be written:
\beq
\|\rVbt_{\repsilon}^{-\frac{1}{2}} \left(\bgb - \bHb\rfb\right) \|=
C +
\| \rVbt_{\repsilon}^{-\frac{1}{2}} \bHb^{\bj}\|^2\rf_\rj^2
-2
\bHb^{\bj T} \rVbt_{\repsilon}^{-\frac{1}{2}} \left(\bgb - \bHb^{-\bj}\rfb^{-\rj} \right)\rf_\rj
\label{Eq:q1_Integral_First_Normbis}
\eeq
where $\bHb^{\bj}$ represents the column $j$ of the matrix $\bHb$, $\bHb^{-\bj}$ represents the matrix $\bHb$ except the column $j$, $\bHb^{\bj}$ and $\rfb^{-\rj}$ represents the vector $\rfb$ except the element $\rf_\rj$. 
Introducing the notation
\beq
\rft_\rj 
= \int \rf_\rj q_{1j}(\rf_{\rj})\d \rf_{\rj}
\;\;;\;\;
\rfbt^{-\rj}=
\begin{bmatrix}
\rft_\red{1} \; \ldots \; \rft_\red{j-1} \; \rft_\red{j+1} \; \ldots \; \rft_{\rM}
\end{bmatrix}^T
\label{Eq:q1_Integral_Not1bis}
\eeq
the expectancy of the first norm becomes:
\beq
\biggl\langle
\|\rVbt_{\repsilon}^{-\frac{1}{2}} \left(\bgb - \bHb\rfb\right) \|
\biggr\rangle_{q_{1-j}(\rf_\rj)}
=
C +
\| \rVbt_{\repsilon}^{-\frac{1}{2}} \bHb^{\bj}\|^2\rf_\rj^2
-2
\bHb^{\bj T} \rVbt_{\repsilon}^{-\frac{1}{2}} \left(\bgb - \bHb^{-\bj}\rfbt^{-\rj} \right)\rf_\rj
\label{Eq:q1_Integral_First_Norm_Exbis}
\eeq
Considering all the free $\rf_\rj$ terms as constants, the expectancy for the second norm becomes:  
\beq
\biggl\langle
\| \rVbt_{\rf}^{-\frac{1}{2}} \rfb \|^2
\biggr\rangle_{q_{1-j}(\rf_\rj)}
 = C + \rvt_{\rf_\rj}^{-1} \rf_\rj^2
\label{Eq:q1_Integral_Second_Norm_Exbis}
\eeq
From Equation~\eqref{Eq:VBA_Proportionalities1bis}, ~\eqref{Eq:q1_Integral_1bis1}, ~\eqref{Eq:q1_Integral_First_Norm_Exbis}, and ~\eqref{Eq:q1_Integral_Second_Norm_Exbis} the proportionality for $q_{1j}(\rf_\rj)$ becomes:
\beq
q_{1j}(\rf_\rj) \propto 
\exp \left\lbrace 
\left( 
\| \rVbt_{\repsilon}^{-\frac{1}{2}} \bHb^{\bj}\|^2 + \rvt_{\rf_\rj}^{-1} 
\right)\rf_\rj^2
-2
\bHb^{\bj T} \rVbt_{\repsilon}^{-1} \left(\bgb - \bHb^{-\bj}\rfb^{-\rj} \right)\rf_\rj
\right\rbrace
\label{Eq:q1_Proportionality}
\eeq
Considering the criterion $J(\rf_\rj) = \left( \| \rVbt_{\repsilon}^{-\frac{1}{2}} \bHb^{\bj}\|^2 + \rvt_{\rf_\rj}^{-1} \right)\rf_\rj^2 -2 \bHb^{\bj T} \rVbt_{\repsilon}^{-1} \left(\bgb - \bHb^{-\bj}\rfb^{-\rj} \right)\rf_\rj$ which is quadratic, we conclude $q_{1j}(\rf_\rj)$ is a Normal distribution.
For computing the mean of the Normal distribution, it is sufficient to compute the solution that minimizes the criterion $J(\rf_\rj)$:
\beq
\frac{\partial J(\rf_\rj)}{\partial \rf_\rj}=0 
\Leftrightarrow
\rfh_\rj = \frac{\bHb^{\bj T} \rVbt_{\repsilon}^{-1} \left( \bgb - \bHb^{-\bj} \rfb^{-\rj} \right)}{\| \rVbt_{\repsilon}^{-\frac{1}{2}}  \bHb^{\bj}\| + \rvt_{\rf}^{-1}}.
\label{Eq:q1_Criterion_Minimization}
\eeq
The variance can be obtained by identification. The analytical expressions for the mean and the variance corresponding to the Normal distributions, $q_1(\rf_\rj)$ are presented in Equation~\eqref{Eq:q1_Analytical_1bis}.
\beq
q_1(\rf_\rj)=\Nc\left(\rf_\rj | \rfh_\rj, \widehat{\mbox{var}}_j \right),
\left\{\barr{ll}
\rfh_\rj = \frac{\bHb^{\bj T} \rVbt_{\repsilon}^{-1} \left(\bgb - \bHb^{-\bj} \rfb^{-\rj} \right)}{\| \rVbt_{\repsilon}^{-\frac{1}{2}} \bHb^{\bj}\|^2 + \rvt_{\rf_\rj}^{-1}}
\\[12pt]
\widehat{\mbox{var}}_j=\frac{1}{\| \rVbt_{\repsilon}^{-\frac{1}{2}} \bHb^{\bj}\|^2 + \rvt_{\rf_\rj}^{-1}}
\earr\right.
,j \in \left\lbrace 1, 2, \ldots, M \right\rbrace
\label{Eq:q1_Analytical_1bis}
\eeq
\paragraph{Computation of the analytical expression of $q_{2j}(\rv_{\rf_\rj})$.}
The proportionality relation corresponding to $q_{2j}(\rv_{\rf_\rj})$ established in Equation~\eqref{Eq:VBA_Proportionalities2bis} refers to $\rv_{\rf_\rj}$, so in the expression of $\ln p\left(\rfb, \rvb_\rf, \rvb_\repsilon | \bgb, \alpha_{f}, \beta_{f}, \alpha_{\epsilon}, \beta_{\epsilon} \right)$ all the terms free of $\rv_{\rf_\rj}$ can be regarded as constants,
\beq
\ln p\left(\rfb, \rvb_\rf, \rvb_\repsilon | \bgb, \alpha_{f}, \beta_{f}, \alpha_{\epsilon}, \beta_{\epsilon} \right) =
C 
-\frac{1}{2}\ln \rv_{\rf_\rj}
-\frac{1}{2} \left\langle \rf_\rj^2\right\rangle_{q_{1j}(\rf_\rj)} \rv_{\rf_\rj}^{-1}
-\left( \alpha_f + 1 \right) \ln \rv_{\rf_\rj}
-\beta_f \rv_{\rf_\rj}^{-1},
\label{Eq:q2_Integral_1bis}
\eeq
so the integral of the logarithm becomes:
\beq
\left\langle \ln p\left(\rfb, \rvb_\rf, \rvb_\repsilon | \bgb, \alpha_{f}, \beta_{f}, \alpha_{\epsilon}, \beta_{\epsilon} \right) \right\rangle_{q_1(\rfb) \; q_{2-j}(\rv_{\rf_\rj}) \; q_3(\rvb_{\repsilon})}
= 
C
- \left(\alpha_f + \frac{3}{2} \right) \ln \rv_{\rf_\rj} 
-\left[ 
\beta_f
+\frac{1}{2} 
\left(
\rfh_\rj^2 + \widehat{\mbox{var}}_j
\right)
\right]
\rv_{\rf_\rj}^{-1}.
\label{Eq:q2_Integral_1bis1}
\eeq
Equation~\eqref{Eq:q2_Integral_1bis1} leads to the conclusion that $q_{2j}(\rv_{\rf_\rj})$ is an Inverse Gamma distribution. Equation~\eqref{Eq:q2_Analytical_1bis} presents the analytical expressions for to the shape and scale parameters corresponding to the Inverse Gamma distribution.
\beq
q_{2j}(\rv_{\rf_\rj})=\Ic\Gc\left(\rv_{\rf_\rj}|\alphah_{f_j},\betah_{f_j}\right),
\left\{\barr{ll}
\alphah_{f_j} = \alpha_f + \frac{1}{2}
\\[10pt]
\betah_{f_j} = \beta_f + \frac{1}{2} 
\left(
\rfh_{\rj}^2 + \widehat{\mbox{var}}_j
\right)
\earr\right.
,j \in \left\lbrace 1, 2, \ldots, M \right\rbrace
\label{Eq:q2_Analytical_1bis}
\eeq
\paragraph{Computation of the analytical expression of $q_{3i}(\rv_{\repsilon_\ri})$.} 
The proportionality relation corresponding to $q_{3i}(\rv_{\repsilon_\ri})$ established in Equation~\eqref{Eq:VBA_Proportionalities3bis} refers to $\rv_{\repsilon_\ri}$ so in the expression of $\ln p\left(\rfb, \rvb_\rf, \rvb_\repsilon | \bgb, \alpha_{f}, \beta_{f}, \alpha_{\epsilon}, \beta_{\epsilon} \right)$ all the terms free of $\rv_{\repsilon_\ri}$ can be regarded as constants:
\beq
\ln p\left(\rfb, \rvb_\rf, \rvb_\repsilon | \bgb, \alpha_{f}, \beta_{f}, \alpha_{\epsilon}, \beta_{\epsilon} \right) = C 
-\left( \alpha_\epsilon + \frac{3}{2} \right)\ln \rv_{\repsilon_\ri}
-\left(
\beta_\epsilon + \frac{1}{2}\left(\bg_\bi - \bHb_\bi\rfb\right)^2
\right)
\rv_{\repsilon_\ri}^{-1}.
\label{Eq:q3_Integral_1bis}
\eeq
Introducing the notation
\beq
\biggl\langle \rfb \biggr\rangle_{q_1(\rfb)}
=
\begin{bmatrix}
\rfh_{\red{1}}
\ldots
\rfh_\rj
\ldots 
\rfh_\rM
\end{bmatrix}^T
\stackrel{Not}{=}
\rfbh
\;\; ; \;\;
\Sigmabh = \diag {\widehat{\mbox{var}}_j}
\label{Eq:q3_Integral_Not1bis}
\eeq
the expectancy of the logarithm becomes
\beq
\begin{split}
\left\langle
\ln p\left(\rfb, \rvb_\rf, \rvb_\repsilon | \bgb, \alpha_{f}, \beta_{f}, \alpha_{\epsilon}, \beta_{\epsilon} \right) 
\right\rangle_{q_1(\rfb) \; q_2(\rvb_\rf) \; q_{3-i}(\rv_{\repsilon_\ri})}
= C & 
-\left( \alpha_\epsilon + 1 + \frac{1}{2} \right)\ln \rv_{\repsilon_\ri}
\\
&-\left(
\beta_\epsilon + \frac{1}{2}\left[ \bHb_\bi \Sigmabh \bHb_\bi^T
+
\left( \bg_\bi - \bHb_\bi \rfbh \right)^2\right]
\right)
\rv_{\repsilon_\ri}^{-1},
\end{split}
\label{Eq:q3_Integral_1bis1}
\eeq
so and the proportionality relation for $q_{3i}(\rv_{\repsilon_\ri})$ from Equation~\eqref{Eq:VBA_Proportionalities3bis} can be written:
\beq
q_{3i}(\rv_{\repsilon_\ri})
\propto
\rv_{\repsilon_\ri}^{-\left( \alpha_\epsilon + \frac{3}{2} \right)}
\exp
\left\lbrace
-\left(
\beta_\epsilon + \frac{1}{2}\left[ \bHb_\bi \Sigmabh \bHb_\bi^T
+
\left( \bg_\bi - \bHb_\bi \rfbh \right)^2\right]
\right)\rv_{\repsilon_\ri}^{-1}
\right\rbrace
\label{Eq:q3_Proportionality}
\eeq
Equation~\eqref{Eq:q3_Proportionality} shows that $q_{3i}(\rv_{\repsilon_\ri})$ are Inverse Gamma distributions. The analytical expressions of the corresponding parameters are presented in Equation~\eqref{Eq:q3_Analytical_1bis}.
\beq
q_{3i}(\rv_{\repsilon_\ri})=\Ic\Gc\left(\rv_{\repsilon_\ri}|\alphah_{\epsilon_i},\betah_{\epsilon_i}\right),
\left\{\barr{ll}
\alphah_{\epsilon_i} = \alpha_\epsilon + \frac{1}{2}
\\[10pt]
\betah_{\epsilon_i} = \beta_\epsilon + \frac{1}{2}\left[ \bHb_\bi \Sigmabh \bHb_\bi^T
+
\left( \bg_\bi - \bHb_\bi \rfbh \right)^2\right]
\earr\right.
,i \in \left\lbrace 1, 2, \ldots, N \right\rbrace
\label{Eq:q3_Analytical_1bis}
\eeq
Since $q_2(\rv_{\rf_\rj}), j \in \left\lbrace 1, 2, \ldots, M \right\rbrace$ and $q_{3i}(\rv_{\repsilon_\ri}), i \in \left\lbrace 1, 2, \ldots, N \right\rbrace $ are Inverse Gamma distributions, it is easy to obtain analytical expressions for $\rVbt_{\repsilon}^{-1}$, defined in Equation~\eqref{Eq:q1_Integral_Not1} and $ \rvt_{\rf_\rj}^{-1}, j \in \left\lbrace 1, 2, \ldots, M \right\rbrace $, obtaining the same expressions as in Equation~\eqref{Eq:V_Expectancies}. Using Equation~\eqref{Eq:V_Expectancies} and Equations~\eqref{Eq:q1_Analytical_1bis},~\eqref{Eq:q2_Analytical_1bis} and~\eqref{Eq:q3_Analytical_1bis}, the final analytical expressions of the separable distributions $q_i$ are presented in Equations~\eqref{Eq:q1_Analytical_2bis},~\eqref{Eq:q2_Analytical_2bis} and~\eqref{Eq:q3_Analytical_2bis}.
\begin{subequations}
\begin{align}
&q_1(\rf_\rj)=\Nc\left(\rf_\rj | \rfh_\rj, \widehat{\mbox{var}}_j \right),\;\;\;\;\;\;
\left\{\barr{ll}
\rfh_\rj = \frac{\bHb^{\bj T} \rVbt_{\repsilon}^{-1} \left(\bgb - \bHb^{-\bj} \rfb^{-\rj} \right)}{\| \rVbt_{\repsilon}^{-\frac{1}{2}} \bHb^{\bj}\|^2 + \rvt_{\rf_\rj}^{-1}}
\\[12pt]
\widehat{\mbox{var}}_j=\frac{1}{\| \rVbt_{\repsilon}^{-\frac{1}{2}} \bHb^{\bj}\|^2 + \rvt_{\rf_\rj}^{-1}}
\earr\right.
,j \in \left\lbrace 1, 2, \ldots, M \right\rbrace
\label{Eq:q1_Analytical_2bis}
\\[4pt]
&q_{2j}(\rv_{\rf_\rj})=\Ic\Gc\left(\rv_{\rf_\rj}|\alphah_{f_j},\betah_{f_j}\right),
\left\{\barr{ll}
\alphah_{f_j} = \alpha_f + \frac{1}{2}
\\[10pt]
\betah_{f_j} = \beta_f + \frac{1}{2} 
\left(
\rfh_{\rj}^2 + \widehat{\mbox{var}}_j
\right)
\earr\right.
,j \in \left\lbrace 1, 2, \ldots, M \right\rbrace
\label{Eq:q2_Analytical_2bis}
\\[4pt]
&q_{3i}(\rv_{\repsilon_\ri})=\Ic\Gc\left(\rv_{\repsilon_\ri}|\alphah_{\epsilon_i},\betah_{\epsilon_i}\right),\;\;\;
\left\{\barr{ll}
\alphah_{\epsilon_i} = \alpha_\epsilon + \frac{1}{2}
\\[10pt]
\betah_{\epsilon_i} = \beta_\epsilon + \frac{1}{2}\left[ \bHb_\bi \Sigmabh \bHb_\bi^T
+
\left( \bg_\bi - \bHb_\bi \rfbh \right)^2\right]
\earr\right.
,i \in \left\lbrace 1, 2, \ldots, N \right\rbrace
\label{Eq:q3_Analytical_2bis}
\end{align}
\end{subequations}
Equations~\eqref{Eq:q1_Analytical_2bis},~\eqref{Eq:q2_Analytical_2bis} and~\eqref{Eq:q3_Analytical_2bis} establish dependencies between the parameters of the distributions, very similar to the one presented in Figures~\eqref{Fig:Dependency_Scheme_1},~\eqref{Fig:Dependency_Scheme_2} and~\eqref{Fig:Dependency_Scheme_3}. The iterative algorithm obtained via PM estimation with full separability, is presented Figure~\eqref{Fig:IA_PM_St_nsStL_unbis}.
\tikzstyle{cloud} = [draw=blue!30!green,fill=orange!12, ellipse, thick, node distance=6em, text width=12em, text centered, minimum height=4em, minimum width=12em]
\tikzstyle{boxbig} = [draw=blue!30!green, fill=orange!12, rectangle, rounded corners, thick, node distance=8em, text width=25em, text centered, minimum height=5em, minimum width=25em]
\tikzstyle{box} = [draw=blue!30!green, fill=orange!12, rectangle, rounded corners, thick, node distance=6.5em, text width=21em, text centered, minimum height=5em, minimum width=21em]
\tikzstyle{boxsmall} = [draw=blue!30!green, fill=orange!12, rectangle, rounded corners, thick, node distance=8.5em, text width=11em, text centered, minimum height=5em, minimum width=11em]
\tikzstyle{line} = [draw=blue!30!green, -latex']
\begin{figure}[!htb]
\centering
\begin{center}
\begin{tikzpicture}[auto]
    \node [boxbig, scale=0.9] (f) {\Large{$ \rfh_\rj = \frac{\bHb^{\bj T} \rVbt_{\repsilon}^{-1} \left(\bgb - \bHb^{-\bj} \rfb^{-\rj} \right)}{\| \rVbt_{\repsilon}^{-\frac{1}{2}} \bHb^{\bj}\|^2 + \rvt_{\rf_\rj}^{-1}}$}
\\[4pt]
\Large{$\widehat{\mbox{var}}_j=\frac{1}{\| \rVbt_{\repsilon}^{-\frac{1}{2}} \bHb^{\bj}\|^2 + \rvt_{\rf_\rj}^{-1}}$}
\\[8pt]
\normalsize{(a) - update estimated $\rf_\rj$ and the corespnding variance $\mbox{var}_j$} 
};

    \node [box, below of=f, node distance=11em, scale=0.9] (vf)
{\Large{$ \alphah_{f_j} = \alpha_f + \frac{1}{2} $}
\normalsize{$\leftarrow$ \textit{ct. during iterations}}
\\[4pt]
\Large{$\betah_{f_j} = \beta_f + \frac{1}{2} \left( \rfh_{\rj}^2 + \widehat{\mbox{var}}_j \right)$}
\\[8pt]
\normalsize{(b) - update estimated $\Ic\Gc$ parameters modelling the variances $\rv_{\rf_\rj}$} 
};

    \node [boxbig, below of=vf, node distance=11em, scale=0.9] (veps) {\Large{$ \alphah_{\epsilon_i} = \alpha_\epsilon + \frac{1}{2} $}
\normalsize{$\leftarrow$ \textit{ct. during iterations}}
\\[4pt]
\Large{$\betah_{\epsilon_i} = \beta_\epsilon + \frac{1}{2}\left[ \bHb_\bi \Sigmabh \bHb_\bi^T + \left( \bg_\bi - \bHb_\bi \rfbh \right)^2\right]$}
\\[8pt]
\normalsize{(c) - update estimated $\Ic\Gc$ parameters modelling the uncertainties variances $\rv_{\repsilon_\ri}$}     
};

    \node [boxsmall, above right = -0.15cm and 0.9cm of vf, node distance=17em, scale=0.9] (Vf) {\Large{$\rVbh_\rf = \diag {\frac{\alphah_{f_j}}{\betah_{f_j}}}$}
\\[8pt]
\normalsize{(d)} 
};

    \node [boxsmall, below left = -4.4cm and 0.3cm of veps, node distance=17em, scale=0.9] (Veps) {\Large{$\rVbh_\repsilon = \diag {\frac{\alphah_{\epsilon_j}}{\betah_{\epsilon_j}}}$}
\\[8pt]
\normalsize{(e)} 
};

	\node [cloud, above of=f, node distance=8em, scale=0.9] (init)
{\Large{Initialization}};  

    \path [line] (f) -- (vf)[near start];
    \path [line] (Vf) |- (f);
    \path [line] (vf) -| (Vf);
    \path [line] (vf) -- (veps);
    \path [line] (Veps) |- (f);
    \path [line] (veps) -| (Veps);
    \path [line] (init) -- (f);    
\end{tikzpicture}
\end{center}
\caption{Iterative algorithm corresponding to PM estimation via VBA - full separability for Student-t hierarchical model, non-stationary Student-t uncertainties model}
\label{Fig:IA_PM_St_nsStL_unbis}
\end{figure}
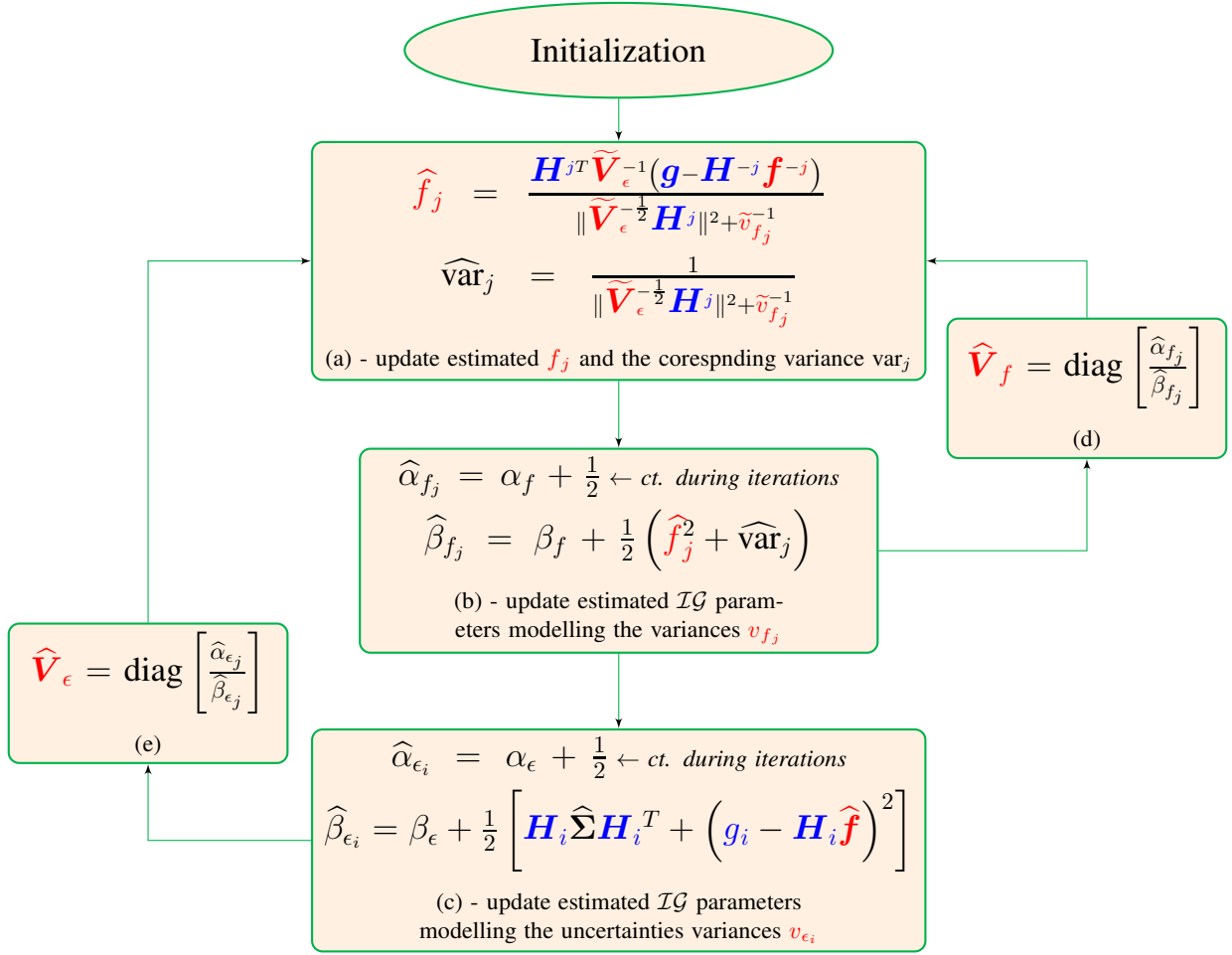

\subsection{IS: Student-t hierarchical model: non-stationary Student-t uncertainties model, unknown uncertainties variances}
\label{Subsec:St_nsStL_un_IS}
\begin{itemize}
\item the hierarchical model is using as a \textbf{prior} the \textbf{Student-t} distribution;
\item the Student-t prior distribution is expressed via \textbf{StPM}, Equation~\eqref{Eq:StPM1}, considering the variance $\rvb_\rf$ as unknown;
\item the \textbf{likelihood} is derived from the distribution proposed for modelling the uncertainties vector $\repsilonb$;
\item for the uncertainties vector $\repsilonb$ a \textbf{non-stationary Student-t uncertainties model} is proposed, i.e. a multivariate Student-t distribution expressed via \textbf{StPM} is used under the following assumption:\\
a) the variance vector $\rvb_{\repsilon}$ is \textbf{unknown}; 
\end{itemize}
\beq
\rotatebox{90}{\hspace{-1.55cm}Student-t nsStL UNK} \;
\vline
\left.\barr{ll}
Likelihood: \textbf{nsStL:}\;
\left\{\barr{ll}
\begin{split}
p\left( \bgb | \rfb, \rvb_{\repsilon} \right) 
=
\Nc\left( \bgb | \bHb\rfb, \rVb_{\repsilon} \right) 
\propto 
\prod_{i=1}^{N} & \rv_{\repsilon_\ri}^{-\frac{1}{2}}
\exp \left\lbrace - \frac{1}{2}\| \rVb_{\repsilon}^{-\frac{1}{2}} \left( \bgb - \bHb\rfb \right) \|^2 \right\rbrace, 
\end{split} 
\\[7pt]
p\left( \rvb_{\repsilon} | \alpha_{\epsilon}, \beta_{\epsilon} \right)
= \prod_{i=1}^{N} \Ic\Gc\left( \rv_{\repsilon_\ri} | \alpha_{\epsilon}, \beta_{\epsilon} \right) \propto  \prod_{i=1}^{N} \rv_{\repsilon_\ri}^{-\alpha_{\epsilon}-1} \; \exp \left\lbrace - \sum_{i=1}^{N}\frac{\beta_{\epsilon}}{\rv_{\repsilon_\ri}} \right\rbrace,
\\[7pt]
\hspace{5.55cm}
\rVb_{\repsilon} = \diag{\rvb_{\repsilon}},\; \rvb_{\repsilon} = \left[\ldots,\rv_{\repsilon_{\ri}}, \ldots\right],
\\[5pt]
\earr\right.
\\[37pt]
Likelihood: \textbf{nsStL:}\;
\left\{\barr{ll}
p(\rfb|\rzb,\rvb_{\rxi}) = \Nc(\rfb| \bDb\rzb, \; \rVb_{\rxi}) \propto \det{\rVb_{\rxi}}^{-\frac{1}{2}} \; \exp \left\lbrace - \frac{1}{2}\| \rVb_{\rxi}^{-\frac{1}{2}} \left( \rfb - \bDb\rzb \right) \|^2 \right\rbrace,
\\[6.7pt]
p(\rvb_{\rxi}|\alpha_{f}, \beta_{f}) = \prod_{j=1}^{M} \Ic\Gc(\rv_{\rxi_\rj}|\alpha_{\xi}, \beta_{\xi}) \propto  \prod_{j=1}^{M} \rv_{\rxi_\rj}^{-\alpha_{\xi}-1} \; \exp \left\lbrace - \sum_{j=1}^{M}\frac{\beta_{\xi}}{\rv_{\rxi_\rj}} \right\rbrace,
\\[7pt]
\hspace{5.65cm}
\rVb_{\rxi} = \diag{\rvb_{\rxi}},\; \rvb_{\rxi} = \left[\ldots,\rv_{\rxi_{\rj}}, \ldots\right],
\\[5pt]
\earr\right.
\\[37pt]
Prior:\;\;\;\;\;\;\;\textbf{StPM:}\;\;
\left\{\barr{ll}
p(\rzb|\zerob,\rvb_{\rz}) = \Nc(\rzb| \zerob, \; \rVb_{\rz}) \propto \det{\rVb_{\rz}}^{-\frac{1}{2}} \; \exp \left\lbrace - \frac{1}{2}\| \rVb_{\rz}^{-\frac{1}{2}} \rzb \|^2 \right\rbrace,
\\[6.7pt]
p(\rvb_{\rz}|\alpha_{z}, \beta_{z}) = \prod_{j=1}^{M} \Ic\Gc(\rv_{\rz_\rj}|\alpha_{z}, \beta_{z}) \propto  \prod_{j=1}^{M} \rv_{\rz_\rj}^{-\alpha_{z}-1} \; \exp \left\lbrace - \sum_{j=1}^{M}\frac{\beta_{z}}{\rv_{\rz_\rj}} \right\rbrace,\\[7pt]
\hspace{5.65cm}
\rVb_{\rz} = \diag{\rvb_{\rz}},\; \rvb_{\rz} = \left[\ldots,\rv_{\rz_{\rj}}, \ldots\right].\\[4pt]
\earr\right.
\earr\right.
\label{Eq:St_nsStL_un_IS}
\eeq
The hierarchical model showed in Figure~\eqref{Fig:FromForwardModelHierarchicalModel2}, which is build over the linear forward model, Equation~\eqref{Eq:LinearModel} and Equation~\eqref{Eq:SparsityViaTransformation}, using as a prior for $\rzb$ a Student-t distribution, expressed via the Student-t Prior Model (StPM), Equation~\eqref{Eq:StPM1} and modelling the uncertainties of the model $\repsilonb$ and $\rxib$ using the non-stationary Student-t Uncertainties Model (nsStUM), Equation~\eqref{Eq:nsStUM}, is presented in Equation~\eqref{Eq:St_nsStL_un_IS}. The posterior distribution is obtained via the Bayes rule, Equation~\eqref{Eq:Post_St_nsStL_un_IS}:
\beq
\begin{split}
p(\rfb, \rzb, \rvb_\repsilon, \rvb_\rxi, \rvb_\rz | \bgb, & \alpha_{\epsilon}, \beta_{\epsilon}, \alpha_{\xi}, \beta_{\xi}, \alpha_{z}, \beta_{z})
\propto  
p\left( \bgb | \rfb, \rvb_{\repsilon} \right) 
p\left( \rvb_{\repsilon} | \alpha_{\epsilon}  \beta_{\epsilon} \right) 
p(\rfb|\rzb,\rvb_{\rxi})
p(\rvb_{\rxi}|\alpha_{f}, \beta_{f})
p(\rzb|\zerob,\rvb_{\rz})
p(\rvb_{\rz}|\alpha_{z}, \beta_{z})
\\& 
\propto 
\;
\prod_{i=1}^{N} \rv_{\repsilon_\ri}^{-\frac{1}{2}}\exp \left\lbrace - \frac{1}{2}\| \rVb_{\repsilon}^{-\frac{1}{2}} \left( \bgb - \bHb\rfb \right) \|^2 \right\rbrace 
\;
\prod_{i=1}^{N} \rv_{\repsilon_\ri}^{-\alpha_{\epsilon}-1} \; \exp \left\lbrace - \sum_{i=1}^{N}\frac{\beta_{\epsilon}}{\rv_{\repsilon_\ri}} \right\rbrace
\\&
\;\;\;\;\;
\det{\rVb_{\rxi}}^{-\frac{1}{2}} \; \exp \left\lbrace - \frac{1}{2}\| \rVb_{\rxi}^{-\frac{1}{2}} \left( \rfb - \bDb\rzb \right) \|^2 \right\rbrace
\;
\prod_{j=1}^{M} \rv_{\rxi_\rj}^{-\alpha_{\xi}-1} \; \exp \left\lbrace - \sum_{j=1}^{M}\frac{\beta_{\xi}}{\rv_{\rxi_\rj}} \right\rbrace
\\&
\;\;\;\;\;
\det{\rVb_{\rz}}^{-\frac{1}{2}} \; \exp \left\lbrace - \frac{1}{2}\| \rVb_{\rz}^{-\frac{1}{2}} \rzb \|^2 \right\rbrace
\;
\prod_{j=1}^{M} \rv_{\rz_\rj}^{-\alpha_{z}-1} \; \exp \left\lbrace - \sum_{j=1}^{M}\frac{\beta_{z}}{\rv_{\rz_\rj}} \right\rbrace
\end{split}
\label{Eq:Post_St_nsStL_un_IS}
\eeq
The goal is to estimate the unknowns of the hierarchical model, namely $\rfb$, the main unknown of the linear forward model expressed in Equation~\eqref{Eq:LinearModel}, $\rzb$, the main unknown of Equation~\eqref{Eq:SparsityViaTransformation}, supposed sparse and consequently modelled using the Student-t distribution and the three variances appearing in the hierarchical model, Equation~\eqref{Eq:St_nsStL_un_IS}, the variance corresponding to the sparse structure $\rzb$, namely $\rvb_\rz$ and the two variances corresponding to the uncertainties of model $\repsilonb$, and $\rxib$, namely $\rvb_\repsilon$ and $\rvb_\rxi$. 
\subsubsection{Joint MAP estimation}
\label{Subsubsec:JMAP_St_nsStL_un_IS}
First, the Joint Maximum A Posterior (JMAP) estimation is considered: the unknowns are estimated on the basis of the available data $\bgb$, by maximizing the posterior distribution:
\beq
\begin{split}
\left( \rfbh, \; \rzbh, \; \rvbh_\rxi, \; \rvbh_\repsilon, \; \rvbh_\rz \right)
&=
\operatorname*{arg\,max}_{\left( \rfb, \; \rzb, \; \rvb_\rxi, \; \rvb_\repsilon, \; \rvb_\rz \right)} p(\rfb, \; \rzb, \; \rvb_\rxi, \; \rvb_\repsilon, \; \rvb_\rz | \bgb, \; \alpha_{\xi}, \; \beta_{\xi}, \; \alpha_{\epsilon}, \; \beta_{\epsilon}, \alpha_{z}, \; \beta_{z})
\\&
=
\operatorname*{arg\,min}_{\left( \rfb, \; \rzb, \; \rvb_\rxi, \; \rvb_\repsilon, \; \rvb_\rz \right)} \Lc \left( \rfb, \; \rzb, \; \rvb_\rxi, \; \rvb_\repsilon, \; \rvb_\rz \right),
\end{split}
\eeq 
where for the second equality the criterion $\Lc \left( \rfb, \; \rzb, \; \rvb_\rxi, \; \rvb_\repsilon \; \rvb_\rz \right)$ is defined as:
\beq
\Lc \left( \rfb, \; \rzb, \; \rvb_\rxi, \; \rvb_\repsilon, \; \rvb_\rz \right) = -\ln  p \left( \rfb, \; \rzb, \; \rvb_\rxi, \; \rvb_\repsilon, \; \rvb_\rz | \bgb, \; \alpha_{\xi}, \; \beta_{\xi}, \; \alpha_{\epsilon}, \; \beta_{\epsilon}, \alpha_{z}, \; \beta_{z} \right)
\label{Eq:Def_L_Criterion_IS}
\eeq
The MAP estimation corresponds to the solution minimizing the criterion $\Lc \left( \rfb, \; \rzb, \; \rvb_\rxi, \; \rvb_\repsilon, \; \rvb_\rz \right)$.
From the analytical expression of the posterior distribution, Equation~\eqref{Eq:Post_St_nsStL_un_IS} and the definition of the criterion $\Lc$,  Equation~\eqref{Eq:Def_L_Criterion_IS}, we obtain:
\beq
\begin{split}
\Lc \left( \rfb, \; \rzb, \; \rvb_\rxi, \; \rvb_\repsilon, \; \rvb_\rz \right)
&
=
-\ln  p \left( \rfb, \; \rzb, \; \rvb_\rxi, \; \rvb_\repsilon, \; \rvb_\rz | \bgb, \; \alpha_{\xi}, \; \beta_{\xi}, \; \alpha_{\epsilon}, \; \beta_{\epsilon}, \alpha_{z}, \; \beta_{z} \right)
\\&
+
\frac{1}{2}\| \rVb_{\repsilon}^{-\frac{1}{2}} \left( \bgb - \bHb\rfb \right) \|^2
+
\left( \alpha_{\epsilon} + \frac{3}{2} \right) \sum_{i=1}^{N} \ln \rv_{\repsilon_\ri} 
+
\sum_{i=1}^{N}\frac{\beta_{\epsilon}}{\rv_{\repsilon_\ri}} 
\\&
+
\frac{1}{2}\| \rVb_{\rxi}^{-\frac{1}{2}} \left( \rfb - \bDb\rzb \right) \|^2
+
\left( \alpha_{\xi} + \frac{3}{2} \right) \sum_{i=1}^{N} \ln \rv_{\rxi_\ri} 
+
\sum_{i=1}^{N}\frac{\beta_{\xi}}{\rv_{\rxi_\ri}}
\\&
+
\frac{1}{2}\| \rVb_{\rz}^{-\frac{1}{2}} \rzb \|^2
+
\left(\alpha_{z} + \frac{3}{2} \right) \sum_{j=1}^{M} \ln \rv_{\rz_\rj}
+
\sum_{j=1}^{M}\frac{\beta_{f}}{\rv_{\rz_\rj}}
\end{split}
\label{Eq:L_Criterion_IS}
\eeq
One of the simplest optimisation algorithm that can be used is an alternate optimization of the criterion $\Lc \left( \rfb, \; \rzb, \; \rvb_\rxi, \; \rvb_\repsilon, \; \rvb_\rz \right)
$ with respect to the each unknown:
\begin{itemize}
\item With respect to $\rfb$:
\beq
\begin{split}
\frac{\partial \Lc \left( \rfb, \; \rzb, \; \rvb_\rxi, \; \rvb_\repsilon, \; \rvb_\rz \right)}{\partial \rfb}=0
&\Leftrightarrow
\frac{\partial}{\partial \rfb}
\left(
\| \rVb_\repsilon^{-\frac{1}{2}} \left( \bgb - \bHb\rfb \right) \|^2
+
\| \rVb_{\rxi}^{-\frac{1}{2}} \left( \rfb - \bDb\rzb \right) \|^2
\right) = 0
\\
&\Leftrightarrow
-\bHb^T \rVb_\repsilon^{-1} \left( \bgb - \bHb \rfb \right) 
+ 
\rVb_\rxi^{-1} \left( \rfb - \bDb\rzb \right) 
= 0
\\
&
\Leftrightarrow 
\left( \bHb^T \rVb_\repsilon^{-1} \bHb + \rVb_\rxi^{-1} \right) \rfb  = 
\bHb^T \rVb_\repsilon^{-1} \bgb + \rVb_\rxi^{-1} \bDb\rzb
\\
&
\Rightarrow
\rfbh 
=
\left( \bHb^T \rVb_\repsilon^{-1} \bHb + \rVb_\rxi^{-1} \right)^{-1} 
\left( \bHb^T \rVb_\repsilon^{-1} \bgb + \rVb_\rxi^{-1} \bDb\rzb \right)
\end{split}
\nonumber
\eeq
\item With respect to $\rzb$:
\beq
\begin{split}
\frac{\partial \Lc \left( \rfb, \; \rzb, \; \rvb_\rxi, \; \rvb_\repsilon, \; \rvb_\rz \right)}{\partial \rzb}=0
&\Leftrightarrow
\frac{\partial}{\partial \rfb}
\left(
\| \rVb_{\rxi}^{-\frac{1}{2}} \left( \rfb - \bDb\rzb \right) \|^2
+
\| \rVb_{\rz}^{-\frac{1}{2}} \rzb \|^2
\right)= 0
\\
&\Leftrightarrow
-\bDb^T \rVb_\rxi^{-1} \left( \rfb - \bDb\rzb \right) 
+ 
\rVb_\rz^{-1} \rzb  
= 0
\\
&
\Leftrightarrow 
\left( \bDb^T \rVb_\rxi^{-1} \bDb + \rVb_\rz^{-1} \right) \rzb  = 
\bDb^T \rVb_\rxi^{-1} \rfb
\\
&
\Rightarrow
\rzbh 
=
\left( \bDb^T \rVb_\rxi^{-1} \bDb + \rVb_\rz^{-1} \right)^{-1} \bDb^T \rVb_\rxi^{-1} \rfb
\end{split}
\nonumber
\eeq
\item With respect to $\rv_{\repsilon_\ri}$, $i \in \left\lbrace 1,2,\ldots, N \right\rbrace$:\\
First, we develop the norm $\| \rVb_\repsilon^{-\frac{1}{2}} \left( \bgb - \bHb\rfb \right) \|^2$:
\beq
\begin{split}
\| \rVb_\repsilon^{-\frac{1}{2}} \left( \bgb - \bHb\rfb \right) \|^2
&= 
\bgb^{T} \rVb_\repsilon^{-1} \bgb
-
2 \bgb^{T} \rVb_\repsilon^{-1} \bHb \rfb
+
\bHb^{T} \rfb^{T} \rVb_\repsilon^{-1} \bHb \rfb
\\
&=
\sum_{i=1}^{N} \rv_{\repsilon_\ri}^{-1} \bg_\bi^{2}
-
2 \sum_{i=1}^{N} \rv_{\repsilon_\ri}^{-1} \bg_\bi \bH_\bi \rfb 
+
\sum_{i=1}^{N} \rv_{\repsilon_\ri}^{-1} \rfb^{T} \bH_\bi^{T} \bH_\bi \rfb,
\nonumber
\end{split}
\eeq
where $\bHb_\bi$ denotes the line i of the matrix $\bHb$, $i\in \left\lbrace 1,2,\ldots,N \right\rbrace$, i.e. $\bHb_\bi = \left[ \bh_{\bi\blue{1}}, \bh_{\bi\blue{1}}, \ldots, \bh_{\bi\bM} \right]$. 
\beq
\begin{split}
\frac{\partial \Lc \left( \rfb, \; \rzb, \; \rvb_\rxi, \; \rvb_\repsilon, \; \rvb_\rz \right)}{\partial \rv_{\repsilon_\ri}}=0
&
\Leftrightarrow
\frac{\partial}{\partial \rv_{\repsilon_\ri}}
\left[
\left( \alpha_{\epsilon} + \frac{3}{2} \right)\ln \rv_{\repsilon_\ri}
+
\left(\beta_{\epsilon} + \frac{1}{2} \left( \bg_\bi^{2} - 2 \bg_\bi \bHb_\bi \rfb  + \rfb^{T} \bHb_\bi^{T} \bHb_\bi \rfb \right)
\right) \rv_{\repsilon_\ri}^{-1}
\right]
=0
\\
&
\Leftrightarrow
\left( \alpha_{\epsilon} + \frac{3}{2} \right) \rv_{\repsilon_\ri}
-
\left(\beta_{\epsilon} + \frac{1}{2} \left( \bg_\bi - \bHb_\bi \rfb  \right)^2 \right)=0
\\
&
\Rightarrow
\rvh_{\repsilon_\ri} = \frac{\beta_{\epsilon} + \frac{1}{2} \left( \bg_\bi - \bHb_\bi \rfb \right)^2}{\alpha_{\epsilon} + \frac{3}{2}}
\end{split}
\nonumber
\eeq
\item With respect to $\rv_{\rxi_\rj}$, $j \in \left\lbrace 1,2,\ldots, M \right\rbrace$:\\
First, we develop the norm $\| \rVb_\rxi^{-\frac{1}{2}} \left( \rfb - \bDb\rzb \right) \|^2$:
\beq
\begin{split}
\| \rVb_\rxi^{-\frac{1}{2}} \left( \rfb - \bDb\rzb \right) \|^2
&= 
\rfb^{T} \rVb_\rxi^{-1} \rfb
-
2 \rfb^{T} \rVb_\rxi^{-1} \bDb \rzb
+
\bDb^{T} \rzb^{T} \rVb_\rxi^{-1} \bDb \rzb
\\
&=
\sum_{j=1}^{M} \rv_{\rxi_\rj}^{-1} \rf_\rj^{2}
-
2 \sum_{j=1}^{M} \rv_{\rxi_\rj}^{-1} \rf_\rj \bD_\bj \rzb 
+
\sum_{j=1}^{M} \rv_{\rxi_\rj}^{-1} \rzb^{T} \bD_\bj^{T} \bD_\bj \rzb,
\nonumber
\end{split}
\eeq
where $\bDb_\bj$ denotes the line j of the matrix $\bDb$, $j\in \left\lbrace 1, 2, \ldots, M \right\rbrace$, i.e. $\bDb_\bj = \left[ \bd_{\bj\blue{1}}, \bd_{\bj\blue{1}}, \ldots, \bd_{\bj\bM} \right]$. 
\beq
\begin{split}
\frac{\partial \Lc \left( \rfb, \; \rzb, \; \rvb_\rxi, \; \rvb_\repsilon, \; \rvb_\rz \right)}{\partial \rv_{\rxi_\rj}}=0
&
\Leftrightarrow
\frac{\partial}{\partial \rv_{\rxi_\rj}}
\left[
\left( \alpha_{\xi} + \frac{3}{2} \right)\ln \rv_{\rxi_\rj}
+
\left(\beta_{\xi} + \frac{1}{2} \left( \rf_\rj^{2} - 2 \rf_\rj \bDb_\bj \rzb  + \rzb^{T} \bDb_\bj^{T} \bHb_\bj \rzb \right)
\right) \rv_{\rxi_\rj}^{-1}
\right]
=0
\\
&
\Leftrightarrow
\left( \alpha_{\xi} + \frac{3}{2} \right) \rv_{\rxi_\rj}
-
\left(\beta_{\xi} + \frac{1}{2} \left( \rf_\rj - \bDb_\bi \rzb  \right)^2 \right)=0
\\
&
\Rightarrow
\rvh_{\rxi_\rj} = \frac{\beta_{\xi} + \frac{1}{2} \left( \rf_\rj - \bDb_\bj \rzb \right)^2}{\alpha_{\xi} + \frac{3}{2}}
\end{split}
\nonumber
\eeq
\item With respect to $\rv_\rz_\rj$, $j \in \left\lbrace 1,2,\ldots, M \right\rbrace$:
\beq
\begin{split}
\frac{\partial \Lc \left( \rfb, \; \rzb, \; \rvb_\rxi, \; \rvb_\repsilon, \; \rvb_\rz \right)}{\partial \rv_{\rz_\rj}}=0
&
\Leftrightarrow
\frac{\partial}{\partial \rv_{\rz_\rj}}
\left[
\left( \alpha_z + \frac{3}{2}  \right)\ln \rv_{\rz_\rj}
+
\left( \beta_z + \frac{1}{2} \rz_\rj^2 \right)
\rv_{\rz_\rj}^{-1}
\right]
=0
\\
&
\Leftrightarrow
\left( \alpha_z + \frac{3}{2}  \right) \rv_{\rz_\rj}
-
\left( \beta_z + \frac{1}{2} \rz_\rj^2 \right)
=0
\\
&
\Rightarrow
\rvh_{\rz_\rj} = \frac{\beta_z + \frac{1}{2} \rz_\rj^2}{\alpha_z + \frac{3}{2}}
\end{split}
\nonumber
\eeq
\end{itemize}
The iterative algorithm obtained via JMAP estimation is presented Figure~\eqref{Fig:IA_JMAP_St_nsStL_un_IS}.
\tikzstyle{cloud} = [draw=blue!30!green,fill=orange!12, ellipse, thick, node distance=6em, text width=12em, text centered, minimum height=4em, minimum width=12em]
\tikzstyle{boxbig} = [draw=blue!30!green, fill=orange!12, rectangle, rounded corners, thick, node distance=6.5em, text width=25em, text centered, minimum height=5em, minimum width=25em]
\tikzstyle{boxhuge} = [draw=blue!30!green, fill=orange!12, rectangle, rounded corners, thick, node distance=4.5em, text width=34em, text centered, minimum height=5em, minimum width=34em]
\tikzstyle{box} = [draw=blue!30!green, fill=orange!12, rectangle, rounded corners, thick, node distance=7.5em, text width=21em, text centered, minimum height=5em, minimum width=21em]
\tikzstyle{boxsmall} = [draw=blue!30!green, fill=orange!12, rectangle, rounded corners, thick, node distance=9em, text width=9em, text centered, minimum height=4em, minimum width=9em]
\tikzstyle{line} = [draw=blue!30!green, -latex']
\begin{figure}[!htb]
\centering
\begin{center}
\begin{tikzpicture}[auto]
    \node [boxhuge, scale=0.9] (f) {\Large{$\rfbh = \left( \bHb^T \rVbh_\repsilon^{-1} \bHb + \rVbh_\rxi^{-1} \right)^{-1} \left( \bHb^T \rVbh_\repsilon^{-1} \bgb + \rVbh_\rxi^{-1} \bDb\rzbh \right)$}
\\[8pt]
\normalsize{(a) - update estimated $\rfb$} 
};

	\node [boxbig, below of=f, node distance=9.6em, scale=0.9] (z) {\Large{$\rzbh =\left( \bDb^T \rVbh_\rxi^{-1} \bDb + \rVbh_\rz^{-1} \right)^{-1} \bDb^T \rVbh_\rxi^{-1} \rfbh$}
\\[5pt]
\normalsize{(b) - update estimated $\rzb$} 
};

    \node [box, below of=z, node distance=9.6em, scale=0.9] (vxi)
{\Large{$ \rvh_{\rxi_\rj} = \frac{\beta_{\xi} + \frac{1}{2} \left( \rfh_\rj - \bDb_\bj \rzbh \right)^2}{\alpha_{\xi} + \frac{3}{2}} $}
\\[5pt]
\normalsize{(c) - update estimated variances $\rv_{\rf_\rj}$} 
};

    \node [box, below of=vxi, node distance=9.6em, scale=0.9] (veps) {\Large{$\rvh_{\repsilon_\ri} = \frac{\beta_{\epsilon} + \frac{1}{2} \left( \bg_\bi - \bHb_\bi \rfbh \right)^2}{\alpha_{\epsilon} + \frac{3}{2}}$}
\\[5pt]
\normalsize{(d) - update estimated uncertainties variances $\rv_{\repsilon_\ri}$}     
};

    \node [box, below of=veps, node distance=9.6em, scale=0.9] (vz) {\Large{$\rvh_{\rz_\rj} = \frac{\beta_z + \frac{1}{2} \rz_\rj^2}{\alpha_z + \frac{3}{2}}$}
\\[5pt]
\normalsize{(e) - update estimated uncertainties variances $\rv_{\repsilon_\ri}$}     
};

    \node [boxsmall, above right = 0.1cm and 0.7cm of vxi, node distance=17em, scale=0.9] (Vxi) {\Large{$\rVbh_\rxi = \diag {\rvbh_{\rxi}}$}
\\[5pt]
\normalsize{(f)}
};

    \node [boxsmall, below left = -3.4cm and 0.7cm of veps, node distance=17em, scale=0.9] (Veps) {\Large{$\rVbh_\repsilon = \diag {\rvbh_{\repsilon}}$}
\\[5pt]
\normalsize{(g)} 
};

    \node [boxsmall, above right = 0.1cm and 2.4cm of vz, node distance=17em, scale=0.9] (Vz) {\Large{$\rVbh_\rz = \diag {\rvbh_{\rz}}$}
\\[5pt]
\normalsize{(e)} 
};

\node [cloud, above of=f, scale=0.9] (init)
{\Large{Initialization}};  

    \path [line] (f) -- (z)[near start];
    \path [line] (z) -- (vxi);
    \path [line] (vxi) -- (veps);
    \path [line] (veps) -- (vz);    
    \path [line] (Vxi) |- (f);
    \path [line] (Vxi) |- (z);
    \path [line] (Veps) |- (f);
    \path [line] (Vz) |- (z);    
    \path [line] (vxi) -| (Vxi);
    \path [line] (veps) -| (Veps);
    \path [line] (vz) -| (Vz);
    \path [line] (z) -| (f);
    \path [line] (init) -- (f);   
\end{tikzpicture}
\end{center}
\caption{Indirect sparsity (via $\rzb$) - Iterative algorithm corresponding to Joint MAP estimation for Student-t hierarchical model, non-stationary Student-t uncertainties model}
\label{Fig:IA_JMAP_St_nsStL_un_IS}
\end{figure}
\begin{algorithm}
\caption{Joint MAP - Student-t hierarchical model, non-stationary Student-t uncertainties model}
\begin{algorithmic}[1]
\Ensure INITIALIZATION $\rfb^{(0)},\rzb^{(0)}$ 




\Function{JMAP}{$\alpha_{\xi},\beta_{\xi},\alpha_{\epsilon},\beta_{\epsilon},\alpha_{z},\beta_{z},\bgb,\bHb, \bDb, \rfb^{(0)}, \rzb^{(0)}, M, N, NoIter$} 

\For{$n = 0$ to ${IterNumb}$}

\For{$j = 1$ to ${M}$}
\State $\rvh_{\rxi_\rj}^{(n)} = \frac{\beta_{\xi} + \frac{1}{2} \left( \rfh_\rj^{(n)} - \bDb_\bj \rzbh^{(n)} \right)^2}{\alpha_{\xi} + \frac{3}{2}} $
\EndFor

\For{$i = 1$ to ${N}$}
\State $\rvh_{\repsilon_\ri}^{(n)} = \frac{\beta_{\epsilon} + \frac{1}{2} \left( \bg_\bi - \bHb_\bi \rfbh^{(n)} \right)^2}{\alpha_{\epsilon} + \frac{3}{2}}$
\EndFor

\For{$j = 1$ to ${M}$}
\State $\rvh_{\rz_\rj}^{(n)} = \frac{\beta_z + \frac{1}{2} \left( \rz_\rj^{(n)} \right)^2}{\alpha_z + \frac{3}{2}}$
\EndFor

\State $\rfbh^{(n+1)} = \left( \bHb^T  \diag{\left(\rvh_{\repsilon_\ri}^{(n)}\right)^{-1}}  \bHb + \diag{\left(\rvh_{\rxi_\rj}^{(n)}\right)^{-1}} \right)^{-1} \left( \bHb^T \diag{\left(\rvh_{\repsilon_\ri}^{(n)}\right)^{-1}} \bgb + \diag{\left(\rvh_{\rxi_\rj}^{(n)}\right)^{-1}} \bDb\rzbh \right)$ 

\State $\rzbh^{(n+1)} =\left( \bDb^T \diag{\left(\rvh_{\rxi_\rj}^{(n)}\right)^{-1}} \bDb + \diag{\left(\rvh_{\rz_\rj}^{(n)}\right)^{-1}} \right)^{-1} \bDb^T \diag{\left( \rvh_{\rxi_\rj}^{(n)}\right)^{-1}} \rfbh^{(n+1)}$ 

\EndFor
\Return $\left( \rfbh^{(n+1)}, \rzbh^{(n+1)}, \rvh_{\rxi_\rj}^{(n)}, \rvh_{\repsilon_\ri}^{(n)},\rvh_{\rz_\rj}^{(n)} \right)$ $n = NoIter$
\EndFunction
\end{algorithmic}
\label{Alg:IA_JMAP_St_nsStL_un_bis_IS}
\end{algorithm}

\subsubsection{Posterior Mean estimation via VBA, partial separability}
\label{Subsubsec:PM_PS_St_nsStL_un_IS}
In this subsection, the Posterior Mean (PM) estimation is considered. The Joint MAP computes the mod of the posterior distribution. The PM computes the mean of the posterior distribution. One of the advantages of this estimator is that it minimizes the Mean Square Error (MSE). Computing the posterior means of any unknown needs great dimensional integration. For example, the mean corresponding to $\rfb$ can be computed from the posterior distribution using Equation~\eqref{Eq:PM_Computation_F_IS},
\beq
E_p \left\lbrace \rfb \right\rbrace = \iiint \rfb \; p(\rfb, \rzb, \rvb_\repsilon, \rvb_\rxi, \rvb_\rz | \bgb, \alpha_{\epsilon}, \beta_{\epsilon}, \alpha_{\xi}, \beta_{\xi}, \alpha_{z}, \beta_{z}) \d\rfb  \d\rzb \d\rvb_{\rxi} \d\rvb_{\repsilon} \d\rvb_{\rz}.
\label{Eq:PM_Computation_F_IS}
\eeq
In general, these computations are not easy. One way to obtain approximate estimates is to approximate $p(\rfb, \rzb, \rvb_\repsilon, \rvb_\rxi, \rvb_\rz | \bgb, \alpha_{\epsilon}, \beta_{\epsilon}, \alpha_{\xi}, \beta_{\xi}, \alpha_{z}, \beta_{z})$ by a separable one $q(\rfb, \rzb, \rvb_\repsilon, \rvb_\rxi, \rvb_\rz | \bgb, \alpha_{\epsilon}, \beta_{\epsilon}, \alpha_{\xi}, \beta_{\xi}, \alpha_{z}, \beta_{z}) = q_1(\rfb) q_2(\rzb)  q_3(\rvb_{\rxi}) q_4(\rvb_{\repsilon}) q_5(\rvb_{\rz}) $, then computing the posterior means using the separability. The mean corresponding to $\rfb$ is computed using the corresponding separable distribution $q_1(\rfb)$, Equation~\eqref{Eq:PM_Computation_F_Separable}, 
\beq
E_{q_{1}} \left\lbrace \rfb \right\rbrace = \int \rfb \; q_1(\rfb) \d\rfb.
\label{Eq:PM_Computation_F_Separable_IS}
\eeq
If the approximation of the posterior distribution with a separable one can be done in such a way that conserves the mean, i.e. Equation~\eqref{Eq:PM_Conservation_IS},
\beq
E_q \left\lbrace x \right\rbrace = E_p \left\lbrace x \right\rbrace, 
\label{Eq:PM_Conservation_IS}
\eeq
for all the unknowns of the model, a great amount of computational cost is gained. In particular, for the proposed hierarchical model, Equation~\eqref{Eq:St_nsStL_un_IS}, the posterior distribution, Equation~\eqref{Eq:Post_St_nsStL_un_IS}, is not a separable one, making the analytical computations of the PM very difficult. One way the compute the PM in this case is to first approximate the posterior law $p(\rfb, \rzb, \rvb_\repsilon, \rvb_\rxi, \rvb_\rz | \bgb, \alpha_{\epsilon}, \beta_{\epsilon}, \alpha_{\xi}, \beta_{\xi}, \alpha_{z}, \beta_{z})$ with a separable law $q(\rfb, \rzb, \rvb_\repsilon, \rvb_\rxi, \rvb_\rz | \bgb, \alpha_{\epsilon}, \beta_{\epsilon}, \alpha_{\xi}, \beta_{\xi}, \alpha_{z}, \beta_{z})$, Equation~\eqref{Eq:Posterior_Approximation_IS},
\beq
\begin{split}
p(\rfb, \rzb, \rvb_\repsilon, \rvb_\rxi, \rvb_\rz | \bgb, \alpha_{\epsilon}, \beta_{\epsilon}, \alpha_{\xi}, \beta_{\xi}, \alpha_{z}, \beta_{z})
&\approx 
q(\rfb, \rzb, \rvb_\repsilon, \rvb_\rxi, \rvb_\rz | \bgb, \alpha_{\epsilon}, \beta_{\epsilon}, \alpha_{\xi}, \beta_{\xi}, \alpha_{z}, \beta_{z})
\\&
=
q_1(\rfb) \; q_2(\rzb) \; q_3(\rvb_{\rxi}) \; q_4(\rvb_{\repsilon}) \; q_5(\rvb_{\rz})\label{Eq:Posterior_Approximation_IS}
\end{split}
\eeq
where the notations from Equation~\eqref{Eq:Posterior_Approximation_Notations1_IS} are used
\beq
q_{3}(\rvb_{\rxi}) = \prod_{j=1}^{M} q_{2j}(\rv_{\rxi_\rj}),
\;\;;\;\;
q_{4}(\rvb_{\repsilon}) = \prod_{i=1}^{N} q_{3i}(\rv_{\repsilon_\ri})
\;\;;\;\;
q_{5}(\rvb_{\rz}) = \prod_{j=1}^{M} q_{3i}(\rv_{\rz_\rj}) 
\label{Eq:Posterior_Approximation_Notations1_IS}
\eeq
by minimizing of the Kullback-Leibler divergence, defined as:
\small
\beq
\begin{split}
\mbox{KL} & \left( q(\rfb, \rzb, \rvb_\repsilon, \rvb_\rxi, \rvb_\rz | \bgb, \alpha_{\epsilon}, \beta_{\epsilon}, \alpha_{\xi}, \beta_{\xi}, \alpha_{z}, \beta_{z})
 : p(\rfb, \rzb, \rvb_\repsilon, \rvb_\rxi, \rvb_\rz | \bgb, \alpha_{\epsilon}, \beta_{\epsilon}, \alpha_{\xi}, \beta_{\xi}, \alpha_{z}, \beta_{z}) \right) =
\\
&=\iint \ldots \int q(\rfb, \rzb, \rvb_\repsilon, \rvb_\rxi, \rvb_\rz | \bgb, \alpha_{\epsilon}, \beta_{\epsilon}, \alpha_{\xi}, \beta_{\xi}, \alpha_{z}, \beta_{z}) \; \ln\frac{q(\rfb, \rzb, \rvb_\repsilon, \rvb_\rxi, \rvb_\rz | \bgb, \alpha_{\epsilon}, \beta_{\epsilon}, \alpha_{\xi}, \beta_{\xi}, \alpha_{z}, \beta_{z})} {p(\rfb, \rzb, \rvb_\repsilon, \rvb_\rxi, \rvb_\rz | \bgb, \alpha_{\epsilon}, \beta_{\epsilon}, \alpha_{\xi}, \beta_{\xi}, \alpha_{z}, \beta_{z})} \d\rfb  \d\rzb \d\rvb_{\rxi} \d\rvb_{\repsilon} \d\rvb_{\rz},
\label{Eq:Kullback-Leibler_IS}
\end{split}
\eeq
\normalsize
where the notations from Equation~\eqref{Eq:Posterior_Approximation_Notations2_IS} are used
\beq
\d \rvb_{\rxi} = \prod_{j=1}^{M} \d \rv_{\rxi_\rj}
\;\;;\;\;
\d \rvb_{\repsilon} = \prod_{i=1}^{N} \d \rv_{\repsilon_\ri}
\;\;;\;\;
\d \rvb_{\rz} = \prod_{j=1}^{M} \d \rv_{\rz_\rj}.
\label{Eq:Posterior_Approximation_Notations2_IS}
\eeq
Equation~\eqref{Eq:Posterior_Approximation_Notations1_IS} is selecting a partial separability for the approximated posterior distribution $q(\rfb, \rzb, \rvb_\repsilon, \rvb_\rxi, \rvb_\rz | \bgb, \alpha_{\epsilon}, \beta_{\epsilon}, \alpha_{\xi}, \beta_{\xi}, \alpha_{z}, \beta_{z})$ in the sense that a total separability is imposed for the distributions corresponding to the three variances appearing in the hierarchical model, $q_3\left( \rvb_{\rxi} \right)$, $q_4\left( \rvb_{\repsilon} \right)$ and $q_5\left( \rvb_{\rz} \right)$ but not for the distribution corresponding to $\rfb$ and $\rzb$. Evidently, a full separability can be imposed, by adding the supplementary conditions $q_{1}(\rfb) = \prod_{j=1}^{M} q_{1j}(\rf_\rj)$ and $q_{3}(\rzb) = \prod_{j=1}^{M} q_{2j}(\rz_\rj)$ in Equation~\eqref{Eq:Posterior_Approximation_Notations1_IS}. This case is considered in Subsection~\eqref{Subsubsec:PM_FS_St_nsStL_un_IS}. The minimization can be done via alternate optimization resulting the following proportionalities from Equations~\eqref{Eq:VBA_Proportionalities1_IS},~\eqref{Eq:VBA_Proportionalities2_IS},~\eqref{Eq:VBA_Proportionalities3_IS},~\eqref{Eq:VBA_Proportionalities4_IS} and ~\eqref{Eq:VBA_Proportionalities5_IS},  
\begin{subequations}
\begin{align}
&q_1(\rfb) \;\;\; \propto \; \exp \biggl\lbrace \left\langle \ln p(\rfb, \rzb, \rvb_\repsilon, \rvb_\rxi, \rvb_\rz | \bgb, \alpha_{\epsilon}, \beta_{\epsilon}, \alpha_{\xi}, \beta_{\xi}, \alpha_{z}, \beta_{z}) \biggr\rangle_{q_2(\rzb) \; q_3(\rvb_{\rxi}) \; q_4(\rvb_{\repsilon}) \; q_5(\rvb_{\rz})} \right\rbrace,
\label{Eq:VBA_Proportionalities1_IS}
\\[4pt]
&q_2(\rzb) \;\;\; \propto \; \exp \biggl\lbrace \left\langle \ln p(\rfb, \rzb, \rvb_\repsilon, \rvb_\rxi, \rvb_\rz | \bgb, \alpha_{\epsilon}, \beta_{\epsilon}, \alpha_{\xi}, \beta_{\xi}, \alpha_{z}, \beta_{z}) \biggr\rangle_{q_1(\rfb) \; q_3(\rvb_{\rxi}) \; q_4(\rvb_{\repsilon}) \; q_5(\rvb_{\rz})} \right\rbrace,
\label{Eq:VBA_Proportionalities2_IS}
\\[4pt]  
&q_{3j}(\rv_{\rxi_\red{j}}) \propto \; \exp \biggl\lbrace \left\langle \ln p(\rfb, \rzb, \rvb_\repsilon, \rvb_\rxi, \rvb_\rz | \bgb, \alpha_{\epsilon}, \beta_{\epsilon}, \alpha_{\xi}, \beta_{\xi}, \alpha_{z}, \beta_{z}) \biggr\rangle_{q_1(\rfb) \; q_2(\rzb) \; q_{3-j}(\rvb_{\rxi_\rj}) \; q_4(\rvb_{\repsilon}) \; q_5(\rvb_{\rz})} \right\rbrace,
\nonumber 
\\&
\hspace{341pt}j \in \left\lbrace 1,2 \ldots, M \right\rbrace,  
\label{Eq:VBA_Proportionalities3_IS}
\\[4pt]
&q_{4i}(\rv_{\repsilon_\ri}) \; \propto \; \exp \biggl\lbrace \left\langle \ln p(\rfb, \rzb, \rvb_\repsilon, \rvb_\rxi, \rvb_\rz | \bgb, \alpha_{\epsilon}, \beta_{\epsilon}, \alpha_{\xi}, \beta_{\xi}, \alpha_{z}, \beta_{z}) \biggr\rangle_{q_1(\rfb) \; q_2(\rzb) \; q_3(\rvb_{\rxi}) \; q_{4-i}(\rvb_{\repsilon_\ri}) \; q_5(\rvb_{\rz})} \right\rbrace,
\nonumber 
\\&
\hspace{342pt}i \in \left\lbrace 1,2 \ldots, N \right\rbrace,
\label{Eq:VBA_Proportionalities4_IS}
\\[4pt]
&q_{5j}(\rv_{\rz_\red{j}}) \propto \; \exp \biggl\lbrace \left\langle \ln p(\rfb, \rzb, \rvb_\repsilon, \rvb_\rxi, \rvb_\rz | \bgb, \alpha_{\epsilon}, \beta_{\epsilon}, \alpha_{\xi}, \beta_{\xi}, \alpha_{z}, \beta_{z}) \biggr\rangle_{q_1(\rfb) \; q_2(\rzb) \; q_3(\rvb_{\rxi}) \; q_4(\rvb_{\repsilon}) \; q_{5-j}(\rvb_{\rz_\rj})} \right\rbrace,
\nonumber 
\\&
\hspace{341pt}j \in \left\lbrace 1,2 \ldots, M \right\rbrace,
\label{Eq:VBA_Proportionalities5_IS}
\end{align}
\end{subequations}
using the notations:
\beq
q_{3-j}(\rv_{\rxi_\red{j}})=\displaystyle \prod_{k=1,k \neq j}^{M} q_{3k}(\rv_{\rxi_\red{k}})
\;\;\;;\;\;\;
q_{4-i}(\rv_{\repsilon_\ri})=\displaystyle \prod_{k=1,k \neq i}^{N} q_{4k}(\rv_{\repsilon_\rk})
\;\;\;;\;\;\;
q_{5-j}(\rv_{\rz_\red{j}})=\displaystyle \prod_{k=1,k \neq j}^{M} q_{5k}(\rv_{\rz_\red{k}})
\label{Eq:Def_q_Minus_IS}
\eeq
and also
\beq
\biggl\langle u(x) \biggr\rangle_{v(y)}= \displaystyle \int u(x) v(y) \d y. 
\label{Eq:Def_Integral_IS}
\eeq
Via Equation~\eqref{Eq:Def_L_Criterion_IS} and Equation~\eqref{Eq:L_Criterion_IS}, the analytical expression of logarithm of the posterior distribution is obtained, Equation~\eqref{Eq:Log_Posterior_IS}:
\beq
\begin{split}
\ln  p \left( \rfb, \rzb, \rvb_\rxi, \rvb_\repsilon, \rvb_\rz | \bgb, \alpha_{\xi}, \beta_{\xi}, \alpha_{\epsilon}, \beta_{\epsilon}, \alpha_{z}, \beta_{z} \right)
=&
-
\frac{1}{2}\| \rVb_{\repsilon}^{-\frac{1}{2}} \left( \bgb - \bHb\rfb \right) \|^2
-
\left( \alpha_{\epsilon} + \frac{3}{2} \right) \sum_{i=1}^{N} \ln \rv_{\repsilon_\ri} 
-
\sum_{i=1}^{N}\frac{\beta_{\epsilon}}{\rv_{\repsilon_\ri}} 
\\&
-
\frac{1}{2}\| \rVb_{\rxi}^{-\frac{1}{2}} \left( \rfb - \bDb\rzb \right) \|^2
-
\left( \alpha_{\xi} + \frac{3}{2} \right) \sum_{j=1}^{M} \ln \rv_{\rxi_\rj} 
-
\sum_{j=1}^{M}\frac{\beta_{\xi}}{\rv_{\rxi_\rj}}
\\&
-
\frac{1}{2}\| \rVb_{\rz}^{-\frac{1}{2}} \rzb \|^2
-
\left(\alpha_{z} + \frac{3}{2} \right) \sum_{j=1}^{M} \ln \rv_{\rz_\rj}
-
\sum_{j=1}^{M}\frac{\beta_{f}}{\rv_{\rz_\rj}}
\end{split}
\label{Eq:Log_Posterior_IS}
\eeq
\paragraph{Computation of the analytical expression of $q_1(\rfb)$ (first part).}
The proportionality relation corresponding to $q_1(\rfb)$ is established in Equation~\eqref{Eq:VBA_Proportionalities1_IS}. In the expression of $p \left( \rfb, \rzb, \rvb_\rxi, \rvb_\repsilon, \rvb_\rz | \bgb, \alpha_{\xi}, \beta_{\xi}, \alpha_{\epsilon}, \beta_{\epsilon}, \alpha_{z}, \beta_{z} \right)$ all the terms free of $\rfb$ can be regarded as constants. Via Equation~\eqref{Eq:Log_Posterior_IS} the integral $\left\langle \right\rangle$ defined in Equation~\eqref{Eq:Def_Integral_IS} becomes:
\beq
\begin{split}
\biggl\langle p \left( \rfb, \rzb, \rvb_\rxi, \rvb_\repsilon, \rvb_\rz | \bgb, \alpha_{\xi}, \beta_{\xi}, \alpha_{\epsilon}, \beta_{\epsilon}, \alpha_{z}, \beta_{z} \right) \biggr\rangle_{q_2(\rzb) \; q_3(\rvb_{\rxi}) \; q_4(\rvb_{\repsilon})}
=&
-\frac{1}{2}
\biggl\langle 
\|\rVb_{\repsilon}^{-\frac{1}{2}} \left(\bgb - \bHb\rfb\right) \|^2
\biggr\rangle_{ q_4(\rvb_{\repsilon}) }
\\&
-\frac{1}{2}
\biggl\langle 
\| \rVb_{\rxi}^{-\frac{1}{2}} \left( \rfb - \bDb\rzb \right) \|^2
\biggr\rangle_{ q_2(\rzb) q_3(\rvb_{\rxi}) }.
\end{split}
\label{Eq:q1_Integral_1_IS}
\eeq
Introducing the notations:
\beq
\begin{split}
\rvt_{\rxi_\rj} = \biggl\langle \rv_{\rxi_\rj}^{-1} \biggr\rangle_{q_{3j}\left( \rv_{\rxi_\rj} \right)}
\;\;;\;\;
\rvbt_{\rxi}\;&=\;
\begin{bmatrix}
\rvt_{\rxi_\red{1}} \ldots 
\rvt_{\rxi_\rj} \ldots 
\rvt_{\rxi_\rM}
\end{bmatrix}^T
\;\;;\;\;
\rVbt_{\rxi} = \diag {\rvbt_{\rxi}}\;
\\
\rvt_{\repsilon_\ri}
=
\biggl\langle \rv_{\repsilon_\ri}^{-1} \biggr\rangle_{q_{4i}\left( \rv_{\repsilon_\ri} \right)}
\;\;;\;\;
\rvbt_\repsilon\;&=\;
\begin{bmatrix}
\rvt_{\repsilon_\red{1}}
\ldots 
\rvt_{\repsilon_\ri}
\ldots 
\rvt_{\repsilon_\rN}
\end{bmatrix}^T
\;\;;\;\;
\rVbt_\repsilon = \diag {\rvbt_\repsilon}
\end{split}
\label{Eq:q1_Integral_Not1_IS}
\eeq
the integral from Equation~\eqref{Eq:q1_Integral_1_IS} becomes:
\beq
\begin{split}
\biggl\langle p \left( \rfb, \rzb, \rvb_\rxi, \rvb_\repsilon, \rvb_\rz | \bgb, \alpha_{\xi}, \beta_{\xi}, \alpha_{\epsilon}, \beta_{\epsilon}, \alpha_{z}, \beta_{z} \right) \biggr\rangle_{q_2(\rzb) \;q_3(\rvb_{\rxi}) \; q_4(\rvb_{\repsilon})}
=&
-\frac{1}{2} \|\rVbt_{\repsilon}^{\frac{1}{2}} \left(\bgb - \bHb\rfb\right) \|^2
\\&
- 
\frac{1}{2} \biggl\langle \| \rVbt_{\rxi}^{\frac{1}{2}} \left( \rfb - \bDb\rzb \right) \|^2\biggr\rangle_{q_2(\rzb)}.
\end{split}
\label{Eq:q1_Integral_2_IS}
\eeq
Evidently, $\biggl\langle p \left( \rfb, \rzb, \rvb_\rxi, \rvb_\repsilon, \rvb_\rz | \bgb, \alpha_{\xi}, \beta_{\xi}, \alpha_{\epsilon}, \beta_{\epsilon}, \alpha_{z}, \beta_{z} \right) \biggr\rangle_{q_2(\rzb) \;q_3(\rvb_{\rxi}) \; q_4(\rvb_{\repsilon})}$ is a quadratic form with respect to $\rfb$. The proportionality from Equation~\eqref{Eq:VBA_Proportionalities1_IS} leads to the following corollary:
\begin{corollary}
$q_1 \left( \rfb \right)$ is a multivariate Normal distribution.
\label{Co:q1}
\end{corollary} 
\paragraph{Computation of the analytical expression of $q_2(\rzb)$.}
The proportionality relation corresponding to $q_2(\rzb)$ is established in Equation~\eqref{Eq:VBA_Proportionalities2_IS}. In the expression of $p \left( \rfb, \rzb, \rvb_\rxi, \rvb_\repsilon, \rvb_\rz | \bgb, \alpha_{\xi}, \beta_{\xi}, \alpha_{\epsilon}, \beta_{\epsilon}, \alpha_{z}, \beta_{z} \right)$ all the terms free of $\rzb$ can be regarded as constants. Via Equation~\eqref{Eq:Log_Posterior_IS} the integral $\left\langle \right\rangle$ defined in Equation~\eqref{Eq:Def_Integral_IS} becomes:
\beq
\begin{split}
\biggl\langle p \left( \rfb, \rzb, \rvb_\rxi, \rvb_\repsilon, \rvb_\rz | \bgb, \alpha_{\xi}, \beta_{\xi}, \alpha_{\epsilon}, \beta_{\epsilon}, \alpha_{z}, \beta_{z} \right) \biggr\rangle_{q_1(\rfb) \; q_3(\rvb_{\rxi}) \; q_5(\rvb_{\rz})}
=&
-\frac{1}{2}
\biggl\langle 
\| \rVb_{\rxi}^{-\frac{1}{2}} \left( \rfb - \bDb\rzb \right) \|^2
\biggr\rangle_{q_1(\rfb) q_3(\rvb_{\rxi}) }
\\&
-\frac{1}{2}
\biggl\langle 
\| \rVb_{\rz}^{-\frac{1}{2}} \rzb \|^2
\biggr\rangle_{ q_5(\rvb_{\rz}) }.
\end{split}
\label{Eq:q2_Integral_1_IS}
\eeq
Using the notations introduced in Equation~\eqref{Eq:q1_Integral_Not1_IS} and introducing the notations:
\beq
\begin{split}
\rvt_{\rz_\rj} = \biggl\langle \rv_{\rz_\rj}^{-1} \biggr\rangle_{q_{5j}\left( \rv_{\rz_\rj} \right)}
\;\;;\;\;
\rvbt_{\rz}\;&=\;
\begin{bmatrix}
\rvt_{\rz_\red{1}} \ldots 
\rvt_{\rz_\rj} \ldots 
\rvt_{\rz_\rM}
\end{bmatrix}^T
\;\;;\;\;
\rVbt_{\rz} = \diag {\rvbt_{\rz}}\;
\end{split}
\label{Eq:q2_Integral_Not1_IS}
\eeq
the integral from Equation~\eqref{Eq:q2_Integral_1_IS} becomes:
\beq
\begin{split}
\biggl\langle p \left( \rfb, \rzb, \rvb_\rxi, \rvb_\repsilon, \rvb_\rz | \bgb, \alpha_{\xi}, \beta_{\xi}, \alpha_{\epsilon}, \beta_{\epsilon}, \alpha_{z}, \beta_{z} \right) \biggr\rangle_{q_2(\rzb) \;q_3(\rvb_{\rxi}) \; q_4(\rvb_{\repsilon})}
=&
-\frac{1}{2}
\biggl\langle 
\| \rVbt_{\rxi}^{\frac{1}{2}} \left( \rfb - \bDb\rzb \right) \|^2
\biggr\rangle_{q_1(\rfb)}
\\&
-\frac{1}{2}
\| \rVbt_{\rz}^{\frac{1}{2}} \rzb \|^2.
\end{split}
\label{Eq:q2_Integral_2_IS}
\eeq
The norm from the first term in Equation~\eqref{Eq:q2_Integral_2_IS} can be developed as it follows:
\beq
\| \rVbt_{\rxi}^{\frac{1}{2}} \left( \rfb - \bDb\rzb \right) \|^2
=
\rfb^T \rVbt_{\rxi} \rfb 
-
2 \rzb^T \bDb^T \rVbt_{\rxi} \rfb
+ \rzb^T \bDb^T \rVbt_{\rxi} \bDb \rzb
=
\| \rVbt_{\rxi}^{\frac{1}{2}} \rfb \|^2
-
2 \rzb^T \bDb^T \rVbt_{\rxi} \rfb 
+
\| \rVbt_{\rxi}^{\frac{1}{2}} \bDb \rzb \|^2,
\label{Eq:Norm_Development_IS}
\eeq  
so Equation~\eqref{Eq:q2_Integral_2_IS} can be developed as it follows:
\beq
\biggl\langle 
\| \rVbt_{\rxi}^{\frac{1}{2}} \left( \rfb - \bDb\rzb \right) \|^2
\biggr\rangle_{q_1(\rfb)}
=
\biggl\langle 
\| \rVbt_{\rxi}^{\frac{1}{2}} \rfb \|^2
\biggr\rangle_{q_1(\rfb)}
-
2 \rzb^T \bDb^T \rVbt_{\rxi} \biggl\langle \rfb \biggr\rangle_{q_1(\rfb)}
+
\| \rVbt_{\rxi}^{\frac{1}{2}} \bDb \rzb \|^2.
\label{Eq:q2_Integral_3_IS}
\eeq  
Using Corollary~\eqref{Co:q1}, we can easily compute $\left\langle 
\| \rVbt_{\rxi}^{\frac{1}{2}} \rfb \|^2
\right\rangle_{q_1(\rfb)}$ and $\left\langle \rfb \right\rangle_{q_1(\rfb)}$. For the multivariate Normal distribution $q_1(\rfb)$ we will denote $\rfbh$ the corresponding mean and $\Sigmabh_{f}$, obtaining:
\beq
\biggl\langle \rfb \biggr\rangle_{q_1(\rfb)} 
=
\rfbh
\;\;\;\;\; ; \;\;\;\;\;
\biggl\langle 
\| \rVbt_{\rxi}^{\frac{1}{2}} \rfb \|^2
\biggr\rangle_{q_1(\rfb)}
=
\| \rVbt_{\rxi}^{\frac{1}{2}} \rfbh \|^2 
+
\Tr{\rVbt_{\rxi}\Sigmabh_{f}} 
\label{Eq:q2_Integral_4_IS}
\eeq
From Equations~\eqref{Eq:q2_Integral_2_IS},~\eqref{Eq:q2_Integral_3_IS},~\eqref{Eq:q2_Integral_4_IS} we obtain:
\beq
\begin{split}
\biggl\langle p \left( \rfb, \rzb, \rvb_\rxi, \rvb_\repsilon, \rvb_\rz | \bgb, \alpha_{\xi}, \beta_{\xi}, \alpha_{\epsilon}, \beta_{\epsilon}, \alpha_{z}, \beta_{z} \right) \biggr\rangle_{q_2(\rzb) \;q_3(\rvb_{\rxi}) \; q_4(\rvb_{\repsilon})}
=&
-\frac{1}{2} \| \rVbt_{\rxi}^{\frac{1}{2}} \rfbh \|^2 
+
-\frac{1}{2} \Tr{\rVbt_{\rxi}\Sigmabh_{f}}
+ \rzb^T \bDb^T \rVbt_{\rxi} \rfbh
\\&
-\frac{1}{2} \| \rVbt_{\rxi}^{\frac{1}{2}} \bDb \rzb \|^2
-\frac{1}{2} \| \rVbt_{\rz}^{\frac{1}{2}} \rzb \|^2
\end{split}
\label{Eq:q2_Integral_5_IS}
\eeq 
Evidently, $\biggl\langle p \left( \rfb, \rzb, \rvb_\rxi, \rvb_\repsilon, \rvb_\rz | \bgb, \alpha_{\xi}, \beta_{\xi}, \alpha_{\epsilon}, \beta_{\epsilon}, \alpha_{z}, \beta_{z} \right) \biggr\rangle_{q_2(\rzb) \;q_3(\rvb_{\rxi}) \; q_4(\rvb_{\repsilon})}$. The proportionality from Equation~\eqref{Eq:VBA_Proportionalities2_IS} leads to the following corollary:
\begin{corollary}
$q_2 \left( \rzb \right)$ is a multivariate Normal distribution.
\label{Co:q2}
\end{corollary}
\hspace{-14pt}Minimizing the criterion
\beq
J(\rzb) 
=
-\frac{1}{2} \| \rVbt_{\rxi}^{\frac{1}{2}} \rfbh \|^2
+
-\frac{1}{2} \Tr{\rVbt_{\rxi}\Sigmah_{\rfb}}
+ \rzb^T \bDb^T \rVbt_{\rxi} \rfbh
-\frac{1}{2} \| \rVbt_{\rxi}^{\frac{1}{2}} \bDb \rzb \|^2
-\frac{1}{2} \| \rVbt_{\rz}^{\frac{1}{2}} \rzb \|^2,
\label{Criterion_q2_IS}
\eeq
leads to the analytical expression of the corresponding mean, denoted $\rzbh$:
\beq
\begin{split}
\frac{\partial J(\rzb) }{\partial \rzb} = 0
&\Leftrightarrow
\frac{\partial }{\partial \rzb} \rfbh^T \rVbt_{\rxi}^T \bDb \rzb
-\frac{1}{2} \frac{\partial }{\partial \rzb} \| \rVbt_{\rxi}^{\frac{1}{2}} \bDb \rzb \|^2
-\frac{1}{2} \frac{\partial }{\partial \rzb} \| \rVbt_{\rz}^{\frac{1}{2}} \rzb \|^2
=0
\Leftrightarrow
\bDb^T \rVbt_{\rxi} \rfbh - \bDb^T \rVbt_{\rxi} \bDb \rzb - \rVbt_{\rz} \rzb
=0
\\&
\Leftrightarrow
\left( \bDb^T \rVbt_{\rxi} \bDb + \rVbt_{\rz} \right) \rzb = \bDb^T \rVbt_{\rxi} \rfbh
\Rightarrow
\rzbh = \left( \bDb^T \rVbt_{\rxi} \bDb + \rVbt_{\rz} \right)^{-1} \bDb^T \rVbt_{\rxi} \rfbh
\end{split}
\label{MeanComputation_q2_IS}
\eeq
The variance can be obtained by identification:
\beq
\Sigmabh_{z} = \left( \bDb^T \rVbt_{\rxi} \bDb + \rVbt_{\rz} \right)^{-1}
\eeq
Finally, we conclude that $q_2(\rzb)$ is a multivariate Normal distribution with the following parameters:
\beq
q_2(\rzb) = \Nc\left( \rzb | \rzbh, \Sigmabh_{z} \right),
\left\{\barr{ll}
\rzbh = \left( \bDb^T \rVbt_{\rxi} \bDb + \rVbt_{\rz} \right)^{-1} \bDb^T \rVbt_{\rxi} \rfbh,
\\[10pt]
\Sigmabh_{z} = \left( \bDb^T \rVbt_{\rxi} \bDb + \rVbt_{\rz} \right)^{-1}.
\earr\right.
\label{Eq:q2_Analytical_1_IS}
\eeq
We note that both the expressions of the mean $\rzbh$ and variance $\Sigmabh_{z}$ depend on expectancies corresponding to the three variances of the hierarchical model. 
\paragraph{Computation of the analytical expression of $q_1(\rfb)$ (second part).}
We come back for computing the parameters of the multivariate Normal distribution $q_1(\rfb)$. Using the fact that $q_2(\rzb)$ was proved a multivariate Normal distribution, the norm from the latter term in Equation~\eqref{Eq:q1_Integral_2_IS} can be computed. The norm was developed in Equation~\eqref{Eq:Norm_Development_IS} and the computation of the integral uses the particular form of $q_2(\rzb)$:
\beq
\begin{split}
\biggl\langle 
\| \rVbt_{\rxi}^{\frac{1}{2}} \left( \rfb - \bDb\rzb \right) \|^2
\biggr\rangle_{q_2(\rzb)}
&=
\| \rVbt_{\rxi}^{\frac{1}{2}} \rfb \|^2
-
2 \rfb^T \rVbt_{\rxi}^T \bDb \biggl\langle \rzb \biggr\rangle_{q_2(\rzb)}
+
\biggl\langle 
\| \rVbt_{\rxi}^{\frac{1}{2}} \bDb \rzb \|^2
\biggr\rangle_{q_2(\rzb)}
\\&
=
\| \rVbt_{\rxi}^{\frac{1}{2}} \rfb \|^2
-
2 \rzbh \bDb^T \rVbt_{\rxi} \rfb
+  
\| \rVbt_{\rxi}^{\frac{1}{2}} \bDb \rzbh \|^2
+
\Tr{\bDb^T \rVbt_{\rxi} \bDb \Sigmabh_{z}}.
\end{split}
\label{Eq:q1_Integral_3_IS}
\eeq
Minimizing the criterion
\beq
J(\rfb) 
=
-\frac{1}{2} \| \rVbt_{\repsilon}^{\frac{1}{2}} \left( \bgb - \bHb \rfb \right) \|^2 
-\frac{1}{2}\| \rVbt_{\rxi}^{\frac{1}{2}} \rfb \|^2
+ \rzbh \bDb^T \rVbt_{\rxi} \rfb
-\frac{1}{2} \| \rVbt_{\rxi}^{\frac{1}{2}} \bDb \rzbh \|^2
-\frac{1}{2} \Tr{\bDb^T \rVbt_{\rxi} \bDb \Sigmabh_{z}}
\label{Criterion_q1_IS}
\eeq
leads to the analytical expression of the corresponding mean, denoted $\rzbh$:
\beq
\begin{split}
\frac{\partial J(\rfb) }{\partial \rfb} = 0
&\Leftrightarrow
-\frac{1}{2} \frac{\partial }{\partial \rfb} \| \rVbt_{\repsilon}^{\frac{1}{2}} \left( \bgb - \bHb \rfb \right) \|^2
-\frac{1}{2} \frac{\partial }{\partial \rfb} \| \rVbt_{\rxi}^{\frac{1}{2}} \rfb \|^2
+ \frac{\partial }{\partial \rfb} \rzbh^T \bDb^T \rVbt_{\rxi} \rfb 
=0
\\&
\Leftrightarrow
\bHb^T \rVbt_{\repsilon} \bgb
- \bHb^T \rVbt_{\repsilon} \bHb \rfb
- \rVbt_{\rxi} \rfb
+ \rVbt_{\rxi}^T \bDb \rzbh  
=0
\\&
\Leftrightarrow
\left( \bHb^T \rVbt_{\repsilon} \bHb + \rVbt_{\rxi} \right) \rfb = \rVbt_{\rxi}^T \bDb \rzbh + \bHb^T \rVbt_{\repsilon} \bgb
\Rightarrow
\rfbh = \left( \bHb^T \rVbt_{\repsilon} \bHb + \rVbt_{\rxi} \right)^{-1} \left( \rVbt_{\rxi}^T \bDb \rzbh + \bHb^T \rVbt_{\repsilon} \bgb \right)
\end{split}
\label{MeanComputation_q1_IS}
\eeq
The variance can be obtained by identification:
\beq
\Sigmabh_{f} = \left( \bHb^T \rVbt_{\repsilon} \bHb + \rVbt_{\rxi} \right)^{-1}
\eeq
Finally, we conclude that $q_1(\rfb)$ is a multivariate Normal distribution with the following parameters:
\beq
q_1(\rfb) = \Nc\left( \rfb | \rfbh, \Sigmabh_{f} \right),
\left\{\barr{ll}
\rfbh = \left( \bHb^T \rVbt_{\repsilon} \bHb + \rVbt_{\rxi} \right)^{-1} \left( \rVbt_{\rxi}^T \bDb \rzbh + \bHb^T \rVbt_{\repsilon} \bgb \right),
\\[10pt]
\Sigmabh_{f} = \left( \bHb^T \rVbt_{\repsilon} \bHb + \rVbt_{\rxi} \right)^{-1}.
\earr\right.
\label{Eq:q1_Analytical_1_IS}
\eeq
We note that both the expressions of the mean $\rfbh$ and variance $\Sigmabh_{f}$ depend on expectancies corresponding to the three variances of the hierarchical model. 
\paragraph{Computation of the analytical expression of $q_{3j}(\rv_{\rxi_\rj})$.}
The proportionality relation corresponding to $q_{3j}(\rv_{\rxi_\rj})$ is presented in established in Equation~\eqref{Eq:VBA_Proportionalities3_IS}. In the expression of $p \left( \rfb, \rzb, \rvb_\rxi, \rvb_\repsilon, \rvb_\rz | \bgb, \alpha_{\xi}, \beta_{\xi}, \alpha_{\epsilon}, \beta_{\epsilon}, \alpha_{z}, \beta_{z} \right)$ all the terms free of $\rv_{\rxi_\rj}$ can be regarded as constants. Via Equation~\eqref{Eq:Log_Posterior_IS} the integral defined in Equation~\eqref{Eq:Def_Integral_IS} becomes:
\beq
\begin{split}
\biggl\langle \ln p(\rfb, \rzb, \rvb_\repsilon, \rvb_\rxi, \rvb_\rz | \bgb, \alpha_{\epsilon}, \beta_{\epsilon}, \alpha_{\xi}, \beta_{\xi}, \alpha_{z}, \beta_{z}) \biggr\rangle_{q_1(\rfb) \; q_2(\rzb) \; q_{3-j}(\rvb_{\rxi_\rj})}
&=
-
\frac{1}{2} \biggl\langle \| \rVb_{\rxi}^{-\frac{1}{2}} \left( \rfb - \bDb\rzb \right) \|^2 \biggr\rangle_{q_1(\rfb) \; q_2(\rzb) \; q_{3-j}(\rvb_{\rxi_\rj})}
\\&
-
\left( \alpha_{\xi} + \frac{3}{2} \right) \ln \rv_{\rxi_\rj} 
-
\beta_{\xi} \rv_{\rxi_\rj}^{-1}
\end{split}
\label{Eq:q3_Integral_1_IS}
\eeq
Introducing the notations:
\beq
\rvbt_{\rxi_{-\rj}}^{-1}=
\begin{bmatrix}
\rvt_{\rxi_\red{1}}^{-1} \;
\ldots \;
\rvt_{\rxi_{\rj-\red{1}}}^{-1} \;
\rv_{\rxi_\rj}^{-1}  \;
\rvt_{\rxi_{\rj+\red{1}}}^{-1}  \;
\ldots \;
\rvt_{\rxi_\rM}^{-1}
\end{bmatrix}^T
\;\; ; \;\;
\rVbt_{\rxi_{-\rj}}=
\mbox{diag}\left(\rvbt_{\rxi_{-\rj}}^{-1} \right)
\label{Eq:q3_Integral_Not1_IS}
\eeq
the integral $\biggl\langle \| \rVb_{\rxi}^{-\frac{1}{2}} \left( \rfb - \bDb\rzb \right) \|^2 \biggr\rangle_{q_1(\rfb) \; q_2(\rzb) \; q_{3-j}(\rvb_{\rxi_\rj})}$ can be written:
\beq
\biggl\langle \| \rVb_{\rxi}^{-\frac{1}{2}} \left( \rfb - \bDb\rzb \right) \|^2 \biggr\rangle_{q_1(\rfb) \; q_2(\rzb) \; q_{3-j}(\rvb_{\rxi_\rj})}
= 
\biggl\langle \| \rVbt_{\rxi_{-\rj}}^{\frac{1}{2}} \left( \rfb - \bDb\rzb \right) \|^2 \biggr\rangle_{q_1(\rfb) \; q_2(\rzb)}
\eeq 
Considering that $q_1(\rfb)$ and $q_2(\rzb)$ are multivariate Normal distributions, Equations~\eqref{Eq:q1_Analytical_1_IS} and~\eqref{Eq:q2_Analytical_1_IS} and considering the development of the norm Equations~\eqref{Eq:Norm_Development_IS}, we have:  
\beq
\begin{split}
\biggl\langle \| \rVbt_{\rxi_{-\rj}}^{\frac{1}{2}} \left( \rfb - \bDb\rzb \right) \|^2 & \biggr\rangle_{q_1(\rfb) \; q_2(\rzb)}
=
\biggl\langle \| \rVbt_{\rxi_{-\rj}}^{\frac{1}{2}} \rfb \|^2 \biggr\rangle_{q_1(\rfb)}
- 2 \biggl\langle \rzb^T \bDb^T \rVbt_{\rxi_{-\rj}} \rfb \biggr\rangle_{q_1(\rfb) \; q_2(\rzb)}
+ \biggl\langle \| \rVbt_{\rxi_{-\rj}}^{\frac{1}{2}} \bDb \rzb \|^2 \biggr\rangle_{q_2(\rzb)}=
\\&
=
\| \rVbt_{\rxi_{-\rj}}^{\frac{1}{2}} \rfbh \|^2
+\Tr{\rVbt_{\rxi_{-\rj}}\Sigmabh_{f}}
-2\rzbh^T \bDb^T \rVbt_{\rxi_{-\rj}} \rfbh
+\| \rVbt_{\rxi_{-\rj}}^{\frac{1}{2}} \bDb \rzbh \|^2
+\Tr{\bDb^T \rVbt_{\rxi_{-\rj}} \bDb \Sigmabh_{z}}
=
\\&
=
C
+\rv_{\rxi_{\rj}}^{-1} \rfh_{\rj}^2 
+\rv_{\rxi_{\rj}}^{-1} \Sigmah_{f_{jj}}
-2\rv_{\rxi_{\rj}}^{-1} \rzbh^T \bDb_{\bj}^T \rfh_{\rj}
+\rv_{\rxi_{\rj}}^{-1} \| \bDb_{\bj} \rzbh \|^2
+\rv_{\rxi_{\rj}}^{-1} \sum_{k=1}^{M} \sum_{i=1}^{M} \bd_{\bj\bk} \bd_{\bj\bi} \Sigmah_{z_{ik}} 
\label{Eq:q3_Integral_2_IS}
\end{split}
\eeq
where $C$ denotes terms not containing $\rv_{\rxi_{\rj}}$ and $\bDb_{\bj}$ denotes the line $j$ of the matrix $\bDb$. From Equation~\eqref{Eq:q3_Integral_1_IS} and Equation~\eqref{Eq:q3_Integral_2_IS}: 
\beq
\begin{split}
\biggl\langle \ln p(\rfb, \rzb, \rvb_\repsilon, \rvb_\rxi, \rvb_\rz | \bgb, \alpha_{\epsilon}, \beta_{\epsilon}, \alpha_{\xi},& \beta_{\xi}, \alpha_{z}, \beta_{z}) \biggr\rangle_{q_1(\rfb) \; q_2(\rzb) \; q_{3-j}(\rvb_{\rxi_\rj})}
=
- \left( \alpha_{\xi} + \frac{3}{2} \right) \ln \rv_{\rxi_\rj} 
\\&
- \left( \beta_{\xi} + \frac{1}{2} \left[ \rfh_{\rj}^2 
+ \Sigmah_{f_{jj}} - \rzbh^T \bDb_{\bj}^T \rfh_{\rj}
+ \| \bDb_{\bj} \rzbh \|^2 + \sum_{k=1}^{M} \sum_{i=1}^{M} \bd_{\bj\bk} \bd_{\bj\bi} \Sigmah_{z_{ik}} \right] \right) \rv_{\rxi_\rj}^{-1}
\end{split}
\eeq
from which it can establish the proportionality corresponding to $q_{3j}(\rv_{\rxi_\rj})$:
\beq
q_{3j}(\rv_{\rxi_\rj})
\propto
\rv_{\rxi_\rj}^{-\left( \alpha_{\xi} + \frac{3}{2} \right)}
\exp
\left\lbrace 
-
\frac{\beta_{\xi} + \frac{1}{2} \left( \rfh_{\rj}^2 
+ \Sigmah_{f_{jj}} - 2 \rzbh^T \bDb_{\bj}^T \rfh_{\rj}
+ \| \bDb_{\bj} \rzbh \|^2 + \sum_{k=1}^{M} \sum_{i=1}^{M} \bd_{\bj\bk} \bd_{\bj\bi} \Sigmah_{z_{ik}} \right)}{\rv_{\rxi_\rj}}
\right\rbrace,
\label{Eq:q3_Integral_3_IS}
\eeq
leading to the following corollary:
\begin{corollary}
$q_{3j} \left( \rv_{\rxi_\rj} \right)$ is an Inverse Gamma distribution.
\label{Co:q3}
\end{corollary} 
\hspace{-14pt}The shape and scale parameters are obtained by identification, from Equation~\eqref{Eq:q3_Integral_3_IS}:
\beq
q_{3j}(\rv_{\rxi_\rj})=\Ic\Gc\left(\rv_{\rxi_\rj}|\alphah_{\xi_j},\betah_{\xi_j}\right),
\left\{\barr{ll}
\alphah_{\xi_j} = \alpha_{\xi} + \frac{3}{2}
\\[10pt]
\betah_{\xi_j} = \beta_{\xi} + \frac{1}{2} \left( \rfh_{\rj}^2 
+ \Sigmah_{f_{jj}} - 2\rzbh^T \bDb_{\bj}^T \rfh_{\rj}
+ \| \bDb_{\bj} \rzbh \|^2 + \sum_{k=1}^{M} \sum_{i=1}^{M} \bd_{\bj\bk} \bd_{\bj\bi} \Sigmah_{z_{ik}} \right)
\earr\right.
\label{Eq:q3_Analytical_1_IS}
\eeq
\paragraph{Computation of the analytical expression of $q_{4i}(\rv_{\repsilon_\ri})$.}
The proportionality relation corresponding to $q_{4i}(\rv_{\repsilon_\ri})$ is presented in established in Equation~\eqref{Eq:VBA_Proportionalities4_IS}. In the expression of $p \left( \rfb, \rzb, \rvb_\rxi, \rvb_\repsilon, \rvb_\rz | \bgb, \alpha_{\xi}, \beta_{\xi}, \alpha_{\epsilon}, \beta_{\epsilon}, \alpha_{z}, \beta_{z} \right)$ all the terms free of $\rv_{\repsilon_\ri}$ can be regarded as constants. Via Equation~\eqref{Eq:Log_Posterior_IS} the integral defined in Equation~\eqref{Eq:Def_Integral_IS} becomes:
\beq
\begin{split}
\biggl\langle \ln p(\rfb, \rzb, \rvb_\repsilon, \rvb_\rxi, \rvb_\rz | \bgb, \alpha_{\epsilon}, \beta_{\epsilon}, \alpha_{\xi}, \beta_{\xi}, \alpha_{z}, \beta_{z}) \biggr\rangle_{q_1(\rfb) \; q_{4-i}(\rvb_{\repsilon_\ri})}
=&
-
\frac{1}{2} \biggl\langle \| \rVb_{\repsilon}^{-\frac{1}{2}} \left( \bgb - \bHb\rfb \right) \|^2 \biggr\rangle_{q_1(\rfb) \; q_{4-i}(\rvb_{\repsilon_\ri})}
\\&
-
\left( \alpha_{\epsilon} + \frac{3}{2} \right)  \ln \rv_{\repsilon_\ri} 
-
\frac{\beta_{\epsilon}}{\rv_{\repsilon_\ri}} 
\end{split}
\label{Eq:q4_Integral_1_IS}
\eeq
Introducing the notations:
\beq
\rvbt_{\repsilon_{-\ri}}^{-1}=
\begin{bmatrix}
\rvt_{\repsilon_\red{1}}^{-1} \;
\ldots \;
\rvt_{\repsilon_{\ri-\red{1}}}^{-1} \;
\rv_{\repsilon_\ri}^{-1}  \;
\rvt_{\repsilon_{\ri+\red{1}}}^{-1}  \;
\ldots \;
\rvt_{\repsilon_\rN}^{-1}
\end{bmatrix}^T
\;\; ; \;\;
\rVbt_{\repsilon_{-\ri}}=
\mbox{diag}\left(\rvbt_{\repsilon_{-\ri}}^{-1} \right)
\label{Eq:q4_Integral_Not1_IS}
\eeq
the integral $\biggl\langle \| \rVb_{\repsilon}^{-\frac{1}{2}} \left( \bgb - \bHb\rfb \right) \|^2 \biggr\rangle_{q_1(\rfb) \; q_{4-i}(\rvb_{\repsilon_\ri})}$ can be written:
\beq
\biggl\langle \| \rVb_{\repsilon}^{-\frac{1}{2}} \left( \bgb - \bHb\rfb \right) \|^2 \biggr\rangle_{q_1(\rfb) \; q_{4-i}(\rvb_{\repsilon_\ri})}
= 
\biggl\langle \| \rVbt_{\repsilon_{-\ri}}^{\frac{1}{2}} \left( \bgb - \bHb\rfb \right) \|^2 \biggr\rangle_{q_1(\rfb)}
\eeq 
Considering that $q_1(\rfb)$ is a multivariate Normal distribution, Equations~\eqref{Eq:q1_Analytical_1_IS} we have:  
\beq
\begin{split}
\biggl\langle \| \rVbt_{\repsilon_{-\ri}}^{\frac{1}{2}} \left( \bgb - \bHb\rfb \right) \|^2 \biggr\rangle_{q_1(\rfb)}
&=
\| \rVbt_{\repsilon_{-\ri}}^{\frac{1}{2}} \left( \bgb - \bHb\rfbh \right) \|^2
+ \Tr{\bHb^T \rVbt_{\repsilon_{-\ri}} \bHb \Sigmabh_{f}}
\\&
=C +
\rv_{\repsilon_\ri}^{-1} \left( \bg_\bi - \bHb_\bi \rfbh \right)^2
+ 
\rv_{\repsilon_\ri}^{-1}\bHb_\bi \Sigmabh \bHb_\bi^T
\label{Eq:q4_Integral_2_IS}
\end{split}
\eeq
where $C$ denotes terms not containing $\rv_{\repsilon_{\ri}}$ and $\bHb_{\bi}$ denotes the line $i$ of the matrix $\bHb$. From Equation~\eqref{Eq:q4_Integral_1_IS} and Equation~\eqref{Eq:q4_Integral_2_IS}: 
\beq
\begin{split}
\biggl\langle \ln p(\rfb, \rzb, \rvb_\repsilon, \rvb_\rxi, \rvb_\rz | \bgb, \alpha_{\epsilon}, \beta_{\epsilon}, \alpha_{\xi}, \beta_{\xi}, \alpha_{z}, \beta_{z}) \biggr\rangle_{q_1(\rfb) \; q_{4-i}(\rvb_{\repsilon_\ri})}
=
C &- \left( \alpha_{\epsilon} + \frac{3}{2} \right) \ln \rv_{\repsilon_\ri} 
\\&
-\left( \beta_{\epsilon} + \frac{1}{2} \left[ 
\left( \bg_\bi - \bHb_\bi \rfbh \right)^2 + \bHb_\bi \Sigmabh_{f} \bHb_\bi^T \right] \right) \rv_{\repsilon_\ri}^{-1}
\end{split}
\eeq
from which it can establish the proportionality corresponding to $q_{4i}(\rv_{\repsilon_\ri})$:
\beq
q_{4i}(\rv_{\repsilon_\ri})
\propto
\rv_{\repsilon_\ri}^{-\left( \alpha_{\xi} + \frac{3}{2} \right)}
\exp
\left\lbrace 
-
\frac{ \beta_{\epsilon} + \frac{1}{2} \left[ 
\left( \bg_\bi - \bHb_\bi \rfbh \right)^2 + \bHb_\bi \Sigmabh_{f} \bHb_\bi^T \right] }{\rv_{\repsilon_\ri}}
\right\rbrace,
\label{Eq:q4_Integral_3_IS}
\eeq
leading to the following corollary:
\begin{corollary}
$q_{4i} \left( \rv_{\repsilon_\ri} \right)$ is an Inverse Gamma distribution.
\label{Co:q4}
\end{corollary} 
\hspace{-14pt}The shape and scale parameters are obtained by identification, from Equation~\eqref{Eq:q4_Integral_3_IS}:
\beq
q_{4i}(\rv_{\repsilon_\ri})=\Ic\Gc\left(\rv_{\repsilon_\ri}|\alphah_{\epsilon_i},\betah_{\epsilon_i}\right),
\left\{\barr{ll}
\alphah_{\epsilon_i} = \alpha_{\epsilon} + \frac{3}{2}
\\[10pt]
\betah_{\epsilon_i} = \beta_{\epsilon} + \frac{1}{2} \left[ 
\left( \bg_\bi - \bHb_\bi \rfbh \right)^2 + \bHb_\bi \Sigmabh_{f} \bHb_\bi^T \right]
\earr\right.
\label{Eq:q4_Analytical_1_IS}
\eeq
\paragraph{Computation of the analytical expression of $q_{5j}(\rv_{\rz_\rj})$.}
The proportionality relation corresponding to $q_{5j}(\rv_{\rz_\rj})$ is presented in established in Equation~\eqref{Eq:VBA_Proportionalities4_IS}. In the expression of $p \left( \rfb, \rzb, \rvb_\rxi, \rvb_\repsilon, \rvb_\rz | \bgb, \alpha_{\xi}, \beta_{\xi}, \alpha_{\epsilon}, \beta_{\epsilon}, \alpha_{z}, \beta_{z} \right)$ all the terms free of $\rv_{\rz_\rj}$ can be regarded as constants. Via Equation~\eqref{Eq:Log_Posterior_IS} the integral defined in Equation~\eqref{Eq:Def_Integral_IS} becomes:
\beq
\begin{split}
\biggl\langle \ln p(\rfb, \rzb, \rvb_\repsilon, \rvb_\rxi, \rvb_\rz | \bgb, \alpha_{\epsilon}, \beta_{\epsilon}, \alpha_{\xi}, \beta_{\xi}, \alpha_{z}, \beta_{z}) \biggr\rangle_{q_2(\rzb) \; q_{5-j}(\rvb_{\rz_\rj})}
=&
-
\frac{1}{2} \biggl\langle \| \rVb_{\rz}^{-\frac{1}{2}} \rzb \|^2 \biggr\rangle_{q_2(\rzb) \; q_{5-j}(\rvb_{\rz_\rj})}
\\&
-
\left(\alpha_{z} + \frac{3}{2} \right) \ln \rv_{\rz_\rj}
-
\frac{\beta_{f}}{\rv_{\rz_\rj}}
\end{split}
\label{Eq:q5_Integral_1_IS}
\eeq
Introducing the notations:
\beq
\rvbt_{\rz_{-\rj}}^{-1}=
\begin{bmatrix}
\rvt_{\rz_\red{1}}^{-1} \;
\ldots \;
\rvt_{\rz_{\rj-\red{1}}}^{-1} \;
\rv_{\rz_\rj}^{-1}  \;
\rvt_{\rz_{\rj+\red{1}}}^{-1}  \;
\ldots \;
\rvt_{\rz_\rM}^{-1}
\end{bmatrix}^T
\;\; ; \;\;
\rVbt_{\rz_{-\rj}}=
\mbox{diag}\left(\rvbt_{\rz_{-\rj}}^{-1} \right)
\label{Eq:q5_Integral_Not1_IS}
\eeq
the integral $\biggl\langle \| \rVb_{\rz}^{-\frac{1}{2}} \rzb \|^2 \biggr\rangle_{q_2(\rzb) \; q_{5-j}(\rvb_{\rz_\rj})}$ can be written:
\beq
\biggl\langle \| \rVb_{\rz}^{-\frac{1}{2}} \rzb \|^2\biggr\rangle_{q_2(\rzb) \; q_{5-j}(\rvb_{\rz_\rj})}
= 
\biggl\langle \| \rVbt_{\rz_{-\rj}}^{\frac{1}{2}} \rzb \|^2 \biggr\rangle_{q_2(\rzb)}
\eeq 
Considering that $q_2(\rzb)$ is a multivariate Normal distribution, Equations~\eqref{Eq:q2_Analytical_1_IS} we have:  
\beq
\begin{split}
\biggl\langle \| \rVbt_{\rz_{-\rj}}^{\frac{1}{2}} \rzb \|^2 \biggr\rangle_{q_2(\rzb)}
&=
\| \rVbt_{\rz_{-\rj}}^{\frac{1}{2}} \rzbh \|^2
+ 
\Tr{\rVbt_{\rz_{-\rj}} \Sigmabh_{z}}
=
C +
\rv_{\rz_\rj}^{-1} \rz_{\rj}^2
+ 
\rv_{\rz_\rj}^{-1} \Sigmah_{z_{jj}}
\label{Eq:q5_Integral_2_IS}
\end{split}
\eeq
From Equation~\eqref{Eq:q5_Integral_1_IS} and Equation~\eqref{Eq:q5_Integral_2_IS}: 
\beq
\begin{split}
\biggl\langle \ln p(\rfb, \rzb, \rvb_\repsilon, \rvb_\rxi, \rvb_\rz | \bgb, \alpha_{\epsilon}, \beta_{\epsilon}, \alpha_{\xi}, \beta_{\xi}, \alpha_{z}, \beta_{z}) \biggr\rangle_{q_2(\rzb) \; q_{5-j}(\rvb_{\rz_\rj})}
=
C &- \left( \alpha_{z} + \frac{3}{2} \right) \ln \rv_{\rz_\rj} 
\\&
-\left( \beta_{z} + \frac{1}{2} \left[ \rz_{\rj}^2 + \Sigmah_{z_{jj}} \right] \right) \rv_{\rz_\rj}^{-1}
\end{split}
\eeq
from which it can establish the proportionality corresponding to $q_{5j}(\rv_{\rz_\rj})$:
\beq
q_{5j}(\rv_{\rz_\rj})
\propto
\rv_{\rz_\rj}^{-\left( \alpha_{z} + \frac{3}{2} \right)}
\exp
\left\lbrace 
-
\frac{ \beta_{z} + \frac{1}{2} \left[ \rz_{\rj}^2 + \Sigmah_{z_{jj}} \right]}{\rv_{\rz_\rj}}
\right\rbrace,
\label{Eq:q5_Integral_3_IS}
\eeq
leading to the following corollary:
\begin{corollary}
$q_{5j} \left( \rv_{\rz_\rj} \right)$ is an Inverse Gamma distribution.
\label{Co:q5}
\hspace{-14pt}\end{corollary} 
The shape and scale parameters are obtained by identification, from Equation~\eqref{Eq:q5_Integral_3_IS}:
\beq
q_{5j}(\rv_{\rz_\rj})=\Ic\Gc\left(\rv_{\repsilon_\ri}|\alphah_{z_j},\betah_{z_j}\right),
\left\{\barr{ll}
\alphah_{z_j} = \alpha_{z} + \frac{3}{2}
\\[10pt]
\betah_{z_j} = \beta_{z} + \frac{1}{2} \left[ \rz_{\rj}^2 + \Sigmah_{z_{jj}} \right].
\earr\right.
\label{Eq:q5_Analytical_1_IS}
\eeq
Equations~\eqref{Eq:q1_Analytical_1_IS},~\eqref{Eq:q2_Analytical_1_IS},~\eqref{Eq:q3_Analytical_1_IS},~\eqref{Eq:q4_Analytical_1_IS} and~\eqref{Eq:q5_Analytical_1_IS} resume the distributions families and the corresponding parameters for $q_1(\rfb)$, $q_2(\rzb)$ ( multivariate Normal distribution), $q_{3j}(\rv_{\rxi_\rj})$, $j\in\left\lbrace 1, 2, \ldots, M \right\rbrace$, $q_{4i}(\rv_{\repsilon_\ri})$, $i\in\left\lbrace 1, 2, \ldots, N \right\rbrace$ and $q_{5j}(\rv_{\rz_\rj})$, $j\in\left\lbrace 1, 2, \ldots, M \right\rbrace$ (Inverse Gamma distributions). However, the parameters corresponding to the multivariate Normal distributions are expressed via $\rVbt_{\rxi}$, $\rVbt_{\repsilon}$ and $\rVbt_{\rz}$ (and by extension all elements forming the three matrices $ \rvt_{\rxi_\rj}$, $j \in \left\lbrace 1, 2, \ldots, M \right\rbrace $, $ \rvt_{\repsilon_\ri}$, $i \in \left\lbrace 1, 2, \ldots, N \right\rbrace $, both defined in Equation~\eqref{Eq:q1_Integral_Not1_IS} and $ \rvt_{\rz_\rj}$, $j \in \left\lbrace 1, 2, \ldots, M \right\rbrace $, defined in Equation~\eqref{Eq:q2_Integral_Not1_IS}.
\paragraph{Computation of the analytical expressions of $\rVbt_{\rxi}$, $\rVbt_{\repsilon}$ and $\rVbt_{\rz}$.}
For an Inverse Gamma distribution with scale and shape parameters $\alpha$ and $\beta$, $\Ic\Gc\left(x|\alpha, \beta \right)$, the following relation holds:
\beq
\biggl\langle x^{-1} \biggr\rangle_{\Ic\Gc(x|\alpha,\beta)} 
=
\frac{\alpha}{\beta}
\label{Eq:Inverse_Gamma_Integral_IS}
\eeq
The prove of the above relation is done by direct computation, using the analytical expression of the Inverse Gamma Distribution: 
\beq
\begin{split}
\biggl\langle x^{-1} \biggr\rangle_{\Ic\Gc(x|\alpha,\beta)} 
&= 
\int 
x^{-1}
\frac{{\beta}^{\alpha}}{\Gamma(\alpha)}
x^{-\alpha-1}
\exp
\left\lbrace
- \frac{\beta}{x}
\right\rbrace
\d x
=
\frac{{\beta}^{\alpha}}{\Gamma(\alpha)}
\frac{\Gamma(\alpha+1)}{{\beta}^{\alpha+1}}
\int
\frac{{\beta}^{\alpha+1}}{\Gamma(\alpha+1)}
x^{-(\alpha+1)-1}
\exp\left\lbrace -\frac{\beta}{x}\right\rbrace
\d x
=\\
&=
\frac{\alpha}{\beta}
\underbrace{
\int \Ic\Gc(x|\alpha+1,\beta)
}_{1}
\d x
=
\frac{\alpha}{\beta}
\end{split}
\nonumber
\eeq
Since $q_{3j}(\rv_{\rxi_\rj})$, $j\in\left\lbrace 1, 2, \ldots, M \right\rbrace$, $q_{4i}(\rv_{\repsilon_\ri})$, $i\in\left\lbrace 1, 2, \ldots, N \right\rbrace$ and $q_{5j}(\rv_{\rz_\rj})$, $j\in\left\lbrace 1, 2, \ldots, M \right\rbrace$ are Inverse Gamma distributions, with the corresponding parameters $\alphah_{\xi_j}$ and $\betah_{\xi_j}$, $j\in\left\lbrace 1, 2, \ldots, M \right\rbrace$, $\alphah_{\epsilon_j}$ and $\betah_{\epsilon_j}$, $j\in\left\lbrace 1, 2, \ldots, N \right\rbrace$ respectively $\alphah_{z_j}$ and $\betah_{z_j}$, $j\in\left\lbrace 1, 2, \ldots, M \right\rbrace$ the expectancies $\rvt_{\rxi_\rj}$, $\rvt_{\repsilon_\ri}$ and $\rvt_{\rz_\rj}$ can be expressed via the parameters of the two Inverse Gamma distributions using Equation~\eqref{Eq:Inverse_Gamma_Integral_IS}:
\beq
\rvt_{\rxi_\rj}
=
\frac{\alphah_{\xi_j}}{\betah_{\xi_j}}
\;\;\;;\;\;\;
\rvt_{\repsilon_\ri}
=
\frac{\alphah_{\epsilon_i}}{\betah_{\epsilon_i}}
\;\;\;;\;\;\;
\rvt_{\rz_\rj}
=
\frac{\alphah_{z_j}}{\betah_{z_j}}
\label{Eq:VepsVfExpectanciesIGSMGen_IS}
\eeq
Using the notation introduced in \eqref{Eq:q1_Integral_Not1_IS}:
\beq
\rVbt_{\rxi}=
\begin{bmatrix}
\frac{\alphah_{\xi_1}}{\betah_{\xi_1}} \ldots 0 \ldots 0 \\
\vdots \ddots \vdots \ddots \vdots \\
0 \ldots \frac{\alphah_{\xi_j}}{\betah_{\xi_j}} \ldots 0 \\
\vdots \ddots \vdots \ddots \vdots \\
0 \ldots 0 \ldots \frac{\alphah_{\xi_M}}{\betah_{\xi_M}} \\
\end{bmatrix}
=
\rVbh_{\rf}
\;\;;\;\;
\rVbt_{\repsilon}=
\begin{bmatrix}
\frac{\alphah_{\epsilon_1}}{\betah_{\epsilon_1}} \ldots 0 \ldots 0 \\
\vdots \ddots \vdots \ddots \vdots \\
0 \ldots \frac{\alphah_{\epsilon_i}}{\betah_{\epsilon_i}} \ldots 0 \\
\vdots \ddots \vdots \ddots \vdots \\
0 \ldots 0 \ldots \frac{\alphah_{\epsilon_N}}{\betah_{\epsilon_N}} \\
\end{bmatrix}
=
\rVbh_{\repsilon}
\;\;;\;\;
\rVbt_{\rz}=
\begin{bmatrix}
\frac{\alphah_{z_1}}{\betah_{z_1}} \ldots 0 \ldots 0 \\
\vdots \ddots \vdots \ddots \vdots \\
0 \ldots \frac{\alphah_{z_j}}{\betah_{z_j}} \ldots 0 \\
\vdots \ddots \vdots \ddots \vdots \\
0 \ldots 0 \ldots \frac{\alphah_{z_M}}{\betah_{z_M}} \\
\end{bmatrix}
=
\rVbh_{\rz}
\label{Eq:V_Expectancies_IS}
\eeq
In Equation~\eqref{Eq:V_Expectancies_IS} other notations are introduced for $\rVbt_{\rxi}$,  $\rVbt_{\repsilon}$ and $\rVbt_{\rz}$. Both values were expressed during the model via unknown expectancies, but via Equation~\eqref{Eq:V_Expectancies_IS} those values don't contain any more integrals to be computed. Therefore, the new notations represent the final analytical expressions used for expressing the density functions $q_i$.
Using Equation~\eqref{Eq:V_Expectancies_IS} and Equations~\eqref{Eq:q1_Analytical_1_IS},~\eqref{Eq:q2_Analytical_1_IS},~\eqref{Eq:q3_Analytical_1_IS},~
\eqref{Eq:q4_Analytical_1_IS}, and~\eqref{Eq:q5_Analytical_1_IS}, the final analytical expressions of the separable distributions $q_i$ are presented in Equations~\eqref{Eq:q1_Analytical_2_IS},~\eqref{Eq:q2_Analytical_2_IS},~\eqref{Eq:q3_Analytical_2_IS},~\eqref{Eq:q4_Analytical_2_IS} and~\eqref{Eq:q5_Analytical_2_IS}.
\begin{subequations}
\begin{align}
&q_1(\rfb) = \Nc\left( \rfb | \rfbh, \Sigmabh_{f} \right), \;\;\;\;\;\;\;\;\;
\left\{\barr{ll}
\rfbh = \left( \bHb^T \rVbh_{\repsilon} \bHb + \rVbh_{\rxi} \right)^{-1} \left( \rVbh_{\rxi}^T \bDb \rzbh + \bHb^T \rVbh_{\repsilon} \bgb \right),
\\[10pt]
\Sigmabh_{f} = \left( \bHb^T \rVbh_{\repsilon} \bHb + \rVbh_{\rxi} \right)^{-1}.
\earr\right.,
\label{Eq:q1_Analytical_2_IS}
\\[4pt]
&q_2(\rzb) = \Nc\left( \rzb | \rzbh, \Sigmabh_{z} \right), \;\;\;\;\;\;\;\;\;\;
\left\{\barr{ll}
\rzbh = \left( \bDb^T \rVbh_{\rxi} \bDb + \rVbh_{\rz} \right)^{-1} \bDb^T \rVbh_{\rxi} \rfbh,
\\[10pt]
\Sigmabh_{z} = \left( \bDb^T \rVbh_{\rxi} \bDb + \rVbh_{\rz} \right)^{-1}.
\earr\right.,
\label{Eq:q2_Analytical_2_IS}
\\[4pt]
&q_{3j}(\rv_{\rxi_\rj})=\Ic\Gc\left(\rv_{\rxi_\rj}|\alphah_{\xi_j},\betah_{\xi_j}\right),
\left\{\barr{ll}
\alphah_{\xi_j} = \alpha_{\xi} + \frac{3}{2}
\\[10pt]
\betah_{\xi_j} = \beta_{\xi} + \frac{1}{2} \left[ \left( \rfh_{\rj} - \bDb_{\bj} \rzbh \right)^2  + \bDb_\bj \Sigmabh_{z} \bDb_\bj^T \right]
\earr\right., j\in\left\lbrace 1, 2, \ldots, M \right\rbrace,
\label{Eq:q3_Analytical_2_IS}
\\[4pt]
&q_{4i}(\rv_{\repsilon_\ri})=\Ic\Gc\left(\rv_{\repsilon_\ri}|\alphah_{\epsilon_i},\betah_{\epsilon_i}\right), \;\;
\left\{\barr{ll}
\alphah_{\epsilon_i} = \alpha_{\epsilon} + \frac{3}{2}
\\[10pt]
\betah_{\epsilon_i} = \beta_{\epsilon} + \frac{1}{2} \left[ 
\left( \bg_\bi - \bHb_\bi \rfbh \right)^2 + \bHb_\bi \Sigmabh_{f} \bHb_\bi^T \right]
\earr\right.,i\in\left\lbrace 1, 2, \ldots, N \right\rbrace,
\label{Eq:q4_Analytical_2_IS}
\\[4pt]
&q_{5j}(\rv_{\rz_\rj})=\Ic\Gc\left(\rv_{\repsilon_\ri}|\alphah_{z_j},\betah_{z_j}\right),
\left\{\barr{ll}
\alphah_{z_j} = \alpha_{z} + \frac{3}{2}
\\[10pt]
\betah_{z_j} = \beta_{z} + \frac{1}{2} \left[ \rz_{\rj}^2 + \Sigmah_{z_{jj}} \right].
\earr\right.,j\in\left\lbrace 1, 2, \ldots, M \right\rbrace.
\label{Eq:q5_Analytical_2_IS}
\end{align}
\end{subequations}
Equation~\eqref{Eq:q1_Analytical_2_IS} establishes the dependency between the parameters corresponding to the multivariate Normal distribution $q_1(\rfb)$ and the others parameters involved in the hierarchical model: the mean $\rfbh$ and the covariance matrix $\Sigmabh_{f}$ depend on $\rVbh_{\repsilon}$, $\rVbh_{\rxi}$ which, via Equation~\eqref{Eq:V_Expectancies_IS} are defined using $\left\lbrace \alphah_{\xi_j},\betah_{\xi_j}\right\rbrace, j\in \left\lbrace 1, 2, \ldots, M \right\rbrace $ and $\left\lbrace \alphah_{\epsilon_i},\betah_{\epsilon_i}\right\rbrace, i\in \left\lbrace 1, 2, \ldots, N \right\rbrace $ and $\rzbh$. The dependency between the parameters of the multivariate Normal distribution $q_1(\rfb)$ and the parameters of the Inverse Gamma distributions $q_{3j}(\rvb_{\rxi_\rj}) ,j \in \left\lbrace 1,2,\ldots,M \right\rbrace$ and $q_{4i}(\rvb_{\repsilon_\ri}), i \in \left\lbrace 1,2,\ldots,N \right\rbrace$ and the multivariate Normal distribution $q_2(\rzb)$ is presented in Figure~\eqref{Fig:Dependency_Scheme_1_IS}.
\begin{figure}[!htb]
\center
\begin{tabular}{c}
\begin{picture}(220,30)
\put(0,0){\framebox(140,26){$\rzbh, \left\lbrace \alphah_{\xi_j},\betah_{\xi_j}\right\rbrace,\left\lbrace \alphah_{\epsilon_i},\betah_{\epsilon_i}\right\rbrace$}}
\put(142,13){\vector(1,0){26}}
\put(168,0){\framebox(56,26){$\rfbh \; , \; \Sigmabh_{f}$}}
\end{picture}
\end{tabular}
\caption{Dependency between $q_1(\rfb)$ parameters and $q_{2}(\rzb)$, $q_{3j}(\rvb_{\rxi_\rj})$ and $q_{4i}(\rvb_{\repsilon_\ri})$ parameters}
\label{Fig:Dependency_Scheme_1_IS}
\end{figure}
Equation~\eqref{Eq:q2_Analytical_2_IS} establishes the dependency between the parameters corresponding to the multivariate Normal distribution $q_2(\rzb)$ and the others parameters involved in the hierarchical model: the mean $\rzbh$ and the covariance matrix $\Sigmabh_{z}$ depend on $\rVbh_{\rxi}$, $\rVbh_{\rz}$ which, via Equation~\eqref{Eq:V_Expectancies_IS} are defined using $\left\lbrace \alphah_{\epsilon_i},\betah_{\epsilon_i}\right\rbrace, i \in \left\lbrace 1, 2, \ldots, N \right\rbrace $ and $\left\lbrace \alphah_{z_j},\betah_{z_j}\right\rbrace, j \in \left\lbrace 1, 2, \ldots, M \right\rbrace $ and $\rfbh$. The dependency between the parameters of the multivariate Normal distribution $q_2(\rzb)$ and the parameters of the Inverse Gamma distributions $q_{4i}(\rvb_{\repsilon_\ri}), i \in \left\lbrace 1,2,\ldots,N \right\rbrace$ and $q_{5j}(\rvb_{\rz_\rj}) ,j \in \left\lbrace 1,2,\ldots,M \right\rbrace$ and the multivariate Normal distribution $q_1(\rfb)$ is presented in Figure~\eqref{Fig:Dependency_Scheme_2_IS}.
\begin{figure}[!htb]
\center
\begin{tabular}{c}
\begin{picture}(220,30)
\put(0,0){\framebox(140,26){$\rfbh, \left\lbrace \alphah_{\epsilon_i},\betah_{\epsilon_i}\right\rbrace,\left\lbrace \alphah_{z_j},\betah_{z_j}\right\rbrace$}}
\put(142,13){\vector(1,0){26}}
\put(168,0){\framebox(56,26){$\rzbh \; , \; \Sigmabh_{z}$}}
\end{picture}
\end{tabular}
\caption{Dependency between $q_2(\rzb)$ parameters and $q_{1}(\rfb)$, $q_{4i}(\rvb_{\repsilon_\ri})$ and $q_{5j}(\rvb_{\rz_\rj})$ parameters}
\label{Fig:Dependency_Scheme_2_IS}
\end{figure}
Equation~\eqref{Eq:q3_Analytical_2_IS} establishes the dependency between the parameters corresponding to the Inverse Gamma distributions $q_{3j}(\rvb_{\rxi_\rj}) ,j \in \left\lbrace 1, 2, \ldots, M \right\rbrace$ and the others parameters involved in the hierarchical model: the shape and scale parameters $\left\lbrace \alphah_{\xi_j}, \betah_{\xi_j} \right\rbrace, j \in \left\lbrace 1, 2, \ldots, M \right\rbrace$ depend on the element $j$ of the mean $\rfbh$ of the multivariate Normal distribution $q_1(\rfb)$, i.e. $\rfh_{\rj}$ and the mean $\rzbh$ and the covariance matrix $\Sigmabh_{z}$ of the multivariate Normal distribution $q_2(\rzb)$, Figure~\eqref{Fig:Dependency_Scheme_3_IS}.
\begin{figure}[!htb]
\center
\begin{tabular}{c}
\begin{picture}(120,30)
\put(0,0){\framebox(46,26){$\rfh_{\rj}, \; \rzbh, \; \Sigmabh_{z}$}}
\put(48,13){\vector(1,0){26}}
\put(76,0){\framebox(56,26){$\left\lbrace \alphah_{\xi_j},\betah_{\xi_j}\right\rbrace $}}
\end{picture}
\end{tabular}
\caption{Dependency between $q_{3j}(\rvb_{\rxi_\rj})$ parameters and $q_{1}(\rfb)$ and $q_2(\rzb)$ parameters}
\label{Fig:Dependency_Scheme_3_IS}
\end{figure}
Equation~\eqref{Eq:q4_Analytical_2_IS} establishes the dependency between the parameters corresponding to the Inverse Gamma distributions $q_{4i}(\rvb_{\repsilon_\ri}) ,i \in \left\lbrace 1,2,\ldots,N \right\rbrace$ and the others parameters involved in the hierarchical model: the shape and scale parameters $\left\lbrace \alphah_{\epsilon_i}, \betah_{\epsilon_i} \right\rbrace, i \in \left\lbrace 1, 2, \ldots, N \right\rbrace$ depend on the mean $\rfbh$ and the covariance matrix $\Sigmabh_{f}$ of the multivariate Normal distribution $q_1(\rfb)$, Figure~\eqref{Fig:Dependency_Scheme_4_IS}.
\begin{figure}[!htb]
\center
\begin{tabular}{c}
\begin{picture}(120,30)
\put(0,0){\framebox(46,26){$ \rfbh \; , \; \Sigmabh_{f}$}}
\put(48,13){\vector(1,0){26}}
\put(76,0){\framebox(56,26){$\left\lbrace \alphah_{\epsilon_i},\betah_{\epsilon_i}\right\rbrace$}}
\end{picture}
\end{tabular}
\caption{Dependency between $q_{4i}(\rvb_{\repsilon_\ri})$ parameters and $q_1(\rfb)$ parameters}
\label{Fig:Dependency_Scheme_4_IS}
\end{figure}
Equation~\eqref{Eq:q5_Analytical_2_IS} establishes the dependency between the parameters corresponding to the Inverse Gamma distributions $q_{5j}(\rvb_{\rz_\rj}) ,j \in \left\lbrace 1,2,\ldots,M \right\rbrace$ and the others parameters involved in the hierarchical model: the shape and scale parameters $\left\lbrace \alphah_{z_j}, \betah_{z_j} \right\rbrace, j \in \left\lbrace 1, 2, \ldots, M \right\rbrace$ depend on the element $j$ of the mean $\rzbh$ and the element $jj$ of the covariance matrix $\Sigmabh_{z}$ corresponding to the multivariate Normal distribution $q_2(\rzb)$, Figure~\eqref{Fig:Dependency_Scheme_5_IS}.
\begin{figure}[!htb]
\center
\begin{tabular}{c}
\begin{picture}(120,30)
\put(0,0){\framebox(46,26){$ \rzh_{\rj} \; , \; \Sigmah_{z_{jj}}$}}
\put(48,13){\vector(1,0){26}}
\put(76,0){\framebox(56,26){$\left\lbrace \alphah_{z_j},\betah_{z_j}\right\rbrace$}}
\end{picture}
\end{tabular}
\caption{Dependency between $q_{5j}(\rvb_{\rz_\rj})$ parameters and $q_2(\rzb)$ parameters}
\label{Fig:Dependency_Scheme_5_IS}
\end{figure}
\newline
The iterative algorithm obtained via PM estimation is presented Figure~\eqref{Fig:IA_PM_St_nsStL_un_IS}.
\tikzstyle{cloud} = [draw=blue!30!green,fill=orange!12, ellipse, thick, node distance=7em, text width=12em, text centered, minimum height=4em, minimum width=12em]
\tikzstyle{boxbig} = [draw=blue!30!green, fill=orange!12, rectangle, rounded corners, thick, node distance=8.5em, text width=26em, text centered, minimum height=3.6em, minimum width=26em]
\tikzstyle{boxhuge} = [draw=blue!30!green, fill=orange!12, rectangle, rounded corners, thick, node distance=12.6em, text width=29em, text centered, minimum height=3.2em, minimum width=29em]
\tikzstyle{boxsmall} = [draw=blue!30!green, fill=orange!12, rectangle, rounded corners, thick, node distance=9em, text width=9em, text centered, minimum height=4em, minimum width=9em]
\tikzstyle{line} = [draw=blue!30!green, -latex']
\begin{figure}[!htb]
\centering
\begin{center}
\begin{tikzpicture}[auto]
    \node [boxhuge, scale=0.9] (f) {\Large{$\rfbh = \left( \bHb^T \rVbh_{\repsilon} \bHb + \rVbh_{\rxi} \right)^{-1} \left( \rVbh_{\rxi}^T \bDb \rzbh + \bHb^T \rVbh_{\repsilon} \bgb \right)$}
\\[4.4pt]
\Large{$\Sigmabh_{f} = \left( \bHb^T \rVbh_{\repsilon} \bHb + \rVbh_{\rxi} \right)^{-1}$}    
\\[4.4pt]
\normalsize{(a) - update estimated $\rfb$ and covariance matrix $\Sigmabh_{f}$} 
};

	\node [boxbig, below of=f, scale=0.9] (z) {\Large{$\rzbh = \left( \bDb^T \rVbh_{\rxi} \bDb + \rVbh_{\rz} \right)^{-1} \bDb^T \rVbh_{\rxi} \rfbh$}
\\[4.4pt]
\Large{$\Sigmabh_{z} = \left( \bDb^T \rVbh_{\rxi} \bDb + \rVbh_{\rz} \right)^{-1}
$}    
\\[4.4pt]
\normalsize{(b) - update estimated $\rzb$ and covariance matrix $\Sigmabh_{z}$} 
};

    \node [boxbig, below of=z, node distance=11.6em, scale=0.9] (vxi)
{\Large{$ \alphah_{\xi_j} = \alpha_{\xi} + \frac{3}{2} $}
\\[5.1pt]
\Large{$\betah_{\xi_j} = \beta_{\xi} + \frac{1}{2} \left[ \left( \rfh_{\rj} - \bDb_{\bj} \rzbh \right)^2  + \bDb_\bj \Sigmabh_{z} \bDb_\bj^T \right]$}
\\[5.1pt]
\normalsize{(c) - update estimated $\Ic\Gc$ parameters corresponding to $\rv_{\rxi_\rj}$} 
};

    \node [boxbig, below of=vxi, node distance=11.6em, scale=0.9] (veps) {\Large{$\alphah_{\epsilon_i} = \alpha_{\epsilon} + \frac{3}{2}$}
\\[5.1pt]
\Large{$\betah_{\epsilon_i} = \beta_{\epsilon} + \frac{1}{2} \left[ 
\left( \bg_\bi - \bHb_\bi \rfbh \right)^2 + \bHb_\bi \Sigmabh_{f} \bHb_\bi^T \right]$}
\\[5.1pt]
\normalsize{(d) - update estimated $\Ic\Gc$ parameters corresponding to $\rv_{\repsilon_\ri}$}     
};

    \node [boxbig, below of=veps, node distance=9em, scale=0.9] (vz) {\Large{$\alphah_{z_j} = \alpha_{z} + \frac{3}{2}$}
\\[5.1pt]
\Large{$\betah_{z_j} = \beta_{z} + \frac{1}{2} \left[ \rz_{\rj}^2 + \Sigmah_{z_{jj}} \right]$}
\\[5.1pt]
\normalsize{(e) - update estimated $\Ic\Gc$ parameters corresponding to $\rv_{\rz_\rj}$}     
};

    \node [boxsmall, above right = 0.095cm and -0.7cm of vxi, node distance=17em, scale=0.9] (Vxi) {\Large{$\rVbh_\rxi = \diag {\rvbh_{\rxi}}$}
\\[5pt]
\normalsize{(f)}
};

    \node [boxsmall, below left = -3.95cm and -0.7cm of veps, node distance=17em, scale=0.9] (Veps) {\Large{$\rVbh_\repsilon = \diag {\rvbh_{\repsilon}}$}
\\[5pt]
\normalsize{(g)} 
};

    \node [boxsmall, above right = -0.08cm and 0.86cm of vz, node distance=17em, scale=0.9] (Vz) {\Large{$\rVbh_\rz = \diag {\rvbh_{\rz}}$}
\\[5pt]
\normalsize{(h)} 
};

\node [cloud, above of=f, scale=0.9] (init)
{\Large{Initialization}};  

    \path [line] (f) -- (z)[near start];
    \path [line] (z) -- (vxi);
    \path [line] (vxi) -- (veps);
    \path [line] (veps) -- (vz);    
    \path [line] (Vxi) |- (f);
    \path [line] (Vxi) |- (z);
    \path [line] (Veps) |- (f);
    \path [line] (Vz) |- (z);    
    \path [line] (vxi) -| (Vxi);
    \path [line] (veps) -| (Veps);
    \path [line] (vz) -| (Vz);
    \path [line] (z) -| (f);
    \path [line] (init) -- (f);   
\end{tikzpicture}
\end{center}
\caption{Indirect sparsity (via $\rzb$) - Iterative algorithm corresponding to partial separability PM estimation for Student-t hierarchical model, non-stationary Student-t uncertainties model}
\label{Fig:IA_PM_St_nsStL_un_IS}
\end{figure}
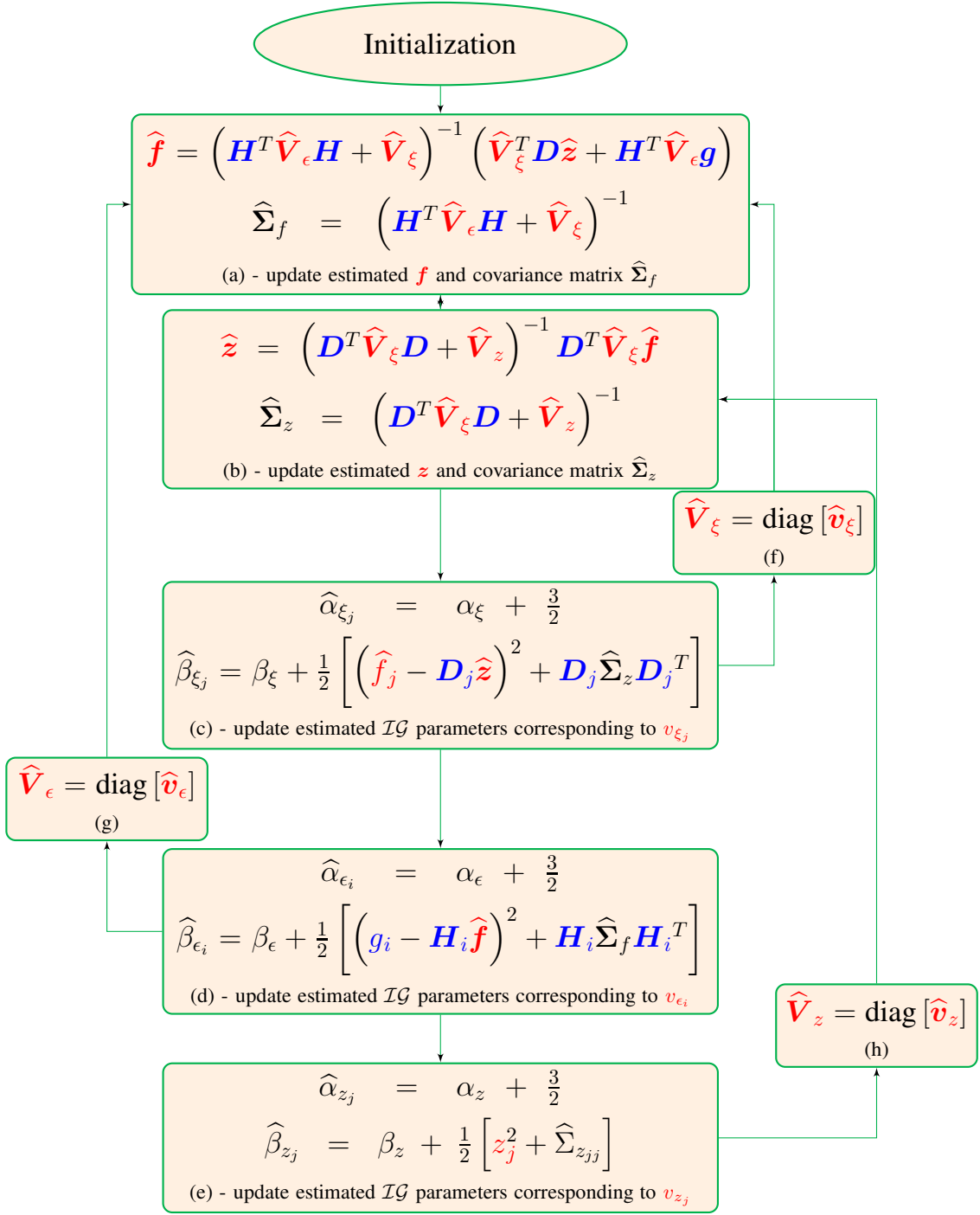
More formal, the iterative algorithm obtained via PM estimation is presented Algorithm~\eqref{Alg:IA_PM_St_nsStL_un_IS}.
\begin{algorithm}
\caption{PM via VBA partial sep. - Student-t hierarchical model, non-stationary Student-t uncertainties model}
\begin{algorithmic}[1]
\Ensure INITIALIZATION $\alphah_{\xi_j}^{(0)},\betah_{\xi_j}^{(0)},\alphah_{\epsilon_i}^{(0)},\betah_{\epsilon_i}^{(0)},\alpha_{z_j}^{(0)},\beta_{z_j}^{(0)}$ 

\Comment{Where $\alphah_{\xi_j}^{(0)}=\alpha_{\xi},
\betah_{\xi_j}^{(0)}=\beta_{\xi},
\alphah_{\epsilon_i}^{(0)}=\alpha_{\epsilon},
\betah_{\epsilon_i}^{(0)}=\beta_{\epsilon},
\alpha_{z_j}^{(0)}=\alpha_{z},
\beta_{z_j}^{(0)}=\beta_{z}$}

\Comment{Leads to $\rVbh_\rxi^{(0)} = \frac{\alpha_{\xi}}{\beta_{\xi}}\Ib,
\rVbh_\repsilon^{(0)} = \frac{\alpha_{\epsilon}}{\beta_{\epsilon}}\Ib,
\rVbh_\rz^{(0)} = \frac{\alpha_{z}}{\beta_{z}} \Ib$}

\Comment{$\alpha_{\xi}, \beta_{\xi},\alpha_{\epsilon}, \beta_{\epsilon}, \alpha_{z}, \beta_{z}$ are set in modeling phase, Eq.~\eqref{Eq:St_nsStL_un_IS}}

\Function{PMviaVBAps}{$\alpha_{\xi},\beta_{\xi},\alpha_{\epsilon},\beta_{\epsilon},\alpha_{z},\beta_{z},\bgb,\bHb, \bDb, \rzb^{(0)}, M, N, NoIter$} 

\For{$n = 0$ to ${IterNumb}$}

\State $\Sigmabh_{f}^{(n)} = \left( \bHb^T \diag {\frac{\alpha_{\epsilon}+\frac{3}{2}}{\betah_{\epsilon_i}^{(n)}}} \bHb + \diag{\frac{\alpha_{\xi}+\frac{3}{2}}{\betah_{\xi_j}^{(n+1)}}} \right)^{-1}$ \Comment{New value $\Sigmabh_{f}^{(n)}$}

\State $\rfbh^{(n)} = \Sigmabh_{f}^{(n)} \left( \diag{\frac{\alpha_{\xi}+\frac{3}{2}}{\betah_{\xi_j}^{(n+1)}}}
^T \bDb \rzbh^{(n)} + \bHb^T \diag {\frac{\alpha_{\epsilon}+\frac{3}{2}}{\betah_{\epsilon_i}^{(n)}}} \bgb \right)$ \Comment{New value $\rfbh^{(n)}$}

\For{$j = 1$ to ${M}$}

\State $\betah_{\xi_j}^{(n+1)} = \beta_{\xi} + \frac{1}{2} \left[ \left( \rfh_{\rj}^{(n)} - \bDb_{\bj} \rzbh^{(n+1)} \right)^2  + \bDb_\bj \Sigmabh_{z}^{(n+1)} \bDb_{\bj}^T \right]$

\EndFor


\For{$i = 1$ to ${N}$}

\State $\betah_{\epsilon_i}^{(n+1)} = \beta_{\epsilon} + \frac{1}{2} \left[ \left( \bg_\bi - \bHb_\bi \rfbh^{(n)} \right)^2 + \bHb_\bi \Sigmabh_{f}^{(n+1)} \bHb_\bi^T \right]$ 

\EndFor


\For{$j = 1$ to ${M}$}

\State $\betah_{z_j}^{(n+1)} = \beta_{z} + \frac{1}{2} \left[ \left( \rz_{\rj}^{(n+1)} \right)^2 + \Sigmah_{z_{jj}}^{(n+1)} \right]$

\EndFor


\State $\Sigmabh_{z}^{(n+1)} = \left( \bDb^T \diag{\frac{\alpha_{\xi}+\frac{3}{2}}{\betah_{\xi_j}^{(n+1)}}} \bDb + \diag{\frac{\alpha_{z}+\frac{3}{2}}{\betah_{z_j}^{(n+1)}}} \right)^{-1}$ \Comment{New value $\Sigmabh_{z}^{(n+1)}$}

\State $\rzbh^{(n+1)} = \Sigmabh_{z}^{(n+1)} \bDb^T \diag{\frac{\alpha_{\xi}+\frac{3}{2}}{\betah_{\xi_j}^{(n+1)}}} \rfbh^{(n)}$ \Comment{New value $\rzbh^{(n+1)}$}

\EndFor

\Return $\left( \rfbh^{(n)}, \Sigmabh_{f}^{(n)}, \rzbh^{(n+1)}, \Sigmabh_{z}^{(n+1)}, \betah_{\xi_j}^{(n+1)}, \betah_{\epsilon_i}^{(n+1)}, \betah_{z_j}^{(n+1)} \right)$ $n = NoIter$
\EndFunction
\end{algorithmic}
\label{Alg:IA_PM_St_nsStL_un_IS}
\end{algorithm}
\newpage
\subsubsection{Posterior Mean estimation via VBA, full separability}
\label{Subsubsec:PM_FS_St_nsStL_un_IS}
In this subsection, the Posterior Mean (PM) estimation is again considered, but via a full separable approximation. The posterior distribution is approximated by a full separable distribution $q\left( \rfb, \rzb,  \rvb_{\rxi}, \rvb_{\repsilon}, \rvb_{\rz} \right)$, i.e. a supplementary condition is added in Equation~\eqref{Eq:Posterior_Approximation_Notations1_IS}:
\beq
q_{1}(\rfb) = \prod_{j=1}^{M} q_{1j}(\rf_{\rj})
\;;\;
q_{2}(\rzb) = \prod_{j=1}^{M} q_{2j}(\rz_{\rj})
\;;\;
q_{3}(\rvb_{\rxi}) = \prod_{j=1}^{M} q_{3j}(\rv_{\rxi_\rj})
\;;\;
q_{4}(\rvb_{\repsilon}) = \prod_{i=1}^{N} q_{4i}(\rv_{\repsilon_\ri})
\;;\;
q_{5}(\rvb_{\rz}) = \prod_{j=1}^{M} q_{5j}(\rv_{\rz_\rj})
\label{Eq:Posterior_Approximation_Notations1bis_IS}
\eeq
As in Subsection~\eqref{Subsubsec:PM_PS_St_nsStL_un_IS}, the approximation is done by minimizing of the Kullback-Leibler divergence, Equation~\eqref{Eq:Kullback-Leibler_IS}, via alternate optimization resulting the following proportionalities from Equations~\eqref{Eq:VBA_Proportionalities1bis_IS},~\eqref{Eq:VBA_Proportionalities2bis_IS},~\eqref{Eq:VBA_Proportionalities3bis_IS},~\eqref{Eq:VBA_Proportionalities4bis_IS} and ~\eqref{Eq:VBA_Proportionalities5bis_IS},  
\begin{subequations}
\begin{align}
&q_{1j}(\rf_{\rj}) \;\;\; \propto \; \exp \biggl\lbrace \left\langle \ln p(\rfb, \rzb, \rvb_\repsilon, \rvb_\rxi, \rvb_\rz | \bgb, \alpha_{\epsilon}, \beta_{\epsilon}, \alpha_{\xi}, \beta_{\xi}, \alpha_{z}, \beta_{z}) \biggr\rangle_{q_{1-j}(\rfb) \; q_2(\rzb) \; q_3(\rvb_{\rxi}) \; q_4(\rvb_{\repsilon}) \; q_5(\rvb_{\rz})} \right\rbrace,
\nonumber 
\\&
\hspace{341pt}j \in \left\lbrace 1,2 \ldots, M \right\rbrace,  
\label{Eq:VBA_Proportionalities1bis_IS}
\\[4pt]
&q_{2j}(\rz_{\rj}) \;\;\; \propto \; \exp \biggl\lbrace \left\langle \ln p(\rfb, \rzb, \rvb_\repsilon, \rvb_\rxi, \rvb_\rz | \bgb, \alpha_{\epsilon}, \beta_{\epsilon}, \alpha_{\xi}, \beta_{\xi}, \alpha_{z}, \beta_{z}) \biggr\rangle_{q_1(\rfb) \; q_{2-j}(\rzb) \; q_3(\rvb_{\rxi}) \; q_4(\rvb_{\repsilon}) \; q_5(\rvb_{\rz})} \right\rbrace,
\nonumber 
\\&
\hspace{341pt}j \in \left\lbrace 1,2 \ldots, M \right\rbrace,  
\label{Eq:VBA_Proportionalities2bis_IS}
\\[4pt]  
&q_{3j}(\rv_{\rxi_\red{j}}) \propto \; \exp \biggl\lbrace \left\langle \ln p(\rfb, \rzb, \rvb_\repsilon, \rvb_\rxi, \rvb_\rz | \bgb, \alpha_{\epsilon}, \beta_{\epsilon}, \alpha_{\xi}, \beta_{\xi}, \alpha_{z}, \beta_{z}) \biggr\rangle_{q_1(\rfb) \; q_2(\rzb) \; q_{3-j}(\rvb_{\rxi_\rj}) \; q_4(\rvb_{\repsilon}) \; q_5(\rvb_{\rz})} \right\rbrace,
\nonumber 
\\&
\hspace{341pt}j \in \left\lbrace 1,2 \ldots, M \right\rbrace,  
\label{Eq:VBA_Proportionalities3bis_IS}
\\[4pt]
&q_{4i}(\rv_{\repsilon_\ri}) \; \propto \; \exp \biggl\lbrace \left\langle \ln p(\rfb, \rzb, \rvb_\repsilon, \rvb_\rxi, \rvb_\rz | \bgb, \alpha_{\epsilon}, \beta_{\epsilon}, \alpha_{\xi}, \beta_{\xi}, \alpha_{z}, \beta_{z}) \biggr\rangle_{q_1(\rfb) \; q_2(\rzb) \; q_3(\rvb_{\rxi}) \; q_{4-i}(\rvb_{\repsilon_\ri}) \; q_5(\rvb_{\rz})} \right\rbrace,
\nonumber 
\\&
\hspace{342pt}i \in \left\lbrace 1,2 \ldots, N \right\rbrace,
\label{Eq:VBA_Proportionalities4bis_IS}
\\[4pt]
&q_{5j}(\rv_{\rz_\red{j}}) \propto \; \exp \biggl\lbrace \left\langle \ln p(\rfb, \rzb, \rvb_\repsilon, \rvb_\rxi, \rvb_\rz | \bgb, \alpha_{\epsilon}, \beta_{\epsilon}, \alpha_{\xi}, \beta_{\xi}, \alpha_{z}, \beta_{z}) \biggr\rangle_{q_1(\rfb) \; q_2(\rzb) \; q_3(\rvb_{\rxi}) \; q_4(\rvb_{\repsilon}) \; q_{5-j}(\rvb_{\rz_\rj})} \right\rbrace,
\nonumber 
\\&
\hspace{341pt}j \in \left\lbrace 1,2 \ldots, M \right\rbrace,
\label{Eq:VBA_Proportionalities5bis_IS}
\end{align}
\end{subequations}
using the notations introduced in Equation~\eqref{Eq:Def_q_Minus_IS}, Equation~\eqref{Eq:Def_Integral_IS} and Equation~\eqref{Eq:Def_q_Minusbis_IS}.
\beq
q_{1-j}(\rf_{\rj})=\displaystyle \prod_{k=1,k \neq j}^{M} q_{1k}(\rf_\rk)
\;\;;\;\;
q_{2-j}(\rz_{\rj})=\displaystyle \prod_{k=1,k \neq j}^{M} q_{2k}(\rz_\rk)
\label{Eq:Def_q_Minusbis_IS}
\eeq
The analytical expression of logarithm of the posterior distribution $\ln p(\rfb, \rzb, \rvb_\repsilon, \rvb_\rxi, \rvb_\rz | \bgb, \alpha_{\epsilon}, \beta_{\epsilon}, \alpha_{\xi}, \beta_{\xi}, \alpha_{z}, \beta_{z})$ is obtained in Equation~\eqref{Eq:Log_Posterior_IS}.
\paragraph{Computation of the analytical expression of $q_{1j}(\rf_{\rj})$.}
The proportionality relation corresponding to $q_{1j}(\rf_{\rj})$ is presented in established in Equation~\eqref{Eq:VBA_Proportionalities1bis_IS}. In the expression of $\ln p\left(\rfb, \rvb_\rf, \rvb_\repsilon | \bgb, \alpha_{f}, \beta_{f}, \alpha_{\epsilon}, \beta_{\epsilon} \right)$ all the terms free of $\rf_\rj$ can be regarded as constants:
\beq
\begin{split}
\biggl\langle \ln p & \left( \rfb, \rzb, \rvb_\rxi, \rvb_\repsilon, \rvb_\rz | \bgb, \alpha_{\xi}, \beta_{\xi}, \alpha_{\epsilon}, \beta_{\epsilon}, \alpha_{z}, \beta_{z} \right) \biggr\rangle_{q_{1-j}(\rfb) \; q_2(\rzb) \; q_3(\rvb_{\rxi}) \; q_4(\rvb_{\repsilon}) \; q_5(\rvb_{\rz})}
=
\\&
=
-\frac{1}{2}
\biggl\langle 
\|\rVb_{\repsilon}^{-\frac{1}{2}} \left(\bgb - \bHb\rfb\right) \|^2
\biggr\rangle_{q_{1-j}(\rfb) \; q_4(\rvb_{\repsilon}) }
-\frac{1}{2}
\biggl\langle 
\| \rVb_{\rxi}^{-\frac{1}{2}} \left( \rfb - \bDb\rzb \right) \|^2
\biggr\rangle_{q_{1-j}(\rfb) q_2(\rzb) q_3(\rvb_{\rxi})}.
\end{split}
\label{Eq:q1_Integral_1bis_IS}
\eeq
Using the notations introduced in Equation~\eqref{Eq:q1_Integral_Not1_IS}, the integral from Equation~\eqref{Eq:q1_Integral_1bis_IS} becomes:
\beq
\begin{split}
\biggl\langle \ln p & \left( \rfb, \rzb, \rvb_\rxi, \rvb_\repsilon, \rvb_\rz | \bgb, \alpha_{\xi}, \beta_{\xi}, \alpha_{\epsilon}, \beta_{\epsilon}, \alpha_{z}, \beta_{z} \right) \biggr\rangle_{q_{1-j}(\rfb) \; q_2(\rzb) \;q_3(\rvb_{\rxi}) \; q_4(\rvb_{\repsilon}) \; q_5(\rvb_{\rz})}
=
\\&=
-\frac{1}{2} \biggl\langle \|\rVbt_{\repsilon}^{\frac{1}{2}} \left(\bgb - \bHb\rfb\right) \|^2 \biggr\rangle_{q_{1-j}(\rfb)} 
- 
\frac{1}{2} \biggl\langle \| \rVbt_{\rxi}^{\frac{1}{2}} \left( \rfb - \bDb\rzb \right) \|^2 \biggr\rangle_{q_{1-j}(\rfb) q_2(\rzb)}.
\end{split}
\label{Eq:q1_Integral_2bis_IS}
\eeq
Introducing the notations
\beq
\rft_{\rj} = \biggl\langle \rf_{\rj} \biggr\rangle_{q_{1j}(\rf_{\rj})}
\;\; ; \;\;
\rfbt_{{-\rj}}
=
\biggl\langle \rfb \biggr\rangle_{q_{1-j}(\rfb)}
=
\begin{bmatrix}
\rft_{\red{1}} \;
\ldots \;
\rft_{{\rj-\red{1}}} \;
\rf_{\rj}  \;
\rft_{{\rj+\red{1}}}  \;
\ldots \;
\rft_{\rM}
\end{bmatrix}^T
\label{Eq:q1_Integral_Not1bis_IS}
\eeq
\beq
\rfb^{-\rj}
=
\begin{bmatrix}
\rf_{\red{1}} \;
\ldots \;
\rf_{\rj-\red{1}} \;
\rf_{\rj+\red{1}}  \;
\ldots \;
\rf_{\rM}
\end{bmatrix}^T
\;\; ; \;\;
\rfbt^{-\rj}
=
\biggl\langle \rfb^{-\rj} \biggr\rangle_{q_{1-j}(\rfb)}
=
\begin{bmatrix}
\rft_{\red{1}} \;
\ldots \;
\rft_{{\rj-\red{1}}} \;
\rft_{{\rj+\red{1}}} \;
\ldots \;
\rft_{\rM}
\end{bmatrix}^T
\label{Eq:q1_Integral_Not2bis_IS}
\eeq
and denoting $\bHb^{\bj}$ the column $j$ of matrix $\bHb$ and $\bHb^{-\bj}$ the matrix $\bHb$ without column j, the development of the norm from the first term if Equation~\eqref{Eq:q1_Integral_2bis_IS} is
\beq
\|\rVbt_{\repsilon}^{\frac{1}{2}} \left(\bgb - \bHb\rfb\right) \|^2
=
\| \rVbt_{\repsilon}^{\frac{1}{2}} \bgb \|^2
-2 \bgb^T \rVbt_{\repsilon} \bHb\rfb 
+ \| \rVbt_{\repsilon}^{\frac{1}{2}} \bHb \rfb \|^2.
\label{Eq:q1_Integral_Eq0bis_IS}
\eeq 
Using the equality $ \bHb \rfb = \bHb^{-\bj} \rfb^{-\rj} + \bHb^{\bj} \rf_{\rj} $ we establish the equalities from Equation~\eqref{Eq:q1_Integral_Eq1bis_IS} and Equation~\eqref{Eq:q1_Integral_Eq2bis_IS}:
\beq
\begin{split}
\bgb^T \rVbt_{\repsilon} \bHb\rfb
&=
\bgb^T \rVbt_{\repsilon} \left(\bHb^{-\bj} \rfb^{-\rj} + \bHb^{\bj} \rf_{\rj}  \right)
=
\bgb^T \rVbt_{\repsilon} \bHb^{-\bj} \rfb^{-\rj} +
\bgb^T \rVbt_{\repsilon} \bHb^{\bj} \rf_{\rj}
\end{split}
\label{Eq:q1_Integral_Eq1bis_IS}
\eeq
\beq
\begin{split}
\| \rVbt_{\repsilon}^{\frac{1}{2}} \bHb \rfb \|^2
&=
\| \rVbt_{\repsilon}^{\frac{1}{2}} \left(\bHb^{-\bj} \rfb^{-\rj} + \bHb^{\bj} \rf_{\rj}  \right) \|^2
=
\left(\rfb^{-\rj T} \bHb^{-\bj T}  + \bHb^{\bj T} \rf_{\rj}  \right) \rVbt_{\repsilon} \left(\bHb^{-\bj} \rfb^{-\rj} + \bHb^{\bj} \rf_{\rj}  \right)
=
\\&
=
\rfb^{-\rj T} \bHb^{-\bj T} \rVbt_{\repsilon} \bHb^{-\bj} \rfb^{-\rj}
+2\bHb^{\bj T} \rVbt_{\repsilon} \bHb^{-\bj} \rfb^{-\rj} \rf_{\rj}
+\bHb^{\bj T}  \rVbt_{\repsilon} \bHb^{\bj} \rf_{\rj}^2
\end{split}
\label{Eq:q1_Integral_Eq2bis_IS}
\eeq
so the corresponding integrals are:
\beq
\begin{split}
\biggl\langle \bgb^T \rVbt_{\repsilon} \bHb\rfb \biggr\rangle_{q_{1-j}(\rfb)}
&=
\bgb^T \rVbt_{\repsilon} \bHb^{-\bj} \rfbt^{-\rj} 
+
\bgb^T \rVbt_{\repsilon} \bHb^{\bj} \rf_{\rj}
\\
\biggl\langle \| \rVbt_{\repsilon}^{\frac{1}{2}} \bHb \rfb \|^2 \biggr\rangle_{q_{1-j}(\rfb)}
&=
\biggl\langle \| \rVbt_{\repsilon}^{\frac{1}{2}} \bHb^{-\bj} \rfb^{-\rj} \|^2 \biggr\rangle_{q_{1-j}(\rfb)} 
+2
\bHb^{\bj T} \rVbt_{\repsilon} \bHb^{-\bj} \rfbt^{-\rj} \rf_{\rj} 
+
\| \rVbt_{\repsilon}^{\frac{1}{2}} \bHb^{\bj}\|^2 \rf_{\rj}^2
\label{Eq:q1_Integral_Eq3bis_IS}
\end{split}
\eeq
Considering all the term that don't contain $\rf_{\rj}$ as constants, using Equation~\eqref{Eq:q1_Integral_Eq3bis_IS} via Equation~\eqref{Eq:q1_Integral_Eq0bis_IS}, the integral of the norm from the first term of Equation~\eqref{Eq:q1_Integral_2bis_IS} is:
\beq
\biggl\langle
\| \rVbt_{\repsilon}^{\frac{1}{2}} \left( \bgb - \bHb \rfb \right) \|^2
\biggr\rangle_{q_{1-j}(\rfb)} 
=
C
-2 
\left( 
\bgb^T \rVbt_{\repsilon} \bHb^{\bj} 
- 
\bHb^{\bj T} \rVbt_{\repsilon} \bHb^{-\bj} \rfbt^{-\rj} 
\right) \rf_{\rj} 
+
\| \rVbt_{\repsilon}^{\frac{1}{2}} \bHb^{\bj} \|^2 \rf_{\rj}^2
\label{Eq:q1_Integral_Eq4bis_IS}
\eeq
The norm corresponding to the latter term of Equation~\eqref{Eq:q1_Integral_2bis_IS} is developed considering all the term that don't contain $\rf_{\rj}$ as constants:
\beq
\| \rVbt_{\rxi}^{\frac{1}{2}} \left( \rfb - \bDb\rzb \right) \|^2
=
\sum_{j=1}^{M} \rvt_{\rxi_\rj} 
\left( 
\rf_\rj - \bDb_{\bj} \rzb
\right)^2
=
C
-2 \rvt_{\rxi_\rj} \bDb_{\bj} \rzb \rf_\rj
+ \rvt_{\rxi_\rj} \rf_\rj^2.
\eeq
Introducing the notations
\beq
\rzt_{\rj} = \biggl\langle \rz_{\rj} \biggr\rangle_{q_{2j}(\rz_{\rj})}
\;\; ; \;\;
\rzbt
=
\biggl\langle \rzb \biggr\rangle_{q_{2j}(\rzb)}
=
\begin{bmatrix}
\rzt_{\red{1}} \;
\ldots \;
\rzt_{{\rj-\red{1}}} \;
\rzt_{\rj}  \;
\rzt_{{\rj+\red{1}}}  \;
\ldots \;
\rzt_{\rM}
\end{bmatrix}^T,
\label{Eq:q1_Integral_Not3bis_IS}
\eeq
the integral of the norm from the latter term of Equation~\eqref{Eq:q1_Integral_2bis_IS} is:
\beq
\biggl\langle \| \rVbt_{\rxi}^{\frac{1}{2}} \left( \rfb - \bDb\rzb \right) \|^2 \biggr\rangle_{q_{1-j}(\rfb) q_2(\rzb)} =
C
-2 \rvt_{\rxi_\rj} \bDb_{\bj} \rzbt \rf_\rj
+ \rvt_{\rxi_\rj} \rf_\rj^2.
\label{Eq:q1_Integral_Eq5bis_IS}
\eeq
From Equation~\eqref{Eq:q1_Integral_2bis_IS} via Equation~\eqref{Eq:q1_Integral_Eq4bis_IS} and Equation~\eqref{Eq:q1_Integral_Eq5bis_IS}:
\beq
\begin{split}
\biggl\langle \ln p &\left( \rfb, \rzb, \rvb_\rxi, \rvb_\repsilon, \rvb_\rz | \bgb, \alpha_{\xi}, \beta_{\xi}, \alpha_{\epsilon}, \beta_{\epsilon}, \alpha_{z}, \beta_{z} \right) \biggr\rangle_{q_{1-j}(\rfb) \; q_2(\rzb) \;q_3(\rvb_{\rxi}) \; q_4(\rvb_{\repsilon})\; q_5(\rvb_{\rz})}
=
C-
\\&
- \left(
\bgb^T \rVbt_{\repsilon} \bHb^{\bj} 
- \bHb^{\bj T} \rVbt_{\repsilon} \bHb^{-\bj} \rfbt^{-\rj}
+ \rvt_{\rxi_\rj} \bDb_{\bj} \rzbt 
\right)\rf_\rj
+ \frac{1}{2} \left(
\| \rVbt_{\repsilon}^{\frac{1}{2}} \bHb^{\bj} \|^2 + \rvt_{\rxi_\rj}
\right)\rf_\rj^2,
\end{split}
\label{Eq:q1_Integral_2bis_IS}
\eeq
so $\biggl\langle \ln p \left( \rfb, \rzb, \rvb_\rxi, \rvb_\repsilon, \rvb_\rz | \bgb, \alpha_{\xi}, \beta_{\xi}, \alpha_{\epsilon}, \beta_{\epsilon}, \alpha_{z}, \beta_{z} \right) \biggr\rangle_{q_{1-j}(\rfb) \; q_2(\rzb) \;q_3(\rvb_{\rxi}) \; q_4(\rvb_{\repsilon})\; q_5(\rvb_{\rz})}$ is a quadratic form with respect to $\rf_{\rj}$. The proportionality from Equation~\eqref{Eq:VBA_Proportionalities1bis_IS} leads to the following corollary:
\begin{corollary}
$q_{1j} \left( \rf_{\rj} \right)$ is a Normal distribution.
\label{Co:q1bis}
\end{corollary} 
\hspace{-14pt}Minimizing the criterion
\beq
J(\rf_\rj)
=
C
- \left(
\bgb^T \rVbt_{\repsilon} \bHb^{\bj} 
- \bHb^{\bj T} \rVbt_{\repsilon} \bHb^{-\bj} \rfbt^{-\rj}
+ \rvt_{\rxi_\rj} \bDb_{\bj} \rzbt 
\right)\rf_\rj
+ \frac{1}{2} \left(
\| \rVbt_{\repsilon}^{\frac{1}{2}} \bHb^{\bj} \|^2 + \rvt_{\rxi_\rj}
\right)\rf_\rj^2,
\label{Eq:q1_Criterion_bis_IS}
\eeq
leads to the mean of the Normal distribution $q_{1j} \left( \rf_{\rj} \right)$:
\beq
\begin{split}
\frac{\partial J(\rf_\rj)
}{\partial \rf_{\rj}} = & 0
\Leftrightarrow
- \left(
\bgb^T \rVbt_{\repsilon} \bHb^{\bj} 
- \bHb^{\bj T} \rVbt_{\repsilon} \bHb^{-\bj} \rfbt^{-\rj}
+ \rvt_{\rxi_\rj} \bDb_{\bj} \rzbt 
\right)
+ 
\left(
\| \rVbt_{\repsilon}^{\frac{1}{2}} \bHb^{\bj}\|^2 + \rvt_{\rxi_\rj}
\right)\rf_\rj = 0
\Rightarrow
\\
\Rightarrow &
\rfh_\rj = \left( \| \rVbt_{\repsilon}^{\frac{1}{2}} \bHb^{\bj} \|^2 + \rvt_{\rxi_\rj} \right)^{-1} \left[ \bHb^{\bj T} \rVbt_{\repsilon} \left( \bgb - \bHb^{-\bj} \rfbt^{-\rj} \right) + \rvt_{\rxi_\rj} \bDb_{\bj} \rzbt \right].
\end{split}
\eeq
By identification, the variance can be easily derived:
\beq
\Sigmah_{f_{jj}} = \left( \| \rVbt_{\repsilon}^{\frac{1}{2}} \bHb^{\bj}\|^2 + \rvt_{\rxi_\rj} \right)^{-1}.
\eeq
Finally, we conclude that $q_{1j} \left( \rf_{\rj} \right)$ is a Normal distribution with the following parameters:
\beq
q_{1j}(\rf_{\rj})=\Nc \left( \rf_{\rj} | \rfh_\rj,\Sigmah_{f_{jj}} \right),
\left\{\barr{ll}
\rfh_\rj = \left( \bHb^{\bj T} \rVbt_{\repsilon} \bHb^{\bj} + \rvt_{\rxi_\rj} \right)^{-1} \left[ \bHb^{\bj T} \rVbt_{\repsilon} \left( \bgb - \bHb^{-\bj} \rfbt^{-\rj} \right) + \rvt_{\rxi_\rj} \bDb_{\bj} \rzbt \right]
\\[10pt]
\Sigmah_{f_{jj}} = \left( \bHb^{\bj T} \rVbt_{\repsilon} \bHb^{\bj} + \rvt_{\rxi_\rj} \right)^{-1}
\earr\right.
\label{Eq:q1_Analytical_1bis_IS}
\eeq
\paragraph{Computation of the analytical expression of $q_{2j}(\rz_{\rj})$.}
The proportionality relation corresponding to $q_{2j}(\rz_{\rj})$ is presented in established in Equation~\eqref{Eq:VBA_Proportionalities2bis_IS}. In the expression of $\ln  p \left( \rfb, \rzb, \rvb_\rxi, \rvb_\repsilon, \rvb_\rz | \bgb, \alpha_{\xi}, \beta_{\xi}, \alpha_{\epsilon}, \beta_{\epsilon}, \alpha_{z}, \beta_{z} \right)$ all the terms free of $\rz_\rj$ can be regarded as constants:
\beq
\begin{split}
\biggl\langle \ln & p \left( \rfb, \rzb, \rvb_\rxi, \rvb_\repsilon, \rvb_\rz | \bgb, \alpha_{\xi}, \beta_{\xi}, \alpha_{\epsilon}, \beta_{\epsilon}, \alpha_{z}, \beta_{z} \right) \biggr\rangle_{q_{1}(\rfb) \; q_{2-j}(\rzb) \; q_3(\rvb_{\rxi}) \; q_4(\rvb_{\repsilon})\; q_5(\rvb_{\rz})}
=
\\&=
-\frac{1}{2}
\biggl\langle 
\| \rVb_{\rxi}^{-\frac{1}{2}} \left( \rfb - \bDb\rzb \right) \|^2 \biggr\rangle_{q_{1}(\rfb) \; q_{2-j}(\rzb) \; q_3(\rvb_{\rxi}) }
-\frac{1}{2}
\biggl\langle 
\| \rVb_{\rz}^{-\frac{1}{2}} \rzb \|^2
\biggr\rangle_{q_2(\rzb) q_5(\rvb_{\rz})}.
\end{split}
\label{Eq:q2_Integral_1bis_IS}
\eeq
Using the notations introduced in Equation~\eqref{Eq:q1_Integral_Not1_IS} and Equation~\eqref{Eq:q2_Integral_Not1_IS}, the integral from Equation~\eqref{Eq:q2_Integral_1bis_IS} becomes:
\beq
\begin{split}
\biggl\langle & \ln p \left( \rfb, \rzb, \rvb_\rxi, \rvb_\repsilon, \rvb_\rz | \bgb, \alpha_{\xi}, \beta_{\xi}, \alpha_{\epsilon}, \beta_{\epsilon}, \alpha_{z}, \beta_{z} \right) \biggr\rangle_{q_{1}(\rfb) \; q_{2-j}(\rzb) \;q_3(\rvb_{\rxi}) \; q_4(\rvb_{\repsilon})\; q_5(\rvb_{\rz})}
=
\\&=
-\frac{1}{2} \biggl\langle \| \rVbt_{\rxi}^{\frac{1}{2}} \left( \rfb - \bDb\rzb \right) \|^2 \biggr\rangle_{q_{1}(\rfb) q_{2-j}(\rzb)}
-\frac{1}{2} \biggl\langle \| \rVbt_{\rz}^{\frac{1}{2}} \rzb \|^2 \biggr\rangle_{q_{2-j}(\rzb)}.
\end{split}
\label{Eq:q2_Integral_2bis_IS}
\eeq
Introducing the notations
\beq
\rzt_{\rj} = \biggl\langle \rz_{\rj} \biggr\rangle_{q_{2j}(\rz_{\rj})}
\;\; ; \;\;
\rzbt_{{-\rj}}
=
\biggl\langle \rzb \biggr\rangle_{q_{2-j}(\rzb)}
=
\begin{bmatrix}
\rzt_{\red{1}} \;
\ldots \;
\rzt_{{\rj-\red{1}}} \;
\rz_{\rj}  \;
\rzt_{{\rj+\red{1}}}  \;
\ldots \;
\rzt_{\rM}
\end{bmatrix}^T
\label{Eq:q2_Integral_Not1bis_IS}
\eeq
\beq
\rzb^{-\rj}
=
\begin{bmatrix}
\rz_{\red{1}} \;
\ldots \;
\rz_{\rj-\red{1}} \;
\rz_{\rj+\red{1}}  \;
\ldots \;
\rz_{\rM}
\end{bmatrix}^T
\;\; ; \;\;
\rzbt^{-\rj}
=
\biggl\langle \rzb^{-\rj} \biggr\rangle_{q_{2-j}(\rzb)}
=
\begin{bmatrix}
\rzt_{\red{1}} \;
\ldots \;
\rzt_{{\rj-\red{1}}} \;
\rzt_{{\rj+\red{1}}} \;
\ldots \;
\rzt_{\rM}
\end{bmatrix}^T
\label{Eq:q2_Integral_Not2bis_IS}
\eeq
and denoting $\bDb^{\bj}$ the column $j$ of matrix $\bDb$ and $\bDb^{-\bj}$ the matrix $\bDb$ without column j, the development of the norm from the first term if Equation~\eqref{Eq:q2_Integral_2bis_IS} is
\beq
\|\rVbt_{\rxi}^{\frac{1}{2}} \left(\rfb - \bDb\rzb\right) \|^2
=
\| \rVbt_{\rxi}^{\frac{1}{2}} \rfb \|^2
-2 \rfb^T \rVbt_{\rxi} \bDb\rzb 
+ \| \rVbt_{\rxi}^{\frac{1}{2}} \bDb \rzb \|^2.
\label{Eq:q2_Integral_Eq0bis_IS}
\eeq 
Using the equality $ \bDb \rzb = \bDb^{-\bj} \rzb^{-\rj} + \bDb^{\bj} \rz_{\rj} $ we establish the equalities from Equation~\eqref{Eq:q2_Integral_Eq1bis_IS} and Equation~\eqref{Eq:q2_Integral_Eq2bis_IS}:
\beq
\begin{split}
\rfb^T \rVbt_{\rxi} \bDb\rzb
&=
\rfb^T \rVbt_{\rxi} \left(\bDb^{-\bj} \rzb^{-\rj} + \bDb^{\bj} \rz_{\rj}  \right)
=
\rfb^T \rVbt_{\rxi} \bDb^{-\bj} \rzb^{-\rj} +
\rfb^T \rVbt_{\rxi} \bDb^{\bj} \rz_{\rj}
\end{split}
\label{Eq:q2_Integral_Eq1bis_IS}
\eeq
\beq
\begin{split}
\| \rVbt_{\rxi}^{\frac{1}{2}} \bDb \rzb \|^2
&=
\| \rVbt_{\rxi}^{\frac{1}{2}} \left(\bDb^{-\bj} \rzb^{-\rj} + \bDb^{\bj} \rz_{\rj}  \right) \|^2
=
\left(\rzb^{-\rj T} \bDb^{-\bj T}  + \bDb^{\bj T} \rz_{\rj}  \right) \rVbt_{\rxi} \left(\bDb^{-\bj} \rzb^{-\rj} + \bDb^{\bj} \rz_{\rj}  \right)
=
\\&
=
\rzb^{-\rj T} \bDb^{-\bj T} \rVbt_{\rxi} \bDb^{-\bj} \rzb^{-\rj}
+2\bDb^{\bj T} \rVbt_{\rxi} \bDb^{-\bj} \rzb^{-\rj} \rz_{\rj}
+\bDb^{\bj T}  \rVbt_{\rxi} \bDb^{\bj} \rz_{\rj}^2
\end{split}
\label{Eq:q2_Integral_Eq2bis_IS}
\eeq
so the corresponding integrals are:
\beq
\begin{split}
\biggl\langle \rfb^T \rVbt_{\rxi} \bDb\rzb \biggr\rangle_{q_{1}(\rfb) \; q_{2-j}(\rzb)}
&=
\rfbt^T \rVbt_{\rxi} \bDb^{-\bj} \rzbt^{-\rj} 
+
\rfbt^T \rVbt_{\rxi} \bDb^{\bj} \rz_{\rj}
\\
\biggl\langle \| \rVbt_{\rxi}^{\frac{1}{2}} \bDb \rzb \|^2 \biggr\rangle_{q_{2-j}(\rzb)}
&=
\biggl\langle \| \rVbt_{\rxi}^{\frac{1}{2}} \bDb^{-\bj} \rzb^{-\rj} \|^2 \biggr\rangle_{q_{2-j}(\rzb)} 
+2
\bDb^{\bj T} \rVbt_{\rxi} \bDb^{-\bj} \rzbt^{-\rj} \rz_{\rj} 
+
\| \rVbt_{\rxi}^{\frac{1}{2}} \bDb^{\bj}\|^2 \rz_{\rj}^2
\label{Eq:q2_Integral_Eq3bis_IS}
\end{split}
\eeq
Considering all the term that don't contain $\rz_{\rj}$ as constants, using Equation~\eqref{Eq:q2_Integral_Eq3bis_IS} via Equation~\eqref{Eq:q2_Integral_Eq0bis_IS}, the integral of the norm from the first term of Equation~\eqref{Eq:q2_Integral_2bis_IS} is:
\beq
\biggl\langle
\| \rVbt_{\rxi}^{\frac{1}{2}} \left( \rfb - \bDb \rzb \right) \|^2 \biggr\rangle_{q_{1}(\rfb)\;q_{2-j}(\rzb)} 
=
C
-2 
\left( 
\rfb^T \rVbt_{\rxi} \bDb^{\bj} 
- 
\bDb^{\bj T} \rVbt_{\rxi} \bDb^{-\bj} \rzbt^{-\rj} 
\right) \rz_{\rj} 
+
\| \rVbt_{\rxi}^{\frac{1}{2}} \bDb^{\bj} \|^2 \rz_{\rj}^2
\label{Eq:q2_Integral_Eq4bis_IS}
\eeq
The norm corresponding to the latter term of Equation~\eqref{Eq:q2_Integral_2bis_IS} is developed considering all the term that don't contain $\rz_{\rj}$ as constants:
\beq
\| \rVbt_{\rz}^{\frac{1}{2}} \rzb \|^2
=
\sum_{j=1}^{M} \rvt_{\rz_\rj} 
 \rz_{\rj}^2
=
C
+ \rvt_{\rz_\rj} \rz_\rj^2.
\eeq
so the integral of the norm from the latter term of Equation~\eqref{Eq:q2_Integral_2bis_IS} is:
\beq
\biggl\langle \| \rVbt_{\rz}^{\frac{1}{2}} \rzb \|^2 \biggr\rangle_{ q_{2-j}(\rzb)} =
C + \rvt_{\rz_\rj} \rz_\rj^2.
\label{Eq:q2_Integral_Eq5bis_IS}
\eeq
From Equation~\eqref{Eq:q2_Integral_2bis_IS} via Equation~\eqref{Eq:q2_Integral_Eq4bis_IS} and Equation~\eqref{Eq:q2_Integral_Eq5bis_IS}:
\beq
\begin{split}
\biggl\langle & \ln p \left( \rfb, \rzb, \rvb_\rxi, \rvb_\repsilon, \rvb_\rz | \bgb, \alpha_{\xi}, \beta_{\xi}, \alpha_{\epsilon}, \beta_{\epsilon}, \alpha_{z}, \beta_{z} \right) \biggr\rangle_{q_{1}(\rfb) \; q_{2-j}(\rzb) \;q_3(\rvb_{\rxi}) \; q_4(\rvb_{\repsilon})\; q_5(\rvb_{\rz})}
=
C-
\\&
-\left(
\rfb^T \rVbt_{\rxi} \bDb^{\bj} 
- \bDb^{\bj T} \rVbt_{\rxi} \bDb^{-\bj} \rzbt^{-\rj} 
\right)\rz_\rj
+\frac{1}{2} \left(
\| \rVbt_{\rxi}^{\frac{1}{2}} \bDb^{\bj}\|^2 + \rvt_{\rz_\rj}
\right)\rz_\rj^2,
\end{split}
\label{Eq:q2_Integral_3bis_IS}
\eeq
so $\biggl\langle \ln p \left( \rfb, \rzb, \rvb_\rxi, \rvb_\repsilon, \rvb_\rz | \bgb, \alpha_{\xi}, \beta_{\xi}, \alpha_{\epsilon}, \beta_{\epsilon}, \alpha_{z}, \beta_{z} \right) \biggr\rangle_{q_{1}(\rfb) \; q_{2-j}(\rzb) \;q_3(\rvb_{\rxi}) \; q_4(\rvb_{\repsilon})\; q_5(\rvb_{\rz})}
$. The proportionality from Equation~\eqref{Eq:VBA_Proportionalities2bis_IS} leads to the following corollary:
\begin{corollary}
$q_{2j} \left( \rz_{\rj} \right)$ is a Normal distribution.
\label{Co:q2bis}
\end{corollary} 
\hspace{-14pt}Minimizing the criterion
\beq
J(\rz_\rj)
=
C
-\left(
\rfb^T \rVbt_{\rxi} \bDb^{\bj} 
- \bDb^{\bj T} \rVbt_{\rxi} \bDb^{-\bj} \rzbt^{-\rj} 
\right)\rz_\rj
+\frac{1}{2} \left(
\| \rVbt_{\rxi}^{\frac{1}{2}} \bDb^{\bj}\|^2 + \rvt_{\rz_\rj}
\right)\rz_\rj^2,
\label{Eq:q2_Criterion_bis_IS}
\eeq
leads to the mean of the Normal distribution $q_{2j} \left( \rz_{\rj} \right)$:
\beq
\begin{split}
\frac{\partial J(\rz_\rj)
}{\partial \rz_{\rj}} = & 0
\Leftrightarrow
- \left(
\rfb^T \rVbt_{\rxi} \bDb^{\bj} 
- \bDb^{\bj T} \rVbt_{\rxi} \bDb^{-\bj} \rzbt^{-\rj} 
\right)
+ 
\left(
\| \rVbt_{\rxi}^{\frac{1}{2}} \bDb^{\bj}\|^2 + \rvt_{\rz_\rj}
\right)\rz_\rj = 0
\Rightarrow
\\
\Rightarrow &
\rzh_\rj = \left( \| \rVbt_{\rxi}^{\frac{1}{2}} \bDb^{\bj}\|^2 + \rvt_{\rz_\rj} \right)^{-1} \left[ \rfb^T \rVbt_{\rxi} \bDb^{\bj} 
- \bDb^{\bj T} \rVbt_{\rxi} \bDb^{-\bj} \rzbt^{-\rj}  \right].
\end{split}
\eeq
By identification, the variance can be easily derived:
\beq
\Sigmah_{z_{jj}} = \left( \| \rVbt_{\rxi}^{\frac{1}{2}} \bDb^{\bj}\|^2 + \rvt_{\rz_\rj} \right)^{-1}.
\eeq
Finally, we conclude that $q_{2j} \left( \rz_{\rj} \right)$ is a Normal distribution with the following parameters:
\beq
q_{2j}(\rz_{\rj})=\Nc \left( \rz_{\rj} | \rzh_\rj,\Sigmah_{z_{jj}} \right),
\left\{\barr{ll}
\rzh_\rj = \left( \bDb^{\bj T} \rVbt_{\rxi} \bDb^{\bj} + \rvt_{\rz_\rj} \right)^{-1} \left[ \bDb^{\bj T}  \rVbt_{\rxi} \left( \rfb - \bDb^{-\bj} \rzbt^{-\rj} \right)  \right]
\\[10pt]
\Sigmah_{z_{jj}} = \left( \bDb^{\bj T} \rVbt_{\rxi} \bDb^{\bj} + \rvt_{\rz_\rj} \right)^{-1}
\earr\right.
\label{Eq:q2_Analytical_1bis_IS}
\eeq
\paragraph{Computation of the analytical expression of $q_{3j}(\rvb_{\rxi_\rj})$.}
The proportionality relation corresponding to $q_{3j}(\rvb_{\rxi_\rj})$ is presented in established in Equation~\eqref{Eq:VBA_Proportionalities3bis_IS}. In the expression of $\ln  p \left( \rfb, \rzb, \rvb_\rxi, \rvb_\repsilon, \rvb_\rz | \bgb, \alpha_{\xi}, \beta_{\xi}, \alpha_{\epsilon}, \beta_{\epsilon}, \alpha_{z}, \beta_{z} \right)$ all the terms free of $\rvb_{\rxi_\rj}$ can be regarded as constants:
\beq
\begin{split}
\biggl\langle \ln & p \left( \rfb, \rzb, \rvb_\rxi, \rvb_\repsilon, \rvb_\rz | \bgb, \alpha_{\xi}, \beta_{\xi}, \alpha_{\epsilon}, \beta_{\epsilon}, \alpha_{z}, \beta_{z} \right) \biggr\rangle_{q_{1}(\rfb) \; q_2(\rzb) \; q_{3-j}(\rvb_{\rxi}) \; q_4(\rvb_{\repsilon})\; q_5(\rvb_{\rz})}
=
\\&=
-\frac{1}{2}
\biggl\langle 
\| \rVb_{\rxi}^{-\frac{1}{2}} \left( \rfb - \bDb\rzb \right) \|^2 \biggr\rangle_{q_{1}(\rfb) \; q_{2}(\rzb) \; q_{3-j}(\rvb_{\rxi}) }
-
\left( \alpha_{\xi} + \frac{3}{2} \right) \ln \rv_{\rxi_\rj} 
-\frac{\beta_{\xi}}{\rv_{\rxi_\rj}}.
\end{split}
\label{Eq:q3_Integral_1bis_IS}
\eeq
Using the notations introduced in Equation~\eqref{Eq:q3_Integral_Not1_IS}, Corollary~\eqref{Co:q1bis} and Corollary~\eqref{Co:q2bis}, the integral from Equation~\eqref{Eq:q3_Integral_1bis_IS} becomes:
\beq
\begin{split}
\biggl\langle & \ln p \left( \rfb, \rzb, \rvb_\rxi, \rvb_\repsilon, \rvb_\rz | \bgb, \alpha_{\xi}, \beta_{\xi}, \alpha_{\epsilon}, \beta_{\epsilon}, \alpha_{z}, \beta_{z} \right) \biggr\rangle_{q_{1}(\rfb) \; q_{2}(\rzb) \;q_{3-j}(\rvb_{\rxi}) \; q_4(\rvb_{\repsilon})\; q_5(\rvb_{\rz})}
=
\\&=
- 
\frac{1}{2} \rv_{\rxi_\rj}^{-1} \biggl\langle \left( \rf_{\rj} - \bDb_{\bj}\rzb \right)^2 \biggr\rangle_{q_{1j}(\rf_{\rj}) q_{2}(\rzb)}
-\left( \alpha_{\xi} + \frac{3}{2} \right) \ln \rv_{\rxi_\rj} 
-\frac{\beta_{\xi}}{\rv_{\rxi_\rj}}
=
\\&=
- 
\frac{1}{2} \rv_{\rxi_\rj}^{-1} \biggl\langle \rf_{\rj}^2 - 2 \rf_{\rj} \bDb_{\bj} \rzb + \left( \bDb_{\bj}\rzb  \right)^2 \biggr\rangle_{q_{1j}(\rf_{\rj}) q_{2}(\rzb)}
-\left( \alpha_{\xi} + \frac{3}{2} \right) \ln \rv_{\rxi_\rj} - \frac{\beta_{\xi}}{\rv_{\rxi_\rj}}
=
\\&=
-\left( \alpha_{\xi} + \frac{3}{2} \right) \ln \rv_{\rxi_\rj} - \frac{\beta_{\xi} + \frac{1}{2} \left[ \left( \rfh_{\rj}  -  \bDb_{\bj}\rzbh \right)^2 + \Sigmah_{f_{jj}} + \bDb_{\bj}^{T} \Sigmabh_{z} \bDb_{\bj} \right]}{\rv_{\rxi_\rj}}.
\end{split}
\label{Eq:q3_Integral_2bis_IS}
\eeq
The proportionality from Equation~\eqref{Eq:VBA_Proportionalities3bis_IS} via Equation~\eqref{Eq:q3_Integral_2bis_IS} leads to the following corollary:
\begin{corollary}
$q_{3j} \left( \rv_{\rxi_\rj} \right)$ is an Inverse Gamma distribution.
\label{Co:q3bis}
\end{corollary} 
\hspace{-14pt}By identification we conclude that $q_{3j} \left( \rv_{\rxi_\rj} \right)$ is an Inverse Gamma distribution with the following parameters:
\beq
q_{3j} \left( \rv_{\rxi_\rj} \right)= \Ic\Gc \left( \rv_{\rxi_\rj} | \alphah_{\xi_j}, \betah_{\xi_j} \right),
\left\{\barr{ll}
\alphah_{\xi_j} = \alpha_{\xi} + \frac{3}{2}
\\[10pt]
\betah_{\xi_j} = \beta_{\xi} + \frac{1}{2} \left[ \left( \rfh_{\rj}  -  \bDb_{\bj}\rzbh \right)^2 + \Sigmah_{f_{jj}} + \bDb_{\bj}^{T} \Sigmabh_{z} \bDb_{\bj} \right]
\earr\right.
\label{Eq:q3_Analytical_1bis_IS}
\eeq
\paragraph{Computation of the analytical expression of $q_{4i}(\rvb_{\repsilon_\ri})$.}
The proportionality relation corresponding to $q_{4j}(\rvb_{\repsilon_\ri})$ is presented in established in Equation~\eqref{Eq:VBA_Proportionalities4bis_IS}. In the expression of $\ln  p \left( \rfb, \rzb, \rvb_\rxi, \rvb_\repsilon, \rvb_\rz | \bgb, \alpha_{\xi}, \beta_{\xi}, \alpha_{\epsilon}, \beta_{\epsilon}, \alpha_{z}, \beta_{z} \right)$ all the terms free of $\rvb_{\rxi_\rj}$ can be regarded as constants:
\beq
\begin{split}
\biggl\langle \ln & p \left( \rfb, \rzb, \rvb_\rxi, \rvb_\repsilon, \rvb_\rz | \bgb, \alpha_{\xi}, \beta_{\xi}, \alpha_{\epsilon}, \beta_{\epsilon}, \alpha_{z}, \beta_{z} \right) \biggr\rangle_{q_{1}(\rfb) \; q_2(\rzb) \; q_{j}(\rvb_{\rxi}) \; q_{4-i}(\rvb_{\repsilon})\; q_5(\rvb_{\rz})}
=
\\&=
-\frac{1}{2}
\biggl\langle 
\| \rVb_{\repsilon}^{-\frac{1}{2}} \left( \bgb - \bHb\rfb \right) \|^2 \biggr\rangle_{q_{1}(\rfb) \; q_{4-i}(\rvb_{\repsilon}) }
-
\left( \alpha_{\epsilon} + \frac{3}{2} \right) \ln \rv_{\repsilon_\ri} 
-\frac{\beta_{\epsilon}}{\rv_{\repsilon_\ri}}.
\end{split}
\label{Eq:q4_Integral_1bis_IS}
\eeq
Using the notations introduced in Equation~\eqref{Eq:q4_Integral_Not1_IS} and Corollary~\eqref{Co:q1bis}, the integral from Equation~\eqref{Eq:q4_Integral_1bis_IS} becomes:
\beq
\begin{split}
\biggl\langle \ln & p \left( \rfb, \rzb, \rvb_\rxi, \rvb_\repsilon, \rvb_\rz | \bgb, \alpha_{\xi}, \beta_{\xi}, \alpha_{\epsilon}, \beta_{\epsilon}, \alpha_{z}, \beta_{z} \right) \biggr\rangle_{q_{1}(\rfb) \; q_2(\rzb) \; q_{j}(\rvb_{\rxi}) \; q_{4-i}(\rvb_{\repsilon})\; q_5(\rvb_{\rz})}
=
\\&=
- 
\frac{1}{2} \rv_{\repsilon_\ri}^{-1} \biggl\langle \left( \bg_{\bi} - \bHb_{\bi}\rfb \right)^2 \biggr\rangle_{q_{1}(\rfb)}
-\left( \alpha_{\epsilon} + \frac{3}{2} \right) \ln \rv_{\repsilon_\ri} 
-\frac{\beta_{\epsilon}}{\rv_{\repsilon_\ri}}
=
\\&=
- 
\frac{1}{2} \rv_{\repsilon_\ri}^{-1} \biggl\langle \bg_{\bi}^2 - 2 \bg_{\bi} \bHb_{\bi} \rzb + \left( \bHb_{\bi} \rfb \right)^2 \biggr\rangle_{q_{1}(\rfb)}
-\left( \alpha_{\epsilon} + \frac{3}{2} \right) \ln \rv_{\repsilon_\ri} - \frac{\beta_{\epsilon}}{\rv_{\repsilon_\ri}}
=
\\&=
-\left( \alpha_{\epsilon} + \frac{3}{2} \right) \ln \rv_{\repsilon_\ri} - \frac{\beta_{\epsilon} + \frac{1}{2} \left[ \left( \bg_{\bi} - \bHb_{\bi}\rfbh \right)^2 + \bHb_{\bi}^{T} \Sigmabh_{f} \bHb_{\bi} \right]}{\rv_{\repsilon_\ri}}.
\end{split}
\label{Eq:q4_Integral_2bis_IS}
\eeq
The proportionality from Equation~\eqref{Eq:VBA_Proportionalities4bis_IS} via Equation~\eqref{Eq:q4_Integral_2bis_IS} leads to the following corollary:
\begin{corollary}
$q_{4i} \left( \rv_{\repsilon_\ri} \right)$ is an Inverse Gamma distribution.
\label{Co:q4bis}
\end{corollary} 
\hspace{-14pt}By identification we conclude that $q_{4i} \left( \rv_{\repsilon_\ri} \right)$ is an Inverse Gamma distribution with the following parameters:
\beq
q_{4i} \left( \rv_{\repsilon_\ri} \right)= \Ic\Gc \left( \rv_{\repsilon_\ri} | \alphah_{\epsilon_i}, \betah_{\epsilon_i} \right),
\left\{\barr{ll}
\alphah_{\epsilon_i} = \alpha_{\epsilon} + \frac{3}{2}
\\[10pt]
\betah_{\epsilon_i} = \beta_{\epsilon} + \frac{1}{2} \left[ \left( \bg_{\bi} - \bHb_{\bi}\rfbh \right)^2 + \bHb_{\bi}^{T} \Sigmabh_{f} \bHb_{\bi} \right]
\earr\right.
\label{Eq:q4_Analytical_1bis_IS}
\eeq
\paragraph{Computation of the analytical expression of $q_{5j}(\rvb_{\rz_\rj})$.}
The proportionality relation corresponding to $q_{5j}(\rvb_{\rz_\rj})$ is presented in established in Equation~\eqref{Eq:VBA_Proportionalities5bis_IS}. In the expression of $\ln  p \left( \rfb, \rzb, \rvb_\rxi, \rvb_\repsilon, \rvb_\rz | \bgb, \alpha_{\xi}, \beta_{\xi}, \alpha_{\epsilon}, \beta_{\epsilon}, \alpha_{z}, \beta_{z} \right)$ all the terms free of $\rvb_{\rz_\rj}$ can be regarded as constants:
\beq
\begin{split}
\biggl\langle \ln & p \left( \rfb, \rzb, \rvb_\rxi, \rvb_\repsilon, \rvb_\rz | \bgb, \alpha_{\xi}, \beta_{\xi}, \alpha_{\epsilon}, \beta_{\epsilon}, \alpha_{z}, \beta_{z} \right) \biggr\rangle_{q_{1}(\rfb) \; q_2(\rzb) \; q_{3}(\rvb_{\rxi}) \; q_4(\rvb_{\repsilon})\; q_{5-j}(\rvb_{\rz})}
=
\\&=
-\frac{1}{2}
\biggl\langle 
\| \rVb_{\rz}^{-\frac{1}{2}} \rzb \|^2 \biggr\rangle_{ q_{2}(\rzb) \; q_{5-j}(\rvb_{\rz})}
-
\left( \alpha_{z} + \frac{3}{2} \right) \ln \rv_{\rz_\rj} 
-\frac{\beta_{z}}{\rv_{\rz_\rj}}.
\end{split}
\label{Eq:q5_Integral_1bis_IS}
\eeq
Using the notations introduced in Equation~\eqref{Eq:q5_Integral_Not1_IS} and Corollary~\eqref{Co:q2bis}, the integral from Equation~\eqref{Eq:q5_Integral_1bis_IS} becomes:
\beq
\begin{split}
\biggl\langle & \ln p \left( \rfb, \rzb, \rvb_\rxi, \rvb_\repsilon, \rvb_\rz | \bgb, \alpha_{\xi}, \beta_{\xi}, \alpha_{\epsilon}, \beta_{\epsilon}, \alpha_{z}, \beta_{z} \right) \biggr\rangle_{q_{1}(\rfb) \; q_{2}(\rzb) \;q_3(\rvb_{\rxi}) \; q_4(\rvb_{\repsilon})\; q_{5-j}(\rvb_{\rz})}
=
\\&=
- 
\frac{1}{2} \rv_{\rz_\rj}^{-1} \biggl\langle \rz_{\rj}^2  \biggr\rangle_{q_{2}(\rzb)}
-\left( \alpha_{z} + \frac{3}{2} \right) \ln \rv_{\rz_\rj} 
-\frac{\beta_{z}}{\rv_{\rz_\rj}}
=
\\&=
- 
\frac{1}{2} \rv_{\rz_\rj}^{-1} \left( \rzh_{\rj}^2 + \Sigmah_{z_{jj}} \right)
-\left( \alpha_{z} + \frac{3}{2} \right) \ln \rv_{\rz_\rj} 
-\frac{\beta_{z}}{\rv_{\rz_\rj}}
=
\\&=
-\left( \alpha_{z} + \frac{3}{2} \right) \ln \rv_{\rz_\rj} -\frac{\beta_{z} + \frac{1}{2} \left( \rzh_{\rj}^2 + \Sigmah_{z_{jj}} \right)}{\rv_{\rz_\rj}}.
\end{split}
\label{Eq:q5_Integral_2bis_IS}
\eeq
The proportionality from Equation~\eqref{Eq:VBA_Proportionalities5bis_IS} via Equation~\eqref{Eq:q5_Integral_2bis_IS} leads to the following corollary:
\begin{corollary}
$q_{5j} \left( \rv_{\rxi_\rj} \right)$ is an Inverse Gamma distribution.
\label{Co:q5bis}
\end{corollary}
\hspace{-14pt}By identification we conclude that $q_{5j} \left( \rv_{\rz_\rj} \right)$ is an Inverse Gamma distribution with the following parameters:
\beq
q_{5j} \left( \rv_{\rz_\rj} \right)= \Ic\Gc \left( \rv_{\rz_\rj} | \alphah_{z_j}, \betah_{z_j} \right),
\left\{\barr{ll}
\alphah_{z_j} = \alpha_{z} + \frac{3}{2}
\\[10pt]
\betah_{z_j} = \beta_{z} + \frac{1}{2} \left( \rzh_{\rj}^2 + \Sigmah_{z_{jj}} \right)
\earr\right.
\label{Eq:q5_Analytical_1bis_IS}
\eeq
Equations~\eqref{Eq:q1_Analytical_1bis_IS},~\eqref{Eq:q2_Analytical_1bis_IS},~\eqref{Eq:q3_Analytical_1bis_IS},~\eqref{Eq:q4_Analytical_1bis_IS} and~\eqref{Eq:q5_Analytical_1bis_IS} resume the distributions families and the corresponding parameters for $q_{1j}(\rf_{\rj})$, $j\in\left\lbrace 1, 2, \ldots, M \right\rbrace$, $q_{2j}(\rz_{\rj})$, $j\in\left\lbrace 1, 2, \ldots, M \right\rbrace$ (Normal distribution), $q_{3j}(\rv_{\rxi_\rj})$, $j\in\left\lbrace 1, 2, \ldots, M \right\rbrace$, $q_{4i}(\rv_{\repsilon_\ri})$, $i\in\left\lbrace 1, 2, \ldots, N \right\rbrace$ and $q_{5j}(\rv_{\rz_\rj})$, $j\in\left\lbrace 1, 2, \ldots, M \right\rbrace$ (Inverse Gamma distributions). However, the parameters corresponding to the Normal distributions are expressed via $\rVbt_{\rxi}$, $\rVbt_{\repsilon}$ and $\rVbt_{\rz}$ (and by extension all elements forming the three matrices $ \rvt_{\rxi_\rj}$, $j \in \left\lbrace 1, 2, \ldots, M \right\rbrace $, $ \rvt_{\repsilon_\ri}$, $i \in \left\lbrace 1, 2, \ldots, N \right\rbrace $, both defined in Equation~\eqref{Eq:q1_Integral_Not1_IS} and $ \rvt_{\rz_\rj}$, $j \in \left\lbrace 1, 2, \ldots, M \right\rbrace $, defined in Equation~\eqref{Eq:q2_Integral_Not1_IS}.
\paragraph{Computation of the analytical expressions of $\rVbt_{\rxi}$, $\rVbt_{\repsilon}$ and $\rVbt_{\rz}$.}
For an Inverse Gamma distribution with scale and shape parameters $\alpha$ and $\beta$, $\Ic\Gc\left(x|\alpha, \beta \right)$, the following relation holds:
\beq
\biggl\langle x^{-1} \biggr\rangle_{\Ic\Gc(x|\alpha,\beta)} 
=
\frac{\alpha}{\beta}
\label{Eq:Inverse_Gamma_Integral_bis_IS}
\eeq
The prove of the above relation is done by direct computation, using the analytical expression of the Inverse Gamma Distribution: 
\beq
\begin{split}
\biggl\langle x^{-1} \biggr\rangle_{\Ic\Gc(x|\alpha,\beta)} 
&= 
\int 
x^{-1}
\frac{{\beta}^{\alpha}}{\Gamma(\alpha)}
x^{-\alpha-1}
\exp
\left\lbrace
- \frac{\beta}{x}
\right\rbrace
\d x
=
\frac{{\beta}^{\alpha}}{\Gamma(\alpha)}
\frac{\Gamma(\alpha+1)}{{\beta}^{\alpha+1}}
\int
\frac{{\beta}^{\alpha+1}}{\Gamma(\alpha+1)}
x^{-(\alpha+1)-1}
\exp\left\lbrace -\frac{\beta}{x}\right\rbrace
\d x
=\\
&=
\frac{\alpha}{\beta}
\underbrace{
\int \Ic\Gc(x|\alpha+1,\beta)
}_{1}
\d x
=
\frac{\alpha}{\beta}
\end{split}
\nonumber
\eeq
Since $q_{3j}(\rv_{\rxi_\rj})$, $j\in\left\lbrace 1, 2, \ldots, M \right\rbrace$, $q_{4i}(\rv_{\repsilon_\ri})$, $i\in\left\lbrace 1, 2, \ldots, N \right\rbrace$ and $q_{5j}(\rv_{\rz_\rj})$, $j\in\left\lbrace 1, 2, \ldots, M \right\rbrace$ are Inverse Gamma distributions, with the corresponding parameters $\alphah_{\xi_j}$ and $\betah_{\xi_j}$, $j\in\left\lbrace 1, 2, \ldots, M \right\rbrace$, $\alphah_{\epsilon_j}$ and $\betah_{\epsilon_j}$, $j\in\left\lbrace 1, 2, \ldots, N \right\rbrace$ respectively $\alphah_{z_j}$ and $\betah_{z_j}$, $j\in\left\lbrace 1, 2, \ldots, M \right\rbrace$ the expectancies $\rvt_{\rxi_\rj}$, $\rvt_{\repsilon_\ri}$ and $\rvt_{\rz_\rj}$ can be expressed via the parameters of the two Inverse Gamma distributions using Equation~\eqref{Eq:Inverse_Gamma_Integral_bis_IS}:
\beq
\rvt_{\rxi_\rj}
=
\frac{\alphah_{\xi_j}}{\betah_{\xi_j}}
=
\rvh_{\rxi_\rj}
\;\;\;;\;\;\;
\rvt_{\repsilon_\ri}
=
\frac{\alphah_{\epsilon_i}}{\betah_{\epsilon_i}}
=
\rvh_{\repsilon_\ri}
\;\;\;;\;\;\;
\rvt_{\rz_\rj}
=
\frac{\alphah_{z_j}}{\betah_{z_j}}
=
\rvh_{\rz_\rj}
\label{Eq:VepsVfExpectanciesIGSMGen_bis_IS}
\eeq
Using the notation introduced in \eqref{Eq:q1_Integral_Not1_IS}:
\beq
\rVbt_{\rxi}=
\begin{bmatrix}
\frac{\alphah_{\xi_1}}{\betah_{\xi_1}} \ldots 0 \ldots 0 \\
\vdots \ddots \vdots \ddots \vdots \\
0 \ldots \frac{\alphah_{\xi_j}}{\betah_{\xi_j}} \ldots 0 \\
\vdots \ddots \vdots \ddots \vdots \\
0 \ldots 0 \ldots \frac{\alphah_{\xi_M}}{\betah_{\xi_M}} \\
\end{bmatrix}
=
\rVbh_{\rf}
\;\;;\;\;
\rVbt_{\repsilon}=
\begin{bmatrix}
\frac{\alphah_{\epsilon_1}}{\betah_{\epsilon_1}} \ldots 0 \ldots 0 \\
\vdots \ddots \vdots \ddots \vdots \\
0 \ldots \frac{\alphah_{\epsilon_i}}{\betah_{\epsilon_i}} \ldots 0 \\
\vdots \ddots \vdots \ddots \vdots \\
0 \ldots 0 \ldots \frac{\alphah_{\epsilon_N}}{\betah_{\epsilon_N}} \\
\end{bmatrix}
=
\rVbh_{\repsilon}
\;\;;\;\;
\rVbt_{\rz}=
\begin{bmatrix}
\frac{\alphah_{z_1}}{\betah_{z_1}} \ldots 0 \ldots 0 \\
\vdots \ddots \vdots \ddots \vdots \\
0 \ldots \frac{\alphah_{z_j}}{\betah_{z_j}} \ldots 0 \\
\vdots \ddots \vdots \ddots \vdots \\
0 \ldots 0 \ldots \frac{\alphah_{z_M}}{\betah_{z_M}} \\
\end{bmatrix}
=
\rVbh_{\rz}
\label{Eq:V_Expectancies_bis_IS}
\eeq
In Equation~\eqref{Eq:V_Expectancies_bis_IS} other notations are introduced for $\rVbt_{\rxi}$,  $\rVbt_{\repsilon}$ and $\rVbt_{\rz}$. Both values were expressed during the model via unknown expectancies, but via Equation~\eqref{Eq:V_Expectancies_bis_IS} those values don't contain any more integrals to be computed. Therefore, the new notations represent the final analytical expressions used for expressing the density functions $q_i$.
Using Equation~\eqref{Eq:V_Expectancies_bis_IS} and Equations~\eqref{Eq:q1_Analytical_1bis_IS},~\eqref{Eq:q2_Analytical_1bis_IS},~\eqref{Eq:q3_Analytical_1bis_IS},~\eqref{Eq:q4_Analytical_1bis_IS} and~\eqref{Eq:q5_Analytical_1bis_IS}, the final analytical expressions of the separable distributions $q_i$ are presented in Equations~\eqref{Eq:q1_Analytical_2bis_IS},~\eqref{Eq:q2_Analytical_2bis_IS},~\eqref{Eq:q3_Analytical_2bis_IS},~\eqref{Eq:q4_Analytical_2bis_IS} and~\eqref{Eq:q5_Analytical_2bis_IS}.
\begin{subequations}
\begin{align}
&q_{1j}(\rf_{\rj})=\Nc \left( \rf_{\rj} | \rfh_\rj,\Sigmah_{f_{jj}} \right),
\;\;\;
\left\{\barr{ll}
\rfh_\rj = \left( \bHb^{\bj T} \rVbh_{\repsilon} \bHb^{\bj} + \frac{\alphah_{\xi_j}}{\betah_{\xi_j}} \right)^{-1} \left[ \bHb^{\bj T} \rVbh_{\repsilon} \left( \bgb - \bHb^{-\bj} \rfbh^{-\rj} \right) + \frac{\alphah_{\xi_j}}{\betah_{\xi_j}} \bDb_{\bj} \rzbh \right]
\\[10pt]
\Sigmah_{f_{jj}} = \left( \bHb^{\bj T} \rVbh_{\repsilon} \bHb^{\bj} + \frac{\alphah_{\xi_j}}{\betah_{\xi_j}} \right)^{-1}
\earr\right.,
\label{Eq:q1_Analytical_2bis_IS}
\\[4pt]
&q_{2j}(\rz_{\rj})=\Nc \left( \rz_{\rj} | \rzh_\rj,\Sigmah_{z_{jj}} \right),
\;\;\;\;
\left\{\barr{ll}
\rzh_\rj = \left( \bDb^{\bj T} \rVbh_{\rxi} \bDb^{\bj} + \frac{\alphah_{z_j}}{\betah_{z_j}} \right)^{-1} \left[ \bDb^{\bj T}  \rVbh_{\rxi} \left( \rfbh - \bDb^{-\bj} \rzbh^{-\rj} \right)  \right]
\\[10pt]
\Sigmah_{z_{jj}} = \left( \bDb^{\bj T} \rVbh_{\rxi} \bDb^{\bj} + \frac{\alphah_{z_j}}{\betah_{z_j}} \right)^{-1}
\earr\right.
,
\label{Eq:q2_Analytical_2bis_IS}
\\[4pt]
&q_{3j}(\rv_{\rxi_\rj})=\Ic\Gc\left(\rv_{\rxi_\rj}|\alphah_{\xi_j},\betah_{\xi_j}\right),
\left\{\barr{ll}
\alphah_{\xi_j} = \alpha_{\xi} + \frac{3}{2}
\\[10pt]
\betah_{\xi_j} = \beta_{\xi} + \frac{1}{2} \left[ \left( \rfh_{\rj} - \bDb_{\bj} \rzbh \right)^2  + \bDb_\bj \Sigmabh_{z} \bDb_\bj^T \right]
\earr\right., j\in\left\lbrace 1, 2, \ldots, M \right\rbrace,
\label{Eq:q3_Analytical_2bis_IS}
\\[4pt]
&q_{4i}(\rv_{\repsilon_\ri})=\Ic\Gc\left(\rv_{\repsilon_\ri}|\alphah_{\epsilon_i},\betah_{\epsilon_i}\right), \;\;
\left\{\barr{ll}
\alphah_{\epsilon_i} = \alpha_{\epsilon} + \frac{3}{2}
\\[10pt]
\betah_{\epsilon_i} = \beta_{\epsilon} + \frac{1}{2} \left[ 
\left( \bg_\bi - \bHb_\bi \rfbh \right)^2 + \bHb_\bi \Sigmabh_{f} \bHb_\bi^T \right]
\earr\right.,i\in\left\lbrace 1, 2, \ldots, N \right\rbrace,
\label{Eq:q4_Analytical_2bis_IS}
\\[4pt]
&q_{5j}(\rv_{\rz_\rj})=\Ic\Gc\left(\rv_{\repsilon_\ri}|\alphah_{z_j},\betah_{z_j}\right),
\left\{\barr{ll}
\alphah_{z_j} = \alpha_{z} + \frac{3}{2}
\\[10pt]
\betah_{z_j} = \beta_{z} + \frac{1}{2} \left[ \rzh_{\rj}^2 + \Sigmah_{z_{jj}} \right].
\earr\right.,j\in\left\lbrace 1, 2, \ldots, M \right\rbrace.
\label{Eq:q5_Analytical_2bis_IS}
\end{align}
\end{subequations}
Equation~\eqref{Eq:q1_Analytical_2bis_IS} establishes the dependency between the parameters corresponding to the Normal distributions $q_{1j}(\rf_\rj)$ and the others parameters involved in the hierarchical model: the mean $\rfh_\rj$ and the variance $\Sigmah_{f_{jj}}$ depend on $\rVbh_{\repsilon}$, $\rVbh_{\rxi}$ which, via Equation~\eqref{Eq:V_Expectancies_bis_IS} are defined using $\left\lbrace \alphah_{\xi_j},\betah_{\xi_j}\right\rbrace, j\in \left\lbrace 1, 2, \ldots, M \right\rbrace $ and $\left\lbrace \alphah_{\epsilon_i},\betah_{\epsilon_i}\right\rbrace, i\in \left\lbrace 1, 2, \ldots, N \right\rbrace $ and $\rzbh$. The dependency between the parameters of the Normal distribution $q_{1j}(\rf_\rj)$ and the parameters of the Inverse Gamma distributions $q_{3j}(\rv_{\rxi_\rj}) ,j \in \left\lbrace 1,2,\ldots,M \right\rbrace$ and $q_{4i}(\rv_{\repsilon_\ri}), i \in \left\lbrace 1,2,\ldots,N \right\rbrace$ and the Normal distributions $q_{2j}(\rz_\rj)$ is presented in Figure~\eqref{Fig:Dependency_Scheme_bis_1_IS}.
\begin{figure}[!htb]
\center
\begin{tabular}{c}
\begin{picture}(220,30)
\put(0,0){\framebox(140,26){$\rzbh, \rfbh^{-\rj}, \alphah_{\xi_j}, \betah_{\xi_j}, \left\lbrace \alphah_{\epsilon_i},\betah_{\epsilon_i}\right\rbrace$}}
\put(142,13){\vector(1,0){26}}
\put(168,0){\framebox(56,26){$\rfh_\rj \; , \; \Sigmah_{f_{jj}}$}}
\end{picture}
\end{tabular}
\caption{Dependency between $q_{1j}(\rf_\rj)$ parameters and $q_{2j}(\rz_\rj)$, $q_{3j}(\rvb_{\rxi_\rj})$ and $q_{4i}(\rvb_{\repsilon_\ri})$ parameters}
\label{Fig:Dependency_Scheme_bis_1_IS}
\end{figure}
Equation~\eqref{Eq:q2_Analytical_2bis_IS} establishes the dependency between the parameters corresponding to the Normal distributions $q_{2j}(\rz_\rj)$ and the others parameters involved in the hierarchical model: the mean $\rzh_\rj$ and the variance $\Sigmah_{z_{jj}}$ depend on $\rVbh_{\rxi}$, $\rVbh_{\rz}$ which, via Equation~\eqref{Eq:V_Expectancies_bis_IS} are defined using $\left\lbrace \alphah_{\epsilon_i},\betah_{\epsilon_i}\right\rbrace, i \in \left\lbrace 1, 2, \ldots, N \right\rbrace $ and $\left\lbrace \alphah_{z_j},\betah_{z_j}\right\rbrace, j \in \left\lbrace 1, 2, \ldots, M \right\rbrace $ and $\rfbh$. The dependency between the parameters of the Normal distribution $q_{2j}(\rz_\rj)$ and the parameters of the Inverse Gamma distributions $q_{4i}(\rvb_{\repsilon_\ri}), i \in \left\lbrace 1,2,\ldots,N \right\rbrace$ and $q_{5j}(\rvb_{\rz_\rj}) ,j \in \left\lbrace 1,2,\ldots,M \right\rbrace$ and the Normal distributions $q_{1j}(\rf_\rj)$ is presented in Figure~\eqref{Fig:Dependency_Scheme_bis_2_IS}.
\begin{figure}[!htb]
\center
\begin{tabular}{c}
\begin{picture}(220,30)
\put(0,0){\framebox(140,26){$\rfbh,\rzbh^{-\rj}, \left\lbrace \alphah_{\epsilon_i},\betah_{\epsilon_i}\right\rbrace, \alphah_{z_j},\betah_{z_j}$}}
\put(142,13){\vector(1,0){26}}
\put(168,0){\framebox(56,26){$\rzh_{\rj} \; , \; \Sigmah_{z_{jj}}$}}
\end{picture}
\end{tabular}
\caption{Dependency between $q_{2j}(\rz_\rj)$ parameters and $q_{1j}(\rf_\rj)$, $q_{4i}(\rvb_{\repsilon_\ri})$ and $q_{5j}(\rvb_{\rz_\rj})$ parameters}
\label{Fig:Dependency_Scheme_bis_2_IS}
\end{figure}
Equation~\eqref{Eq:q3_Analytical_2bis_IS} establishes the dependency between the parameters corresponding to the Inverse Gamma distributions $q_{3j}(\rvb_{\rxi_\rj}) ,j \in \left\lbrace 1, 2, \ldots, M \right\rbrace$ and the others parameters involved in the hierarchical model: the shape and scale parameters $\left\lbrace \alphah_{\xi_j}, \betah_{\xi_j} \right\rbrace, j \in \left\lbrace 1, 2, \ldots, M \right\rbrace$ depend on the of the Normal distribution $q_{1j}(\rf_\rj)$, i.e. $\rfh_{\rj}$ and the mean $\rzb_\rj$ and the variance $\Sigmah_{z_{jj}}$ of the Normal distribution $q_{2j}(\rz_\rj)$, Figure~\eqref{Fig:Dependency_Scheme_bis_3_IS}.
\begin{figure}[!htb]
\center
\begin{tabular}{c}
\begin{picture}(120,30)
\put(0,0){\framebox(46,26){$\rfh_{\rj}, \; \rzbh, \; \Sigmabh_{z}$}}
\put(48,13){\vector(1,0){26}}
\put(76,0){\framebox(56,26){$ \alphah_{\xi_j},\betah_{\xi_j}$}}
\end{picture}
\end{tabular}
\caption{Dependency between $q_{3j}(\rvb_{\rxi_\rj})$ parameters and $q_{1j}(\rf_\rj)$ and $q_{2j}(\rz_\rj)$ parameters}
\label{Fig:Dependency_Scheme_bis_3_IS}
\end{figure}
Equation~\eqref{Eq:q4_Analytical_2bis_IS} establishes the dependency between the parameters corresponding to the Inverse Gamma distributions $q_{4i}(\rvb_{\repsilon_\ri}) ,i \in \left\lbrace 1,2,\ldots,N \right\rbrace$ and the others parameters involved in the hierarchical model: the shape and scale parameters $\left\lbrace \alphah_{\epsilon_i}, \betah_{\epsilon_i} \right\rbrace, i \in \left\lbrace 1, 2, \ldots, N \right\rbrace$ depend on the mean $\rfh_\rj$ and the variance $\Sigmah_{f_{jj}}$ of the Normal distribution $q_{1j}(\rf_\rj)$, Figure~\eqref{Fig:Dependency_Scheme_bis_4_IS}.
\begin{figure}[!htb]
\center
\begin{tabular}{c}
\begin{picture}(120,30)
\put(0,0){\framebox(46,26){$ \rfbh \; , \; \Sigmabh_{f}$}}
\put(48,13){\vector(1,0){26}}
\put(76,0){\framebox(56,26){$ \alphah_{\epsilon_i},\betah_{\epsilon_i}$}}
\end{picture}
\end{tabular}
\caption{Dependency between $q_{4i}(\rvb_{\repsilon_\ri})$ parameters and $q_{1j}(\rf_\rj)$ parameters}
\label{Fig:Dependency_Scheme_bis_4_IS}
\end{figure}
Equation~\eqref{Eq:q5_Analytical_2bis_IS} establishes the dependency between the parameters corresponding to the Inverse Gamma distributions $q_{5j}(\rvb_{\rz_\rj}) ,j \in \left\lbrace 1,2,\ldots,M \right\rbrace$ and the others parameters involved in the hierarchical model: the shape and scale parameters $\left\lbrace \alphah_{z_j}, \betah_{z_j} \right\rbrace, j \in \left\lbrace 1, 2, \ldots, M \right\rbrace$ depend on the mean $\rzh_\rj$ and the variance $\Sigmah_{z_{jj}}$ corresponding to the Normal distribution $q_{2j}(\rz_\rj)$, Figure~\eqref{Fig:Dependency_Scheme_5_IS}.
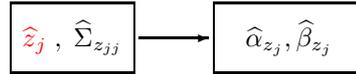
\begin{figure}[!htb]
\center
\begin{tabular}{c}
\begin{picture}(120,30)
\put(0,0){\framebox(46,26){$ \rzh_{\rj} \; , \; \Sigmah_{z_{jj}}$}}
\put(48,13){\vector(1,0){26}}
\put(76,0){\framebox(56,26){$\alphah_{z_j},\betah_{z_j}$}}
\end{picture}
\end{tabular}
\caption{Dependency between $q_{5j}(\rvb_{\rz_\rj})$ parameters and $q_{2j}(\rz_\rj)$ parameters}
\label{Fig:Dependency_Scheme_bis_5_IS}
\end{figure}
\newline
The iterative algorithm obtained via PM estimation is presented Figure~\eqref{Fig:IA_PM_St_nsStL_un_bis_IS}.
\tikzstyle{cloud} = [draw=blue!30!green,fill=orange!12, ellipse, thick, node distance=7em, text width=12em, text centered, minimum height=4em, minimum width=12em]
\tikzstyle{boxbig} = [draw=blue!30!green, fill=orange!12, rectangle, rounded corners, thick, node distance=8.5em, text width=32em, text centered, minimum height=3.6em, minimum width=32em]
\tikzstyle{boxhuge} = [draw=blue!30!green, fill=orange!12, rectangle, rounded corners, thick, node distance=12.6em, text width=38em, text centered, minimum height=3.2em, minimum width=38em]
\tikzstyle{boxsmall} = [draw=blue!30!green, fill=orange!12, rectangle, rounded corners, thick, node distance=9em, text width=9em, text centered, minimum height=4em, minimum width=9em]
\tikzstyle{line} = [draw=blue!30!green, -latex']
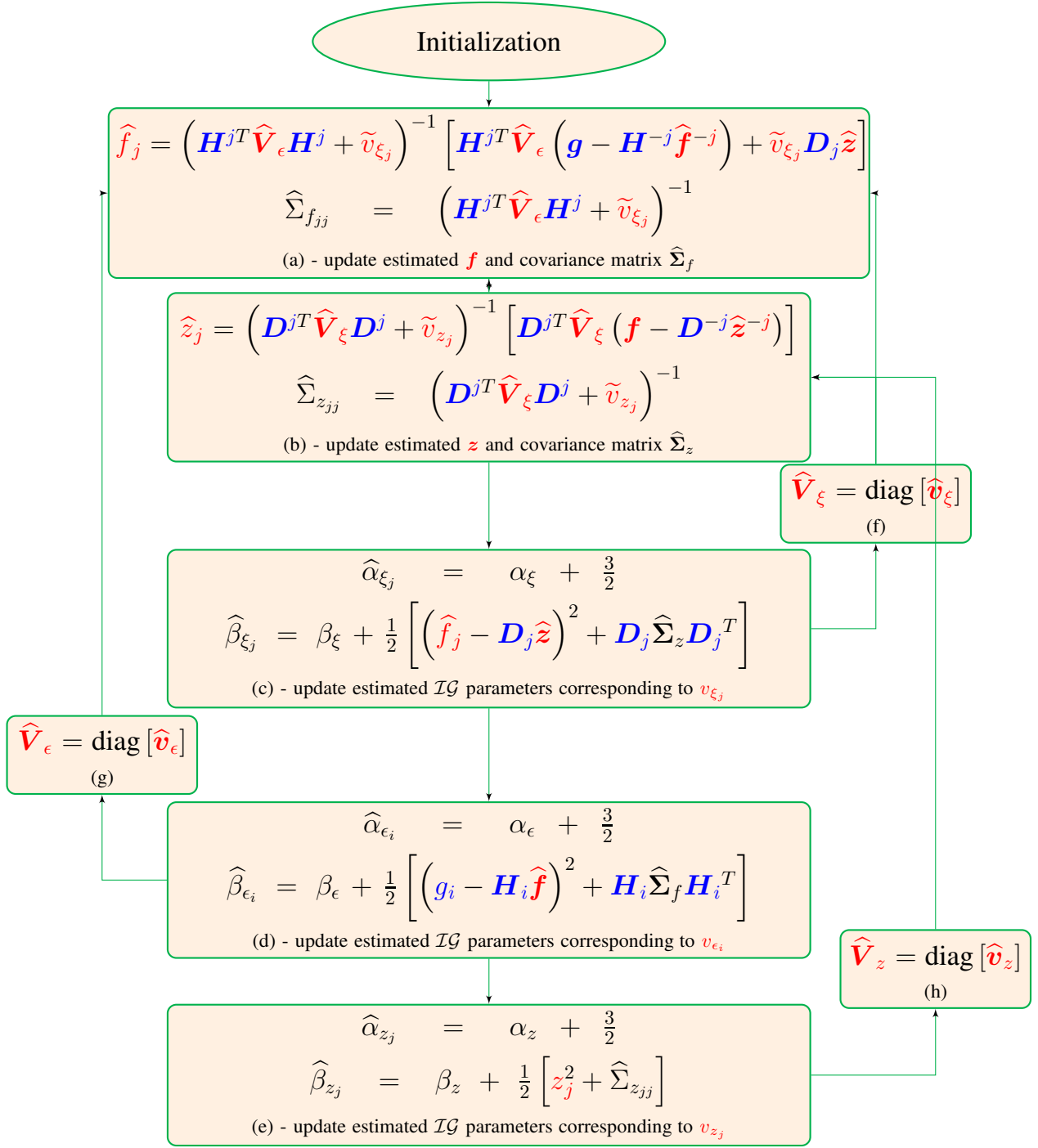
\begin{figure}[!htb]
\centering
\begin{center}
\begin{tikzpicture}[auto]
    \node [boxhuge, scale=0.9] (f) {\Large{$\rfh_\rj = \left( \bHb^{\bj T} \rVbh_{\repsilon} \bHb^{\bj} + \rvt_{\rxi_\rj} \right)^{-1} \left[ \bHb^{\bj T} \rVbh_{\repsilon} \left( \bgb - \bHb^{-\bj} \rfbh^{-\rj} \right) + \rvt_{\rxi_\rj} \bDb_{\bj} \rzbh \right]$}
\\[4.4pt]
\Large{$\Sigmah_{f_{jj}} = \left( \bHb^{\bj T} \rVbh_{\repsilon} \bHb^{\bj} + \rvt_{\rxi_\rj} \right)^{-1}$}    
\\[4.4pt]
\normalsize{(a) - update estimated $\rfb$ and covariance matrix $\Sigmabh_{f}$} 
};

	\node [boxbig, below of=f, scale=0.9] (z) {\Large{$\rzh_\rj = \left( \bDb^{\bj T} \rVbh_{\rxi} \bDb^{\bj} + \rvt_{\rz_\rj} \right)^{-1} \left[ \bDb^{\bj T}  \rVbh_{\rxi} \left( \rfb - \bDb^{-\bj} \rzbh^{-\rj} \right)  \right]$}
\\[4.4pt]
\Large{$\Sigmah_{z_{jj}} = \left( \bDb^{\bj T} \rVbh_{\rxi} \bDb^{\bj} + \rvt_{\rz_\rj} \right)^{-1}$}    
\\[4.4pt]
\normalsize{(b) - update estimated $\rzb$ and covariance matrix $\Sigmabh_{z}$} 
};

    \node [boxbig, below of=z, node distance=11.6em, scale=0.9] (vxi)
{\Large{$ \alphah_{\xi_j} = \alpha_{\xi} + \frac{3}{2} $}
\\[5.1pt]
\Large{$\betah_{\xi_j} = \beta_{\xi} + \frac{1}{2} \left[ \left( \rfh_{\rj} - \bDb_{\bj} \rzbh \right)^2  + \bDb_\bj \Sigmabh_{z} \bDb_\bj^T \right]$}
\\[5.1pt]
\normalsize{(c) - update estimated $\Ic\Gc$ parameters corresponding to $\rv_{\rxi_\rj}$} 
};

    \node [boxbig, below of=vxi, node distance=11.6em, scale=0.9] (veps) {\Large{$\alphah_{\epsilon_i} = \alpha_{\epsilon} + \frac{3}{2}$}
\\[5.1pt]
\Large{$\betah_{\epsilon_i} = \beta_{\epsilon} + \frac{1}{2} \left[ 
\left( \bg_\bi - \bHb_\bi \rfbh \right)^2 + \bHb_\bi \Sigmabh_{f} \bHb_\bi^T \right]$}
\\[5.1pt]
\normalsize{(d) - update estimated $\Ic\Gc$ parameters corresponding to $\rv_{\repsilon_\ri}$}     
};

    \node [boxbig, below of=veps, node distance=9em, scale=0.9] (vz) {\Large{$\alphah_{z_j} = \alpha_{z} + \frac{3}{2}$}
\\[5.1pt]
\Large{$\betah_{z_j} = \beta_{z} + \frac{1}{2} \left[ \rz_{\rj}^2 + \Sigmah_{z_{jj}} \right]$}
\\[5.1pt]
\normalsize{(e) - update estimated $\Ic\Gc$ parameters corresponding to $\rv_{\rz_\rj}$}     
};

    \node [boxsmall, above right = 0.09cm and -0.5cm of vxi, node distance=17em, scale=0.9] (Vxi) {\Large{$\rVbh_\rxi = \diag {\rvbh_{\rxi}}$}
\\[5pt]
\normalsize{(f)}
};

    \node [boxsmall, below left = -3.95cm and -0.5cm of veps, node distance=17em, scale=0.9] (Veps) {\Large{$\rVbh_\repsilon = \diag {\rvbh_{\repsilon}}$}
\\[5pt]
\normalsize{(g)} 
};

    \node [boxsmall, above right = -0.08cm and 0.46cm of vz, node distance=17em, scale=0.9] (Vz) {\Large{$\rVbh_\rz = \diag {\rvbh_{\rz}}$}
\\[5pt]
\normalsize{(h)} 
};

\node [cloud, above of=f, scale=0.9] (init)
{\Large{Initialization}};  

    \path [line] (f) -- (z)[near start];
    \path [line] (z) -- (vxi);
    \path [line] (vxi) -- (veps);
    \path [line] (veps) -- (vz);    
    \path [line] (Vxi) |- (f);
    \path [line] (Vxi) |- (z);
    \path [line] (Veps) |- (f);
    \path [line] (Vz) |- (z);    
    \path [line] (vxi) -| (Vxi);
    \path [line] (veps) -| (Veps);
    \path [line] (vz) -| (Vz);
    \path [line] (z) -| (f);
    \path [line] (init) -- (f);   
\end{tikzpicture}
\end{center}
\caption{Indirect sparsity (via $\rzb$) - Iterative algorithm corresponding to full separability PM estimation for Student-t hierarchical model, non-stationary Student-t uncertainties model}
\label{Fig:IA_PM_St_nsStL_un_bis_IS}
\end{figure}
More formal, the iterative algorithm obtained via PM estimation is presented Algorithm~\eqref{Alg:IA_PM_St_nsStL_un_bis_IS}.
\begin{algorithm}
\caption{PM via VBA full sep. - Student-t hierarchical model, non-stationary Student-t uncertainties model}
\begin{algorithmic}[1]
\Ensure INITIALIZATION $\alphah_{\xi_j}^{(0)},\betah_{\xi_j}^{(0)},\alphah_{\epsilon_i}^{(0)},\betah_{\epsilon_i}^{(0)},\alpha_{z_j}^{(0)},\beta_{z_j}^{(0)}$ 

\Comment{Where $\alphah_{\xi_j}^{(0)}=\alpha_{\xi},
\betah_{\xi_j}^{(0)}=\beta_{\xi},
\alphah_{\epsilon_i}^{(0)}=\alpha_{\epsilon},
\betah_{\epsilon_i}^{(0)}=\beta_{\epsilon},
\alpha_{z_j}^{(0)}=\alpha_{z},
\beta_{z_j}^{(0)}=\beta_{z}$}

\Comment{Leads to $\rVbh_\rxi^{(0)} = \frac{\alpha_{\xi}}{\beta_{\xi}}\Ib,
\rVbh_\repsilon^{(0)} = \frac{\alpha_{\epsilon}}{\beta_{\epsilon}}\Ib,
\rVbh_\rz^{(0)} = \frac{\alpha_{z}}{\beta_{z}} \Ib$}

\Comment{$\alpha_{\xi}, \beta_{\xi},\alpha_{\epsilon}, \beta_{\epsilon}, \alpha_{z}, \beta_{z}$ are set in modeling phase, Eq.~\eqref{Eq:St_nsStL_un_IS}}

\Function{PMviaVBAfs}{$\alpha_{\xi},\beta_{\xi},\alpha_{\epsilon},\beta_{\epsilon},\alpha_{z},\beta_{z},\bgb,\bHb, \bDb, \rzb^{(0)}, \rfb^{(0)}, M, N, NoIter$} 

\For{$n = 0$ to ${IterNumb}$}

\For{$j = 1$ to ${M}$}

\State $\Sigmah_{f_{jj}}^{(n+1)} = \left( \bHb^{\bj T} \diag {\frac{\alpha_{\epsilon}+\frac{3}{2}}{\betah_{\epsilon_i}^{(n)}}} \bHb^{\bj} + \frac{\alphah_{\xi}+\frac{3}{2}}{\betah_{\xi_j}^{(n)}} \right)^{-1}$ \Comment{New value $\Sigmah_{f_{jj}}^{(n+1)}$}

\State $\rfh_\rj^{(n+1)} = \Sigmah_{f_{jj}}^{(n+1)} \left[ \bHb^{\bj T} \diag {\frac{\alpha_{\epsilon}+\frac{3}{2}}{\betah_{\epsilon_i}^{(n)}}} \left( \bgb - \bHb^{-\bj} \left( \rfbh^{-\rj} \right)^{(n)} \right) + \frac{\alphah_{\xi}+\frac{3}{2}}{\betah_{\xi_j}^{(n)}} \bDb_{\bj} \rzbh^{(n)} \right]$\Comment{New value $\rfh_\rj^{(n+1)}$}

\EndFor

\For{$j = 1$ to ${M}$}

\State $\Sigmah_{z_{jj}}^{(n+1)} = \left( \bDb^{\bj T} \diag {\frac{\alpha_{\xi}+\frac{3}{2}}{\betah_{\xi_i}^{(n)}}} \bDb^{\bj} + \frac{\alphah_{z}+\frac{3}{2}}{\betah_{z_j}^{(n)}} \right)^{-1}$ \Comment{New value $\Sigmah_{z_{jj}}^{(n+1)}$}

\State $\rzh_\rj^{(n+1)} = \Sigmah_{z_{jj}}^{(n+1)} \left[ \bDb^{\bj T}  \diag {\frac{\alpha_{\xi}+\frac{3}{2}}{\betah_{\xi_i}^{(n)}}} \left( \rfbh^{(n+1)} - \bDb^{-\bj} \left( \rzbh^{-\rj} \right)^{(n)} \right)  \right]$ \Comment{New value $\rzh_\rj^{(n+1)}$}

\EndFor

\For{$j = 1$ to ${M}$}

\State $\betah_{\xi_j}^{(n+1)} = \beta_{\xi} + \frac{1}{2} \left[ \left( \rfh_{\rj}^{(n+1)} - \bDb_{\bj} \rzbh^{(n+1)} \right)^2  + \bDb_\bj \Sigmabh_{z}^{(n+1)} \bDb_{\bj}^T \right]$

\EndFor


\For{$i = 1$ to ${N}$}

\State $\betah_{\epsilon_i}^{(n+1)} = \beta_{\epsilon} + \frac{1}{2} \left[ \left( \bg_\bi - \bHb_\bi \rfbh^{(n+1)} \right)^2 + \bHb_\bi \Sigmabh_{f}^{(n+1)} \bHb_\bi^T \right]$ 

\EndFor


\For{$j = 1$ to ${M}$}

\State $\betah_{z_j}^{(n+1)} = \beta_{z} + \frac{1}{2} \left[ \left( \rzh_{\rj}^{(n+1)} \right)^2 + \Sigmah_{z_{jj}}^{(n+1)} \right]$

\EndFor


\EndFor

\Return $\left( \rfbh^{(n+1)}, \Sigmabh_{f}^{(n+1)}, \rzbh^{(n+1)}, \Sigmabh_{z}^{(n+1)}, , \betah_{\xi_j}^{(n+1)}, \betah_{\epsilon_i}^{(n+1)},\betah_{z_j}^{(n+1)} \right)$ , $n = NoIter$
\EndFunction
\end{algorithmic}
\label{Alg:IA_PM_St_nsStL_un_bis_IS}
\end{algorithm}

\subsection{Student-t hierarchical model: stationary Laplace uncertainties model, unknown uncertainties variance}
\label{Subsec:St_sLL_un}
\begin{itemize}
\item the hierarchical model is using as a \textbf{prior} the \textbf{Student-t} distribution;
\item the Student-t prior distribution is expressed via \textbf{StPM}, Equation~\eqref{Eq:StPM1}, considering the variance $\rvb_\rf$ as unknown;
\item the \textbf{likelihood} is derived from the distribution proposed for modelling the uncertainties vector $\repsilonb$;
\item for the uncertainties vector $\repsilonb$ a \textbf{stationary Laplace uncertainties model} is proposed, i.e. a multivariate Laplace distribution expressed via \textbf{LPM} is used under the following two assumptions:\\
a) each element of the uncertainties vector has the same \textbf{variance}, $\rv_{\repsilon}$;\\
b) the variance $\rv_{\repsilon}$ is \textbf{unknown}; 
\end{itemize}
\beq
\rotatebox{90}{\hspace{-1.35cm}Student-t sLL UNK} \;
\vline
\left.\barr{ll}
Likelihood: \textbf{sLL:}\;\;
\left\{\barr{ll}
p\left( \bgb | \rfb, \rv_{\repsilon} \right) = \Nc \left( \bgb | \bHb\rfb, \rv_{\repsilon}^{-1} \Ib \right), 
\\[7pt]
p\left( \rv_{\repsilon} | 1, \frac{b_{\epsilon}}{2} \right)
= \Ic\Gc\left( \rv_{\repsilon} | 1, \frac{b_{\epsilon}}{2} \right).\\[4pt]
\earr\right.
\\[19pt]
Prior:\;\;\;\;\;\;\;\textbf{StPM:}\;\;
\left\{\barr{ll}
p(\rfb|0,\rvb_{\rf}) = \Nc(\rfb|0, \; \rVb_{\rf}) \propto \det{\rVb_{\rf}}^{-\frac{1}{2}} \; \exp \left\lbrace - \frac{1}{2}\| \rVb_{\rf}^{-\frac{1}{2}} \rfb \| \right\rbrace,
\\[7pt]
p(\rvb_{\rf}|\alpha_{f}, \beta_{f}) = \prod_{j=1}^{M} \Ic\Gc(\rv_{\rf_\rj}|\alpha_{f}, \beta_{f}) \propto \prod_{j=1}^{M}\rv_{\rf_\rj}^{-\alpha_{f}-1} \; \exp \left\lbrace -\sum_{j=1}^{M}\frac{\beta_{f}}{\rv_{\rf_\rj}} \right\rbrace,\\[7pt]
\hspace{5.6cm}
\rVb_{\rf} = \diag{\rvb_{\rf}},\; \rvb_{\rf} = \left[\ldots,\rv_{\rf_{\rj}}, \ldots\right].\\[4pt]
\earr\right.
\earr\right.
\label{Eq:St_sLL_un}
\eeq

\subsection{Student-t hierarchical model: non-stationary Laplace uncertainties model, unknown uncertainties variances}
\label{Subsec:St_nsLL_un}
\begin{itemize}
\item the hierarchical model is using as a \textbf{prior} the \textbf{Student-t} distribution;
\item the Student-t prior distribution is expressed via \textbf{StPM}, Equation~\eqref{Eq:StPM1}, considering the variance $\rvb_\rf$ as unknown;
\item the \textbf{likelihood} is derived from the distribution proposed for modelling the uncertainties vector $\repsilonb$;
\item for the uncertainties vector $\repsilonb$ a \textbf{non-stationary Laplace uncertainties model} is proposed, i.e. a multivariate Laplace distribution expressed via \textbf{LPM} is used under the following assumption:\\
a) the variance vector $\rvb_{\repsilon}$ is \textbf{unknown}; 
\end{itemize}
\beq
\rotatebox{90}{\hspace{-1.55cm}Student-t nsLL UNK} \;
\vline
\left.\barr{ll}
Likelihood: \textbf{nsLL:}\;
\left\{\barr{ll}
p\left( \bgb | \rfb, \rvb_{\repsilon} \right) = \Nc \left( \bgb | \bHb\rfb, \rVb_{\repsilon}^{-1} \right), 
\\[7pt]
p\left( \rvb_{\repsilon} | 1, \frac{b_{\epsilon}}{2} \right)
= \prod_{i=1}^{N} \Ic\Gc\left( \rv_{\repsilon_\ri} | 1, \frac{b_{\epsilon}}{2} \right),\\[7pt]
\rVb_{\repsilon} = \diag{\rvb_{\repsilon}},\; \rvb_{\repsilon} = \left[\ldots,\rv_{\repsilon_{\ri}}, \ldots\right],\\[4pt]
\earr\right.
\\[27pt]
Prior:\;\;\;\;\;\;\;\textbf{StPM:}\;\;\;
\left\{\barr{ll}
p(\rfb|0,\rvb_{\rf}) = \Nc(\rfb|0, \; \rVb_{\rf}) \propto \det{\rVb_{\rf}}^{-\frac{1}{2}} \; \exp \left\lbrace - \frac{1}{2}\| \rVb_{\rf}^{-\frac{1}{2}} \rfb \| \right\rbrace,
\\[6.7pt]
p(\rvb_{\rf}|\alpha_{f}, \beta_{f}) = \prod_{j=1}^{M} \Ic\Gc(\rv_{\rf_\rj}|\alpha_{f}, \beta_{f}) \propto  \prod_{j=1}^{M} \rv_{\rf_\rj}^{-\alpha_{f}-1} \; \exp \left\lbrace - \sum_{j=1}^{M}\frac{\beta_{f}}{\rv_{\rf_\rj}} \right\rbrace,\\[7pt]
\hspace{5.6cm}
\rVb_{\rf} = \diag{\rvb_{\rf}},\; \rvb_{\rf} = \left[\ldots,\rv_{\rf_{\rj}}, \ldots\right].\\[4pt]
\earr\right.
\earr\right.
\label{Eq:St_nsLL_un}
\eeq

\section{Hierarchical Models with Laplace prior via LPM}
Considering the Laplace distribution for modelling the sparse structure of $\rfb$ in linear forward model, Equation~\eqref{Eq:LinearModel}, expressed via the conjugate priors LPM, depending on the model proposed for the uncertainties of the model, $\repsilonb$, and implicitly the corresponding likelihood, sGL or nsGL, different hierarchical models can be proposed:

\subsection{Laplace hierarchical model: stationary Gaussian uncertainties model, known uncertainties variance}
\label{Subsec:L_sGL_k}
\begin{itemize}
\item the hierarchical model is using as a \textbf{prior} the \textbf{Laplace} distribution;
\item the Laplace prior distribution is expressed via \textbf{LPM}, Equation~\eqref{Eq:LPM1}, considering the variance $\rvb_\rf$ unknown;
\item the \textbf{likelihood} is derived from the distribution proposed for modelling the uncertainties vector $\repsilonb$;
\item for the uncertainties vector $\repsilonb$ a \textbf{stationary Gaussian uncertainties model} is proposed, i.e. a multivariate Gaussian distribution is used under the following two assumptions:\\
a) each element of the uncertainties vector has the same \textbf{variance}, $v_{\epsilon}$;\\
b) the variance $v_{\epsilon}$ is \textbf{known}; 
\end{itemize}
\beq
\rotatebox{90}{\hspace{-1.1cm}Laplace sGL K} \;
\vline
\left.\barr{ll}
Likelihood: \textbf{sGL:}\;\;\;\;\;\;\;
p\left( \bgb | \rfb, v_{\epsilon} \right)
=
\Nc\left( \bgb | \bHb\rfb, v_{\epsilon} \Ib \right)
\propto v_{\epsilon}^{-\frac{N}{2}} \; \exp \left\lbrace - \frac{1}{2 v_{\epsilon}}\| \left( \bgb - \bHb\rfb \right) \| \right\rbrace.
\\[17pt]
Prior:\;\;\;\;\;\;\;\textbf{LPM:}\;\;
\left\{\barr{ll}
p(\rfb|0,\rvb_{\rf}) = \Nc(\rfb|0, \; \rVb_{\rf}^{-1}) \propto \det{\rVb_{\rf}}^{\frac{1}{2}} \; \exp \left\lbrace - \frac{1}{2}\| \rVb_{\rf}^{\frac{1}{2}} \rfb \| \right\rbrace,
\\[7pt]
p(\rvb_{\rf}|b_f) = \prod_{j=1}^{M} \Ic\Gc(\rv_{\rf_\rj}|1, \; \frac{b_f}{2}) \propto \prod_{j=1}^{M} \rv_{\rf_\rj}^{-2} \; \exp \left\lbrace - \sum_{j=1}^{M} \frac{b_f}{2\rv_{\rf_\rj}} \right\rbrace, \\[7pt]
\hspace{4.4cm}
\rVb_{\rf} = \diag{\rvb_{\rf}},\; \rvb_{\rf} = \left[\ldots,\rv_{\rf_{\rj}}, \ldots\right].\\[4pt]
\earr\right.
\earr\right.
\label{Eq:L_sGL_k}
\eeq

\subsection{Laplace hierarchical model: non-stationary Gaussian uncertainties model, known uncertainties variances}
\label{Subsec:L_nsGL_k}
\begin{itemize}
\item the hierarchical model is using as a \textbf{prior} the \textbf{Laplace} distribution;
\item the Laplace prior distribution is expressed via \textbf{LPM}, Equation~\eqref{Eq:LPM1}, considering the variance $\rvb_\rf$ unknown;
\item the \textbf{likelihood} is derived from the distribution proposed for modelling the uncertainties vector $\repsilonb$;
\item for the uncertainties vector $\repsilonb$ a \textbf{stationary Gaussian uncertainties model} is proposed, i.e. a multivariate Gaussian distribution is used under the following assumption:\\
a) the variance vector $\vb_{\epsilon}$ is \textbf{known}; 
\end{itemize}
\beq
\rotatebox{90}{\hspace{-1.15cm}Laplace nsGL K} \;
\vline
\left.\barr{ll}
Likelihood: \textbf{nsGL:}\;
\left\{\barr{ll}
\begin{split}
p\left( \bgb | \rfb, \vb_{\epsilon} \right) 
=
\Nc\left( \bgb | \bHb\rfb, \Vb_{\epsilon} \right) 
\propto 
\prod_{i=1}^{N} & v_{\epsilon_i}^{-\frac{1}{2}}
\exp \left\lbrace - \frac{1}{2}\| \Vb_{\epsilon}^{-\frac{1}{2}} \left( \bgb - \bHb\rfb \right) \| \right\rbrace,
\\
&
\Vb_{\epsilon} = \diag{\vb_{\epsilon}},\; \vb_{\epsilon} = \left[\ldots v_{\epsilon_{i}}, \ldots\right]. 
\end{split}
\earr\right.
\\[25pt]
Prior:\;\;\;\;\;\;\;\textbf{LPM:}\;\;
\left\{\barr{ll}
p(\rfb|0,\rvb_{\rf}) = \Nc(\rfb|0, \; \rVb_{\rf}^{-1}) \propto \det{\rVb_{\rf}}^{\frac{1}{2}} \; \exp \left\lbrace - \frac{1}{2}\| \rVb_{\rf}^{\frac{1}{2}} \rfb \| \right\rbrace,
\\[7pt]
p(\rvb_{\rf}|b_f) = \prod_{j=1}^{M} \Ic\Gc(\rv_{\rf_\rj}|1, \; \frac{b_f}{2}) \propto \prod_{j=1}^{M} \rv_{\rf_\rj}^{-2} \; \exp \left\lbrace - \sum_{j=1}^{M} \frac{b_f}{2\rv_{\rf_\rj}} \right\rbrace, \\[7pt]
\hspace{4.4cm}
\rVb_{\rf} = \diag{\rvb_{\rf}},\; \rvb_{\rf} = \left[\ldots,\rv_{\rf_{\rj}}, \ldots\right].\\[4pt]
\earr\right.
\earr\right.
\label{Eq:L_nsGL_k}
\eeq

\subsection{Laplace hierarchical model: stationary Gaussian uncertainties model, unknown uncertainties variance}
\label{Subsec:L_sGL_un}
\begin{itemize}
\item the hierarchical model is using as a \textbf{prior} the \textbf{Laplace} distribution;
\item the Laplace prior distribution is expressed via \textbf{LPM}, Equation~\eqref{Eq:LPM1}, considering the variance $\rvb_\rf$ as unknown;
\item the \textbf{likelihood} is derived from the distribution proposed for modelling the uncertainties vector $\repsilonb$;
\item for the uncertainties vector $\repsilonb$ a \textbf{stationary Gaussian uncertainties model} is proposed, i.e. a multivariate Gaussian distribution is used under the following two assumptions:\\
a) each element of the uncertainties vector has the same \textbf{variance}, $\rv_{\repsilon}$;\\
b) the variance $\rv_{\repsilon}$ is \textbf{unknown}; 
\end{itemize}
\beq
\rotatebox{90}{\hspace{-1.35cm}Laplace sGL UNK} \;
\vline
\left.\barr{ll}
Likelihood: \textbf{sGL:}\;\;\;\;\;\;\;
p\left( \bgb | \rfb, \rv_{\repsilon} \right)
=
\Nc\left( \bgb | \bHb\rfb, \rv_{\repsilon} \Ib \right)
\propto \rv_{\repsilon}^{-\frac{N}{2}} \; \exp \left\lbrace - \frac{1}{2\rv_{\repsilon}}\| \left( \bgb - \bHb\rfb \right) \| \right\rbrace.
\\[17pt]
Prior:\;\;\;\;\;\;\;\textbf{LPM:}\;\;
\left\{\barr{ll}
p(\rfb|0,\rvb_{\rf}) = \Nc(\rfb|0, \; \rVb_{\rf}^{-1}) \propto \det{\rVb_{\rf}}^{\frac{1}{2}} \; \exp \left\lbrace - \frac{1}{2}\| \rVb_{\rf}^{\frac{1}{2}} \rfb \| \right\rbrace,
\\[7pt]
p(\rvb_{\rf}|b_f) = \prod_{j=1}^{M} \Ic\Gc(\rv_{\rf_\rj}|1, \; \frac{b_f}{2}) \propto \prod_{j=1}^{M} \rv_{\rf_\rj}^{-2} \; \exp \left\lbrace - \sum_{j=1}^{M} \frac{b_f}{2\rv_{\rf_\rj}} \right\rbrace, \\[7pt]
\hspace{4.4cm}
\rVb_{\rf} = \diag{\rvb_{\rf}},\; \rvb_{\rf} = \left[\ldots,\rv_{\rf_{\rj}}, \ldots\right].\\[4pt]
\earr\right.
\earr\right.
\label{Eq:L_sGL_un}
\eeq

\subsection{Laplace hierarchical model: non-stationary Gaussian uncertainties model, unknown uncertainties variances}
\label{Subsec:L_nsGL_un}
\begin{itemize}
\item the hierarchical model is using as a \textbf{prior} the \textbf{Laplace} distribution;
\item the Laplace prior distribution is expressed via \textbf{LPM}, Equation~\eqref{Eq:LPM1}, considering the variance $\rvb_\rf$ as unknown;
\item the \textbf{likelihood} is derived from the distribution proposed for modelling the uncertainties vector $\repsilonb$;
\item for the uncertainties vector $\repsilonb$ a \textbf{stationary Gaussian uncertainties model} is proposed, i.e. a multivariate Gaussian distribution is used under the following assumption:\\
a) the variance vector $\rvb_{\repsilon}$ is \textbf{unknown}; 
\end{itemize}
\beq
\rotatebox{90}{\hspace{-1.55cm}Laplace nsGL UNK} \;
\vline
\left.\barr{ll}
Likelihood: \textbf{nsGL:}\;
\left\{\barr{ll}
\begin{split}
p\left( \bgb | \rfb, \rv_{\repsilon} \right) 
=
\Nc\left( \bgb | \bHb\rfb, \rVb_{\repsilon} \right) 
\propto 
\prod_{i=1}^{N} & \rv_{\repsilon_\ri}^{-\frac{1}{2}}
\exp \left\lbrace - \frac{1}{2}\| \rVb_{\repsilon}^{-\frac{1}{2}} \left( \bgb - \bHb\rfb \right) \| \right\rbrace,
\\
&
\rVb_{\repsilon} = \diag{\rvb_{\repsilon}},\; \rvb_{\repsilon} = \left[\ldots,\rv_{\repsilon_{\ri}}, \ldots\right]. 
\end{split}
\earr\right.
\\[25pt]
Prior:\;\;\;\;\;\;\;\textbf{LPM:}\;\;
\left\{\barr{ll}
p(\rfb|0,\rvb_{\rf}) = \Nc(\rfb|0, \; \rVb_{\rf}^{-1}) \propto \det{\rVb_{\rf}}^{\frac{1}{2}} \; \exp \left\lbrace - \frac{1}{2}\| \rVb_{\rf}^{\frac{1}{2}} \rfb \| \right\rbrace,
\\[7pt]
p(\rvb_{\rf}|b_f) = \prod_{j=1}^{M} \Ic\Gc(\rv_{\rf_\rj}|1, \; \frac{b_f}{2}) \propto \prod_{j=1}^{M} \rv_{\rf_\rj}^{-2} \; \exp \left\lbrace - \sum_{j=1}^{M} \frac{b_f}{2\rv_{\rf_\rj}} \right\rbrace, \\[7pt]
\hspace{4.4cm}
\rVb_{\rf} = \diag{\rvb_{\rf}},\; \rvb_{\rf} = \left[\ldots,\rv_{\rf_{\rj}}, \ldots\right].\\[4pt]
\earr\right.
\earr\right.
\label{Eq:L_nsGL_un}
\eeq

\subsection{Laplace hierarchical model: stationary Student-t uncertainties model, unknown uncertainties variance}
\label{Subsec:L_sStL_un}
\begin{itemize}
\item the hierarchical model is using as a \textbf{prior} the \textbf{Laplace} distribution;
\item the Laplace prior distribution is expressed via \textbf{LPM}, Equation~\eqref{Eq:LPM1}, considering the variance $\rvb_\rf$ as unknown;
\item the \textbf{likelihood} is derived from the distribution proposed for modelling the uncertainties vector $\repsilonb$;
\item for the uncertainties vector $\repsilonb$ a \textbf{stationary Student-t uncertainties model} is proposed, i.e. a multivariate Student distribution expressed via \textbf{StPM} is used under the following two assumptions:\\
a) each element of the uncertainties vector has the same \textbf{variance}, $\rv_{\repsilon}$;\\
b) the variance $\rv_{\repsilon}$ is \textbf{unknown}; 
\end{itemize}
\beq
\rotatebox{90}{\hspace{-1.35cm}Laplace sStL UNK} \;
\vline
\left.\barr{ll}
Likelihood: \textbf{sStL:}\;
\left\{\barr{ll}
p\left( \bgb | \rfb, \rv_{\repsilon} \right)
=
\Nc\left( \bgb | \bHb\rfb, \rv_{\repsilon} \Ib \right)
\propto \rv_{\repsilon}^{-\frac{N}{2}} \; \exp \left\lbrace - \frac{1}{2\rv_{\repsilon}}\| \left( \bgb - \bHb\rfb \right) \| \right\rbrace,
\\[7pt]
p\left( \rv_{\repsilon} | \alpha_{\epsilon}, \beta_{\epsilon} \right) = 
\Ic\Gc\left( \rv_{\repsilon} | \alpha_{\epsilon}, \beta_{\epsilon} \right) \propto \rv_{\repsilon}^{-\alpha_{\epsilon}-1} \; \exp \left\lbrace \frac{\beta_{\epsilon}}{\rv_{\repsilon}} \right\rbrace,\\[4pt]
\earr\right.
\\[25pt]
Prior:\;\;\;\;\;\;\;\textbf{LPM:}\;\;
\left\{\barr{ll}
p(\rfb|0,\rvb_{\rf}) = \Nc(\rfb|0, \; \rVb_{\rf}^{-1}) \propto \det{\rVb_{\rf}}^{\frac{1}{2}} \; \exp \left\lbrace - \frac{1}{2}\| \rVb_{\rf}^{\frac{1}{2}} \rfb \| \right\rbrace,
\\[7pt]
p(\rvb_{\rf}|b_f) = \prod_{j=1}^{M} \Ic\Gc(\rv_{\rf_\rj}|1, \; \frac{b_f}{2}) \propto \prod_{j=1}^{M} \rv_{\rf_\rj}^{-2} \; \exp \left\lbrace - \sum_{j=1}^{M} \frac{b_f}{2\rv_{\rf_\rj}} \right\rbrace, \\[7pt]
\hspace{4.4cm}
\rVb_{\rf} = \diag{\rvb_{\rf}},\; \rvb_{\rf} = \left[\ldots,\rv_{\rf_{\rj}}, \ldots\right].\\[4pt]
\earr\right.
\earr\right.
\label{Eq:L_sStL_un}
\eeq

\subsection{Laplace hierarchical model: non-stationary Student-t uncertainties model, unknown uncertainties variances}
\label{Subsec:L_nsStL_un}
\begin{itemize}
\item the hierarchical model is using as a \textbf{prior} the \textbf{Laplace} distribution;
\item the Laplace prior distribution is expressed via \textbf{LPM}, Equation~\eqref{Eq:LPM1}, considering the variance $\rvb_\rf$ as unknown;
\item the \textbf{likelihood} is derived from the distribution proposed for modelling the uncertainties vector $\repsilonb$;
\item for the uncertainties vector $\repsilonb$ a \textbf{non-stationary Student-t uncertainties model} is proposed, i.e. a multivariate Student-t distribution expressed via \textbf{StPM} is used under the following assumption:\\
a) the variance vector $\rvb_{\repsilon}$ is \textbf{unknown}; 
\end{itemize}
\beq
\rotatebox{90}{\hspace{-1.55cm}Laplace nsStL UNK} \;
\vline
\left.\barr{ll}
Likelihood: \textbf{nsStL:}\;
\left\{\barr{ll}
\begin{split}
p\left( \bgb | \rfb, \rv_{\repsilon} \right) 
=
\Nc\left( \bgb | \bHb\rfb, \rVb_{\repsilon} \right) 
\propto 
\prod_{i=1}^{N} & \rv_{\repsilon_\ri}^{-\frac{1}{2}}
\exp \left\lbrace - \frac{1}{2}\| \rVb_{\repsilon}^{-\frac{1}{2}} \left( \bgb - \bHb\rfb \right) \| \right\rbrace, 
\end{split} 
\\[7pt]
p\left( \rvb_{\repsilon} | \alpha_{\epsilon}, \beta_{\epsilon} \right)
= \prod_{i=1}^{N} \Ic\Gc\left( \rv_{\repsilon_\ri} | \alpha_{\epsilon}, \beta_{\epsilon} \right) \propto  \prod_{i=1}^{N} \rv_{\repsilon_\ri}^{-\alpha_{\epsilon}-1} \; \exp \left\lbrace - \sum_{i=1}^{N}\frac{\beta_{\epsilon}}{\rv_{\repsilon_\ri}} \right\rbrace,\\[7pt]
\hspace{5.55cm}
\rVb_{\repsilon} = \diag{\rvb_{\repsilon}},\; \rvb_{\repsilon} = \left[\ldots,\rv_{\repsilon_{\ri}}, \ldots\right],\\[4pt]
\earr\right.
\\[33pt]
Prior:\;\;\;\;\;\;\;\textbf{LPM:}\;\;
\left\{\barr{ll}
p(\rfb|0,\rvb_{\rf}) = \Nc(\rfb|0, \; \rVb_{\rf}^{-1}) \propto \det{\rVb_{\rf}}^{\frac{1}{2}} \; \exp \left\lbrace - \frac{1}{2}\| \rVb_{\rf}^{\frac{1}{2}} \rfb \| \right\rbrace,
\\[7pt]
p(\rvb_{\rf}|\beta_f) = \prod_{j=1}^{M} \Ic\Gc(\rv_{\rf_\rj}|1, \; \frac{\beta_f}{2}) \propto \prod_{j=1}^{M} \rv_{\rf_\rj}^{-2} \; \exp \left\lbrace - \sum_{j=1}^{M} \frac{\beta_f}{2\rv_{\rf_\rj}} \right\rbrace, \\[7pt]
\hspace{4.4cm}
\rVb_{\rf} = \diag{\rvb_{\rf}},\; \rvb_{\rf} = \left[\ldots,\rv_{\rf_{\rj}}, \ldots\right].\\[4pt]
\earr\right.
\earr\right.
\label{Eq:L_nsStL_un}
\eeq
The hierarchical model build over the linear forward model, Equation~\eqref{Eq:LinearModel}, using as a prior for $\rfb$ a Laplace distribution, expressed via the Laplace Prior Model (LPM), Equation~\eqref{Eq:LPM1} and modelling the uncertainties of the model $\repsilonb$ using the non-stationary Student Uncertainties Model (nsStUM), Equation~\eqref{Eq:nsStUM}, is presented in Equation~\eqref{Eq:L_nsStL_un}. The posterior distribution is obtained via the Bayes rule, Equation~\eqref{Eq:Post_L_nsStL_un}:
\beq
\begin{split}
p(\rfb, \rvb_\rf, \rvb_\repsilon | \bgb, \beta_{f}, \alpha_{\epsilon} \beta_{\epsilon})
& \propto 
\; 
p\left( \bgb | \rfb, \rvb_{\repsilon} \right) 
\;
p\left( \rvb_{\repsilon} | \beta_{\epsilon} \right) 
\;
p(\rfb|0,\rvb_{\rf}) 
\;
p(\rvb_{\rf} | \beta_{f})\\
& \propto 
\;
\prod_{i=1}^{N}  \rv_{\repsilon_\ri}^{-\frac{1}{2}}
\;
\exp \left\lbrace - \frac{1}{2}\| \rVb_{\repsilon}^{-\frac{1}{2}} \left( \bgb - \bHb\rfb \right) \| \right\rbrace 
\;
\prod_{i=1}^{N} \rv_{\repsilon_\ri}^{-\left( \alpha_{\epsilon} + 1 \right)} 
\;
\exp \left\lbrace - \sum_{i=1}^{N}\frac{\beta_{\epsilon}}{\rv_{\repsilon_\ri}} \right\rbrace
\\
&
\;\;\;\;\;
\prod_{j=1}^{M}  \rv_{\rf_\ri}^{\frac{1}{2}} 
\;
\exp \left\lbrace - \frac{1}{2}\| \rVb_{\rf}^{\frac{1}{2}} \rfb \| \right\rbrace
\;
\prod_{j=1}^{M} \rv_{\rf_\rj}^{-2}
\;
\exp \left\lbrace - \frac{1}{2} \sum_{j=1}^{M}\frac{\beta_{f}}{\rv_{\rf_\rj}} \right\rbrace
\end{split}
\label{Eq:Post_L_nsStL_un}
\eeq
The goal is to estimate the unknowns of the hierarchical model, namely $\rfb$, the main unknown of the linear forward model, Equation~\eqref{Eq:LinearModel} which was suppose sparse, and consequently modelled via the Laplace distribution and the two variances appearing in the hierarchical model, Equation~\eqref{Eq:L_nsStL_un}, the variance corresponding to the sparse structure $\rfb$, namely $\rvb_\rf$ and the variance corresponding to uncertainties of model $\repsilonb$, namely $\rvb_\repsilon$. 
\subsubsection{Joint MAP estimation}
\label{Subsubsec:JMAP_L_nsStL_un}
First, the Joint Maximum A Posterior (JMAP) estimation is considered: the unknowns are estimated on the basis of the available data $\bgb$, by maximizing the posterior distribution:
\beq
\left( \rfbh, \; \rvbh_\rf, \; \rvbh_\repsilon \right)
=
\operatorname*{arg\,max}_{\left( \rfb, \; \rvb_\rf, \; \rvb_\repsilon \right)} p(\rfb, \; \rvb_\rf, \; \rvb_\repsilon | \bgb, \; \beta_{f}, \; \alpha_{\epsilon}, \; \beta_{\epsilon})
=
\operatorname*{arg\,min}_{\left( \rfb, \; \rvb_\rf, \; \rvb_\repsilon \right)} \Lc(\rfb, \; \rvb_\rf, \; \rvb_\repsilon),
\eeq 
where for the second equality the criterion $\Lc(\rfb, \rvb_{\rf}, \rvb_{\repsilon})$ is defined as:
\beq
\Lc(\rfb, \rvb_{\rf}, , \rvb_{\repsilon})= -\ln  p(\rfb, \; \rvb_\rf, \; \rvb_\repsilon | \bgb, \; \beta_{f}, \; \alpha_{\epsilon}, \; \beta_{\epsilon})
\label{Eq:Def_L_Crit_L_nsStL_un}
\eeq
The MAP estimation corresponds to the solution minimizing the criterion $\Lc(\rfb, \rvb_{\rf}, \rvb_{\repsilon})$.
From the analytical expression of the posterior distribution, Equation~\eqref{Eq:Post_L_nsStL_un} and the definition of the criterion $\Lc$, Equation~\eqref{Eq:Def_L_Crit_L_nsStL_un}, we obtain:
\beq
\begin{split}
\Lc(\rfb, \rvb_{\repsilon}, \rvb_{\rf})
=
-\ln  p(\rfb, \rvb_{\repsilon}, \rvb_{\rf}|\bgb \; \beta_{f}, \; \alpha_{\epsilon}, \; \beta_{\epsilon})
&
=
\frac{1}{2} \| \rVb_{\repsilon}^{-\frac{1}{2}} \left( \bgb - \bHb\rfb \right) \|
+
\left( \alpha_{\epsilon} + \frac{3}{2} \right) \sum_{i=1}^{N} \ln \rv_{\repsilon_\ri} 
+
\sum_{i=1}^{N}\frac{\beta_{\epsilon}}{\rv_{\repsilon_\ri}} 
\\
&
+
\frac{1}{2} \| \rVb_{\rf}^{\frac{1}{2}} \rfb \|
+
\frac{3}{2} \sum_{j=1}^{M} \ln \rv_{\rf_\rj}
+
\frac{1}{2} \sum_{j=1}^{M}\frac{\beta_{f}}{\rv_{\rf_\rj}}
\end{split}
\label{Eq:L_Crit_L_nsStL_un}
\eeq
One of the simplest optimisation algorithm that can be used is an alternate optimization of the criterion $\Lc(\rfb, \rvb_{\repsilon}, \rvb_{\rf})$ with respect to the each unknown:
\begin{itemize}
\item With respect to $\rfb$:
\beq
\begin{split}
\frac{\partial \Lc(\rfb, \rvb_{\rf}, \rvb_{\repsilon})}{\partial \rfb}=0
&\Leftrightarrow
\frac{\partial}{\partial \rfb}
\left(
\| \rVb_\repsilon^{-\frac{1}{2}} \left( \bgb - \bHb\rfb \right) \|
+
\|\rVb_\rf^{\frac{1}{2}}\rfb\|
\right)
=
-\bHb^T \rVb_\repsilon^{-1} \left( \bgb - \bHb \rfb \right) + \rVb_\rf \rfb = 0
\\
&
\Leftrightarrow 
\left( \bHb^T \rVb_\repsilon^{-1} \bHb + \rVb_\rf \right) \rfb  = 
\bHb^T \rVb_\repsilon^{-1} \bgb
\\
&
\Rightarrow
\rfbh 
=
\left( \bHb^T \rVb_\repsilon^{-1} \bHb + \rVb_\rf \right)^{-1} \bHb^T \rVb_\repsilon^{-1} \bgb
\end{split}
\nonumber
\eeq
\item With respect to $\rv_\rf$, $j \in \left\lbrace 1,2,\ldots, M \right\rbrace$:
\beq
\begin{split}
\frac{\partial \Lc(\rfb, \rvb_{\rf}, \rvb_{\rf})}{\partial \rv_{\rf_\rj}}=0
&
\Leftrightarrow
\frac{\partial}{\partial \rv_{\rf_\rj}}
\left(
3 \ln \rv_{\rf_\rj}
+
\beta_f \rv_{\rf_\rj}^{-1} 
+ 
\rf_\rj^2 \rv_{\rf_\rj}
\right)
=0
\\
&
\Leftrightarrow
\rf_\rj^2 \rv_{\rf_\rj}^{2}
+ 
3 \rv_{\rf_\rj}
-
\beta_f  
=0
\\
&
\Rightarrow
\rvh_{\rf_\rj} =
\frac{-3 \pm \sqrt{9 - 4 \rf_\rj^2 \beta_f}}{2 \rf_\rj^2}
\end{split}
\nonumber
\eeq
\item With respect to $\rv_{\repsilon_\ri}$, $i \in \left\lbrace 1,2,\ldots, N \right\rbrace$:\\
First, we develop the norm $\| \rVb_\repsilon^{-\frac{1}{2}} \left( \bgb - \bHb\rfb \right) \|$:
\beq
\begin{split}
\| \rVb_\repsilon^{-\frac{1}{2}} \left( \bgb - \bHb\rfb \right) \|
&= 
\bgb^{T} \rVb_\repsilon^{-1} \bgb
-
2 \bgb^{T} \rVb_\repsilon^{-1} \bHb \rfb
+
\bHb^{T} \rfb^{T} \rVb_\repsilon^{-1} \bHb \rfb
\\
&=
\sum_{i=1}^{N} \rv_{\repsilon_\ri}^{-1} \bg_\bi^{2}
-
2 \sum_{i=1}^{N} \rv_{\repsilon_\ri}^{-1} \bg_\bi \bH_\bi \rfb 
+
\sum_{i=1}^{N} \rv_{\repsilon_\ri}^{-1} \rfb^{T} \bH_\bi^{T} \bH_\bi \rfb,
\nonumber
\end{split}
\eeq
where $\bHb_\bi$ denotes the line i of the matrix $\bHb$, $i\in \left\lbrace 1,2,\ldots,N \right\rbrace$, i.e. $\bHb_\bi = \left[ \bh_{\bi\blue{1}}, \bh_{\bi\blue{1}}, \ldots, \bh_{\bi\bM} \right]$. 
\beq
\begin{split}
\frac{\partial \Lc(\rfb, \rvb_{\rf}, \rvb_{\repsilon})}{\partial \rv_{\repsilon_\ri}}=0
&
\Leftrightarrow
\frac{\partial}{\partial \rv_{\repsilon_\ri}}
\left[
\left( \alpha_{\epsilon} + \frac{3}{2} \right)\ln \rv_{\repsilon_\ri}
+
\left(\beta_{\epsilon} + \frac{1}{2} \left( \bg_\bi^{2} - 2 \bg_\bi \bHb_\bi \rfb  + \rfb^{T} \bHb_\bi^{T} \bHb_\bi \rfb \right)
\right) \rv_{\repsilon_\ri}^{-1}
\right]
=0
\\
&
\Leftrightarrow
\left( \alpha_{\epsilon} + \frac{3}{2} \right) \rv_{\repsilon_\ri}
-
\left(\beta_{\epsilon} + \frac{1}{2} \left( \bg_\bi - \bHb_\bi \rfb  \right)^2 \right)=0
\\
&
\Rightarrow
\rvh_{\repsilon_\ri} = \frac{\beta_{\epsilon} + \frac{1}{2} \left( \bg_\bi - \bHb_\bi \rfb \right)^2}{\alpha_{\epsilon} + \frac{3}{2}}
\end{split}
\nonumber
\eeq
\end{itemize}
The iterative algorithm obtained via JMAP estimation is presented Figure~\eqref{Fig:IA_JMAP_L_nsStL_un}.
\tikzstyle{cloud} = [draw=blue!30!green,fill=orange!12, ellipse, thick, node distance=6em, text width=12em, text centered, minimum height=4em, minimum width=12em]
\tikzstyle{boxbig} = [draw=blue!30!green, fill=orange!12, rectangle, rounded corners, thick, node distance=6.5em, text width=25em, text centered, minimum height=5em, minimum width=25em]
\tikzstyle{box} = [draw=blue!30!green, fill=orange!12, rectangle, rounded corners, thick, node distance=6.5em, text width=21em, text centered, minimum height=5em, minimum width=21em]
\tikzstyle{boxsmall} = [draw=blue!30!green, fill=orange!12, rectangle, rounded corners, thick, node distance=8.5em, text width=9em, text centered, minimum height=5em, minimum width=9em]
\tikzstyle{line} = [draw=blue!30!green, -latex']
\begin{figure}[!htb]
\centering
\begin{center}
\begin{tikzpicture}[auto]
    \node [boxbig, scale=0.9] (f) {\Large{$\rfbh 
=
\left( \bHb^T \rVbh_\repsilon^{-1} \bHb + \rVbh_\rf \right)^{-1} \bHb^T \rVbh_\repsilon^{-1} \bgb$}
\\[8pt]
\normalsize{(a) - update estimated $\rfb$} 
};

    \node [box, below of=f, node distance=9.6em, scale=0.9] (vf)
{\Large{$ \rvh_{\rf_\rj} = \frac{-3 \pm \sqrt{9 - 4 \rf_\rj^2 \beta_f}}{2 \rf_\rj^2} $}
\\[8pt]
\normalsize{(b) - update estimated variances $\rv_{\rf_\rj}$} 
};

    \node [box, below of=vf, node distance=9.6em, scale=0.9] (veps) {\Large{$\rvh_{\repsilon_\ri} = \frac{\beta_{\epsilon} + \frac{1}{2} \left( \bg_\bi - \bHb_\bi \rfb \right)^2}{\alpha_{\epsilon} + \frac{3}{2}}$}
\\[8pt]
\normalsize{(c) - update estimated uncertainties variances $\rv_{\repsilon_\ri}$}     
};

    \node [boxsmall, above right = 0.05cm and 0.5cm of vf, node distance=17em, scale=0.9] (Vf) {\Large{$\rVbh_\rf = \diag {\rvbh_{\rf}}$}
\\[8pt]
(d) 
};

    \node [boxsmall, below left = -3.4cm and 0.5cm of veps, node distance=17em, scale=0.9] (Veps) {\Large{$\rVbh_\repsilon = \diag {\rvbh_{\repsilon}}$}
\\[8pt]
(e) 
};

\node [cloud, above of=f, scale=0.9] (init)
{\Large{Initialization}};  

    \path [line] (f) -- (vf)[near start];
    \path [line] (Vf) |- (f);
    \path [line] (vf) -| (Vf);
    \path [line] (vf) -- (veps);
    \path [line] (Veps) |- (f);
    \path [line] (veps) -| (Veps);
    \path [line] (init) -- (f);
    
\end{tikzpicture}
\end{center}
\caption{Iterative algorithm corresponding to Joint MAP estimation for Laplace hierarchical model, non-stationary Student-t uncertainties model}
\label{Fig:IA_JMAP_L_nsStL_un}
\end{figure}
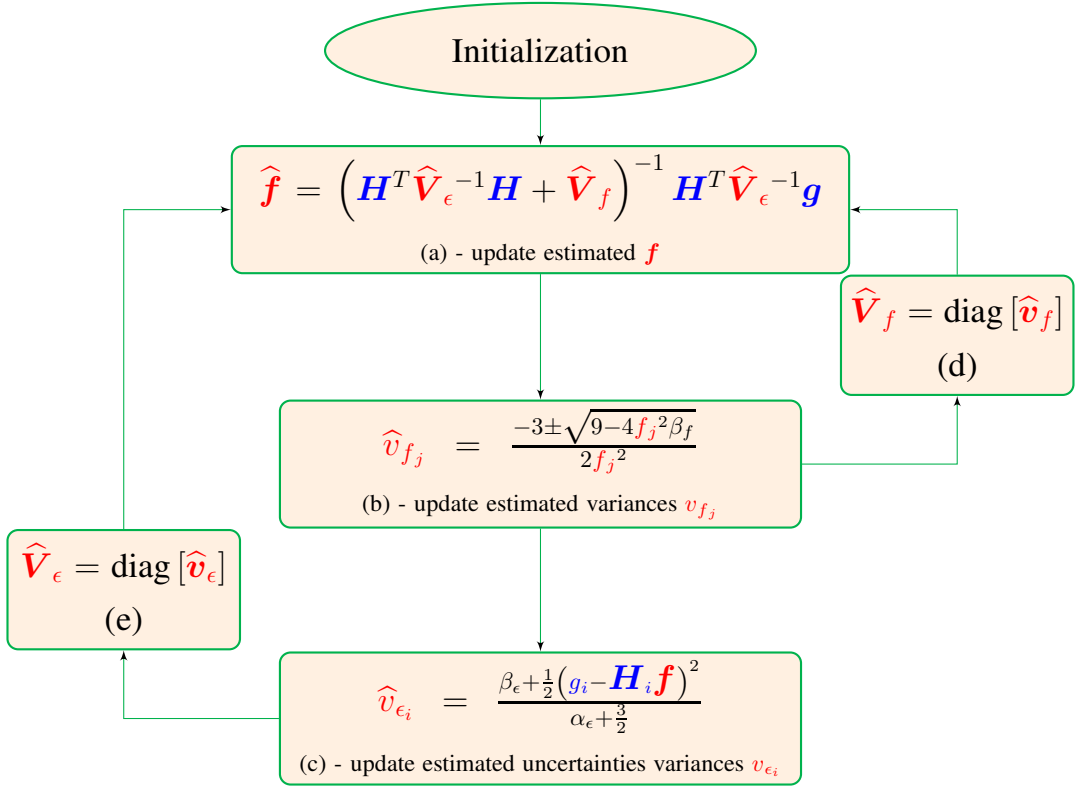
\subsubsection{Posterior Mean estimation via VBA, partial separability}
\label{Subsubsec:PM_PS_L_nsStL_un}
In this subsection, the Posterior Mean (PM) estimation is considered. The Joint MAP computes the mod of the posterior distribution. The PM computes the mean of the posterior distribution. One of the advantages of this estimator is that it minimizes the Mean Square Error (MSE). Computing the posterior means of any unknown needs great dimensional integration. For example, the mean corresponding to $\rfb$ can be computed from the posterior distribution using Equation~\eqref{Eq:PM_Comput_F_L_nsStL_un},
\beq
E_p \left\lbrace \rfb \right\rbrace = \iiint \rfb \; p(\rfb, \rvb_\rf, \rvb_\repsilon | \bgb, \beta_{f}, \alpha_{\epsilon}, \beta_{\epsilon}) \d\rfb \d\rvb_{\rf} \d\rvb_{\repsilon}.
\label{Eq:PM_Comput_F_L_nsStL_un}
\eeq
In general, these computations are not easy. One way to obtain approximate estimates is to approximate $p(\rfb, \rvb_\rf, \rvb_\repsilon | \bgb, \beta_{f}, \beta_{\epsilon})$ by a separable one $q(\rfb, \rvb_\rf, \rvb_\repsilon | \bgb, \beta_{f} , \alpha_{\epsilon}, \beta_{\epsilon}) = q_1(\rfb) \; q_2(\rvb_{\rf}) \; q_3(\rvb_{\repsilon}) $, then computing the posterior means using the separability. The mean corresponding to $\rfb$ is computed using the corresponding separable distribution $q_1(\rfb)$, Equation~\eqref{Eq:PM_Comput_F_L_nsStL_un_Separable}, 
\beq
E_{q_{1}} \left\lbrace \rfb \right\rbrace = \int \rfb \; q_1(\rfb) \d\rfb.
\label{Eq:PM_Comput_F_L_nsStL_un_Separable}
\eeq
If the approximation of the posterior distribution with a separable one can be done in such a way that conserves the mean, i.e. Equation~\eqref{Eq:PM_Conservation_L_nsStL_un},
\beq
E_q \left\lbrace x \right\rbrace = E_p \left\lbrace x \right\rbrace, 
\label{Eq:PM_Conservation_L_nsStL_un}
\eeq
for all the unknowns of the model, a great amount of computational cost is gained. In particular, for the proposed hierarchical model, Equation~\eqref{Eq:St_nsStL_un}, the posterior distribution, Equation~\eqref{Eq:Post_L_nsStL_un}, is not a separable one, making the analytical computations of the PM very difficult. One way the compute the PM in this case is to first approximate the posterior law $p(\rfb, \rvb_\rf, \rvb_\repsilon | \bgb, \beta_{f} ,\alpha_{\epsilon}, \beta_{\epsilon})$ with a separable law $q(\rfb, \rvb_\rf, \rvb_\repsilon | \bgb, \beta_{f}, \alpha_{\epsilon}, \beta_{\epsilon})$, Equation~\eqref{Eq:Post_Approx_L_nsStL_un},
\beq
p(\rfb, \rvb_\rf, \rvb_\repsilon | \bgb, \beta_{f}, \alpha_{\epsilon} ,  \beta_{\epsilon}) 
\approx 
q(\rfb, \rvb_\rf, \rvb_\repsilon | \bgb, \beta_{f}, \alpha_{\epsilon} , \beta_{\epsilon})
=
q_1(\rfb) \; q_2(\rvb_{\rf}) \; q_3(\rvb_{\repsilon})\label{Eq:Post_Approx_L_nsStL_un}
\eeq
where the notations from Equation~\eqref{Eq:Post_Approx_L_nsStL_un_Notations1} are used
\beq
q_{2}(\rvb_{\rf}) = \prod_{j=1}^{M} q_{2j}(\rv_{\rf_\rj}),
\;\;;\;\;
q_{3}(\rvb_{\repsilon}) = \prod_{i=1}^{N} q_{3i}(\rv_{\repsilon_\ri})
\label{Eq:Post_Approx_L_nsStL_un_Notations1}
\eeq
by minimizing of the Kullback-Leibler divergence, defined as:
\beq
\begin{split}
\mbox{KL} & \left( q(\rfb, \rvb_\rf, \rvb_\repsilon | \bgb, \beta_{f}, \alpha_{\epsilon} , \beta_{\epsilon}) : p(\rfb, \rvb_\rf, \rvb_\repsilon | \bgb, \beta_{f}, \alpha_{\epsilon} , \beta_{\epsilon}) \right) =
\\
&=\iint \ldots \int q(\rfb, \rvb_\rf, \rvb_\repsilon | \bgb, \beta_{f}, \alpha_{\epsilon} , \beta_{\epsilon}) \; \ln\frac{q(\rfb, \rvb_\rf, \rvb_\repsilon | \bgb, \beta_{f}, \alpha_{\epsilon} , \beta_{\epsilon})} {p(\rfb, \rvb_\rf, \rvb_\repsilon | \bgb, \beta_{f}, \alpha_{\epsilon} , \beta_{\epsilon})} \d \rfb \d \rvb_{\repsilon} \d \rvb_{\rf}
\label{Eq:Kull-Leib_L_nsStL_un}
\end{split}
\eeq
where the notations from Equation~\eqref{Eq:Post_Approx_L_nsStL_un_Notations2} are used
\beq
\d \rvb_{\rf} = \prod_{j=1}^{M} \d \rv_{\rf_\rj}
\;\;;\;\;
\d \rvb_{\repsilon} = \prod_{i=1}^{N} \d \rv_{\repsilon_\ri}.
\label{Eq:Post_Approx_L_nsStL_un_Notations2}
\eeq
Equation~\eqref{Eq:Post_Approx_L_nsStL_un_Notations1} is selecting a partial separability for the approximated posterior distribution $q(\rfb, \rvb_\rf, \rvb_\repsilon | \bgb,  \beta_{f}, \alpha_{\epsilon}, \beta_{\epsilon})$ in the sense that a total separability is imposed for the distributions corresponding to the two variances appearing in the hierarchical model, $q_2\left( \rvb_{\rf} \right)$ and $q_3\left( \rvb_{\repsilon} \right)$ but not for the distribution corresponding to $\rfb$. Evidently, a full separability can be imposed, by adding the supplementary condition $q_{1}(\rfb) = \prod_{j=1}^{M} q_{1j}(\rf_\rj)$ in Equation~\eqref{Eq:Post_Approx_L_nsStL_un_Notations1}. This case is considered in Subsection~\eqref{Subsubsec:PM_FS_L_nsStL_un}. The minimization can be done via alternate optimization resulting the following proportionalities from Equations~\eqref{Eq:VBA_Proport1_L_nsStL_un},~\eqref{Eq:VBA_Proport2_L_nsStL_un} and~\eqref{Eq:VBA_Proport3_L_nsStL_un},  
\begin{subequations}
\begin{align}
&q_1(\rfb) \;\;\; \propto \; \exp \left\lbrace \biggl\langle \ln p(\rfb, \rvb_\rf, \rvb_\repsilon | \bgb, \beta_{f}, \alpha_{\epsilon}, \beta_{\epsilon}) \biggr\rangle_{ q_2(\rvb_{\rf}) \; q_3(\rvb_{\repsilon})} \right\rbrace,
\label{Eq:VBA_Proport1_L_nsStL_un}
\\[4pt]
&q_{2j}(\rv_{\rf_\red{j}}) \propto \; \exp \left\lbrace \biggl\langle \ln p(\rfb, \rvb_\rf, \rvb_\repsilon | \bgb, \beta_{f}, \alpha_{\epsilon}, \beta_{\epsilon}) \biggr\rangle_{q_1(\rfb) \; q_{2-j}(\rv_{\rf_\red{j}}) \; q_3(\rvb_{\repsilon})} \right\rbrace,\;\;j \in \left\lbrace 1,2 \ldots, M \right\rbrace,
\label{Eq:VBA_Proport2_L_nsStL_un}
\\[4pt]
&q_{3i}(\rv_{\repsilon_\ri}) \; \propto \; \exp \left\lbrace \biggl\langle \ln p(\rfb, \rvb_\rf, \rvb_\repsilon | \bgb, \beta_{f}, \alpha_{\epsilon}, \beta_{\epsilon}) \biggr\rangle_{q_1(\rfb) \; q_2(\rvb_{\rf}) \; q_{3-i}(\rv_{\repsilon_\ri})} \right\rbrace,\;\;i \in \left\lbrace 1,2 \ldots, N \right\rbrace,
\label{Eq:VBA_Proport3_L_nsStL_un}
\end{align}
\end{subequations}
using the notations:
\beq
q_{2-j}(\rv_{\rf_\red{j}})=\displaystyle \prod_{k=1,k \neq j}^{M} q_{2k}(\rv_{\rf_\red{k}})
\;\;\;;\;\;\;
q_{3-i}(\rv_{\repsilon_\ri})=\displaystyle \prod_{k=1,k \neq i}^{N} q_{3k}(\rv_{\repsilon_\rk})
\label{Eq:Def_q_Min_L_nsStL_un}
\eeq
and also
\beq
\biggl\langle u(x) \biggr\rangle_{v(y)}= \displaystyle \int u(x) v(y) \d y. 
\label{Eq:Def_Integ_L_nsStL_un}
\eeq
Via Equation~\eqref{Eq:Def_L_Crit_L_nsStL_un} and Equation~\eqref{Eq:L_Crit_L_nsStL_un}, the analytical expression of logarithm of the posterior distribution is obtained, Equation~\eqref{Eq:Log_Post_L_nsStL_un}:
\beq
\begin{split}
\ln  p(\rfb, \rvb_{\repsilon}, \rvb_{\rf}|\bgb, \beta_{f}, \alpha_{\epsilon}, \beta_{\epsilon})
=
&
-\frac{1}{2}\| \rVb_{\repsilon}^{-\frac{1}{2}} \left( \bgb - \bHb\rfb \right) \|
-
\left(\alpha_{\epsilon} + \frac{3}{2} \right) \sum_{i=1}^{N} \ln \rv_{\repsilon_\ri} 
-
\sum_{i=1}^{N}\frac{\beta_{\epsilon}}{\rv_{\repsilon_\ri}} 
\\
&
-
\frac{1}{2}\| \rVb_{\rf}^{\frac{1}{2}} \rfb \|
-
\frac{3}{2} \sum_{j=1}^{M} \ln \rv_{\rf_\rj}
-
\frac{1}{2} \sum_{j=1}^{M}\frac{\beta_{f}}{\rv_{\rf_\rj}}
\end{split}
\label{Eq:Log_Post_L_nsStL_un}
\eeq
\paragraph{Computation of the analytical expression of $q_1(\rfb)$.}
The proportionality relation corresponding to $q_1(\rfb)$ is presented in established in Equation~\eqref{Eq:VBA_Proport1_L_nsStL_un}. In the expression of $\ln p\left(\rfb, \rvb_\rf, \rvb_\repsilon | \bgb, \beta_{f}, \alpha_{\epsilon}, \beta_{\epsilon} \right)$ all the terms free of $\rfb$ can be regarded as constants. Via Equation~\eqref{Eq:Log_Post_L_nsStL_un} the integral defined in Equation~\eqref{Eq:Def_Integ_L_nsStL_un} becomes:
\beq
\begin{split}
\biggl\langle \ln p(\rfb, \rvb_\rf, \rvb_\repsilon | \bgb, \beta_{f}, \alpha_{\epsilon}, \beta_{\epsilon}) \biggr\rangle_{q_2(\rvb_{\rf}) \; q_3(\rvb_{\repsilon})}
&
=
-\frac{1}{2}
\biggl\langle 
\|\rVb_{\repsilon}^{-\frac{1}{2}} \left(\bgb - \bHb\rfb\right) \|
\biggr\rangle_{ q_3(\rvb_{\repsilon}) }
-\frac{1}{2}
\biggl\langle 
\| \rVb_{\rf}^{\frac{1}{2}} \rfb \|
\biggl\rangle_{ q_2(\rvb_{\rf}) }.
\end{split}
\label{Eq:q1_Integ_1_L_nsStL_un}
\eeq
Introducing the notations:
\beq
\begin{split}
\rvt_{\rf_\rj} = \biggl\langle \rv_{\rf_\rj} \biggr\rangle_{q_{2j}\left( \rv_{\rf_\rj} \right)}
\;\;;\;\;
\rvbt_{\rf}\;&=\;
\begin{bmatrix}
\rvt_{\rf_\red{1}} \ldots 
\rvt_{\rf_\rj} \ldots 
\rvt_{\rf_\rM}
\end{bmatrix}^T
\;\;;\;\;
\rVbt_{\rf} = \diag {\rvbt_{\rf}}\;
\\
\rvt_{\repsilon_\ri}^{-1}
=
\biggl\langle \rv_{\repsilon_\ri}^{-1} \biggr\rangle_{q_{3i}\left( \rv_{\repsilon_\ri} \right)}
\;\;;\;\;
\rvbt_\repsilon^{-1}\;&=\;
\begin{bmatrix}
\rvt_{\repsilon_\red{1}}^{-1}
\ldots 
\rvt_{\repsilon_\ri}^{-1}
\ldots 
\rvt_{\repsilon_\rN}^{-1}
\end{bmatrix}^T
\;\;;\;\;
\rVbt_\repsilon^{-1} = \diag {\rvbt_\repsilon^{-1}}
\end{split}
\label{Eq:q1_Integ_Not1_L_nsStL_un}
\eeq
the integral from Equation~\eqref{Eq:q1_Integ_1_L_nsStL_un} becomes:
\beq
\begin{split}
\biggl\langle \ln p(\rfb, \rvb_\rf, \rvb_\repsilon | \bgb,  \beta_{f}, \alpha_{\epsilon}, \beta_{\epsilon}) \biggr\rangle_{q_2(\rvb_{\rf}) \; q_3(\rvb_{\repsilon})}
&
= -\frac{1}{2} \|\rVbt_{\repsilon}^{-\frac{1}{2}} \left(\bgb - \bHb\rfb\right) \| - \frac{1}{2} \| \rVbt_{\rf}^{\frac{1}{2}} \rfb \|.
\end{split}
\eeq
Noting that $\biggl\langle \ln p(\rfb, \rvb_\rf, \rvb_\repsilon | \bgb, \alpha_{\epsilon}, \beta_{f}, \beta_{\epsilon}) \biggr\rangle_{q_2(\rvb_{\rf}) \; q_3(\rvb_{\repsilon})}
$ is a quadratic criterion and considering the proportionality from Equation~\eqref{Eq:VBA_Proport1_L_nsStL_un} it can be concluded that $q_1 \left( \rfb \right)$  is a multivariate Normal distribution. Minimizing the criterion leads to the analytical expression of the corresponding mean. The variance is obtained by identification: 
\beq
q_1(\rfb) = \Nc\left( \rfb | \rfbh, \Sigmabh \right),
\left\{\barr{ll}
\rfbh = 
\left( \bHb^T  \rVbt_\repsilon^{-1} \bHb + \rVbt_{\rf} \right)^{-1}
\bHb^T \rVbt_\repsilon^{-1} \bgb,
\\[10pt]
\Sigmabh = \left( \bHb_\blue{1}^T  \rVbt_\repsilon^{-1} \bHb + \rVbt_{\rf} \right)^{-1}.
\earr\right.
\label{Eq:q1_Anal_1_L_nsStL_un}
\eeq
We note that both the expressions of the mean and variance depend on expectancies corresponding to two variances of the hierarchical model.
\paragraph{Computation of the analytical expression of $q_{2j}(\rv_{\rf_\rj})$.}
The proportionality relation corresponding to $q_{2j}(\rv_{\rf_\rj})$ is presented in established in Equation~\eqref{Eq:VBA_Proport2_L_nsStL_un}. In the expression of $\ln p\left(\rfb, \rvb_\rf, \rvb_\repsilon | \bgb, \beta_{f}, \alpha_{\epsilon}, \beta_{\epsilon} \right)$ all the terms free of $\rv_{\rf_\rj}$ can be regarded as constants. Via Equation~\eqref{Eq:Log_Post_L_nsStL_un} the integral defined in Equation~\eqref{Eq:Def_Integ_L_nsStL_un} becomes:
\beq
\begin{split}
\biggl\langle \ln p(\rfb, \rvb_\rf, \rvb_\repsilon | \bgb, \beta_{f}, \alpha_{\epsilon}, \beta_{\epsilon}) \biggr\rangle_{q_1(\rfb) \; q_{2-j}(\rv_{\rf_\red{j}}) \; q_3(\rvb_{\repsilon})}
&
=
-
\frac{1}{2}
\biggl\langle
\| \rVb_{\rf}^{\frac{1}{2}} \rfb \|
\biggr\rangle_{q_1(\rfb) \; q_{2-j}(\rv_{\rf_\red{j}})}
-
\frac{3}{2} \ln \rv_{\rf_\rj}
-
\frac{1}{2} \frac{\beta_{f}}{\rv_{\rf_\rj}}
\end{split}
\label{Eq:q2_Integ_1_L_nsStL_un}
\eeq
Introducing the notations:
\beq
\rvbt_{\rf_{-\ri}}=
\begin{bmatrix}
\rvt_{\rf_\red{1}} \;
\ldots \;
\rvt_{\rf_{\ri-\red{1}}} \;
\rv_{\rf_\ri}  \;
\rvt_{\rf_{\ri+\red{1}}} \;
\ldots \;
\rvt_{\rf_\rN}
\end{bmatrix}^T
\;\;;\;\;
\rVbt_{\rf_{-\ri}}=
\mbox{diag}\left(\rvbt_{\rf_{-\ri}} \right)
\label{Eq:q2_Integ_Not1_L_nsStL_un}
\eeq
the integral $\biggl\langle
\| \rVb_{\rf}^{\frac{1}{2}} \rfb \|
\biggr\rangle_{q_1(\rfb) \; q_{2-j}(\rv_{\rf_\red{j}})}
$ can be written:
\beq
\biggl\langle
\| \rVb_{\rf}^{\frac{1}{2}} \rfb \|
\biggr\rangle_{q_1(\rfb) \; q_{2-j}(\rv_{\rf_\red{j}})}
= 
\biggl\langle \| \rVbt_{\rf_{-\ri}}^{\frac{1}{2}} \rfb \|^2 \biggr\rangle_{q_1(\rfb)}
\eeq 
Considering that $q_1(\rfb)$ is a multivariate Normal distribution, Equation~\eqref{Eq:q1_Anal_1_L_nsStL_un}:
\beq
\biggl\langle \|  \rVbt_{\rf_{-\ri}}^{\frac{1}{2}}  \rfb \|^2 \biggr\rangle_{q_1(\rfb)}
=
\| \rVbt_{\rf_{-\ri}}^{\frac{1}{2}} \rfbh \|^2 + \mbox{Tr}\left(\rVbt_{\rf_{-\ri}} \Sigmabh \right) 
=
C + \rv_{\rf_\ri} \left( \rfh_\rj^2 + \Sigmabh_{jj} \right)
\label{Eq:q2_Integ_2_L_nsStL_un}
\eeq
From Equation~\eqref{Eq:q2_Integ_1_L_nsStL_un} and Equation~\eqref{Eq:q2_Integ_2_L_nsStL_un}: 
\beq
\begin{split}
\biggl\langle \ln p(\rfb, \rvb_\rf, \rvb_\repsilon | \bgb, \beta_{f}, \alpha_{\epsilon}, \beta_{\epsilon}) \biggr\rangle_{q_1(\rfb) \; q_{2-j}(\rv_{\rf_\red{j}}) \; q_3(\rvb_{\repsilon})}
=
- \frac{3}{2} \ln \rv_{\rf_\rj}
- \frac{1}{2} \beta_{f} \rv_{\rf_\rj}^{-1} 
- \frac{1}{2} \left( \rfh_\rj^2 + \Sigmabh_{jj} \right) \rv_{\rf_\rj}
\end{split}
\eeq
from which it can establish the proportionality corresponding to
$q_{2j}(\rv_{\rf_\rj})$:
\beq
q_{2j}(\rv_{\rf_\rj})
\propto
\rv_{\rf_\rj}^{-\frac{3}{2}}
\exp
\biggl\lbrace 
-\frac{1}{2}
\left[ \left( \rfh_\rj^2 + \Sigmabh_{jj} \right) \rv_{\rf_\rj} + \beta_{f} \rv_{\rf_\rj}^{-1} \right]
\biggr\rbrace,
\eeq
leading to the conclusion that $q_{2j}(\rv_{\rf_\rj})$ is a generalized inverse Gaussian distribution (see Equation~\eqref{Eq:GenInvGau}) with the following parameters:
\beq
q_{2j}(\rv_{\rf_\rj})=\Gc\Ic\Gc\left(\rv_{\rf_\rj}|\aht_{f_j},\bht_{f_j}, \cht_{f_j} \right),
\left\{\barr{ll}
\aht_{f_j} = \rfh_\rj^2 + \Sigmabh_{jj}
\\[10pt]
\bht_{f_j} = \beta_f 
\\[10pt]
\cht_{f_j} = -\frac{1}{2}
\earr\right.
\label{Eq:q2_Anal_1_L_nsStL_un}
\eeq
\paragraph{Computation of the analytical expression of $q_{3i}(\rv_{\repsilon_\ri})$.}
The proportionality relation corresponding to $q_{3i}(\rv_{\repsilon_\ri})$ is presented in established in Equation~\eqref{Eq:VBA_Proport3_L_nsStL_un}. In the expression of $\ln p\left(\rfb, \rvb_\rf, \rvb_\repsilon | \bgb, \beta_{f}, \alpha_{\epsilon},  \beta_{\epsilon} \right)$ all the terms free of $\rv_{\repsilon_\ri}$ can be regarded as constants. Via Equation~\eqref{Eq:Log_Post_L_nsStL_un} the integral defined in Equation~\eqref{Eq:Def_Integ_L_nsStL_un} becomes:
\beq
\begin{split}
\biggl\langle \ln p(\rfb, \rvb_\rf, \rvb_\repsilon | \bgb, \beta_{f}, \alpha_{\epsilon}, \beta_{\epsilon}) \biggr\rangle_{q_1(\rfb) \; q_{2}(\rvb_{\rf}) \; q_{3-i}(\rv_{\repsilon_\ri})}
=
&
-\frac{1}{2} \biggl\langle \| \rVb_{\repsilon}^{-\frac{1}{2}} \left( \bgb - \bHb \rfb \right) \| \biggr\rangle_{q_1(\rfb) \; q_{3-i}(\rv_{\repsilon_\ri})}
\\
&
-
\left( \alpha_{\epsilon} + \frac{3}{2} \right) \ln \rv_{\repsilon_\ri}
-
\frac{\beta_{\epsilon}}{\rv_{\repsilon_\ri}}
\end{split}
\label{Eq:q3_Integ_1_L_nsStL_un}
\eeq
Introducing the notations:
\beq
\rvbt_{\repsilon_{-\ri}}^{-1}=
\begin{bmatrix}
\rvt_{\repsilon_\red{1}}^{-1} \;
\ldots \;
\rvt_{\repsilon_{\ri-\red{1}}}^{-1} \;
\rv_{\repsilon_\ri}^{-1}  \;
\rvt_{\repsilon_{\ri+\red{1}}}^{-1} \;
\ldots \;
\rvt_{\repsilon_\rN}^{-1}
\end{bmatrix}^T
\;\;;\;\;
\rVbt_{\repsilon_{-\ri}}^{-1}=
\mbox{diag}\left(\rvbt_{\repsilon_{-\ri}}^{-1} \right)
\label{Eq:q3_Integ_Not1_L_nsStL_un}
\eeq
the integral $\biggl\langle \| \rVb_{\repsilon}^{-\frac{1}{2}} \left( \bgb - \bHb \rfb \right) \| \biggr\rangle_{q_1(\rfb) \; q_{3-i}(\rv_{\repsilon_\ri})}
$ can be written:
\beq
\biggl\langle \| \rVb_{\repsilon}^{-\frac{1}{2}} \left( \bgb - \bHb \rfb \right) \| \biggr\rangle_{q_1(\rfb) \; q_{3-i}(\rv_{\repsilon_\ri})}
= 
\biggl\langle \| \rVbt_{\repsilon_{-\ri}}^{-\frac{1}{2}} \left( \bgb - \bHb \rfb \right) \|^2 \biggr\rangle_{q_1(\rfb)}
\eeq 
Considering that $q_1(\rfb)$ is a multivariate Normal distribution, Equation~\eqref{Eq:q1_Anal_1_L_nsStL_un}:
\beq
\biggl\langle
\| \rVbt_{\repsilon_{-\ri}}^{-\frac{1}{2}} \left( \bgb - \bHb \rfb \right) \|^2
\biggr\rangle_{q_1(\rfb)}
=
\| \rVbt_{\repsilon_{-\ri}}^{-\frac{1}{2}}
\left( \bgb - \bHb \rfbh \right) \|^2
+
\mbox{Tr}\left( \bHb^T \rVb_{\repsilon_{-\ri}}^{-1} \bHb \Sigmabh \right)
\label{Eq:q3_Integ_2_L_nsStL_un}
\eeq
and considering as constants all terms free of $\rv_{\repsilon_\ri}$:
\beq
\| \rVbt_{\repsilon_{-\ri}}^{-\frac{1}{2}}
\left( \bgb - \bHb \rfbh \right) \|^2
=
C +
\rv_{\repsilon_\ri}^{-1} \left( \bg_\bi - \bHb_\bi \rfbh \right)^2
\;\;;\;\;
\mbox{Tr}\left( \bHb^T \rVbt_{\repsilon_{-\ri}}^{-1} \bHb \Sigmabh \right)
=
C + 
\rv_{\repsilon_\ri}^{-1} \bHb_\bi \Sigmabh \bHb_\bi^T
\eeq
where $\bHb_\bi$ is the line i of the matrix $\bHb$, so we can conclude:
\beq
\biggl\langle
\| \rVb_{\repsilon}^{-\frac{1}{2}}\left( \bgb - \bHb\rfb \right)\|^2
\biggr\rangle_{q_1(\rfb) \; q_{3-i}(\rv_{\repsilon_\ri})}
=
C + 
\left(
\bHb_\bi \Sigmabh \bHb_\bi^T
+
\left( \bg_\bi - \bHb_\bi \rfbh \right)^2
\right)
\rv_{\repsilon_\ri}^{-1}
\label{Eq:q3_Integ_3_L_nsStL_un}
\eeq
From Equation~\eqref{Eq:q3_Integ_1_L_nsStL_un} and Equation~\eqref{Eq:q3_Integ_3_L_nsStL_un}: 
\beq
\begin{split}
\biggl\langle \ln p(\rfb, \rvb_\rf, \rvb_\repsilon | \bgb, \beta_{f}, \alpha_{\epsilon}, \beta_{\epsilon}) \biggr\rangle_{q_1(\rfb) \; q_{2}(\rvb_{\rf}) \; q_{3-i}(\rv_{\repsilon_\ri})}
=
&
-
\left( \alpha_{\epsilon} + \frac{1}{2} \right) \ln \rv_{\repsilon_\ri}
\\
&
-  
\left(
\beta_{\epsilon}
+ 
\frac{1}{2}
\left[ \bHb_\bi \Sigmabh \bHb_\bi^T +
\left( \bg_\bi - \bHb_\bi \rfbh \right)^2 \right] 
\right)
\rv_{\repsilon_\ri}^{-1}
\end{split}
\eeq
from which it can establish the proportionality corresponding to
$q_{3i}(\rv_{\repsilon_\ri})$:
\beq
q_{3i}(\rv_{\repsilon_\ri})
\propto
\rv_{\repsilon_\ri}^{ - \left( \alpha_{\epsilon} + \frac{3}{2} \right) }
\exp
\left\lbrace 
-  
\left(
\beta_{\epsilon}
+ 
\frac{1}{2}
\left[ \bHb_\bi \Sigmabh \bHb_\bi^T +
\left( \bg_\bi - \bHb_\bi \rfbh \right)^2 \right] 
\right)
\rv_{\repsilon_\ri}^{-1}
\right\rbrace,
\eeq
leading to the conclusion that $q_{3i}(\rv_{\repsilon_\ri})$ are Inverse Gamma distributions with the following parameters:
\beq
q_{3i}(\rv_{\repsilon_\ri})=\Ic\Gc \left( \rv_{\repsilon_\ri} | \alphah_{\epsilon_i}, \betah_{\epsilon_i} \right),
\left\{\barr{ll}
\alphah_{\epsilon_i} = \alpha_{\epsilon} + \frac{3}{2}
\\[10pt]
\betah_{\epsilon_i} = \beta_{\epsilon} + \frac{1}{2} \left[ \bHb_\bi \Sigmabh \bHb_\bi^T + \left( \bg_\bi - \bHb_\bi \rfbh \right)^2 \right] 
\earr\right.
\label{Eq:q3_Anal_1_L_nsStL_un}
\eeq
Equations~\eqref{Eq:q1_Anal_1_L_nsStL_un},~\eqref{Eq:q2_Anal_1_L_nsStL_un} and~\eqref{Eq:q3_Anal_1_L_nsStL_un} resume the distributions families and the corresponding parameters for $q_1(\rfb)$, a multivariate Normal distribution, $q_{2j}(\rv_{\rf_\rj})$, $j\in\left\lbrace 1, 2, \ldots, M \right\rbrace$ Inverse Gamma distributions and $q_{3i}(\rv_{\repsilon_\ri})$, $i\in\left\lbrace 1, 2, \ldots, N \right\rbrace$, generalized inverse Gaussian distributions. However, the parameters corresponding to the multivariate Normal distribution are expressed via $\rVbt_{\repsilon}^{-1}$ and $\rVbt_{\rf}$ (and by extension all elements forming the three matrices $\rvt_{\repsilon_\ri}^{-1}$, $i\in\left\lbrace 1, 2, \ldots, N \right\rbrace$ and $\rvt_{\rf_\rj}$, $j\in\left\lbrace 1, 2, \ldots, M \right\rbrace$).
\paragraph{Computation of the analytical expression $\rVbt_{\rf}$.}
For an generalized inverse Gaussian distribution with parameters $a$, $b$ and $-\frac{1}{2}$, $\Gc\Ic\Gc\left( x | a, b, c \right)$, the following relation holds:
\beq
\biggl\langle x \biggr\rangle_{\Gc\Ic\Gc(x|a,b,-\frac{1}{2})} 
=
\left( \frac{a}{b} \right)^{-\frac{1}{2}}
\frac{\Kc_{\frac{1}{2}}\left( \sqrt{ab} \right)}{\Kc_{-\frac{1}{2}}\left( \sqrt{ab} \right)},
\label{Eq:Inv_Gam_Integ_L_nsStL_un}
\eeq
where $\Kc_{p}$ represents the modified Bessel function of the second kind.\\
To prove the above relation we consider the direct computation, using the analytical expression of the generalized inverse Gaussian distribution: 
\beq
\begin{split}
\biggl\langle x \biggr\rangle_{\Gc\Ic\Gc(x|a,b,-\frac{1}{2})} 
= 
&
\int x \; \Gc\Ic\Gc(x|a,b,-\frac{1}{2}) \d x
=
\int 
x
\frac{\left( \frac{a}{b} \right)^{-\frac{1}{2}}}{\Kc_{-\frac{1}{2}}\left( \sqrt{ab} \right)}
x^{-\frac{1}{2}-1}
\exp
\left\lbrace -\frac{1}{2} \left( ax + bx^{-1} \right) \right\rbrace
\d x
\\
= 
&
\frac{\left( \frac{a}{b} \right)^{-\frac{1}{2}}}{\Kc_{-\frac{1}{2}}\left( \sqrt{ab} \right)}
\int 
x^{\frac{1}{2}-1}
\exp
\left\lbrace -\frac{1}{2} \left( ax + bx^{-1} \right) \right\rbrace
\d x
\\
=
&
\frac{\left( \frac{a}{b} \right)^{-\frac{1}{2}}}{\Kc_{-\frac{1}{2}}\left( \sqrt{ab} \right)}
\frac{\Kc_{\frac{1}{2}}\left( \sqrt{ab} \right)}{\left( \frac{a}{b} \right)^{\frac{1}{2}}}
\int 
\frac{\left( \frac{a}{b} \right)^{\frac{1}{2}}}{\Kc_{\frac{1}{2}}\left( \sqrt{ab} \right)}
x^{\frac{1}{2}-1}
\exp
\left\lbrace -\frac{1}{2} \left( ax + bx^{-1} \right) \right\rbrace
\d x
\\
=
&
\left( \frac{a}{b} \right)^{-1}
\underbrace{\frac{\Kc_{\frac{1}{2}}\left( \sqrt{ab} \right)}{\Kc_{-\frac{1}{2}}\left( \sqrt{ab} \right)}}_{1}
\underbrace{\int \Gc\Ic\Gc(x|a,b,\frac{1}{2}) \d x}_{1}
=\frac{b}{a}
\end{split}
\nonumber
\eeq
The fact that the integral of the generalized inverse Gaussian distribution is obvious. Proving that the ratio between the two 
modified Bessel functions of the second kind is $1$, i.e. that $\Kc_{\frac{1}{2}}\left( \sqrt{ab} \right) = \Kc_{-\frac{1}{2}}\left( \sqrt{ab} \right)$ comes from expressing the modified Bessel function of the second kind $\Kc_{\alpha}\left( x \right)$ via the modified Bessel function of the first kind $\Ic_{\alpha}\left( x \right)$:
\beq
\Kc_{\alpha}\left( x \right) 
=
\frac{\pi}{2}
\frac{\Ic_{-\alpha}\left( x \right)-\Ic_{\alpha}\left(x\right)}{\sin\left( \alpha \pi \right)}
=
\frac{\pi}{2}
\frac{\Ic_{\alpha}\left( x \right)-\Ic_{-\alpha}\left(x\right)}{-\sin\left( \alpha \pi \right)}
=
\frac{\pi}{2}
\frac{\Ic_{\alpha}\left( x \right)-\Ic_{-\alpha}\left(x\right)}{\sin\left( -\alpha \pi \right)}
=
\Kc_{-\alpha}\left( x \right)
\eeq
Since $q_{2j}(\rv_{\rf_\rj})$, $j\in\left\lbrace 1, 2, \ldots, M \right\rbrace$ are generalized inverse Gaussian distributions, with the corresponding parameters $\aht_{f_j}$, $\bht_{f_j}$ and $\cht_{f_j}$, $j\in\left\lbrace 1, 2, \ldots, M \right\rbrace$, the expectancies $\rvt_{\rf_\rj}$ can be expressed via the parameters of the generalized inverse Gaussian distributions using Equation~\eqref{Eq:Inv_Gam_Integ_L_nsStL_un}:
\beq
\rvt_{\rf_\rj}
=
\frac{\bht_{f_j}}{\aht_{f_j}}
\label{Eq:VepsVfExpectIGSMGen_L_nsStL_un}
\eeq
Using the notation introduced in \eqref{Eq:q1_Integ_Not1_L_nsStL_un}:
\beq
\rVbt_{\rf}=
\begin{bmatrix}
\frac{\bht_{f_1}}{\aht_{f_1}} \ldots 0 \ldots 0 \\
\vdots \ddots \vdots \ddots \vdots \\
0 \ldots \frac{\bht_{f_j}}{\aht_{f_j}} \ldots 0 \\
\vdots \ddots \vdots \ddots \vdots \\
0 \ldots 0 \ldots \frac{\bht_{f_M}}{\aht_{f_M}} \\
\end{bmatrix}
=
\rVbh_{\rf}
\label{Eq:V_Expect_L_nsStL_un}
\eeq
In Equation~\eqref{Eq:V_Expect_L_nsStL_un} other notation is introduced for $\rVbt_{\rf}$. The value was expressed during the model via unknown expectancies, but via Equation~\eqref{Eq:V_Expect_L_nsStL_un} this value doesn't contain any more integrals to be computed. Therefore, the new notation represents the final analytical expression used for expressing the density functions $q_i$.
\paragraph{Computation of the analytical expression $\rVbt_{\repsilon}^{-1}$.}
For an Inverse Gamma distribution with scale and shape parameters $\alpha$ and $\beta$, $\Ic\Gc\left(x|\alpha, \beta \right)$, the following relation holds:
\beq
\biggl\langle x^{-1} \biggr\rangle_{\Ic\Gc(x|\alpha,\beta)} 
=
\frac{\alpha}{\beta}
\label{Eq:Inverse_Gamma_Integral_L_nsStL_un_Second}
\eeq
The prove of the above relation is done by direct computation, using the analytical expression of the Inverse Gamma Distribution: 
\beq
\begin{split}
\biggl\langle x^{-1} \biggr\rangle_{\Ic\Gc(x|\alpha,\beta)} 
&= 
\int 
x^{-1}
\frac{{\beta}^{\alpha}}{\Gamma(\alpha)}
x^{-\alpha-1}
\exp
\left\lbrace
- \frac{\beta}{x}
\right\rbrace
\d x
=
\frac{{\beta}^{\alpha}}{\Gamma(\alpha)}
\frac{\Gamma(\alpha+1)}{{\beta}^{\alpha+1}}
\int
\frac{{\beta}^{\alpha+1}}{\Gamma(\alpha+1)}
x^{-(\alpha+1)-1}
\exp\left\lbrace -\frac{\beta}{x}\right\rbrace
\d x
=\\
&=
\frac{\alpha}{\beta}
\underbrace{
\int \Ic\Gc(x|\alpha+1,\beta)
}_{1}
\d x
=
\frac{\alpha}{\beta}
\end{split}
\nonumber
\eeq
Since $q_{3i}(\rv_{\repsilon_\ri})$, $i\in\left\lbrace 1, 2, \ldots, N \right\rbrace$ are Inverse Gamma distributions, with the corresponding parameters $\alphah_{\epsilon_i}$ and $\betah_{\epsilon_i}$, $i\in\left\lbrace 1, 2, \ldots, N \right\rbrace$ the expectancies $\rvt_{\repsilon_\ri}^{-1}$ can be expressed via the parameters of the Inverse Gamma distributions using Equation~\eqref{Eq:Inverse_Gamma_Integral_L_nsStL_un_Second}:
\beq
\rvt_{\repsilon_\ri}^{-1}
=
\frac{\alphah_{\epsilon_i}}{\betah_{\epsilon_i}}
\label{Eq:VepsVfExpectanciesIGSMGen_L_nsStL_un_Second}
\eeq
Using the notation introduced in \eqref{Eq:q1_Integ_Not1_L_nsStL_un}:
\beq
\rVbt_{\repsilon}^{-1}=
\begin{bmatrix}
\frac{\alphah_{\epsilon_1}}{\betah_{\epsilon_1}} \ldots 0 \ldots 0 \\
\vdots \ddots \vdots \ddots \vdots \\
0 \ldots \frac{\alphah_{\epsilon_i}}{\betah_{\epsilon_i}} \ldots 0 \\
\vdots \ddots \vdots \ddots \vdots \\
0 \ldots 0 \ldots \frac{\alphah_{\epsilon_N}}{\betah_{\epsilon_N}} \\
\end{bmatrix}
=
\rVbh_{\repsilon}^{-1}
\label{Eq:V_Expectancies_L_nsStL_un_Second}
\eeq
In Equation~\eqref{Eq:V_Expectancies_L_nsStL_un_Second} other notation is introduced for $\rVbt_{\repsilon}^{-1}$. The value was expressed during the model via unknown expectancies, but via Equation~\eqref{Eq:V_Expectancies_L_nsStL_un_Second} this value doesn't contain any more integrals to be computed. Therefore, the new notation represents the final analytical expressions used for expressing the density functions $q_i$.
Using Equation~\eqref{Eq:V_Expect_L_nsStL_un} and Equations~\eqref{Eq:q1_Anal_1_L_nsStL_un},~\eqref{Eq:q2_Anal_1_L_nsStL_un} and~\eqref{Eq:q3_Anal_1_L_nsStL_un}, the final analytical expressions of the separable distributions $q_i$ are presented in Equations~\eqref{Eq:q1_Anal_2_L_nsStL_un},~\eqref{Eq:q2_Anal_2_L_nsStL_un} and~\eqref{Eq:q3_Anal_2_L_nsStL_un}.
\begin{subequations}
\begin{align}
&q_1(\rfb) = \Nc\left( \rfb | \rfbh, \Sigmabh \right),
\;\;\;\;\;\;\;\;\;\;\;\;\;\;\;\;\;\;\;\;
\left\{\barr{ll}
\rfbh = 
\left( \bHb^T  \rVbh_\repsilon \bHb + \rVbh_{\rf} \right)^{-1}
\bHb^T \rVbh_\repsilon \bgb,
\\[10pt]
\Sigmabh = \left( \bHb_\blue{1}^T  \rVbh_\repsilon \bHb + \rVbh_{\rf} \right)^{-1}
\earr\right.,
\label{Eq:q1_Anal_2_L_nsStL_un}
\\[4pt]
&q_{2j}(\rv_{\rf_\rj}) = \Gc\Ic\Gc \left( \rv_{\rf_\rj} | \aht_{f_j}, \bht_{f_j}, \cht_{f_j} \right),
\left\{\barr{ll}
\aht_{f_j} = \rfh_\rj^2 + \Sigmabh_{jj}
\\[10pt]
\bht_{f_j} = \beta_f 
\\[10pt]
\cht_{f_j} = -\frac{1}{2}
\earr\right.
,j\in\left\lbrace 1, 2, \ldots, M \right\rbrace,
\label{Eq:q2_Anal_2_L_nsStL_un}
\\[4pt]
&q_{3i}(\rv_{\repsilon_\ri})=\Ic\Gc \left( \rv_{\repsilon_\ri} | \alphah_{\epsilon_i}, \betah_{\epsilon_i} \right),
\;\;\;\;\;\;\;\;\;\;
\left\{\barr{ll}
\alphah_{\epsilon_i} = \alpha_{\epsilon} + \frac{3}{2}
\\[10pt]
\betah_{\epsilon_i} = \beta_{\epsilon} + \frac{1}{2} \left[ \bHb_\bi \Sigmabh \bHb_\bi^T + \left( \bg_\bi - \bHb_\bi \rfbh \right)^2 \right] 
\earr\right.
,i\in\left\lbrace 1, 2, \ldots, N \right\rbrace.
\label{Eq:q3_Anal_2_L_nsStL_un}
\end{align}
\end{subequations}
Equation~\eqref{Eq:q1_Anal_2_L_nsStL_un} establishes the dependency between the parameters corresponding to the multivariate Normal distribution $q_1(\rfb)$ and the others parameters involved in the hierarchical model: the mean $\rfbh$ and the covariance matrix $\Sigmabh$ depend on $\rVbh_{\repsilon}$ and $\rVbh_{\rf}$ which, via Equation~\eqref{Eq:V_Expect_L_nsStL_un} are defined using $\left\lbrace \aht_{f_j},\bht_{f_j}\right\rbrace, j\in \left\lbrace 1, 2, \ldots, M \right\rbrace $ and $\left\lbrace \aht_{\epsilon_i},\bht_{\epsilon_i}\right\rbrace, i\in \left\lbrace 1, 2, \ldots, N \right\rbrace $. The dependency between the parameters of the multivariate Normal distribution $q_1(\rfb)$ and the parameters of the generalized inverse Gaussian distributions $q_{2j}(\rvb_{\rf_\rj}) ,j \in \left\lbrace 1,2,\ldots,M \right\rbrace$ and $q_{3i}(\rvb_{\repsilon_\ri}), i \in \left\lbrace 1,2,\ldots,N \right\rbrace$ is presented in Figure~\eqref{Fig:Depend_Sch_1_L_nsStL_un}.
\begin{figure}[!htb]
\center
\begin{tabular}{c}
\begin{picture}(180,30)
\put(0,0){\framebox(100,26){$\left\lbrace \aht_{f_j},\bht_{f_j}\right\rbrace,\left\lbrace \aht_{\epsilon_j},\bht_{\epsilon_j}\right\rbrace$}}
\put(102,13){\vector(1,0){26}}
\put(128,0){\framebox(56,26){$\rfbh \; , \; \Sigmabh$}}
\end{picture}
\end{tabular}
\caption{Dependency between $q_1(\rfb)$ parameters and $q_{2j}(\rvb_{\rf_\rj})$ and $q_{3i}(\rvb_{\repsilon_\ri})$ parameters}
\label{Fig:Depend_Sch_1_L_nsStL_un}
\end{figure}
Equation~\eqref{Eq:q2_Anal_2_L_nsStL_un} establishes the dependency between the parameters corresponding to the Inverse Gamma distributions $q_{2j}(\rvb_{\rf_\rj}) ,j \in \left\lbrace 1,2,\ldots,M \right\rbrace$ and the others parameters involved in the hierarchical model: the shape and scale parameters $\left\lbrace \aht_{f_j}, \bht_{f_j} \right\rbrace, j \in \left\lbrace 1,2,\ldots,M \right\rbrace$ depend on the mean $\rfbh$ and the covariance matrix $\Sigmabh$ of the multivariate Normal distribution $q_1(\rfb)$, Figure~\eqref{Fig:Depend_Sch_2_L_nsStL_un}.
\begin{figure}[!htb]
\center
\begin{tabular}{c}
\begin{picture}(120,30)
\put(0,0){\framebox(46,26){$ \rfbh \; , \; \Sigmabh$}}
\put(48,13){\vector(1,0){26}}
\put(76,0){\framebox(56,26){$\left\lbrace \aht_{f_j},\bht_{f_j}\right\rbrace$}}
\end{picture}
\end{tabular}
\caption{Dependency between $q_{2j}(\rvb_{\rf_\rj})$ parameters and $q_1(\rfb)$ and $q_{3i}(\rvb_{\repsilon_\ri})$ parameters}
\label{Fig:Depend_Sch_2_L_nsStL_un}
\end{figure}
Equation~\eqref{Eq:q3_Anal_2_L_nsStL_un} establishes the dependency between the parameters corresponding to the Inverse Gamma distributions $q_{3i}(\rvb_{\repsilon_\ri}) ,i \in \left\lbrace 1,2,\ldots,N \right\rbrace$ and the others parameters involved in the hierarchical model: the shape and scale parameters $\left\lbrace \aht_{\epsilon_i}, \bht_{\epsilon_i} \right\rbrace, i \in \left\lbrace 1,2,\ldots,N \right\rbrace$ depend on the mean $\rfbh$ and the covariance matrix $\Sigmabh$ of the multivariate Normal distribution $q_1(\rfb)$, Figure~\eqref{Fig:Depend_Sch_3_L_nsStL_un}.
\begin{figure}[!htb]
\center
\begin{tabular}{c}
\begin{picture}(120,30)
\put(0,0){\framebox(46,26){$ \rfbh \; , \; \Sigmabh $}}
\put(48,13){\vector(1,0){26}}
\put(76,0){\framebox(56,26){$\left\lbrace \aht_{\epsilon_j},\bht_{\epsilon_j}\right\rbrace $}}
\end{picture}
\end{tabular}
\caption{Dependency between $q_{3i}(\rvb_{\repsilon_\ri})$ parameters and $q_{2j}(\rvb_{\rf_\rj})$ and $q_1(\rfb)$ parameters}
\label{Fig:Depend_Sch_3_L_nsStL_un}
\end{figure}
\newline
The iterative algorithm obtained via PM estimation is presented Figure~\eqref{Fig:IA_PM_L_nsStL_un}.
\tikzstyle{cloud} = [draw=blue!30!green,fill=orange!12, ellipse, thick, node distance=6em, text width=12em, text centered, minimum height=4em, minimum width=12em]
\tikzstyle{boxbig} = [draw=blue!30!green, fill=orange!12, rectangle, rounded corners, thick, node distance=8em, text width=25em, text centered, minimum height=5em, minimum width=25em]
\tikzstyle{box} = [draw=blue!30!green, fill=orange!12, rectangle, rounded corners, thick, node distance=6.5em, text width=21em, text centered, minimum height=5em, minimum width=21em]
\tikzstyle{boxsmall} = [draw=blue!30!green, fill=orange!12, rectangle, rounded corners, thick, node distance=8.5em, text width=11em, text centered, minimum height=5em, minimum width=11em]
\tikzstyle{line} = [draw=blue!30!green, -latex']
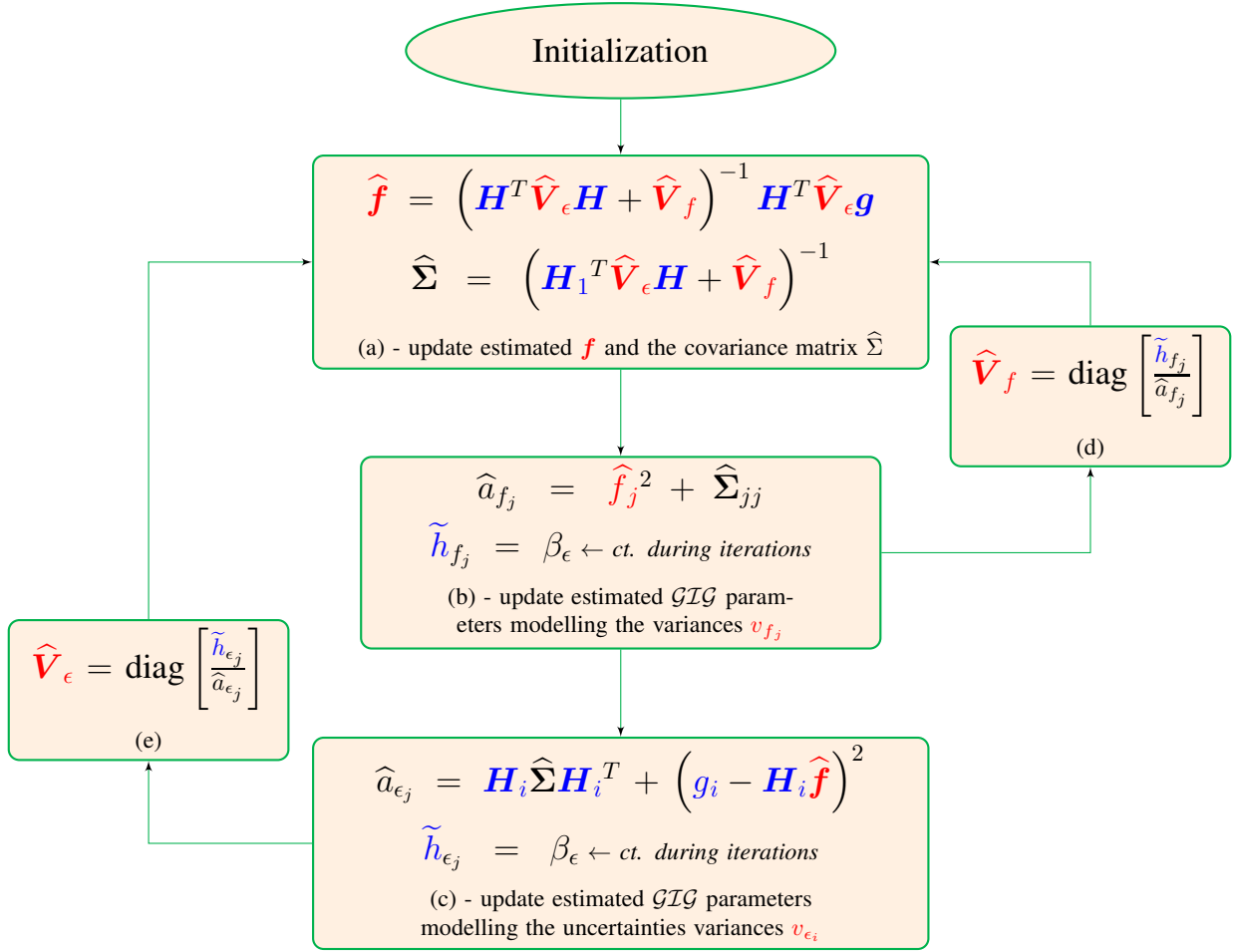
\begin{figure}[!htb]
\centering
\begin{center}
\begin{tikzpicture}[auto]
    \node [boxbig, scale=0.9] (f) {\Large{$\rfbh = 
\left( \bHb^T  \rVbh_\repsilon \bHb + \rVbh_{\rf} \right)^{-1}
\bHb^T \rVbh_\repsilon \bgb$}
\\[4pt]
\Large{$\Sigmabh = \left( \bHb_\blue{1}^T  \rVbh_\repsilon \bHb + \rVbh_{\rf} \right)^{-1}$}
\\[8pt]
\normalsize{(a) - update estimated $\rfb$ and the covariance matrix $\Sigmah$} 
};

    \node [box, below of=f, node distance=11em, scale=0.9] (vf)
{\Large{$ \aht_{f_j} = \rfh_\rj^2 + \Sigmabh_{jj}$}
\\[4pt]
\Large{$\bht_{f_j} = \beta_\epsilon $}
\normalsize{$\leftarrow$ \textit{ct. during iterations}}
\\[8pt]
\normalsize{(b) - update estimated $\Gc\Ic\Gc$ parameters modelling the variances $\rv_{\rf_\rj}$} 
};

    \node [boxbig, below of=vf, node distance=11em, scale=0.9] (veps) {\Large{$\aht_{\epsilon_j} = \bHb_\bi \Sigmabh \bHb_\bi^T
+ \left( \bg_\bi - \bHb_\bi \rfbh \right)^2$}
\\[4pt]
\Large{$\bht_{\epsilon_j} = \beta_\epsilon $}
\normalsize{$\leftarrow$ \textit{ct. during iterations}}
\\[8pt]
\normalsize{(c) - update estimated $\Gc\Ic\Gc$ parameters modelling the uncertainties variances $\rv_{\repsilon_\ri}$}     
};

    \node [boxsmall, above right = -0.15cm and 0.9cm of vf, node distance=17em, scale=0.9] (Vf) {\Large{$\rVbh_\rf = \diag {\frac{\bht_{f_j}}{\aht_{f_j}}}$}
\\[8pt]
\normalsize{(d)} 
};

    \node [boxsmall, below left = -4.4cm and 0.3cm of veps, node distance=17em, scale=0.9] (Veps) {\Large{$\rVbh_\repsilon = \diag {\frac{\bht_{\epsilon_j}}{\aht_{\epsilon_j}}}$}
\\[8pt]
\normalsize{(e)} 
};

\node [cloud, above of=f, node distance=8em, scale=0.9] (init)
{\Large{Initialization}};  

    \path [line] (f) -- (vf)[near start];
    \path [line] (Vf) |- (f);
    \path [line] (vf) -| (Vf);
    \path [line] (vf) -- (veps);
    \path [line] (Veps) |- (f);
    \path [line] (veps) -| (Veps);
    \path [line] (init) -- (f);
    
\end{tikzpicture}
\end{center}
\caption{Iterative algorithm corresponding to PM estimation via VBA - partial separability for Laplace hierarchical model, non-stationary Student-t uncertainties model}
\label{Fig:IA_PM_L_nsStL_un}
\end{figure}
\subsubsection{Posterior Mean estimation via VBA, full separability}
\label{Subsubsec:PM_FS_L_nsStL_un}
In this subsection, the Posterior Mean (PM) estimation is again considered, but via a full separable approximation. The posterior distribution is approximated by a full separable distribution $q\left( \rfb, \rvb_{\rf},\rvb_{\repsilon} \right)$, i.e. a supplementary condition is added in Equation~\eqref{Eq:Post_Approx_L_nsStL_un_Notations1}:
\beq
q_{1}(\rfb) = \prod_{j=1}^{M} q_{1j}(\rf_{\rj}),
\;\;;\;\;
q_{2}(\rvb_{\rf}) = \prod_{j=1}^{M} q_{2j}(\rv_{\rf_\rj}),
\;\;;\;\;
q_{3}(\rvb_{\repsilon}) = \prod_{i=1}^{N} q_{3i}(\rv_{\repsilon_\ri})
\label{Eq:Post_Approx_L_nsStL_un_Notations1bis}
\eeq
As in Subsection~\eqref{Subsubsec:PM_PS_L_nsStL_un}, the approximation is done by minimizing of the Kullback-Leibler divergence, Equation~\eqref{Eq:Kull-Leib_L_nsStL_un}, via alternate optimization resulting the following proportionalities from Equations~\eqref{Eq:VBA_Proport1_L_nsStL_unbis},~\eqref{Eq:VBA_Proport2_L_nsStL_unbis} and ~\eqref{Eq:VBA_Proport3_L_nsStL_unbis},  
\begin{subequations}
\begin{align}
&q_1(\rf_\red{j}) \;\;\; \propto \; \exp \left\lbrace \biggl\langle \ln p(\rfb, \rvb_\rf, \rvb_\repsilon | \bgb, \beta_{f}, \alpha_{\epsilon}, \beta_{\epsilon}) \biggr\rangle_{ q_{1-j}(\rf_\rj) q_2(\rvb_{\rf}) \; q_3(\rvb_{\repsilon})} \right\rbrace,\;\;j \in \left\lbrace 1,2 \ldots, M \right\rbrace,
\label{Eq:VBA_Proport1_L_nsStL_unbis}
\\[4pt]
&q_{2j}(\rv_{\rf_\red{j}}) \propto \; \exp \left\lbrace \biggl\langle \ln p(\rfb, \rvb_\rf, \rvb_\repsilon | \bgb, \beta_{f}, \alpha_{\epsilon}, \beta_{\epsilon}) \biggr\rangle_{q_1(\rfb) \; q_{2-j}(\rv_{\rf_\red{j}}) \; q_3(\rvb_{\repsilon})} \right\rbrace,\;\;j \in \left\lbrace 1,2 \ldots, M \right\rbrace,
\label{Eq:VBA_Proport2_L_nsStL_unbis}
\\[4pt]
&q_{3i}(\rv_{\repsilon_\ri}) \; \propto \; \exp \left\lbrace \biggl\langle \ln p(\rfb, \rvb_\rf, \rvb_\repsilon | \bgb, \beta_{f}, \alpha_{\epsilon}, \beta_{\epsilon}) \biggr\rangle_{q_1(\rfb) \; q_2(\rvb_{\rf}) \; q_{3-i}(\rv_{\repsilon_\ri})} \right\rbrace,\;\;i \in \left\lbrace 1,2 \ldots, N \right\rbrace,
\label{Eq:VBA_Proport3_L_nsStL_unbis}
\end{align}
\end{subequations}
using the notations introduced in Equation~\eqref{Eq:Def_q_Min_L_nsStL_un}, Equation~\eqref{Eq:Def_Integ_L_nsStL_un} and Equation~\eqref{Eq:Def_q_Min_L_nsStL_unbis}:
\beq
q_{1-j}(\rv_{\rj})=\displaystyle \prod_{k=1,k \neq j}^{M} q_{1k}(\rv_\rk)
\label{Eq:Def_q_Min_L_nsStL_unbis}
\eeq
The analytical expression of logarithm of the posterior distribution $\ln p(\rfb, \rvb_\rf, \rvb_\repsilon | \bgb, \beta_{f}, \alpha_{\epsilon}, \beta_{\epsilon})$ is obtained in Equation~\eqref{Eq:Log_Post_L_nsStL_un}.
\paragraph{Computation of the analytical expression of $q_1(\rfb)$.}
The proportionality relation corresponding to $q_1(\rfb)$ is presented in established in Equation~\eqref{Eq:VBA_Proport1_L_nsStL_unbis}. In the expression of $\ln p\left(\rfb, \rvb_\rf, \rvb_\repsilon | \bgb, \beta_{f}, \alpha_{\epsilon},  \beta_{\epsilon} \right)$ all the terms free of $\rf_\ri$ can be regarded as constants:
\beq
\begin{split}
\biggl\langle \ln p(\rfb, \rvb_\rf, \rvb_\repsilon | \bgb, \beta_{f}, \alpha_{\epsilon}, \beta_{\epsilon}) \biggr\rangle_{q_{1-j}(\rf_\rj) \; q_2(\rvb_{\rf}) \; q_3(\rvb_{\repsilon})}
=
&
-\frac{1}{2}
\biggl\langle 
\|\rVb_{\repsilon}^{\frac{1}{2}} \left(\bgb - \bHb\rfb\right) \|
\biggr\rangle_{q_{1-j}(\rf_\rj) \; q_3(\rvb_{\repsilon}) }
\\
&
-\frac{1}{2}
\biggl\langle 
\| \rVb_{\rf}^{\frac{1}{2}} \rfb \|
\biggr\rangle_{q_{1-j}(\rf_\rj) \; q_2(\rvb_{\rf}) }.
\end{split}
\label{Eq:q1_Integ_1_L_nsStL_unbis}
\eeq
Using Equation~\eqref{Eq:q1_Integ_Not1_L_nsStL_un}
the integral from Equation~\eqref{Eq:q1_Integ_1_L_nsStL_unbis} becomes:
\beq
\begin{split}
\left\langle \ln p(\rfb, \rvb_\rf, \rvb_\repsilon | \bgb,  \beta_{f}, \beta_{\epsilon}) \right\rangle_{q_{1-j}(\rf_\rj) \; q_2(\rvb_{\rf}) \; q_3(\rvb_{\repsilon})}
&
= 
-
\frac{1}{2} 
\biggl\langle 
\|\rVbt_{\repsilon}^{\frac{1}{2}} \left(\bgb - \bHb\rfb\right) \| 
\biggr\rangle_{q_{1-j}(\rf_\rj)} 
-
\frac{1}{2} 
\biggl\langle 
\| \rVbt_{\rf}^{\frac{1}{2}} \rfb \| 
\biggr\rangle_{q_{1-j}(\rf_\rj)}.
\label{Eq:q1_Integ_1_L_nsStL_unbis1}
\end{split}
\eeq
Considering all the $\rf_\rj$ free terms as constants, the first norm can be written:
\beq
\|\rVbt_{\repsilon}^{\frac{1}{2}} \left(\bgb - \bHb\rfb\right) \|=
C +
\| \rVbt_{\repsilon}^{\frac{1}{2}} \bHb^{\bj}\|^2\rf_\rj^2
-2
\bHb^{\bj T} \rVbt_{\repsilon}^{\frac{1}{2}} \left(\bgb - \bHb^{-\bj}\rfb^{-\rj} \right)\rf_\rj
\label{Eq:q1_Integ_First_Normbis_L_nsStL_un}
\eeq
where $\bHb^{\bj}$ represents the column $j$ of the matrix $\bHb$, $\bHb^{-\bj}$ represents the matrix $\bHb$ except the column $j$, $\bHb^{\bj}$ and $\rfb^{-\rj}$ represents the vector $\rfb$ except the element $\rf_\rj$. 
Introducing the notation
\beq
\rft_\rj 
= \int \rf_\rj \; q_{1j}(\rf_{\rj})\; \d \rf_{\rj}
\;\;;\;\;
\rfbt^{-\rj}=
\begin{bmatrix}
\rft_\red{1} \; \ldots \; \rft_\red{j-1} \; \rft_\red{j+1} \; \ldots \; \rft_{\rM}
\end{bmatrix}^T
\label{Eq:q1_Integ_Not1_L_nsStL_unbis}
\eeq
the expectancy of the first norm becomes:
\beq
\biggl\langle
\|\rVbt_{\repsilon}^{\frac{1}{2}} \left(\bgb - \bHb\rfb\right) \|
\biggr\rangle_{q_{1-j}(\rf_\rj)}
=
C +
\| \rVbt_{\repsilon}^{\frac{1}{2}} \bHb^{\bj}\|^2\rf_\rj^2
-2
\bHb^{\bj T} \rVbt_{\repsilon}^{\frac{1}{2}} \left(\bgb - \bHb^{-\bj}\rfbt^{-\rj} \right)\rf_\rj
\label{Eq:q1_Integ_First_Norm_Exbis_L_nsStL_un}
\eeq
Considering all the free $\rf_\rj$ terms as constants, the expectancy for the second norm becomes:  
\beq
\biggl\langle
\| \rVbt_{\rf}^{\frac{1}{2}} \rfb \|^2
\biggr\rangle_{q_{1-j}(\rf_\rj)}
 = C + \rvt_{\rf_\rj} \rf_\rj^2
\label{Eq:q1_Integ_Second_Norm_Exbis_L_nsStL_un}
\eeq
From Equation~\eqref{Eq:VBA_Proport1_L_nsStL_unbis}, ~\eqref{Eq:q1_Integ_1_L_nsStL_unbis1}, ~\eqref{Eq:q1_Integ_First_Norm_Exbis_L_nsStL_un}, and ~\eqref{Eq:q1_Integ_Second_Norm_Exbis_L_nsStL_un} the proportionality for $q_{1j}(\rf_\rj)$ becomes:
\beq
q_{1j}(\rf_\rj) \propto 
\exp \left\lbrace 
\left( 
\| \rVbt_{\repsilon}^{\frac{1}{2}} \bHb^{\bj}\|^2 + \rvt_{\rf_\rj} \right) \rf_\rj^2
-2
\bHb^{\bj T} \rVbt_{\repsilon} \left(\bgb - \bHb^{-\bj}\rfb^{-\rj} \right)\rf_\rj
\right\rbrace
\label{Eq:q1_Proport_L_nsStL_un}
\eeq
Considering the criterion $J(\rf_\rj) = \left( \| \rVbt_{\repsilon}^{\frac{1}{2}} \bHb^{\bj}\|^2 + \rvt_{\rf_\rj} \right)\rf_\rj^2 -2 \bHb^{\bj T} \rVbt_{\repsilon} \left(\bgb - \bHb^{-\bj}\rfb^{-\rj} \right)\rf_\rj$ which is quadratic, we conclude $q_{1j}(\rf_\rj)$ is a Normal distribution.
For computing the mean of the Normal distribution, it is sufficient to compute the solution that minimizes the criterion $J(\rf_\rj)$:
\beq
\frac{\partial J(\rf_\rj)}{\partial \rf_\rj}=0 
\Leftrightarrow
\rfh_\rj = \frac{\bHb^{\bj T} \rVbt_{\repsilon} \left( \bgb - \bHb^{-\bj} \rfb^{-\rj} \right)}{\| \rVbt_{\repsilon}^{\frac{1}{2}}  \bHb^{\bj}\| + \rvt_{\rf}}.
\label{Eq:q1_Crit_Minim_L_nsStL_un}
\eeq
The variance can be obtained by identification. The analytical expressions for the mean and the variance corresponding to the Normal distributions, $q_1(\rf_\rj)$ are presented in Equation~\eqref{Eq:q1_Anal_1_L_nsStL_unbis}.
\beq
q_1(\rf_\rj)=\Nc\left(\rf_\rj | \rfh_\rj, \widehat{\mbox{var}}_j \right),
\left\{\barr{ll}
\rfh_\rj = \frac{\bHb^{\bj T} \rVbt_{\repsilon} \left(\bgb - \bHb^{-\bj} \rfb^{-\rj} \right)}{\| \rVbt_{\repsilon}^{\frac{1}{2}} \bHb^{\bj}\| + \rvt_{\rf_\rj}}
\\[12pt]
\widehat{\mbox{var}}_j=\frac{1}{\| \rVbt_{\repsilon}^{\frac{1}{2}} \bHb^{\bj}\| + \rvt_{\rf_\rj}}
\earr\right.
,j \in \left\lbrace 1, 2, \ldots, M \right\rbrace
\label{Eq:q1_Anal_1_L_nsStL_unbis}
\eeq
\paragraph{Computation of the analytical expression of $q_{2j}(\rv_{\rf_\rj})$.}
The proportionality relation corresponding to $q_{2j}(\rv_{\rf_\rj})$ established in Equation~\eqref{Eq:VBA_Proport2_L_nsStL_unbis} refers to $\rv_{\rf_\rj}$, so in the expression of $\ln p\left(\rfb, \rvb_\rf, \rvb_\repsilon | \bgb, \beta_{f}, \alpha_{\epsilon}, \beta_{\epsilon} \right)$ all the terms free of $\rv_{\rf_\rj}$ can be regarded as constants,
\beq
\ln p\left(\rfb, \rvb_\rf, \rvb_\repsilon | \bgb, \beta_{f}, \alpha_{\epsilon}, \beta_{\epsilon} \right) =
C 
+
\frac{1}{2}\ln \rv_{\rf_\rj}
-
\frac{1}{2} \biggl\langle \rf_\rj^2\biggr\rangle_{q_{1j}(\rf_\rj)} \rv_{\rf_\rj}
-
2 \ln \rv_{\rf_\rj}
-
\frac{\beta_f}{2}  \rv_{\rf_\rj}^{-1},
\label{Eq:q2_Integ_1_L_nsStL_unbis}
\eeq
so the integral of the logarithm becomes:
\beq
\biggl\langle \ln p\left(\rfb, \rvb_\rf, \rvb_\repsilon | \bgb, \beta_{f}, \alpha_{\epsilon}, \beta_{\epsilon} \right) \biggr\rangle_{q_1(\rfb) \; q_{2-j}(\rv_{\rf_\rj}) \; q_3(\rvb_{\repsilon})}
= 
C
- 
\frac{3}{2} \ln \rv_{\rf_\rj} 
-
\frac{\beta_f}{2} \rv_{\rf_\rj}^{-1}
-
\frac{1}{2} \left( \rfh_\rj^2 + \widehat{\mbox{var}}_j \right) \rv_{\rf_\rj}.
\label{Eq:q2_Integ_1_L_nsStL_unbis1}
\eeq
Equation~\eqref{Eq:q2_Integ_1_L_nsStL_unbis1} leads to the conclusion that $q_{2j}(\rv_{\rf_\rj})$ is an generalized inverse Gaussian distribution. Equation~\eqref{Eq:q2_Anal_1_L_nsStL_unbis} presents the analytical expressions for the parameters corresponding to the Inverse Gamma distribution.
\beq
q_{2j}(\rv_{\rf_\rj}) = \Gc\Ic\Gc \left( \rv_{\rf_\rj} | \aht_{f_j}, \bht_{f_j}, \cht_{f_j} \right),
\left\{\barr{ll}
\aht_{f_j} = \rfh_\rj^2 + \widehat{\mbox{var}}_j
\\[10pt]
\bht_{f_j} = \beta_f
\\[10pt]
\cht_{f_j} = - \frac{1}{2} 
\earr\right.
,j \in \left\lbrace 1, 2, \ldots, M \right\rbrace
\label{Eq:q2_Anal_1_L_nsStL_unbis}
\eeq
\paragraph{Computation of the analytical expression of $q_{3i}(\rv_{\repsilon_\ri})$.} 
The proportionality relation corresponding to $q_{3i}(\rv_{\repsilon_\ri})$ established in Equation~\eqref{Eq:VBA_Proport3_L_nsStL_unbis} refers to $\rv_{\repsilon_\ri}$ so in the expression of $\ln p\left(\rfb, \rvb_\rf, \rvb_\repsilon | \bgb, \beta_{f}, \alpha_{\epsilon}, \beta_{\epsilon} \right)$ all the terms free of $\rv_{\repsilon_\ri}$ can be regarded as constants:
\beq
\ln p \left( \rfb, \rvb_\rf, \rvb_\repsilon | \bgb, \beta_{f}, \beta_{\epsilon} \right) = C 
-
\frac{3}{2} \ln \rv_{\repsilon_\ri}
-
\frac{\beta_\epsilon}{2}\rv_{\repsilon_\ri}^{-1} 
+ 
\frac{1}{2}\left(\bg_\bi - \bHb_\bi\rfb\right)^2
\rv_{\repsilon_\ri}.
\label{Eq:q3_Integ_1_L_nsStL_unbis}
\eeq
Introducing the notation
\beq
\biggl\langle \rfb \biggr\rangle_{q_1(\rfb)}
=
\begin{bmatrix}
\rfh_{\red{1}}
\ldots
\rfh_\rj
\ldots 
\rfh_\rM
\end{bmatrix}^T
\stackrel{Not}{=}
\rfbh
\;\; ; \;\;
\Sigmabh = \diag {\widehat{\mbox{var}}_j}
\label{Eq:q3_Integ_Not1_L_nsStL_unbis}
\eeq
the expectancy of the logarithm becomes
\beq
\begin{split}
\biggl\langle
\ln p\left(\rfb, \rvb_\rf, \rvb_\repsilon | \bgb, \beta_{f}, \alpha_{\epsilon}, \beta_{\epsilon} \right) 
\biggr\rangle_{q_1(\rfb) \; q_2(\rvb_\rf) \; q_{3-i}(\rv_{\repsilon_\ri})}
= C & 
-\frac{3}{2} \ln \rv_{\repsilon_\ri}
-
\frac{\beta_\epsilon}{2} \rv_{\repsilon_\ri}^{-1}
- 
\frac{1}{2}
\left[ \bHb_\bi \Sigmabh \bHb_\bi^T
- 
\left( \bg_\bi - \bHb_\bi \rfbh \right)^2
\right] \rv_{\repsilon_\ri},
\end{split}
\label{Eq:q3_Integ_1_L_nsStL_unbis1}
\eeq
so the proportionality relation for $q_{3i}(\rv_{\repsilon_\ri})$ from Equation~\eqref{Eq:VBA_Proport3_L_nsStL_unbis} can be written:
\beq
q_{3i}(\rv_{\repsilon_\ri})
\propto
\rv_{\repsilon_\ri}^{-\frac{3}{2}}
\exp
\left\lbrace
-\frac{1}{2}
\left(
\left[ \bHb_\bi \Sigmabh \bHb_\bi^T + \left( \bg_\bi - \bHb_\bi \rfbh \right)^2 \right] \rv_{\repsilon_\ri}
+
\beta_\epsilon \rv_{\repsilon_\ri}^{-1} 
\right)
\right\rbrace
\label{Eq:q3_Proport_L_nsStL_un}
\eeq
Equation~\eqref{Eq:q3_Proport_L_nsStL_un} shows that $q_{3i}(\rv_{\repsilon_\ri})$ are generalized inverse Gaussian distributions. The analytical expressions of the corresponding parameters are presented in Equation~\eqref{Eq:q3_Anal_1_L_nsStL_unbis}.
\beq
q_{3i}(\rv_{\repsilon_\ri}) = \Gc\Ic\Gc \left( \rv_{\repsilon_\ri} | \aht_{\epsilon_i}, \bht_{\epsilon_i}, \cht_{\epsilon_i} \right),
\left\{\barr{ll}
\aht_{\epsilon_i} = \bHb_\bi \Sigmabh \bHb_\bi^T + \left( \bg_\bi - \bHb_\bi \rfbh \right)^2
\\[10pt]
\bht_{\epsilon_i} = \beta_\epsilon
\\[10pt]
\cht_{\epsilon_i} = - \frac{1}{2}
\earr\right.
,i \in \left\lbrace 1, 2, \ldots, N \right\rbrace
\label{Eq:q3_Anal_1_L_nsStL_unbis}
\eeq
Since $q_2(\rv_{\rf_\rj}), j \in \left\lbrace 1, 2, \ldots, M \right\rbrace$ and $q_{3i}(\rv_{\repsilon_\ri}), i \in \left\lbrace 1, 2, \ldots, N \right\rbrace $ are generalized inverse Gaussian distributions, it is easy to obtain analytical expressions for $\rVbt_{\repsilon}$, defined in Equation~\eqref{Eq:q1_Integ_Not1_L_nsStL_un} and $ \rvt_{\rf_\rj}, j \in \left\lbrace 1, 2, \ldots, M \right\rbrace $, obtaining the same expressions as in Equation~\eqref{Eq:V_Expect_L_nsStL_un}. Using Equation~\eqref{Eq:V_Expect_L_nsStL_un} and Equations~\eqref{Eq:q1_Anal_1_L_nsStL_unbis},~\eqref{Eq:q2_Anal_1_L_nsStL_unbis} and~\eqref{Eq:q3_Anal_1_L_nsStL_unbis}, the final analytical expressions of the separable distributions $q_i$ are presented in Equations~\eqref{Eq:q1_Anal_2_L_nsStL_unbis},~\eqref{Eq:q2_Anal_2_L_nsStL_unbis} and~\eqref{Eq:q3_Anal_2_L_nsStL_unbis}.
\begin{subequations}
\begin{align}
&q_1(\rf_\rj)=\Nc\left(\rf_\rj | \rfh_\rj, \widehat{\mbox{var}}_j \right),\;\;\;\;\;\;\;\;\;\;\;\;\;
\left\{\barr{ll}
\rfh_\rj = \frac{\bHb^{\bj T} \rVbh_{\repsilon} \left(\bgb - \bHb^{-\bj} \rfb^{-\rj} \right)}{\| \rVbh_{\repsilon}^{\frac{1}{2}} \bHb^{\bj}\| + \rvh_{\rf_\rj}}
\\[12pt]
\widehat{\mbox{var}}_j=\frac{1}{\| \rVbh_{\repsilon}^{\frac{1}{2}} \bHb^{\bj}\| + \rvh_{\rf_\rj}}
\earr\right.
,j \in \left\lbrace 1, 2, \ldots, M \right\rbrace
\label{Eq:q1_Anal_2_L_nsStL_unbis}
\\[4pt]
&q_{2j}(\rv_{\rf_\rj})=\Gc\Ic\Gc\left(\rv_{\rf_\rj}|\aht_{f_j},\bht_{f_j},\cht_{f_j}\right),
\left\{\barr{ll}
\aht_{f_j} = \rfh_\rj^2 + \widehat{\mbox{var}}_j
\\[10pt]
\bht_{f_j} = \beta_f
\\[10pt]
\cht_{f_j} = - \frac{1}{2} 
\earr\right.
,j \in \left\lbrace 1, 2, \ldots, M \right\rbrace
\label{Eq:q2_Anal_2_L_nsStL_unbis}
\\[4pt]
&q_{3i}(\rv_{\repsilon_\ri})=\Gc\Ic\Gc\left(\rv_{\repsilon_\ri}|\aht_{\epsilon_i},\bht_{\epsilon_i},\cht_{\epsilon_i}\right),\;\;\;
\left\{\barr{ll}
\aht_{\epsilon_i} = \bHb_\bi \Sigmabh \bHb_\bi^T + \left( \bg_\bi - \bHb_\bi \rfbh \right)^2
\\[10pt]
\bht_{\epsilon_i} = \beta_\epsilon
\\[10pt]
\cht_{\epsilon_i} = - \frac{1}{2}
\earr\right.
,i \in \left\lbrace 1, 2, \ldots, N \right\rbrace
\label{Eq:q3_Anal_2_L_nsStL_unbis}
\end{align}
\end{subequations}
Equations~\eqref{Eq:q1_Anal_2_L_nsStL_unbis},~\eqref{Eq:q2_Anal_2_L_nsStL_unbis} and~\eqref{Eq:q3_Anal_2_L_nsStL_unbis} establish dependencies between the parameters of the distributions, very similar to the one presented in Figures~\eqref{Fig:Depend_Sch_1_L_nsStL_un},~\eqref{Fig:Depend_Sch_2_L_nsStL_un} and~\eqref{Fig:Depend_Sch_3_L_nsStL_un}. The iterative algorithm obtained via PM estimation with full separability, is presented Figure~\eqref{Fig:IA_PM_L_nsStL_unbis}.
\tikzstyle{cloud} = [draw=blue!30!green,fill=orange!12, ellipse, thick, node distance=6em, text width=12em, text centered, minimum height=4em, minimum width=12em]
\tikzstyle{boxbig} = [draw=blue!30!green, fill=orange!12, rectangle, rounded corners, thick, node distance=8em, text width=25em, text centered, minimum height=5em, minimum width=25em]
\tikzstyle{box} = [draw=blue!30!green, fill=orange!12, rectangle, rounded corners, thick, node distance=6.5em, text width=21em, text centered, minimum height=5em, minimum width=21em]
\tikzstyle{boxsmall} = [draw=blue!30!green, fill=orange!12, rectangle, rounded corners, thick, node distance=8.5em, text width=11em, text centered, minimum height=5em, minimum width=11em]
\tikzstyle{line} = [draw=blue!30!green, -latex']
\begin{figure}[!htb]
\centering
\begin{center}
\begin{tikzpicture}[auto]
    \node [boxbig, scale=0.9] (f) {\Large{$ \rfh_\rj = \frac{\bHb^{\bj T} \rVbh_{\repsilon} \left(\bgb - \bHb^{-\bj} \rfb^{-\rj} \right)}{\| \rVbh_{\repsilon}^{\frac{1}{2}} \bHb^{\bj}\| + \rvh_{\rf_\rj}}$}
\\[4pt]
\Large{$\widehat{\mbox{var}}_j=\frac{1}{\| \rVbh_{\repsilon}^{\frac{1}{2}} \bHb^{\bj}\| + \rvh_{\rf_\rj}}$}
\\[8pt]
\normalsize{(a) - update estimated $\rf_\rj$ and the coresponding variance $\mbox{var}_j$} 
};

    \node [box, below of=f, node distance=11em, scale=0.9] (vf)
{\Large{$ \aht_{f_j} = \rfh_\rj^2 + \widehat{\mbox{var}}_j $}
\\[4pt]
\Large{$ \bht_{f_j} = \beta_f + \frac{1}{2} $}
\normalsize{$\leftarrow$ \textit{ct. during iterations}}
\\[8pt]
\normalsize{(b) - update estimated $\Ic\Gc$ parameters modelling the variances $\rv_{\rf_\rj}$} 
};

    \node [boxbig, below of=vf, node distance=11em, scale=0.9] (veps) {\Large{$ \aht_{\epsilon_i} = \bHb_\bi \Sigmabh \bHb_\bi^T + \left( \bg_\bi - \bHb_\bi \rfbh \right)^2 $}
\\[4pt]
\Large{$ \bht_{\epsilon_i} = \beta_\epsilon $}
\normalsize{$\leftarrow$ \textit{ct. during iterations}}
\\[8pt]
\normalsize{(c) - update estimated $\Ic\Gc$ parameters modelling the uncertainties variances $\rv_{\repsilon_\ri}$}     
};

    \node [boxsmall, above right = -0.15cm and 0.9cm of vf, node distance=17em, scale=0.9] (Vf) {\Large{$\rVbh_\rf = \diag {\frac{\bht_{f_j}}{\aht_{f_j}}}$}
\\[8pt]
\normalsize{(d)} 
};

    \node [boxsmall, below left = -4.4cm and 0.3cm of veps, node distance=17em, scale=0.9] (Veps) {\Large{$\rVbh_\repsilon = \diag {\frac{\bht_{\epsilon_j}}{\aht_{\epsilon_j}}}$}
\\[8pt]
\normalsize{(e)} 
};

\node [cloud, above of=f, node distance=8em, scale=0.9] (init)
{\Large{Initialization}};  

    \path [line] (f) -- (vf)[near start];
    \path [line] (Vf) |- (f);
    \path [line] (vf) -| (Vf);
    \path [line] (vf) -- (veps);
    \path [line] (Veps) |- (f);
    \path [line] (veps) -| (Veps);
    \path [line] (init) -- (f);
    
\end{tikzpicture}
\end{center}
\caption{Iterative algorithm corresponding to PM estimation via VBA - full separability for Laplace hierarchical model, non-stationary Student-t uncertainties model}
\label{Fig:IA_PM_L_nsStL_unbis}
\end{figure}
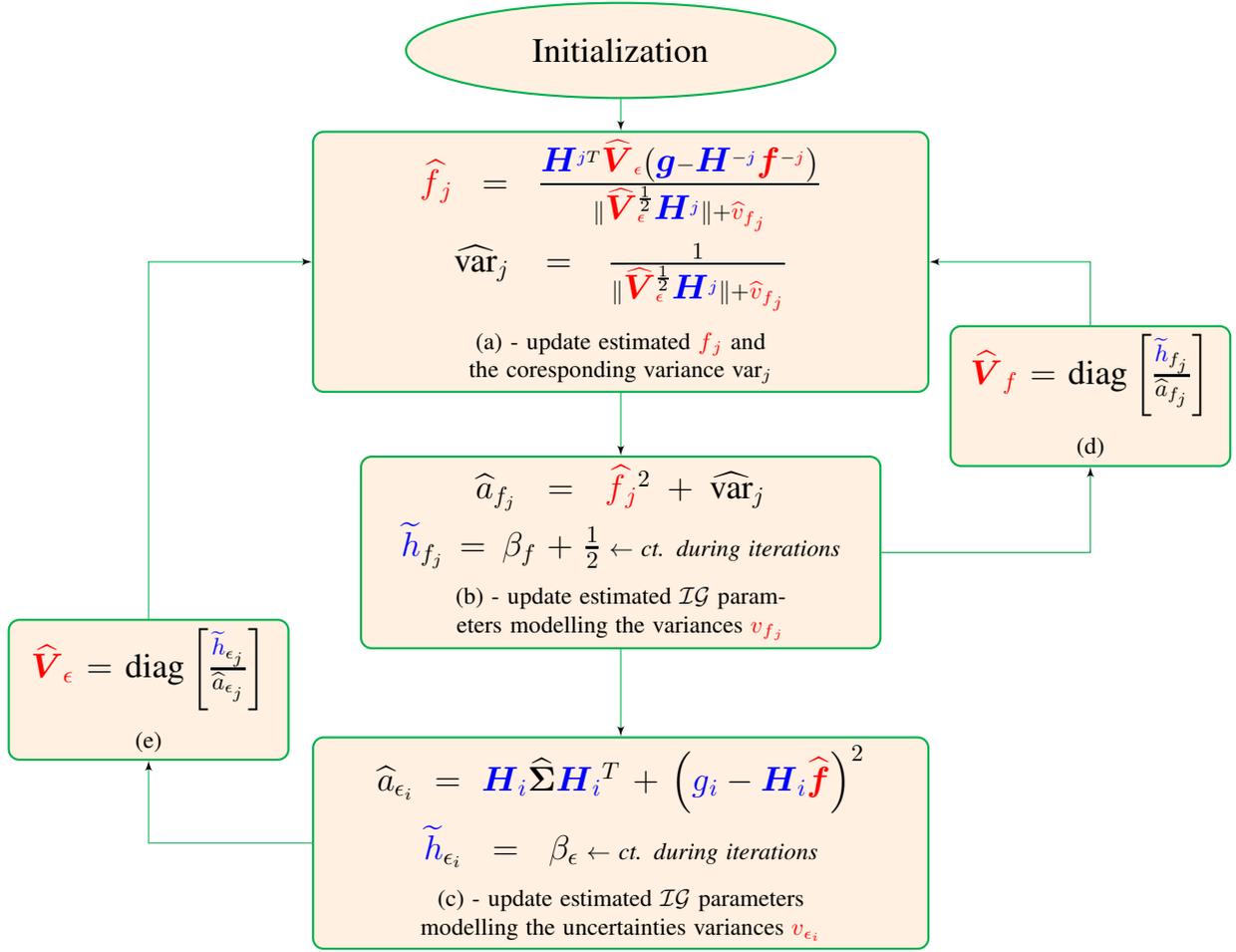

\subsection{Laplace hierarchical model: stationary Laplace uncertainties model, unknown uncertainties variance}
\label{Subsec:L_sLL_un}
\begin{itemize}
\item the hierarchical model is using as a \textbf{prior} the \textbf{Laplace} distribution;
\item the Laplace prior distribution is expressed via \textbf{LPM}, Equation~\eqref{Eq:LPM1}, considering the variance $\rvb_\rf$ as unknown;
\item the \textbf{likelihood} is derived from the distribution proposed for modelling the uncertainties vector $\repsilonb$;
\item for the uncertainties vector $\repsilonb$ a \textbf{stationary Laplace uncertainties model} is proposed, i.e. a multivariate Laplace distribution expressed via \textbf{LPM} is used under the following two assumptions:\\
a) each element of the uncertainties vector has the same \textbf{variance}, $\rv_{\repsilon}$;\\
b) the variance $\rv_{\repsilon}$ is \textbf{unknown}; 
\end{itemize}
\beq
\rotatebox{90}{\hspace{-1.35cm}Laplace sLL UNK} \;
\vline
\left.\barr{ll}
Likelihood: \textbf{sLL:}\;\;\;
\left\{\barr{ll}
p\left( \bgb | \rfb, \rv_{\repsilon} \right) = \Nc \left( \bgb | \bHb\rfb, \rv_{\repsilon}^{-1} \Ib \right) \propto \rv_{\repsilon}^{\frac{N}{2}} \; \exp \left\lbrace   - \frac{\rv_\repsilon}{2}  \| \left( \bgb - \bHb\rfb \right) \| \right\rbrace, 
\\[7pt]
p\left( \rv_{\repsilon} | 1, \frac{b_{\epsilon}}{2} \right)
= \Ic\Gc\left( \rv_{\repsilon} | 1, \frac{b_{\epsilon}}{2} \right) \propto \rv_{\repsilon_\ri}^{-2} \; \exp \left\lbrace - \frac{b_\epsilon}{2\rv_{\repsilon_\rj}} \right\rbrace ,\\[4pt]
\earr\right.
\\[23pt]
Prior:\;\;\;\;\;\;\;\textbf{LPM:}\;\;
\left\{\barr{ll}
p(\rfb|0,\rvb_{\rf}) = \Nc(\rfb|0, \; \rVb_{\rf}^{-1}) \propto \det{\rVb_{\rf}}^{\frac{1}{2}} \; \exp \left\lbrace - \frac{1}{2}\| \rVb_{\rf}^{\frac{1}{2}} \rfb \| \right\rbrace,
\\[7pt]
p(\rvb_{\rf}|b_f) = \prod_{j=1}^{M} \Ic\Gc(\rv_{\rf_\rj}|1, \; \frac{b_f}{2}) \propto \prod_{j=1}^{M} \rv_{\rf_\rj}^{-2} \; \exp \left\lbrace - \sum_{j=1}^{M} \frac{b_f}{2\rv_{\rf_\rj}} \right\rbrace, \\[7pt]
\hspace{4.4cm}
\rVb_{\rf} = \diag{\rvb_{\rf}},\; \rvb_{\rf} = \left[\ldots,\rv_{\rf_{\rj}}, \ldots\right].\\[4pt]
\earr\right.
\earr\right.
\label{Eq:L_sLL_un}
\eeq

\subsection{Laplace hierarchical model: non-stationary Laplace uncertainties model, unknown uncertainties variances}
\label{Subsec:L_nsLL_un}
\begin{itemize}
\item the hierarchical model is using as a \textbf{prior} the \textbf{Laplace} distribution;
\item the Laplace prior distribution is expressed via \textbf{LPM}, Equation~\eqref{Eq:LPM1}, considering the variance $\rvb_\rf$ as unknown;
\item the \textbf{likelihood} is derived from the distribution proposed for modelling the uncertainties vector $\repsilonb$;
\item for the uncertainties vector $\repsilonb$ a \textbf{non-stationary Laplace uncertainties model} is proposed, i.e. a multivariate Laplace distribution expressed via \textbf{LPM} is used under the following assumption:\\
a) the variance vector $\rvb_{\repsilon}$ is \textbf{unknown}; 
\end{itemize}
\beq
\rotatebox{90}{\hspace{-1.55cm}Laplace nsLL UNK} \;
\vline
\left.\barr{ll}
Likelihood: \textbf{nsLL:}\;
\left\{\barr{ll}
p\left( \bgb | \rfb, \rvb_{\repsilon} \right) = \Nc \left( \bgb | \bHb\rfb, \rVb_{\repsilon}^{-1} \right) \propto 
 \det{\rVb_{\repsilon}}^{\frac{1}{2}}
\exp \left\lbrace - \frac{1}{2}\| \rVb_{\repsilon}^{\frac{1}{2}} \left( \bgb - \bHb\rfb \right) \| \right\rbrace, 
\\[7pt]
p\left( \rvb_{\repsilon} | \beta_{\epsilon} \right)
= \prod_{i=1}^{N} \Ic\Gc\left( \rv_{\repsilon_\ri} | 1, \frac{\beta_{\epsilon}}{2} \right) \propto \prod_{i=1}^{N} \rv_{\repsilon_\ri}^{-2} \; \exp \left\lbrace - \sum_{i=1}^{N} \frac{\beta_\epsilon}{2\rv_{\repsilon_\rj}} \right\rbrace ,\\[7pt]
\hspace{5.4cm}
\rVb_{\repsilon} = \diag{\rvb_{\repsilon}},\; \rvb_{\repsilon} = \left[\ldots,\rv_{\repsilon_{\ri}}, \ldots\right],\\[4pt]
\earr\right.
\\[33pt]
Prior:\;\;\;\;\;\;\;\textbf{LPM:}\;\;
\left\{\barr{ll}
p(\rfb|0,\rvb_{\rf}) = \Nc(\rfb|0, \; \rVb_{\rf}^{-1}) \propto \det{\rVb_{\rf}}^{\frac{1}{2}} \; \exp \left\lbrace - \frac{1}{2}\| \rVb_{\rf}^{\frac{1}{2}} \rfb \| \right\rbrace,
\\[7pt]
p(\rvb_{\rf}|\beta_f) = \prod_{j=1}^{M} \Ic\Gc(\rv_{\rf_\rj}|1, \frac{\beta_f}{2}) \propto \prod_{j=1}^{M} \rv_{\rf_\rj}^{-2} \; \exp \left\lbrace - \sum_{j=1}^{M} \frac{\beta_f}{2\rv_{\rf_\rj}} \right\rbrace,\\[7pt]
\hspace{4.4cm}
\rVb_{\rf} = \diag{\rvb_{\rf}},\; \rvb_{\rf} = \left[\ldots,\rv_{\rf_{\rj}}, \ldots\right].\\[4pt]
\earr\right.
\earr\right.
\label{Eq:L_nsLL_un}
\eeq
The hierarchical model build over the linear forward model, Equation~\eqref{Eq:LinearModel}, using as a prior for $\rfb$ a Laplace distribution, expressed via the Laplace Prior Model (LPM), Equation~\eqref{Eq:LPM1} and modelling the uncertainties of the model $\repsilonb$ using the non-stationary Laplace Uncertainties Model (nsLUM), Equation~\eqref{Eq:nsLUM}, is presented in Equation~\eqref{Eq:L_nsLL_un}. The posterior distribution is obtained via the Bayes rule, Equation~\eqref{Eq:Post_L_nsLL_un}:
\beq
\begin{split}
p(\rfb, \rvb_\rf, \rvb_\repsilon | \bgb, \beta_{f}, \beta_{\epsilon})
& \propto 
\; 
p\left( \bgb | \rfb, \rvb_{\repsilon} \right) 
\;
p\left( \rvb_{\repsilon} | \beta_{\epsilon} \right) 
\;
p(\rfb|0,\rvb_{\rf}) 
\;
p(\rvb_{\rf} | \beta_{f})\\
& \propto 
\;
\prod_{i=1}^{N}  \rv_{\repsilon_\ri}^{\frac{1}{2}}
\;
\exp \left\lbrace - \frac{1}{2}\| \rVb_{\repsilon}^{\frac{1}{2}} \left( \bgb - \bHb\rfb \right) \| \right\rbrace 
\;
\prod_{i=1}^{N} \rv_{\repsilon_\ri}^{-2} 
\;
\exp \left\lbrace - \sum_{i=1}^{N}\frac{\beta_{\epsilon}}{\rv_{\repsilon_\ri}} \right\rbrace\\
&
\;\;\;\;\;
\prod_{j=1}^{M}  \rv_{\rf_\ri}^{\frac{1}{2}} 
\;
\exp \left\lbrace - \frac{1}{2}\| \rVb_{\rf}^{\frac{1}{2}} \rfb \| \right\rbrace
\;
\prod_{j=1}^{M} \rv_{\rf_\rj}^{-2}
\;
\exp \left\lbrace - \sum_{j=1}^{M}\frac{\beta_{f}}{\rv_{\rf_\rj}} \right\rbrace
\end{split}
\label{Eq:Post_L_nsLL_un}
\eeq
The goal is to estimate the unknowns of the hierarchical model, namely $\rfb$, the main unknown of the linear forward model, Equation~\eqref{Eq:LinearModel} which was suppose sparse, and consequently modelled via the Laplace distribution and the two variances appearing in the hierarchical model, Equation~\eqref{Eq:L_nsLL_un}, the variance corresponding to the sparse structure $\rfb$, namely $\rvb_\rf$ and the variance corresponding to uncertainties of model $\repsilonb$, namely $\rvb_\repsilon$. 
\subsubsection{Joint MAP estimation}
\label{Subsubsec:JMAP_L_nsLL_un}
First, the Joint Maximum A Posterior (JMAP) estimation is considered: the unknowns are estimated on the basis of the available data $\bgb$, by maximizing the posterior distribution:
\beq
\left( \rfbh, \; \rvbh_\rf, \; \rvbh_\repsilon \right)
=
\operatorname*{arg\,max}_{\left( \rfb, \; \rvb_\rf, \; \rvb_\repsilon \right)} p(\rfb, \; \rvb_\rf, \; \rvb_\repsilon | \bgb, \; \beta_{f}, \; \beta_{\epsilon})
=
\operatorname*{arg\,min}_{\left( \rfb, \; \rvb_\rf, \; \rvb_\repsilon \right)} \Lc(\rfb, \; \rvb_\rf, \; \rvb_\repsilon),
\eeq 
where for the second equality the criterion $\Lc(\rfb, \rvb_{\rf}, \rvb_{\repsilon})$ is defined as:
\beq
\Lc(\rfb, \rvb_{\rf}, , \rvb_{\repsilon})= -\ln  p(\rfb, \; \rvb_\rf, \; \rvb_\repsilon | \bgb, \; \beta_{f}, \; \beta_{\epsilon})
\label{Eq:Def_L_Crit_L_nsLL_un}
\eeq
The MAP estimation corresponds to the solution minimizing the criterion $\Lc(\rfb, \rvb_{\rf}, \rvb_{\repsilon})$.
From the analytical expression of the posterior distribution, Equation~\eqref{Eq:Post_L_nsLL_un} and the definition of the criterion $\Lc$ Equation~\eqref{Eq:Def_L_Crit_L_nsLL_un}, we obtain:
\beq
\begin{split}
\Lc(\rfb, \rvb_{\repsilon}, \rvb_{\rf})
=
-\ln  p(\rfb, \rvb_{\repsilon}, \rvb_{\rf}|\bgb)
&
=
\frac{1}{2} \| \rVb_{\repsilon}^{\frac{1}{2}} \left( \bgb - \bHb\rfb \right) \|
+
\frac{3}{2} \sum_{i=1}^{N} \ln \rv_{\repsilon_\ri} 
+
\frac{1}{2} \sum_{i=1}^{N}\frac{\beta_{\epsilon}}{\rv_{\repsilon_\ri}} 
\\
&
+
\frac{1}{2} \| \rVb_{\rf}^{\frac{1}{2}} \rfb \|
+
\frac{3}{2} \sum_{j=1}^{M} \ln \rv_{\rf_\rj}
+
\frac{1}{2} \sum_{j=1}^{M}\frac{\beta_{f}}{\rv_{\rf_\rj}}
\end{split}
\label{Eq:L_Crit_L_nsLL_un}
\eeq
One of the simplest optimisation algorithm that can be used is an alternate optimization of the criterion $\Lc(\rfb, \rvb_{\repsilon}, \rvb_{\rf})$ with respect to the each unknown:
\begin{itemize}
\item With respect to $\rfb$:
\beq
\begin{split}
\frac{\partial \Lc(\rfb, \rvb_{\rf}, \rvb_{\repsilon})}{\partial \rfb}=0
&\Leftrightarrow
\frac{\partial}{\partial \rfb}
\left(
\| \rVb_\repsilon^{\frac{1}{2}} \left( \bgb - \bHb\rfb \right) \|
+
\|\rVb_\rf^{\frac{1}{2}}\rfb\|
\right)
=
-\bHb^T \rVb_\repsilon \left( \bgb - \bHb \rfb \right) + \rVb_\rf \rfb = 0
\\
&
\Leftrightarrow 
\left( \bHb^T \rVb_\repsilon \bHb + \rVb_\rf \right) \rfb  = 
\bHb^T \rVb_\repsilon \bgb
\\
&
\Rightarrow
\rfbh 
=
\left( \bHb^T \rVb_\repsilon \bHb + \rVb_\rf \right)^{-1} \bHb^T \rVb_\repsilon \bgb
\end{split}
\nonumber
\eeq
\item With respect to $\rv_\rf$, $j \in \left\lbrace 1,2,\ldots, M \right\rbrace$:
\beq
\begin{split}
\frac{\partial \Lc(\rfb, \rvb_{\rf}, \rvb_{\rf})}{\partial \rv_{\rf_\rj}}=0
&
\Leftrightarrow
\frac{\partial}{\partial \rv_{\rf_\rj}}
\left(
3 \ln \rv_{\rf_\rj}
+
\beta_f \rv_{\rf_\rj}^{-1} 
+ 
\rf_\rj^2 \rv_{\rf_\rj}
\right)
=0
\\
&
\Leftrightarrow
\rf_\rj^2 \rv_{\rf_\rj}^{2}
+ 
3 \rv_{\rf_\rj}
-
\beta_f  
=0
\\
&
\Rightarrow
\rvh_{\rf_\rj} =
\frac{-3 \pm \sqrt{9 - 4 \rf_\rj^2 \beta_f}}{2 \rf_\rj^2}
\end{split}
\nonumber
\eeq
\item With respect to $\rv_{\repsilon_\ri}$, $i \in \left\lbrace 1,2,\ldots, N \right\rbrace$:\\
First, we develop the norm $\| \rVb_\repsilon^{\frac{1}{2}} \left( \bgb - \bHb\rfb \right) \|$:
\beq
\begin{split}
\| \rVb_\repsilon^{\frac{1}{2}} \left( \bgb - \bHb\rfb \right) \|
&= 
\bgb^{T} \rVb_\repsilon \bgb
-
2 \bgb^{T} \rVb_\repsilon \bHb \rfb
+
\bHb^{T} \rfb^{T} \rVb_\repsilon \bHb \rfb
\\
&=
\sum_{i=1}^{N} \rv_{\repsilon_\ri} \bg_\bi^{2}
-
2 \sum_{i=1}^{N} \rv_{\repsilon_\ri} \bg_\bi \bH_\bi \rfb 
+
\sum_{i=1}^{N} \rv_{\repsilon_\ri} \rfb^{T} \bH_\bi^{T} \bH_\bi \rfb,
\nonumber
\end{split}
\eeq
where $\bHb_\bi$ denotes the line i of the matrix $\bHb$, $i\in \left\lbrace 1,2,\ldots,N \right\rbrace$, i.e. $\bHb_\bi = \left[ \bh_{\bi\blue{1}}, \bh_{\bi\blue{1}}, \ldots, \bh_{\bi\bM} \right]$. 
\beq
\begin{split}
\frac{\partial \Lc(\rfb, \rvb_{\rf}, \rvb_{\repsilon})}{\partial \rv_{\repsilon_\ri}}=0
&
\Leftrightarrow
\frac{\partial}{\partial \rv_{\repsilon_\ri}}
\left[
3 \ln \rv_{\repsilon_\ri}
+
\beta_{\epsilon} \rv_{\repsilon_\ri}^{-1}
+ 
\left( \bg_\bi^{2} - 2 \bg_\bi \bHb_\bi \rfb  + \rfb^{T} \bHb_\bi^{T} \bHb_\bi \rfb \right) \rv_{\repsilon_\ri}
\right]
=0
\\
&
\Leftrightarrow
\left( \bg_\bi - \bHb_\bi \rfb  \right)^2 \rv_{\repsilon_\ri}^{2}
+
3 \rv_{\repsilon_\ri}
-
\beta_{\epsilon} 
=0
\\
&
\Rightarrow
\rvh_{\repsilon_\ri} = \frac{-3 \pm \sqrt{9 - 4 \left( \bg_\bi - \bHb_\bi \rfb  \right)^2 \beta_\epsilon}}{2 \left( \bg_\bi - \bHb_\bi \rfb  \right)^2 }
\end{split}
\nonumber
\eeq
\end{itemize}
The iterative algorithm obtained via JMAP estimation is presented Figure~\eqref{Fig:IA_JMAP_L_nsLL_un}.
\tikzstyle{cloud} = [draw=blue!30!green,fill=orange!12, ellipse, thick, node distance=6em, text width=12em, text centered, minimum height=4em, minimum width=12em]
\tikzstyle{boxbig} = [draw=blue!30!green, fill=orange!12, rectangle, rounded corners, thick, node distance=6.5em, text width=25em, text centered, minimum height=5em, minimum width=25em]
\tikzstyle{box} = [draw=blue!30!green, fill=orange!12, rectangle, rounded corners, thick, node distance=6.5em, text width=21em, text centered, minimum height=5em, minimum width=21em]
\tikzstyle{boxsmall} = [draw=blue!30!green, fill=orange!12, rectangle, rounded corners, thick, node distance=8.5em, text width=9em, text centered, minimum height=5em, minimum width=9em]
\tikzstyle{line} = [draw=blue!30!green, -latex']
\begin{figure}[!htb]
\centering
\begin{center}
\begin{tikzpicture}[auto]
    \node [boxbig, scale=0.9] (f) {\Large{$\rfbh 
=
\left( \bHb^T \rVbh_\repsilon \bHb + \rVbh_\rf \right)^{-1} \bHb^T \rVbh_\repsilon \bgb$}
\\[8pt]
\normalsize{(a) - update estimated $\rfb$} 
};

    \node [box, below of=f, node distance=9.6em, scale=0.9] (vf)
{\Large{$ \rvh_{\rf_\rj} = \frac{-3 \pm \sqrt{9 - 4 \rf_\rj^2 \beta_f}}{2 \rf_\rj^2} $}
\\[8pt]
\normalsize{(b) - update estimated variances $\rv_{\rf_\rj}$} 
};

    \node [box, below of=vf, node distance=9.6em, scale=0.9] (veps) {\Large{$\rvh_{\repsilon_\ri} = \frac{-3 \pm \sqrt{9 - 4 \left( \bg_\bi - \bHb_\bi \rfb  \right)^2 \beta_\epsilon}}{2 \left( \bg_\bi - \bHb_\bi \rfb  \right)^2 }$}
\\[8pt]
\normalsize{(c) - update estimated uncertainties variances $\rv_{\repsilon_\ri}$}     
};

    \node [boxsmall, above right = 0.05cm and 0.5cm of vf, node distance=17em, scale=0.9] (Vf) {\Large{$\rVbh_\rf = \diag {\rvbh_{\rf}}$}
\\[8pt]
(d) 
};

    \node [boxsmall, below left = -3.4cm and 0.5cm of veps, node distance=17em, scale=0.9] (Veps) {\Large{$\rVbh_\repsilon = \diag {\rvbh_{\repsilon}}$}
\\[8pt]
(e) 
};

\node [cloud, above of=f, scale=0.9] (init)
{\Large{Initialization}};  

    \path [line] (f) -- (vf)[near start];
    \path [line] (Vf) |- (f);
    \path [line] (vf) -| (Vf);
    \path [line] (vf) -- (veps);
    \path [line] (Veps) |- (f);
    \path [line] (veps) -| (Veps);
    \path [line] (init) -- (f);
    
\end{tikzpicture}
\end{center}
\caption{Iterative algorithm corresponding to Joint MAP estimation for Laplace hierarchical model, non-stationary Laplace uncertainties model}
\label{Fig:IA_JMAP_L_nsLL_un}
\end{figure}
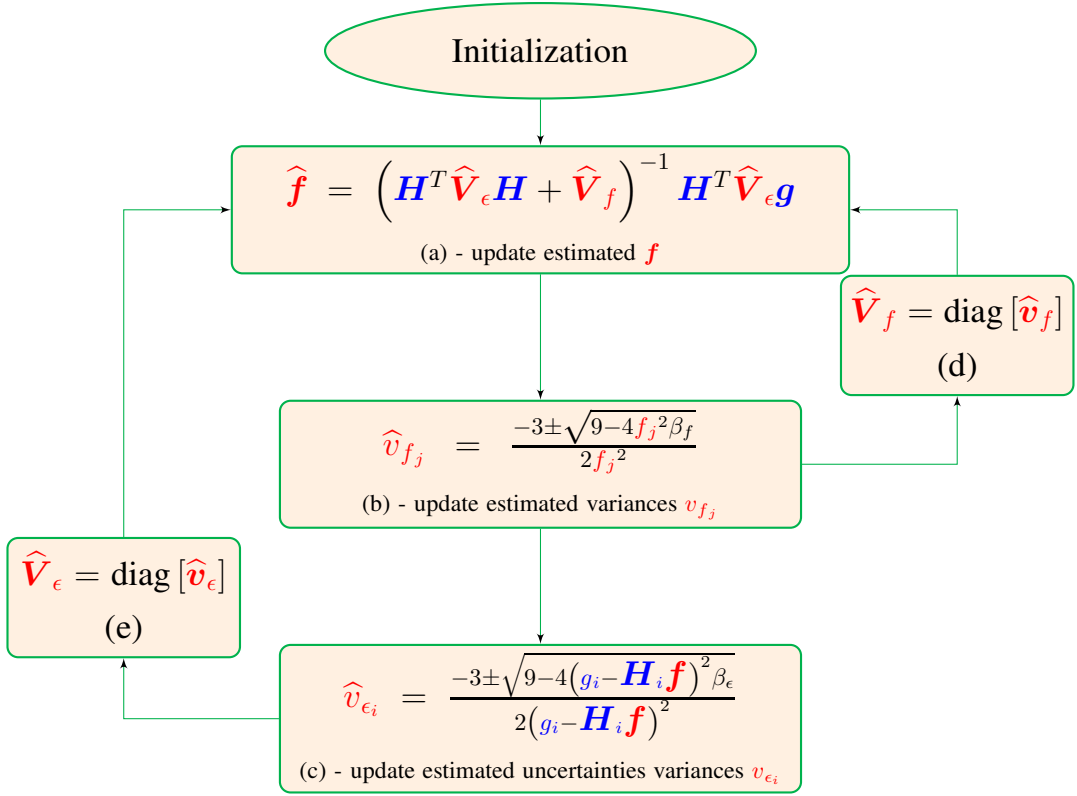
\subsubsection{Posterior Mean estimation via VBA, partial separability}
\label{Subsubsec:PM_PS_L_nsLL_un}
In this subsection, the Posterior Mean (PM) estimation is considered. The Joint MAP computes the mod of the posterior distribution. The PM computes the mean of the posterior distribution. One of the advantages of this estimator is that it minimizes the Mean Square Error (MSE). Computing the posterior means of any unknown needs great dimensional integration. For example, the mean corresponding to $\rfb$ can be computed from the posterior distribution using Equation~\eqref{Eq:PM_Comput_F_L_nsLL_un},
\beq
E_p \left\lbrace \rfb \right\rbrace = \iiint \rfb \; p(\rfb, \rvb_\rf, \rvb_\repsilon | \bgb, \beta_{f}, \beta_{\epsilon}) \d\rfb \d\rvb_{\rf} \d\rvb_{\repsilon}.
\label{Eq:PM_Comput_F_L_nsLL_un}
\eeq
In general, these computations are not easy. One way to obtain approximate estimates is to approximate $p(\rfb, \rvb_\rf, \rvb_\repsilon | \bgb, \beta_{f}, \beta_{\epsilon})$ by a separable one $q(\rfb, \rvb_\rf, \rvb_\repsilon | \bgb, \beta_{f}, \beta_{\epsilon}) = q_1(\rfb) \; q_2(\rvb_{\rf}) \; q_3(\rvb_{\repsilon}) $, then computing the posterior means using the separability. The mean corresponding to $\rfb$ is computed using the corresponding separable distribution $q_1(\rfb)$, Equation~\eqref{Eq:PM_Comput_F_L_nsLL_un_Separable}, 
\beq
E_{q_{1}} \left\lbrace \rfb \right\rbrace = \int \rfb \; q_1(\rfb) \d\rfb.
\label{Eq:PM_Comput_F_L_nsLL_un_Separable}
\eeq
If the approximation of the posterior distribution with a separable one can be done in such a way that conserves the mean, i.e. Equation~\eqref{Eq:PM_Conservation_L_nsLL_un},
\beq
E_q \left\lbrace x \right\rbrace = E_p \left\lbrace x \right\rbrace, 
\label{Eq:PM_Conservation_L_nsLL_un}
\eeq
for all the unknowns of the model, a great amount of computational cost is gained. In particular, for the proposed hierarchical model, Equation~\eqref{Eq:St_nsStL_un}, the posterior distribution, Equation~\eqref{Eq:Post_L_nsLL_un}, is not a separable one, making the analytical computations of the PM very difficult. One way the compute the PM in this case is to first approximate the posterior law $p(\rfb, \rvb_\rf, \rvb_\repsilon | \bgb, \beta_{f}, \beta_{\epsilon})$ with a separable law $q(\rfb, \rvb_\rf, \rvb_\repsilon | \bgb, \beta_{f}, \beta_{\epsilon})$, Equation~\eqref{Eq:Post_Approx_L_nsLL_un},
\beq
p(\rfb, \rvb_\rf, \rvb_\repsilon | \bgb, \beta_{f},  \beta_{\epsilon}) 
\approx 
q(\rfb, \rvb_\rf, \rvb_\repsilon | \bgb, \beta_{f}, \beta_{\epsilon})
=
q_1(\rfb) \; q_2(\rvb_{\rf}) \; q_3(\rvb_{\repsilon})\label{Eq:Post_Approx_L_nsLL_un}
\eeq
where the notations from Equation~\eqref{Eq:Post_Approx_L_nsLL_un_Notations1} are used
\beq
q_{2}(\rvb_{\rf}) = \prod_{j=1}^{M} q_{2j}(\rv_{\rf_\rj}),
\;\;;\;\;
q_{3}(\rvb_{\repsilon}) = \prod_{i=1}^{N} q_{3i}(\rv_{\repsilon_\ri})
\label{Eq:Post_Approx_L_nsLL_un_Notations1}
\eeq
by minimizing of the Kullback-Leibler divergence, defined as:
\beq
\begin{split}
\mbox{KL} & \left( q(\rfb, \rvb_\rf, \rvb_\repsilon | \bgb, \beta_{f}, \beta_{\epsilon}) : p(\rfb, \rvb_\rf, \rvb_\repsilon | \bgb, \beta_{f}, \beta_{\epsilon}) \right) =
\\
&=\iint \ldots \int q(\rfb, \rvb_\rf, \rvb_\repsilon | \bgb, \beta_{f}, \beta_{\epsilon}) \; \ln\frac{q(\rfb, \rvb_\rf, \rvb_\repsilon | \bgb, \beta_{f}, \beta_{\epsilon})} {p(\rfb, \rvb_\rf, \rvb_\repsilon | \bgb, \beta_{f}, \beta_{\epsilon})} \d \rfb \d \rvb_{\repsilon} \d \rvb_{\rf}
\label{Eq:Kull-Leib_L_nsLL_un}
\end{split}
\eeq
where the notations from Equation~\eqref{Eq:Post_Approx_L_nsLL_un_Notations2} are used
\beq
\d \rvb_{\rf} = \prod_{j=1}^{M} \d \rv_{\rf_\rj}
\;\;;\;\;
\d \rvb_{\repsilon} = \prod_{i=1}^{N} \d \rv_{\repsilon_\ri}.
\label{Eq:Post_Approx_L_nsLL_un_Notations2}
\eeq
Equation~\eqref{Eq:Post_Approx_L_nsLL_un_Notations1} is selecting a partial separability for the approximated posterior distribution $q(\rfb, \rvb_\rf, \rvb_\repsilon | \bgb,  \beta_{f}, \beta_{\epsilon})$ in the sense that a total separability is imposed for the distributions corresponding to the two variances appearing in the hierarchical model, $q_2\left( \rvb_{\rf} \right)$ and $q_3\left( \rvb_{\repsilon} \right)$ but not for the distribution corresponding to $\rfb$. Evidently, a full separability can be imposed, by adding the supplementary condition $q_{1}(\rfb) = \prod_{j=1}^{M} q_{1j}(\rf_\rj)$ in Equation~\eqref{Eq:Post_Approx_L_nsLL_un_Notations1}. This case is considered in Subsection~\eqref{Subsubsec:PM_FS_L_nsLL_un}. The minimization can be done via alternate optimization resulting the following proportionalities from Equations~\eqref{Eq:VBA_Proport1_L_nsLL_un},~\eqref{Eq:VBA_Proport2_L_nsLL_un} and~\eqref{Eq:VBA_Proport3_L_nsLL_un},  
\begin{subequations}
\begin{align}
&q_1(\rfb) \;\;\; \propto \; \exp \left\lbrace \biggl\langle \ln p(\rfb, \rvb_\rf, \rvb_\repsilon | \bgb, \beta_{f}, \beta_{\epsilon}) \biggr\rangle_{ q_2(\rvb_{\rf}) \; q_3(\rvb_{\repsilon})} \right\rbrace,
\label{Eq:VBA_Proport1_L_nsLL_un}
\\[4pt]
&q_{2j}(\rv_{\rf_\red{j}}) \propto \; \exp \left\lbrace \biggl\langle \ln p(\rfb, \rvb_\rf, \rvb_\repsilon | \bgb, \beta_{f}, \beta_{\epsilon}) \biggr\rangle_{q_1(\rfb) \; q_{2-j}(\rv_{\rf_\red{j}}) \; q_3(\rvb_{\repsilon})} \right\rbrace,\;\;j \in \left\lbrace 1,2 \ldots, M \right\rbrace,
\label{Eq:VBA_Proport2_L_nsLL_un}
\\[4pt]
&q_{3i}(\rv_{\repsilon_\ri}) \; \propto \; \exp \left\lbrace \biggl\langle \ln p(\rfb, \rvb_\rf, \rvb_\repsilon | \bgb, \beta_{f}, \beta_{\epsilon}) \biggr\rangle_{q_1(\rfb) \; q_2(\rvb_{\rf}) \; q_{3-i}(\rv_{\repsilon_\ri})} \right\rbrace,\;\;i \in \left\lbrace 1,2 \ldots, N \right\rbrace,
\label{Eq:VBA_Proport3_L_nsLL_un}
\end{align}
\end{subequations}
using the notations:
\beq
q_{2-j}(\rv_{\rf_\red{j}})=\displaystyle \prod_{k=1,k \neq j}^{M} q_{2k}(\rv_{\rf_\red{k}})
\;\;\;;\;\;\;
q_{3-i}(\rv_{\repsilon_\ri})=\displaystyle \prod_{k=1,k \neq i}^{N} q_{3k}(\rv_{\repsilon_\rk})
\label{Eq:Def_q_Min_L_nsLL_un}
\eeq
and also
\beq
\biggl\langle u(x) \biggr\rangle_{v(y)}= \displaystyle \int u(x) v(y) \d y. 
\label{Eq:Def_Integ_L_nsLL_un}
\eeq
Via Equation~\eqref{Eq:Def_L_Crit_L_nsLL_un} and Equation~\eqref{Eq:L_Crit_L_nsLL_un}, the analytical expression of logarithm of the posterior distribution is obtained, Equation~\eqref{Eq:Log_Post_L_nsLL_un}:
\beq
\begin{split}
\ln  p(\rfb, \rvb_{\repsilon}, \rvb_{\rf}|\bgb, \beta_{f}, \beta_{\epsilon})
=
&
-\frac{1}{2}\| \rVb_{\repsilon}^{\frac{1}{2}} \left( \bgb - \bHb\rfb \right) \|
-
\frac{3}{2} \sum_{i=1}^{N} \ln \rv_{\repsilon_\ri} 
-
\frac{1}{2} \sum_{i=1}^{N}\frac{\beta_{\epsilon}}{\rv_{\repsilon_\ri}} 
-
\frac{1}{2}\| \rVb_{\rf}^{\frac{1}{2}} \rfb \|
-
\frac{3}{2} \sum_{j=1}^{M} \ln \rv_{\rf_\rj}
-
\frac{1}{2} \sum_{j=1}^{M}\frac{\beta_{f}}{\rv_{\rf_\rj}}
\end{split}
\label{Eq:Log_Post_L_nsLL_un}
\eeq
\paragraph{Computation of the analytical expression of $q_1(\rfb)$.}
The proportionality relation corresponding to $q_1(\rfb)$ is presented in established in Equation~\eqref{Eq:VBA_Proport1_L_nsLL_un}. In the expression of $\ln p\left(\rfb, \rvb_\rf, \rvb_\repsilon | \bgb, \beta_{f}, \beta_{\epsilon} \right)$ all the terms free of $\rfb$ can be regarded as constants. Via Equation~\eqref{Eq:Log_Post_L_nsLL_un} the integral defined in Equation~\eqref{Eq:Def_Integ_L_nsLL_un} becomes:
\beq
\begin{split}
\biggl\langle \ln p(\rfb, \rvb_\rf, \rvb_\repsilon | \bgb, \beta_{f}, \beta_{\epsilon}) \biggr\rangle_{q_2(\rvb_{\rf}) \; q_3(\rvb_{\repsilon})}
&
=
-\frac{1}{2}
\biggl\langle 
\|\rVb_{\repsilon}^{\frac{1}{2}} \left(\bgb - \bHb\rfb\right) \|
\biggr\rangle_{ q_3(\rvb_{\repsilon}) }
-\frac{1}{2}
\biggl\langle 
\| \rVb_{\rf}^{\frac{1}{2}} \rfb \|
\biggl\rangle_{ q_2(\rvb_{\rf}) }.
\end{split}
\label{Eq:q1_Integ_1_L_nsLL_un}
\eeq
Introducing the notations:
\beq
\begin{split}
\rvt_{\rf_\rj} = \biggl\langle \rv_{\rf_\rj} \biggr\rangle_{q_{2j}\left( \rv_{\rf_\rj} \right)}
\;\;;\;\;
\rvbt_{\rf}\;&=\;
\begin{bmatrix}
\rvt_{\rf_\red{1}} \ldots 
\rvt_{\rf_\rj} \ldots 
\rvt_{\rf_\rM}
\end{bmatrix}^T
\;\;;\;\;
\rVbt_{\rf} = \diag {\rvbt_{\rf}}\;
\\
\rvt_{\repsilon_\ri}
=
\biggl\langle \rv_{\repsilon_\ri} \biggr\rangle_{q_{3i}\left( \rv_{\repsilon_\ri} \right)}
\;\;;\;\;
\rvbt_\repsilon\;&=\;
\begin{bmatrix}
\rvt_{\repsilon_\red{1}}
\ldots 
\rvt_{\repsilon_\ri}
\ldots 
\rvt_{\repsilon_\rN}
\end{bmatrix}^T
\;\;;\;\;
\rVbt_\repsilon = \diag {\rvbt_\repsilon}
\end{split}
\label{Eq:q1_Integ_Not1_L_nsLL_un}
\eeq
the integral from Equation~\eqref{Eq:q1_Integ_1_L_nsLL_un} becomes:
\beq
\begin{split}
\biggl\langle \ln p(\rfb, \rvb_\rf, \rvb_\repsilon | \bgb,  \beta_{f}, \beta_{\epsilon}) \biggr\rangle_{q_2(\rvb_{\rf}) \; q_3(\rvb_{\repsilon})}
&
= -\frac{1}{2} \|\rVbt_{\repsilon}^{\frac{1}{2}} \left(\bgb - \bHb\rfb\right) \| - \frac{1}{2} \| \rVbt_{\rf}^{\frac{1}{2}} \rfb \|.
\end{split}
\eeq
Noting that $\biggl\langle \ln p(\rfb, \rvb_\rf, \rvb_\repsilon | \bgb, \beta_{f}, \beta_{\epsilon}) \biggr\rangle_{q_2(\rvb_{\rf}) \; q_3(\rvb_{\repsilon})}
$ is a quadratic criterion and considering the proportionality from Equation~\eqref{Eq:VBA_Proport1_L_nsLL_un} it can be concluded that $q_1 \left( \rfb \right)$  is a multivariate Normal distribution. Minimizing the criterion leads to the analytical expression of the corresponding mean. The variance is obtained by identification: 
\beq
q_1(\rfb) = \Nc\left( \rfb | \rfbh, \Sigmabh \right),
\left\{\barr{ll}
\rfbh = 
\left( \bHb^T  \rVbt_\repsilon \bHb + \rVbt_{\rf} \right)^{-1}
\bHb^T \rVbt_\repsilon \bgb,
\\[10pt]
\Sigmabh = \left( \bHb_\blue{1}^T  \rVbt_\repsilon \bHb + \rVbt_{\rf} \right)^{-1}.
\earr\right.
\label{Eq:q1_Anal_1_L_nsLL_un}
\eeq
We note that both the expressions of the mean and variance depend on expectancies corresponding to two variances of the hierarchical model.
\paragraph{Computation of the analytical expression of $q_{2j}(\rv_{\rf_\rj})$.}
The proportionality relation corresponding to $q_{2j}(\rv_{\rf_\rj})$ is presented in established in Equation~\eqref{Eq:VBA_Proport2_L_nsLL_un}. In the expression of $\ln p\left(\rfb, \rvb_\rf, \rvb_\repsilon | \bgb, \beta_{f}, \beta_{\epsilon} \right)$ all the terms free of $\rv_{\rf_\rj}$ can be regarded as constants. Via Equation~\eqref{Eq:Log_Post_L_nsLL_un} the integral defined in Equation~\eqref{Eq:Def_Integ_L_nsLL_un} becomes:
\beq
\begin{split}
\biggl\langle \ln p(\rfb, \rvb_\rf, \rvb_\repsilon | \bgb, \beta_{f}, \beta_{\epsilon}) \biggr\rangle_{q_1(\rfb) \; q_{2-j}(\rv_{\rf_\red{j}}) \; q_3(\rvb_{\repsilon})}
&
=
-
\frac{1}{2}
\biggl\langle
\| \rVb_{\rf}^{\frac{1}{2}} \rfb \|
\biggr\rangle_{q_1(\rfb) \; q_{2-j}(\rv_{\rf_\red{j}})}
-
\frac{3}{2} \ln \rv_{\rf_\rj}
-
\frac{1}{2} \frac{\beta_{f}}{\rv_{\rf_\rj}}
\end{split}
\label{Eq:q2_Integ_1_L_nsLL_un}
\eeq
Introducing the notations:
\beq
\rvbt_{\rf_{-\ri}}=
\begin{bmatrix}
\rvt_{\rf_\red{1}} \;
\ldots \;
\rvt_{\rf_{\ri-\red{1}}} \;
\rv_{\rf_\ri}  \;
\rvt_{\rf_{\ri+\red{1}}} \;
\ldots \;
\rvt_{\rf_\rN}
\end{bmatrix}^T
\;\;;\;\;
\rVbt_{\rf_{-\ri}}=
\mbox{diag}\left(\rvbt_{\rf_{-\ri}} \right)
\label{Eq:q2_Integ_Not1_L_nsLL_un}
\eeq
the integral $\biggl\langle
\| \rVb_{\rf}^{\frac{1}{2}} \rfb \|
\biggr\rangle_{q_1(\rfb) \; q_{2-j}(\rv_{\rf_\red{j}})}
$ can be written:
\beq
\biggl\langle
\| \rVb_{\rf}^{\frac{1}{2}} \rfb \|
\biggr\rangle_{q_1(\rfb) \; q_{2-j}(\rv_{\rf_\red{j}})}
= 
\biggl\langle \| \rVbt_{\rf_{-\ri}}^{\frac{1}{2}} \rfb \|^2 \biggr\rangle_{q_1(\rfb)}
\eeq 
Considering that $q_1(\rfb)$ is a multivariate Normal distribution, Equation~\eqref{Eq:q1_Anal_1_L_nsLL_un}:
\beq
\biggl\langle \|  \rVbt_{\rf_{-\ri}}^{\frac{1}{2}}  \rfb \|^2 \biggr\rangle_{q_1(\rfb)}
=
\| \rVbt_{\rf_{-\ri}}^{\frac{1}{2}} \rfbh \|^2 + \mbox{Tr}\left(\rVbt_{\rf_{-\ri}} \Sigmabh \right) 
=
C + \rv_{\rf_\ri} \left( \rfh_\rj^2 + \Sigmabh_{jj} \right)
\label{Eq:q2_Integ_2_L_nsLL_un}
\eeq
From Equation~\eqref{Eq:q2_Integ_1_L_nsLL_un} and Equation~\eqref{Eq:q2_Integ_2_L_nsLL_un}: 
\beq
\begin{split}
\biggl\langle \ln p(\rfb, \rvb_\rf, \rvb_\repsilon | \bgb, \beta_{f}, \beta_{\epsilon}) \biggr\rangle_{q_1(\rfb) \; q_{2-j}(\rv_{\rf_\red{j}}) \; q_3(\rvb_{\repsilon})}
=
- \frac{3}{2} \ln \rv_{\rf_\rj}
- \frac{1}{2} \beta_{f} \rv_{\rf_\rj}^{-1} 
- \frac{1}{2} \left( \rfh_\rj^2 + \Sigmabh_{jj} \right) \rv_{\rf_\rj}
\end{split}
\eeq
from which it can establish the proportionality corresponding to
$q_{2j}(\rv_{\rf_\rj})$:
\beq
q_{2j}(\rv_{\rf_\rj})
\propto
\rv_{\rf_\rj}^{-\frac{3}{2}}
\exp
\biggl\lbrace 
-\frac{1}{2}
\left[ \left( \rfh_\rj^2 + \Sigmabh_{jj} \right) \rv_{\rf_\rj} + \beta_{f} \rv_{\rf_\rj}^{-1} \right]
\biggr\rbrace,
\eeq
leading to the conclusion that $q_{2j}(\rv_{\rf_\rj})$ is a generalized inverse Gaussian distribution (see Equation~\eqref{Eq:GenInvGau}) with the following parameters:
\beq
q_{2j}(\rv_{\rf_\rj})=\Gc\Ic\Gc\left(\rv_{\rf_\rj}|\aht_{f_j},\bht_{f_j}, \cht_{f_j} \right),
\left\{\barr{ll}
\aht_{f_j} = \rfh_\rj^2 + \Sigmabh_{jj}
\\[10pt]
\bht_{f_j} = \beta_f 
\\[10pt]
\cht_{f_j} = -\frac{1}{2}
\earr\right.
\label{Eq:q2_Anal_1_L_nsLL_un}
\eeq
\paragraph{Computation of the analytical expression of $q_{3i}(\rv_{\repsilon_\ri})$.}
The proportionality relation corresponding to $q_{3i}(\rv_{\repsilon_\ri})$ is presented in established in Equation~\eqref{Eq:VBA_Proport3_L_nsLL_un}. In the expression of $\ln p\left(\rfb, \rvb_\rf, \rvb_\repsilon | \bgb, \beta_{f},  \beta_{\epsilon} \right)$ all the terms free of $\rv_{\repsilon_\ri}$ can be regarded as constants. Via Equation~\eqref{Eq:Log_Post_L_nsLL_un} the integral defined in Equation~\eqref{Eq:Def_Integ_L_nsLL_un} becomes:
\beq
\begin{split}
\biggl\langle \ln p(\rfb, \rvb_\rf, \rvb_\repsilon | \bgb, \beta_{f}, \beta_{\epsilon}) \biggr\rangle_{q_1(\rfb) \; q_{2}(\rvb_{\rf}) \; q_{3-i}(\rv_{\repsilon_\ri})}
=
&
-\frac{1}{2} \biggl\langle \| \rVb_{\repsilon}^{\frac{1}{2}} \left( \bgb - \bHb \rfb \right) \| \biggr\rangle_{q_1(\rfb) \; q_{3-i}(\rv_{\repsilon_\ri})}
-
\frac{3}{2} \ln \rv_{\repsilon_\ri}
-
\frac{1}{2} \frac{\beta_{\epsilon}}{\rv_{\repsilon_\ri}}
\end{split}
\label{Eq:q3_Integ_1_L_nsLL_un}
\eeq
Introducing the notations:
\beq
\rvbt_{\repsilon_{-\ri}}=
\begin{bmatrix}
\rvt_{\repsilon_\red{1}} \;
\ldots \;
\rvt_{\repsilon_{\ri-\red{1}}} \;
\rv_{\repsilon_\ri}  \;
\rvt_{\repsilon_{\ri+\red{1}}} \;
\ldots \;
\rvt_{\repsilon_\rN}
\end{bmatrix}^T
\;\;;\;\;
\rVbt_{\repsilon_{-\ri}}=
\mbox{diag}\left(\rvbt_{\repsilon_{-\ri}} \right)
\label{Eq:q3_Integ_Not1_L_nsLL_un}
\eeq
the integral $\biggl\langle \| \rVb_{\repsilon}^{\frac{1}{2}} \left( \bgb - \bHb \rfb \right) \| \biggr\rangle_{q_1(\rfb) \; q_{3-i}(\rv_{\repsilon_\ri})}
$ can be written:
\beq
\biggl\langle \| \rVb_{\repsilon}^{\frac{1}{2}} \left( \bgb - \bHb \rfb \right) \| \biggr\rangle_{q_1(\rfb) \; q_{3-i}(\rv_{\repsilon_\ri})}
= 
\biggl\langle \| \rVbt_{\repsilon_{-\ri}}^{\frac{1}{2}} \left( \bgb - \bHb \rfb \right) \|^2 \biggr\rangle_{q_1(\rfb)}
\eeq 
Considering that $q_1(\rfb)$ is a multivariate Normal distribution, Equation~\eqref{Eq:q1_Anal_1_L_nsLL_un}:
\beq
\biggl\langle
\| \rVbt_{\repsilon_{-\ri}}^{\frac{1}{2}} \left( \bgb - \bHb \rfb \right) \|^2
\biggr\rangle_{q_1(\rfb)}
=
\| \rVbt_{\repsilon_{-\ri}}^{\frac{1}{2}}
\left( \bgb - \bHb \rfbh \right) \|^2
+
\mbox{Tr}\left( \bHb^T \rVb_{\repsilon_{-\ri}} \bHb \Sigmabh \right)
\label{Eq:q3_Integ_2_L_nsLL_un}
\eeq
and considering as constants all terms free of $\rv_{\repsilon_\ri}$:
\beq
\| \rVbt_{\repsilon_{-\ri}}^{\frac{1}{2}}
\left( \bgb - \bHb \rfbh \right) \|^2
=
C +
\rv_{\repsilon_\ri} \left( \bg_\bi - \bHb_\bi \rfbh \right)^2
\;\;;\;\;
\mbox{Tr}\left( \bHb^T \rVbt_{\repsilon_{-\ri}} \bHb \Sigmabh \right)
=
C + 
\rv_{\repsilon_\ri} \bHb_\bi \Sigmabh \bHb_\bi^T
\eeq
where $\bHb_\bi$ is the line i of the matrix $\bHb$, so we can conclude:
\beq
\biggl\langle
\| \rVb_{\repsilon}^{\frac{1}{2}}\left( \bgb - \bHb\rfb \right)\|^2
\biggr\rangle_{q_1(\rfb) \; q_{3-i}(\rv_{\repsilon_\ri})}
=
C + 
\left(
\bHb_\bi \Sigmabh \bHb_\bi^T
+
\left( \bg_\bi - \bHb_\bi \rfbh \right)^2
\right)
\rv_{\repsilon_\ri}
\label{Eq:q3_Integ_3_L_nsLL_un}
\eeq
From Equation~\eqref{Eq:q3_Integ_1_L_nsLL_un} and Equation~\eqref{Eq:q3_Integ_3_L_nsLL_un}: 
\beq
\begin{split}
\biggl\langle \ln p(\rfb, \rvb_\rf, \rvb_\repsilon | \bgb, \beta_{f}, \beta_{\epsilon}) \biggr\rangle_{q_1(\rfb) \; q_{2}(\rvb_{\rf}) \; q_{3-i}(\rv_{\repsilon_\ri})}
=
&
- 
\frac{3}{2} \ln \rv_{\repsilon_\ri}
- 
\frac{\beta_{\epsilon}}{2} \rv_{\repsilon_\ri}^{-1}
- 
\frac{1}{2} 
\left[ \bHb_\bi \Sigmabh \bHb_\bi^T +
\left( \bg_\bi - \bHb_\bi \rfbh \right)^2 \right] \rv_{\repsilon_\ri}
\end{split}
\eeq
from which it can establish the proportionality corresponding to
$q_{3i}(\rv_{\repsilon_\ri})$:
\beq
q_{3i}(\rv_{\repsilon_\ri})
\propto
\rv_{\repsilon_\ri}^{ -\frac{3}{2} }
\exp
\left\lbrace 
-\frac{1}{2}
\left(
\left[ \bHb_\bi \Sigmabh \bHb_\bi^T +
\left( \bg_\bi - \bHb_\bi \rfbh \right)^2 \right] \rv_{\repsilon_\ri}
+
\beta_{\epsilon} \rv_{\repsilon_\ri}^{-1}
\right)
\right\rbrace,
\eeq
leading to the conclusion that $q_{3i}(\rv_{\repsilon_\ri})$ is an generalized inverse Gaussian distribution with the following parameters:
\beq
q_{3i}(\rv_{\repsilon_\ri})=\Gc\Ic\Gc \left( \rv_{\repsilon_\ri} | \aht_{\epsilon_i}, \bht_{\epsilon_i}, \cht_{\epsilon_i} \right),
\left\{\barr{ll}
\aht_{\epsilon_j} = \bHb_\bi \Sigmabh \bHb_\bi^T
+ \left( \bg_\bi - \bHb_\bi \rfbh \right)^2
\\[10pt]
\bht_{\epsilon_j} = \beta_\epsilon 
\\[10pt]
\cht_{\epsilon_j} = -\frac{1}{2}
\earr\right.
\label{Eq:q3_Anal_1_L_nsLL_un}
\eeq
Equations~\eqref{Eq:q1_Anal_1_L_nsLL_un},~\eqref{Eq:q2_Anal_1_L_nsLL_un} and~\eqref{Eq:q3_Anal_1_L_nsLL_un} resume the distributions families and the corresponding parameters for $q_1(\rfb)$, a multivariate Normal distribution and  $q_{2j}(\rv_{\rf_\rj})$, $j\in\left\lbrace 1, 2, \ldots, M \right\rbrace$ and $q_{3i}(\rv_{\repsilon_\ri})$, $i\in\left\lbrace 1, 2, \ldots, N \right\rbrace$, generalized inverse Gaussian distributions. However, the parameters corresponding to the multivariate Normal distribution are expressed via $\rVbt_{\repsilon}$ and $\rVbt_{\rf}$ (and by extension all elements forming the three matrices $\rvt_{\repsilon_\ri}$, $i\in\left\lbrace 1, 2, \ldots, N \right\rbrace$ and $\rvt_{\rf_\rj}$, $j\in\left\lbrace 1, 2, \ldots, M \right\rbrace$).
\paragraph{Computation of the analytical expressions of $\rVbt_{\repsilon}$ and $\rVbt_{\rf}$.}
For an generalized inverse Gaussian distribution with parameters $a$, $b$ and $-\frac{1}{2}$, $\Gc\Ic\Gc\left( x | a, b, c \right)$, the following relation holds:
\beq
\biggl\langle x \biggr\rangle_{\Gc\Ic\Gc(x|a,b,-\frac{1}{2})} 
=
\left( \frac{a}{b} \right)^{-\frac{1}{2}}
\frac{\Kc_{\frac{1}{2}}\left( \sqrt{ab} \right)}{\Kc_{-\frac{1}{2}}\left( \sqrt{ab} \right)},
\label{Eq:Inv_Gam_Integ_L_nsLL_un}
\eeq
where $\Kc_{p}$ represents the modified Bessel function of the second kind.\\
To prove the above relation we consider the direct computation, using the analytical expression of the generalized inverse Gaussian distribution: 
\beq
\begin{split}
\biggl\langle x \biggr\rangle_{\Gc\Ic\Gc(x|a,b,-\frac{1}{2})} 
= 
&
\int x \; \Gc\Ic\Gc(x|a,b,-\frac{1}{2}) \d x
=
\int 
x
\frac{\left( \frac{a}{b} \right)^{-\frac{1}{2}}}{\Kc_{-\frac{1}{2}}\left( \sqrt{ab} \right)}
x^{-\frac{1}{2}-1}
\exp
\left\lbrace -\frac{1}{2} \left( ax + bx^{-1} \right) \right\rbrace
\d x
\\
= 
&
\frac{\left( \frac{a}{b} \right)^{-\frac{1}{2}}}{\Kc_{-\frac{1}{2}}\left( \sqrt{ab} \right)}
\int 
x^{\frac{1}{2}-1}
\exp
\left\lbrace -\frac{1}{2} \left( ax + bx^{-1} \right) \right\rbrace
\d x
\\
=
&
\frac{\left( \frac{a}{b} \right)^{-\frac{1}{2}}}{\Kc_{-\frac{1}{2}}\left( \sqrt{ab} \right)}
\frac{\Kc_{\frac{1}{2}}\left( \sqrt{ab} \right)}{\left( \frac{a}{b} \right)^{\frac{1}{2}}}
\int 
\frac{\left( \frac{a}{b} \right)^{\frac{1}{2}}}{\Kc_{\frac{1}{2}}\left( \sqrt{ab} \right)}
x^{\frac{1}{2}-1}
\exp
\left\lbrace -\frac{1}{2} \left( ax + bx^{-1} \right) \right\rbrace
\d x
\\
=
&
\left( \frac{a}{b} \right)^{-1}
\underbrace{\frac{\Kc_{\frac{1}{2}}\left( \sqrt{ab} \right)}{\Kc_{-\frac{1}{2}}\left( \sqrt{ab} \right)}}_{1}
\underbrace{\int \Gc\Ic\Gc(x|a,b,\frac{1}{2}) \d x}_{1}
=\frac{b}{a}
\end{split}
\nonumber
\eeq
The fact that the integral of the generalized inverse Gaussian distribution is obvious. Proving that the ratio between the two 
modified Bessel functions of the second kind is $1$, i.e. that $\Kc_{\frac{1}{2}}\left( \sqrt{ab} \right) = \Kc_{-\frac{1}{2}}\left( \sqrt{ab} \right)$ comes from expressing the modified Bessel function of the second kind $\Kc_{\alpha}\left( x \right)$ via the modified Bessel function of the first kind $\Ic_{\alpha}\left( x \right)$:
\beq
\Kc_{\alpha}\left( x \right) 
=
\frac{\pi}{2}
\frac{\Ic_{-\alpha}\left( x \right)-\Ic_{\alpha}\left(x\right)}{\sin\left( \alpha \pi \right)}
=
\frac{\pi}{2}
\frac{\Ic_{\alpha}\left( x \right)-\Ic_{-\alpha}\left(x\right)}{-\sin\left( \alpha \pi \right)}
=
\frac{\pi}{2}
\frac{\Ic_{\alpha}\left( x \right)-\Ic_{-\alpha}\left(x\right)}{\sin\left( -\alpha \pi \right)}
=
\Kc_{-\alpha}\left( x \right)
\eeq
Since $q_{2j}(\rv_{\rf_\rj})$, $j\in\left\lbrace 1, 2, \ldots, M \right\rbrace$ and $q_{3i}(\rv_{\repsilon_\ri})$, $i\in\left\lbrace 1, 2, \ldots, N \right\rbrace$ are generalized inverse Gaussian distributions, with the corresponding parameters $\aht_{f_j}$, $\bht_{f_j}$ and $\cht_{f_j}$, $j\in\left\lbrace 1, 2, \ldots, M \right\rbrace$ respectively $\aht_{\epsilon_i}$ and $\bht_{\epsilon_i}$ and $\cht_{\epsilon_i}$, $i\in\left\lbrace 1, 2, \ldots, N \right\rbrace$ the expectancies $\rvt_{\rf_\rj}$ and $\rvt_{\repsilon_\ri}$ can be expressed via the parameters of the two generalized inverse Gaussian distributions using Equation~\eqref{Eq:Inv_Gam_Integ_L_nsLL_un}:
\beq
\rvt_{\rf_\rj}
=
\frac{\bht_{f_j}}{\aht_{f_j}}
\;\;\;;\;\;\;
\rvt_{\repsilon_\ri}
=
\frac{\bht_{\epsilon_i}}{\aht_{\epsilon_i}}
\label{Eq:VepsVfExpectIGSMGen_L_nsLL_un}
\eeq
Using the notation introduced in \eqref{Eq:q1_Integ_Not1_L_nsLL_un}:
\beq
\rVbt_{\rf}=
\begin{bmatrix}
\frac{\bht_{f_1}}{\aht_{f_1}} \ldots 0 \ldots 0 \\
\vdots \ddots \vdots \ddots \vdots \\
0 \ldots \frac{\bht_{f_j}}{\aht_{f_j}} \ldots 0 \\
\vdots \ddots \vdots \ddots \vdots \\
0 \ldots 0 \ldots \frac{\bht_{f_M}}{\aht_{f_M}} \\
\end{bmatrix}
=
\rVbh_{\rf}
\;\;;\;\;
\rVbt_{\repsilon}=
\begin{bmatrix}
\frac{\bht_{\epsilon_1}}{\aht_{\epsilon_1}} \ldots 0 \ldots 0 \\
\vdots \ddots \vdots \ddots \vdots \\
0 \ldots \frac{\bht_{\epsilon_i}}{\aht_{\epsilon_i}} \ldots 0 \\
\vdots \ddots \vdots \ddots \vdots \\
0 \ldots 0 \ldots \frac{\bht_{\epsilon_N}}{\aht_{\epsilon_N}} \\
\end{bmatrix}
=
\rVbh_{\repsilon}
\label{Eq:V_Expect_L_nsLL_un}
\eeq
In Equation~\eqref{Eq:V_Expect_L_nsLL_un} other notations are introduced for $\rVbt_{\rf}$ and $\rVbt_{\repsilon}$. Both values were expressed during the model via unknown expectancies, but via Equation~\eqref{Eq:V_Expect_L_nsLL_un} those values don't contain any more integrals to be computed. Therefore, the new notations represent the final analytical expressions used for expressing the density functions $q_i$.
Using Equation~\eqref{Eq:V_Expect_L_nsLL_un} and Equations~\eqref{Eq:q1_Anal_1_L_nsLL_un},~\eqref{Eq:q2_Anal_1_L_nsLL_un} and~\eqref{Eq:q3_Anal_1_L_nsLL_un}, the final analytical expressions of the separable distributions $q_i$ are presented in Equations~\eqref{Eq:q1_Anal_2_L_nsLL_un},~\eqref{Eq:q2_Anal_2_L_nsLL_un} and~\eqref{Eq:q3_Anal_2_L_nsLL_un}.
\begin{subequations}
\begin{align}
&q_1(\rfb) = \Nc\left( \rfb | \rfbh, \Sigmabh \right),
\;\;\;\;\;\;\;\;\;\;\;\;\;\;\;\;\;\;\;\;
\left\{\barr{ll}
\rfbh = 
\left( \bHb^T  \rVbh_\repsilon \bHb + \rVbh_{\rf} \right)^{-1}
\bHb^T \rVbh_\repsilon \bgb,
\\[10pt]
\Sigmabh = \left( \bHb_\blue{1}^T  \rVbh_\repsilon \bHb + \rVbh_{\rf} \right)^{-1}
\earr\right.,
\label{Eq:q1_Anal_2_L_nsLL_un}
\\[4pt]
&q_{2j}(\rv_{\rf_\rj}) = \Gc\Ic\Gc \left( \rv_{\rf_\rj} | \aht_{f_j}, \bht_{f_j}, \cht_{f_j} \right),
\left\{\barr{ll}
\aht_{f_j} = \rfh_\rj^2 + \Sigmabh_{jj}
\\[10pt]
\bht_{f_j} = \beta_f 
\\[10pt]
\cht_{f_j} = -\frac{1}{2}
\earr\right.
,j\in\left\lbrace 1, 2, \ldots, M \right\rbrace,
\label{Eq:q2_Anal_2_L_nsLL_un}
\\[4pt]
&q_{3i}(\rv_{\repsilon_\ri}) = \Gc\Ic\Gc \left( \rv_{\repsilon_\ri} | \aht_{\epsilon_i}, \bht_{\epsilon_i}, \cht_{\epsilon_i} \right),
\;\;\;
\left\{\barr{ll}
\aht_{\epsilon_j} = \bHb_\bi \Sigmabh \bHb_\bi^T
+ \left( \bg_\bi - \bHb_\bi \rfbh \right)^2
\\[10pt]
\bht_{\epsilon_j} = \beta_\epsilon 
\\[10pt]
\cht_{\epsilon_j} = -\frac{1}{2}
\earr\right.
,i\in\left\lbrace 1, 2, \ldots, N \right\rbrace.
\label{Eq:q3_Anal_2_L_nsLL_un}
\end{align}
\end{subequations}
Equation~\eqref{Eq:q1_Anal_2_L_nsLL_un} establishes the dependency between the parameters corresponding to the multivariate Normal distribution $q_1(\rfb)$ and the others parameters involved in the hierarchical model: the mean $\rfbh$ and the covariance matrix $\Sigmabh$ depend on $\rVbh_{\repsilon}$ and $\rVbh_{\rf}$ which, via Equation~\eqref{Eq:V_Expect_L_nsLL_un} are defined using $\left\lbrace \aht_{f_j},\bht_{f_j}\right\rbrace, j\in \left\lbrace 1, 2, \ldots, M \right\rbrace $ and $\left\lbrace \aht_{\epsilon_i},\bht_{\epsilon_i}\right\rbrace, i\in \left\lbrace 1, 2, \ldots, N \right\rbrace $. The dependency between the parameters of the multivariate Normal distribution $q_1(\rfb)$ and the parameters of the generalized inverse Gaussian distributions $q_{2j}(\rvb_{\rf_\rj}) ,j \in \left\lbrace 1,2,\ldots,M \right\rbrace$ and $q_{3i}(\rvb_{\repsilon_\ri}), i \in \left\lbrace 1,2,\ldots,N \right\rbrace$ is presented in Figure~\eqref{Fig:Depend_Sch_1_L_nsLL_un}.
\begin{figure}[!htb]
\center
\begin{tabular}{c}
\begin{picture}(180,30)
\put(0,0){\framebox(100,26){$\left\lbrace \aht_{f_j},\bht_{f_j}\right\rbrace,\left\lbrace \aht_{\epsilon_j},\bht_{\epsilon_j}\right\rbrace$}}
\put(102,13){\vector(1,0){26}}
\put(128,0){\framebox(56,26){$\rfbh \; , \; \Sigmabh$}}
\end{picture}
\end{tabular}
\caption{Dependency between $q_1(\rfb)$ parameters and $q_{2j}(\rvb_{\rf_\rj})$ and $q_{3i}(\rvb_{\repsilon_\ri})$ parameters}
\label{Fig:Depend_Sch_1_L_nsLL_un}
\end{figure}
Equation~\eqref{Eq:q2_Anal_2_L_nsLL_un} establishes the dependency between the parameters corresponding to the Inverse Gamma distributions $q_{2j}(\rvb_{\rf_\rj}) ,j \in \left\lbrace 1,2,\ldots,M \right\rbrace$ and the others parameters involved in the hierarchical model: the shape and scale parameters $\left\lbrace \aht_{f_j}, \bht_{f_j} \right\rbrace, j \in \left\lbrace 1,2,\ldots,M \right\rbrace$ depend on the mean $\rfbh$ and the covariance matrix $\Sigmabh$ of the multivariate Normal distribution $q_1(\rfb)$, Figure~\eqref{Fig:Depend_Sch_2_L_nsLL_un}.
\begin{figure}[!htb]
\center
\begin{tabular}{c}
\begin{picture}(120,30)
\put(0,0){\framebox(46,26){$ \rfbh \; , \; \Sigmabh$}}
\put(48,13){\vector(1,0){26}}
\put(76,0){\framebox(56,26){$\left\lbrace \aht_{f_j},\bht_{f_j}\right\rbrace$}}
\end{picture}
\end{tabular}
\caption{Dependency between $q_{2j}(\rvb_{\rf_\rj})$ parameters and $q_1(\rfb)$ and $q_{3i}(\rvb_{\repsilon_\ri})$ parameters}
\label{Fig:Depend_Sch_2_L_nsLL_un}
\end{figure}
Equation~\eqref{Eq:q3_Anal_2_L_nsLL_un} establishes the dependency between the parameters corresponding to the Inverse Gamma distributions $q_{3i}(\rvb_{\repsilon_\ri}) ,i \in \left\lbrace 1,2,\ldots,N \right\rbrace$ and the others parameters involved in the hierarchical model: the shape and scale parameters $\left\lbrace \aht_{\epsilon_i}, \bht_{\epsilon_i} \right\rbrace, i \in \left\lbrace 1,2,\ldots,N \right\rbrace$ depend on the mean $\rfbh$ and the covariance matrix $\Sigmabh$ of the multivariate Normal distribution $q_1(\rfb)$, Figure~\eqref{Fig:Depend_Sch_3_L_nsLL_un}.
\begin{figure}[!htb]
\center
\begin{tabular}{c}
\begin{picture}(120,30)
\put(0,0){\framebox(46,26){$ \rfbh \; , \; \Sigmabh $}}
\put(48,13){\vector(1,0){26}}
\put(76,0){\framebox(56,26){$\left\lbrace \aht_{\epsilon_j},\bht_{\epsilon_j}\right\rbrace $}}
\end{picture}
\end{tabular}
\caption{Dependency between $q_{3i}(\rvb_{\repsilon_\ri})$ parameters and $q_{2j}(\rvb_{\rf_\rj})$ and $q_1(\rfb)$ parameters}
\label{Fig:Depend_Sch_3_L_nsLL_un}
\end{figure}
\newline
The iterative algorithm obtained via PM estimation is presented Figure~\eqref{Fig:IA_PM_L_nsLL_un}.
\tikzstyle{cloud} = [draw=blue!30!green,fill=orange!12, ellipse, thick, node distance=6em, text width=12em, text centered, minimum height=4em, minimum width=12em]
\tikzstyle{boxbig} = [draw=blue!30!green, fill=orange!12, rectangle, rounded corners, thick, node distance=8em, text width=25em, text centered, minimum height=5em, minimum width=25em]
\tikzstyle{box} = [draw=blue!30!green, fill=orange!12, rectangle, rounded corners, thick, node distance=6.5em, text width=21em, text centered, minimum height=5em, minimum width=21em]
\tikzstyle{boxsmall} = [draw=blue!30!green, fill=orange!12, rectangle, rounded corners, thick, node distance=8.5em, text width=11em, text centered, minimum height=5em, minimum width=11em]
\tikzstyle{line} = [draw=blue!30!green, -latex']
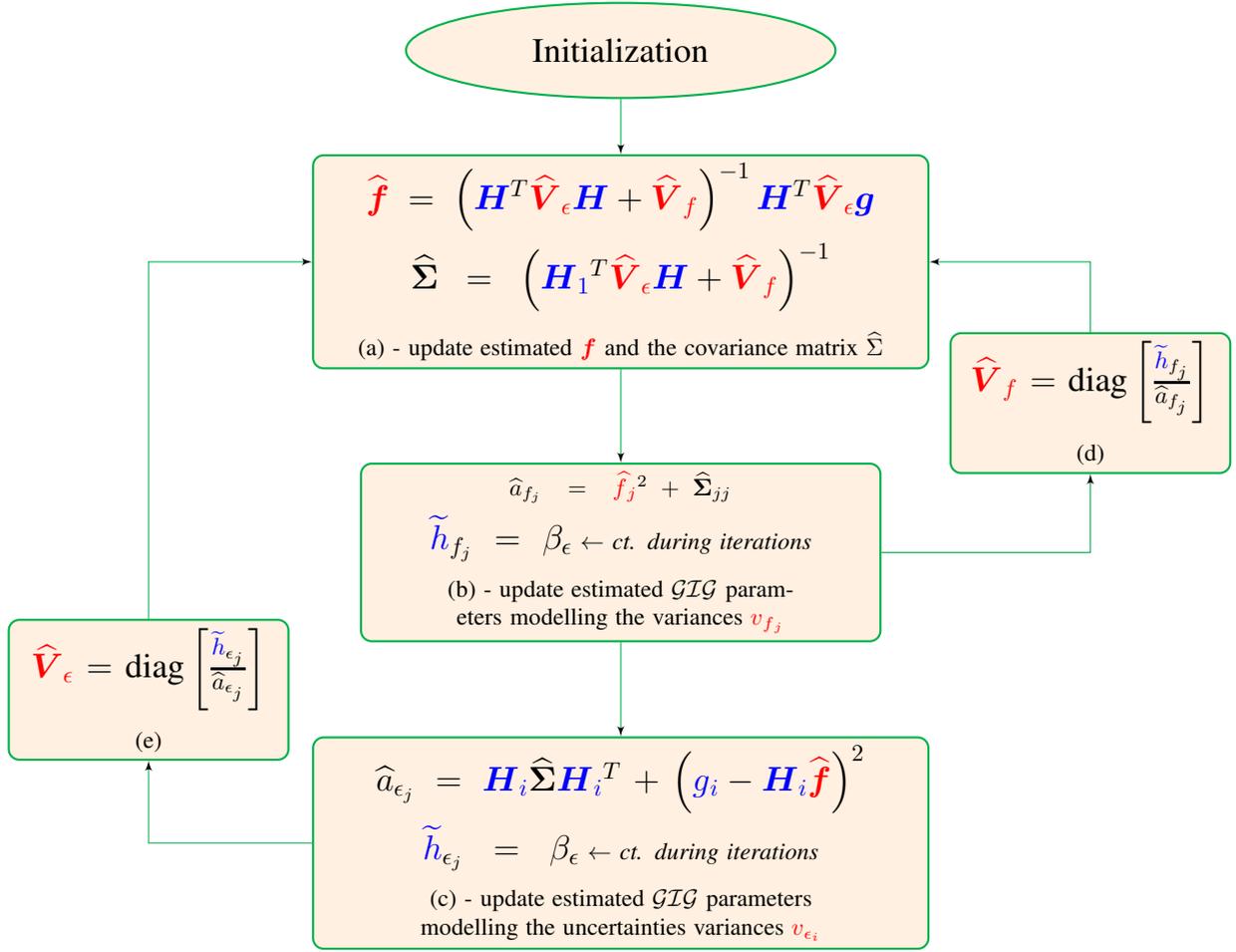
\begin{figure}[!htb]
\centering
\begin{center}
\begin{tikzpicture}[auto]
    \node [boxbig, scale=0.9] (f) {\Large{$\rfbh = 
\left( \bHb^T  \rVbh_\repsilon \bHb + \rVbh_{\rf} \right)^{-1}
\bHb^T \rVbh_\repsilon \bgb$}
\\[4pt]
\Large{$\Sigmabh = \left( \bHb_\blue{1}^T  \rVbh_\repsilon \bHb + \rVbh_{\rf} \right)^{-1}$}
\\[8pt]
\normalsize{(a) - update estimated $\rfb$ and the covariance matrix $\Sigmah$} 
};

    \node [box, below of=f, node distance=11em, scale=0.9] (vf)
{\normalsize{$ \aht_{f_j} = \rfh_\rj^2 + \Sigmabh_{jj}$}
\\[4pt]
\Large{$\bht_{f_j} = \beta_\epsilon $}
\normalsize{$\leftarrow$ \textit{ct. during iterations}}
\\[8pt]
\normalsize{(b) - update estimated $\Gc\Ic\Gc$ parameters modelling the variances $\rv_{\rf_\rj}$} 
};

    \node [boxbig, below of=vf, node distance=11em, scale=0.9] (veps) {\Large{$\aht_{\epsilon_j} = \bHb_\bi \Sigmabh \bHb_\bi^T
+ \left( \bg_\bi - \bHb_\bi \rfbh \right)^2$}
\\[4pt]
\Large{$\bht_{\epsilon_j} = \beta_\epsilon $}
\normalsize{$\leftarrow$ \textit{ct. during iterations}}
\\[8pt]
\normalsize{(c) - update estimated $\Gc\Ic\Gc$ parameters modelling the uncertainties variances $\rv_{\repsilon_\ri}$}     
};

    \node [boxsmall, above right = -0.15cm and 0.9cm of vf, node distance=17em, scale=0.9] (Vf) {\Large{$\rVbh_\rf = \diag {\frac{\bht_{f_j}}{\aht_{f_j}}}$}
\\[8pt]
\normalsize{(d)} 
};

    \node [boxsmall, below left = -4.4cm and 0.3cm of veps, node distance=17em, scale=0.9] (Veps) {\Large{$\rVbh_\repsilon = \diag {\frac{\bht_{\epsilon_j}}{\aht_{\epsilon_j}}}$}
\\[8pt]
\normalsize{(e)} 
};

\node [cloud, above of=f, node distance=8em, scale=0.9] (init)
{\Large{Initialization}};  

    \path [line] (f) -- (vf)[near start];
    \path [line] (Vf) |- (f);
    \path [line] (vf) -| (Vf);
    \path [line] (vf) -- (veps);
    \path [line] (Veps) |- (f);
    \path [line] (veps) -| (Veps);
    \path [line] (init) -- (f);
    
\end{tikzpicture}
\end{center}
\caption{Iterative algorithm corresponding to PM estimation via VBA - partial separability for Laplace hierarchical model, non-stationary Laplace uncertainties model}
\label{Fig:IA_PM_L_nsLL_un}
\end{figure}
\subsubsection{Posterior Mean estimation via VBA, full separability}
\label{Subsubsec:PM_FS_L_nsLL_un}
In this subsection, the Posterior Mean (PM) estimation is again considered, but via a full separable approximation. The posterior distribution is approximated by a full separable distribution $q\left( \rfb, \rvb_{\rf},\rvb_{\repsilon} \right)$, i.e. a supplementary condition is added in Equation~\eqref{Eq:Post_Approx_L_nsLL_un_Notations1}:
\beq
q_{1}(\rfb) = \prod_{j=1}^{M} q_{1j}(\rf_{\rj}),
\;\;;\;\;
q_{2}(\rvb_{\rf}) = \prod_{j=1}^{M} q_{2j}(\rv_{\rf_\rj}),
\;\;;\;\;
q_{3}(\rvb_{\repsilon}) = \prod_{i=1}^{N} q_{3i}(\rv_{\repsilon_\ri})
\label{Eq:Post_Approx_L_nsLL_un_Notations1bis}
\eeq
As in Subsection~\eqref{Subsubsec:PM_PS_L_nsLL_un}, the approximation is done by minimizing of the Kullback-Leibler divergence, Equation~\eqref{Eq:Kull-Leib_L_nsLL_un}, via alternate optimization resulting the following proportionalities from Equations~\eqref{Eq:VBA_Proport1_L_nsLL_unbis},~\eqref{Eq:VBA_Proport2_L_nsLL_unbis} and ~\eqref{Eq:VBA_Proport3_L_nsLL_unbis},  
\begin{subequations}
\begin{align}
&q_1(\rf_\red{j}) \;\;\; \propto \; \exp \left\lbrace \biggl\langle \ln p(\rfb, \rvb_\rf, \rvb_\repsilon | \bgb, \beta_{f}, \beta_{\epsilon}) \biggr\rangle_{ q_{1-j}(\rf_\rj) q_2(\rvb_{\rf}) \; q_3(\rvb_{\repsilon})} \right\rbrace,\;\;j \in \left\lbrace 1,2 \ldots, M \right\rbrace,
\label{Eq:VBA_Proport1_L_nsLL_unbis}
\\[4pt]
&q_{2j}(\rv_{\rf_\red{j}}) \propto \; \exp \left\lbrace \biggl\langle \ln p(\rfb, \rvb_\rf, \rvb_\repsilon | \bgb, \beta_{f}, \beta_{\epsilon}) \biggr\rangle_{q_1(\rfb) \; q_{2-j}(\rv_{\rf_\red{j}}) \; q_3(\rvb_{\repsilon})} \right\rbrace,\;\;j \in \left\lbrace 1,2 \ldots, M \right\rbrace,
\label{Eq:VBA_Proport2_L_nsLL_unbis}
\\[4pt]
&q_{3i}(\rv_{\repsilon_\ri}) \; \propto \; \exp \left\lbrace \biggl\langle \ln p(\rfb, \rvb_\rf, \rvb_\repsilon | \bgb, \beta_{f}, \beta_{\epsilon}) \biggr\rangle_{q_1(\rfb) \; q_2(\rvb_{\rf}) \; q_{3-i}(\rv_{\repsilon_\ri})} \right\rbrace,\;\;i \in \left\lbrace 1,2 \ldots, N \right\rbrace,
\label{Eq:VBA_Proport3_L_nsLL_unbis}
\end{align}
\end{subequations}
using the notations introduced in Equation~\eqref{Eq:Def_q_Min_L_nsLL_un}, Equation~\eqref{Eq:Def_Integ_L_nsLL_un} and Equation~\eqref{Eq:Def_q_Min_L_nsLL_unbis}:
\beq
q_{1-j}(\rv_{\rj})=\displaystyle \prod_{k=1,k \neq j}^{M} q_{1k}(\rv_\rk)
\label{Eq:Def_q_Min_L_nsLL_unbis}
\eeq
The analytical expression of logarithm of the posterior distribution $\ln p(\rfb, \rvb_\rf, \rvb_\repsilon | \bgb, \beta_{f}, \beta_{\epsilon})$ is obtained in Equation~\eqref{Eq:Log_Post_L_nsLL_un}.
\paragraph{Computation of the analytical expression of $q_1(\rfb)$.}
The proportionality relation corresponding to $q_1(\rfb)$ is presented in established in Equation~\eqref{Eq:VBA_Proport1_L_nsLL_unbis}. In the expression of $\ln p\left(\rfb, \rvb_\rf, \rvb_\repsilon | \bgb, \beta_{f},  \beta_{\epsilon} \right)$ all the terms free of $\rf_\ri$ can be regarded as constants:
\beq
\begin{split}
\biggl\langle \ln p(\rfb, \rvb_\rf, \rvb_\repsilon | \bgb, \beta_{f}, \beta_{\epsilon}) \biggr\rangle_{q_{1-j}(\rf_\rj) \; q_2(\rvb_{\rf}) \; q_3(\rvb_{\repsilon})}
=
&
-\frac{1}{2}
\biggl\langle 
\|\rVb_{\repsilon}^{\frac{1}{2}} \left(\bgb - \bHb\rfb\right) \|
\biggr\rangle_{q_{1-j}(\rf_\rj) \; q_3(\rvb_{\repsilon}) }
\\
&
-\frac{1}{2}
\biggl\langle 
\| \rVb_{\rf}^{\frac{1}{2}} \rfb \|
\biggr\rangle_{q_{1-j}(\rf_\rj) \; q_2(\rvb_{\rf}) }.
\end{split}
\label{Eq:q1_Integ_1_L_nsLL_unbis}
\eeq
Using Equation~\eqref{Eq:q1_Integ_Not1_L_nsLL_un}
the integral from Equation~\eqref{Eq:q1_Integ_1_L_nsLL_unbis} becomes:
\beq
\begin{split}
\left\langle \ln p(\rfb, \rvb_\rf, \rvb_\repsilon | \bgb,  \beta_{f}, \beta_{\epsilon}) \right\rangle_{q_{1-j}(\rf_\rj) \; q_2(\rvb_{\rf}) \; q_3(\rvb_{\repsilon})}
&
= 
-
\frac{1}{2} 
\biggl\langle 
\|\rVbt_{\repsilon}^{\frac{1}{2}} \left(\bgb - \bHb\rfb\right) \| 
\biggr\rangle_{q_{1-j}(\rf_\rj)} 
-
\frac{1}{2} 
\biggl\langle 
\| \rVbt_{\rf}^{\frac{1}{2}} \rfb \| 
\biggr\rangle_{q_{1-j}(\rf_\rj)}.
\label{Eq:q1_Integ_1_L_nsLL_unbis1}
\end{split}
\eeq
Considering all the $\rf_\rj$ free terms as constants, the first norm can be written:
\beq
\|\rVbt_{\repsilon}^{\frac{1}{2}} \left(\bgb - \bHb\rfb\right) \|=
C +
\| \rVbt_{\repsilon}^{\frac{1}{2}} \bHb^{\bj}\|^2\rf_\rj^2
-2
\bHb^{\bj T} \rVbt_{\repsilon}^{\frac{1}{2}} \left(\bgb - \bHb^{-\bj}\rfb^{-\rj} \right)\rf_\rj
\label{Eq:q1_Integ_First_Normbis_L_nsLL_un}
\eeq
where $\bHb^{\bj}$ represents the column $j$ of the matrix $\bHb$, $\bHb^{-\bj}$ represents the matrix $\bHb$ except the column $j$, $\bHb^{\bj}$ and $\rfb^{-\rj}$ represents the vector $\rfb$ except the element $\rf_\rj$. 
Introducing the notation
\beq
\rft_\rj 
= \int \rf_\rj \; q_{1j}(\rf_{\rj})\; \d \rf_{\rj}
\;\;;\;\;
\rfbt^{-\rj}=
\begin{bmatrix}
\rft_\red{1} \; \ldots \; \rft_\red{j-1} \; \rft_\red{j+1} \; \ldots \; \rft_{\rM}
\end{bmatrix}^T
\label{Eq:q1_Integ_Not1_L_nsLL_unbis}
\eeq
the expectancy of the first norm becomes:
\beq
\biggl\langle
\|\rVbt_{\repsilon}^{\frac{1}{2}} \left(\bgb - \bHb\rfb\right) \|
\biggr\rangle_{q_{1-j}(\rf_\rj)}
=
C +
\| \rVbt_{\repsilon}^{\frac{1}{2}} \bHb^{\bj}\|^2\rf_\rj^2
-2
\bHb^{\bj T} \rVbt_{\repsilon}^{\frac{1}{2}} \left(\bgb - \bHb^{-\bj}\rfbt^{-\rj} \right)\rf_\rj
\label{Eq:q1_Integ_First_Norm_Exbis_L_nsLL_un}
\eeq
Considering all the free $\rf_\rj$ terms as constants, the expectancy for the second norm becomes:  
\beq
\biggl\langle
\| \rVbt_{\rf}^{\frac{1}{2}} \rfb \|^2
\biggr\rangle_{q_{1-j}(\rf_\rj)}
 = C + \rvt_{\rf_\rj} \rf_\rj^2
\label{Eq:q1_Integ_Second_Norm_Exbis_L_nsLL_un}
\eeq
From Equation~\eqref{Eq:VBA_Proport1_L_nsLL_unbis}, ~\eqref{Eq:q1_Integ_1_L_nsLL_unbis1}, ~\eqref{Eq:q1_Integ_First_Norm_Exbis_L_nsLL_un}, and ~\eqref{Eq:q1_Integ_Second_Norm_Exbis_L_nsLL_un} the proportionality for $q_{1j}(\rf_\rj)$ becomes:
\beq
q_{1j}(\rf_\rj) \propto 
\exp \left\lbrace 
\left( 
\| \rVbt_{\repsilon}^{\frac{1}{2}} \bHb^{\bj}\|^2 + \rvt_{\rf_\rj} \right) \rf_\rj^2
-2
\bHb^{\bj T} \rVbt_{\repsilon} \left(\bgb - \bHb^{-\bj}\rfb^{-\rj} \right)\rf_\rj
\right\rbrace
\label{Eq:q1_Proport_L_nsLL_un}
\eeq
Considering the criterion $J(\rf_\rj) = \left( \| \rVbt_{\repsilon}^{\frac{1}{2}} \bHb^{\bj}\|^2 + \rvt_{\rf_\rj} \right)\rf_\rj^2 -2 \bHb^{\bj T} \rVbt_{\repsilon} \left(\bgb - \bHb^{-\bj}\rfb^{-\rj} \right)\rf_\rj$ which is quadratic, we conclude $q_{1j}(\rf_\rj)$ is a Normal distribution.
For computing the mean of the Normal distribution, it is sufficient to compute the solution that minimizes the criterion $J(\rf_\rj)$:
\beq
\frac{\partial J(\rf_\rj)}{\partial \rf_\rj}=0 
\Leftrightarrow
\rfh_\rj = \frac{\bHb^{\bj T} \rVbt_{\repsilon} \left( \bgb - \bHb^{-\bj} \rfb^{-\rj} \right)}{\| \rVbt_{\repsilon}^{\frac{1}{2}}  \bHb^{\bj}\| + \rvt_{\rf}}.
\label{Eq:q1_Crit_Minim_L_nsLL_un}
\eeq
The variance can be obtained by identification. The analytical expressions for the mean and the variance corresponding to the Normal distributions, $q_1(\rf_\rj)$ are presented in Equation~\eqref{Eq:q1_Anal_1_L_nsLL_unbis}.
\beq
q_1(\rf_\rj)=\Nc\left(\rf_\rj | \rfh_\rj, \widehat{\mbox{var}}_j \right),
\left\{\barr{ll}
\rfh_\rj = \frac{\bHb^{\bj T} \rVbt_{\repsilon} \left(\bgb - \bHb^{-\bj} \rfb^{-\rj} \right)}{\| \rVbt_{\repsilon}^{\frac{1}{2}} \bHb^{\bj}\| + \rvt_{\rf_\rj}}
\\[12pt]
\widehat{\mbox{var}}_j=\frac{1}{\| \rVbt_{\repsilon}^{\frac{1}{2}} \bHb^{\bj}\| + \rvt_{\rf_\rj}}
\earr\right.
,j \in \left\lbrace 1, 2, \ldots, M \right\rbrace
\label{Eq:q1_Anal_1_L_nsLL_unbis}
\eeq
\paragraph{Computation of the analytical expression of $q_{2j}(\rv_{\rf_\rj})$.}
The proportionality relation corresponding to $q_{2j}(\rv_{\rf_\rj})$ established in Equation~\eqref{Eq:VBA_Proport2_L_nsLL_unbis} refers to $\rv_{\rf_\rj}$, so in the expression of $\ln p\left(\rfb, \rvb_\rf, \rvb_\repsilon | \bgb, \beta_{f}, \beta_{\epsilon} \right)$ all the terms free of $\rv_{\rf_\rj}$ can be regarded as constants,
\beq
\ln p\left(\rfb, \rvb_\rf, \rvb_\repsilon | \bgb, \beta_{f}, \beta_{\epsilon} \right) =
C 
+
\frac{1}{2}\ln \rv_{\rf_\rj}
-
\frac{1}{2} \biggl\langle \rf_\rj^2\biggr\rangle_{q_{1j}(\rf_\rj)} \rv_{\rf_\rj}
-
2 \ln \rv_{\rf_\rj}
-
\frac{\beta_f}{2}  \rv_{\rf_\rj}^{-1},
\label{Eq:q2_Integ_1_L_nsLL_unbis}
\eeq
so the integral of the logarithm becomes:
\beq
\biggl\langle \ln p\left(\rfb, \rvb_\rf, \rvb_\repsilon | \bgb, \beta_{f}, \beta_{\epsilon} \right) \biggr\rangle_{q_1(\rfb) \; q_{2-j}(\rv_{\rf_\rj}) \; q_3(\rvb_{\repsilon})}
= 
C
- 
\frac{3}{2} \ln \rv_{\rf_\rj} 
-
\frac{\beta_f}{2} \rv_{\rf_\rj}^{-1}
-
\frac{1}{2} \left( \rfh_\rj^2 + \widehat{\mbox{var}}_j \right) \rv_{\rf_\rj}.
\label{Eq:q2_Integ_1_L_nsLL_unbis1}
\eeq
Equation~\eqref{Eq:q2_Integ_1_L_nsLL_unbis1} leads to the conclusion that $q_{2j}(\rv_{\rf_\rj})$ is an generalized inverse Gaussian distribution. Equation~\eqref{Eq:q2_Anal_1_L_nsLL_unbis} presents the analytical expressions for the parameters corresponding to the Inverse Gamma distribution.
\beq
q_{2j}(\rv_{\rf_\rj}) = \Gc\Ic\Gc \left( \rv_{\rf_\rj} | \aht_{f_j}, \bht_{f_j}, \cht_{f_j} \right),
\left\{\barr{ll}
\aht_{f_j} = \rfh_\rj^2 + \widehat{\mbox{var}}_j
\\[10pt]
\bht_{f_j} = \beta_f
\\[10pt]
\cht_{f_j} = - \frac{1}{2} 
\earr\right.
,j \in \left\lbrace 1, 2, \ldots, M \right\rbrace
\label{Eq:q2_Anal_1_L_nsLL_unbis}
\eeq
\paragraph{Computation of the analytical expression of $q_{3i}(\rv_{\repsilon_\ri})$.} 
The proportionality relation corresponding to $q_{3i}(\rv_{\repsilon_\ri})$ established in Equation~\eqref{Eq:VBA_Proport3_L_nsLL_unbis} refers to $\rv_{\repsilon_\ri}$ so in the expression of $\ln p\left(\rfb, \rvb_\rf, \rvb_\repsilon | \bgb, \beta_{f}, \beta_{\epsilon} \right)$ all the terms free of $\rv_{\repsilon_\ri}$ can be regarded as constants:
\beq
\ln p \left( \rfb, \rvb_\rf, \rvb_\repsilon | \bgb, \beta_{f}, \beta_{\epsilon} \right) = C 
-
\frac{3}{2} \ln \rv_{\repsilon_\ri}
-
\frac{\beta_\epsilon}{2}\rv_{\repsilon_\ri}^{-1} 
+ 
\frac{1}{2}\left(\bg_\bi - \bHb_\bi\rfb\right)^2
\rv_{\repsilon_\ri}.
\label{Eq:q3_Integ_1_L_nsLL_unbis}
\eeq
Introducing the notation
\beq
\biggl\langle \rfb \biggr\rangle_{q_1(\rfb)}
=
\begin{bmatrix}
\rfh_{\red{1}}
\ldots
\rfh_\rj
\ldots 
\rfh_\rM
\end{bmatrix}^T
\stackrel{Not}{=}
\rfbh
\;\; ; \;\;
\Sigmabh = \diag {\widehat{\mbox{var}}_j}
\label{Eq:q3_Integ_Not1_L_nsLL_unbis}
\eeq
the expectancy of the logarithm becomes
\beq
\begin{split}
\biggl\langle
\ln p\left(\rfb, \rvb_\rf, \rvb_\repsilon | \bgb, \beta_{f}, \beta_{\epsilon} \right) 
\biggr\rangle_{q_1(\rfb) \; q_2(\rvb_\rf) \; q_{3-i}(\rv_{\repsilon_\ri})}
= C & 
-\frac{3}{2} \ln \rv_{\repsilon_\ri}
-
\frac{\beta_\epsilon}{2} \rv_{\repsilon_\ri}^{-1}
- 
\frac{1}{2}
\left[ \bHb_\bi \Sigmabh \bHb_\bi^T
- 
\left( \bg_\bi - \bHb_\bi \rfbh \right)^2
\right] \rv_{\repsilon_\ri},
\end{split}
\label{Eq:q3_Integ_1_L_nsLL_unbis1}
\eeq
so the proportionality relation for $q_{3i}(\rv_{\repsilon_\ri})$ from Equation~\eqref{Eq:VBA_Proport3_L_nsLL_unbis} can be written:
\beq
q_{3i}(\rv_{\repsilon_\ri})
\propto
\rv_{\repsilon_\ri}^{-\frac{3}{2}}
\exp
\left\lbrace
-\frac{1}{2}
\left(
\left[ \bHb_\bi \Sigmabh \bHb_\bi^T + \left( \bg_\bi - \bHb_\bi \rfbh \right)^2 \right] \rv_{\repsilon_\ri}
+
\beta_\epsilon \rv_{\repsilon_\ri}^{-1} 
\right)
\right\rbrace
\label{Eq:q3_Proport_L_nsLL_un}
\eeq
Equation~\eqref{Eq:q3_Proport_L_nsLL_un} shows that $q_{3i}(\rv_{\repsilon_\ri})$ are generalized inverse Gaussian distributions. The analytical expressions of the corresponding parameters are presented in Equation~\eqref{Eq:q3_Anal_1_L_nsLL_unbis}.
\beq
q_{3i}(\rv_{\repsilon_\ri}) = \Gc\Ic\Gc \left( \rv_{\repsilon_\ri} | \aht_{\epsilon_i}, \bht_{\epsilon_i}, \cht_{\epsilon_i} \right),
\left\{\barr{ll}
\aht_{\epsilon_i} = \bHb_\bi \Sigmabh \bHb_\bi^T + \left( \bg_\bi - \bHb_\bi \rfbh \right)^2
\\[10pt]
\bht_{\epsilon_i} = \beta_\epsilon
\\[10pt]
\cht_{\epsilon_i} = - \frac{1}{2}
\earr\right.
,i \in \left\lbrace 1, 2, \ldots, N \right\rbrace
\label{Eq:q3_Anal_1_L_nsLL_unbis}
\eeq
Since $q_2(\rv_{\rf_\rj}), j \in \left\lbrace 1, 2, \ldots, M \right\rbrace$ and $q_{3i}(\rv_{\repsilon_\ri}), i \in \left\lbrace 1, 2, \ldots, N \right\rbrace $ are generalized inverse Gaussian distributions, it is easy to obtain analytical expressions for $\rVbt_{\repsilon}$, defined in Equation~\eqref{Eq:q1_Integ_Not1_L_nsLL_un} and $ \rvt_{\rf_\rj}, j \in \left\lbrace 1, 2, \ldots, M \right\rbrace $, obtaining the same expressions as in Equation~\eqref{Eq:V_Expect_L_nsLL_un}. Using Equation~\eqref{Eq:V_Expect_L_nsLL_un} and Equations~\eqref{Eq:q1_Anal_1_L_nsLL_unbis},~\eqref{Eq:q2_Anal_1_L_nsLL_unbis} and~\eqref{Eq:q3_Anal_1_L_nsLL_unbis}, the final analytical expressions of the separable distributions $q_i$ are presented in Equations~\eqref{Eq:q1_Anal_2_L_nsLL_unbis},~\eqref{Eq:q2_Anal_2_L_nsLL_unbis} and~\eqref{Eq:q3_Anal_2_L_nsLL_unbis}.
\begin{subequations}
\begin{align}
&q_1(\rf_\rj)=\Nc\left(\rf_\rj | \rfh_\rj, \widehat{\mbox{var}}_j \right),\;\;\;\;\;\;\;\;\;\;\;\;\;
\left\{\barr{ll}
\rfh_\rj = \frac{\bHb^{\bj T} \rVbh_{\repsilon} \left(\bgb - \bHb^{-\bj} \rfb^{-\rj} \right)}{\| \rVbh_{\repsilon}^{\frac{1}{2}} \bHb^{\bj}\| + \rvh_{\rf_\rj}}
\\[12pt]
\widehat{\mbox{var}}_j=\frac{1}{\| \rVbh_{\repsilon}^{\frac{1}{2}} \bHb^{\bj}\| + \rvh_{\rf_\rj}}
\earr\right.
,j \in \left\lbrace 1, 2, \ldots, M \right\rbrace
\label{Eq:q1_Anal_2_L_nsLL_unbis}
\\[4pt]
&q_{2j}(\rv_{\rf_\rj})=\Gc\Ic\Gc\left(\rv_{\rf_\rj}|\aht_{f_j},\bht_{f_j},\cht_{f_j}\right),
\left\{\barr{ll}
\aht_{f_j} = \rfh_\rj^2 + \widehat{\mbox{var}}_j
\\[10pt]
\bht_{f_j} = \beta_f
\\[10pt]
\cht_{f_j} = - \frac{1}{2} 
\earr\right.
,j \in \left\lbrace 1, 2, \ldots, M \right\rbrace
\label{Eq:q2_Anal_2_L_nsLL_unbis}
\\[4pt]
&q_{3i}(\rv_{\repsilon_\ri})=\Gc\Ic\Gc\left(\rv_{\repsilon_\ri}|\aht_{\epsilon_i},\bht_{\epsilon_i},\cht_{\epsilon_i}\right),\;\;\;
\left\{\barr{ll}
\aht_{\epsilon_i} = \bHb_\bi \Sigmabh \bHb_\bi^T + \left( \bg_\bi - \bHb_\bi \rfbh \right)^2
\\[10pt]
\bht_{\epsilon_i} = \beta_\epsilon
\\[10pt]
\cht_{\epsilon_i} = - \frac{1}{2}
\earr\right.
,i \in \left\lbrace 1, 2, \ldots, N \right\rbrace
\label{Eq:q3_Anal_2_L_nsLL_unbis}
\end{align}
\end{subequations}
Equations~\eqref{Eq:q1_Anal_2_L_nsLL_unbis},~\eqref{Eq:q2_Anal_2_L_nsLL_unbis} and~\eqref{Eq:q3_Anal_2_L_nsLL_unbis} establish dependencies between the parameters of the distributions, very similar to the one presented in Figures~\eqref{Fig:Depend_Sch_1_L_nsLL_un},~\eqref{Fig:Depend_Sch_2_L_nsLL_un} and~\eqref{Fig:Depend_Sch_3_L_nsLL_un}. The iterative algorithm obtained via PM estimation with full separability, is presented Figure~\eqref{Fig:IA_PM_L_nsLL_unbis}.
\tikzstyle{cloud} = [draw=blue!30!green,fill=orange!12, ellipse, thick, node distance=6em, text width=12em, text centered, minimum height=4em, minimum width=12em]
\tikzstyle{boxbig} = [draw=blue!30!green, fill=orange!12, rectangle, rounded corners, thick, node distance=8em, text width=25em, text centered, minimum height=5em, minimum width=25em]
\tikzstyle{box} = [draw=blue!30!green, fill=orange!12, rectangle, rounded corners, thick, node distance=6.5em, text width=21em, text centered, minimum height=5em, minimum width=21em]
\tikzstyle{boxsmall} = [draw=blue!30!green, fill=orange!12, rectangle, rounded corners, thick, node distance=8.5em, text width=11em, text centered, minimum height=5em, minimum width=11em]
\tikzstyle{line} = [draw=blue!30!green, -latex']
\begin{figure}[!htb]
\centering
\begin{center}
\begin{tikzpicture}[auto]
    \node [boxbig, scale=0.9] (f) {\Large{$ \rfh_\rj = \frac{\bHb^{\bj T} \rVbh_{\repsilon} \left(\bgb - \bHb^{-\bj} \rfb^{-\rj} \right)}{\| \rVbh_{\repsilon}^{\frac{1}{2}} \bHb^{\bj}\| + \rvh_{\rf_\rj}}$}
\\[4pt]
\Large{$\widehat{\mbox{var}}_j=\frac{1}{\| \rVbh_{\repsilon}^{\frac{1}{2}} \bHb^{\bj}\| + \rvh_{\rf_\rj}}$}
\\[8pt]
\normalsize{(a) - update estimated $\rf_\rj$ and the coresponding variance $\mbox{var}_j$} 
};

    \node [box, below of=f, node distance=11em, scale=0.9] (vf)
{\Large{$ \aht_{f_j} = \rfh_\rj^2 + \widehat{\mbox{var}}_j $}
\\[4pt]
\Large{$ \bht_{f_j} = \beta_f + \frac{1}{2} $}
\normalsize{$\leftarrow$ \textit{ct. during iterations}}
\\[8pt]
\normalsize{(b) - update estimated $\Ic\Gc$ parameters modelling the variances $\rv_{\rf_\rj}$} 
};

    \node [boxbig, below of=vf, node distance=11em, scale=0.9] (veps) {\Large{$ \aht_{\epsilon_i} = \bHb_\bi \Sigmabh \bHb_\bi^T + \left( \bg_\bi - \bHb_\bi \rfbh \right)^2 $}
\\[4pt]
\Large{$ \bht_{\epsilon_i} = \beta_\epsilon $}
\normalsize{$\leftarrow$ \textit{ct. during iterations}}
\\[8pt]
\normalsize{(c) - update estimated $\Ic\Gc$ parameters modelling the uncertainties variances $\rv_{\repsilon_\ri}$}     
};

    \node [boxsmall, above right = -0.15cm and 0.9cm of vf, node distance=17em, scale=0.9] (Vf) {\Large{$\rVbh_\rf = \diag {\frac{\bht_{f_j}}{\aht_{f_j}}}$}
\\[8pt]
\normalsize{(d)} 
};

    \node [boxsmall, below left = -4.4cm and 0.3cm of veps, node distance=17em, scale=0.9] (Veps) {\Large{$\rVbh_\repsilon = \diag {\frac{\bht_{\epsilon_j}}{\aht_{\epsilon_j}}}$}
\\[8pt]
\normalsize{(e)} 
};

\node [cloud, above of=f, node distance=8em, scale=0.9] (init)
{\Large{Initialization}};  

    \path [line] (f) -- (vf)[near start];
    \path [line] (Vf) |- (f);
    \path [line] (vf) -| (Vf);
    \path [line] (vf) -- (veps);
    \path [line] (Veps) |- (f);
    \path [line] (veps) -| (Veps);
    \path [line] (init) -- (f);
    
\end{tikzpicture}
\end{center}
\caption{Iterative algorithm corresponding to PM estimation via VBA - full separability for Laplace hierarchical model, non-stationary Laplace uncertainties model}
\label{Fig:IA_PM_L_nsLL_unbis}
\end{figure}
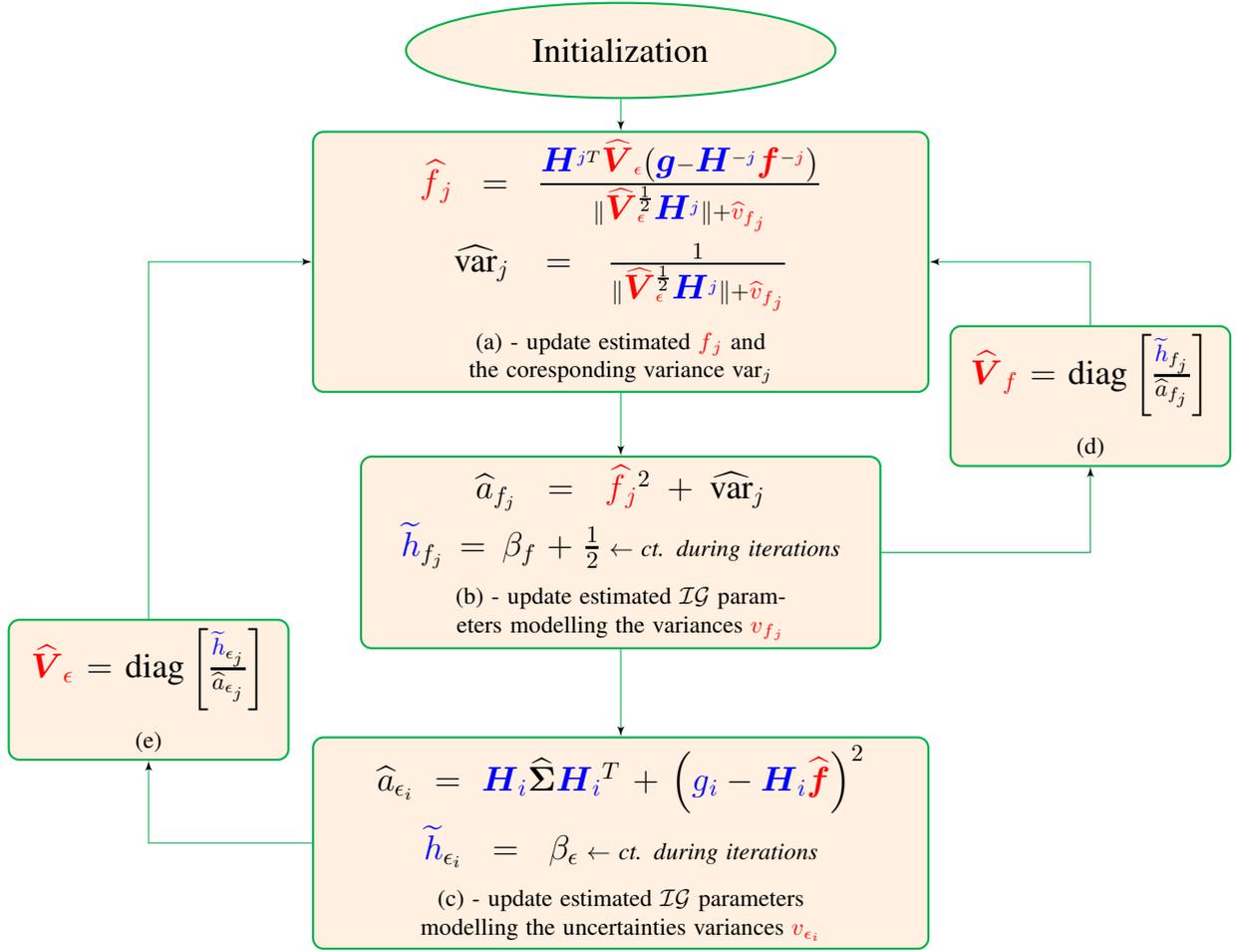

\newpage
\listoffigures
\listoftables

\newpage
\bibliographystyle{alpha}
\bibliography{Biblio}

\end{bmcformat}
\end{document}